\documentclass[12pt,preprint]{aastex}
\usepackage{natbib,graphicx}

\shorttitle{Lobes of Eleven Powerful Radio Galaxies}

\begin{document}

\title{A Detailed Study of the Lobes of Eleven Powerful Radio Galaxies}
\author{\today}
\author{Ruth A. Daly\altaffilmark{1}, Preeti Kharb\altaffilmark{2},
Christopher P. O'Dea\altaffilmark{2}, Stefi A. Baum\altaffilmark{3},
Matthew P. Mory\altaffilmark{1},
Justin McKane\altaffilmark{1},
Christopher Altenderfer\altaffilmark{1},
\& Michael Beury\altaffilmark{1}}

\altaffiltext{1}{Dept. of Physics, Penn State University, Berks Campus}
\altaffiltext{2}{Dept. of Physics, Rochester Institute of Technology }
\altaffiltext{3}{Center for Imaging Science, Rochester Institute of 
Technology }

\begin{abstract}
Radio lobes of a sample of eleven very powerful classical double radio 
galaxies were studied.  Each source was rotated so that the symmetry
axis of the source was horizontal, and vertical cross-sectional cuts
were taken across the source at intervals of one beam size. These were
used to study the cross-sectional surface brightness profiles, the 
width of each slice, radio emissivity as a function 
of position across each slice, the first and second moments, and the 
average surface brightness, minimum energy 
magnetic field strength, and pressure of each 
slice.  Typically, a Gaussian provides a good description of 
the surface brightness profile of cross-sectional
slices. The Gaussian FWHM as a function of distance from the hot
spot first increases and then decreases with increasing distance from
the hot spot.  The width as a function of distance from the
hot spot is generally highly symmetric on each side of the source. 
The radio emissivity is often close to flat across a 
slice, indicating a roughly constant emissivity and pressure 
for that slice. 
Some slices show variations in radio emissivity that indicate an
``edge-peaked'' pressure profile for that slice.  When this occurs
it is generally found in slices near the local maxima of the bridge 
width. The emissivity does not exhibit any signature of emission
from a jet.  
The first moment is generally quite close to zero indicating 
only small excursions of the ridge line 
from the symmetry axis of the source. The second
moment indicates the same source shape as is found using the Gaussian
FWHM.  The average surface brightness is peaked at the hot spot, and 
is fairly flat across most of the radio lobes. The average magnetic
field strength and pressure peak at the hot spot and gradually decrease
with increasing distance from the hot spot,
reaching a roughly constant value at a location that is typically just
before the location of a local maximum of the bridge width.  These
results are interpreted in terms of a heuristic model for the radio lobes.    

\end{abstract}


\section{INTRODUCTION}

Physical processes in powerful active galactic nuclei (AGNs) with radio jets 
that appear to be transporting plasma from the galactic nucleus to the 
intergalactic medium, are not completely 
understood. An understanding of these
processes can provide powerful insights into galactic
nuclei and the medium through which the jets traverse. 
Comprehensive spectral studies of the 
Fanaroff-Riley-type-II (FRII) radio galaxies \citep{FanaroffRiley74}, have 
indicated 
that the charged particles/electrons are injected into the terminal regions
of the jets called hot spots, 
which expand to form radio lobes   
consistent with the spectral index steepening 
that is observed in the FRII radio lobes 
\citep[for example,][]{MS85, 
Alexander87, AL87, LMS89, 
Carilli91, LPR92, M98, P99, M99, 
Blundell99,  K00, GDW00, GKBD04, M07, J08,  
Kharb08, K09}.

A detailed study of the radio bridge in the nearby powerful FRII radio 
galaxy, Cygnus~A, was presented by \citet{Carvalho05}. 
This study showed that the observed surface brightness distribution
perpendicular to the source axis is well 
characterized by a Gaussian function 
over most of the length of the source. The first moment of the
cross-sectional surface brightness profiles traces the wobbling  
of the ridge-line, which could be the result of the wobbling of the
outflow axis, while the Gaussian FWHM provides 
a reasonable estimate of the 
bridge width. The ratio of the Gaussian FWHM to the second moment is fairly
constant along the source, with an average value of about 2.5, indicating
that both these parameters give a consistent measure of the bridge width.
The observed surface brightness, estimated minimum pressure and magnetic field
strength 
decrease with distance from the hot spot.
In this paper, we present a similar study 
on a sample of eleven large and
powerful FRII radio galaxies.

The 11 sources studied here are part of a sample of 13 FRII radio galaxies 
which were observed with multiple configurations of the Very Large Array 
(VLA) at 330~MHz, 1.4, 5 and 8~GHz. The first results from this sample were 
presented by \citet{Kharb08}. The sources had arm-length ratios 
close to unity, small misalignment angles, and weak or no radio jets, 
all of which are 
consistent with the sources lying close to the plane of the sky. 
The results from the spectral aging analysis of this radio galaxy sample
are presented by O'Dea et al. (2009), while the cosmological 
results are presented by Daly et al. (2008, 2009).  
The 11 sources of interest here are the ones that showed extended lobe 
emission. Extended radio lobes were not detected for 
the two most distant sources, 3C13 and 3C470.
The 11 FRII radio galaxies studied here  
have radio luminosities at 178~MHz greater 
than 
$10^{28}$~W~Hz$^{-1}$; angular sizes larger than $27^{\prime\prime}$, and span a  
redshift range of $z\simeq$~0.4--1.3.  A spatially flat 
cosmological model with current
normalized mean mass density $\Omega_m = 0.3$, cosmological constant
$\Omega_{\Lambda} = 0.7$, and Hubble constant 
$H_0 = 70 \rm{~ km ~s^{-1} ~Mpc^{-1}}$ are assumed throughout. 
The analysis and results are presented in section \ref{data}, 
and a summary and conclusions are given in section \ref{summary}.

\section{DATA ANALYSIS AND RESULTS}
\label{data}

The very powerful FRII radio galaxies were observed with the A, B, C and
D-array configurations of the VLA \citep{Napier83} between November, 2002
and November, 2003 at 330~MHz, 1.4, 4.8 and 8.4~GHz.
Details of the data-reduction and analysis are presented 
by \citet{Kharb08}.  

The Astronomical Image Processing System (AIPS) along with auxiliary
programs were used to obtain and study cross-sectional 
slices of the radio surface brightness at evenly spaced intervals 
along the symmetry axis of each radio bridge. 
The FITS files of the final 1.4~GHz and 5~GHz maps convolved with 
a 2$^{\prime\prime}$ or 2.5$^{\prime\prime}$ beam were rotated so that the hot spots 
identified using the 8.4 GHz data 
lay on a horizontal line. The axis joining the two primary hot spots, with
each hot spot 
located at the extremity of each side of the source, 
was taken to be the 
symmetry axis of the source. This seemed a good choice since the 
sources have very small misalignment angles \citep{Kharb08}, 
and is confirmed by the fact that the first moment shows very mild
departures from zero across most sources. 
The data were clipped so as to include only data with 
an intensity
greater than the (mean + 3 $\times$ rms) noise level of each map.
Cross-sectional slices of the source, separated
by one FWHM of the beam, were obtained for each side of each source 
with the requirement that one of the slices intersect the middle of the 
primary hot spot on that side of the source. 
The center of the source was taken to
be the core position identified by \citet{Kharb08}.  
The properties of the cross-sectional
slices were used to study the cross-sectional shape 
of the surface brightness profile
(see section \ref{GaussianFits}), the radio emissivity profile of the 
slice (see section \ref{radioemissivity}), the first and second moments of the 
slice and a comparison of the second moment with the Gaussian FWHM 
(see sections \ref{Firstmoment} and \ref{Secondmoment}), and the 
average surface brightness, minimum energy magnetic field strength,
and average minimum energy pressure of the
plasma in each slice (see section \ref{SBP}). 

 \subsection{Cross-Sectional Slices and Gaussian Fits}
\label{GaussianFits}
The rotated 1.4 GHz map is shown for each source (see the figures).
Each side of each source is studied separately, and is sliced in
the vertical direction, perpendicular to the symmetry axis of the 
source. Distances (in arc seconds) are measured relative to the 
hot spot; $D_{HSL}$ refers to the distance along the symmetry axis
of the source from the left hot spot going toward the center of the source, 
and $D_{HSR}$ refers to that
from the right hot spot of the rotated map toward the center of the source. 

Each of the cross-sectional surface brightness profiles is fit with a
Gaussian.  These profiles and the best fit Gaussian for each is shown
in the figures.  The FWHM of the best fit Gaussian, $W_{Gauss}$ is also shown
for each slice of each source. These are obtained by subtracting the 
FWHM of the observing beam in quadrature: 
$W_{Gauss} = [(W_{G,M})^2 - (\theta_{FW})^2]^{1/2}$, where 
$W_{G,M}$ is the measured FWHM of the Gaussian and $\theta_{FW}$ is
the FWHM of the observing beam. 

Considering the regions of the source with widths indicating a 
size larger than the beamsize,  
a Gaussian typically provides a very good fit to the cross-sectional
surface brightness profiles.
The 1.4 and 5 GHz Gaussian widths are in 
very good agreement. The degree of symmetry of the 
source width as a function of position is noted in the figure captions. 
Many of the sources show a remarkable 
degree of symmetry of source width as
a function of distance from the hot spot between the right 
and left hand sides of the source. This suggests that the properties
of the ambient medium 
vary in the same way with distance from the 
center of the source on each side of the source, and that the beam
power of the twin jets is similar at any given time. 
The widths obtained at 1.4 and 5 GHz
are very similar, as expected. 

In many of the sources, the bridge width initially increases with 
distance from the hot spot, and then decreases.  
Considering the 22 radio 
bridges studied here (two for each of the 11 sources), this behavior
is clearly seen in 16 bridges (the left side of 3C6.1;
3C34; the 5GHz data of 3C44; 3C54; the right side of 3C114; 
3C142.1; 3C172; 3C441; and 3C469.1), 
and is consistent with the properties
of 4 more bridges (the right side of 3C6.1; 
the left side of 3C114; and 3C169.1).  
Interestingly, the similarity of the source width as a function 
of distance from the hot spot for the 
right and left sides of a source typically includes the 
location of the local maxima of the bridge width. 
There is only one source, 
3C41, 
that has a width as a function of position that seems to level
off. One possible explanation 
for the turnover of the bridge width as a function of
distance from the hot spot 
is discussed in section \ref{radioemissivity}.

\subsection{Radio Emissivity} 
\label{radioemissivity}
The emissivity, or energy emitted per unit volume per unit time per unit
frequency, can be obtained from the surface brightness profiles, 
as described in detail in section 2.5 of Carvalho et al. (2005).  
For a given surface brightness profile, the emissivity can be obtained
as a function of distance from the symmetry axis of the source assuming
only that the emissivity is semi-cylindrically symmetric, that is, 
it is cylindrically symmetric over each half of the slice 
(that above the symmetry axis, and that below the symmetry axis) 
separately. So the 
emissivity is a free function of distance from the symmetry axis of the 
source but the emissivity at a given distance $r$ 
is assumed to be cylindrically symmetric at that distance $r$ over that
half of the slice
(see Fig. 14 of Carvalho et al. 2005).
The emissivities thus obtained are shown with the surface brightness
profiles from which they are obtained in the figures. 
The emissivities are computed from the outside of the slice toward
the slice center, starting from each of the two extremities of the 
cross-sectional slice.  When the emissivity obtained at the center of
the slice is continuous, it indicates that the emissivity computed by
stepping in from one side of the slice matches that obtained by 
stepping in from the other side of the slice, suggesting that in 
these cases the assumption of semi-cylindrical symmetry is a valid
assumption.  

The emissivity $\epsilon_{\nu} \equiv dE/(dV~dt~d\nu)$ is 
related to that measured directly from the radio map, 
referred to as $\epsilon_M$, and a normalization factor $K_{\epsilon}$:
$\epsilon_{\nu} = K_{\epsilon} \epsilon_M$.  When the 
measured value, $\epsilon_M$, has units of Jy/(beam arcsec), as
it does for the sources studied here, 
the normalization factor is  
$K_{\epsilon} = 3.17 ~(1+z)^4 ~\rm{[(a_0r)/Gpc]^{-1} 
~(\theta_{FW}/arcsec)^{-2}}
10^{-34} \rm{~erg ~cm^{-3} ~s^{-1} ~Hz^{-1}}$, where $a_0r$ is the coordinate 
distance to the source, and $\theta_{FW}$ is the FWHM of the 
observing beam.  This follows from the following
considerations: the emission coefficient 
$j_{\nu} = \epsilon_{\nu}/(4 \pi)$, and 
the specific intensity produced over a path length $\Delta y$ is 
$I_{\nu} = j_{\nu} \Delta y$, where $I_{\nu} = dE/(dA~dt~d\nu~d\Omega)$,
where $d\Omega$ is a differential solid angle 
(e.g. Rybicki and Lightman 1979).  The observed 
specific intensity is $I_{\nu_0} = I_{\nu} (1+z)^{-3}$ and in this 
study is measured
in Jansky/beam. To convert this to Jy per steradian it must be multiplied
by $(\pi \theta_b^2)^{-1}$, where $\theta_b = \theta_{FW}/[2 
\sqrt(\ln 2)]$. 

Each emissivity figure lists the surface brightness units on the left hand
side of the figure, and the emissivity units, $\epsilon_M$ in 
Jy/(beam arcsec), 
on the right hand side of the figure.  The value of the normalization
$K_{\epsilon}$ for each source is listed in Table 1 for 
$\epsilon_M$ in units of Jy/(beam arcsec), 
and the value
of $\epsilon_{\nu}$ is obtained using the expression
$\epsilon_{\nu} = K_{\epsilon} \epsilon_M$.  Because the resolution for
each source is the same at the two frequencies considered, the normalization
factor is frequency independent. 

The sources have emissivity profiles that are consistent with 
a roughly constant emissivity in each cross-sectional slice for
most slices of a source, 
that is, the emissivity across the slice is
fairly flat.  In addition, in most slices of most sources, the
emissivity obtained near the center of the slice is continuous.
This indicates good agreement of the emissivity 
near the center of the slice 
obtained independently 
by stepping in from each of the two sides of the slice.
Continuity of the emissivity near the center of a slice 
suggests that the assumption of semi-cylindrical symmetry 
is valid for that slice.

There are some very interesting deviations from a flat emissivity 
profile for a given slice. Some of the emissivity profiles suggest 
that the emissivity is largest near the edges of the slice, and 
the emissivity has a minimum near the center of the slice.  
In most of the instances in which this occurs, the emissivity
is continuous near the center of the slice, suggesting that 
the assumption of semi-cylindrical symmetry is valid for that 
slice. Interestingly, 
slices with emissivities that peak near the edge of the 
slice occur when the bridge width is close to a maximum. 
That is, the behavior of the emissivity profile is 
related to the bridge width as a function of distance from 
the hot spot. An emissivity
profile that is ``edge-peaked'' is often found at the 
locations of the source where the bridge width is close to a maximum,
discussed in section \ref{GaussianFits}. 
This can be seen on 
the left side of 3C6.1 (and perhaps also the right side
of that source), 3C34, 3C142.1, 3C169.1, 3C172, 3C441, 
and perhaps the right side of 3C44
(see the 5 GHz results).

The emissivity of synchrotron radiation 
is related to the pressure of the 
relativistic fluid producing the radio emission. This follows
because the minimum energy pressure $P \propto \epsilon_{\nu}^{4/7}$ 
introduced by \citet{Burbidge56} and the pressure can be obtained from the 
minimum energy pressure allowing for offsets from  minimum energy conditions, 
as discussed by Carvalho et al. (2005). 
So the pressure of the relativistic fluid is 
roughly proportional to the square root of the emissivity. 
Slices with the emissivity peaked away from the slice 
center, and with an emissivity minimum near the slice
center, have a pressure profile with similar properties:
the pressure is larger near the edge of the slice than
it is near the center of the slice. And, 
slice to slice variations in the local emissivity and hence
pressure are also apparent in the figures even though
the average pressure of slices tend to become fairly flat 
some distance from the hot spot for many 
sources, as discussed in section \ref{SBP}. 

Both the behavior of the source width as a function of distance
from the hot spot and the existence of ``edge-peaked''
pressure profiles for slices near local maxima in the 
source width are reminiscent of the response of an outflow to 
a sudden drop in external pressure 
such as is experienced by an outflow passing through a nozzle 
(e.g. Courant \& Friedrichs 1948;
Owczarek 1964;  K\"onigl 1980; Daly \& Marscher 1988).  
This suggests that, in  
the rest frame of the hot spot, the hot spot is somewhat like 
the opening of a nozzle through which a relativistic fluid
flows. In response to the series of waves that carry 
information about the pressure of the external medium to 
the relativistic fluid, the system over-expands in some
regions and contracts in others 
causing the width to have a local maxima
(much like those described in section \ref{GaussianFits}), 
and, for slices near those with maximum widths, 
causing the pressure near the edge of the slice to exceed that near
the center. 

Another possibility is that the fluctuations of the minimum energy pressure 
could be due to changes in the magnetic field structure that would
result in changes in the offset of the field strength
from the minimum energy value, such as the field structures discussed by 
\citet{E97} and 
\citet{TJR04}. The fluctuations could also be
an artifact of non-semi-cylindrical symmetry in a particular slice.
This is signalled when the emissivity near the center of the slice
is discontinuous or zero, as discussed earlier. 

There are some slices with pressures that are both ``edge-peaked'' and
``center-peaked'', having three pressure maxima with pressure  
minima between the slice
center and edge.  
These can be seen for example, in 3C44, 3C142.1, and 3C441. 
These types of features are also observed
in the emissivity and pressure slices of Cygnus A 
(Carvalho et al. 2005).

\subsection{The First Moment}
\label{Firstmoment}

The first moment is a measure of the location of the surface 
brightness-weighted central axis of the source. This is obtained 
for each cross-sectional slice 
using the 
expression $\bar{x}=\sum x_i S_i / \sum S_i$, where $S_i$ 
is the surface brightness at the position $x_i$. 
In obtaining the moments, each slice is treated independently.
The locations $x_i$ for each slice are evenly spaced with separations that are 
very small compared with the beamsize.  
The first moment can be
affected by the presence of the radio jet, or a bulk sideways motion 
of the source.  
``Wiggles'' in the first moment could be suggestive of a wobble 
of the outflow axis. Precession of the jet axes could give wiggles in the
opposite sense on the two sides of the source. The wiggles could
also be due to the wandering of the hot spot, as the plasma beam moves 
about along the cavity walls of the cocoon as in the `dentist's drill' model 
\citep{Scheuer82}.
The basic underlying assumption is that the ridge line is due to the 
location of the 
hot spot when that material was laid down, though it could 
later be affected by some 
other process, for example, by turbulence or instabilities in the cocoon or 
interaction with the jet. 

The first moment as a function of the slice distance from the hot spot is
shown in the figures. 
We observe wiggles in the first moment with a typical amplitude of 
1$^{\prime\prime}$ to 2$^{\prime\prime}$ over scales of 10 to 20$^{\prime\prime}$.
This tracks the wiggles of the
source ridge line, which can be seen in the contour plots.

Two obvious processes that could cause the first moment to deviate
systematically from zero are a linear motion of the source 
relative to the ambient gas 
in a 
direction perpendicular to the source symmetry axis, and a 
twisting of the direction of the outflow axis of the AGN. 
The former would cause the first moment of 
both the right and left hand sides of the 
source to deviate from zero in the same sense.  This is seen in 
the first moments of 
3C44, 3C114, 
3C142.1, 3C169.1, 3C172, and 
perhaps in the 5 GHz data of 3C469.1.  For these sources, it is possible to
estimate the magnitude of the velocity of the source in the 
perpendicular direction, $v_{\perp}$, relative to the rate of 
growth of the source $v_L$ along the directions of the hot spots.
The deviation of the first moment from zero, $\Delta(FM)$,
divided by the  
distance over which this occurs, $D_{FM}$, is related to the 
source velocities: $v_{\perp} = (\Delta(FM)/D_{FM})~v_L$. 
For 3C44, the first moment varies by about 1$^{\prime\prime}$ over a 
distance of about 10$^{\prime\prime}$, so $v_{\perp} \simeq 0.1 v_L$ 
$\simeq 2.5 \times 10^3 \rm{ km/s}$, where a value of $v_L$ of 
about 0.05c obtained by O'Dea et al. (2009) for this source was
used.  This value of $v_L$ was obtained assuming an offset, $b$, of the 
magnetic field strength in the radio bridge from minimum 
energy conditions of 0.25. 
For 3C114, the first moment deviates from zero by about
1$^{\prime\prime}$ over 20$^{\prime\prime}$, so 
$v_{\perp} \simeq 0.05 v_L \simeq 750 \rm{km/s}$ for
$v_L \simeq 0.05\rm{c}$, obtained by O'Dea et al. (2009) for 
b = 0.25. 
For 3C142.1, the first moment deviates from zero
by about 2$^{\prime\prime}$ over a distance of about 10$^{\prime\prime}$, so 
$v_{\perp} \simeq 0.2 v_L \simeq 2 \times 10^3 \rm{ km/s}$ for 
$v_L \simeq 0.033\rm{c}$ obtained by O'Dea et al. (2009) assuming 
b = 0.25. For 3C169.1, the first moment deviates by about 3$^{\prime\prime}$ 
over a distance of about 20$^{\prime\prime}$, so $v_{\perp} \simeq 0.15 v_L
\simeq 2 \times 10^3 \rm{km/s}$ for a value of $v_L \simeq 0.044 \rm{c}$
obtained by O'Dea et al. (2009) for b = 0.25. For 3C172, 
the first moment deviates by about 4$^{\prime\prime}$ over 20$^{\prime\prime}$ on the 
left hand side of the source, 
so $v_{\perp} \simeq 0.2 v_L \simeq 3 \times 10^3 \rm{km/s}$ 
for $v_L \simeq 0.058$ for this side of the source for b=0.25
(O'Dea et al. 2009).  The right hand side of the source shows a step function
behavior, so the left hand side only is considered here. 
For 3C469.1, the deviation for the first moment from zero is about
1$^{\prime\prime}$ over a distance of 10$^{\prime\prime}$, so $v_{\perp} \sim 
0.1 v_L \sim 3 \times 10^3 \rm{km/s}$ for $v_L \sim 0.1\rm{c}$ obtained
for b = 0.25 by O'Dea et al. (2009). 

The values
of $v_{\perp} \simeq \rm{few} \times 10^3 \rm{km/s}$ 
obtained seem quite reasonable.  
All of these values of $v_L$ and 
$v_{\perp}$ increase by about a factor of 3 if a value of b=1 
(i.e. minimum energy conditions) is 
adopted. If the systematic deviations of the first moment from zero 
is due to the motion of the AGN relative to the ambient gas, the
reasonable values of $v_{\perp}$ obtained for b = 0.25 suggest 
that the synchrotron ages obtained for these
sources assuming b = 0.25 are reliable.  
Of course, these deviations
of the first moment from zero could be due to other factors, in which
case these estimates of $v_{\perp}$ obtained should be taken as upper bounds. 

A deviation of the first moment from zero that is in the opposite sense
for each side of the source would be produced if the direction 
of the outflow axis of the
AGN is shifting over time. This behavior is seen in the 
first moments of 3C41, 3C54 and 3C441. 3C41 has a first moment
that deviates from zero by about 1$^{\prime\prime}$ over a distance of
5$^{\prime\prime}$, suggesting that the axis of the outflow has shifted
by about 0.2 radians, or about $10^o$.  3C54 has a first moment
that deviates from zero by about 1$^{\prime\prime}$ over a distance of
about 10$^{\prime\prime}$, suggesting that the outflow axis has shifted by 
about 0.1 radians, or $6^o$.  And, 3C441 has a first moment that has
shifted by about 2$^{\prime\prime}$ over a 
distance of about 10$^{\prime\prime}$, 
suggesting the outflow axis has moved by about 0.2 radians or about
$10^o$.  Of course, other processes, such as side flows, etc. 
could have caused these deviations in which case the numbers obtained
should be taken as upper bounds. 
Taking these constraints as bounds, the sources have $v_{\perp}$ less
than about (750 - few $\times 10^3$) km/s, and changes in outflow
angle of less than about (few - 10) degrees. 
 
The remaining two sources, 3C6.1, 3C34 have 
a first moment that appears to be flat on one side of the source,
and deviates from zero on the other side of the source. 
The left side of 3C6.1 has a first moment that deviates by less
than 0.5$^{\prime\prime}$ over a distance of 10$^{\prime\prime}$, so the 
perpendicular velocity must be quite small, 
$v_{\perp} < 0.05 v_L < 750 \rm{km/s}$ for 
$v_L \simeq 0.05 \rm{c}$ obtained by O'Dea et al (2009) for this
source assuming b=0.25.  The outflow axis of the source could not
have shifted by more than about 0.05 radians, or $3^o$.
The structure of 3C34 is more complex, suggesting a sideways drift of
the first moment on the left side of the source, and more chaotic 
motion on the right side of the source. 
This is consistent with other results obtained for this source.
Based 
on optical imaging and spectroscopy, Best et al. (1997) 
concluded that the jet in 3C34 (on the western or right side of the
source) seems to have passed 
through a companion galaxy in the cluster to which 3C34 belongs, at a distance 
of about 120 kpc (17$^{\prime\prime}$) from the core of 3C34, thereby inducing massive 
star formation in the companion galaxy. 
Based on the ageing starburst in the companion galaxy, Best et al. (1997)  
concluded that the jet axis in 3C34 eventually changed direction due to precession, 
and is not currently passing through the companion galaxy. The precession is 
responsible for the new currently active hot spot, which is situated to the south 
of the older hot spot. 
Our first moment analysis shows a sideways drift on the left side of the source 
(east side of 3C34), and a more chaotic motion on the right side 
(west side of 3C34). This is consistent with the conclusions of  
Best et al. (1997).  Interestingly, the source 
seems to be quite symmetric 
in terms of the
width as a function of distance from the hot spot, and
there does not seem to be any sign of the interaction on that
parameter.  The emissivity on the right side of the source seems to 
go to zero near the symmetry axis of the source at distances of 6, 
8, and 18 arcseconds from the hotspot, which may be a signature that 
the assumption of semi-cylindrical symmetry has broken down at these
locations. Distances of 6 and 8 arcseconds from the hotspot are near 
the location of the companion galaxy with which the radio jets may
have interacted.

\subsection{Second Moment and Comparison with Gaussian FWHM}
\label{Secondmoment}

The second moment is another 
measure of the bridge width. It is obtained for each cross-sectional
slice 
using the 
expression $(\sum S_i [x_i - \bar{x}]^2/\sum S_i)^{1/2}$, where $S_i$ 
is the surface brightness at the position $x_i$, and 
$\bar{x}$ is the first moment (see section \ref{Firstmoment}). 
In general, the source width as measured by the second moment
is very similar to that indicated by the Gaussian width $W_G$.  

The ratio of the second moment to the Gaussian FWHM 
remains fairly constant for each side of each source, 
with source to source variations that range from about 
2.0 to 3.0 for the 11 sources studied, 
as can
be seen in the figures.    
The second moment is a measure of the half width of the source, 
so the width indicated by the 
Gaussian FWHM is about 25\% larger than the width indicated by 
twice the second moment. 
For a slice with a constant emissivity, the ratio of the 
Gaussian FWHM to the second moment is expected to be 
about 2.7 (see Table 1 of Carvalho et al. 2005 and account
for the fact that the second moment measures a half width rather
than a full width). This value for the 
ratio is quite close to those obtained empirically 
here, and is consistent with the typical emissivity profiles 
obtained for most slices. This ratio is expected to be slightly
larger when the emissivity is ``edge-peaked'' 
(Carvalho et al. 2005), but the difference is so small that
is it difficult to detect. 

\subsection{Mean Surface Brightness, Magnetic Field Strength, and Pressure}
\label{SBP}

The average surface brightness of a slice, $\bar{S}$, is obtained 
by dividing the area under the surface brightness profile by 
$W_G$, and is shown in the figures; the units of $\bar{S}$ are
Jy/beam. 
The hot spot dominates the surface 
brightness of the lobe.

\begin{deluxetable}{lcccccccc}
\tabletypesize{\small}
\tablecaption{Normalizations for Figures}
\tablewidth{0pt}
\tablehead{
\colhead{(1)}  & \colhead{(2)}   & \colhead{(3)}  & 
\colhead{(4)}& \colhead{(5)}  & \colhead{(6)}& 
\colhead{(7)}   & \colhead{(8)}& \colhead{(9)} \\
\colhead{Source}  & \colhead{$\theta_{FWHM}$}   & \colhead{z}  & 
\colhead{$(a_0r)$}& \colhead{$K_{\epsilon}$}  & \colhead{$K_{B}(1.4)$}& 
\colhead{$K_{B}(5)$}   & \colhead{$K_{P}(1.4)$}& \colhead{$K_{P}(5)$} \\
&{arcsec}&&\colhead{Gpc}&\colhead{$10^{-34}$}&\colhead{$10^{-6}$} 
& \colhead{$10^{-6}$} & \colhead{$10^{-11}$}
& \colhead{$10^{-11}$} \\
&&&& \colhead{$\rm{erg/s/cm^{3}/Hz}$} & \colhead{G} 
& \colhead{G} & \colhead{$\rm{erg/cm^{3}}$}
& \colhead{$\rm{erg/cm^{3}}$} }
\startdata																
3C6.1	&	2	&	0.8404	&	2.896	&	3.14	&	107	&	138	&	35.2	&	58.5	\\
3C34	&	2	&	0.69	&	2.476	&	2.61	&	99.5	&	128	&	30.6	&	50.9	\\
3C41	&	2	&	0.794	&	2.771	&	2.96	&	104	&	135	&	33.7	&	56.0	\\
3C44	&	2.5	&	0.66	&	2.388	&	1.61	&	86.4	&	111	&	23.1	&	38.4	\\
3C54	&	2.5	&	0.8274	&	2.862	&	1.97	&	93.3	&	120	&	26.9	&	44.8	\\
3C114	&	2.5	&	0.815	&	2.828	&	1.94	&	92.8	&	120	&	26.6	&	44.3	\\
3C142.1	&	2	&	0.4061	&	1.573	&	1.97	&	88.5	&	114	&	24.2	&	40.3	\\
3C169.1	&	2.5	&	0.633	&	2.307	&	1.56	&	85.4	&	110	&	22.5	&	37.5	\\
3C172	&	2.5	&	0.5191	&	1.951	&	1.38	&	81.3	&	105	&	20.4	&	34.0	\\
3C441	&	2.5	&	0.707	&	2.526	&	1.70	&	88.3	&	114	&	24.1	&	40.1	\\
3C469.1	&	2.5	&	1.336	&	4.047	&	3.73	&	118	&	152	&	42.7	&	
71.1	\\
\enddata
\tablecomments
{Col.1: Source Name. Col.2: FWHM of the Gaussian observing beam.
Col.3: Source redshift.
Col.4: Coordinate distance to the source in Gigaparsecs. 
Col.5: Normalization to the emissivities $\epsilon_M$ shown in 
figures when $\epsilon_M$ is in units of Jy/(beam arcsec), 
as is the case for all of the sources shown here. 
Col.6: Normalization to the 1.4 GHz magnetic field strengths 
shown 
in the figures.
Col.7: Normalization to the 5 GHz magnetic field strengths shown 
in the figures.
Col.8: Normalization to the 1.4 GHz pressures shown 
in the figures.
Col.9: Normalization to the 5 GHz pressures shown 
in the figures. }
\label{tabsummary}
\end{deluxetable}

The minimum-energy magnetic field strength \citep{Burbidge56}
is obtained for each surface brightness profile slice using the
expression $B_{min} = K_B B_M$ where $B_M$ is the magnetic
field strength shown in the figures, and the normalization 
$K_B$ is listed in Table 1 for $B_M$ in the units shown. 
The magnetic field strength shown
in the figures is obtained using the expression 
$B_M = (\bar{S}/W_G)^{2/7}$.
The normalization factor $K_B$ is obtained using the expression
for the minimum energy magnetic field strength presented by 
Miley (1980) assuming no relativistic protons, a volume filling factor
of unity, and a spectral index of 0.7.  

The minimum-energy pressure \citep{Burbidge56} is obtained for each
surface brightness slice 
using the expression $P_{min} = K_P P_M$ where 
$P_M$ is the pressure shown in the figures. This is obtained using
the expression $P_M=(\bar{S}/W_G)^{4/7}$. The normalization factor
$K_P$ is obtained from $K_B$ noting that the source pressure is 
$P = [(4/3)b^{-1.5} +b^2]B_{min}^2/(24 \pi)$, where $b$ represents
the offset of the magnetic field strength $B$ from the minimum
energy field, $b \equiv B/B_{min}$, taken here to be unity. 
Values of $K_P$ are listed in Table 1 for $P_M$ in the units
shown in the figures. 

All three quantities, $\bar{S}$, $B_{min}$, and $P_{min}$ 
peak in the vicinity of the hot spot, have a region where each  
steadily decreases across the source away from the hot spot, and then
level off to a roughly constant or very slowly declining value.

Interestingly, the location at which the pressure levels off
and becomes roughly constant seems to be related to the location of the 
local maxima of the bridge width discussed in section \ref{GaussianFits}.
Moving away from the hot spot, the
pressure seems to level off just before the local maxima of
the bridge width for 3C6.1, 3C34, 3C41, 3C44, the right side of
3C142.1, the left side of 3C169.1 (perhaps in two places), 
and 3C469.1. The effect is less obvious but may also be 
present for 3C54, 3C114, the left side of 3C142.1, and the left side of
3C441. The effect is not seen in 3C172 or the right side of
3C169.1, and the pressure is not seen to level off for the 
right side of 3C441. 

Thus, it would appear that in many sources the bridge width 
increases with distance from the hot spot and the pressure of the 
relativistic fluid in this region has strong gradients. 
Pressure waves propagating through the fluid tend to bring the
pressure into equilibrium.  Equilibrium is approached or 
reached near the 
location of the local maxima of the radio bridge width 
(and this approach to equilibrium 
is perhaps the reason that the bridge width does not
continue to increase with distance from the hot spot). 
Sometimes the source over expands at the local maxima 
of the bridge width, and
a signature of this over expansion could be emissivity 
and pressure slice profiles
that are ``edge-peaked'' with a minimum near the symmetry axis of the 
lobe. 
Regions downstream of this point, which are further away from 
the hot spot, are in rough pressure equilibrium.  Pressure waves
propagate throughout the radio lobe region as indicated by fluctuations
of the emissivity and pressure 
both within a slice and as seen by comparing emissivities
of adjacent slices. 

Effect which could cause apparent fluctuations of pressure across
a slice are deviations of the magnetic field strength from 
minimum energy conditions or a non-semi-cylindrically symmetric slice,
as discussed in section \ref{radioemissivity}.  It is interesting
that the emissivity obtained independently by stepping in from the 
top and bottom of the slice match in most slices, suggesting that
semi-cylindrical symmetry is a good approximation for most slices, 
and that the 
pressure fluctuations seen across most slices are rather modest.

\section{SUMMARY AND CONCLUSIONS}
\label{summary}

The radio lobes of a 
sample of 11 very powerful classical double radio galaxies were 
studied in detail.  Each source was rotated so that the line 
connecting the hot spots was horizontal.  Vertical cross-sectional
cuts separated by one beam width of the radio surface brightness were 
taken beginning at the hot spot of each side of each source and
moving in toward the source center.  

\begin{enumerate}
\item Gaussian Fits: 
Each cross-sectional slice was fit with a Gaussian and these Gaussian
fits and the FWHM of the Gaussian, $W_G$, were studied as a function of
distance from the hot spot. A Gaussian provides a very good description
of the surface brightness profile for most slices. 
The lobe width $W_G$ as a function of distance from the hot spot
was very similar for each side of a given source, indicating a high
degree of symmetry relative to the hot spot positions. This suggests that 
the properties of the ambient medium are symmetric about the 
source center and the properties of the twin jets at 
any given time are quite similar.  

Many of the widths first increase and then decrease with distance
from the hot spot. The position of the  
local maxima seems to be related to the 
pressure profile of the source 
and is sometimes related to the properties of the radio emissivity.

\item Radio Emissivity: The radio emissivity, $\epsilon_{\nu}$ 
may be obtained by noting $\epsilon_{\nu} \propto S_{\nu}/l$, where
$S_{\nu}$ is the radio surface brightness and $l$ is the path length,
and assuming that at a given distance $r$ from the slice center the 
emissivity is constant over a thin semi-circular shell 
at radius r. By starting at 
the top or bottom of the slice and stepping in, the emissivity 
may be obtained at each distance $r$ for the top and bottom half
of the slice. The results obtained for the top and bottom halves
are independent. For most slices, the 
emissivity near the center of the slice
is continuous, indicating that the emissivity obtained there by stepping
in from the top and bottom of the source match, and suggesting that 
the assumption of semi-cylindrical symmetry is valid for that slice. 

The emissivities of many slices are fairly flat, indicating a region of
constant emissivity.  Some of the slices indicate an ``edge-peaked''
emissivity that can include a minimum emissivity near the source center. 
When these are detected, the slices seen to be located near the 
local maxima of the radio bridge width.  Taken at face value,they
suggest that the pressure of the slice is a minimum at the slice
center and larger near the edge of the slice since the pressure 
goes roughly as the square root of the emissivity. This is the 
type of behavior expected as pressure waves carry information 
about the pressure of the external medium 
into the relativistic fluid.
No signature of a jet-like feature that could produce a central
peak of an emissivity profile that propagates through adjacent slices 
is seen. 

\item The First Moment: The first moment of each slice was obtained
and studied as a function of distance from the hot spot.  A first
moment of zero would indicate that the surface brightness peak of
each slice lies along the symmetry axis of the source. The first moment
is found to wander from zero by very small amounts: typically less
than 1$^{\prime\prime}$ to 2$^{\prime\prime}$ over distances of about 10$^{\prime\prime}$ 
to 20$^{\prime\prime}$.  The first moment may be taken as an indication of
the location of the hot spot at an earlier stage of the source lifetime. 
Deviations of the first moment from the current symmetry axis of
the source could be caused by actual deviations of the 
location of the hot spot from the current symmetry axis, or by many
other processes.  One such process is the 
motion of the source relative to the ambient gas. 
Measurements of the deviations of the first moment from zero were
used to place bounds on the velocity of the source relative to 
the ambient gas in the direction perpendicular to the source symmetry
axis, and the results suggest that this velocity must be less than
or equal to about $10^3$ km/s or so.  Measurements of the first moment
can also be used to place limits on the angle over which the outflow
axis could have shifted relative to the ambient gas, and the
results suggest that this angle must be less than a few to 10 degrees. 

\item The Second Moment: The second moment was obtained and compared with 
the Gaussian width. The two were found to indicate a similar source
shape since their ratio is fairly constant.  The value of this ratio
is typically close to what is expected for the emissivity profiles
obtained. 

\item Mean Surface Brightness, Magnetic Field Strength, and Pressure: 
Average values of the 
surface brightness, minimum energy magnetic field
strength, and pressure were obtained for each slice.
The surface brightness peaks at the location of the hot spot,
and remains fairly constant across the rest of the source.
The magnetic field strength and pressure also peak 
at the hot spot, and decrease in value as the distance from 
the hot spot increases. At some distance from
the hot spot they seem to reach a roughly 
constant value and remain fairly constant for distances
greater than this, suggesting that equilibrium is being 
approached or reached. Interestingly, 
the pressures seem to level off at a distance from the hot spot that
is located just before the local maxima of the bridge width.

The results obtained may be interpreted as the response of the system
to the large pressure difference between the hot spot and the 
ambient gas, as discussed in sections \ref{radioemissivity} and 
\ref{SBP}.  In the rest frame of the hot spot, the problem 
is somewhat like that of an outflow entering a region of low external
pressure, which has been explored for a relativistic fluid by 
 K\"onigl (1980) and Daly \& Marscher (1988).  The fluid expands in the 
lateral direction as information about the external pressure 
propagates into the relativistic fluid.  At the point of
maximal expansion, the pressure is a minimum near the symmetry
axis of the source, and is largest at the source edges.  
In the ideal case studied by  K\"onigl and Daly \& Marscher, this 
pattern continues, but in the sources studied here, the data
suggest that pressure waves are able to bring the source
into rough pressure equilibrium downstream of this location
for most of the sources studied.  

\end{enumerate}

\acknowledgements
This work was supported in part by U. S. National Science Foundation
grant AST-0507465 (RAD). 
The National Radio Astronomy Observatory is a facility of the National
Science Foundation operated under cooperative agreement by Associated
Universities, Inc.
This research has made use of the NASA/IPAC Extragalactic Database (NED)
which is operated by the Jet Propulsion Laboratory, California Institute
of Technology, under contract with the National Aeronautics and Space
Administration.

\begin{figure}
\plotone{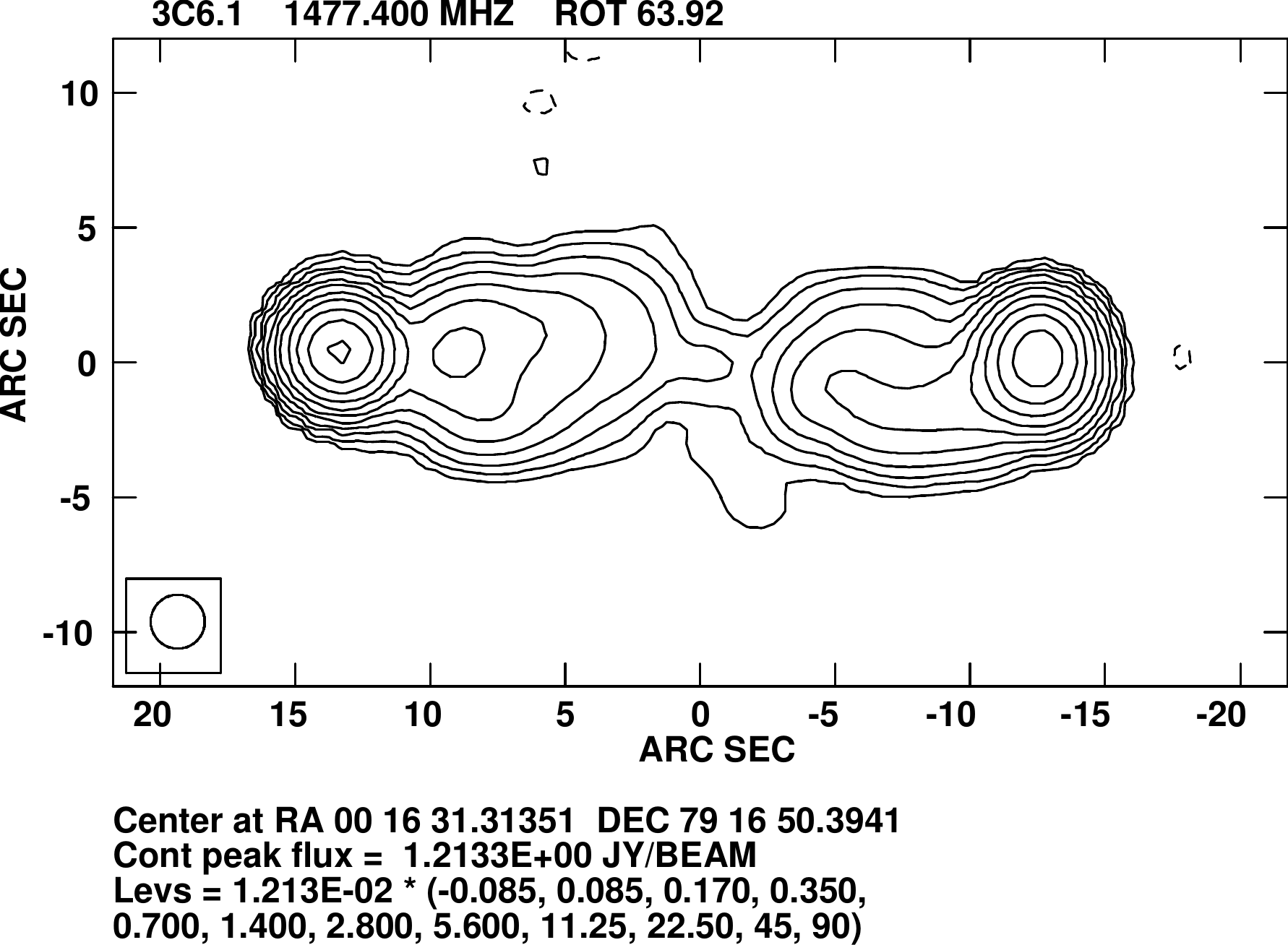}
\caption{3C6.1 at 1.4 GHz and 2 arcsec resolution 
rotated by ${63.92 \deg}$ counter-clockwise so that 
the centroids of the two hot spots lie on a horizontal line. The location
of the core of the source, identified as described in 
\citet{Kharb08},  has
x and y coordinates of 0 arcsec.  Only data with SN greater 
than $(\rm{mean} 
+  3 \times \rm{rms})$ noise level of the map were included in the 
source analysis; all of the results presented here were obtained from 
this image and a similar image at 5 GHz.}
\label{3C6.1rotfig}
\end{figure}

\begin{figure}
\plottwo{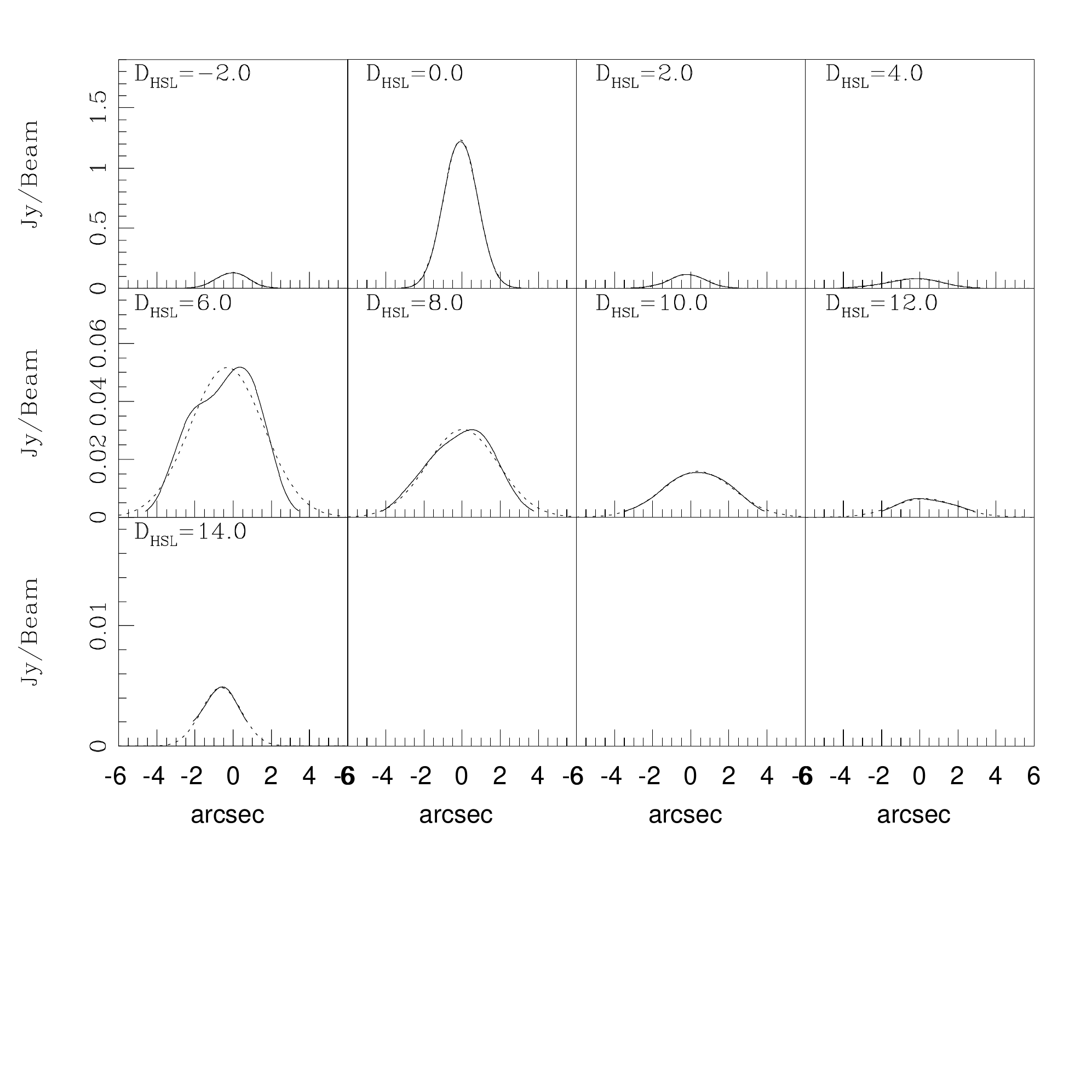}{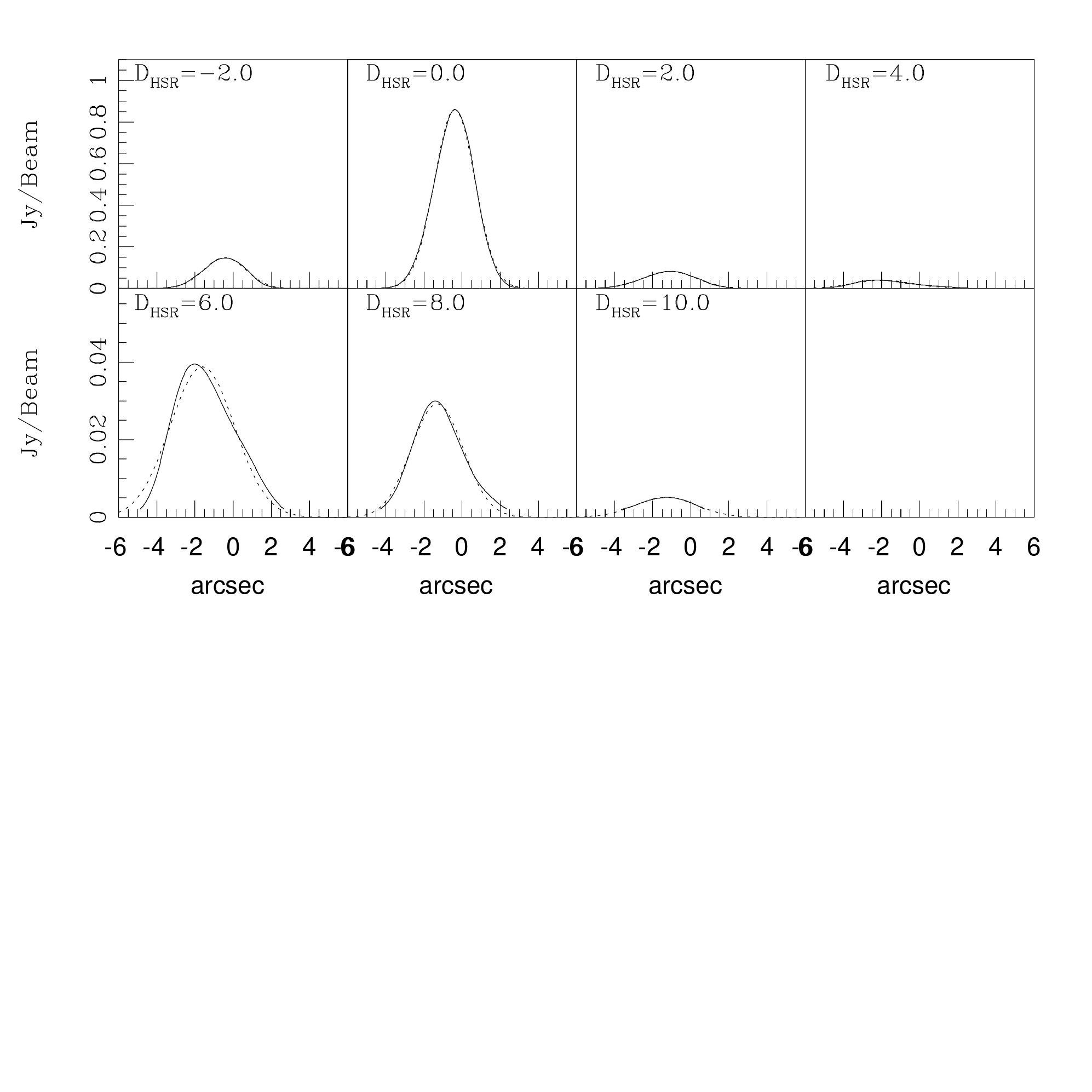}
\caption{3C6.1 surface brightness (solid line) and 
best fit Gaussian (dotted line)  for each cross-sectional
slice of the radio bridge at 1.4 GHz for the left hand side of the
source (left panel) and the right hand side of the source (right panel).
The distance from the hot spot is indicated in the top left hand
corner of each slide; $D_{HSL}$ indicates the distance from the 
left hot spot, and $D_{HSR}$ indicates the distance from the 
right hot spot, where the left and right hand sides of the source
are identified relative to the core (see Fig. \ref{3C6.1rotfig}).  
The results indicate that a Gaussian provides an excellent 
description of the surface brightness profile of each cross-sectional 
slice. Note that the surface brightness scale of each row is different. }
\label{3C6.1SG}
\end{figure}

\begin{figure}
\plottwo{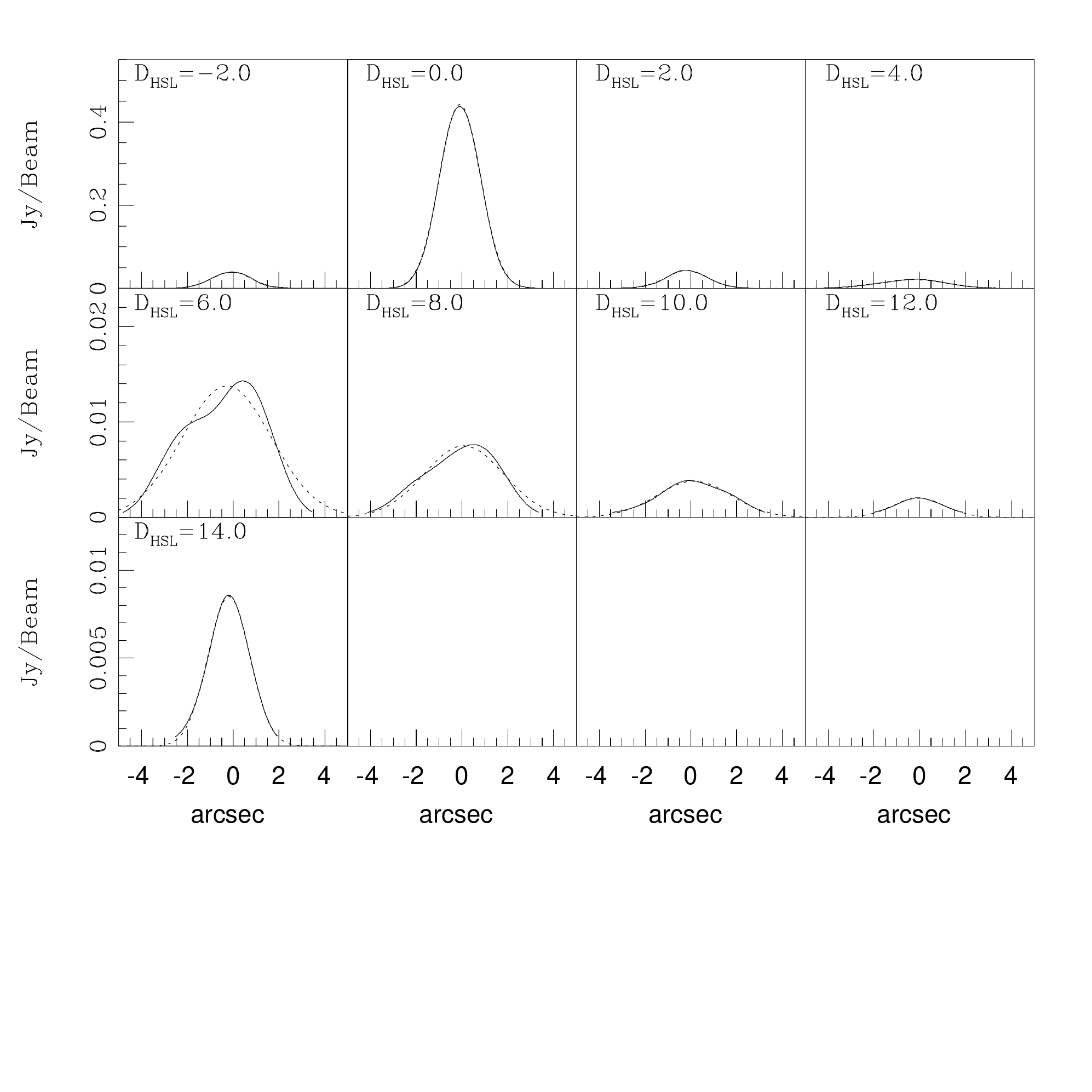}{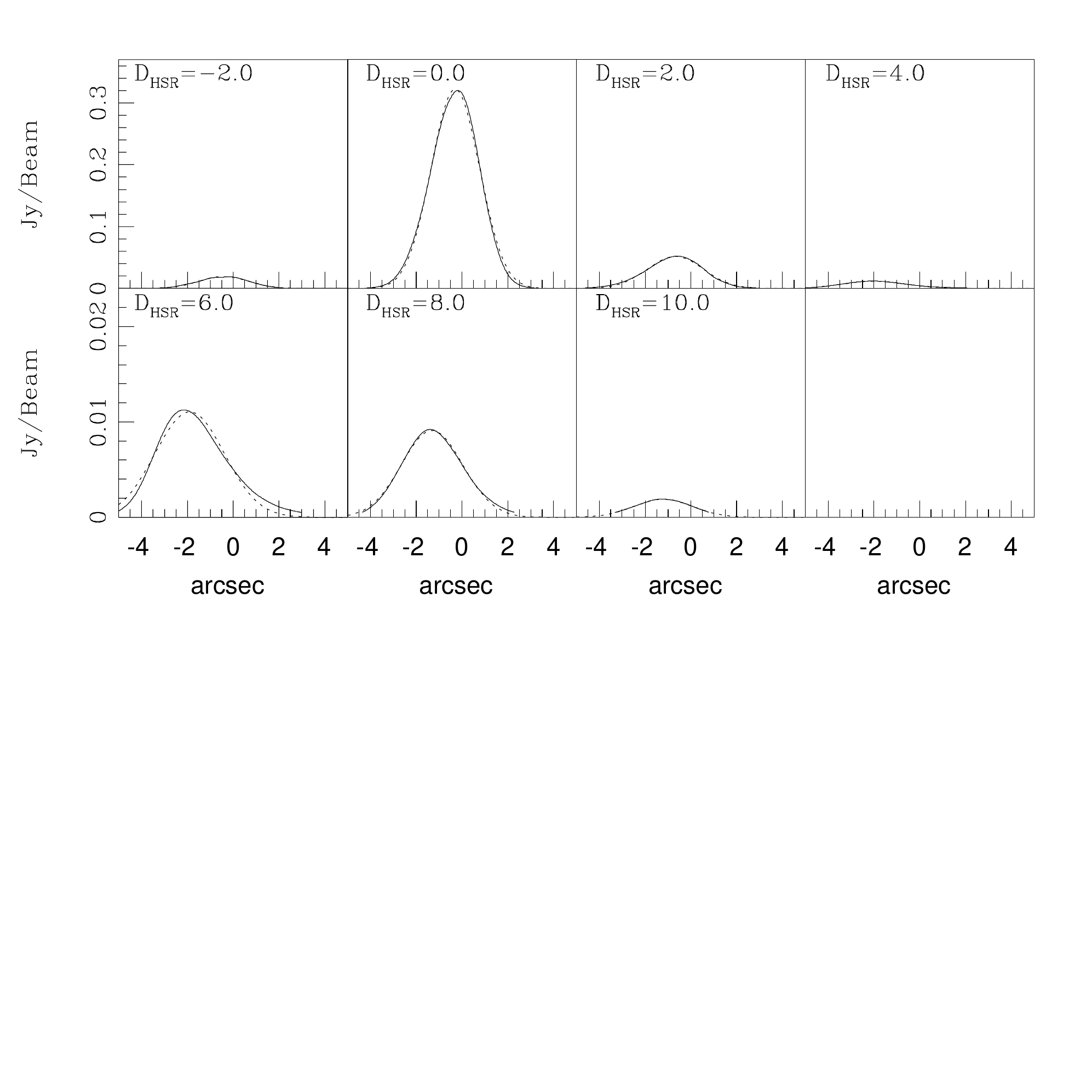}
\caption{3C6.1 surface brightness (solid line) and 
best fit Gaussian (dotted line)  for each cross-sectional
slice of the radio bridge at 5 GHz for the left hand side of the
source (left panel) and the right hand side of the source (right panel),
as in Fig. \ref{3C6.1SG} but at 5 GHz. The results
are very similar to those obtained at 1.4 GHz.}
\end{figure}

\begin{figure}
\plottwo{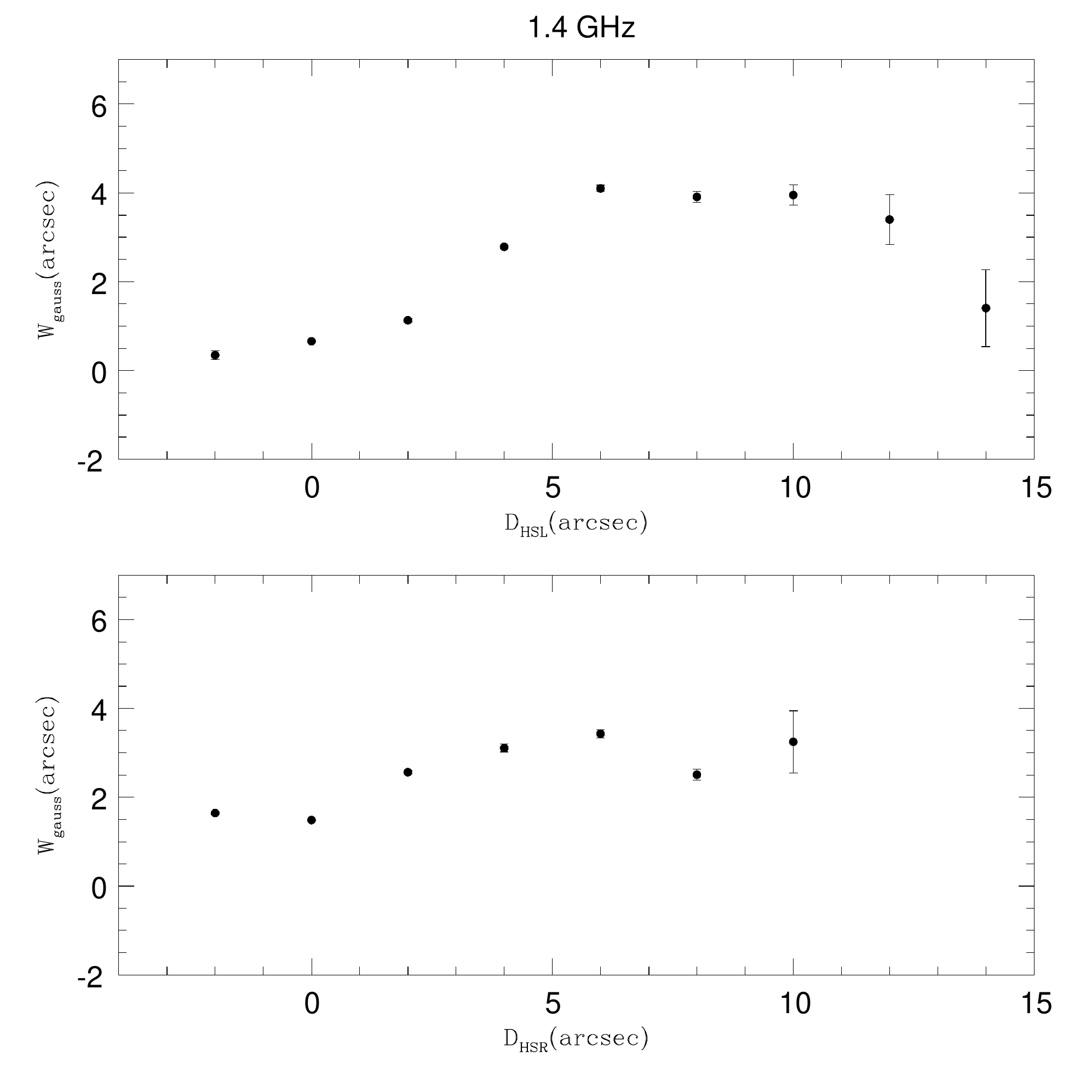}{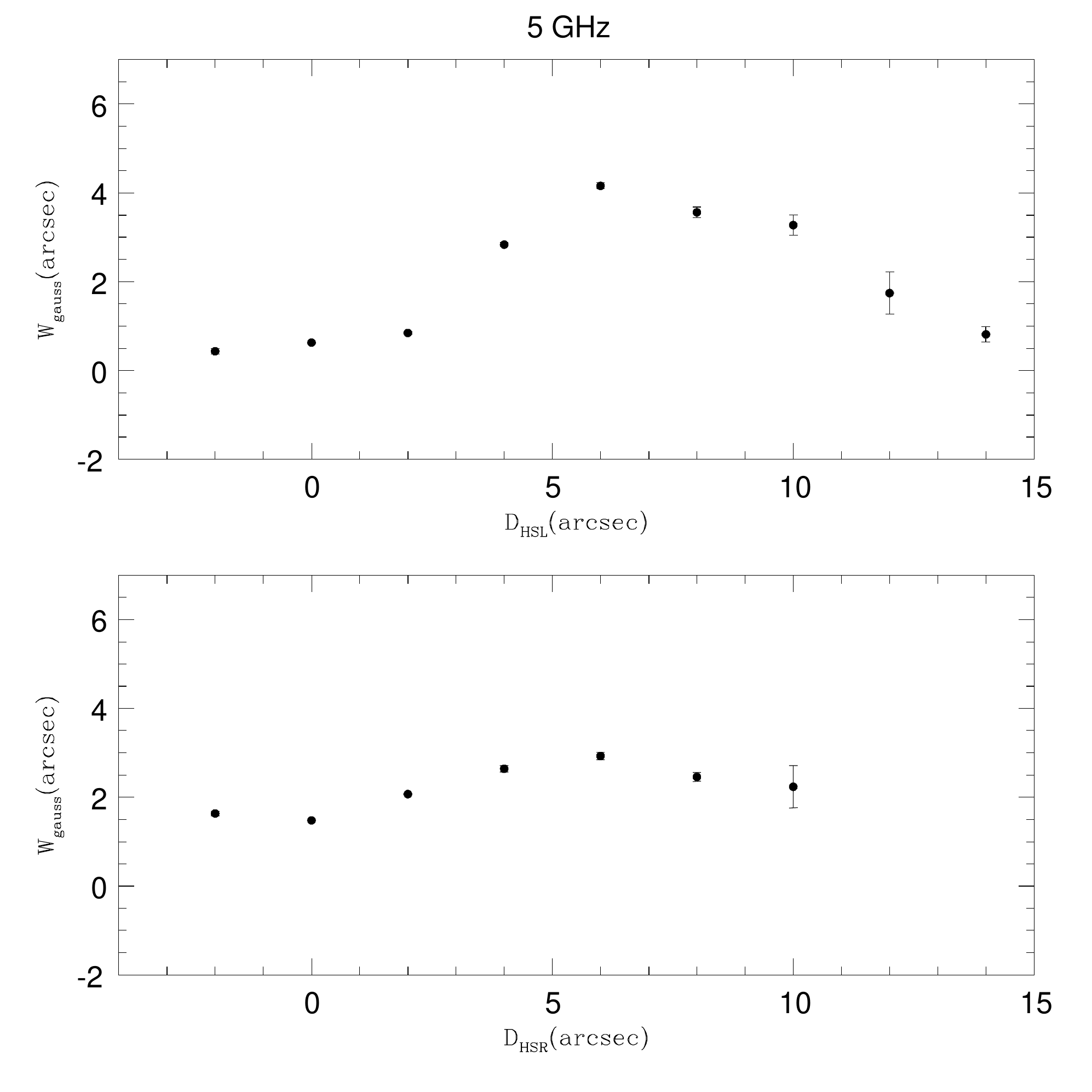}
\caption{3C6.1 Gaussian full width at half maximum 
(FWHM) 
as a function of 
distance from the hot spot at 1.4 GHz and 
5 GHz (left and right panels, respectively) 
for the left and right hand sides of the source (top and bottom
panels, respectively) with one measurement shown 
per 2 arcsec resolution element; angular distances 
are measured from the 
centroid of the hot spot.
This measure of the 
source width indicates that the right and left hand sides of the 
source are highly symmetric in 
terms of width as a function of distance from the hot spot. 
Very similar results are obtained at 1.4 and 5 GHz.  }
\label{3C6.1WG}
\end{figure}

\begin{figure}
\plottwo{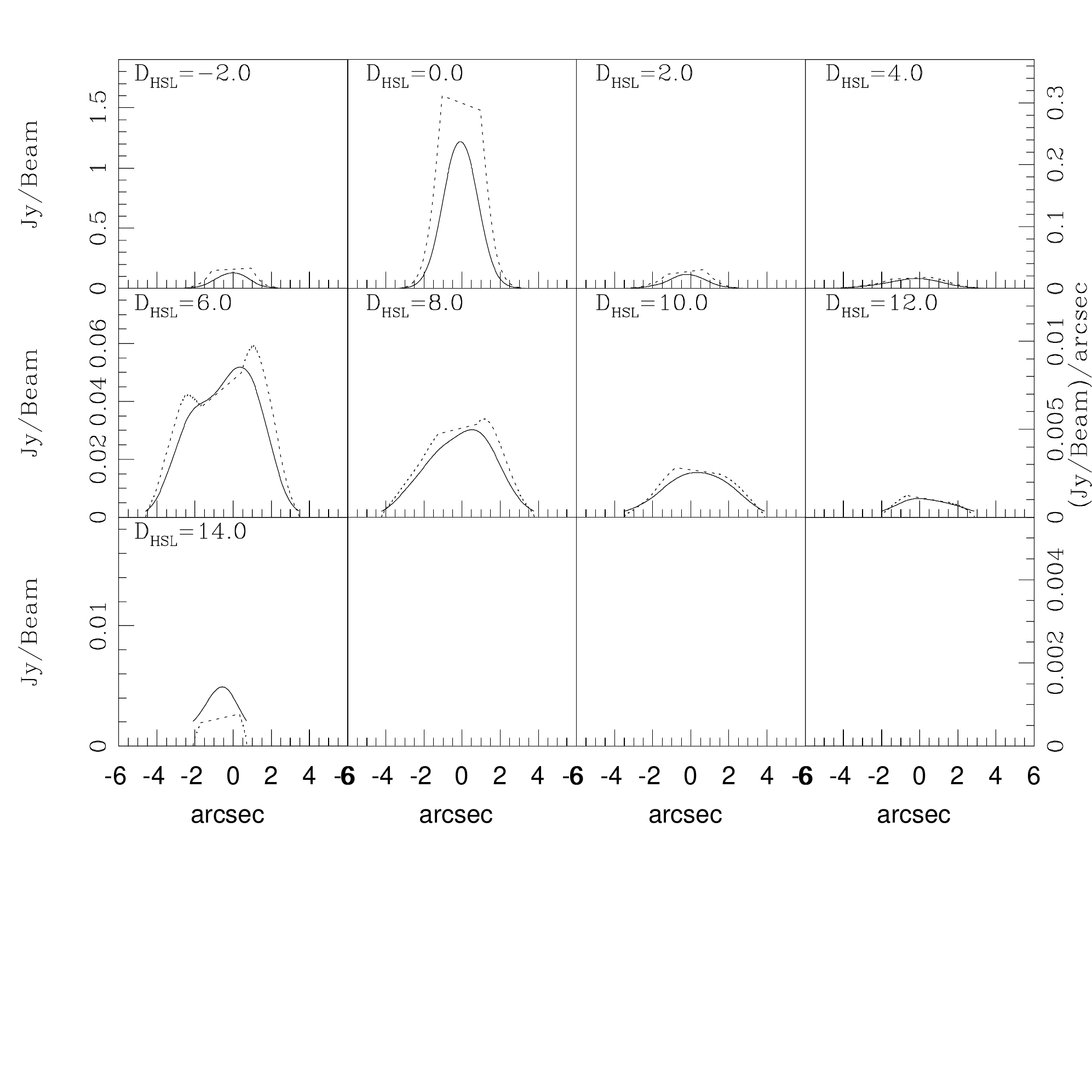}{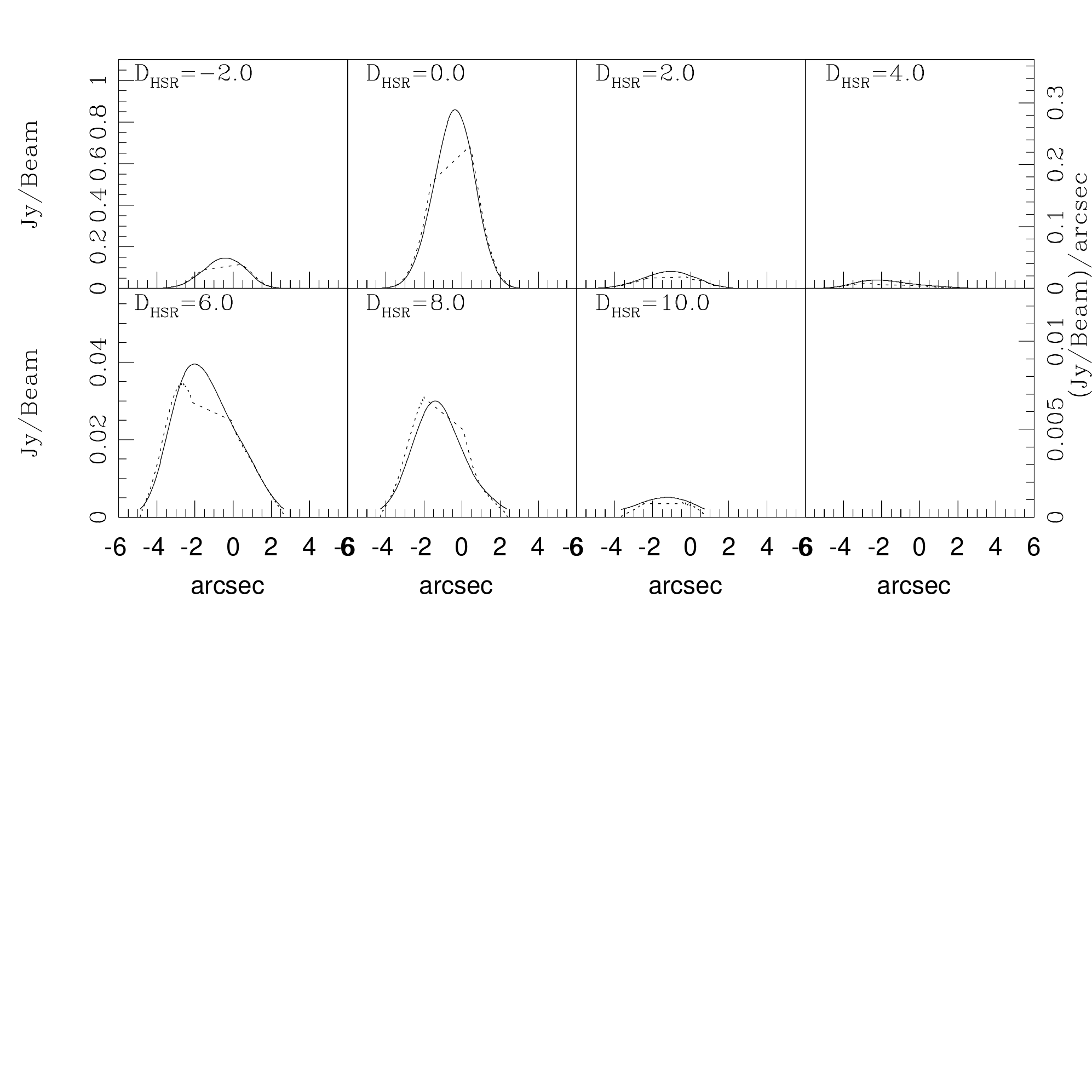}
\caption{3C6.1  emissivity (dotted line) 
and surface brightness (solid line) 
for each cross-sectional
slice of the radio bridge at 1.4 GHz for the left hand side of the
source (left panel) and the right hand side of the source (right panel).
The distance from the hot spot is indicated in the top left hand
corner of each slide; $D_{HSL}$ indicates the distance from the 
left hot spot, and $D_{HSR}$ indicates the distance from the 
right hot spot, where the left and right hand sides of the source
are identified relative to the core (see Fig. \ref{3C6.1rotfig}).  
The units of surface brightness are shown on the left hand side
of each figure, while those for the emissivity are shown on the 
right hand side of each figure. The normalization of the emissivity
is given in Table 1.  
The results indicate that a constant
emissivity per slice provides a reasonable description of the source.
Note that the surface brightness and emissivity scales of each row are different.}
\label{3C6.1SEM}
\end{figure}

\begin{figure}
\plottwo{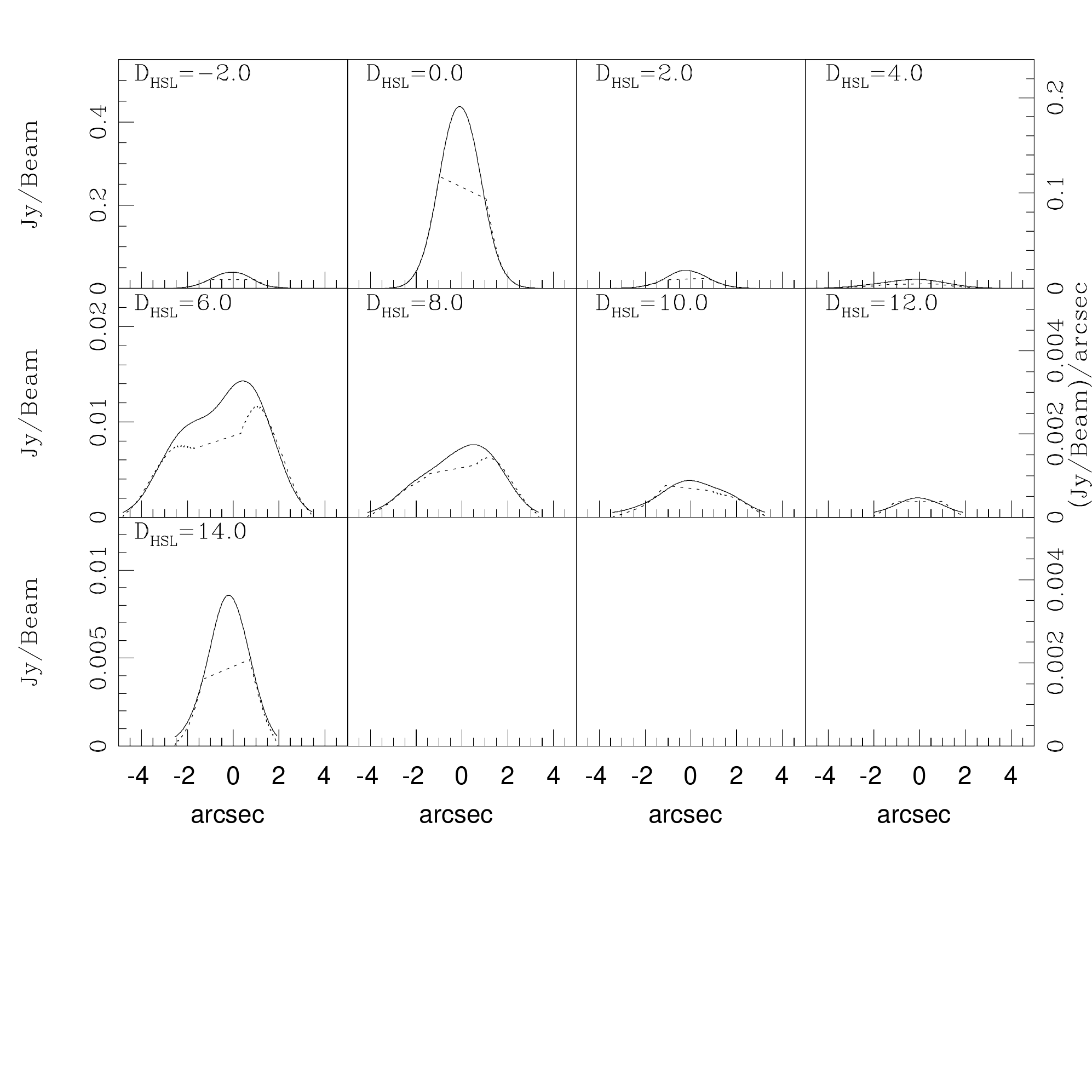}{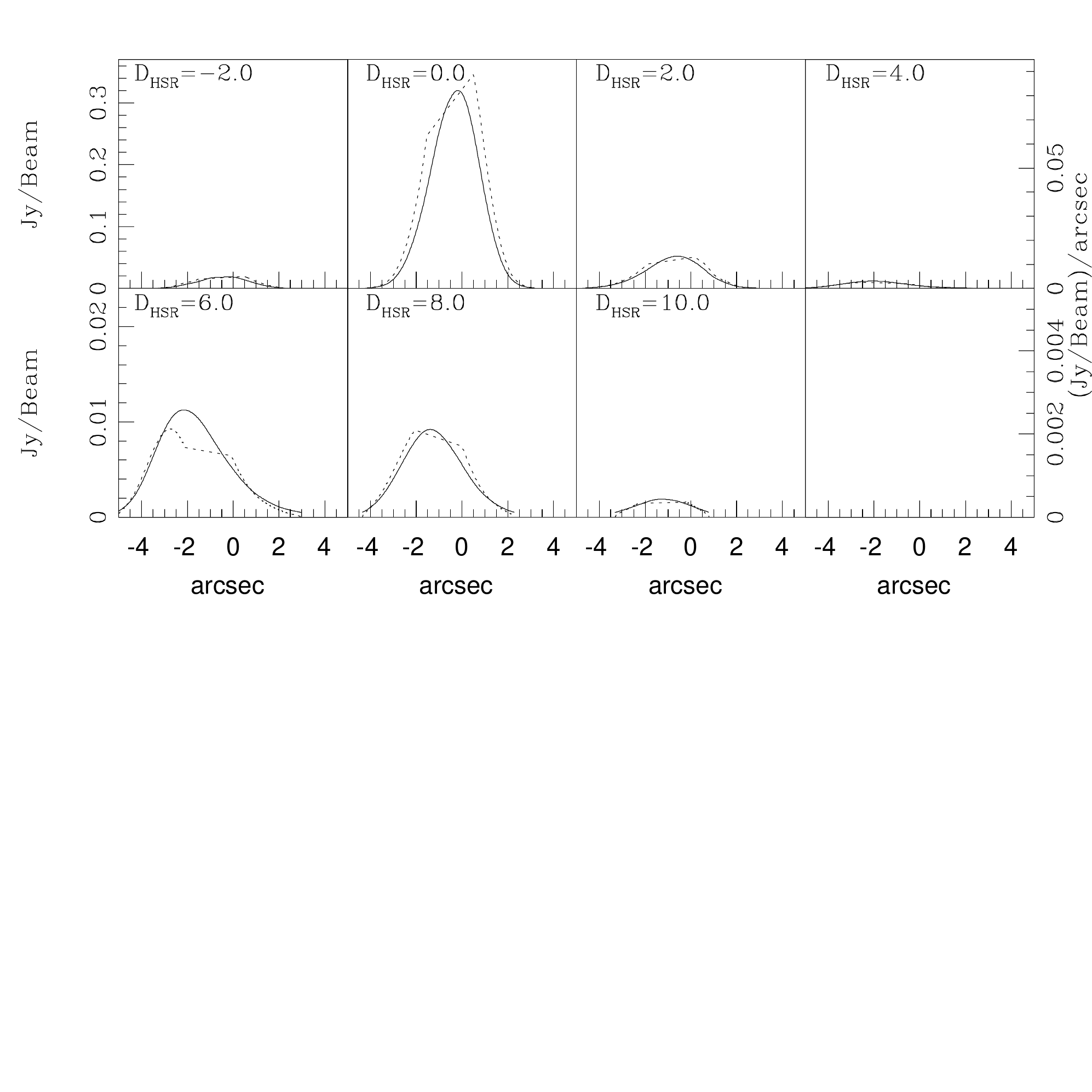}
\caption{3C6.1  emissivity (dotted line) 
and surface brightness (solid line) 
for each cross-sectional
slice of the radio bridge at 5 GHz for the left hand side of the
source (left panel) and the right hand side of the source (right panel),
as in Fig. \ref{3C6.1SEM} but at 5 GHz.  The results
are very similar to those obtained at 1.4 GHz.}
\end{figure}

\begin{figure}
\plottwo{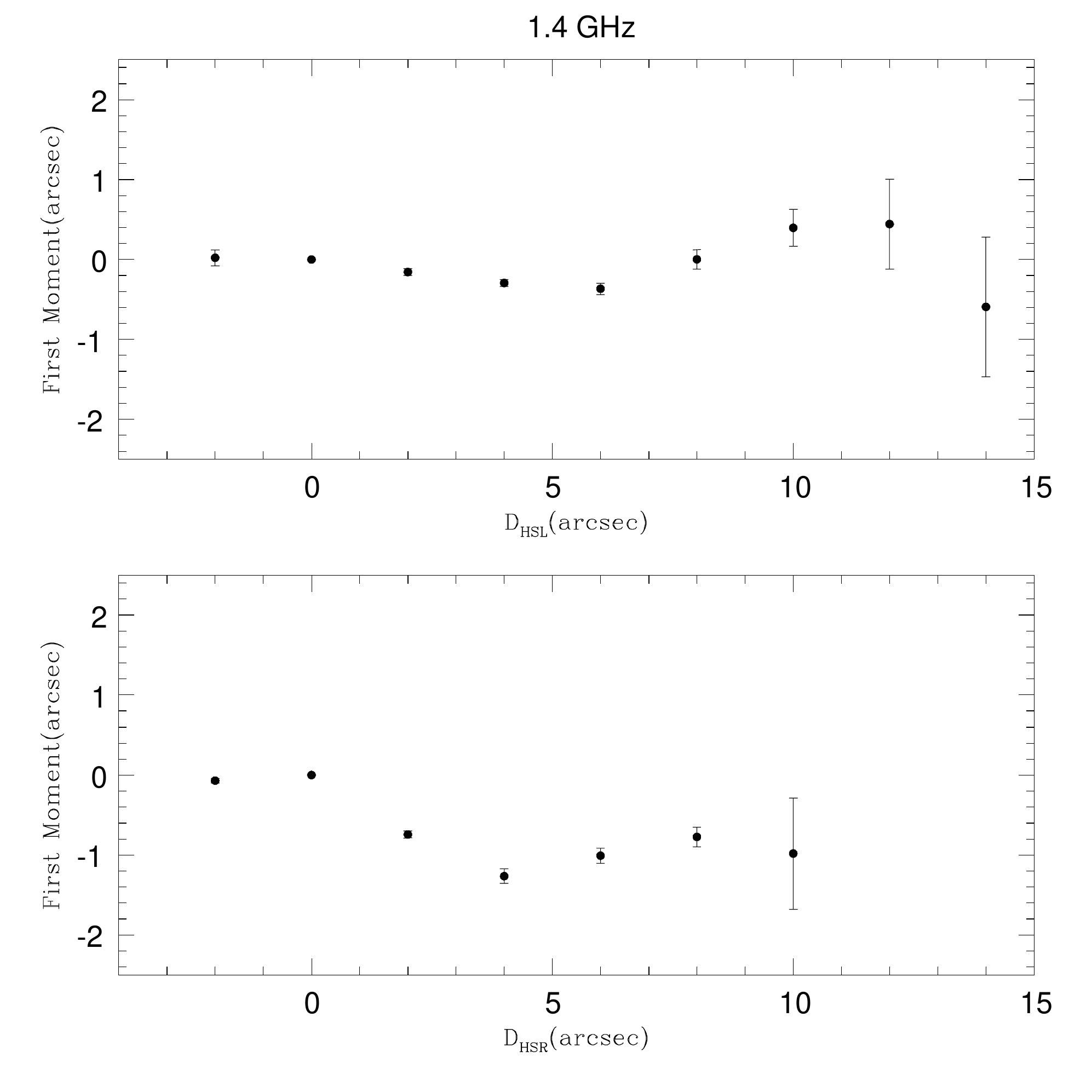}{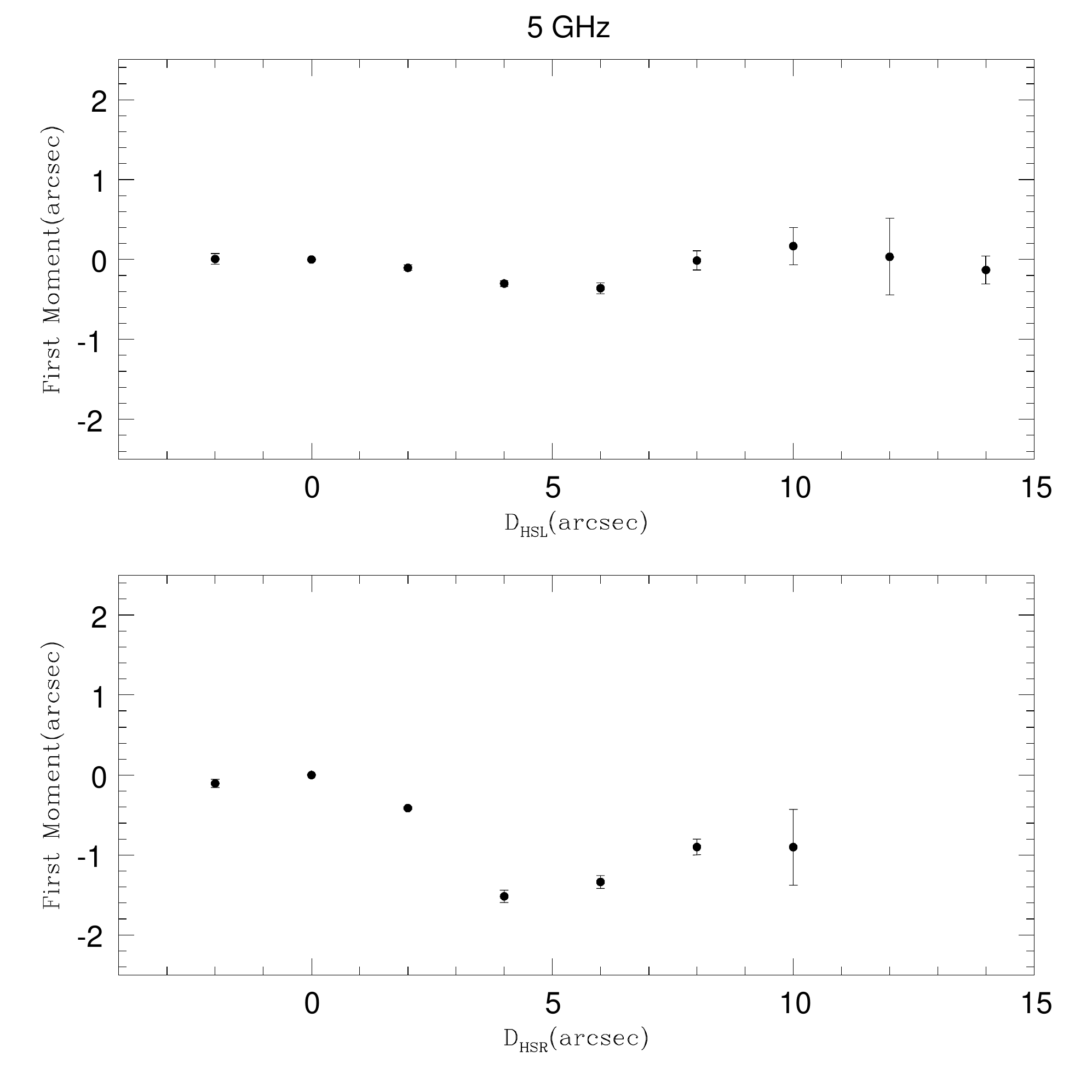}
\caption{3C6.1 first moment as a function of distance from 
the hot spot for 1.4 and 5 GHz (left and right panels, respectively) 
for the left and right hand sides of the source
(top and bottom panels, respectively), 
as in Fig. \ref{3C6.1WG}. The first moment is measured
relative to the horizontal line that connects the two hot spot 
centroids; a value of zero for the first moment means that the
first moment lies along the horizontal line connecting the two
hot spots. Results obtained at 1.4 and 5 GHz are nearly identical.
The left hand side of the source has a first moment that remains
close to zero, while the first moment of the 
right hand side of the source falls slightly below this line by about
1 to 2 arcsec (see Fig. \ref{3C6.1rotfig}). }
\label{3C6.1FM}
\end{figure}

\begin{figure}
\plottwo{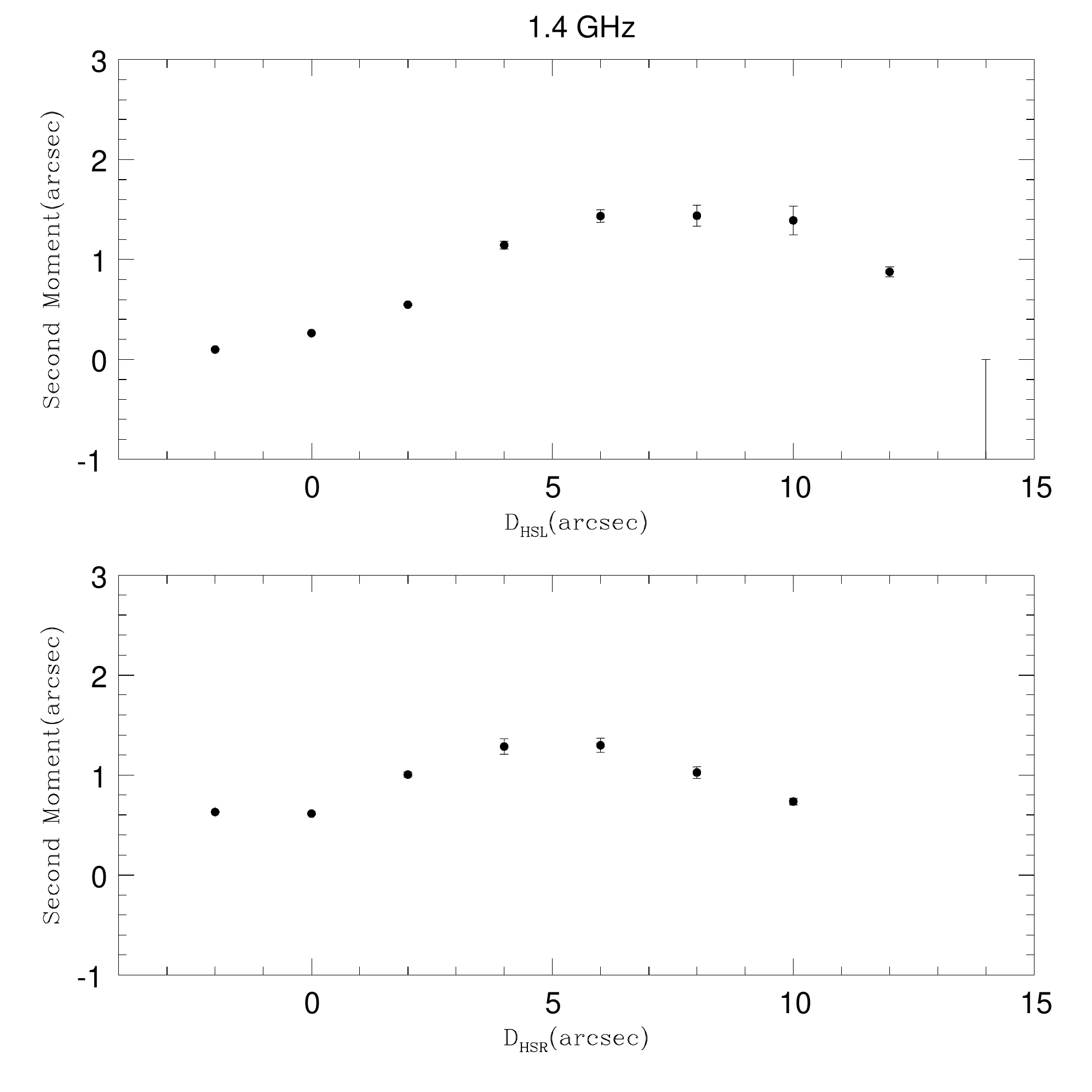}{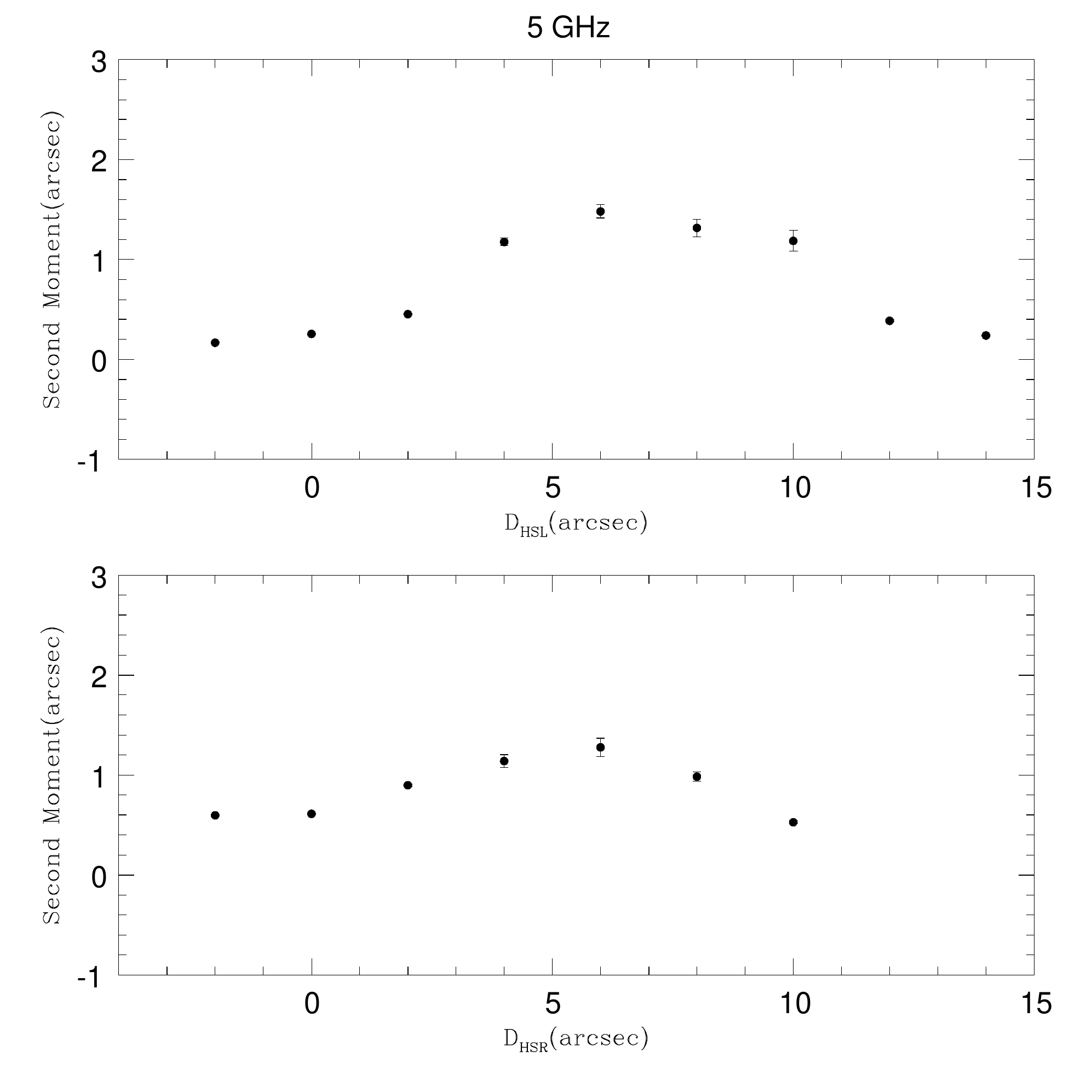}
\caption{3C6.1 second moment as a function of distance from the
hot spot at 1.4 and 5 GHz (left and right panels, respectively) 
for the left and right hand sides of the source (top and bottom
panels, respectively), as in Fig. \ref{3C6.1WG}.  Comparing the two
panels on the left, it is clear that the source is highly symmetric
in terms of the source width as a function of distance from the hot
spot. Results obtained at 1.4 and 5 GHz are nearly identical. }
\label{3C6.1SM}
\end{figure}

\begin{figure}
\plottwo{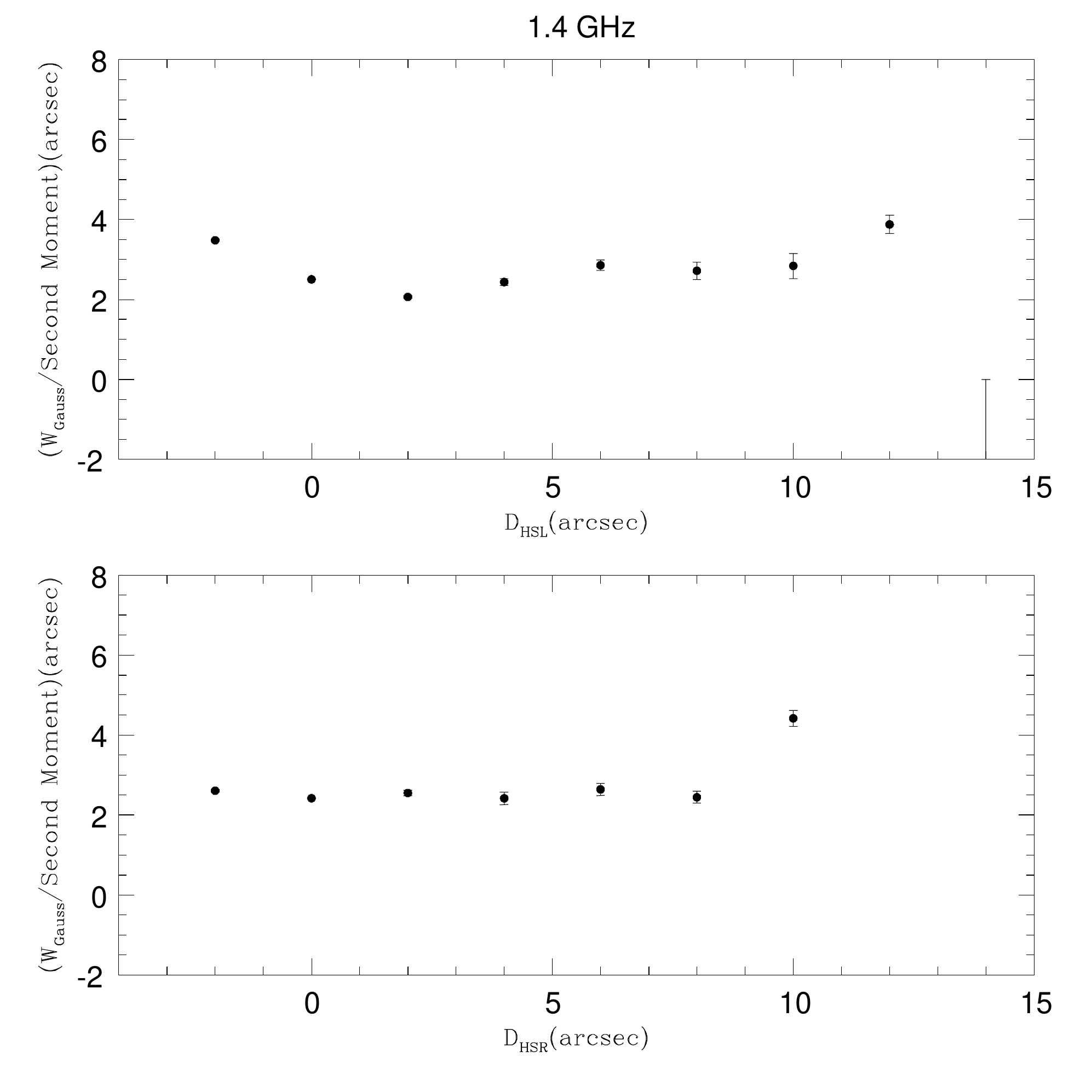}{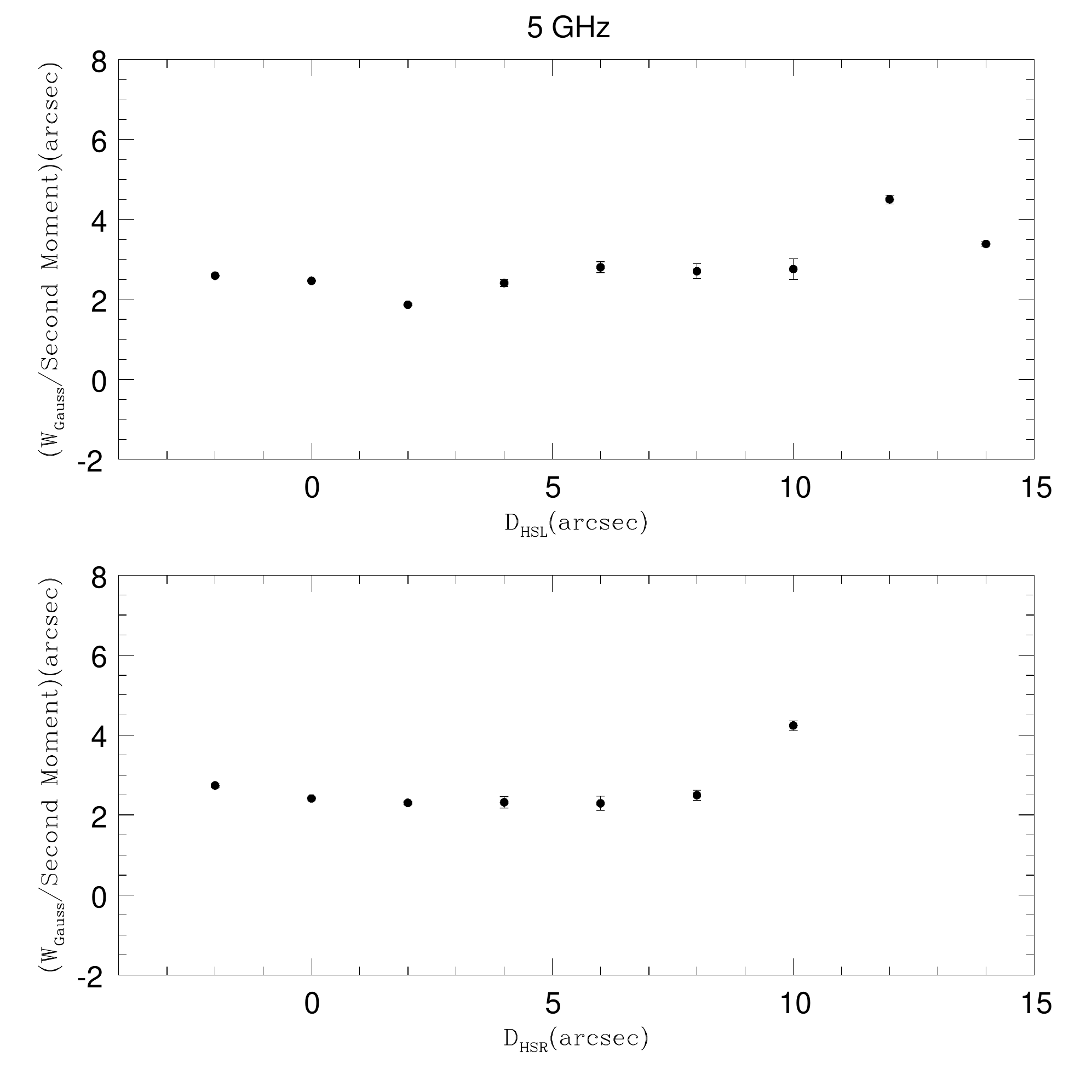}
\caption{3C6.1 ratio of the Gaussian FWHM to the second moment as a
function of distance from the hot spot at 1.4 and 5 GHz (left and
right panels, respectively) for the left and right hand sides of the 
source (top and bottom panels, respectively), as in Fig. \ref{3C6.1WG}.
The value of this ratio is fairly constant for each side of the source,
and has a value of about 2.5 to 3.  Similar results are obtained at
1.4 and 5 GHz.}
\label{3C6.1R}
\end{figure}

\begin{figure}
\plottwo{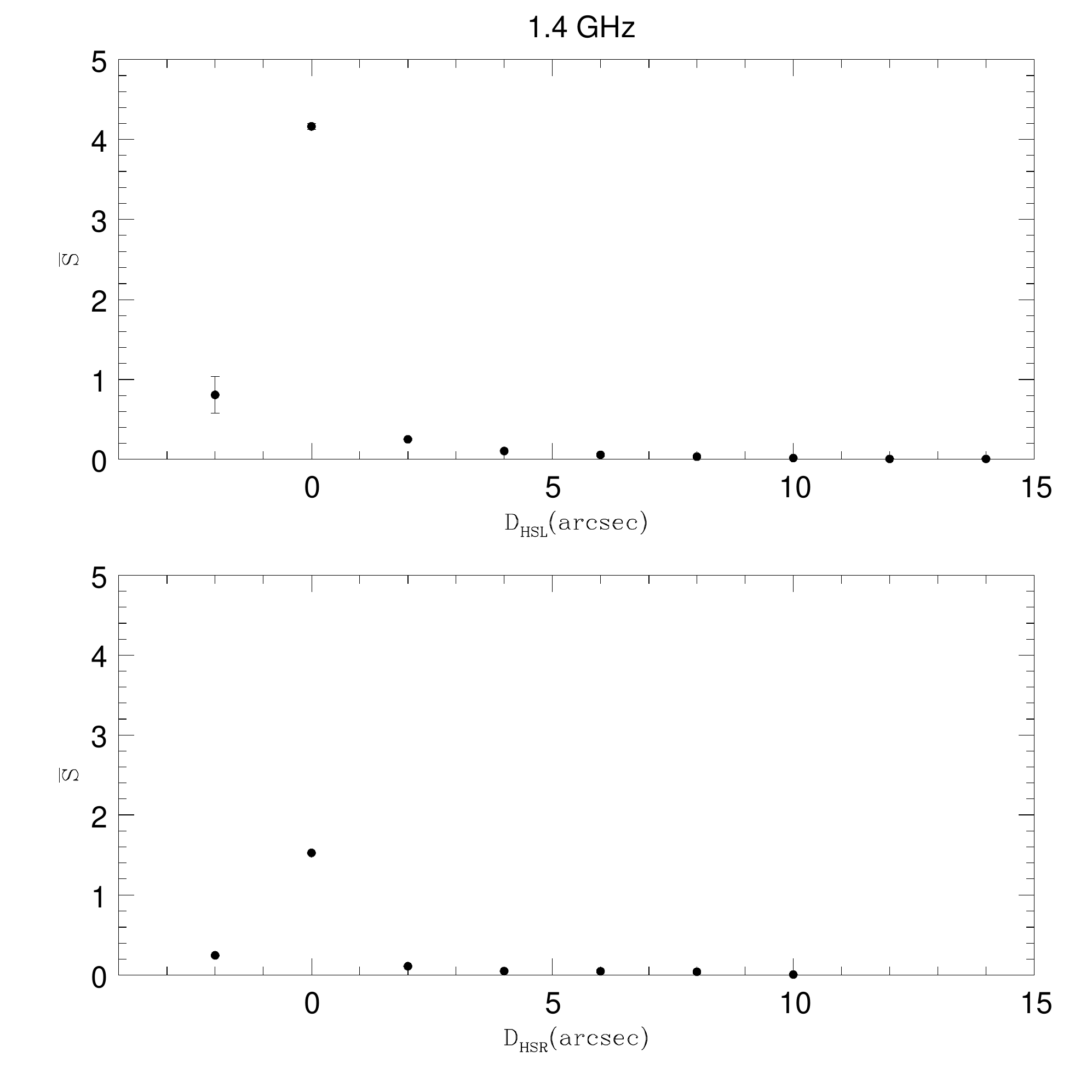}{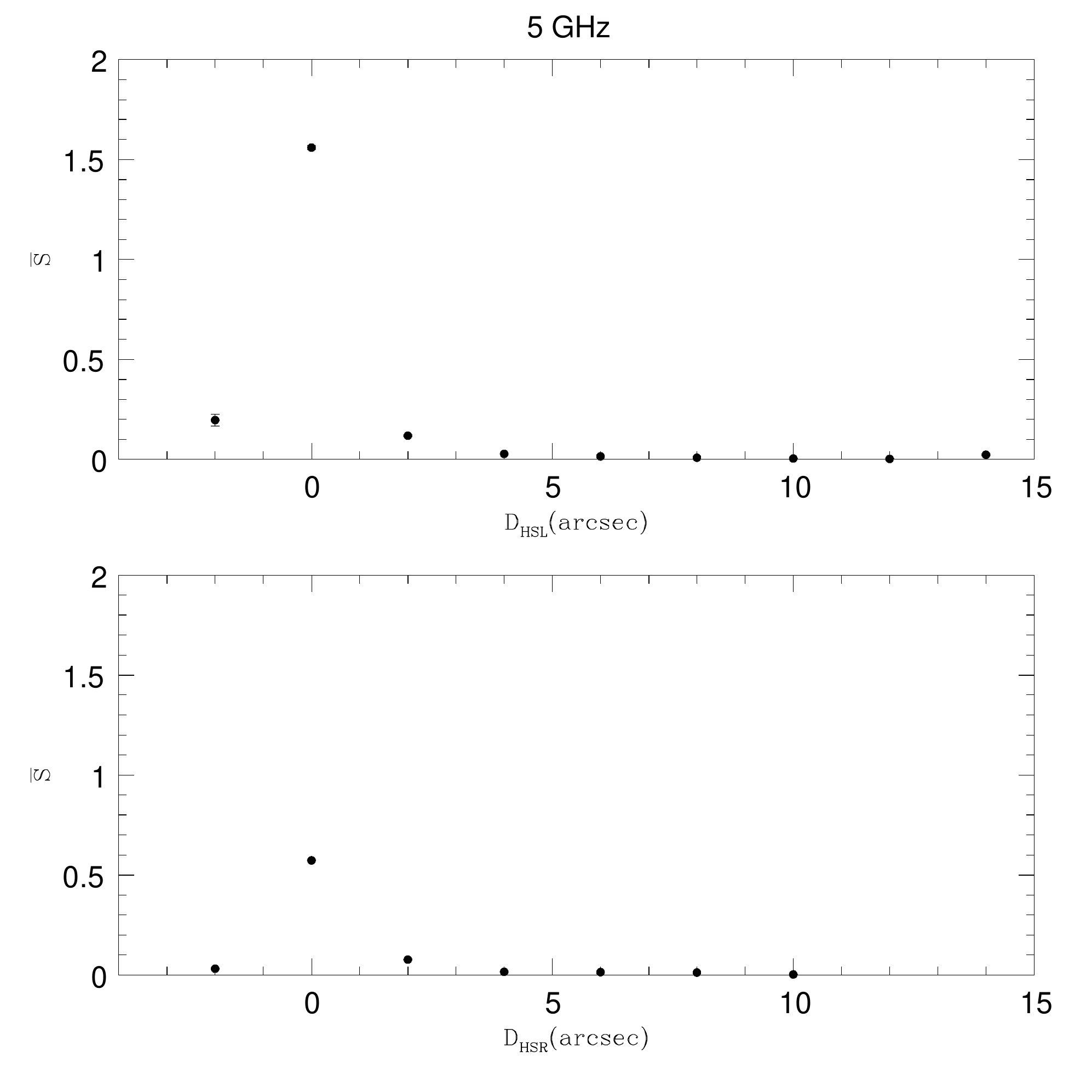}
\caption{3C6.1 average surface brightness in Jy/beam 
as a function of 
distance from the 
hot spot at 1.4 and 5 GHz (left and
right panels, respectively) for the left and right hand sides of the 
source (top and bottom panels, respectively), as in Fig. \ref{3C6.1WG}.
The brightness peak at each
hot spot is clearly visible.  
There is excellent agreement between results obtained at 1.4 GHz
and 5 GHz. }
\label{3C6.1S}
\end{figure}

\begin{figure}
\plottwo{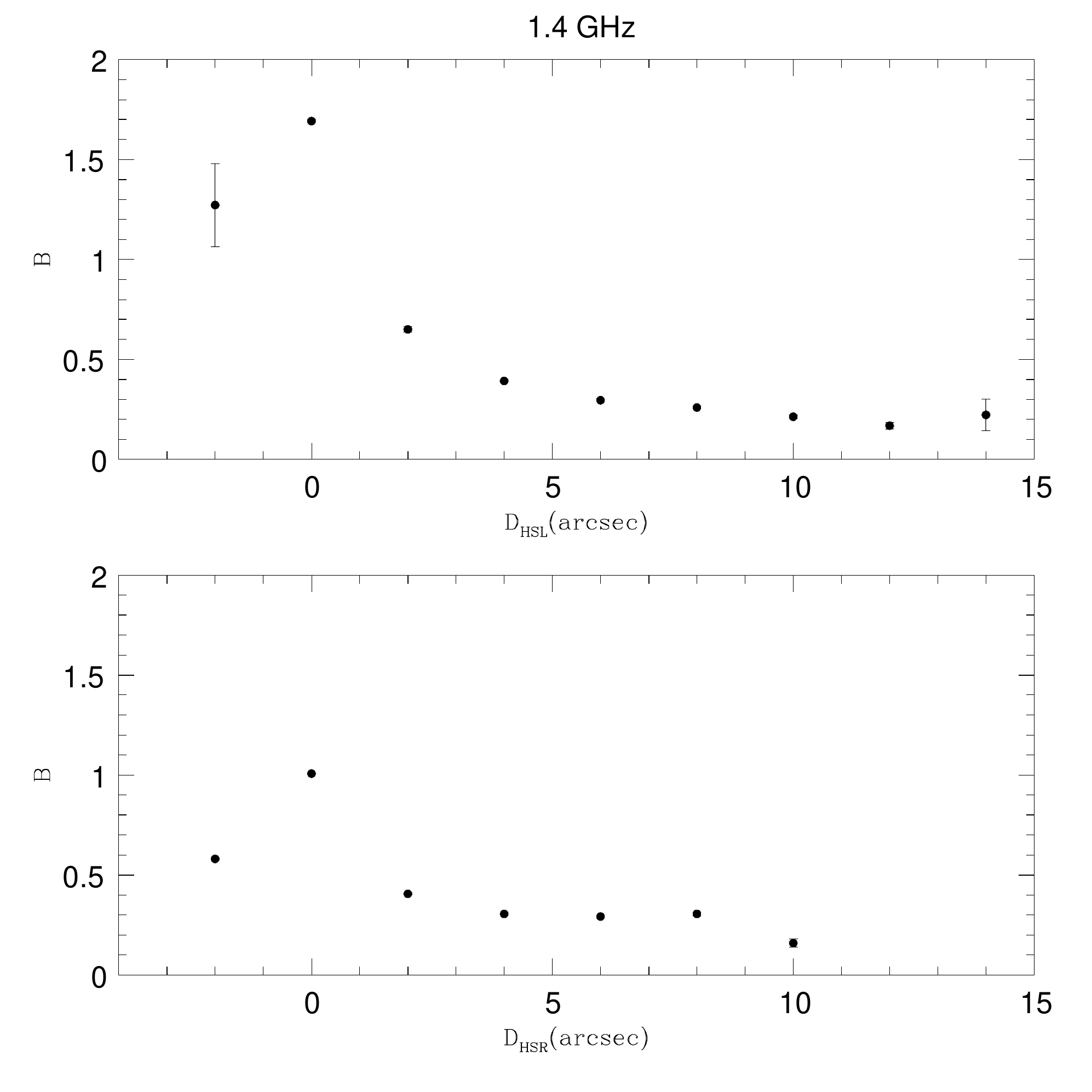}{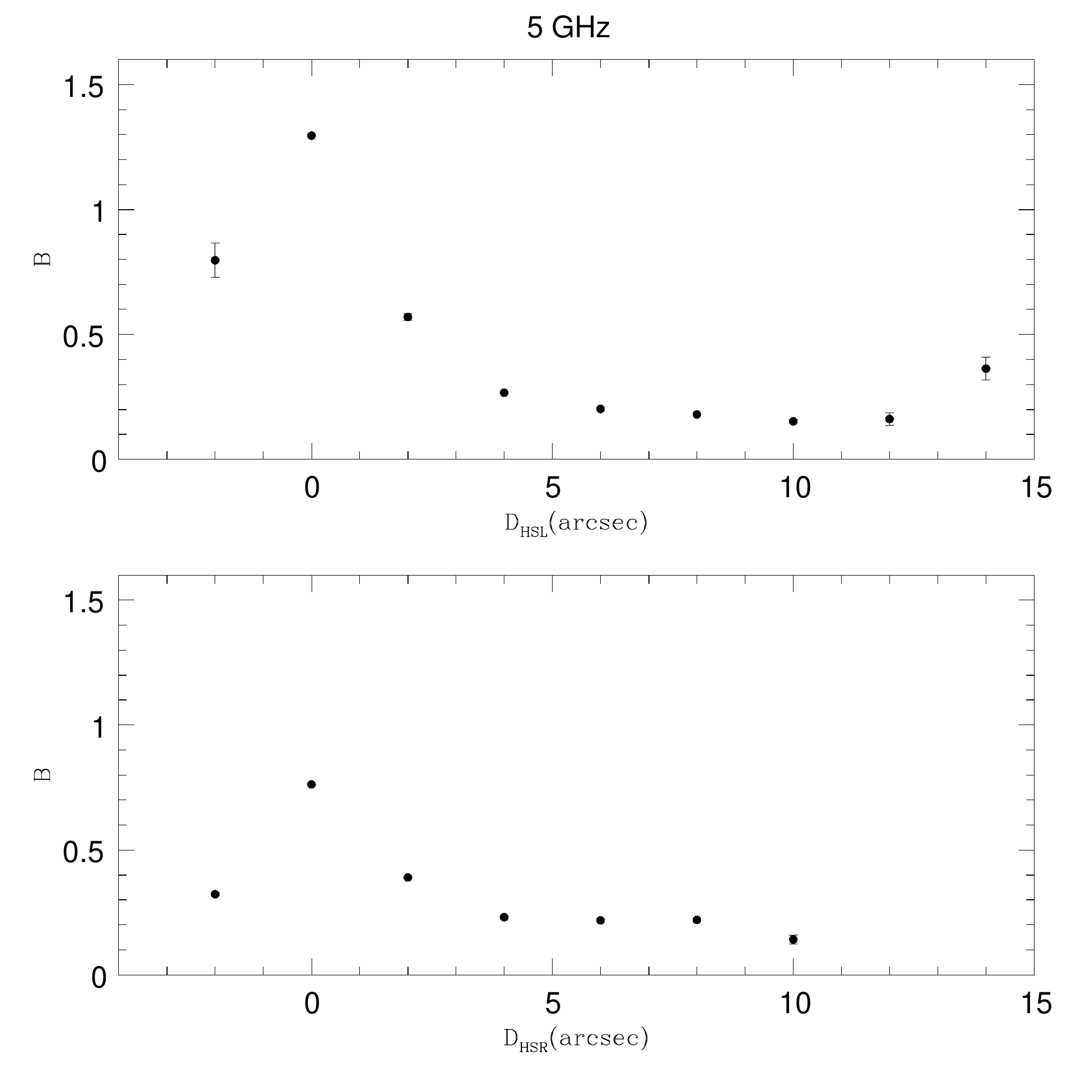}
\caption{3C6.1 minimum energy magnetic field strength as a function of 
distance from the 
hot spot at 1.4 and 5 GHz (left and
right panels, respectively) for the left and right hand sides of the 
source (top and bottom panels, respectively), as in Fig. \ref{3C6.1WG}.
Nearly identical results are obtained
at 1.4 and 5 GHz. The normalization of the magnetic field
strength is given in Table 1.}
\label{3C6.1B}
\end{figure}

\begin{figure}
\plottwo{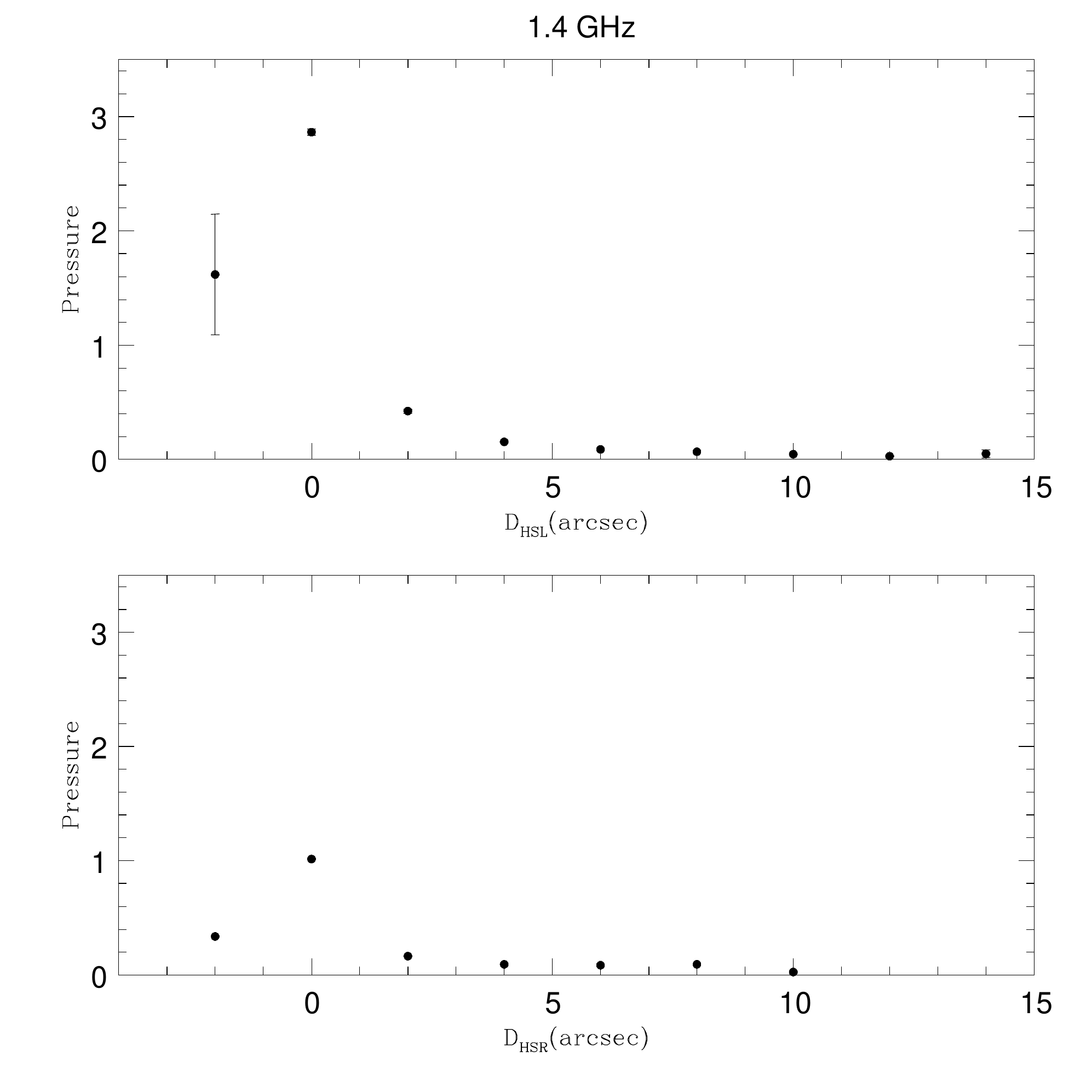}{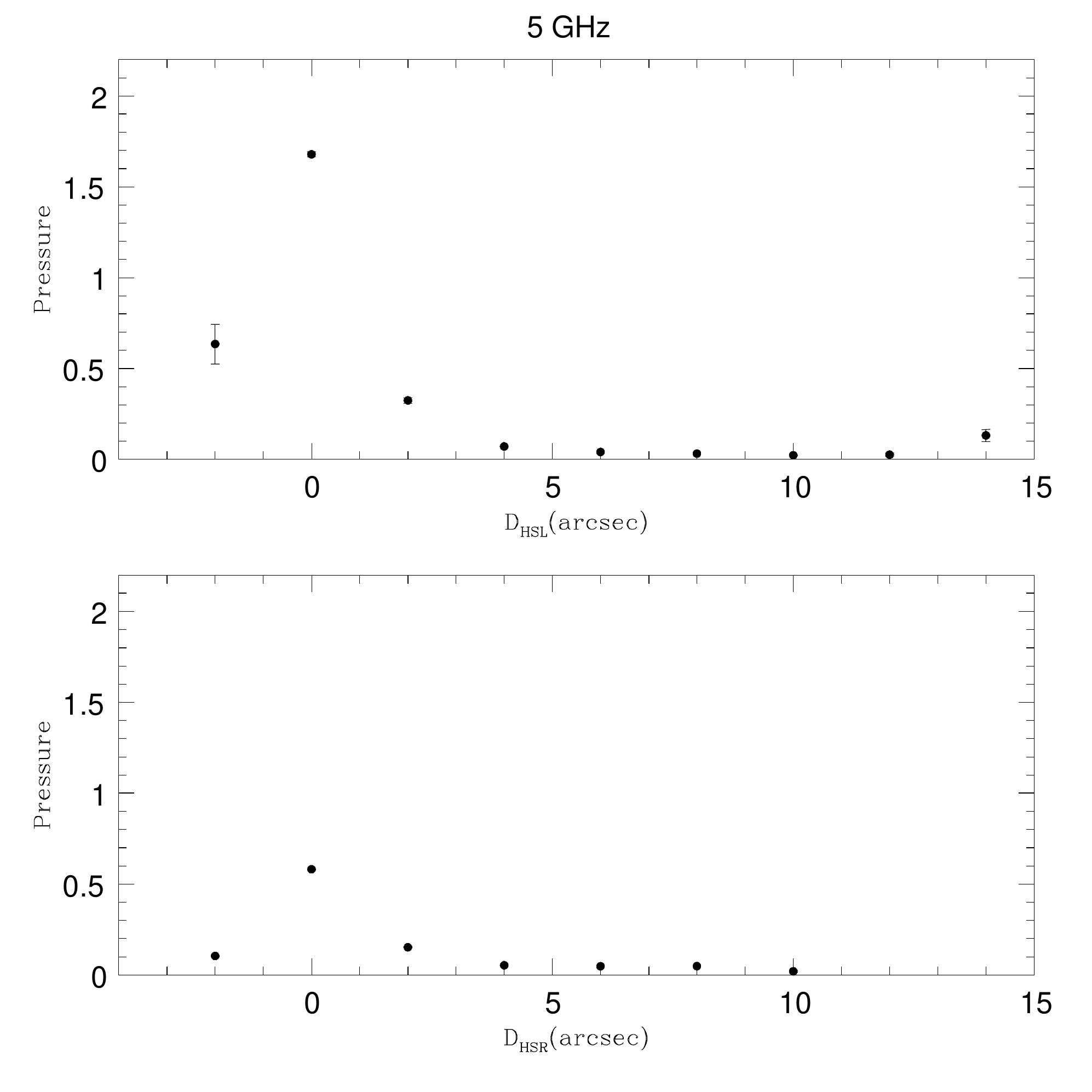}
\caption{3C6.1 minimum energy pressure as a function of 
distance from the 
hot spot at 1.4 and 5 GHz (left and
right panels, respectively) for the left and right hand sides of the 
source (top and bottom panels, respectively), as in Fig. \ref{3C6.1B}.
Nearly identical results are obtained
at 1.4 and 5 GHz. The normalization of the pressure is given 
in Table 1.}
\label{3C6.1P}
\end{figure}

\clearpage
\begin{figure}
\plotone{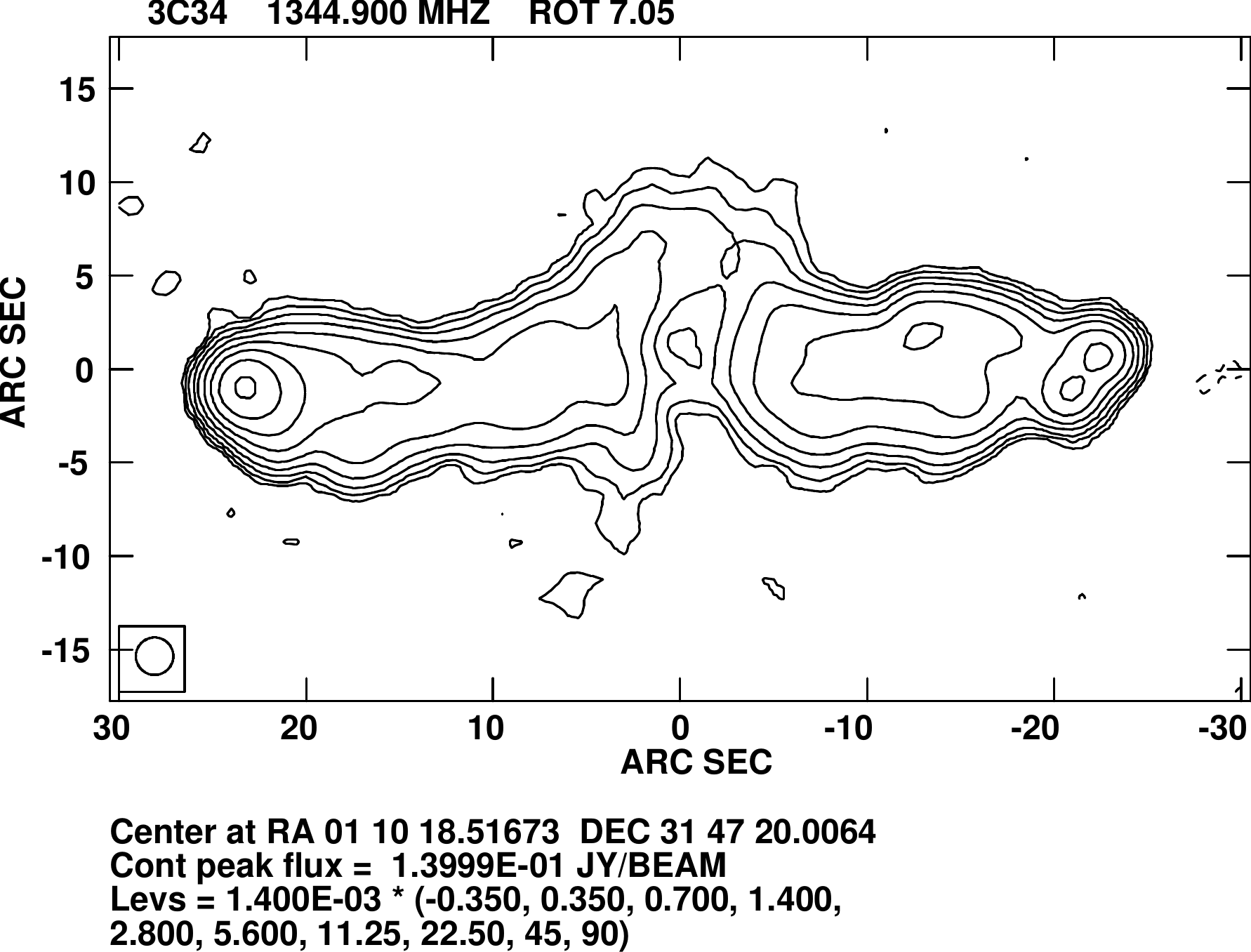}
\caption{3C34 at 1.4 GHz and 2'' resolution rotated by 7.05 deg 
counter-clockwise, as in Fig. \ref{3C6.1rotfig}. }
\end{figure}

\begin{figure}
\plottwo{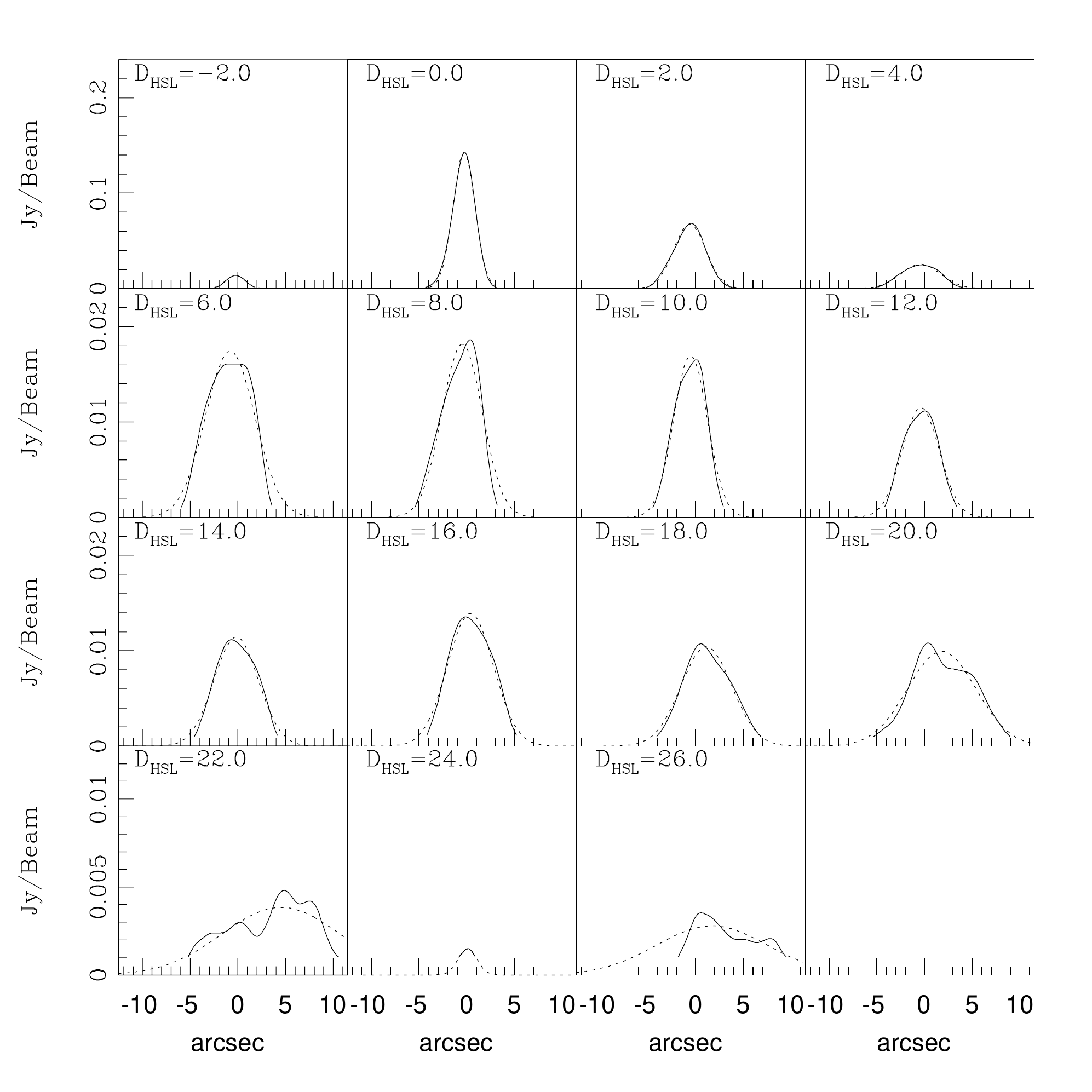}{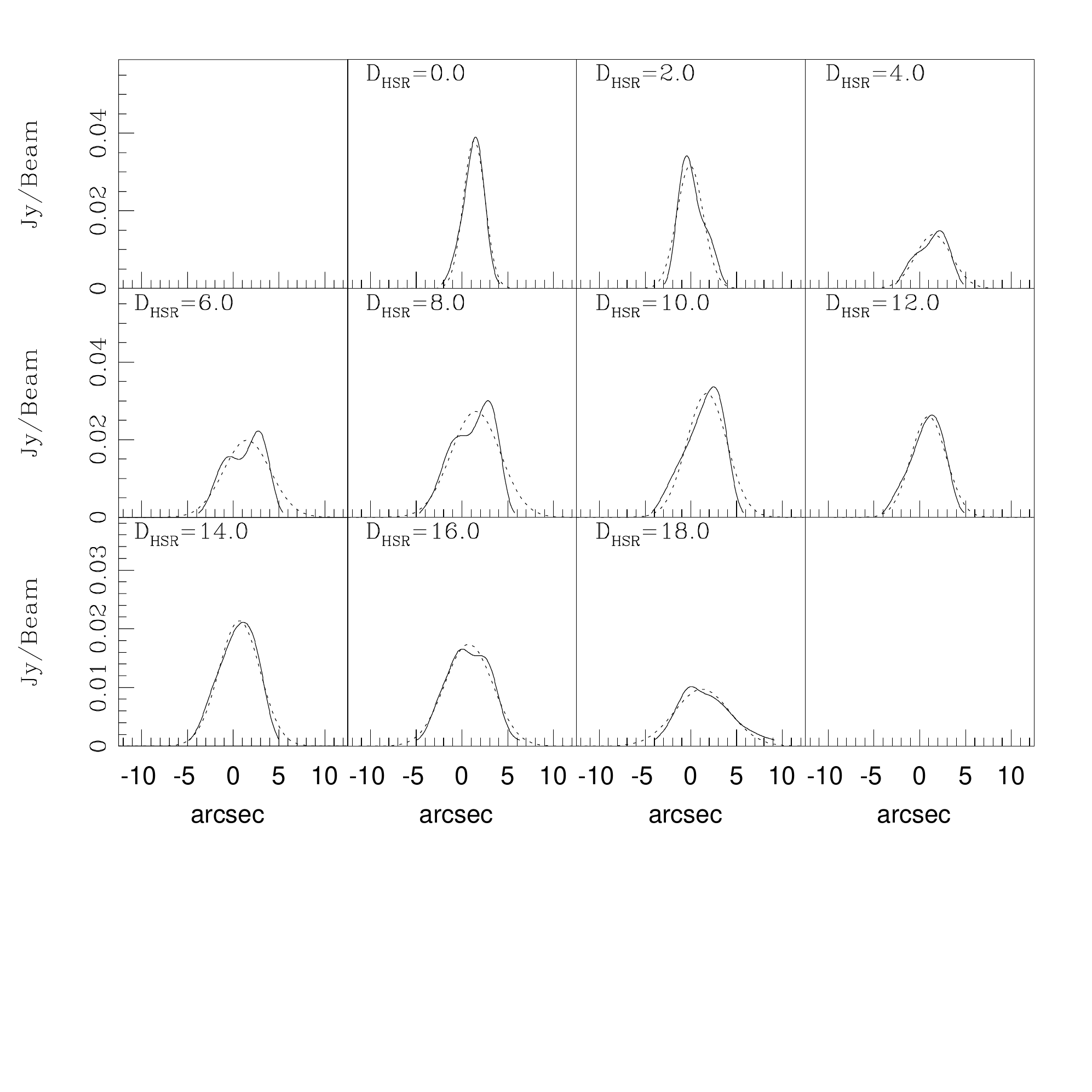}
\caption{3C34 surface brightness (solid line) and best fit 
Gaussian (dotted line) for each cross-sectional slice of the 
radio bridge at 1.4 GHz, as in Fig. \ref{3C6.1SG}.
A Gaussian provides an excellent description of the surface brightness
profile of each cross-sectional slice. }
\label{3C34SG}
\end{figure}

\begin{figure}
\plottwo{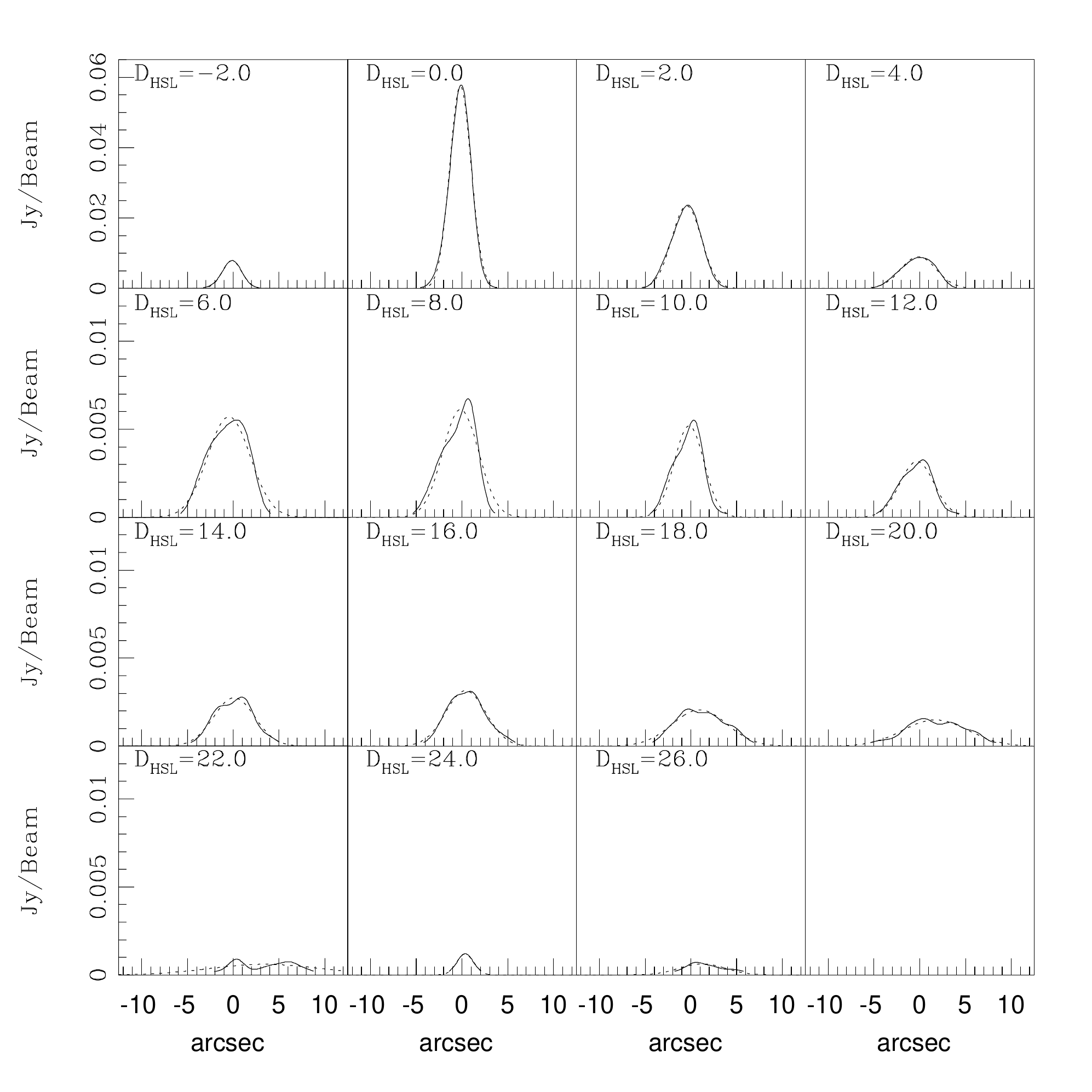}{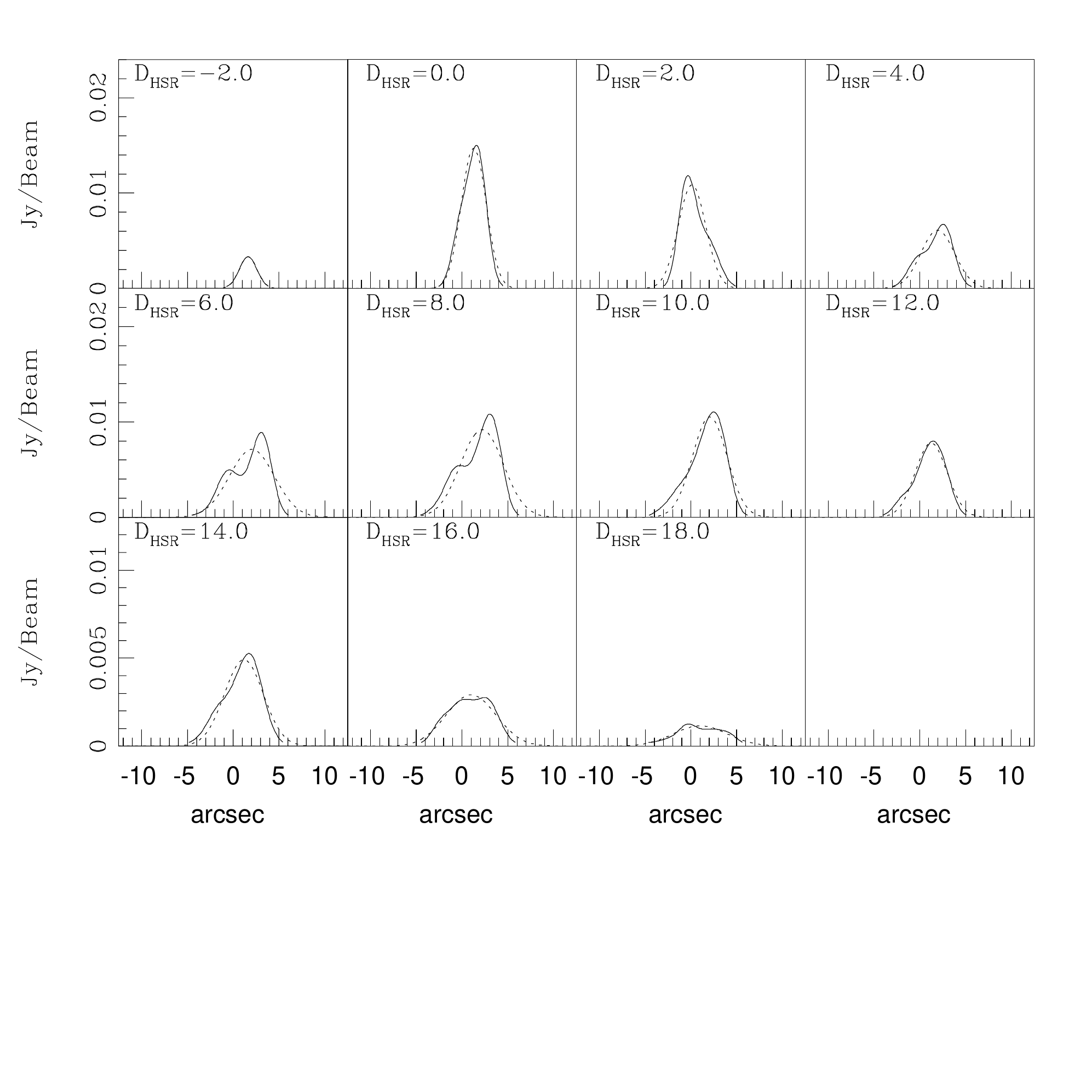}
\caption{3C34 surface brightness (solid line) and best fit 
Gaussian (dotted line) for each cross-sectional slice of the 
radio bridge at 5 GHz, as in 
Fig. \ref{3C34SG} but at 5 GHz. 
Results obtained at 5 GHz are nearly identical to those
obtained at 1.4 GHz.}
\end{figure}

\begin{figure}
\plottwo{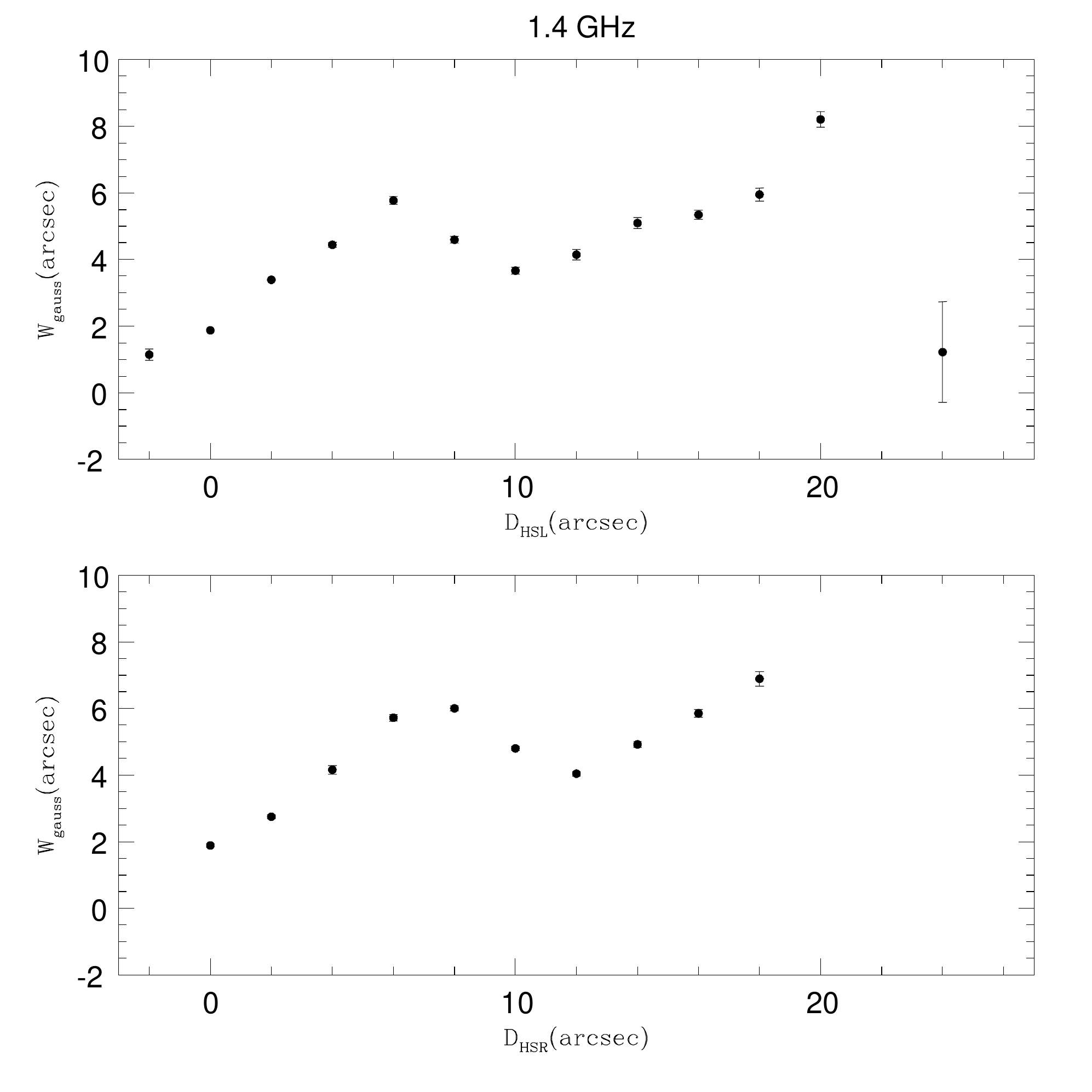}{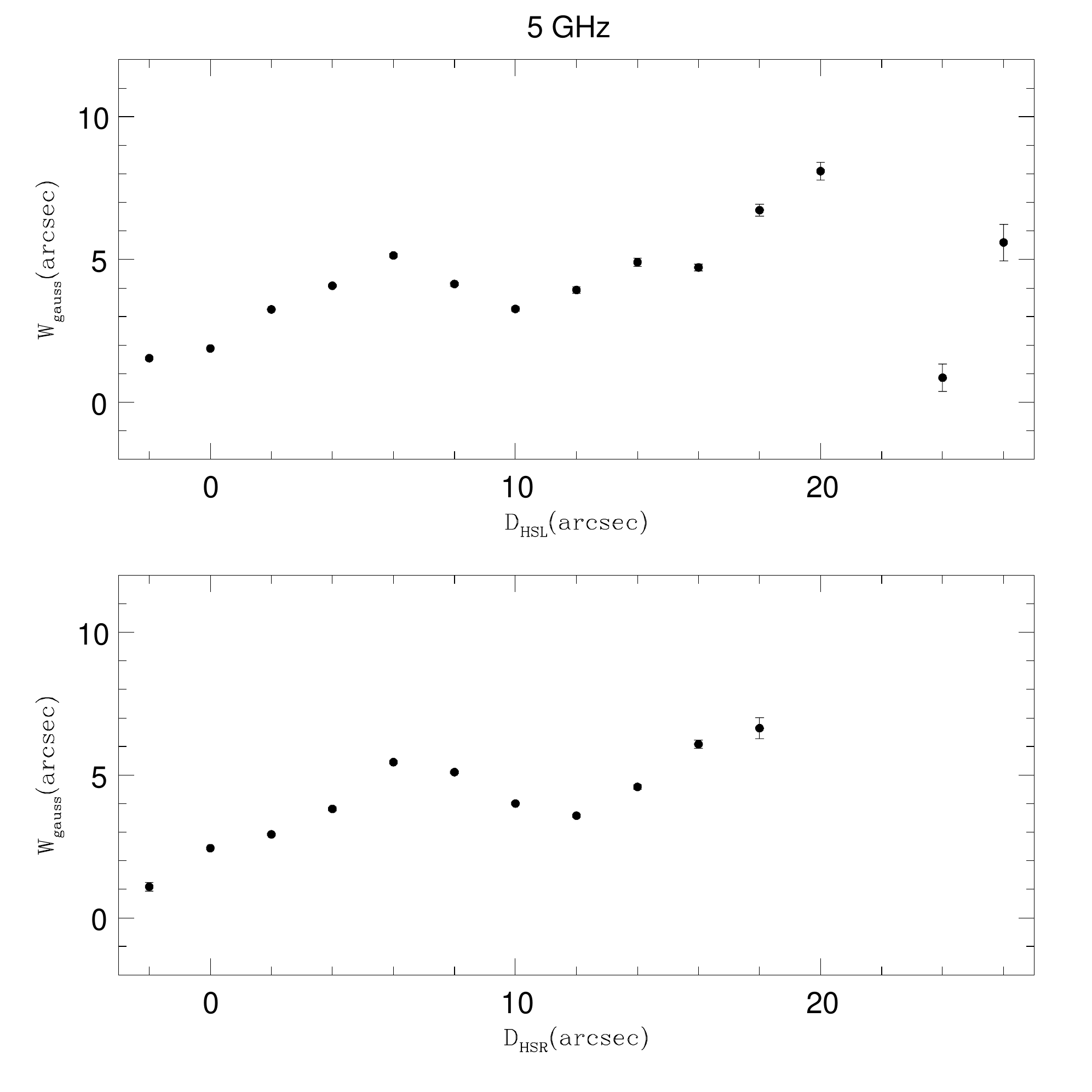}
\caption{3C34 Gaussian FWHM as a function of 
distance from the hot spot  
at 1.4 and 5 GHz (left and
right panels, respectively) for the left and right hand sides of the 
source (top and bottom panels, respectively), as in Fig. \ref{3C6.1WG}.
The right and left hand sides of the source are almost
perfectly symmetric.  Similar results are obtained at 
1.4 and 5 GHz. }
\label{3C34WG}
\end{figure}

\begin{figure}
\plottwo{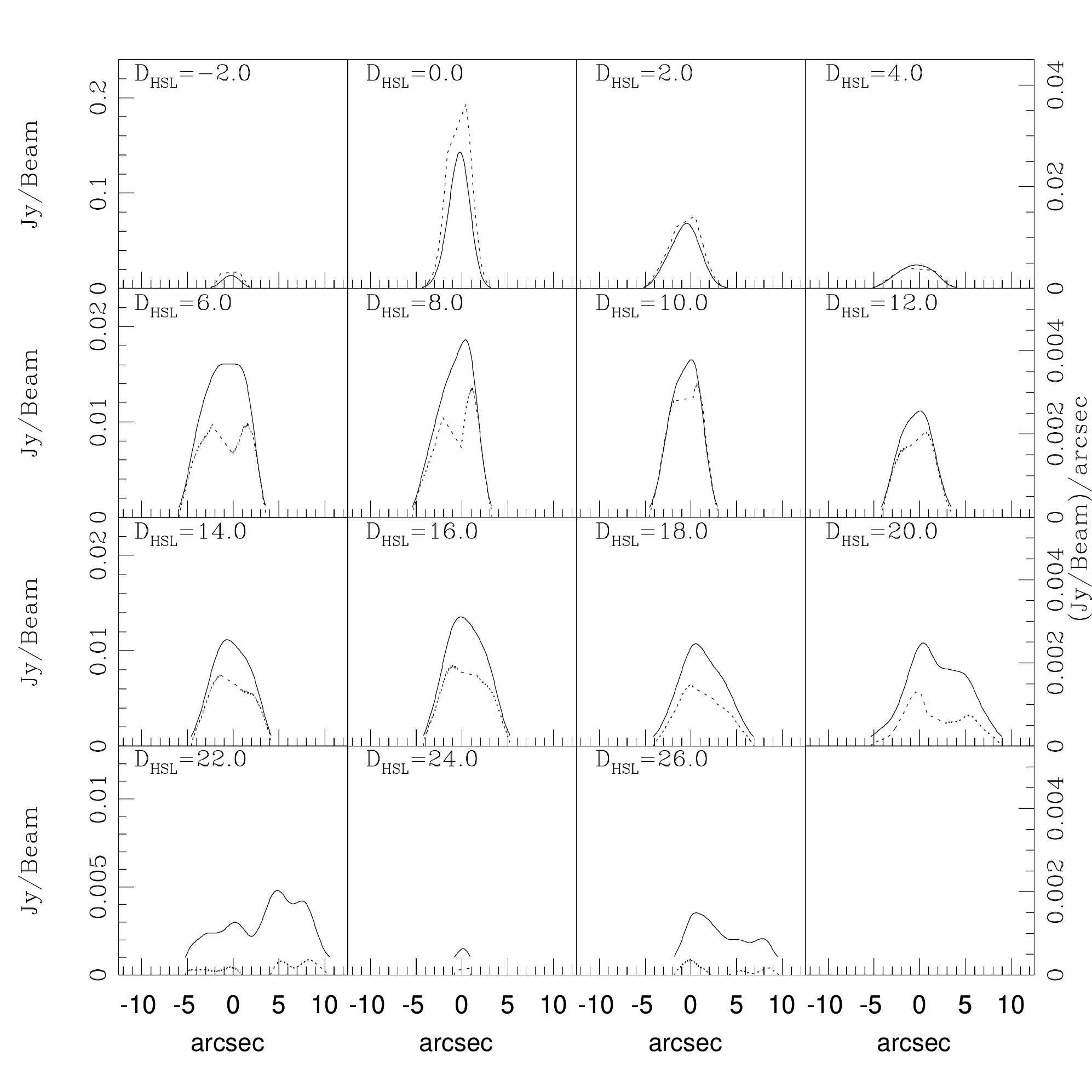}{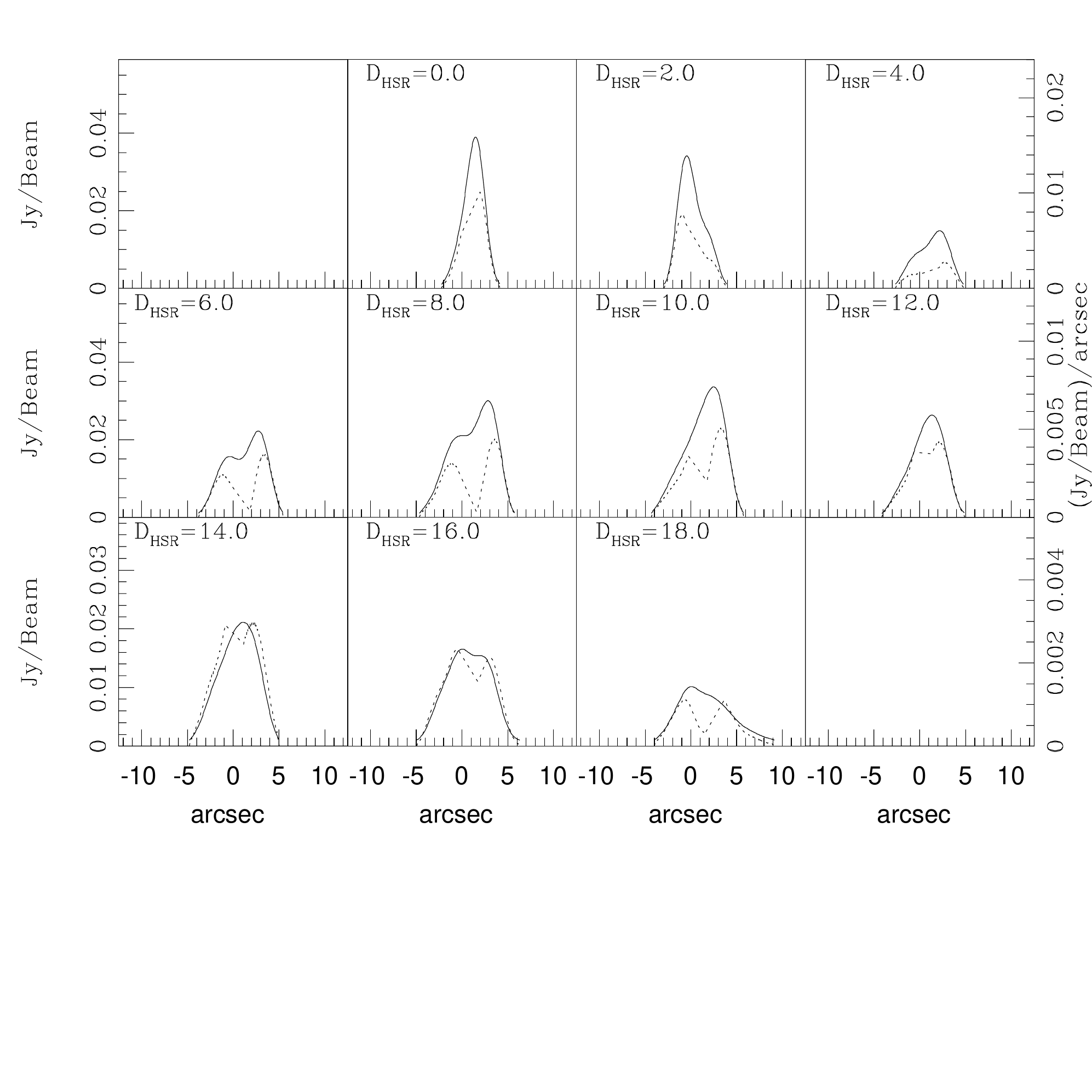}
\caption{3C34 emissivity (dotted line) and surface brightness
(solid line) for each cross-sectional slice of the radio 
bridge at 1.4 GHz, as in Fig. \ref{3C6.1SEM}. A constant volume
emissivity per slice provides a reasonable description of the 
data over most of the source. }
\label{3C34SEM}
\end{figure}

\begin{figure}
\plottwo{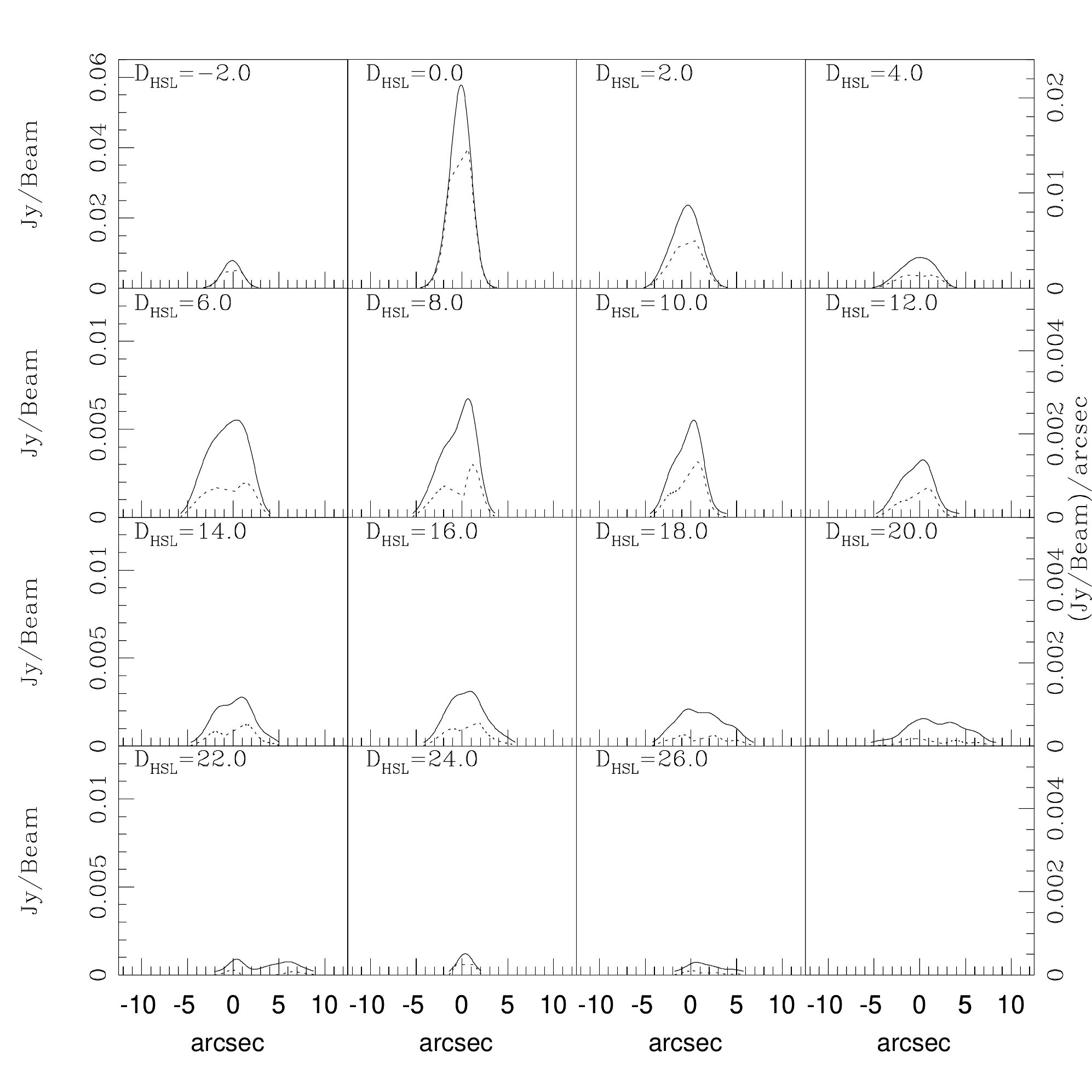}{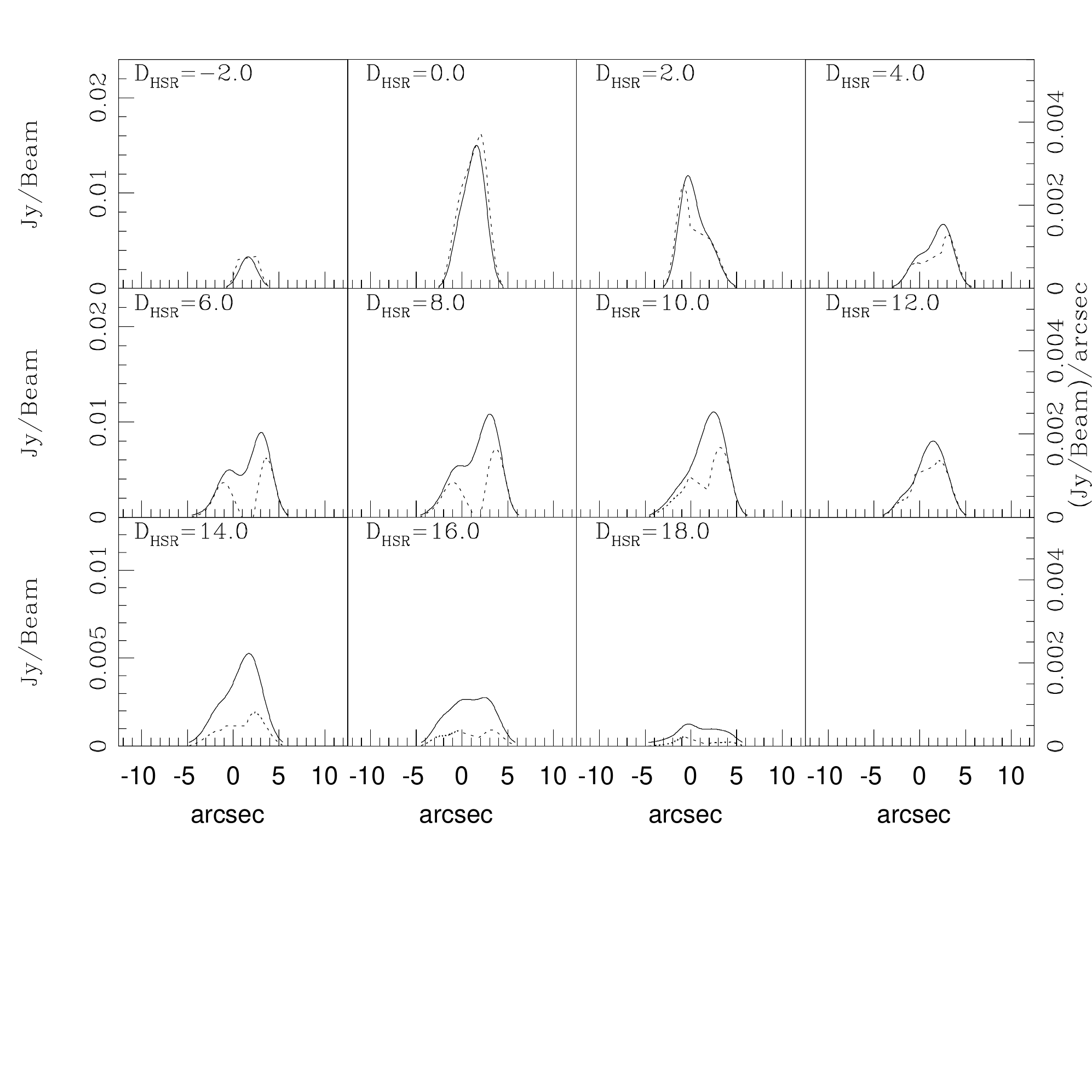}
\caption{3C34 
emissivity (dotted line) and surface brightness
(solid line) for each cross-sectional slice of the radio 
bridge at 5 GHz, as in Fig. \ref{3C34SEM} but at 5 GHz.
Results obtained at 5 GHz are nearly identical to those
obtained at 1.4 GHz.}
\end{figure}

\begin{figure}
\plottwo{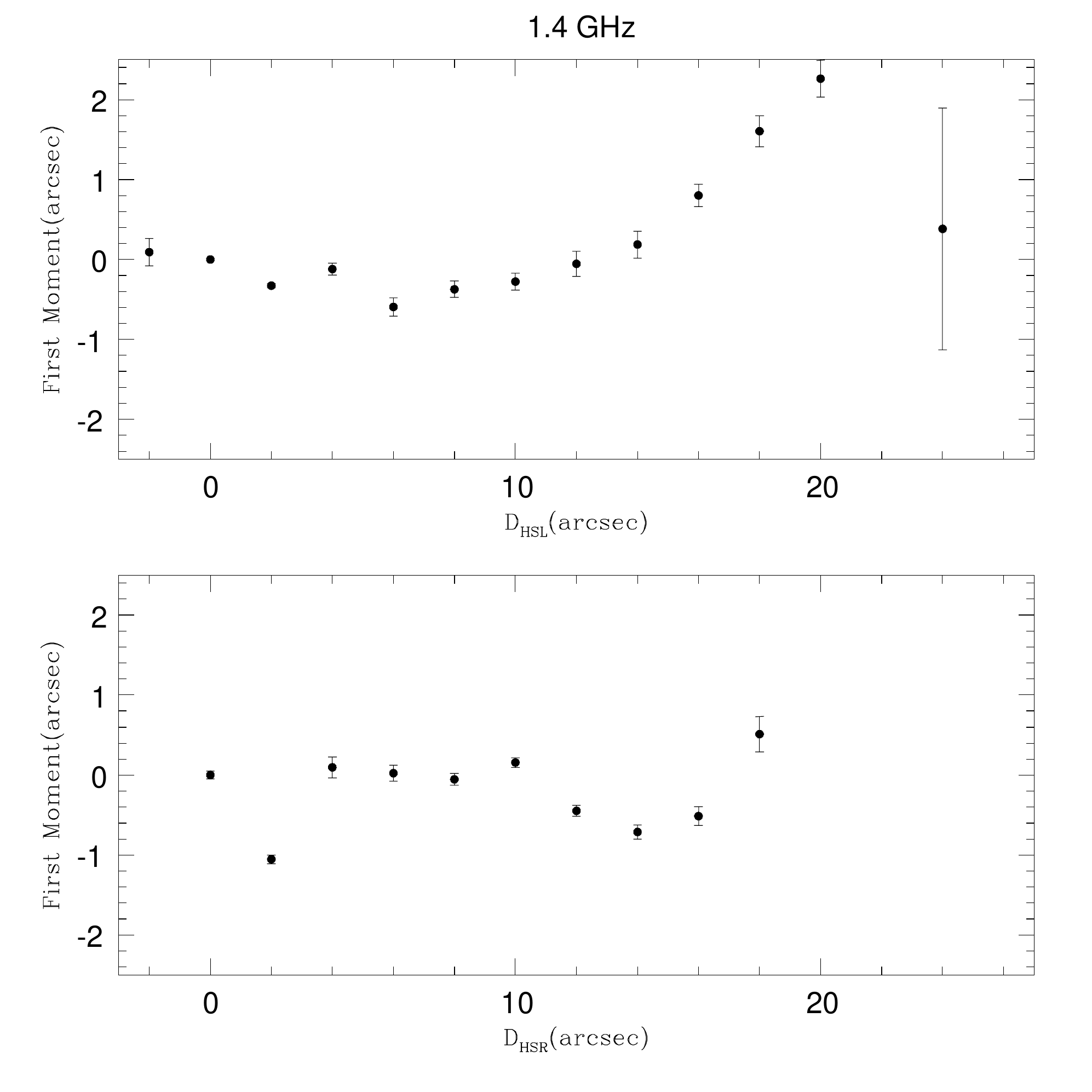}{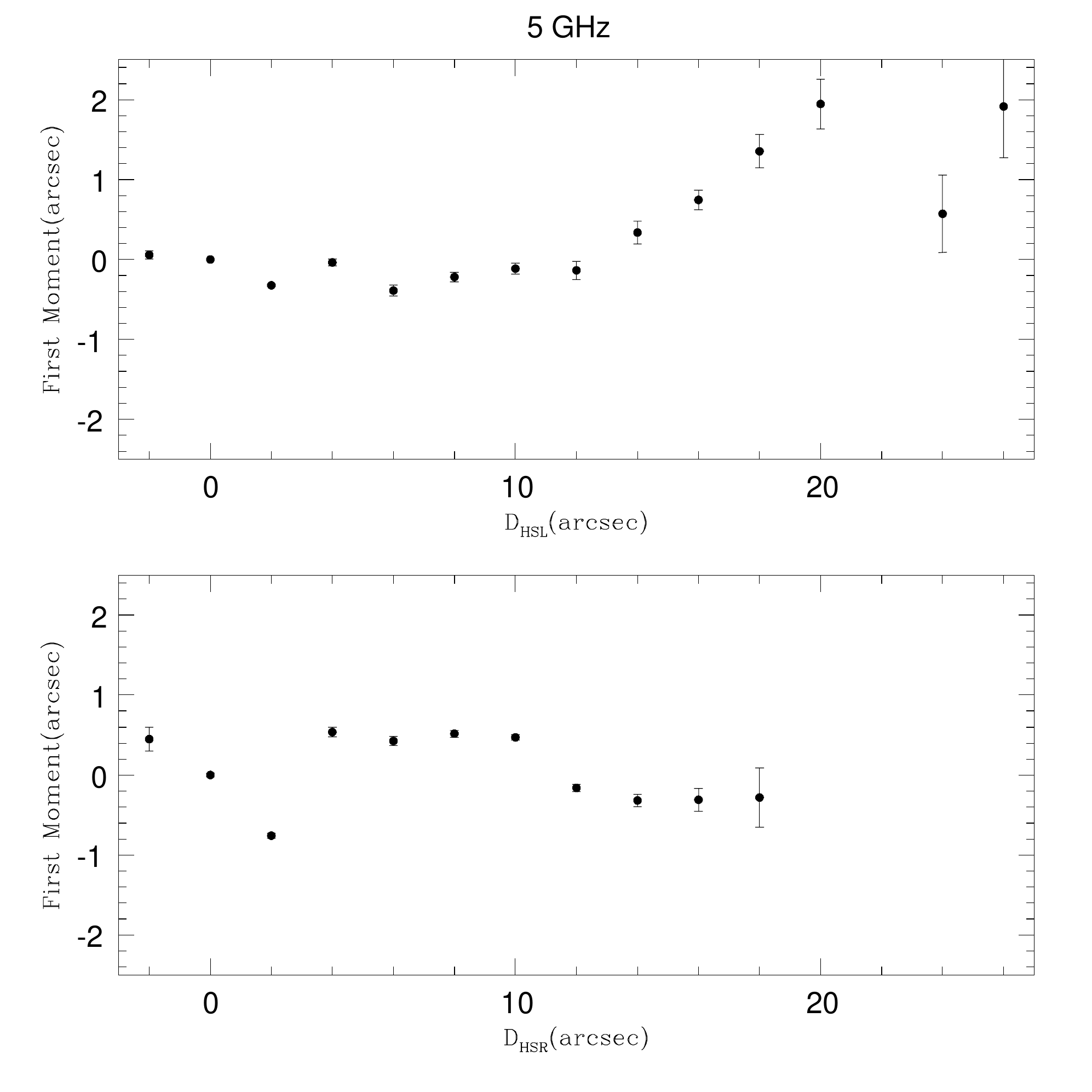}
\caption{3C34 first moment as a function of distance from 
the hot spot at 1.4 GHz and 5 GHz (left and right panels,
respectively) for the left and right hand sides of the source
(top and bottom, respectively), as in Fig. \ref{3C6.1FM}. 
A side flow with an angular size of a few arc seconds 
is seen on the left hand side of the source
near the location of the AGN. This represents the ``oldest''
radio emitting plasma in the source, and could be due to the 
motion of the source relative to the ambient gas, or a 
back flow of plasma into the ambient gas, as discussed, 
for example, by \citet{LW84} and \citet{LMS89}. 
Similar results are obtained at 1.4 and 5 GHz. }
\end{figure}

\begin{figure}
\plottwo{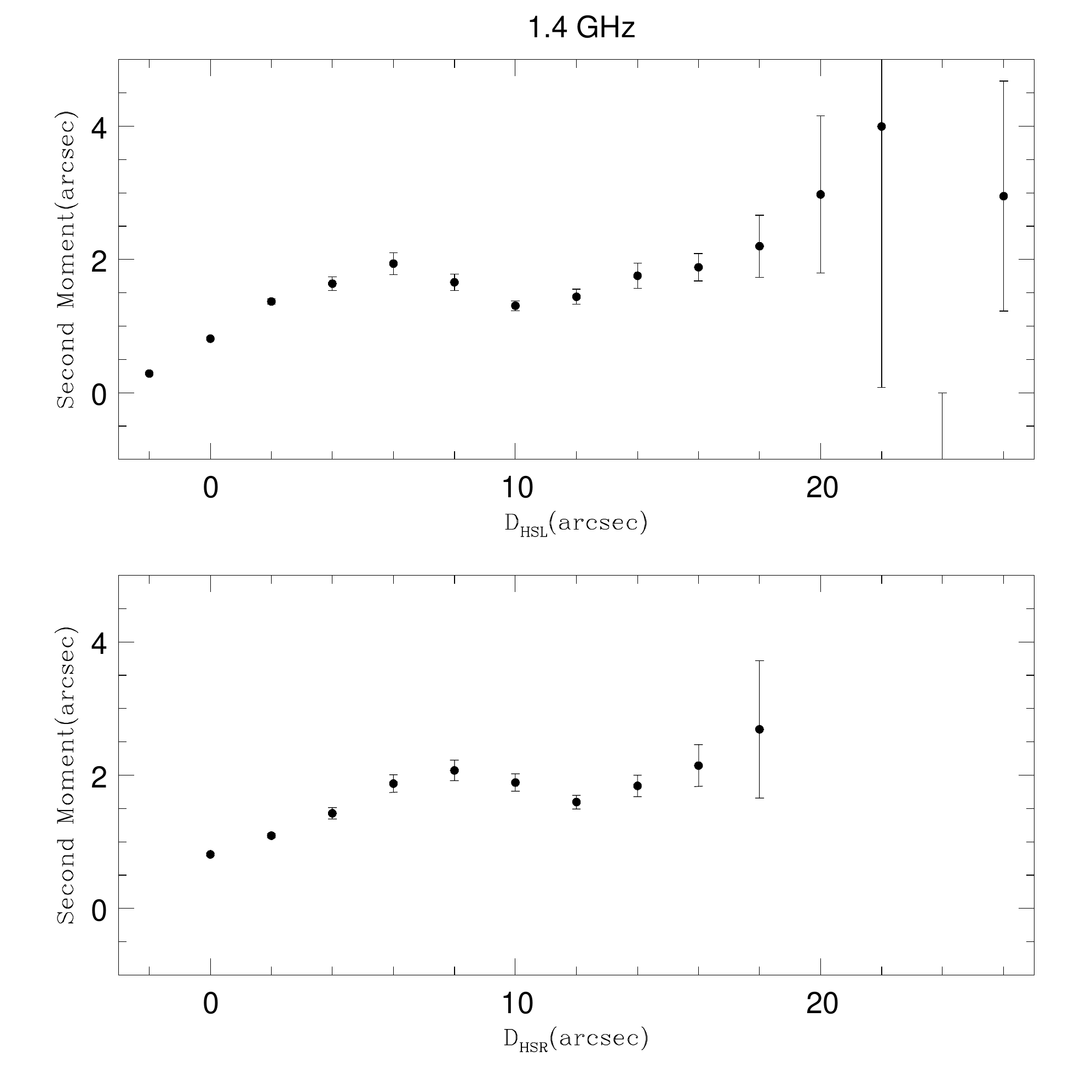}{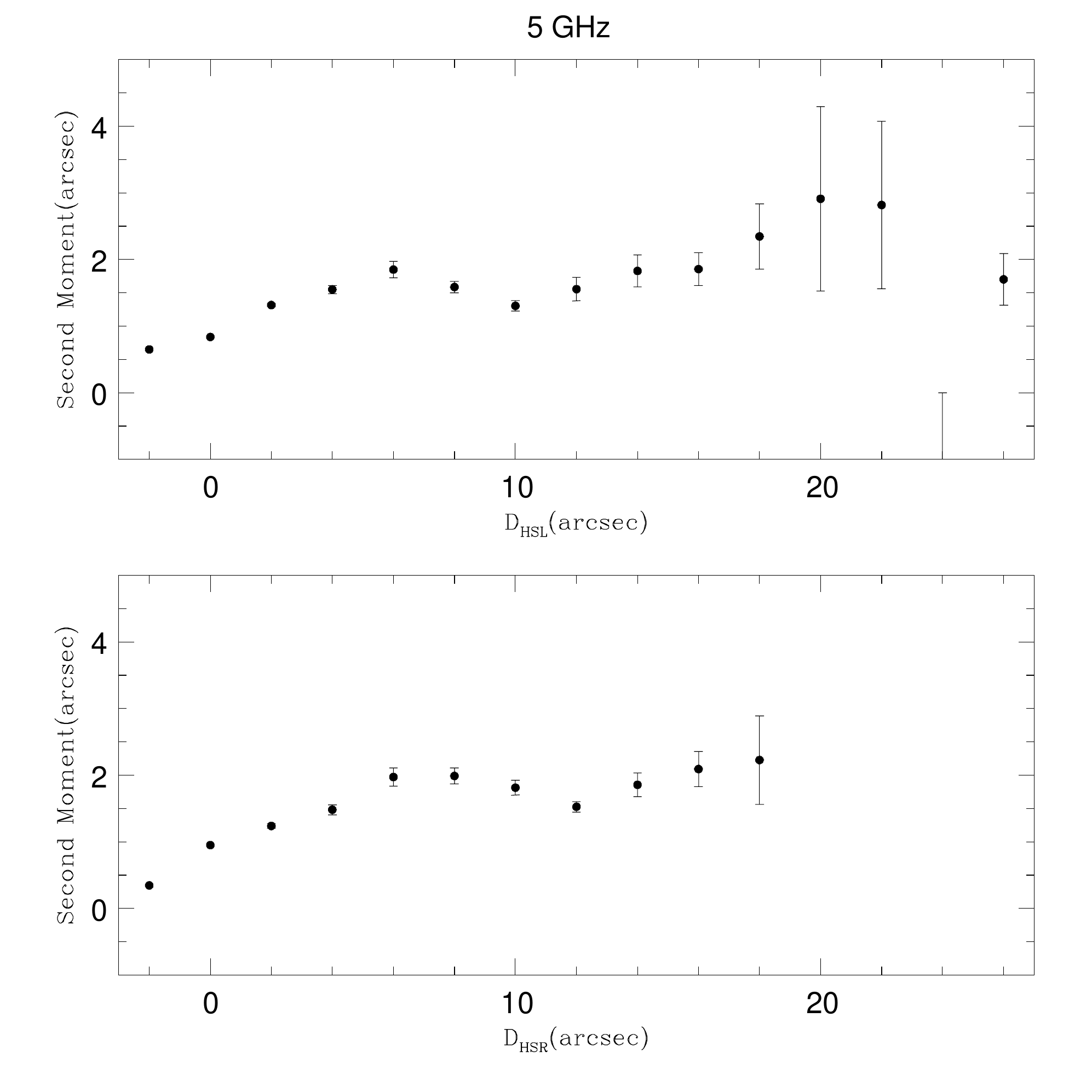}
\caption{3C34 second moment as a function of distance from the
hot spot at 1.4 GHz and 5 GHz (left and right panels, 
respectively) for the left and right sides of the source
(top and bottom panels, respectively), as in Fig. \ref{3C6.1SM}.
The right and left hand sides of the source are highly symmetric,
as seen in Fig. \ref{3C34WG}.  Similar results are obtained at
1.4 and 5 GHz. }
\end{figure}

\begin{figure}
\plottwo{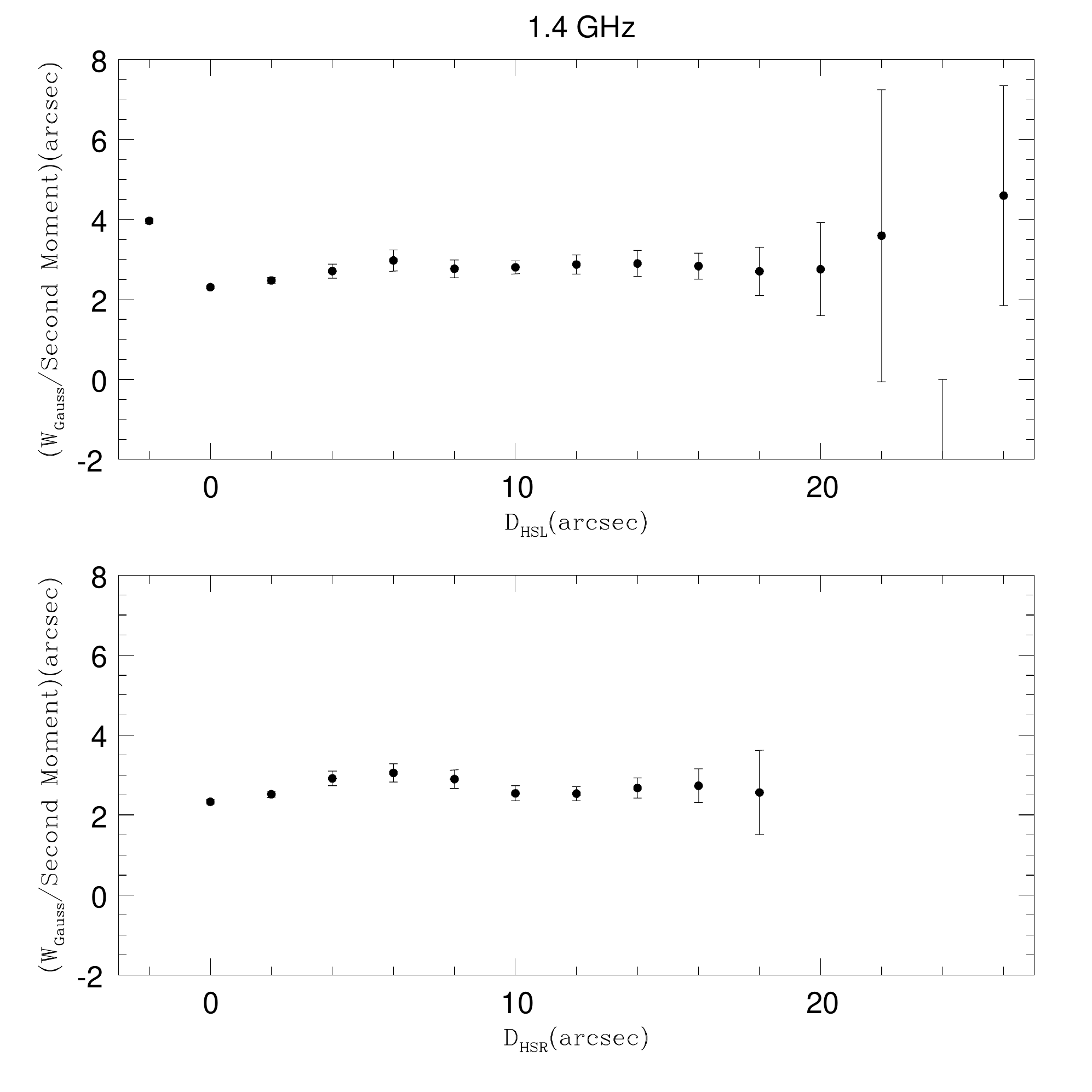}{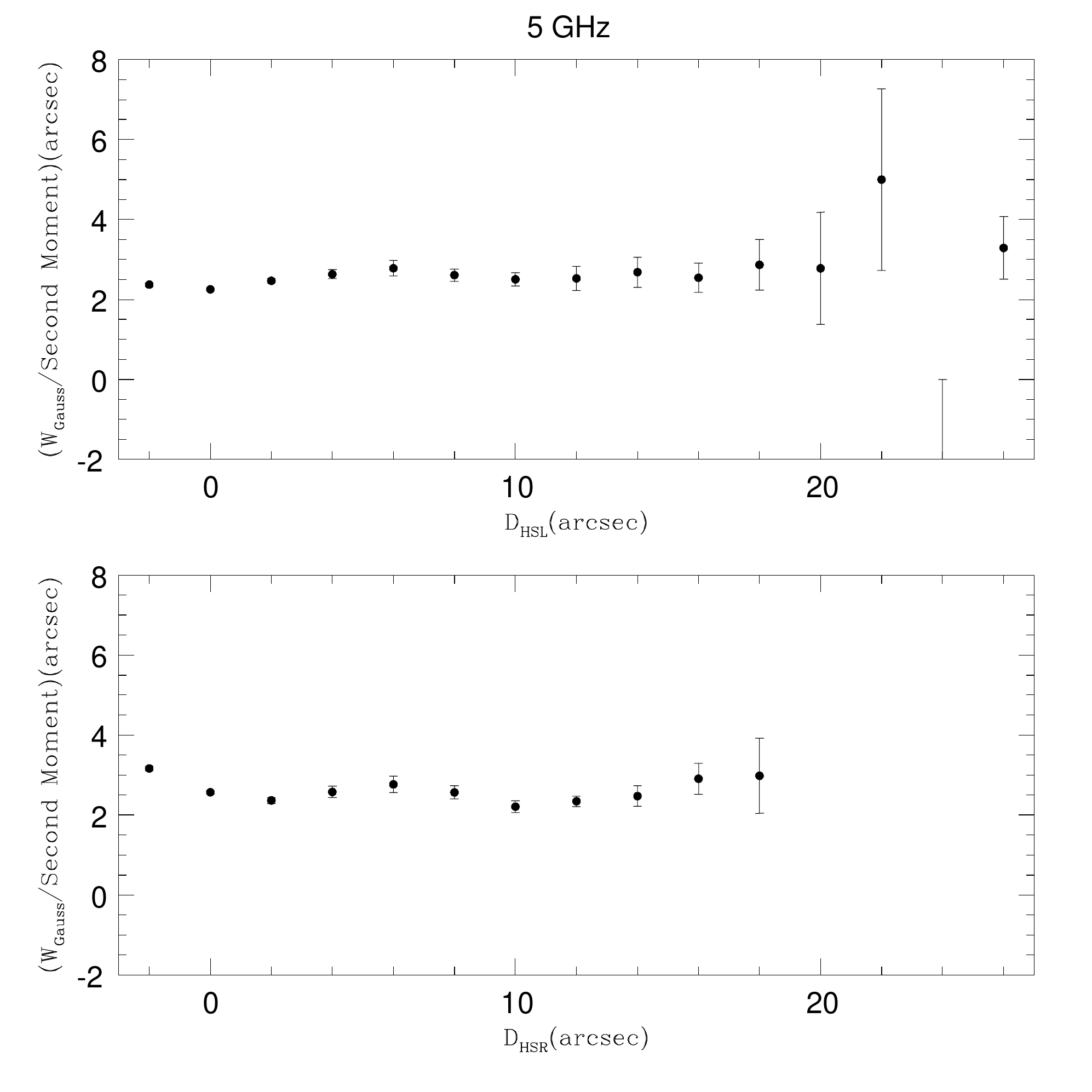}
\caption{3C34 ratio of the Gaussian FWHM to the second moment as a
function of distance from the hot spot at 1.4 and 5 GHz 
(left and right panels, respectively) for the left and right
hand sides of the source (top and bottom panels, respectively), 
as in Fig. \ref{3C6.1R}.  The value of this ratio is fairly constant
for each side of the source, and has a value of about 2.  Similar 
results are obtained at 1.4 and 5 GHz. }
\end{figure}

\begin{figure}
\plottwo{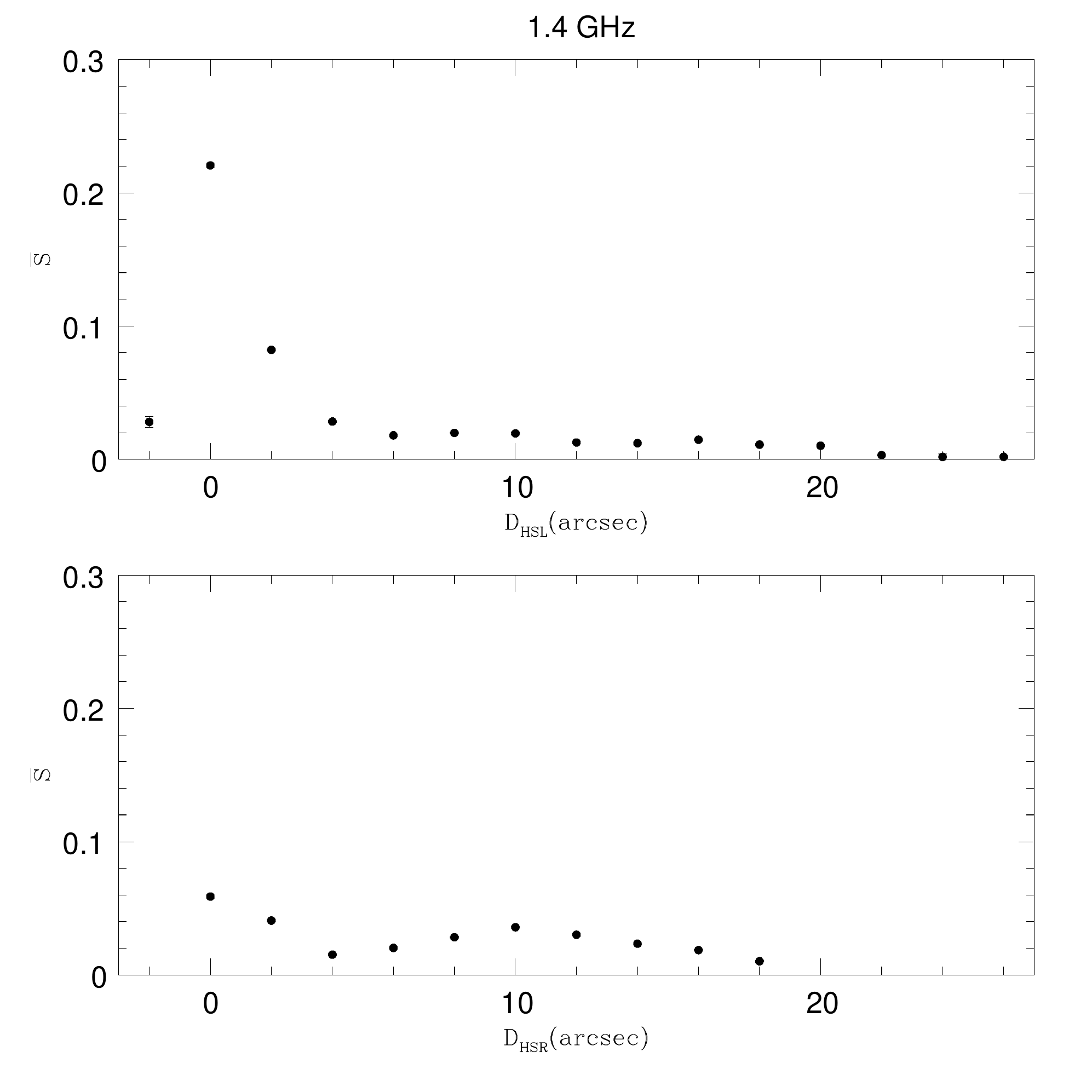}{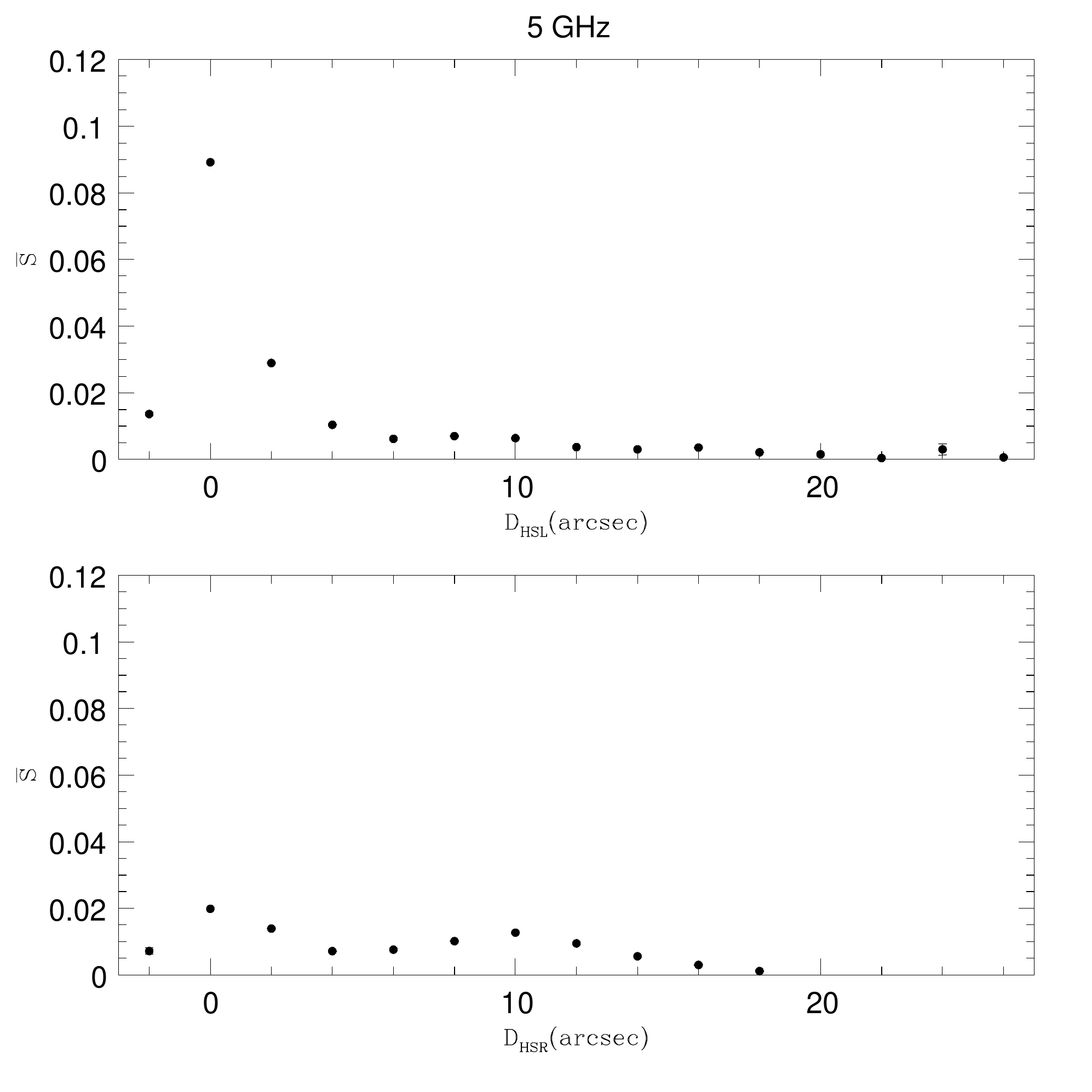}
\caption{3C34 average surface brightness in units of Jy/beam 
as a function of distance from the 
hot spot at 1.4 and 5 GHz 
(left and right panels, respectively) for the left and right
hand sides of the source (top and bottom panels, respectively), 
as in Fig. \ref{3C6.1S}.  }
\end{figure}

\begin{figure}
\plottwo{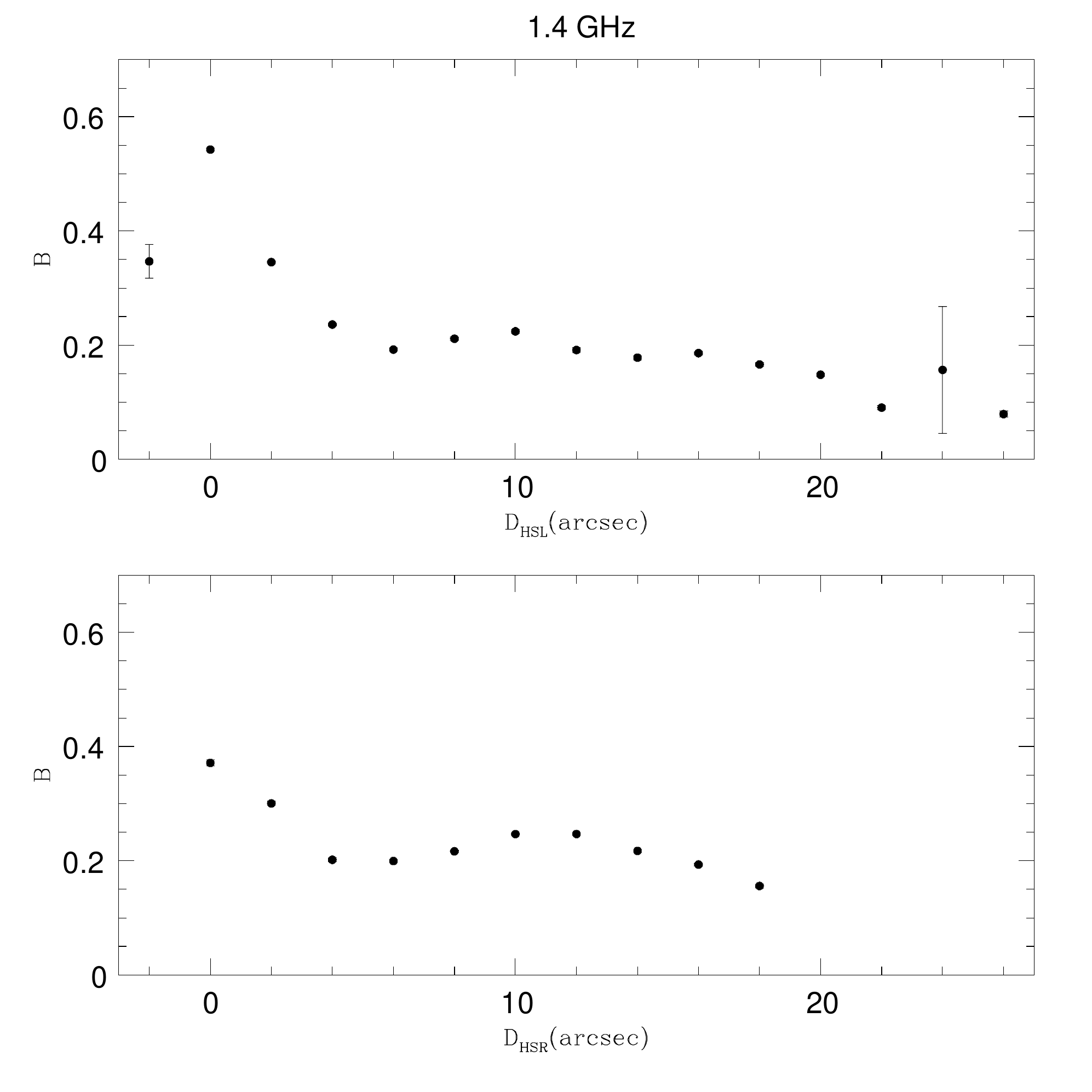}{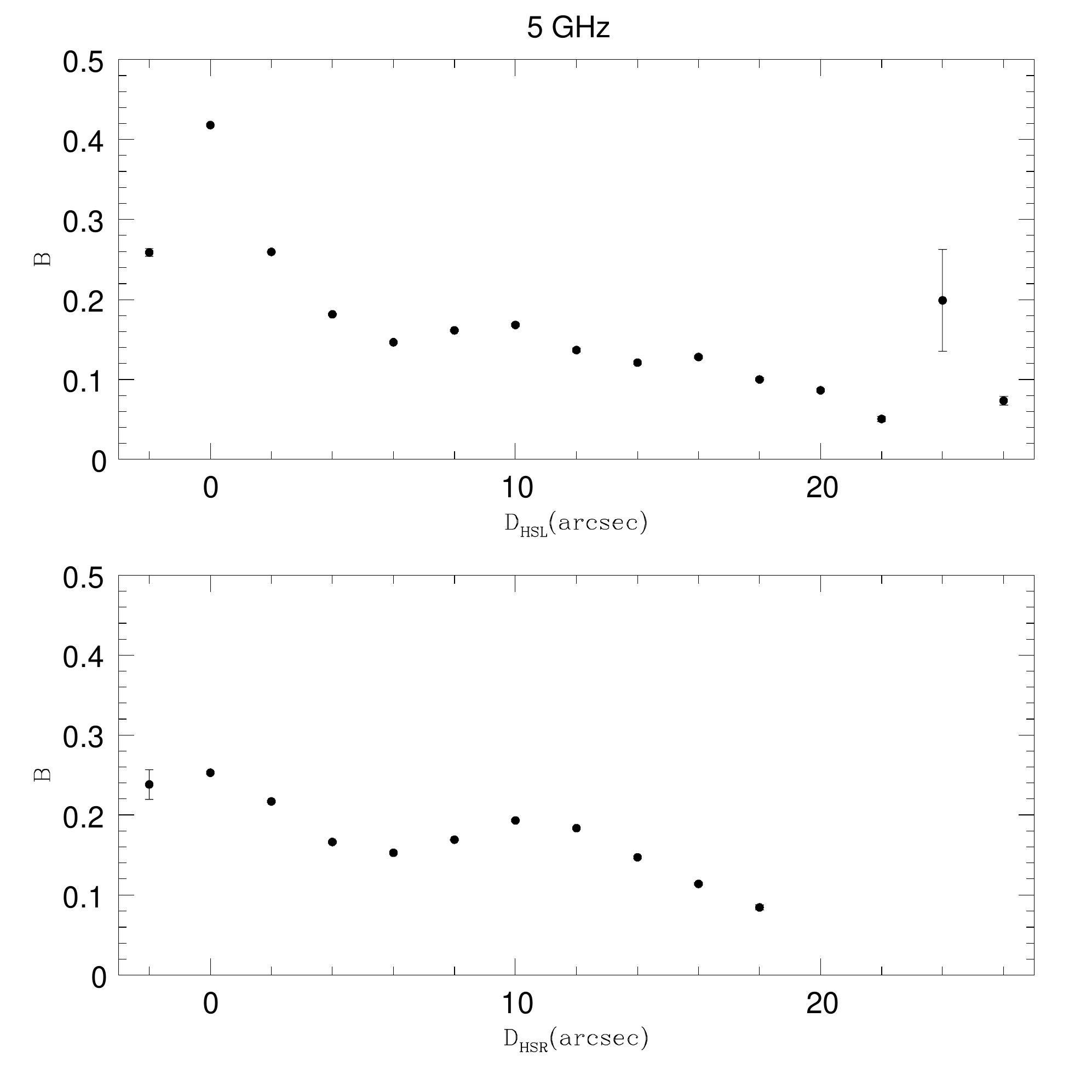}
\caption{3C34 minimum energy magnetic field strength 
as a function of distance from the 
hot spot at 1.4 and 5 GHz 
(left and right panels, respectively) for the left and right
hand sides of the source (top and bottom panels, respectively), 
as in Fig. \ref{3C6.1B}. The normalization is given in Table 1.}
\label{3C34B}
\end{figure}

\begin{figure}
\plottwo{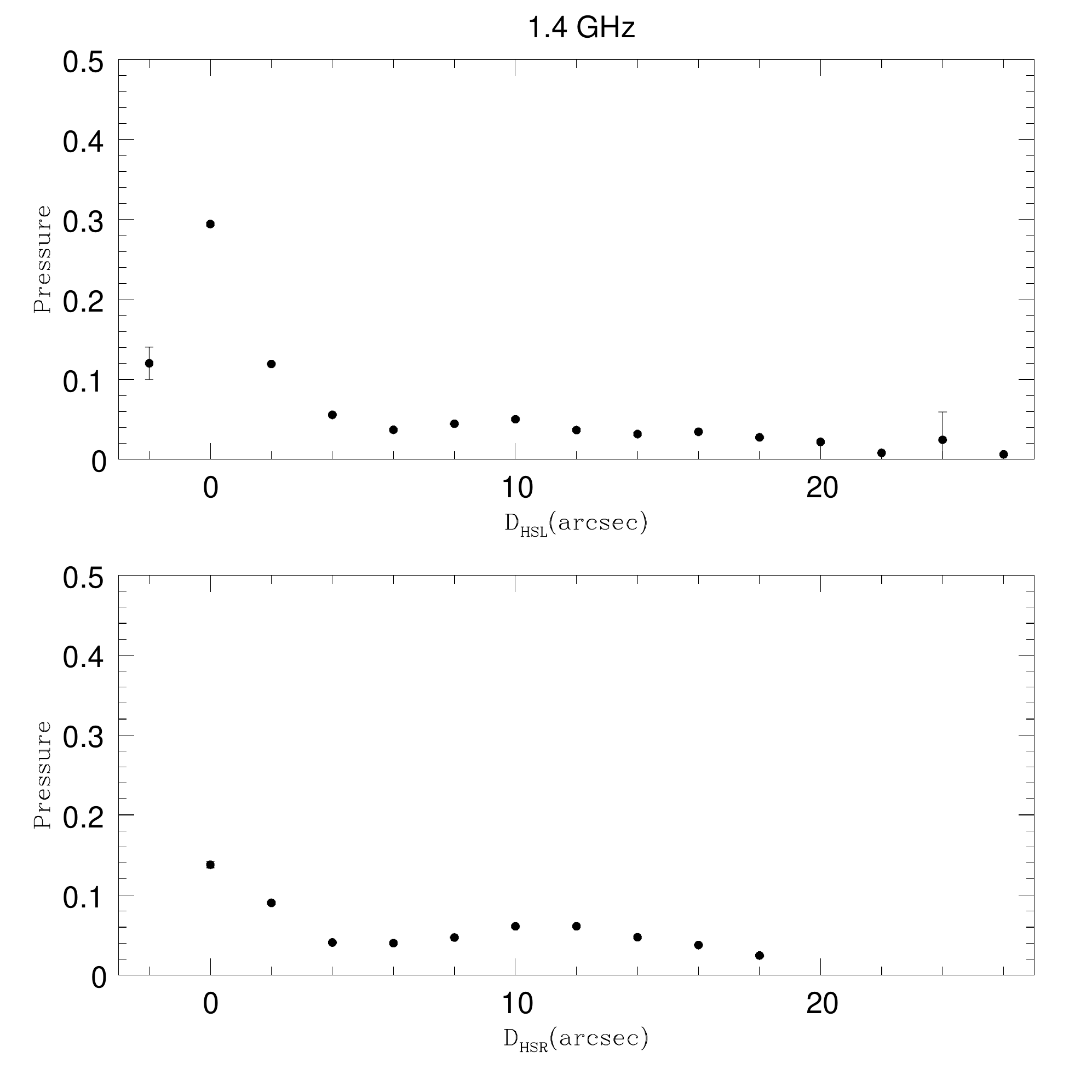}{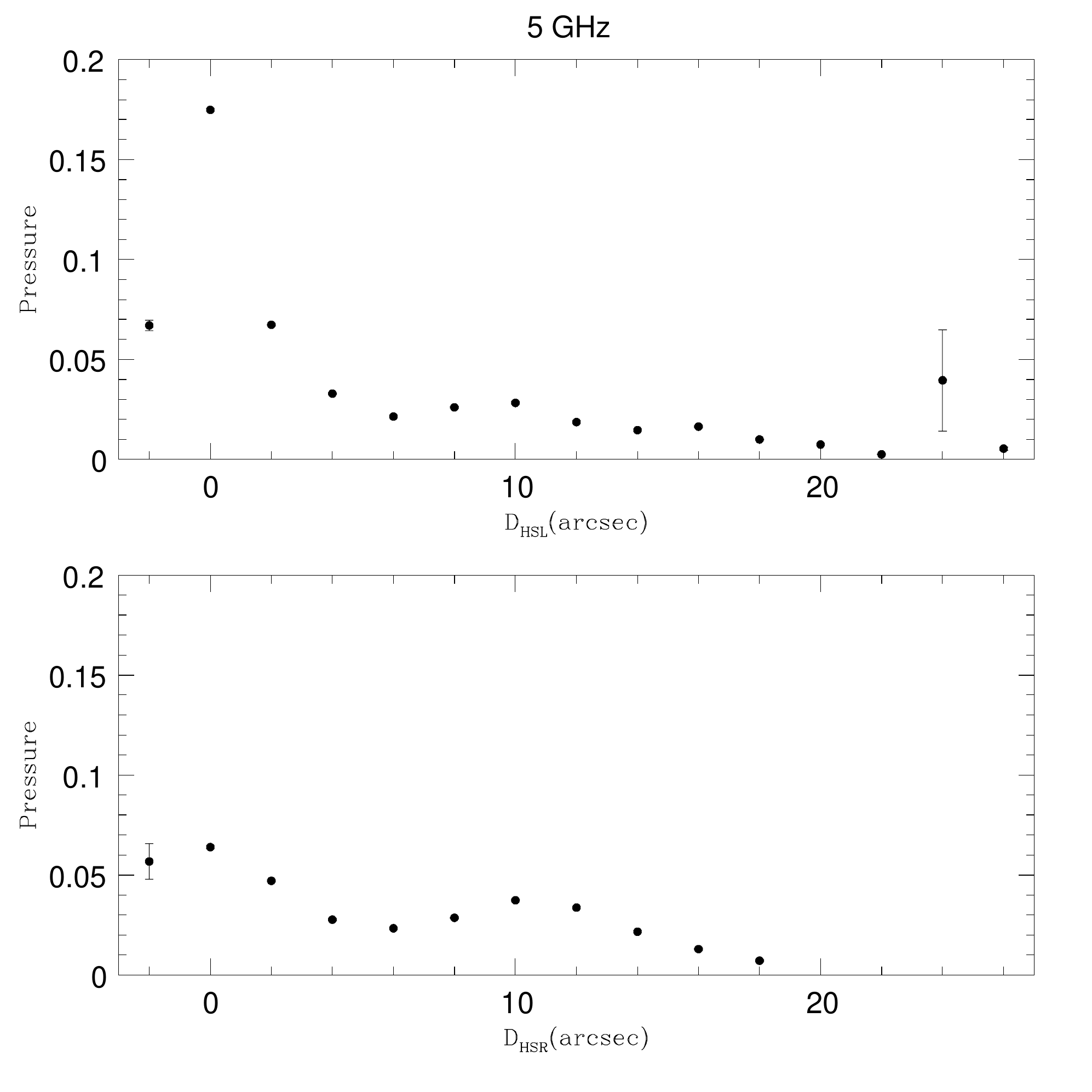}
\caption{3C34 minimum energy pressure as a function of 
distance from the hot spot at 1.4 and 5 GHz 
(left and right panels, respectively) for the left and right
hand sides of the source (top and bottom panels, respectively), 
as in Fig. \ref{3C6.1P}. The normalization is given in Table 1.}
\end{figure}

\clearpage
\begin{figure}
\plotone{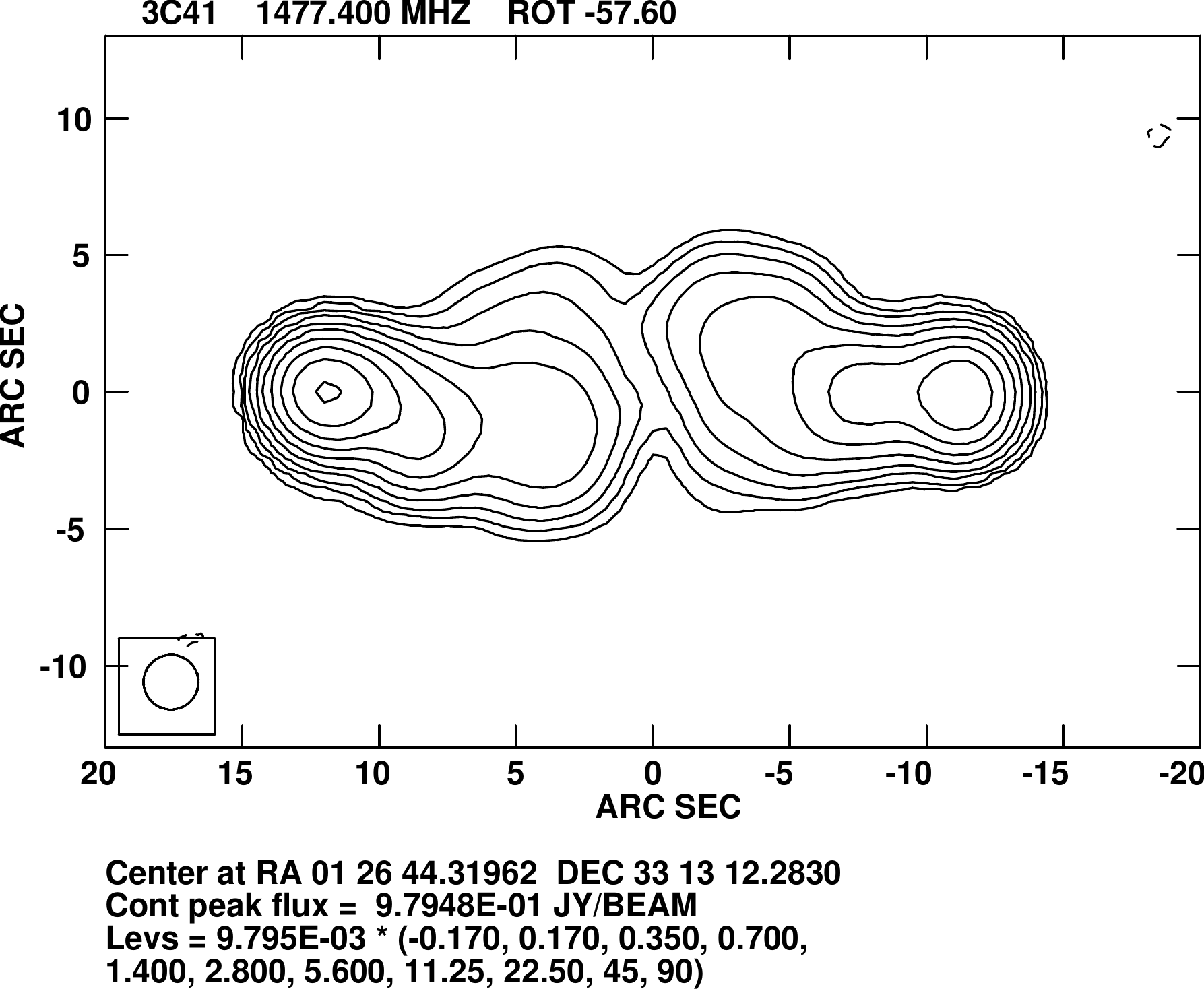}
\caption{3C41 at 1.4 GHz and 2'' resolution rotated by 57.60 deg 
clockwise, as in Fig. \ref{3C6.1rotfig}.}
\end{figure}

\begin{figure}
\plottwo{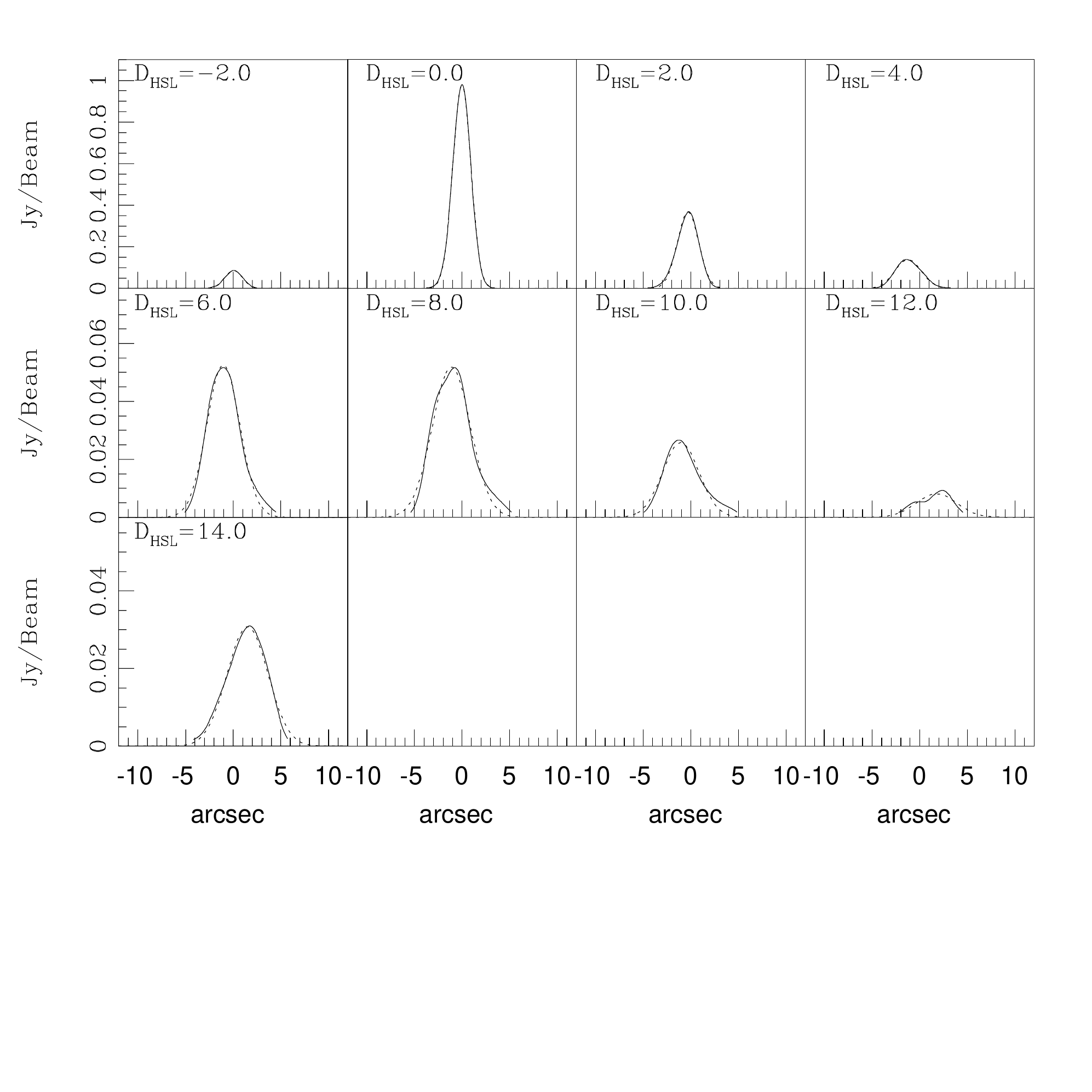}{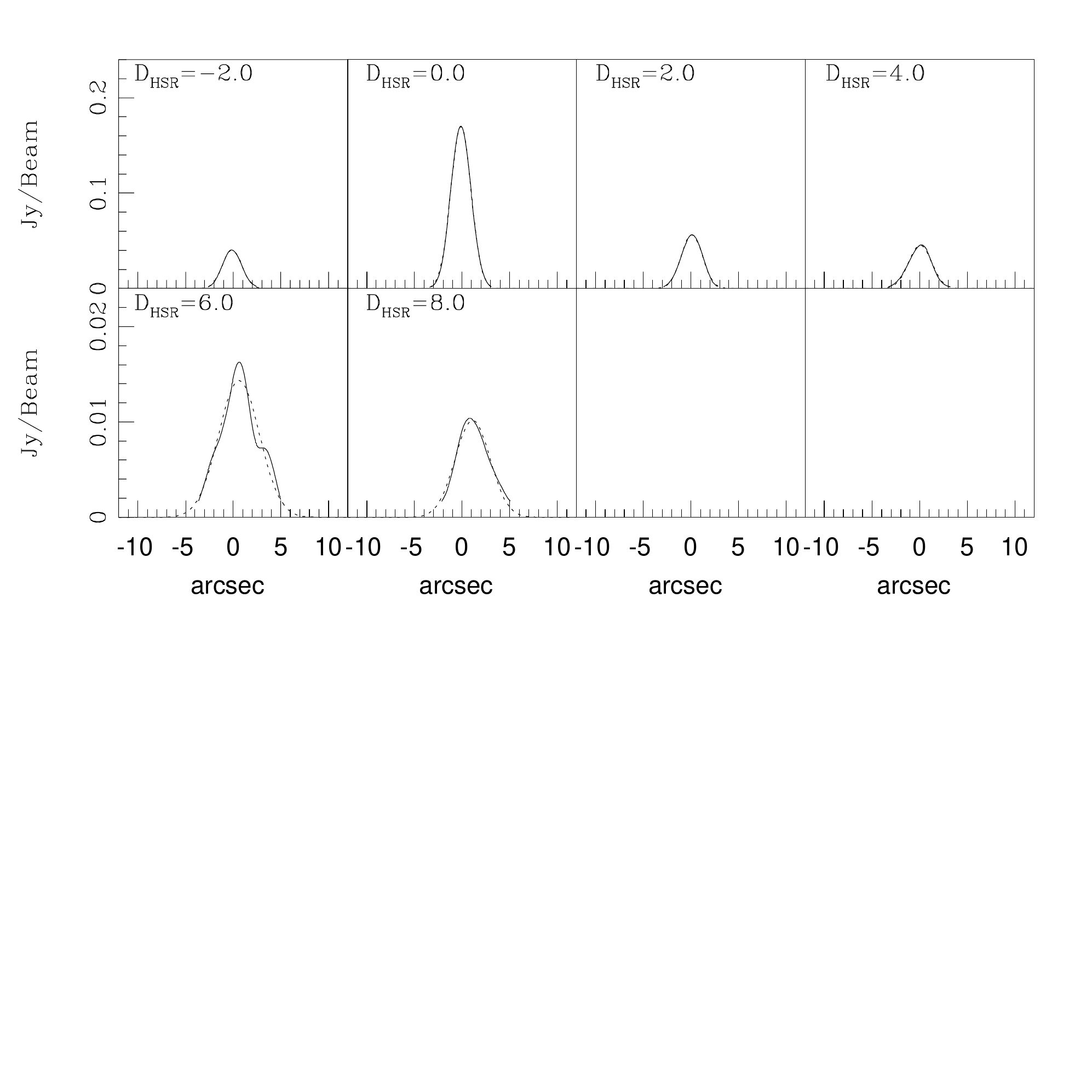}
\caption{3C41 
surface brightness (solid line) and best fit 
Gaussian (dotted line) for each cross-sectional slice of the 
radio bridge at 1.4 GHz, as in Fig. \ref{3C6.1SG}.
A Gaussian provides an excellent description of the surface brightness
profile of each cross-sectional slice. }
\label{3C41SG}
\end{figure}

\begin{figure}
\plottwo{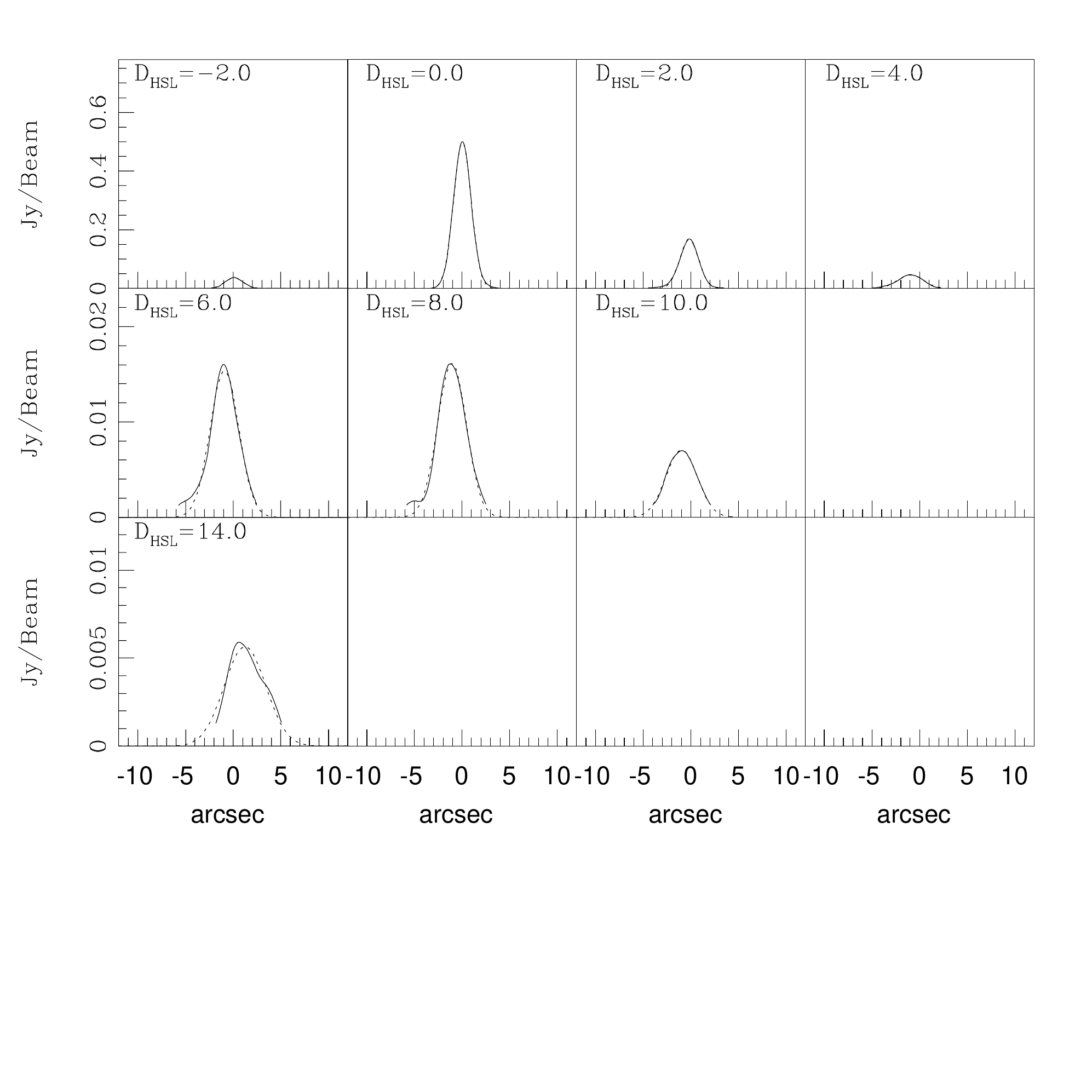}{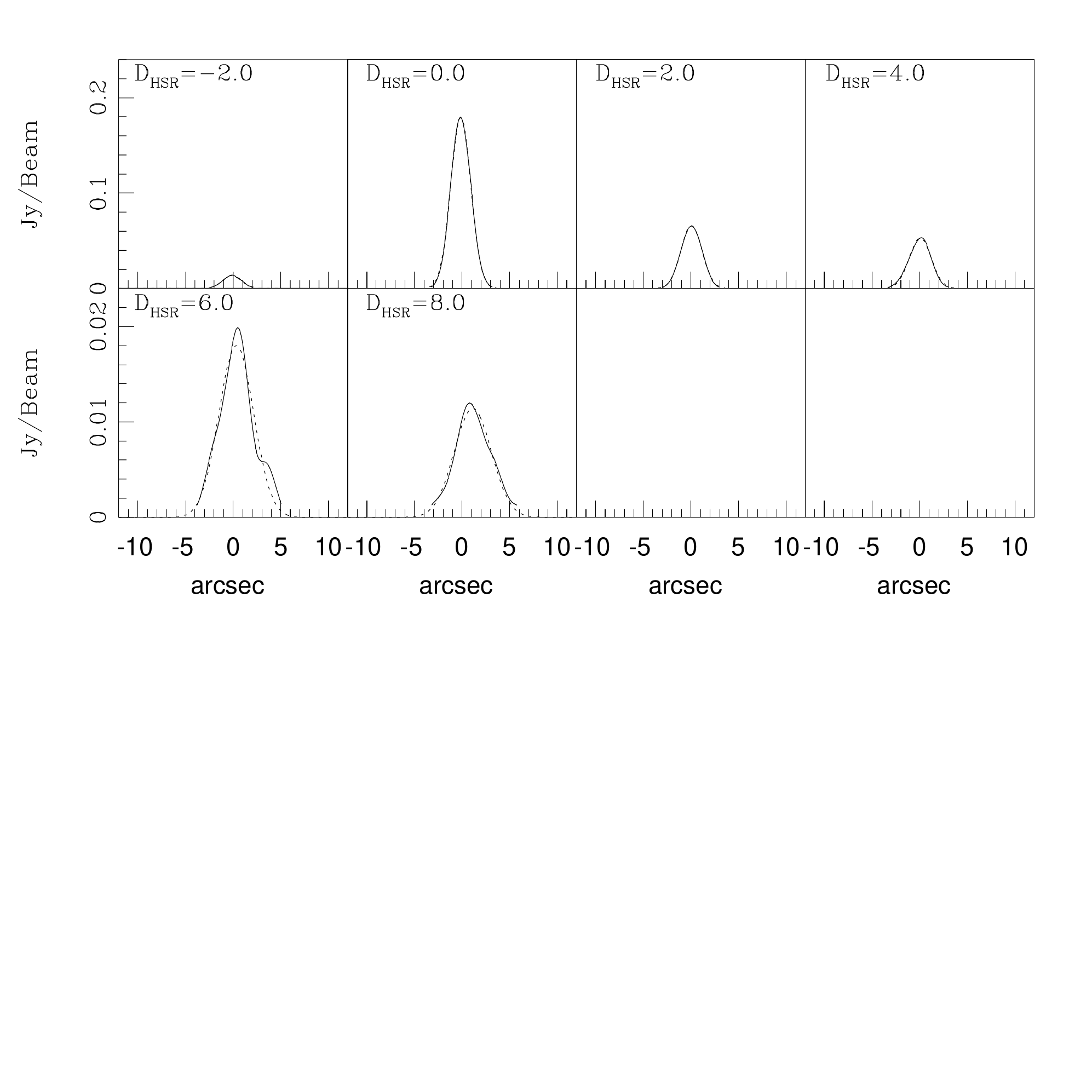}
\caption{3C41 
surface brightness (solid line) and best fit 
Gaussian (dotted line) for each cross-sectional slice of the 
radio bridge at 5 GHz, as in 
Fig. \ref{3C41SG} but at 5 GHz. 
Results obtained at 5 GHz are nearly identical to those
obtained at 1.4 GHz.}
\end{figure}

\begin{figure}
\plottwo{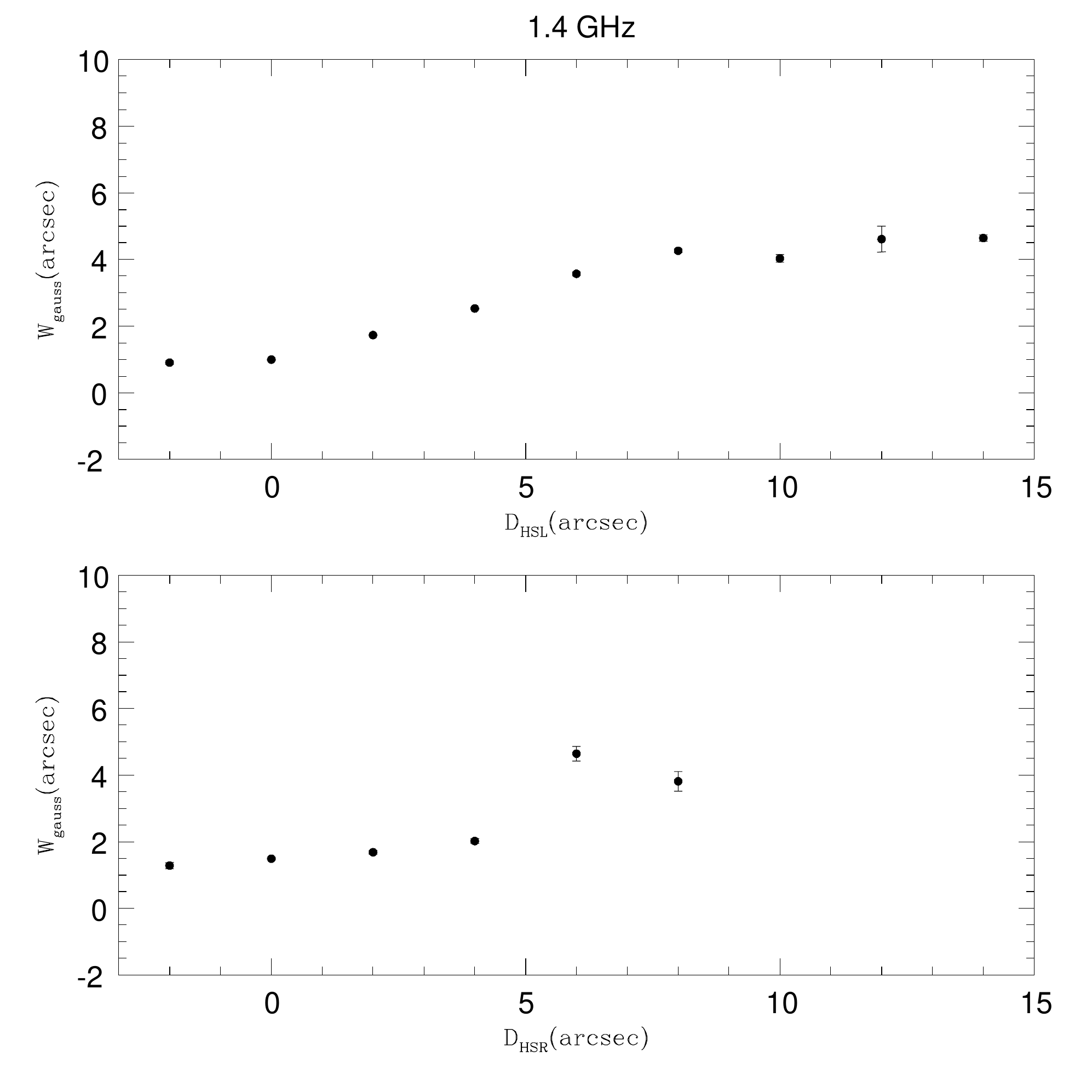}{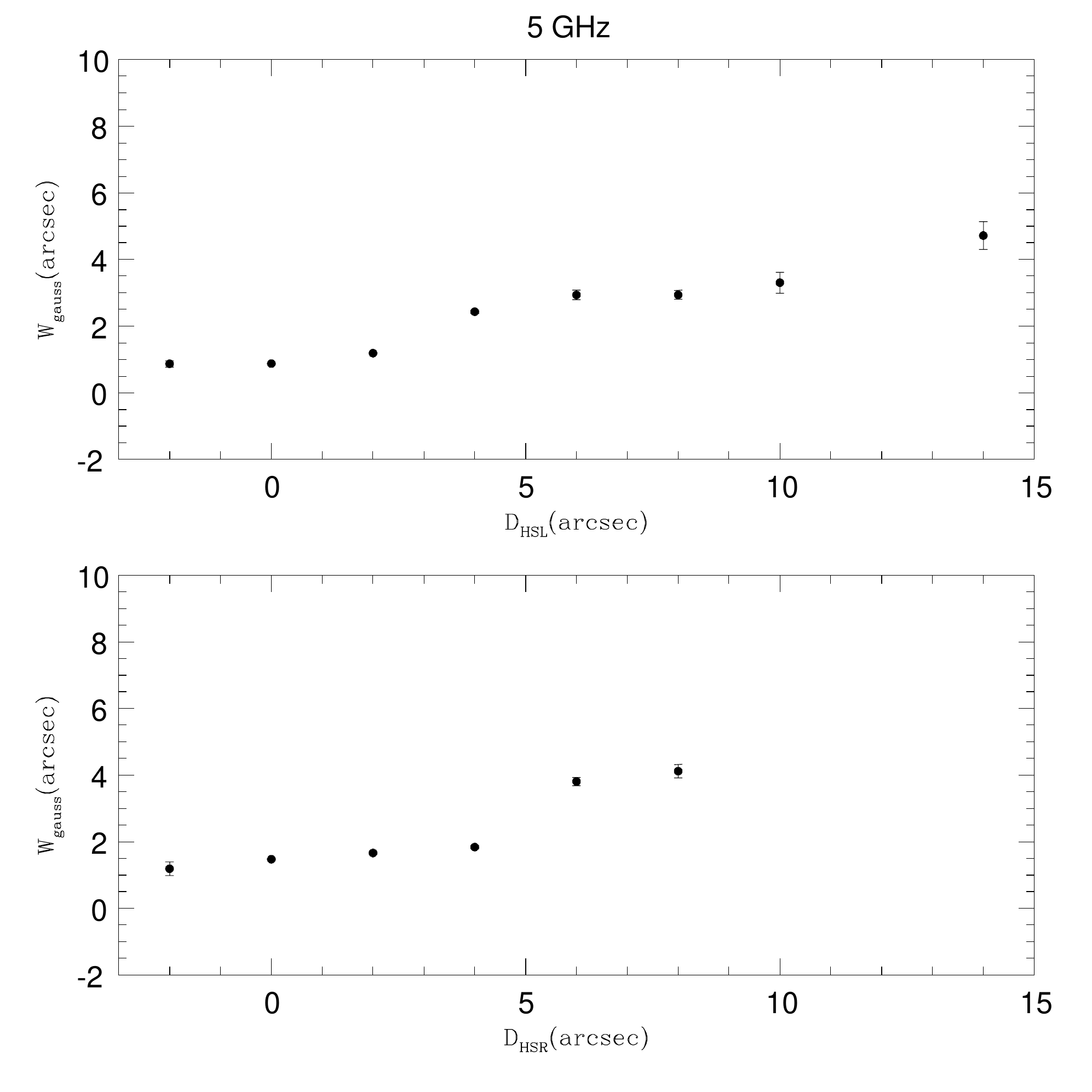}
\caption{3C41  Gaussian FWHM as a function of 
distance from the hot spot 
at 1.4 and 5 GHz (left and
right panels, respectively) for the left and right hand sides of the 
source (top and bottom panels, respectively), as in Fig. \ref{3C6.1WG}.
The right and left hand sides of the source are fairly 
symmetric.  Similar results are obtained at 
1.4 and 5 GHz. }
\end{figure}

\begin{figure}
\plottwo{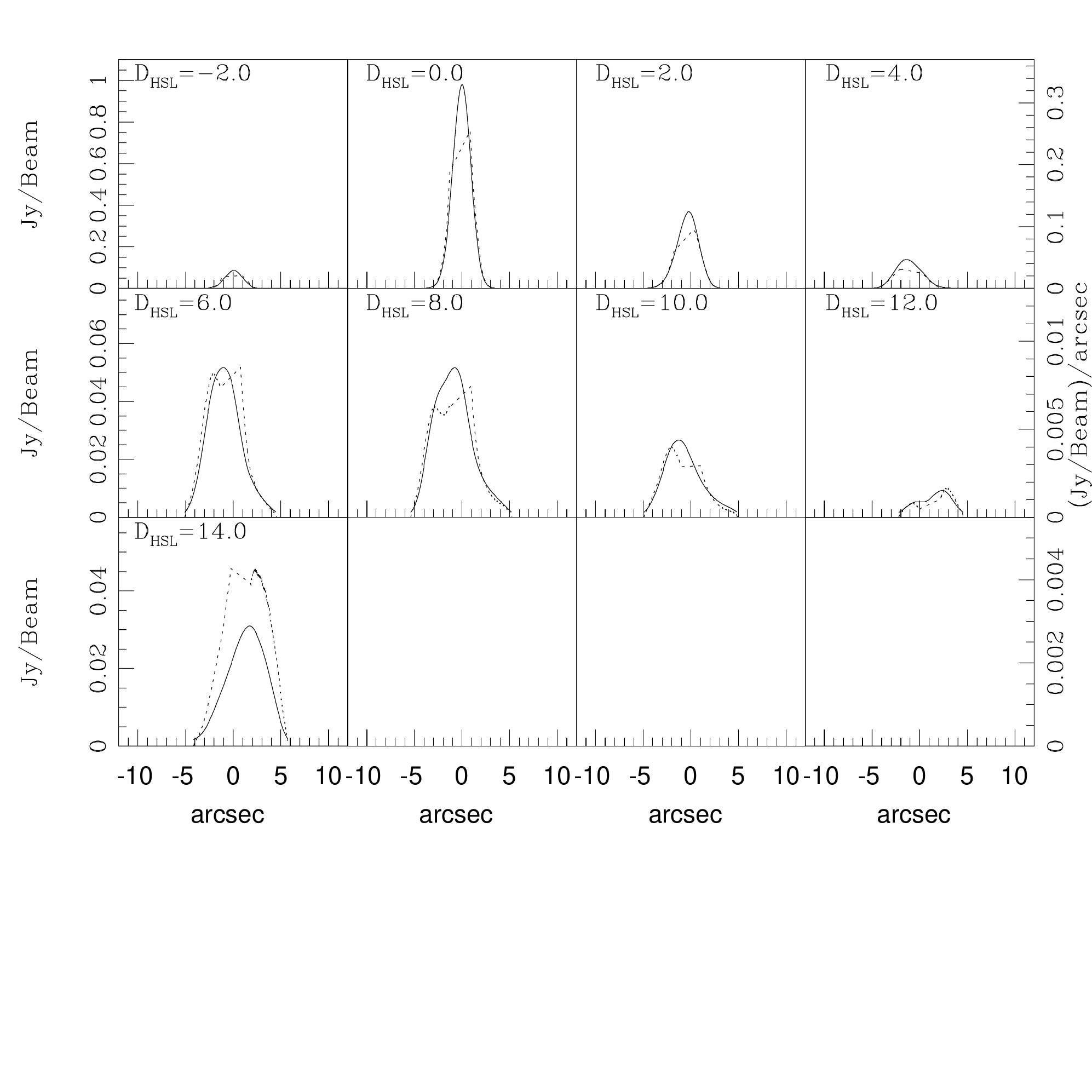}{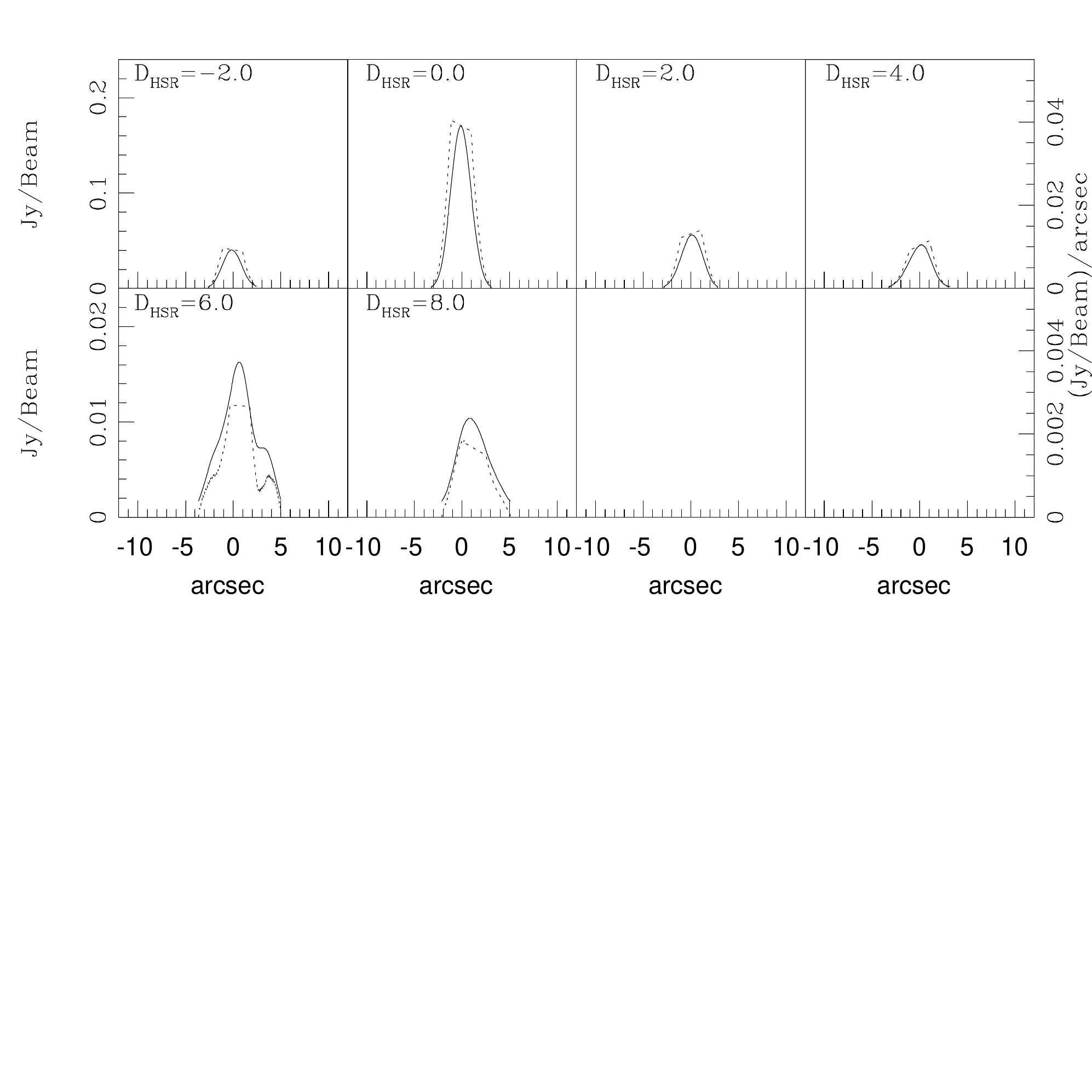}
\caption{3C41 emissivity (dotted line) and surface brightness
(solid line) for each cross-sectional slice of the radio 
bridge at 1.4 GHz, as in Fig. \ref{3C6.1SEM}. A constant volume
emissivity per slice provides a reasonable description of the 
data over most of the source. }
\label{3C41SEM}
\end{figure}

\begin{figure}
\plottwo{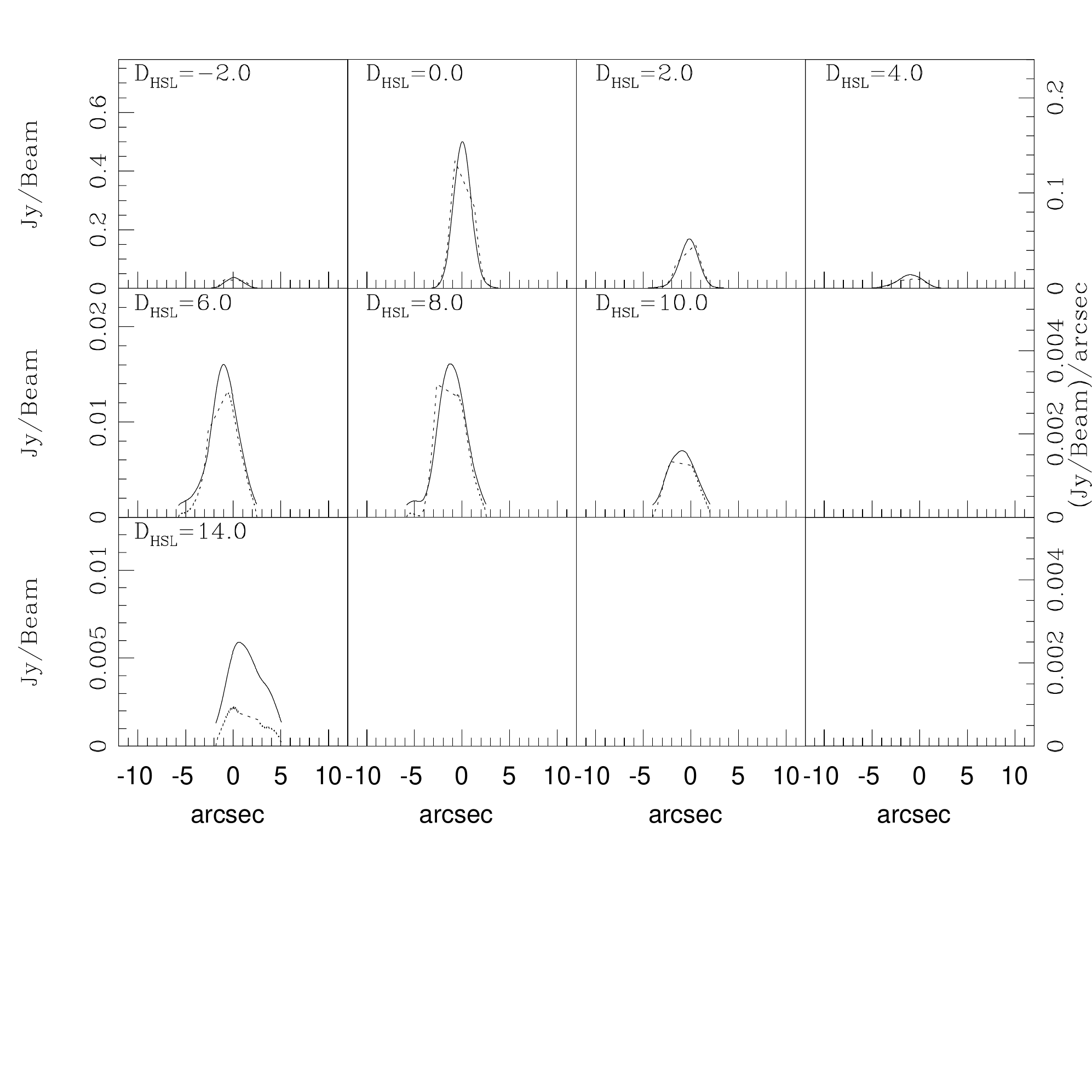}{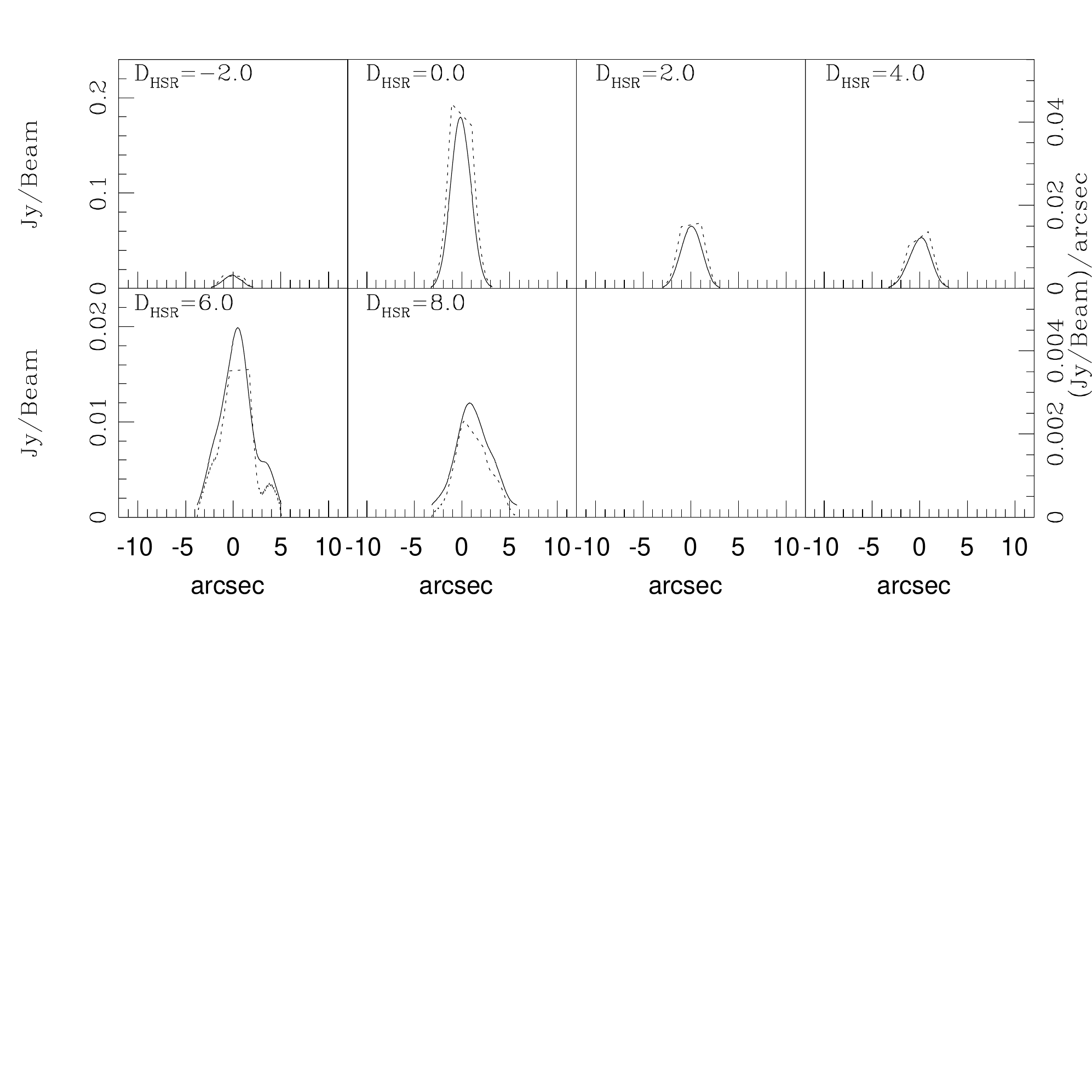}
\caption{3C41 
emissivity (dotted line) and surface brightness
(solid line) for each cross-sectional slice of the radio 
bridge at 5 GHz, as in Fig. \ref{3C41SEM} but at 5 GHz.
Results obtained at 5 GHz are nearly identical to those
obtained at 1.4 GHz.}
\end{figure}

\begin{figure}
\plottwo{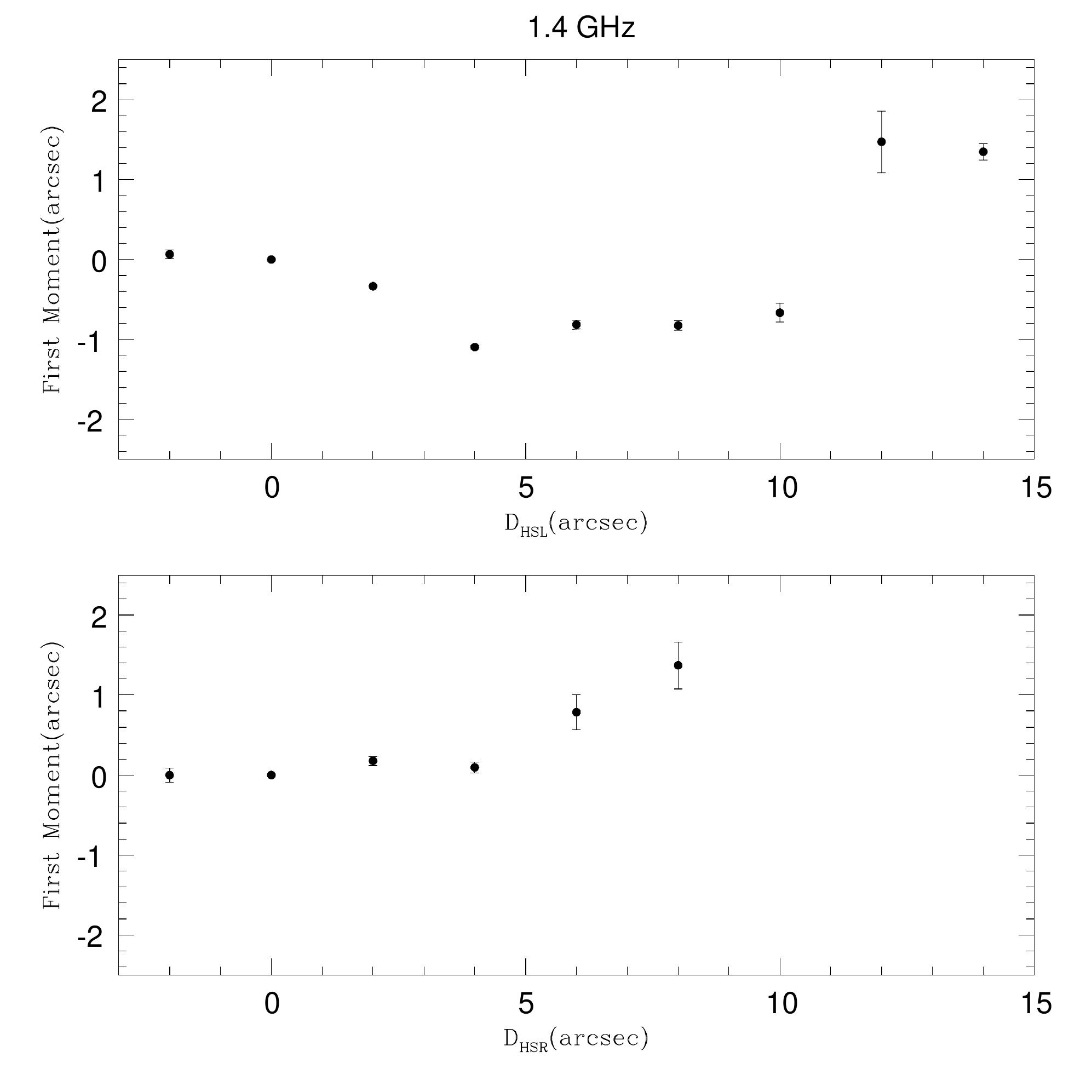}{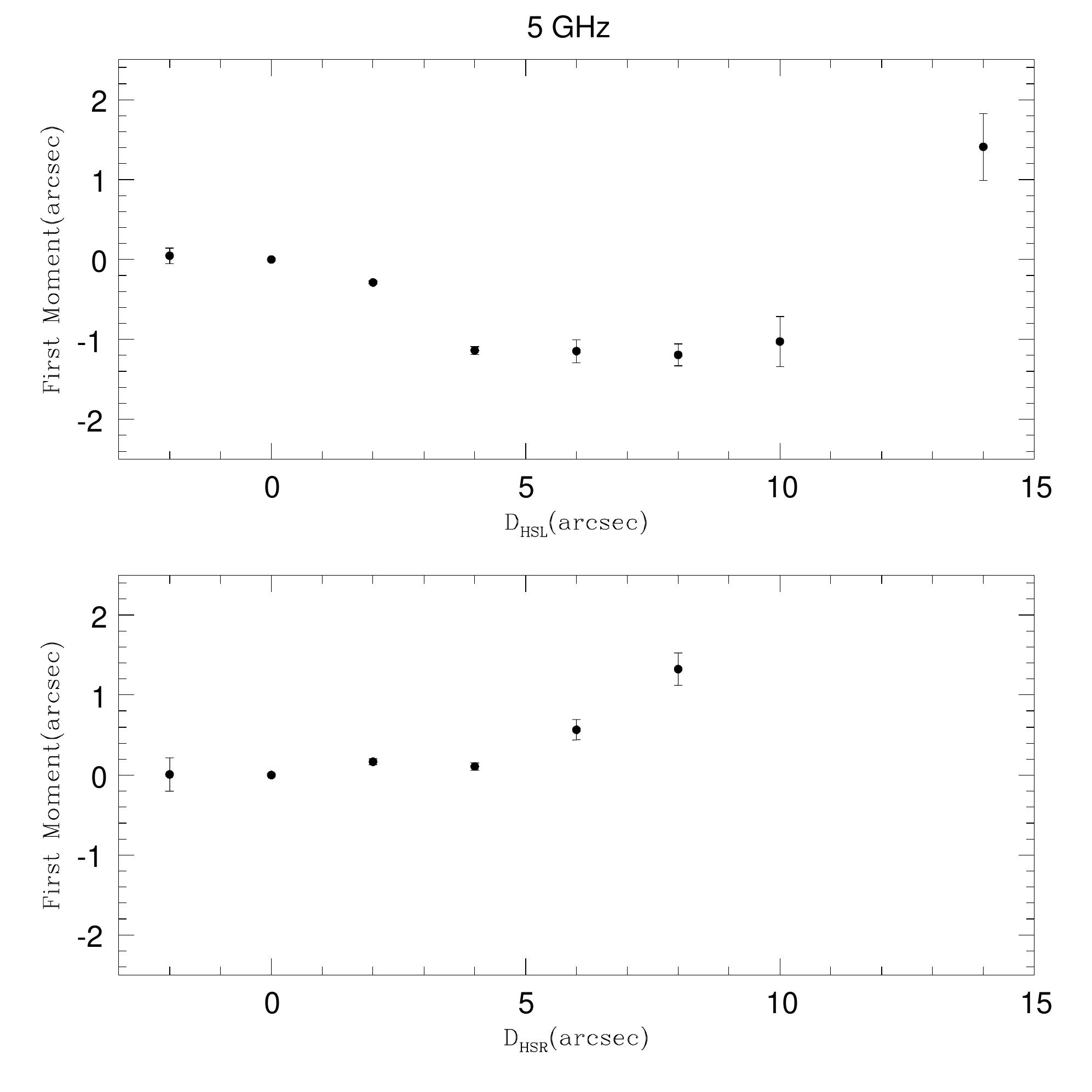}
\caption{3C41 first moment as a function of distance from 
the hot spot at 1.4 GHz and 5 GHz (left and right panels,
respectively) for the left and right hand sides of the source
(top and bottom, respectively).}
\end{figure}

\begin{figure}
\plottwo{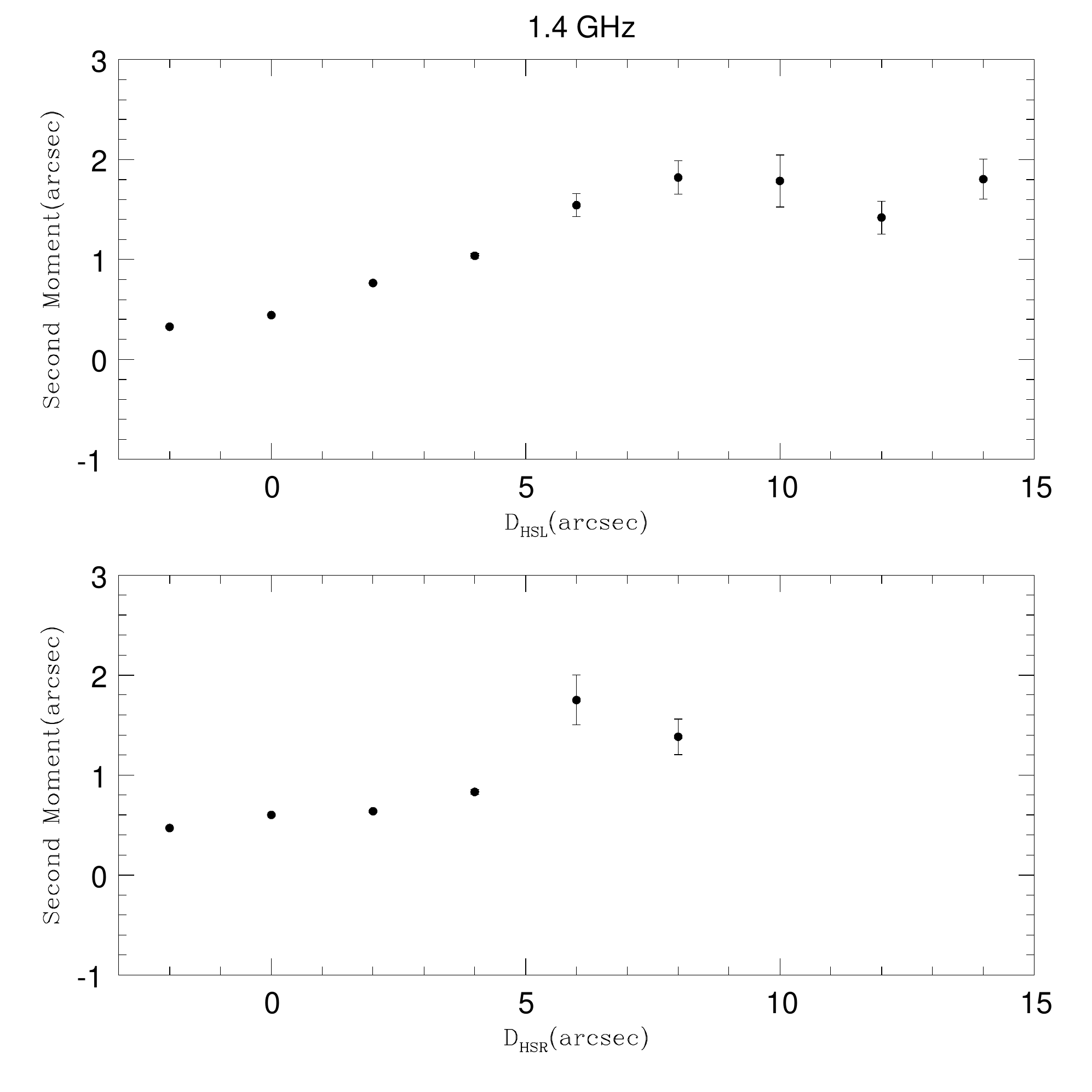}{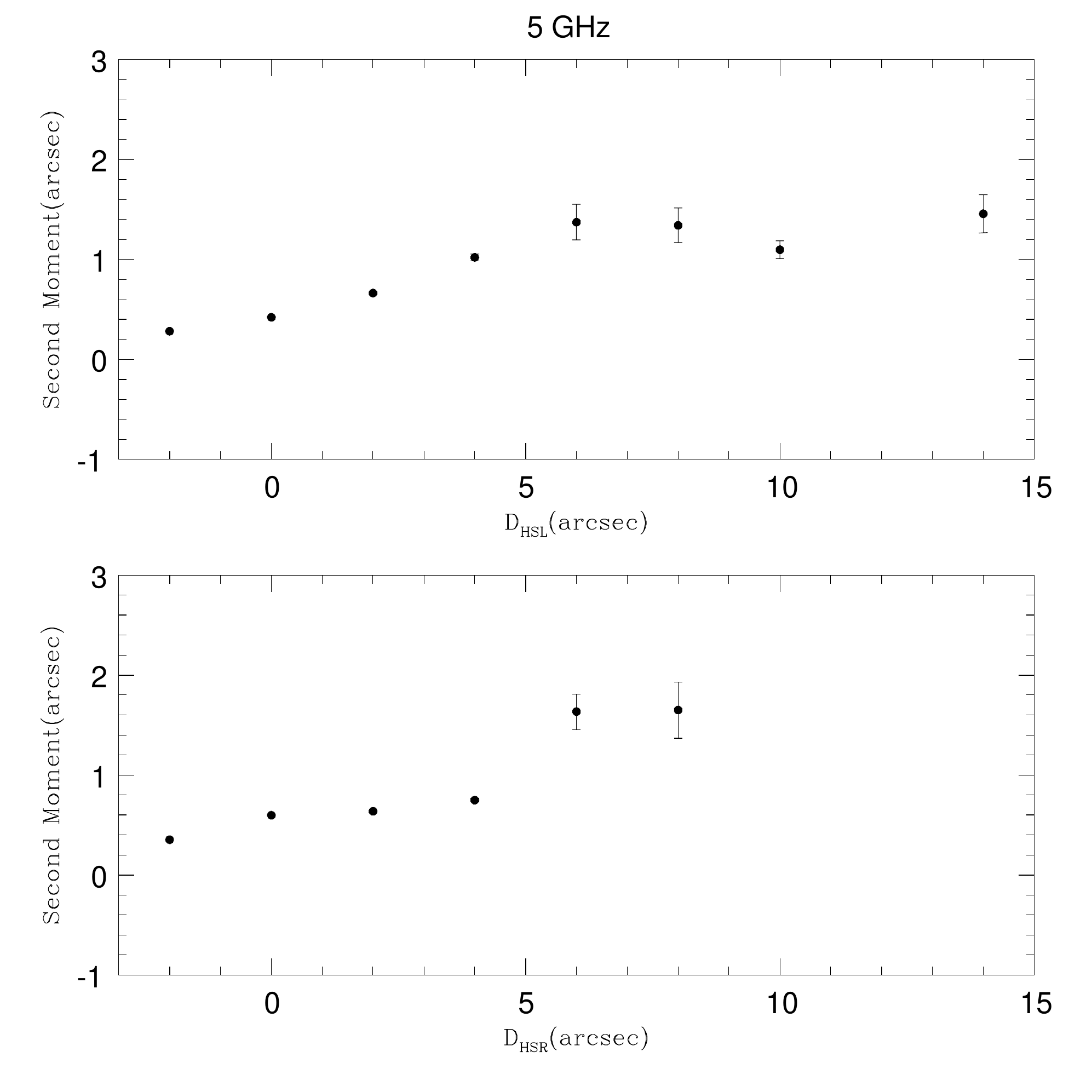}
\caption{3C41 second moment as a function of distance from the
hot spot at 1.4 GHz and 5 GHz (left and right panels, 
respectively) for the left and right sides of the source
(top and bottom panels, respectively).}
\end{figure}

\begin{figure}
\plottwo{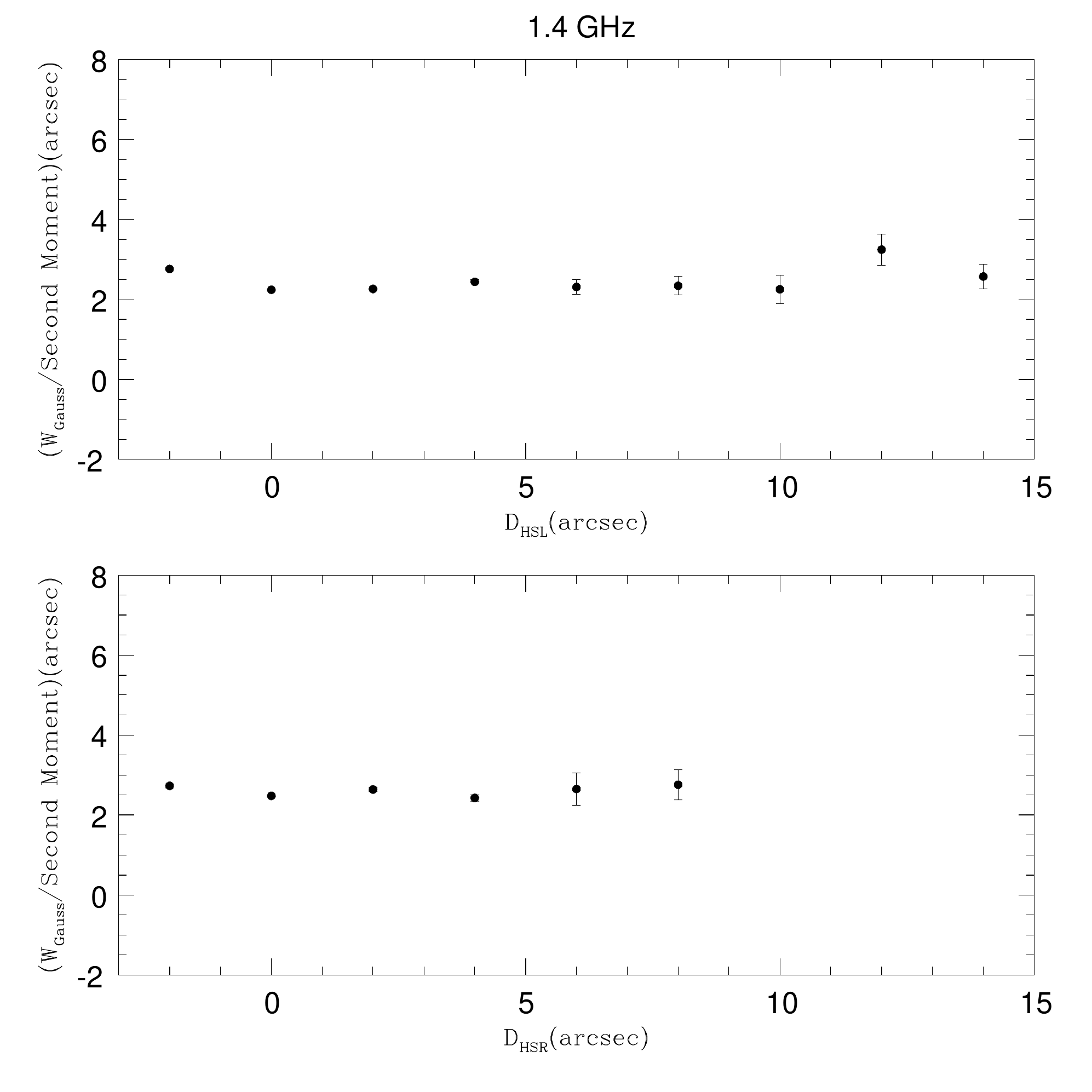}{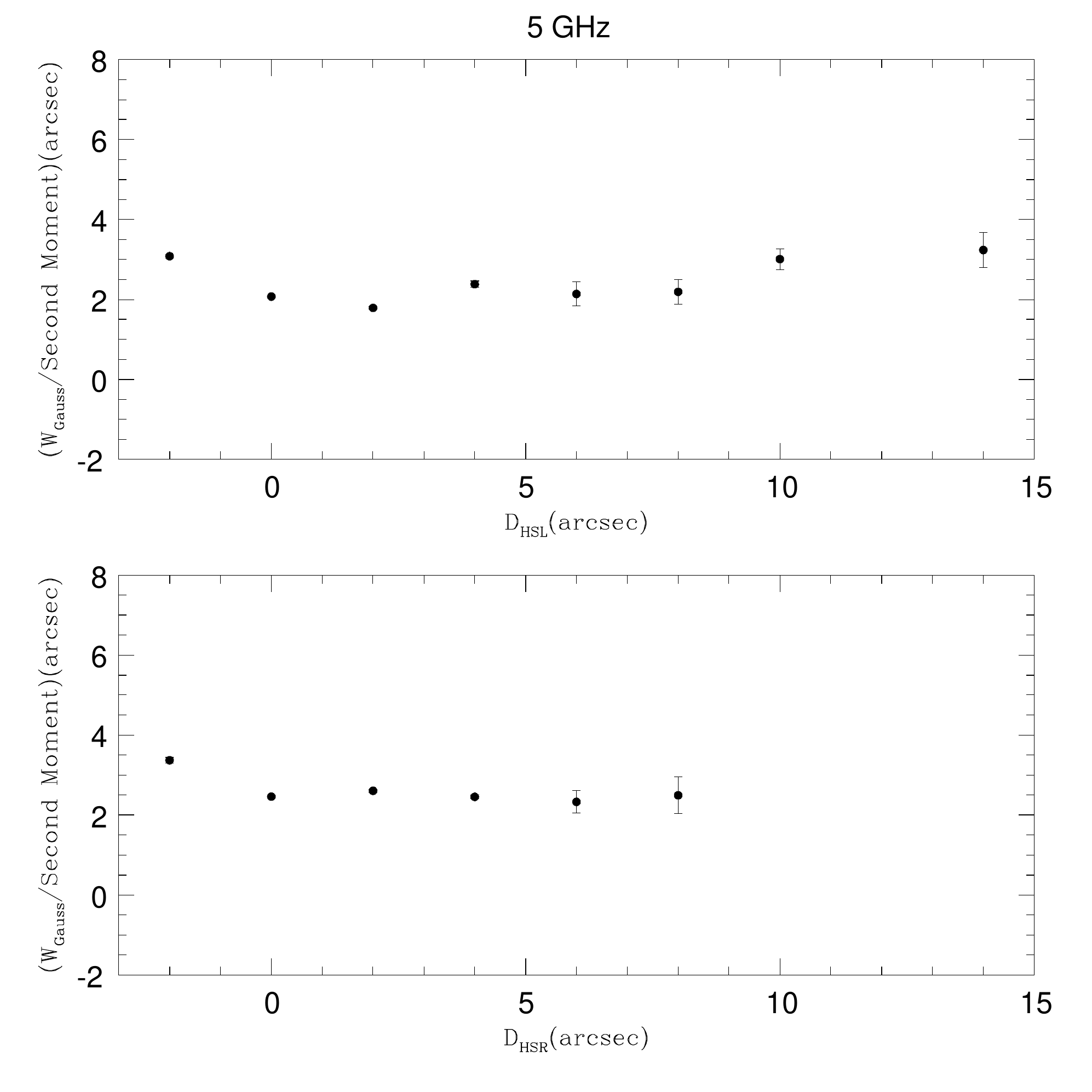}
\caption{3C41 ratio of the Gaussian FWHM to the second moment as a
function of distance from the hot spot at 1.4 and 5 GHz 
(left and right panels, respectively) for the left and right
hand sides of the source (top and bottom panels, respectively). The value of this ratio is fairly constant for each side of the source, and has a value of 
about 2. Similar results are obtained at 1.4 and 5 GHz.} 
\end{figure}

\begin{figure}
\plottwo{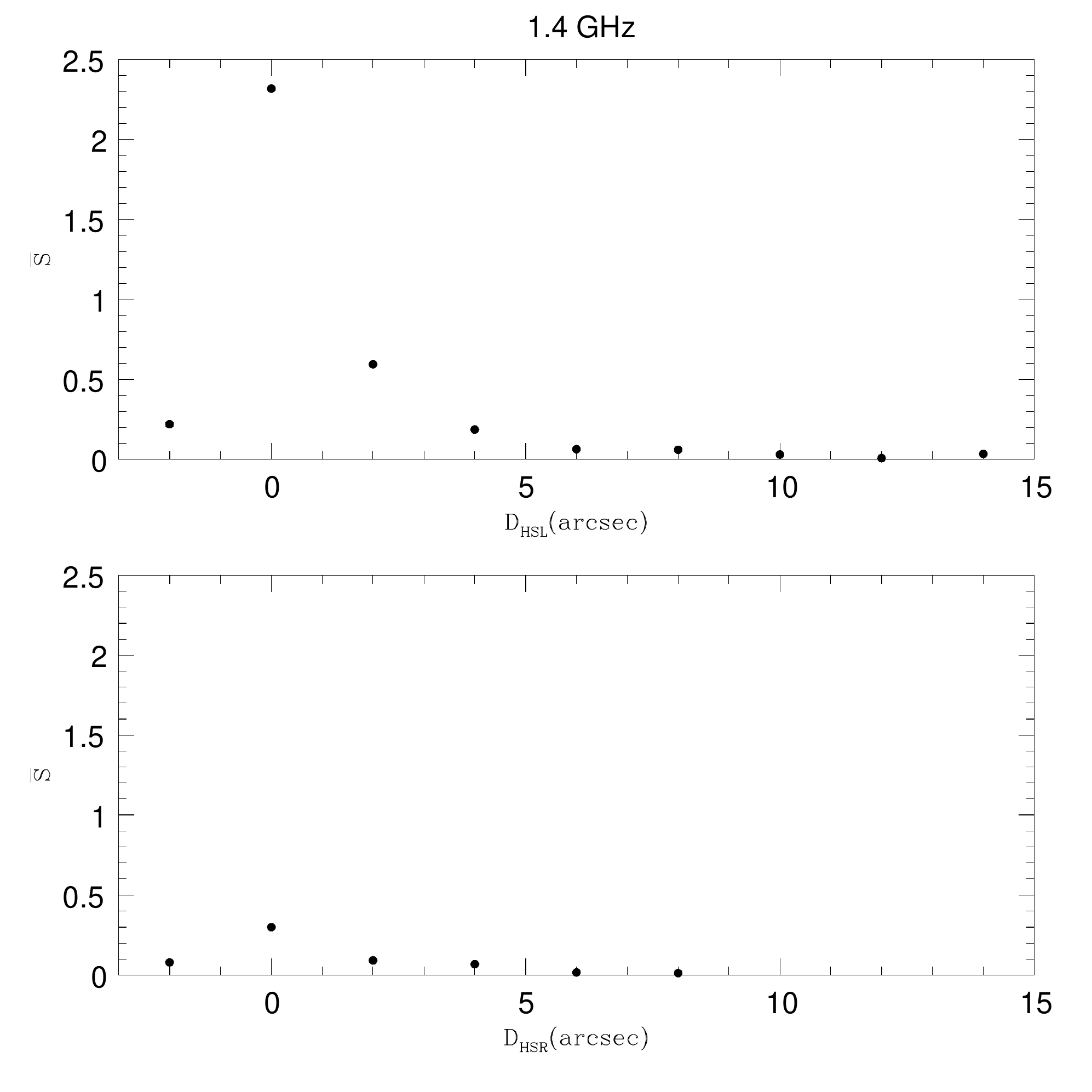}{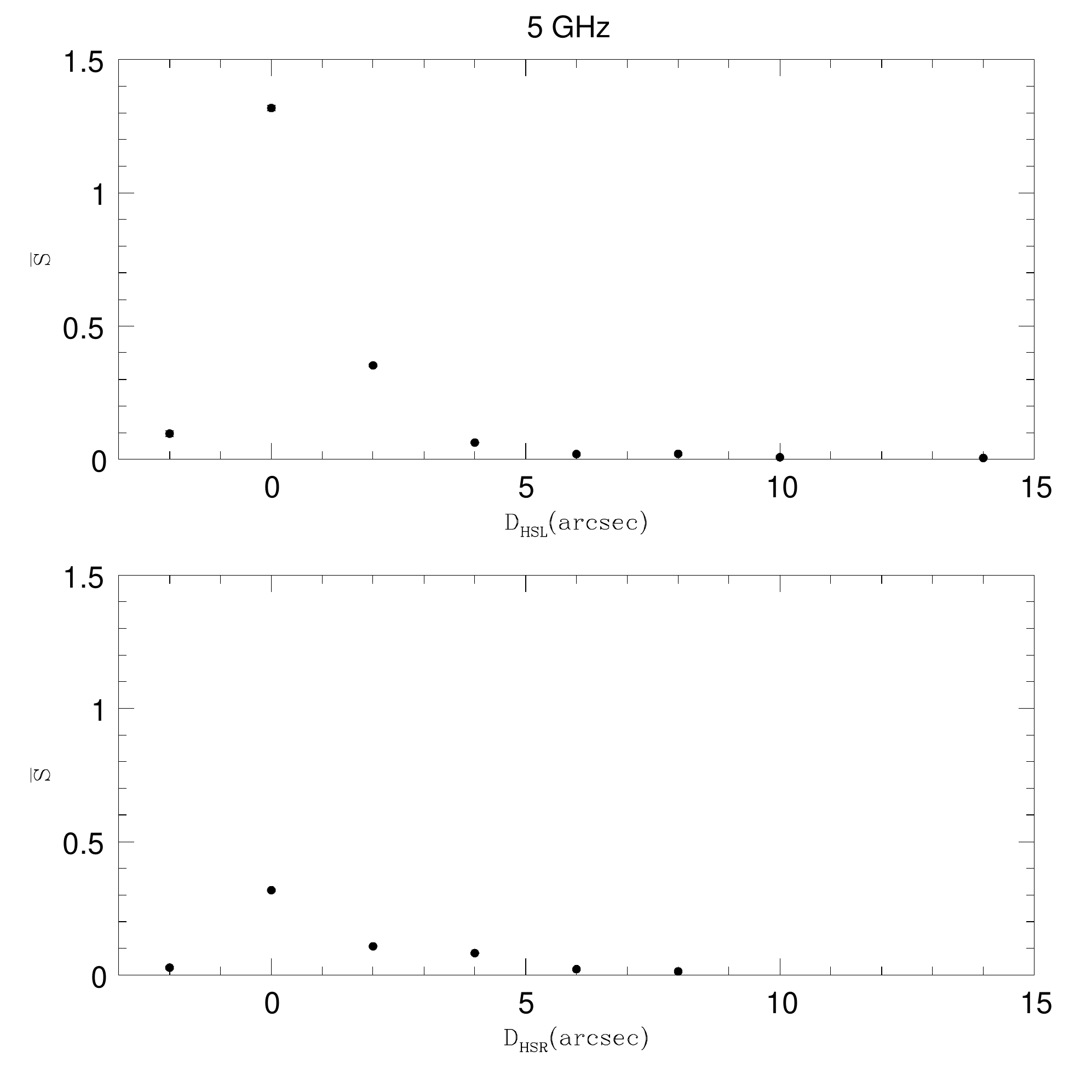}
\caption{3C41 average surface brightness in units of Jy/beam 
as a function of distance from the 
hot spot at 1.4 and 5 GHz 
(left and right panels, respectively) for the left and right
hand sides of the source (top and bottom panels, respectively).}
\end{figure}

\begin{figure}
\plottwo{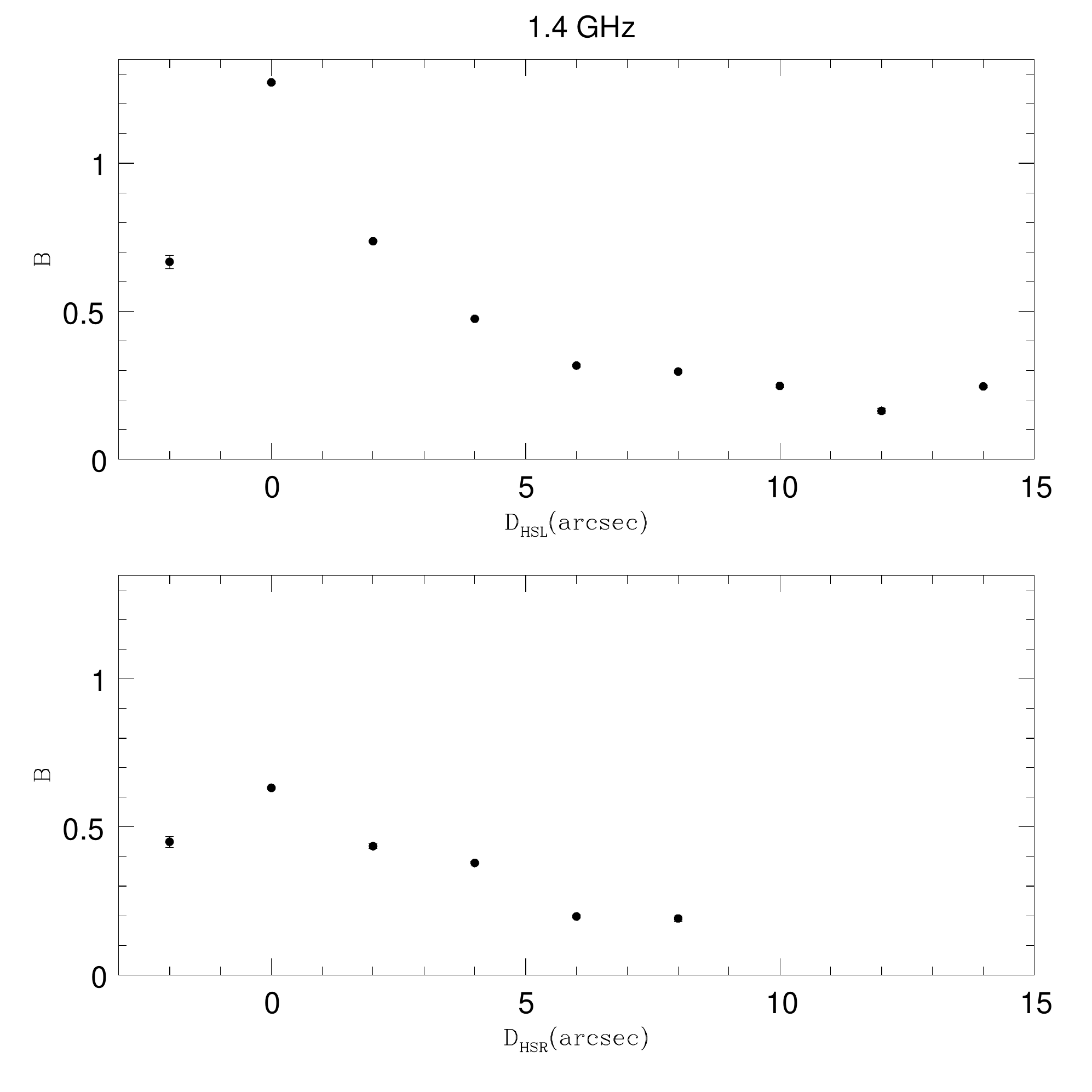}{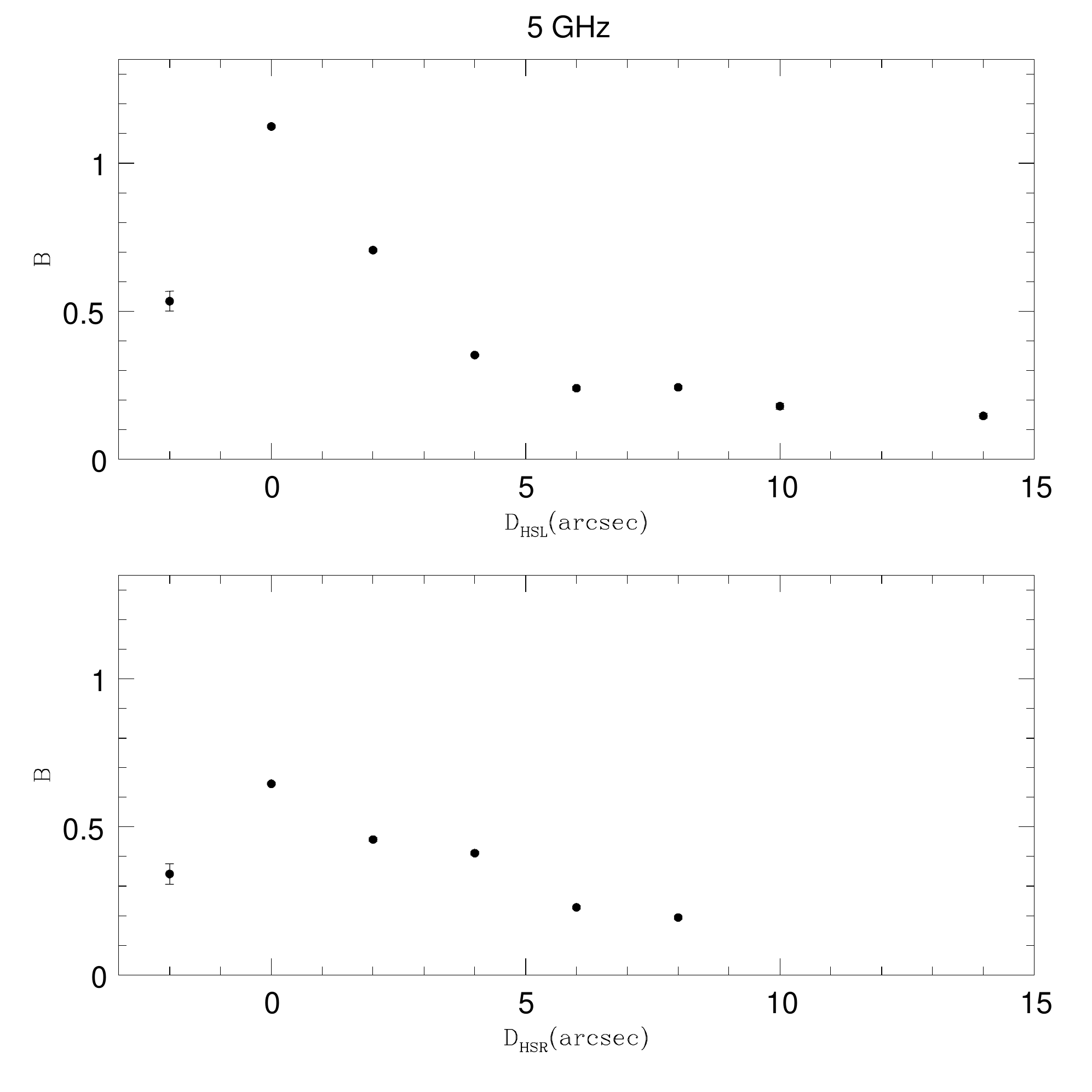}
\caption{3C41 minimum energy magnetic field strength 
as a function of distance from the 
hot spot at 1.4 and 5 GHz 
(left and right panels, respectively) for the left and right
hand sides of the source (top and bottom panels, respectively).
The normalization is given in Table 1. 
}
\end{figure}

\begin{figure}
\plottwo{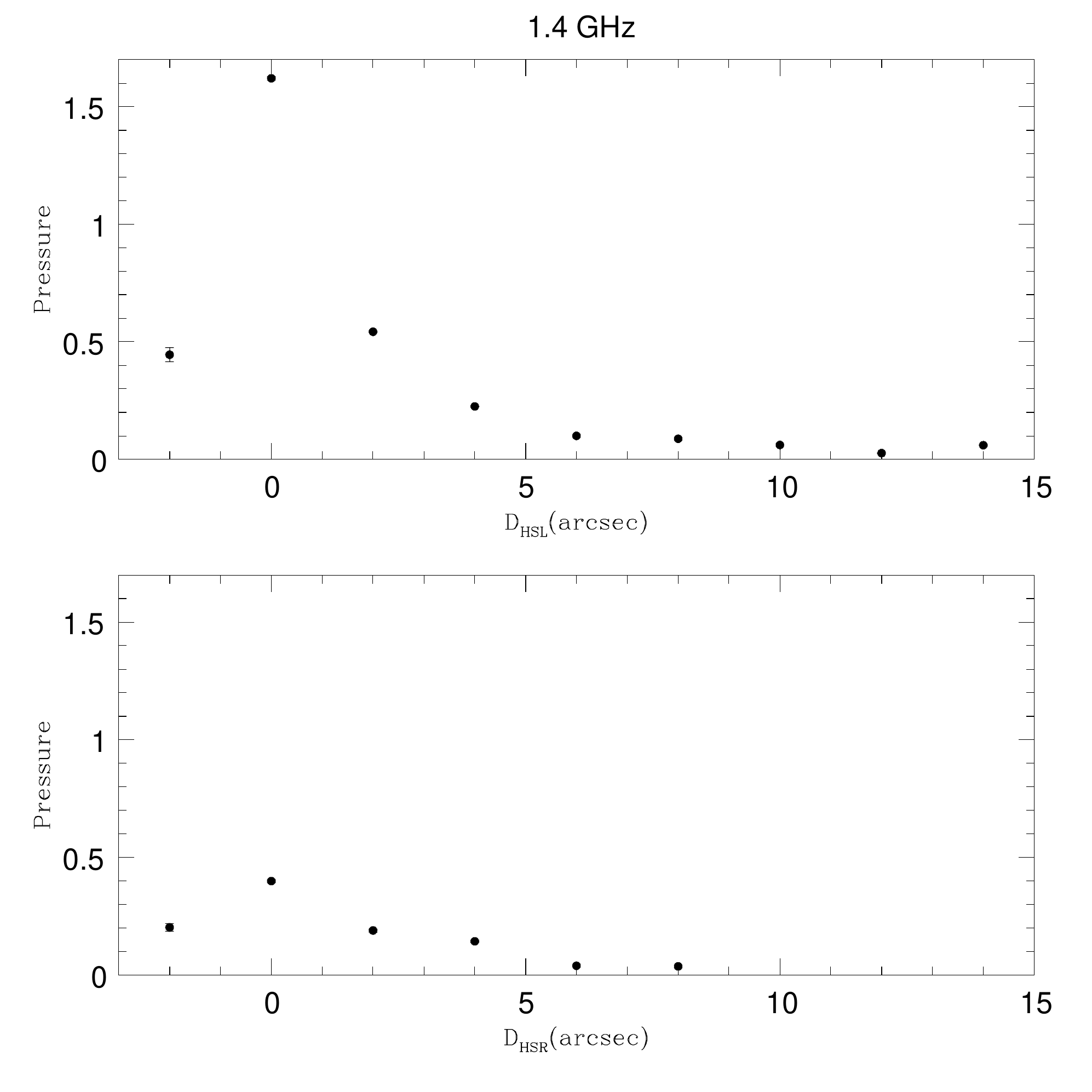}{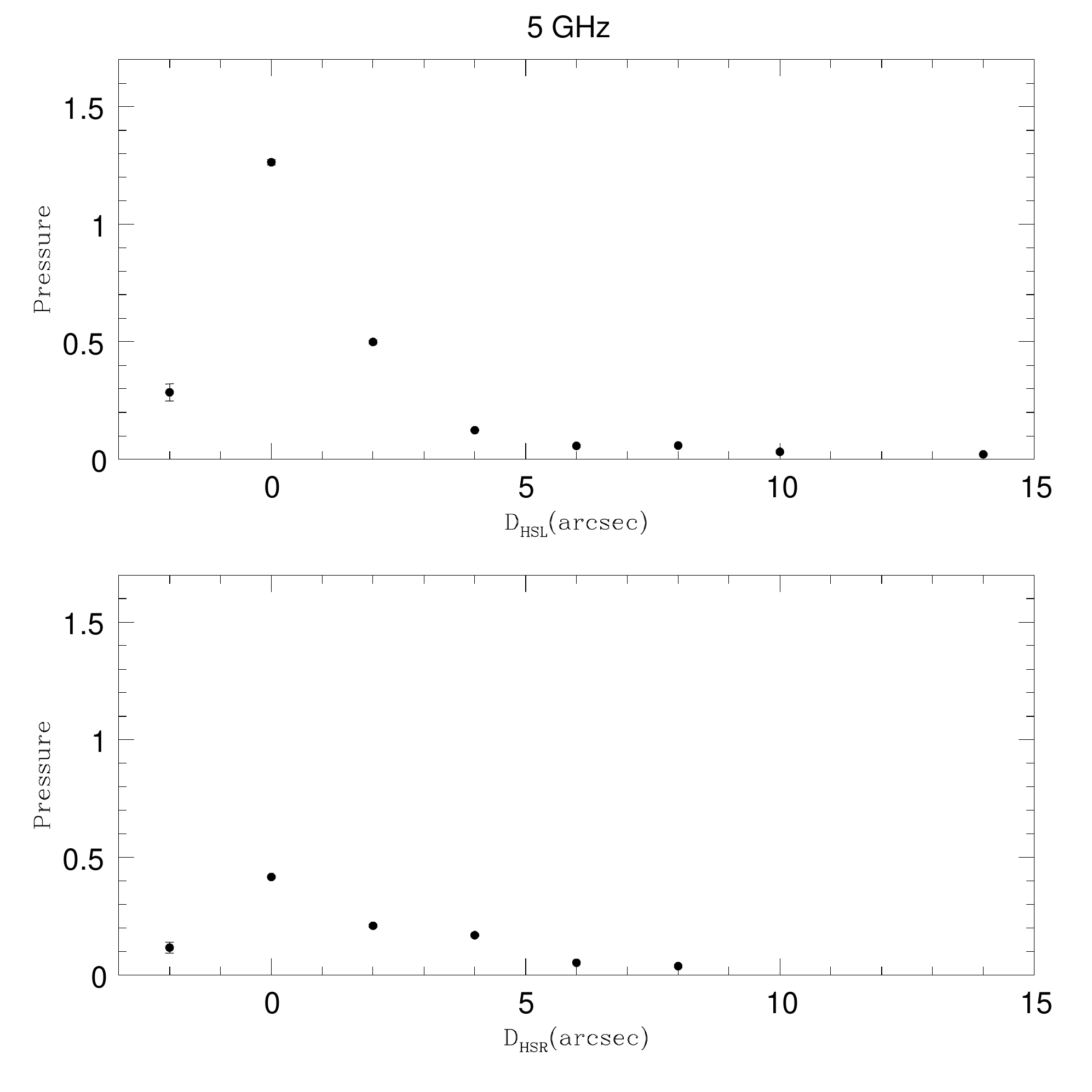}
\caption{3C41 minimum energy pressure as a function of 
distance from the hot spot at 1.4 and 5 GHz 
(left and right panels, respectively) for the left and right
hand sides of the source (top and bottom panels, respectively), 
as in Fig. \ref{3C6.1P}. The normalization is given in Table 1.
}
\end{figure}

\clearpage
\begin{figure}
\plotone{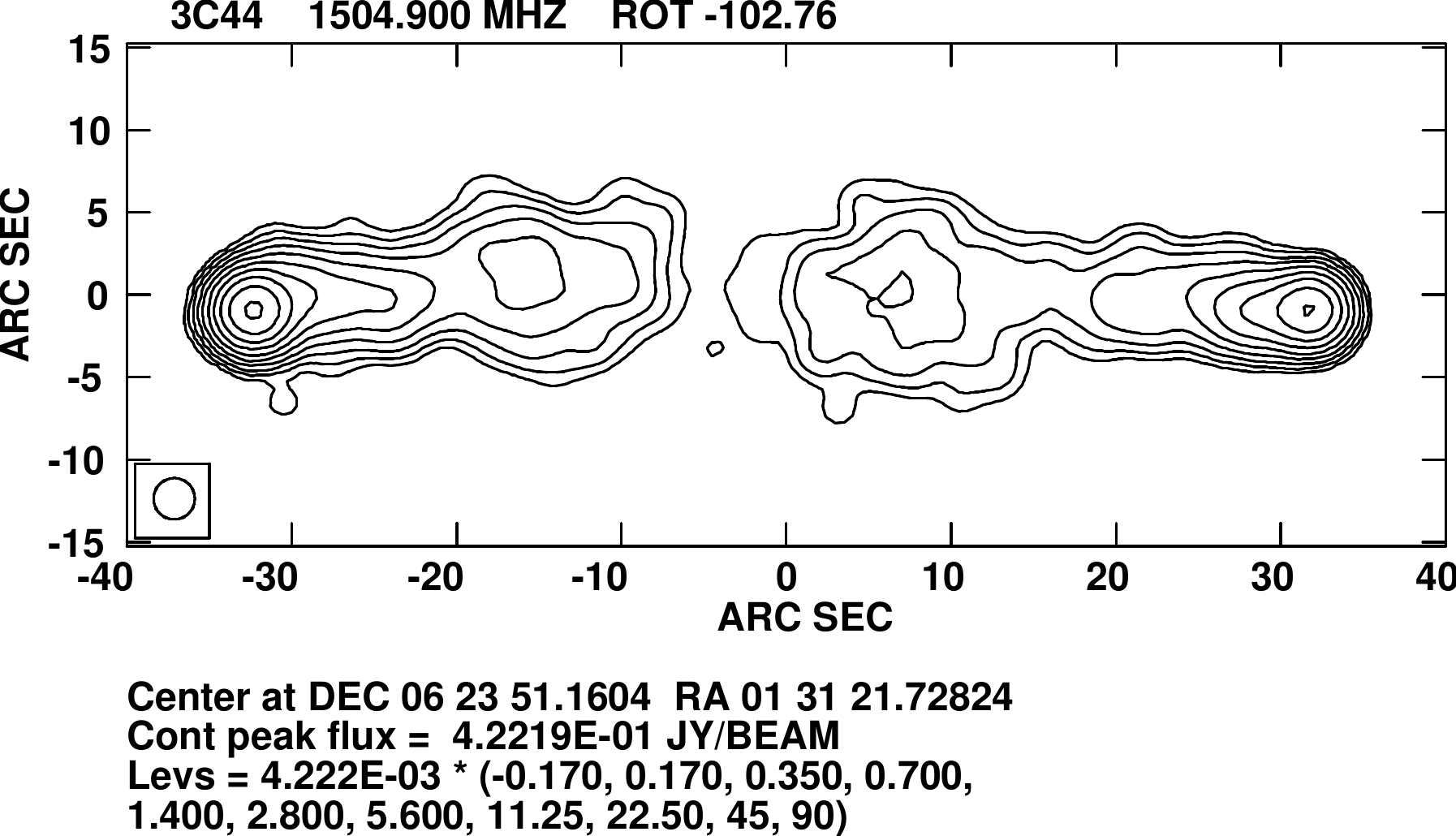}
\caption{3C44 at 1.4 GHz and 2.5'' resolution rotated by 102.76 deg 
clockwise, as in Fig. \ref{3C6.1rotfig}.}
\end{figure}

\begin{figure}
\plottwo{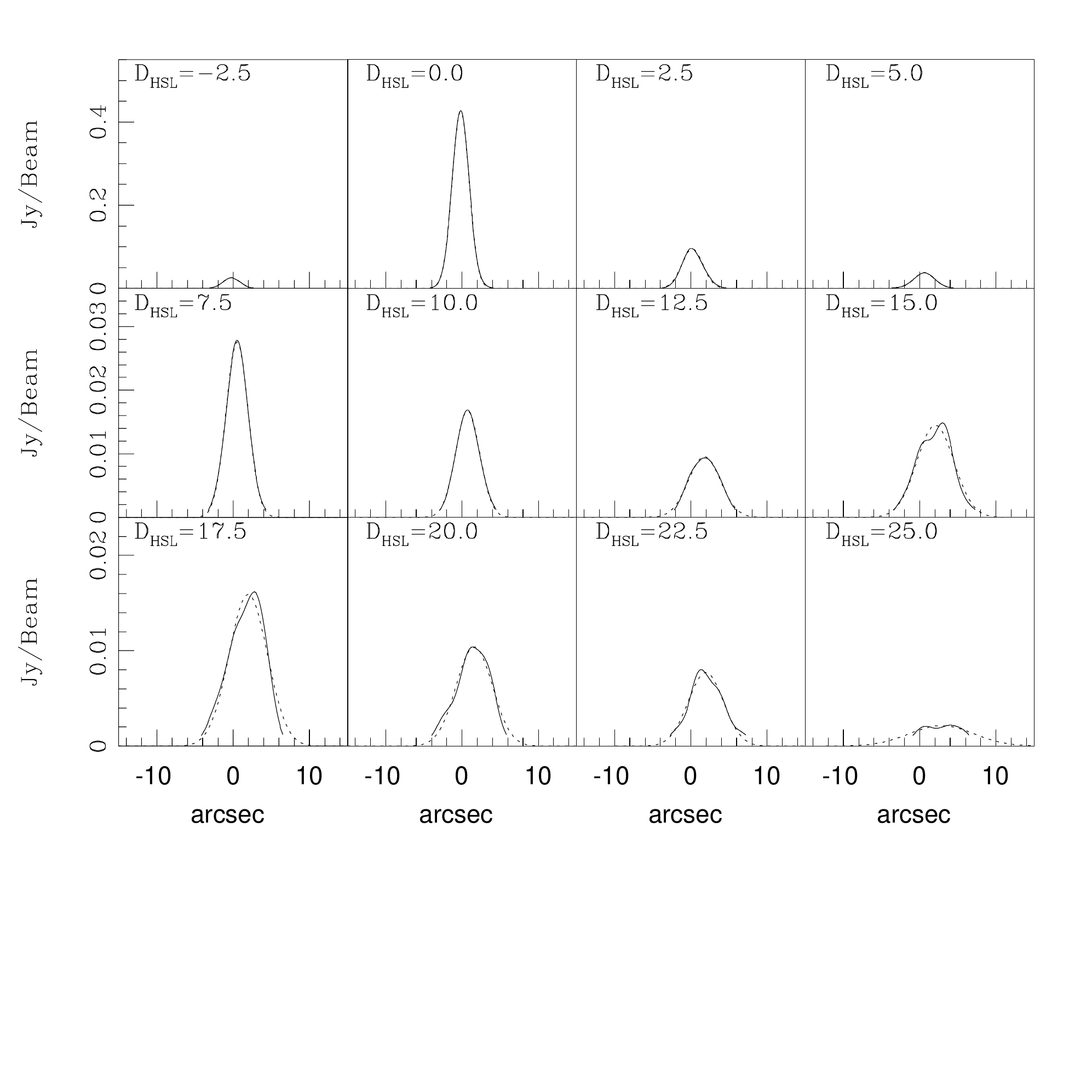}{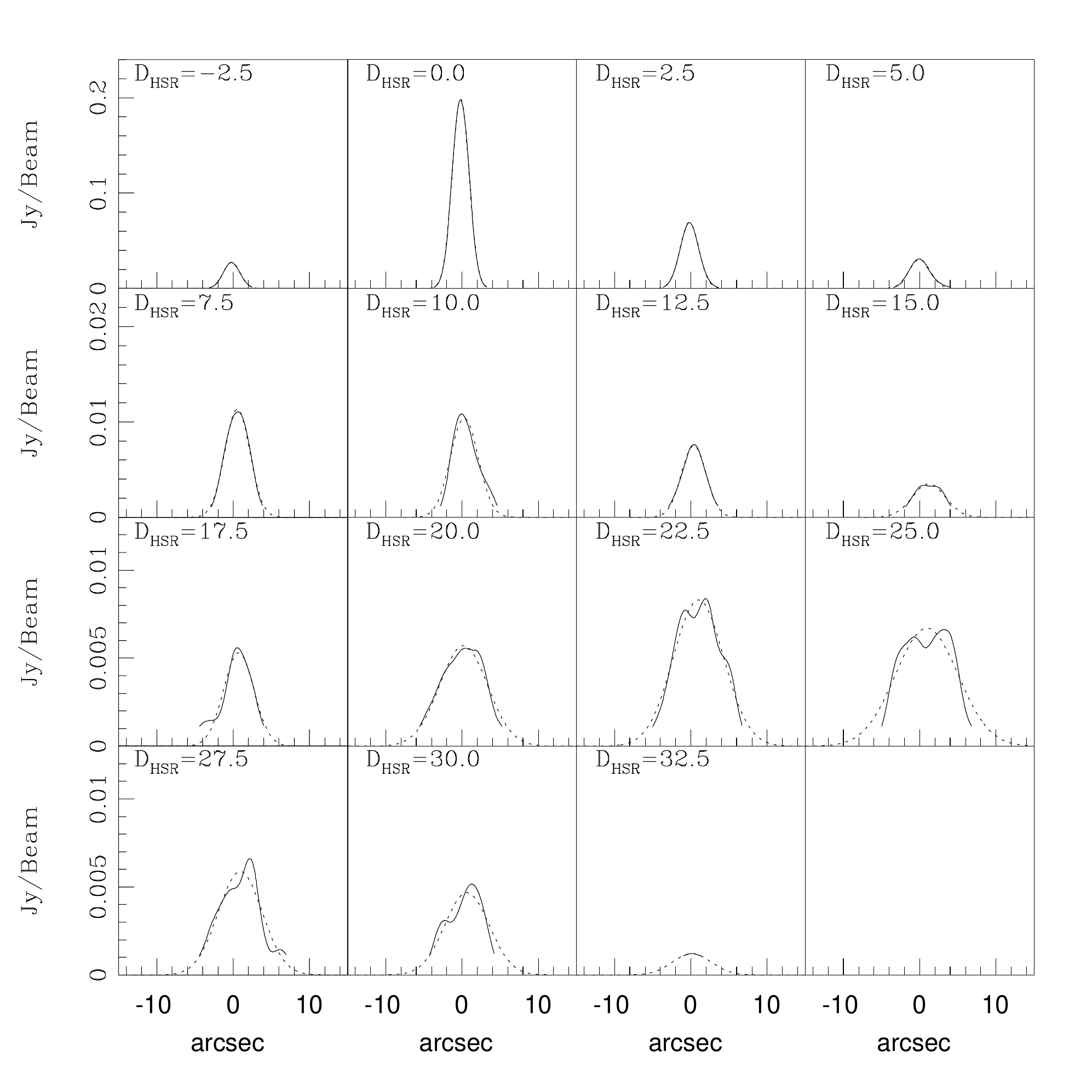}
\caption{3C44 
surface brightness (solid line) and best fit 
Gaussian (dotted line) for each cross-sectional slice of the 
radio bridge at 1.4 GHz, as in Fig. \ref{3C6.1SG}.
A Gaussian provides an excellent description of the surface brightness
profile of each cross-sectional slice. }
\label{3C44SG}
\end{figure}

\begin{figure}
\plottwo{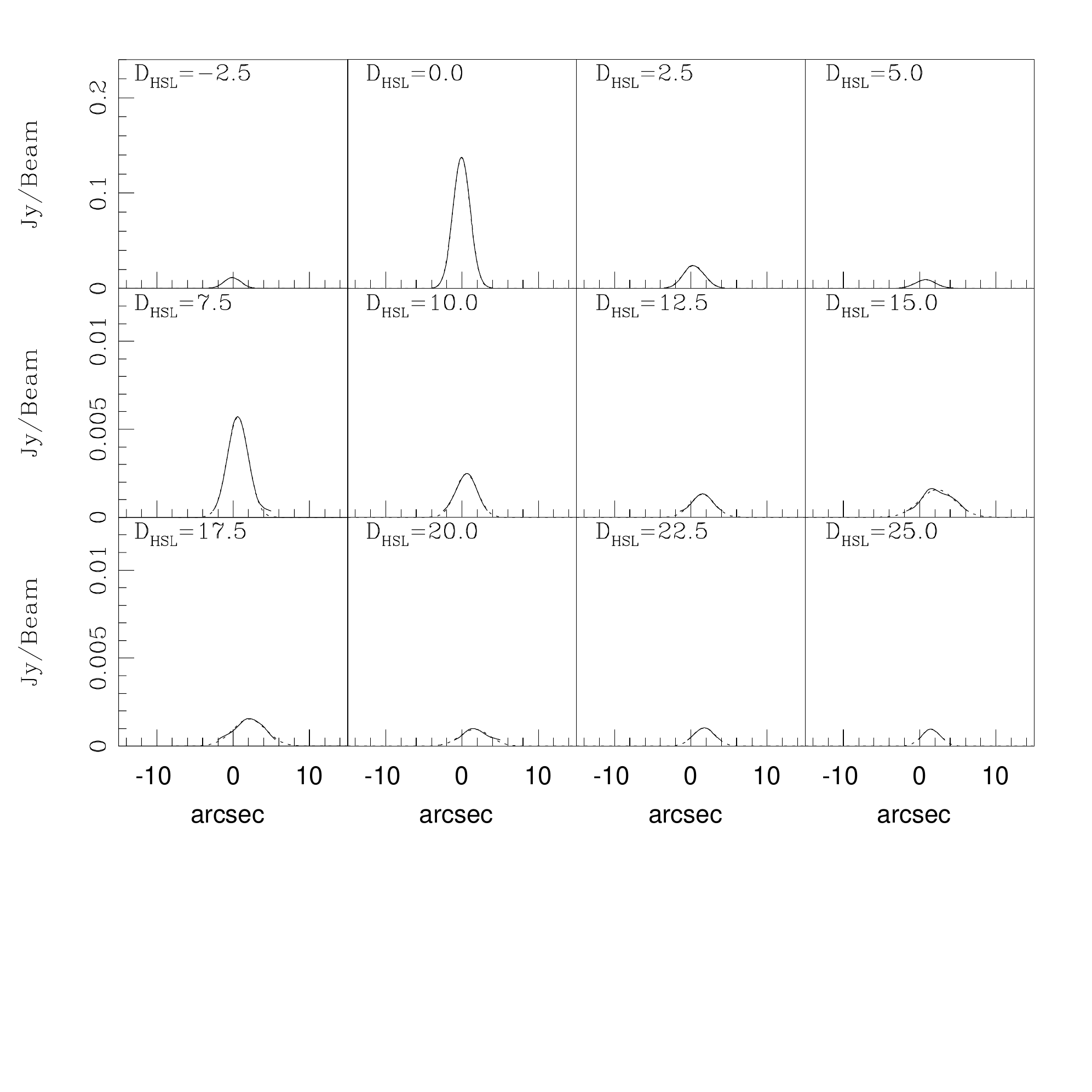}{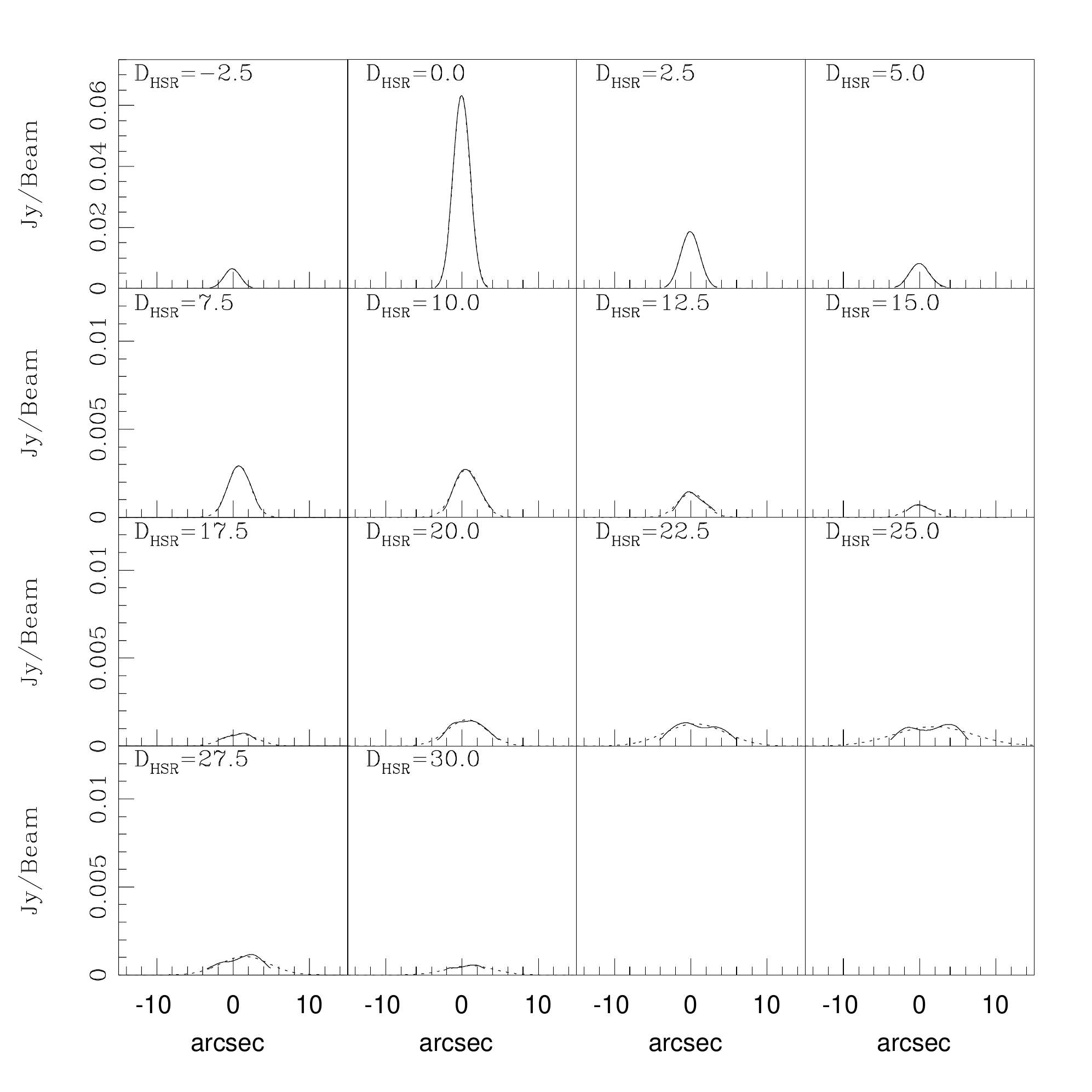}
\caption{3C44 
surface brightness (solid line) and best fit 
Gaussian (dotted line) for each cross-sectional slice of the 
radio bridge at 5 GHz, as in 
Fig. \ref{3C44SG} but at 5 GHz. 
Results obtained at 5 GHz are nearly identical to those
obtained at 1.4 GHz.}
\end{figure}

\begin{figure}
\plottwo{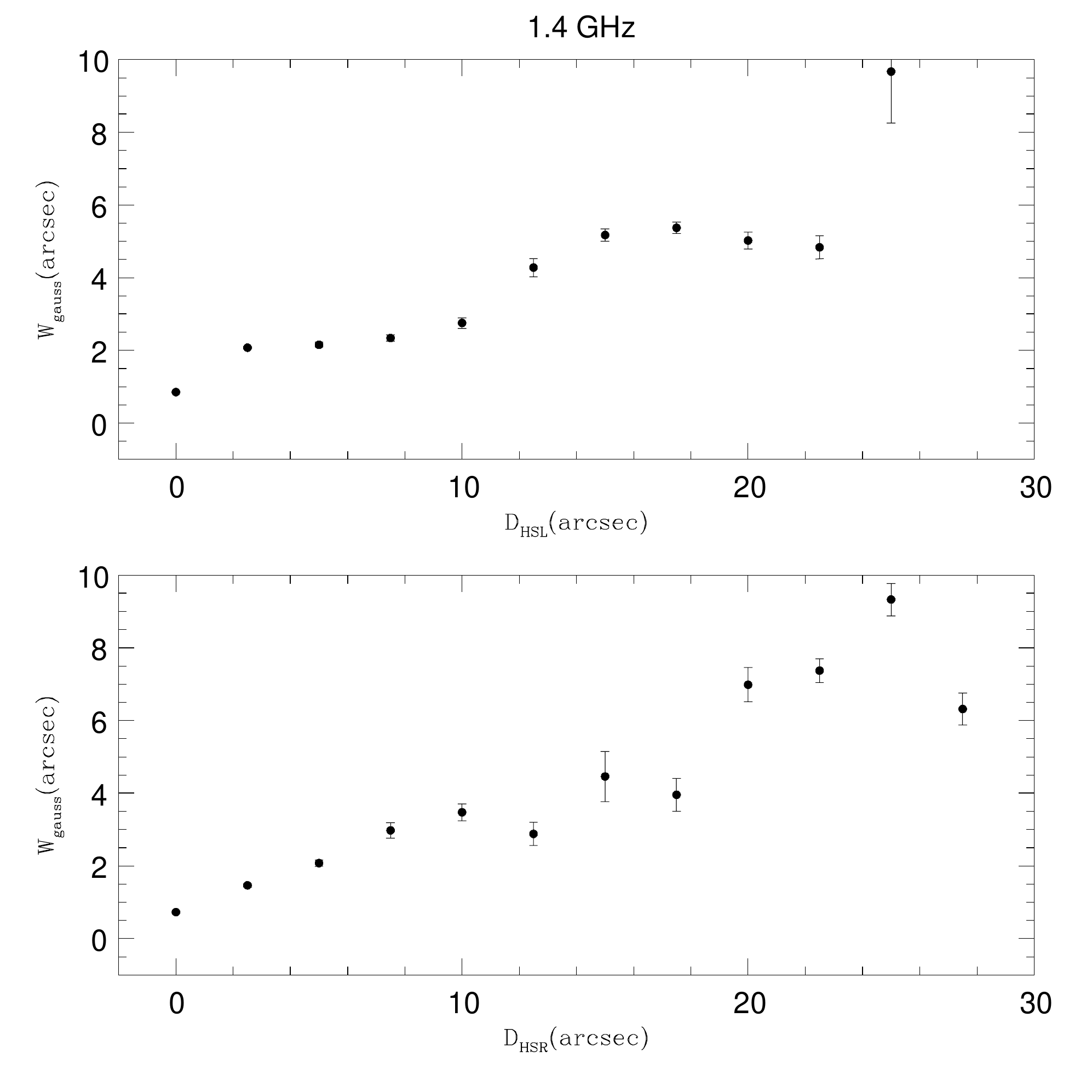}{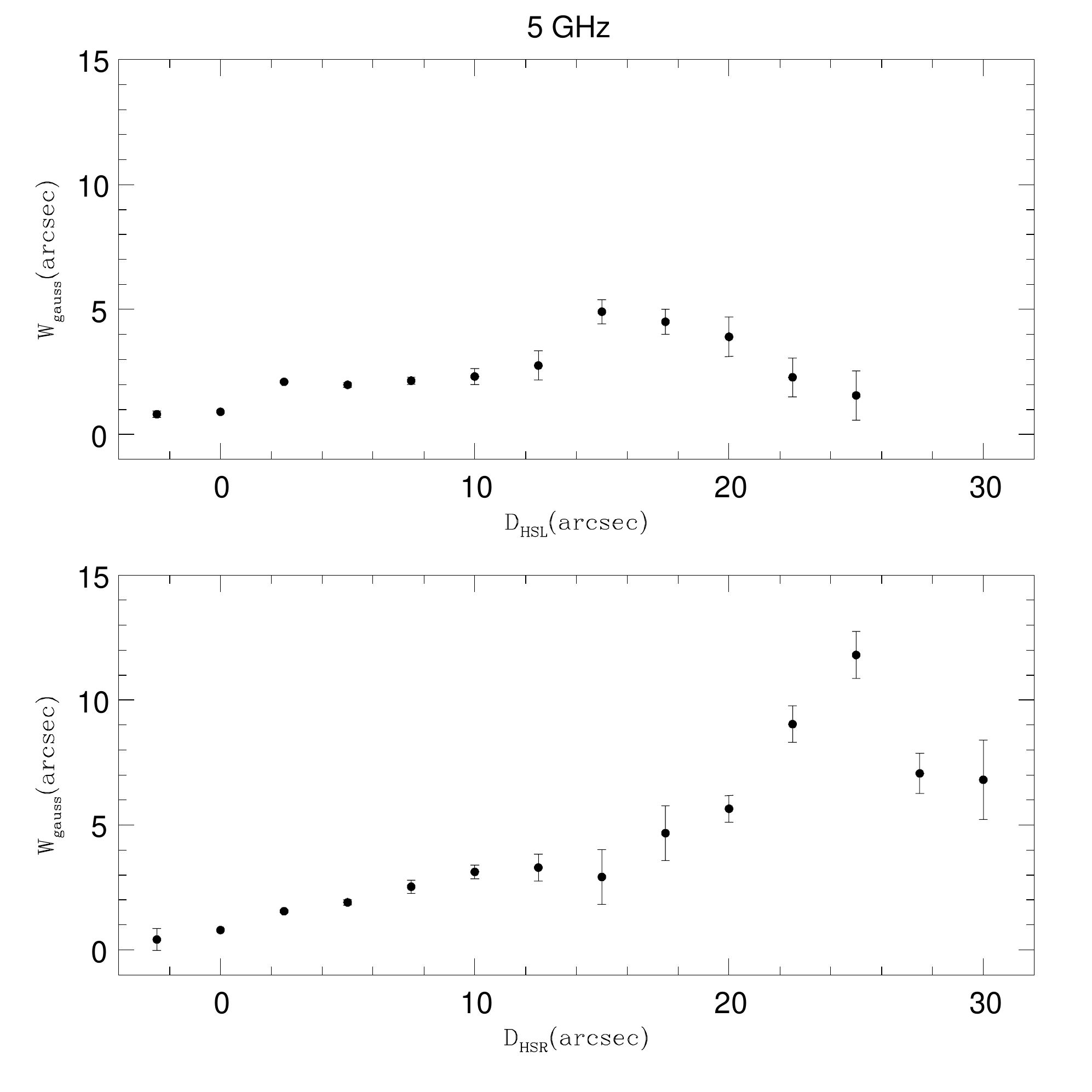}
\caption{3C44 
 Gaussian FWHM as a function of 
distance from the hot spot   
at 1.4 and 5 GHz (left and
right panels, respectively) for the left and right hand sides of the 
source (top and bottom panels, respectively), as in Fig. \ref{3C6.1WG}.
The right and left hand sides of the source are fairly 
symmetric.  Similar results are obtained at 
1.4 and 5 GHz.}
\end{figure}

\begin{figure}
\plottwo{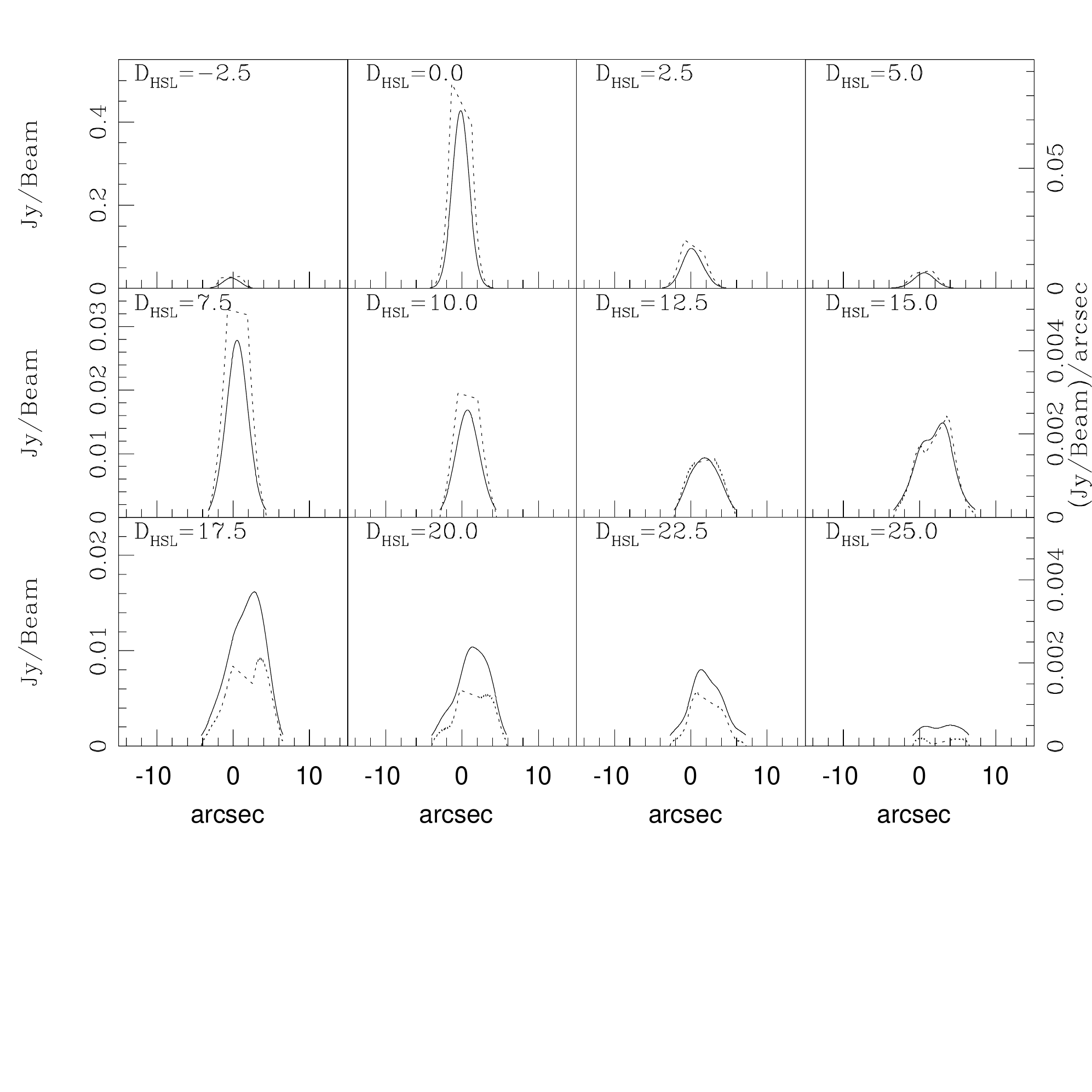}{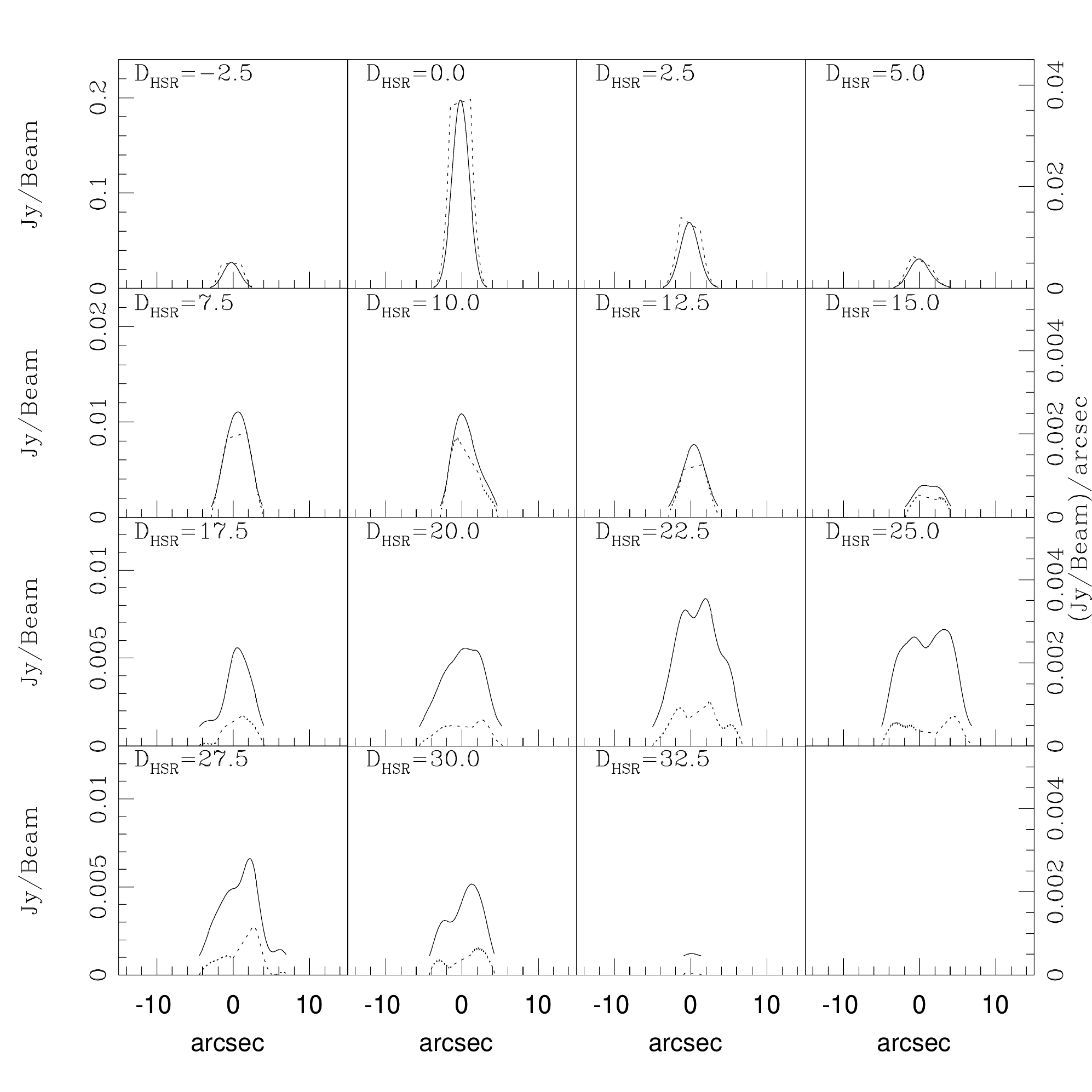}
\caption{3C44 emissivity (dotted line) and surface brightness
(solid line) for each cross-sectional slice of the radio 
bridge at 1.4 GHz, as in Fig. \ref{3C6.1SEM}. A constant volume
emissivity per slice provides a reasonable description of the 
data over most of the source. }
\label{3C44SEM}
\end{figure}

\begin{figure}
\plottwo{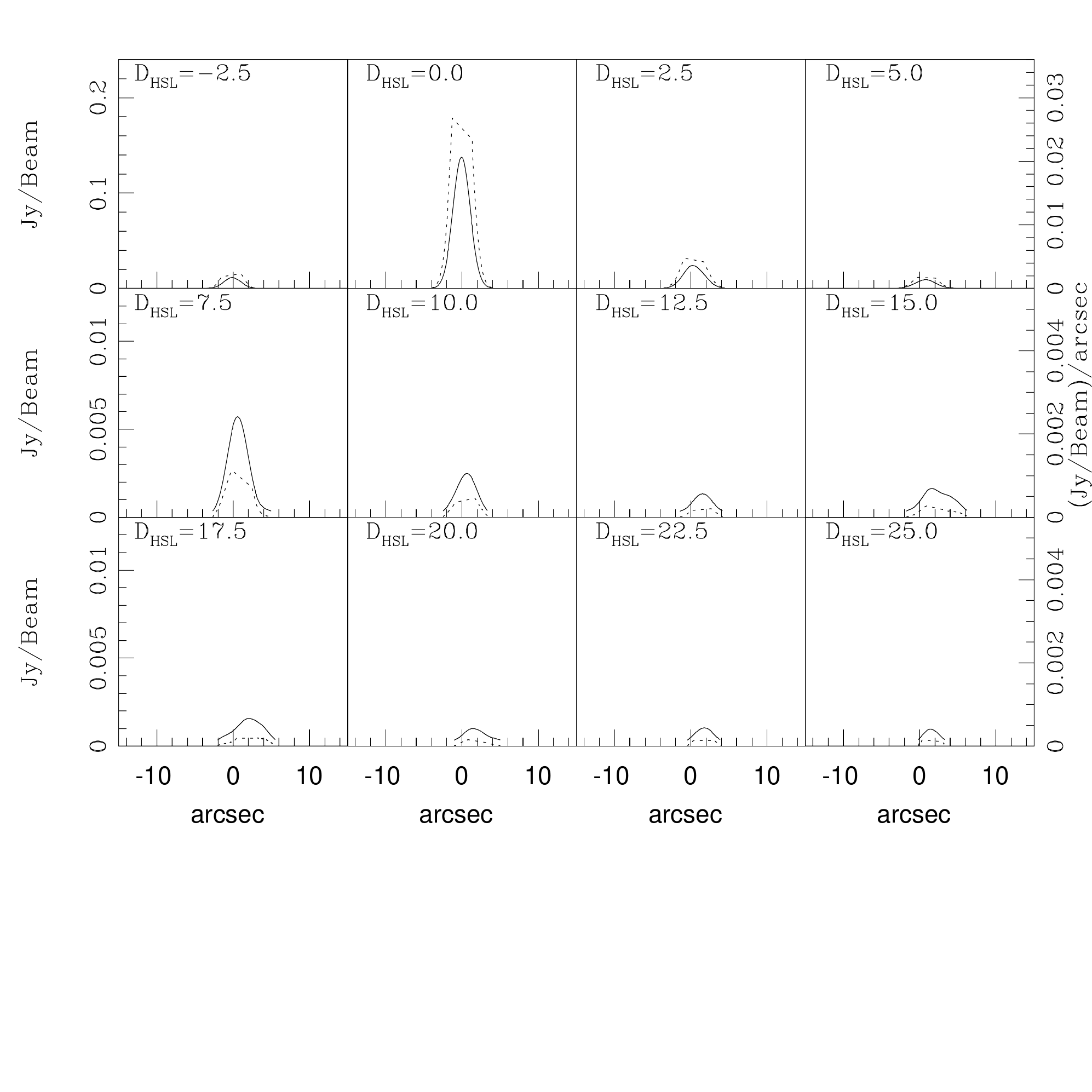}{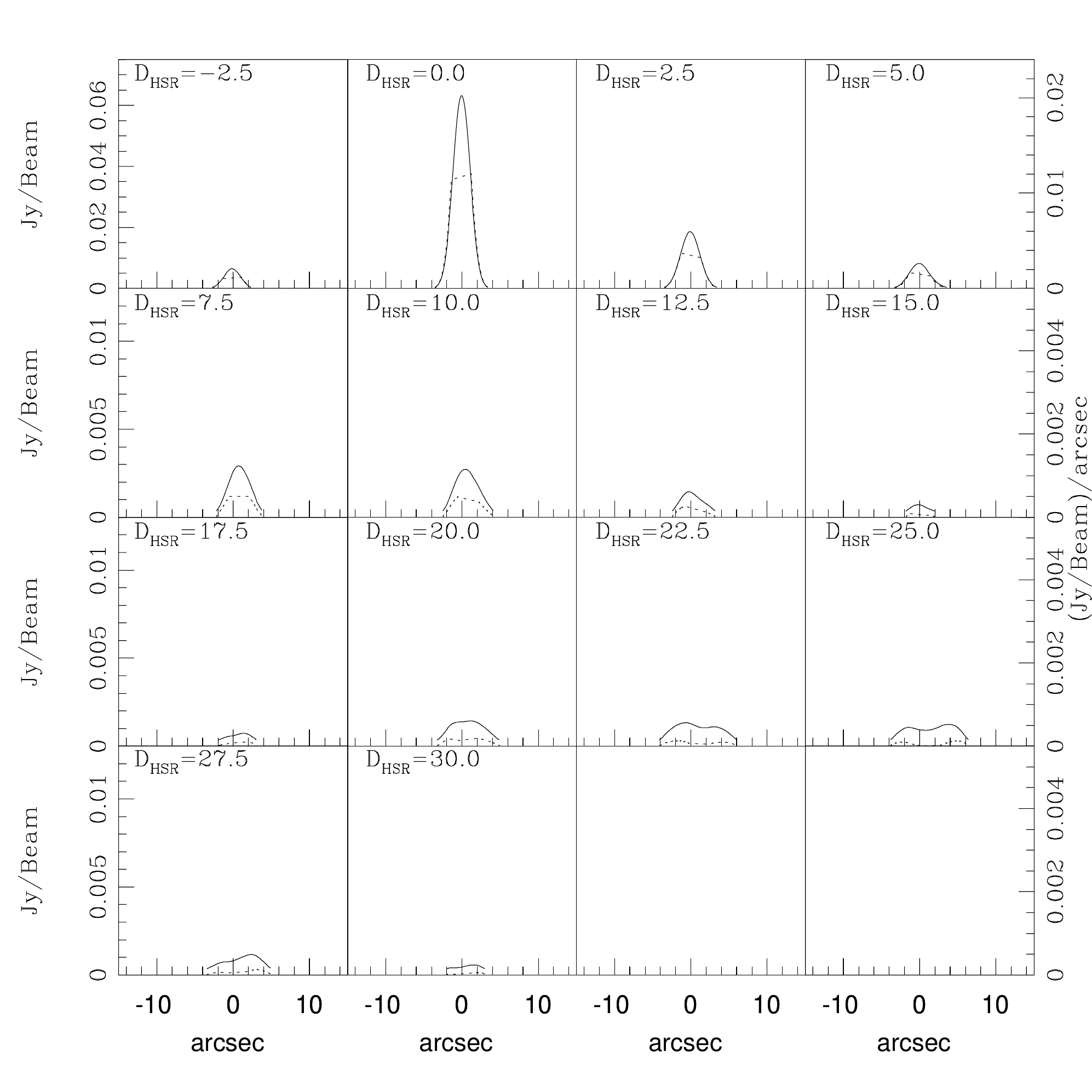}
\caption{3C44
emissivity (dotted line) and surface brightness
(solid line) for each cross-sectional slice of the radio 
bridge at 5 GHz, as in Fig. \ref{3C44SEM} but at 5 GHz.
Results obtained at 5 GHz are nearly identical to those
obtained at 1.4 GHz.}
\end{figure}

\begin{figure}
\plottwo{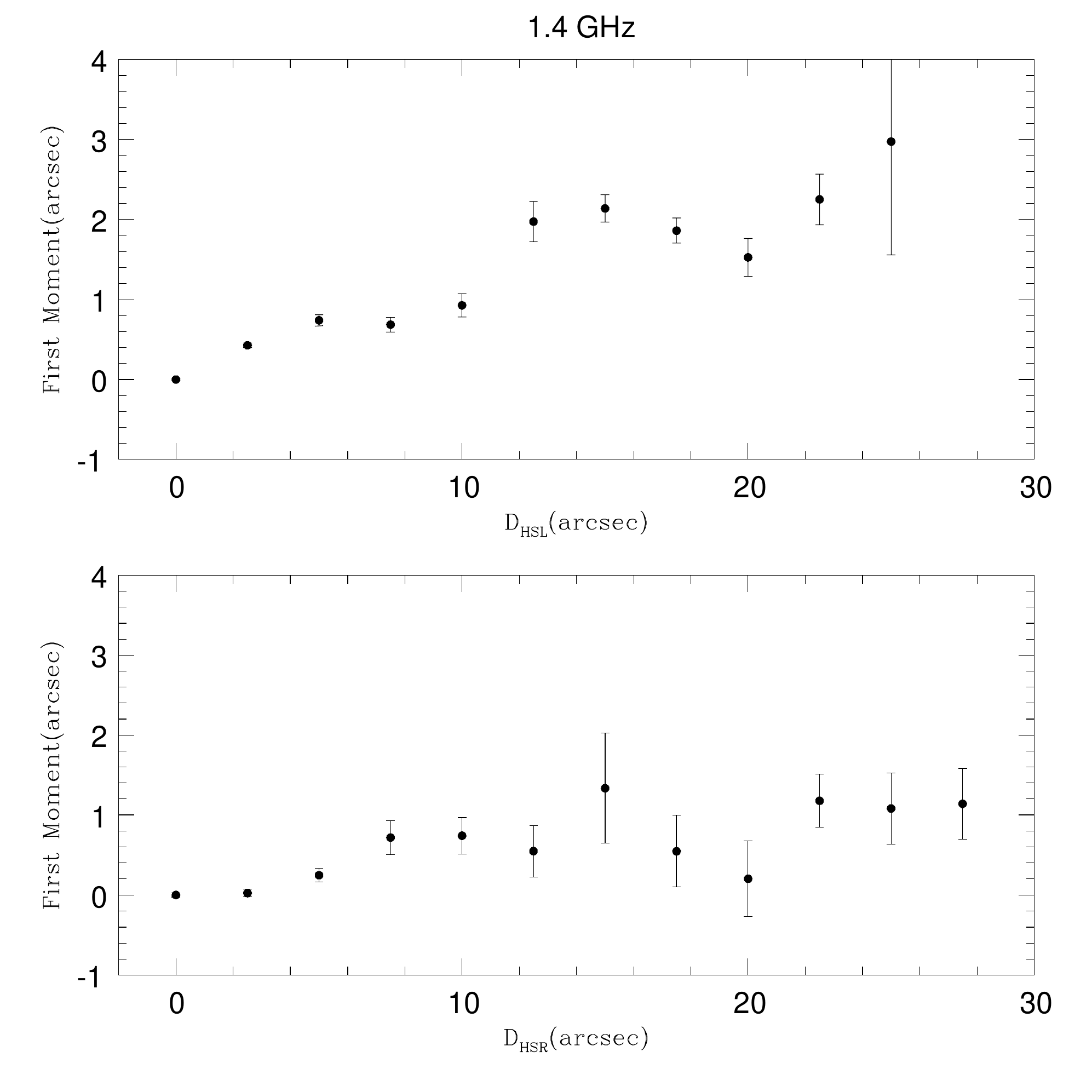}{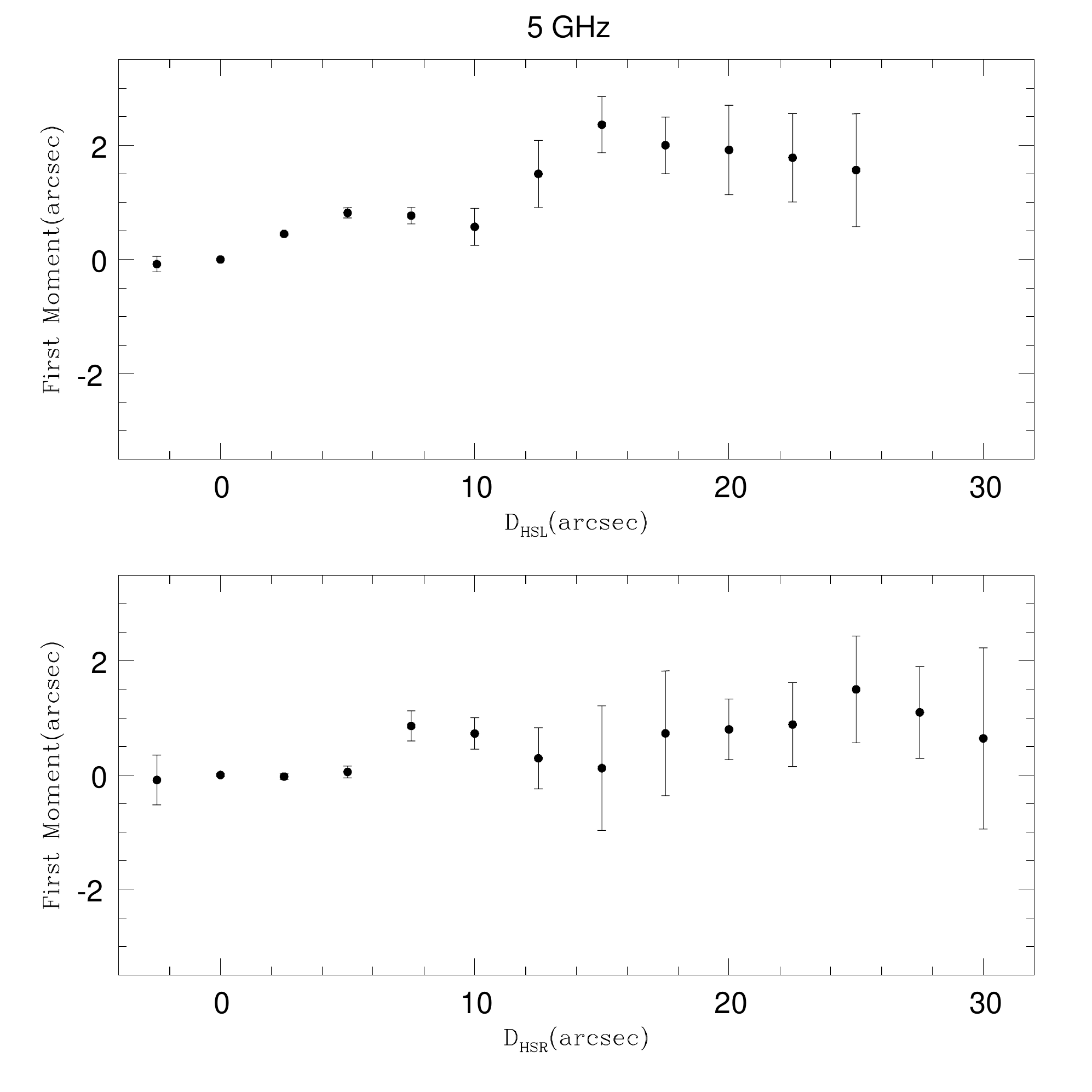}
\caption{3C44 first moment as a function of distance from 
the hot spot at 1.4 GHz and 5 GHz (left and right panels,
respectively) for the left and right hand sides of the source
(top and bottom, respectively).}
\end{figure}

\begin{figure}
\plottwo{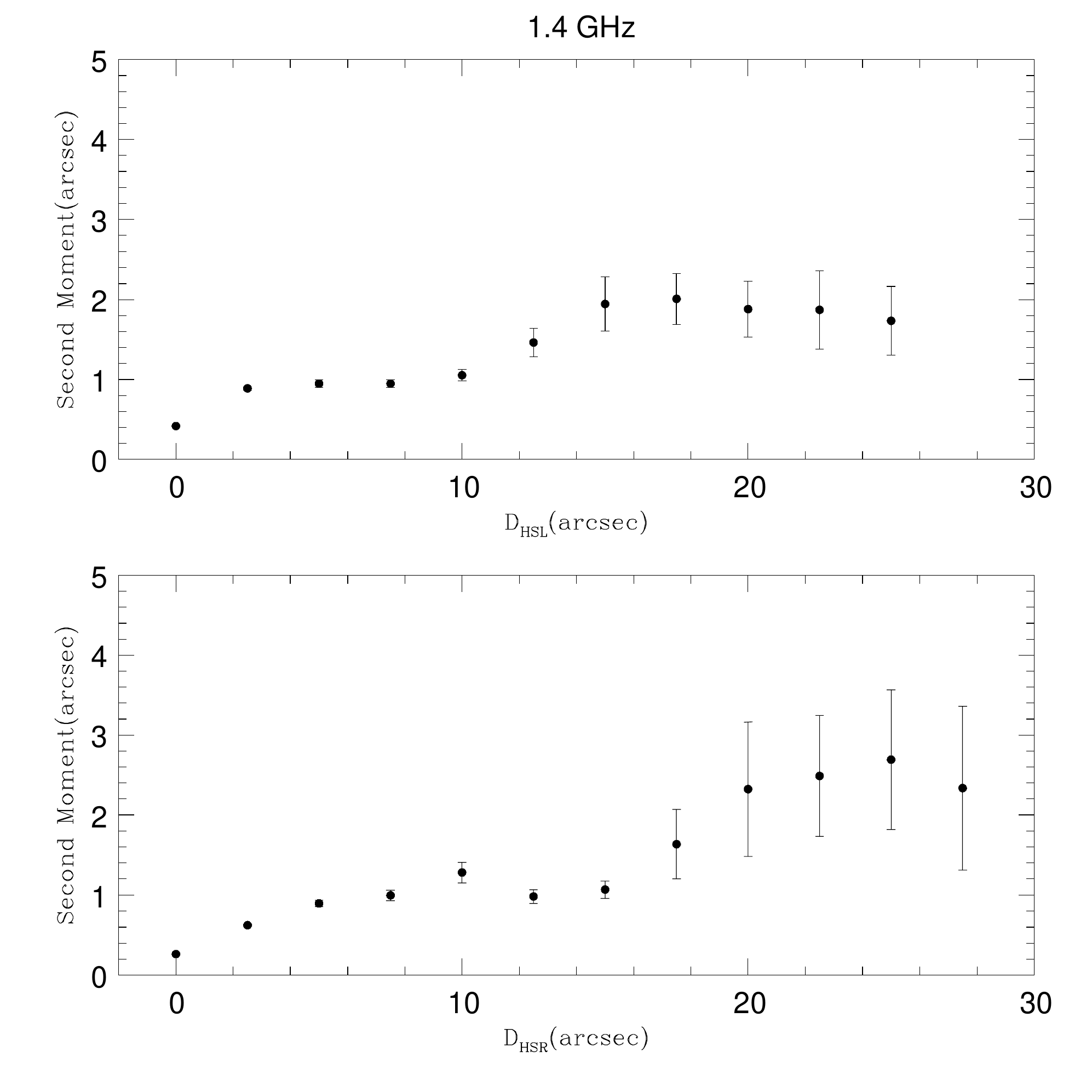}{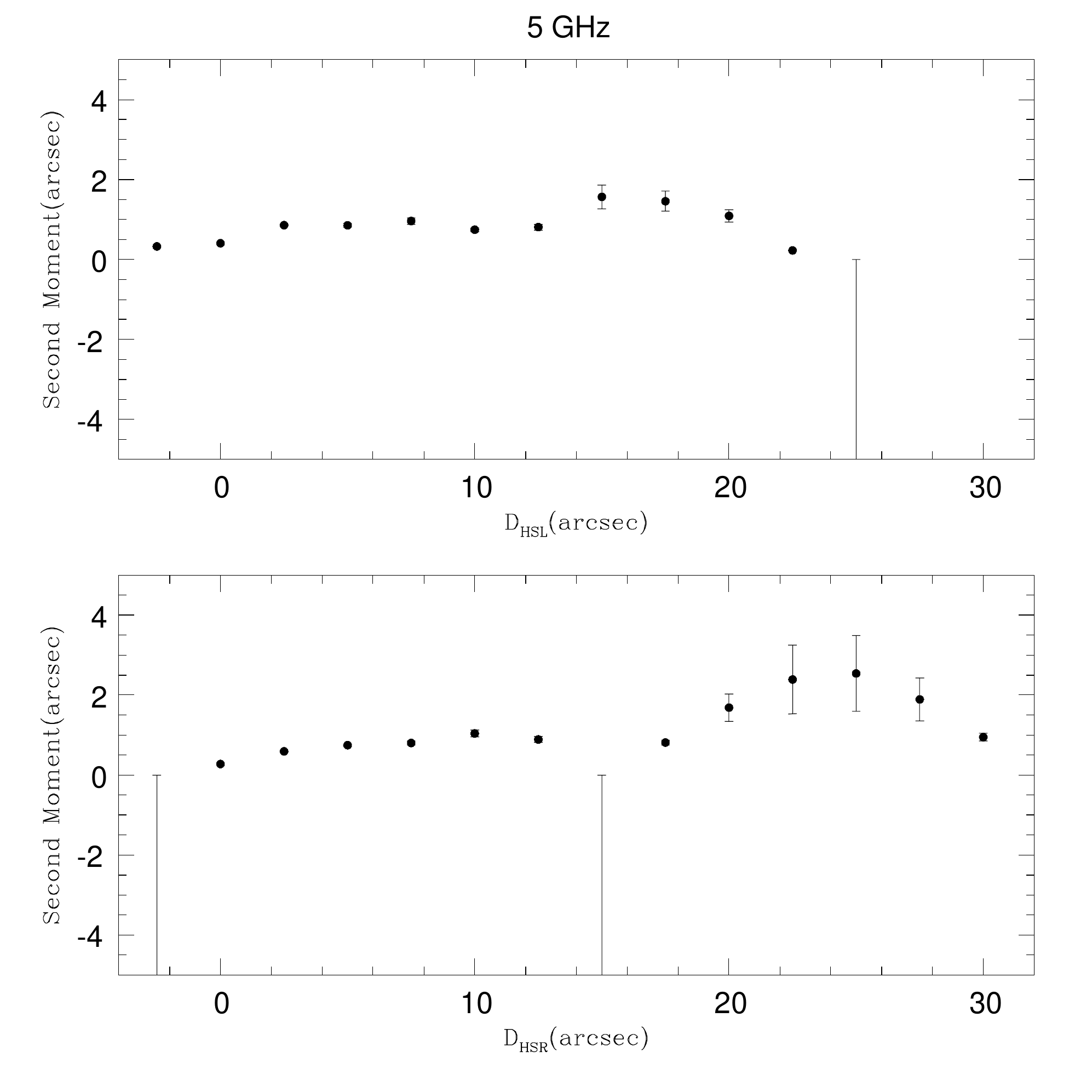}
\caption{3C44 second moment as a function of distance from the
hot spot at 1.4 GHz and 5 GHz (left and right panels, 
respectively) for the left and right sides of the source
(top and bottom panels, respectively).}
\end{figure}

\begin{figure}
\plottwo{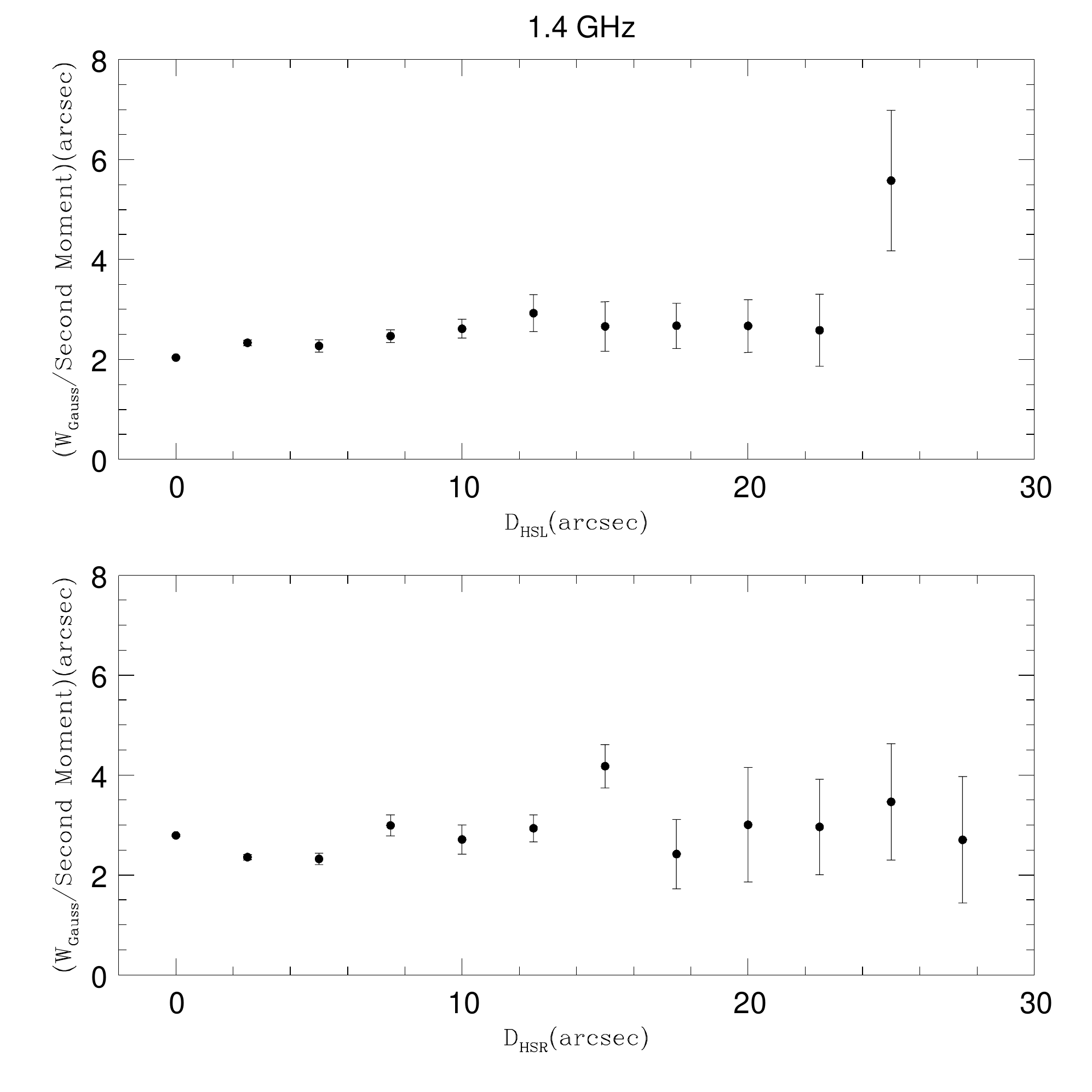}{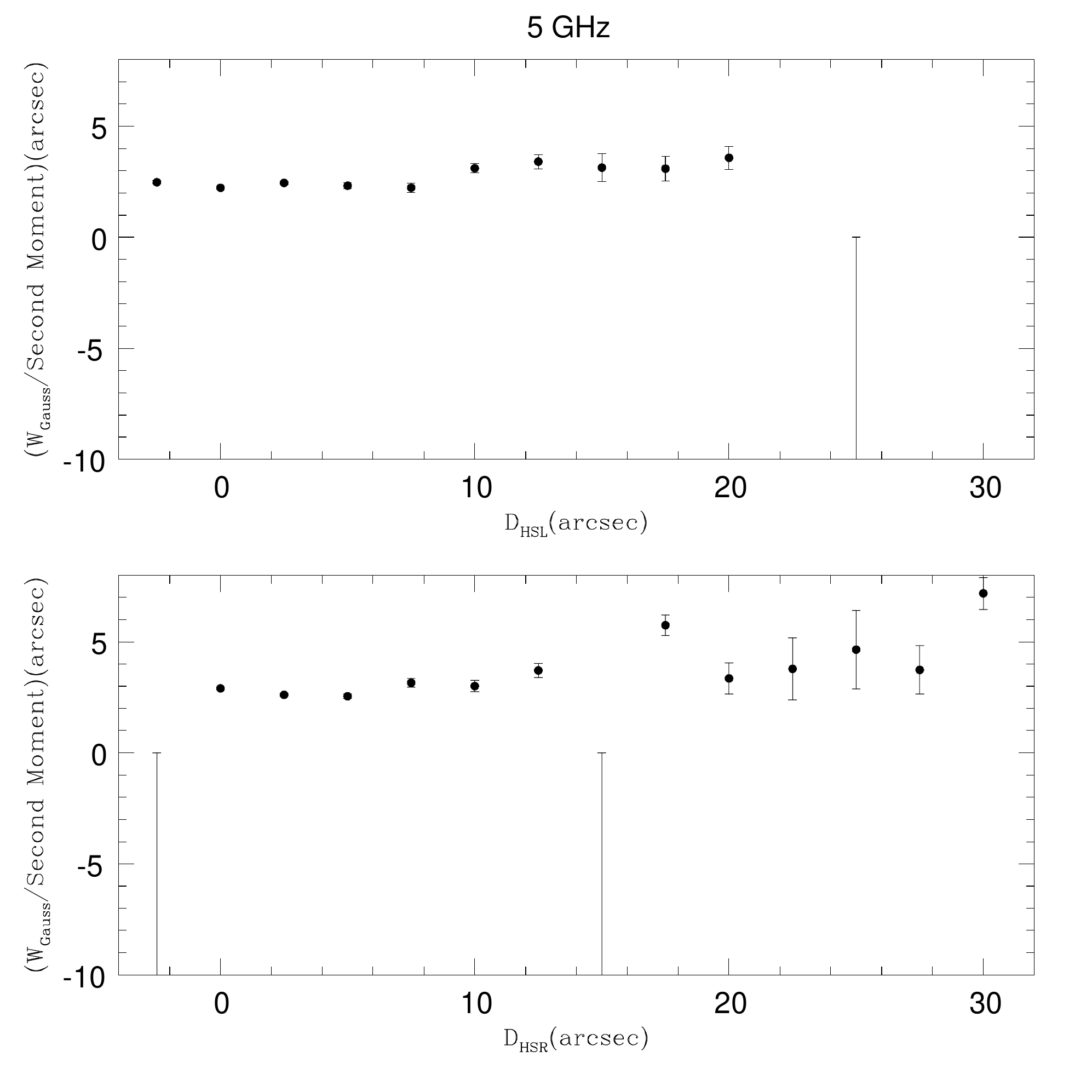}
\caption{3C44 ratio of the Gaussian FWHM to the second moment as a
function of distance from the hot spot at 1.4 and 5 GHz 
(left and right panels, respectively) for the left and right
hand sides of the source (top and bottom panels, respectively). The value of this ratio is fairly constant for each side of the source, and has a value of about 2. Similar results are obtained at 1.4 and 5 GHz.} 
\end{figure}

\begin{figure}
\plottwo{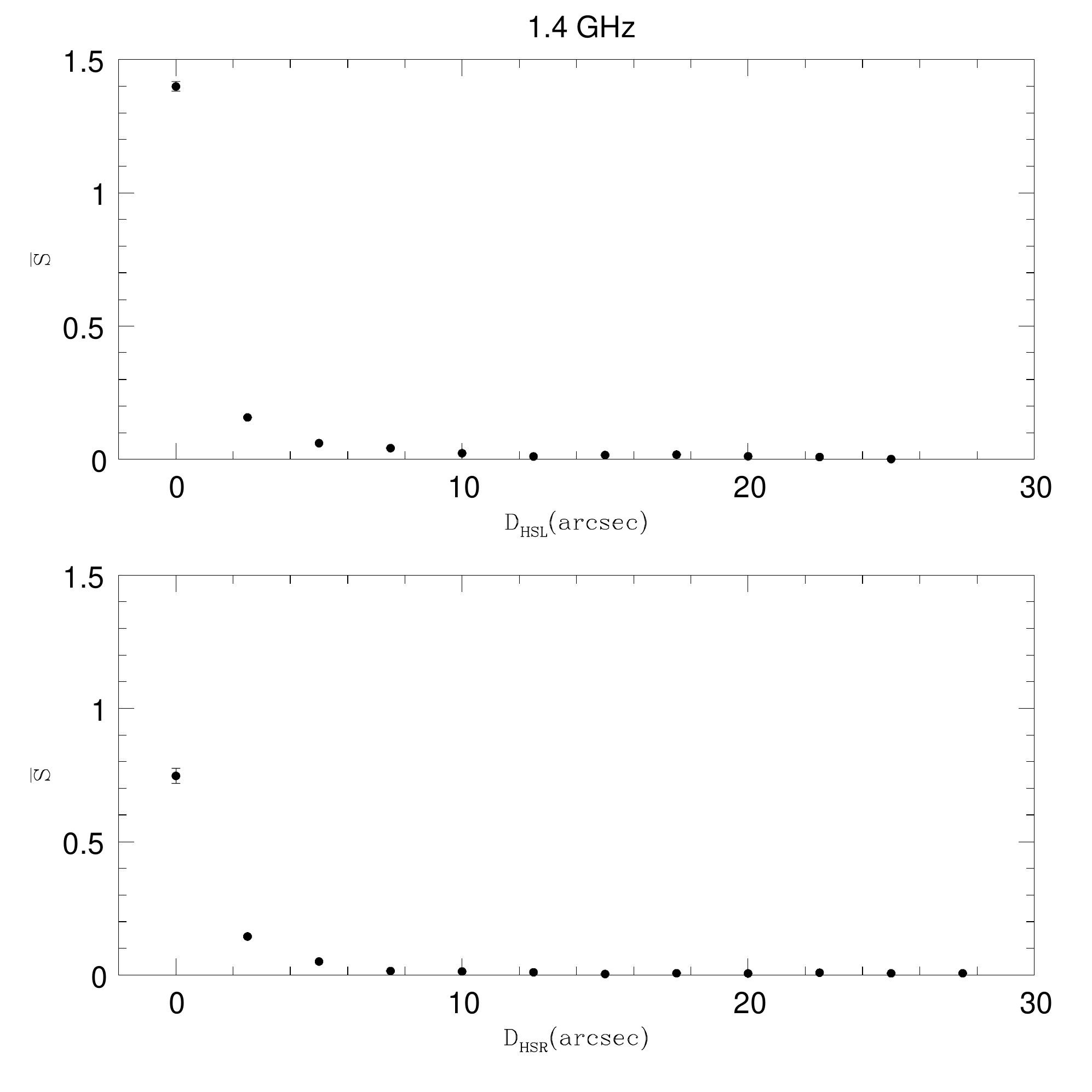}{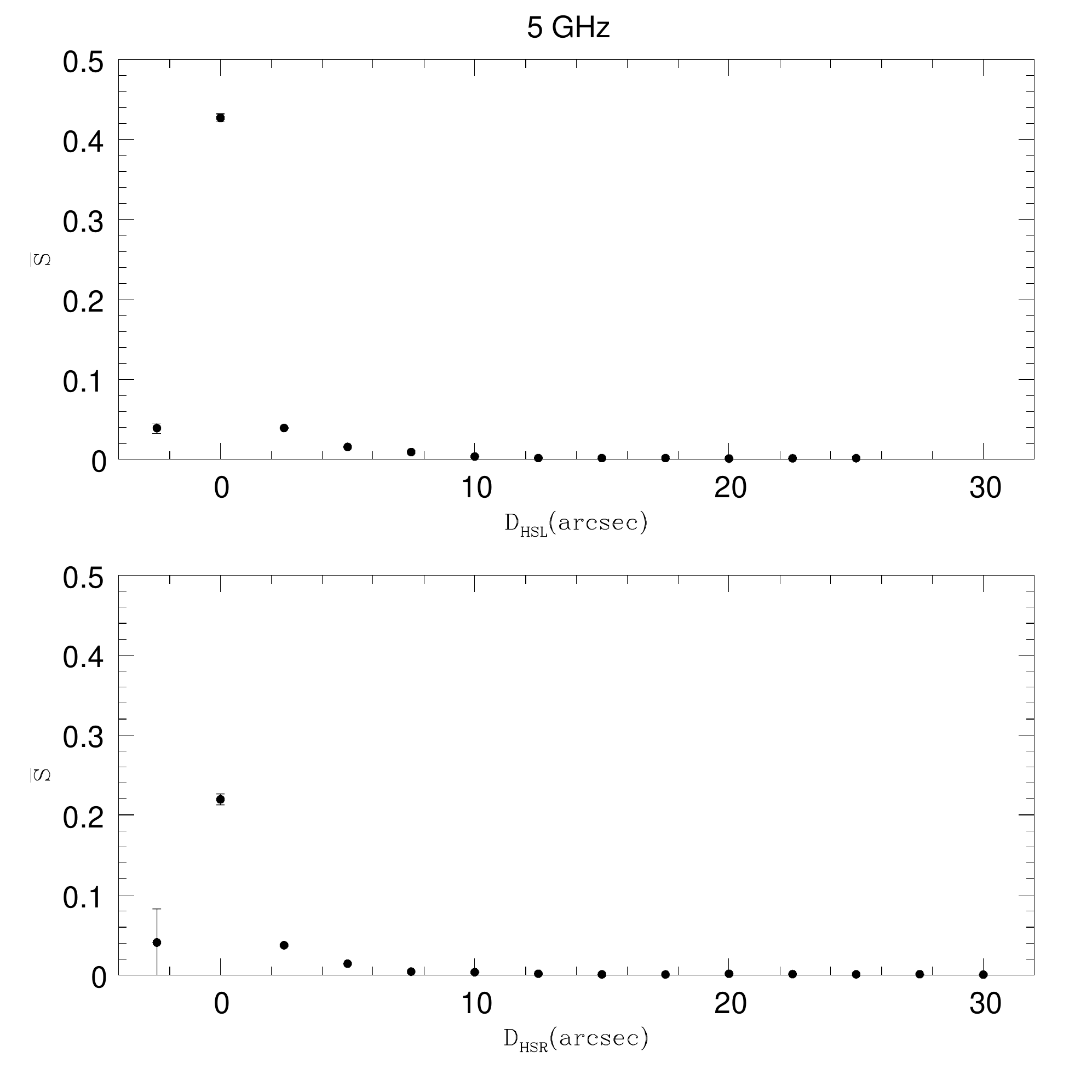}
\caption{3C44 average surface brightness in units of Jy/beam 
as a function of distance from the 
hot spot at 1.4 and 5 GHz 
(left and right panels, respectively) for the left and right
hand sides of the source (top and bottom panels, respectively).}
\end{figure}

\begin{figure}
\plottwo{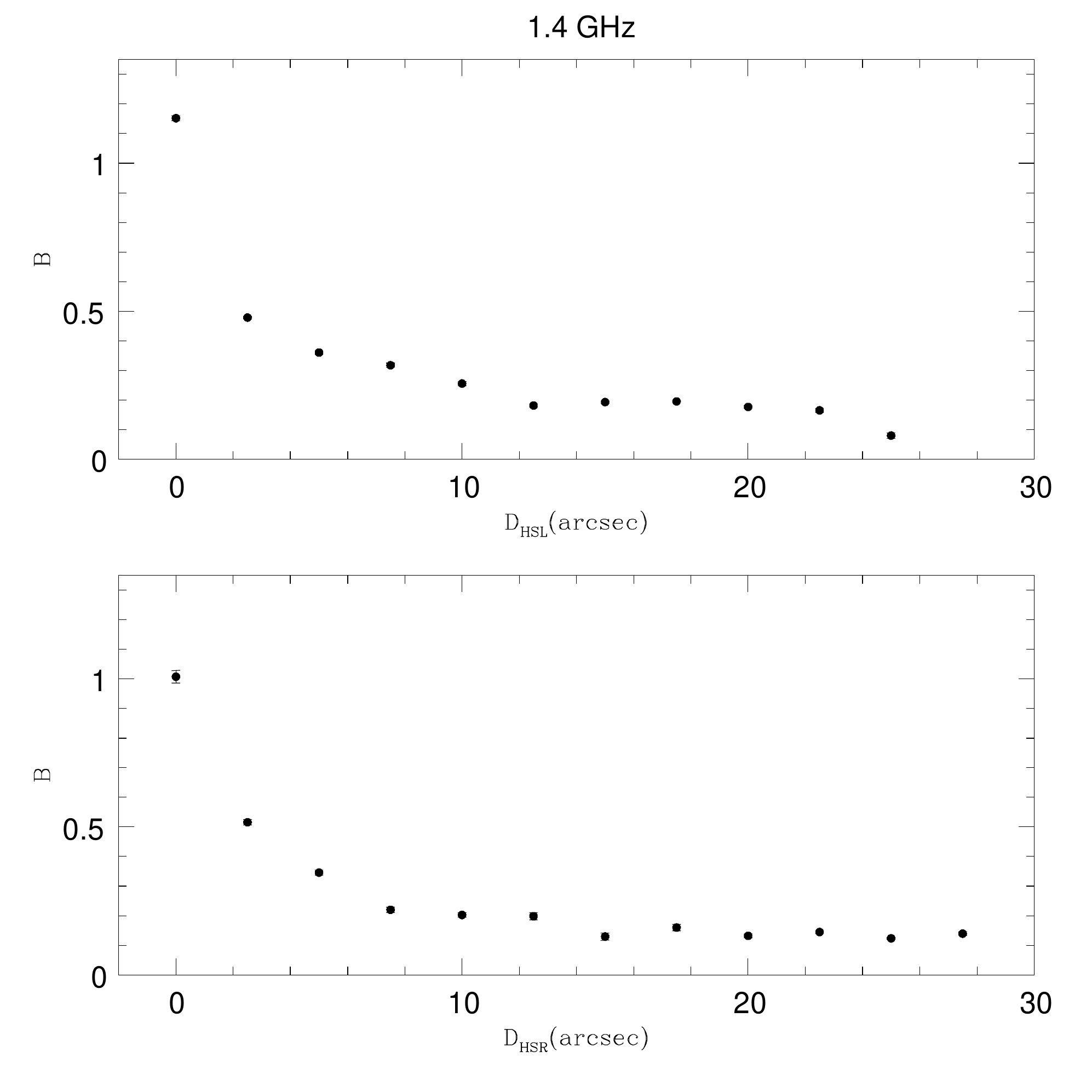}{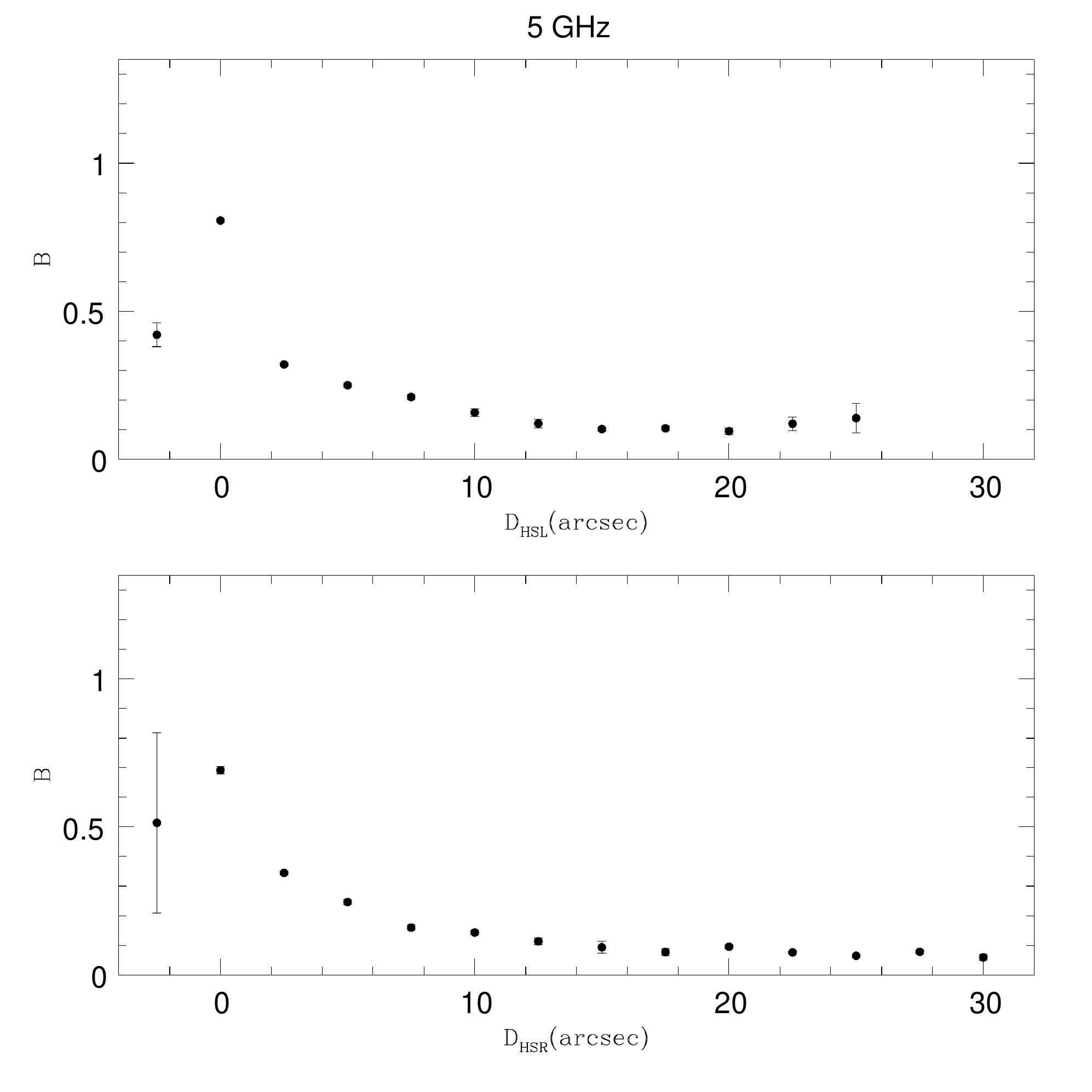}
\caption{3C44 minimum energy magnetic field strength 
as a function of distance from the 
hot spot at 1.4 and 5 GHz 
(left and right panels, respectively) for the left and right
hand sides of the source (top and bottom panels, respectively).
The normalization is given in Table 1. 
}
\end{figure}

\begin{figure}
\plottwo{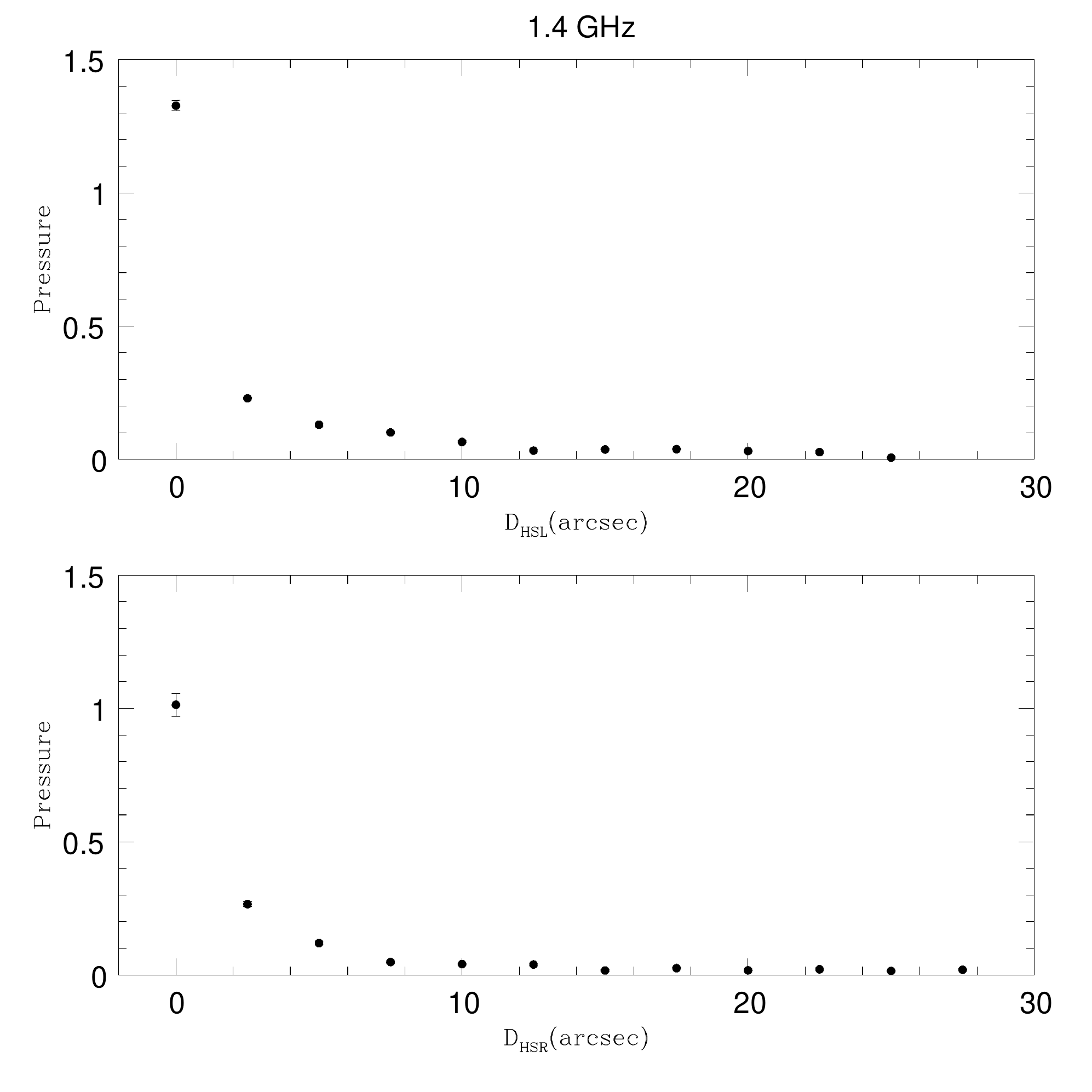}{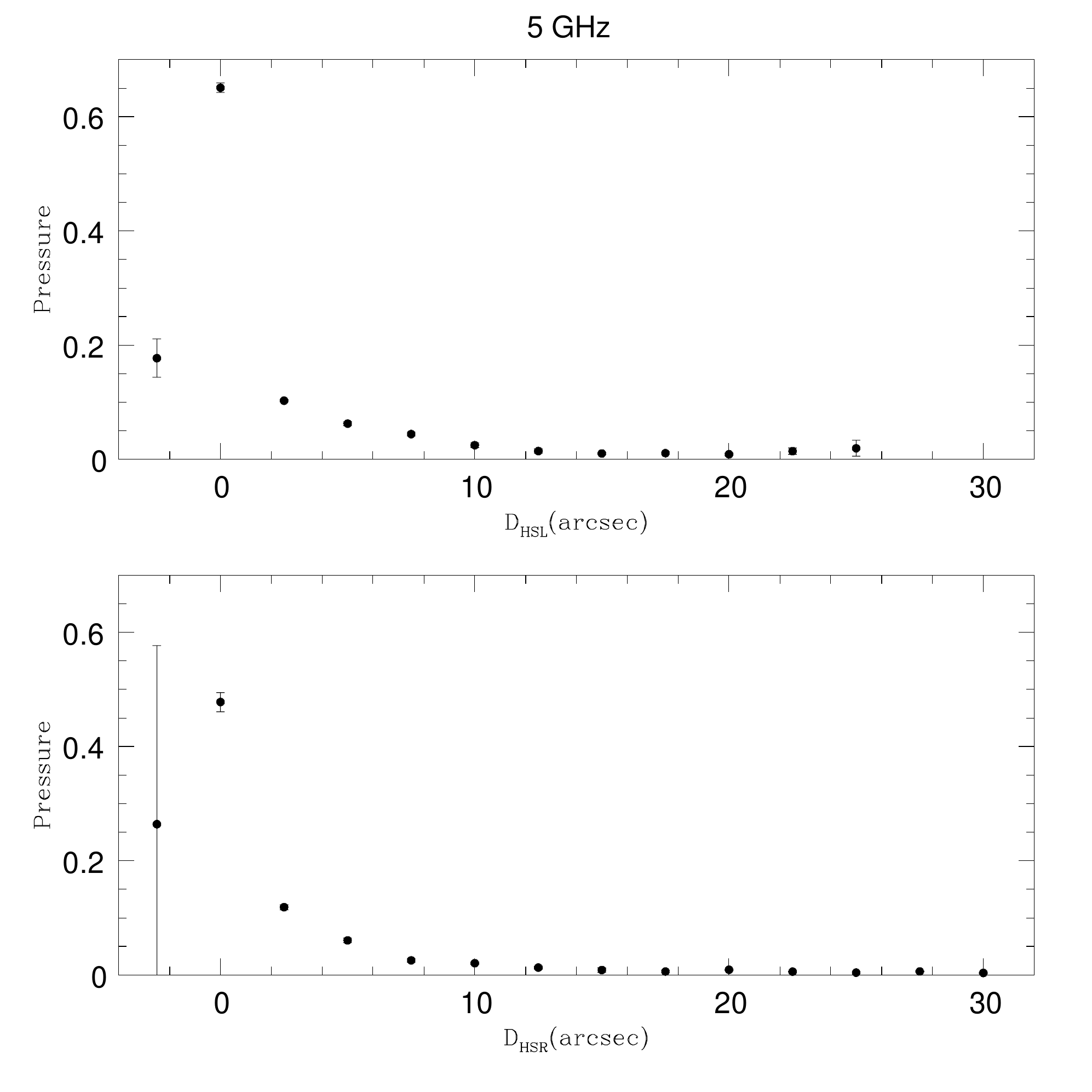}
\caption{3C44 
minimum energy pressure as a function of 
distance from the hot spot at 1.4 and 5 GHz 
(left and right panels, respectively) for the left and right
hand sides of the source (top and bottom panels, respectively), 
as in Fig. \ref{3C6.1P}. The normalization is given in Table 1.
}
\end{figure}

\clearpage
\begin{figure}
\plotone{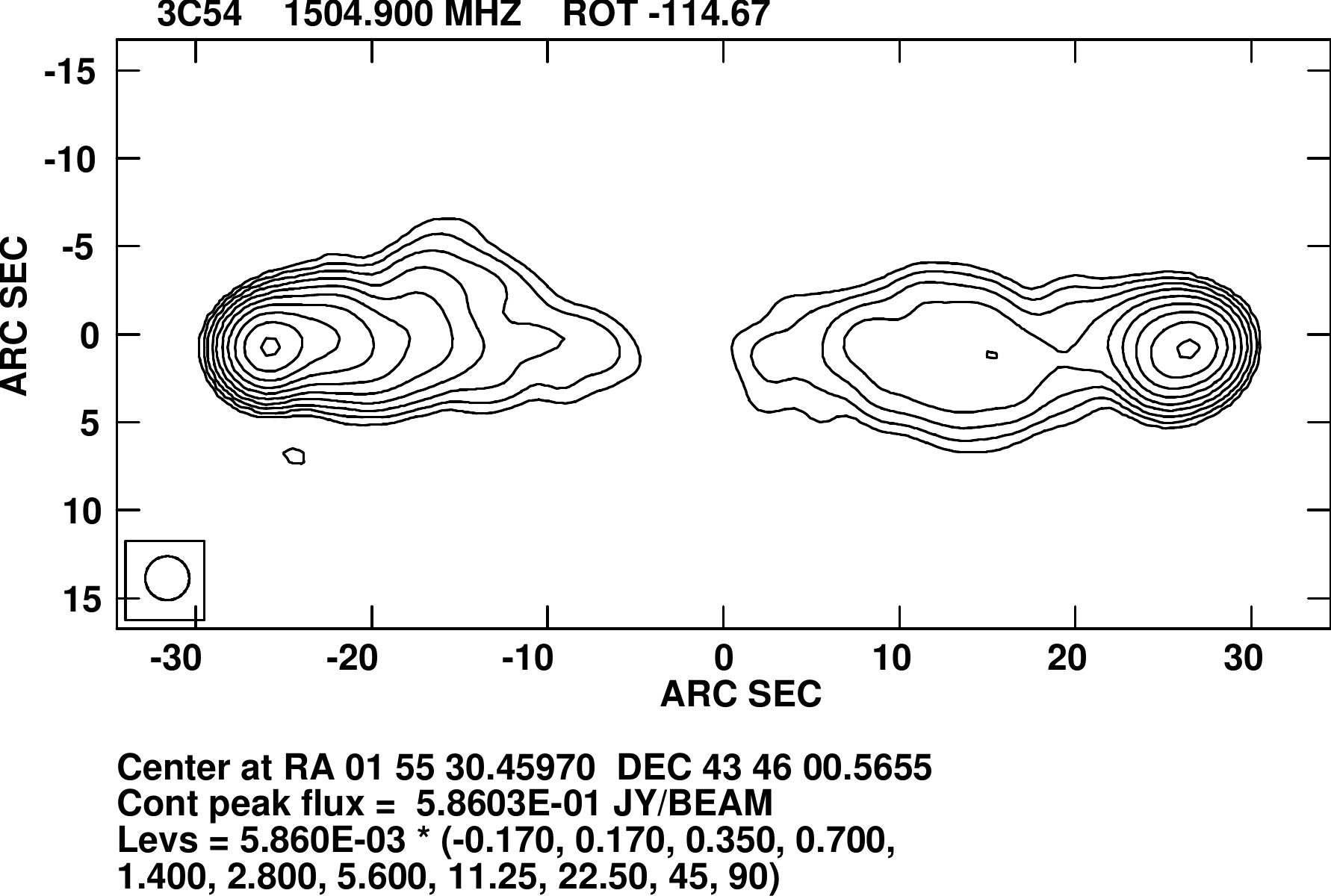}
\caption{3C54 at 1.4 GHz and 2.5'' resolution rotated by 114.67 deg 
clockwise, as in Fig. \ref{3C6.1rotfig}.}
\end{figure}

\begin{figure}
\plottwo{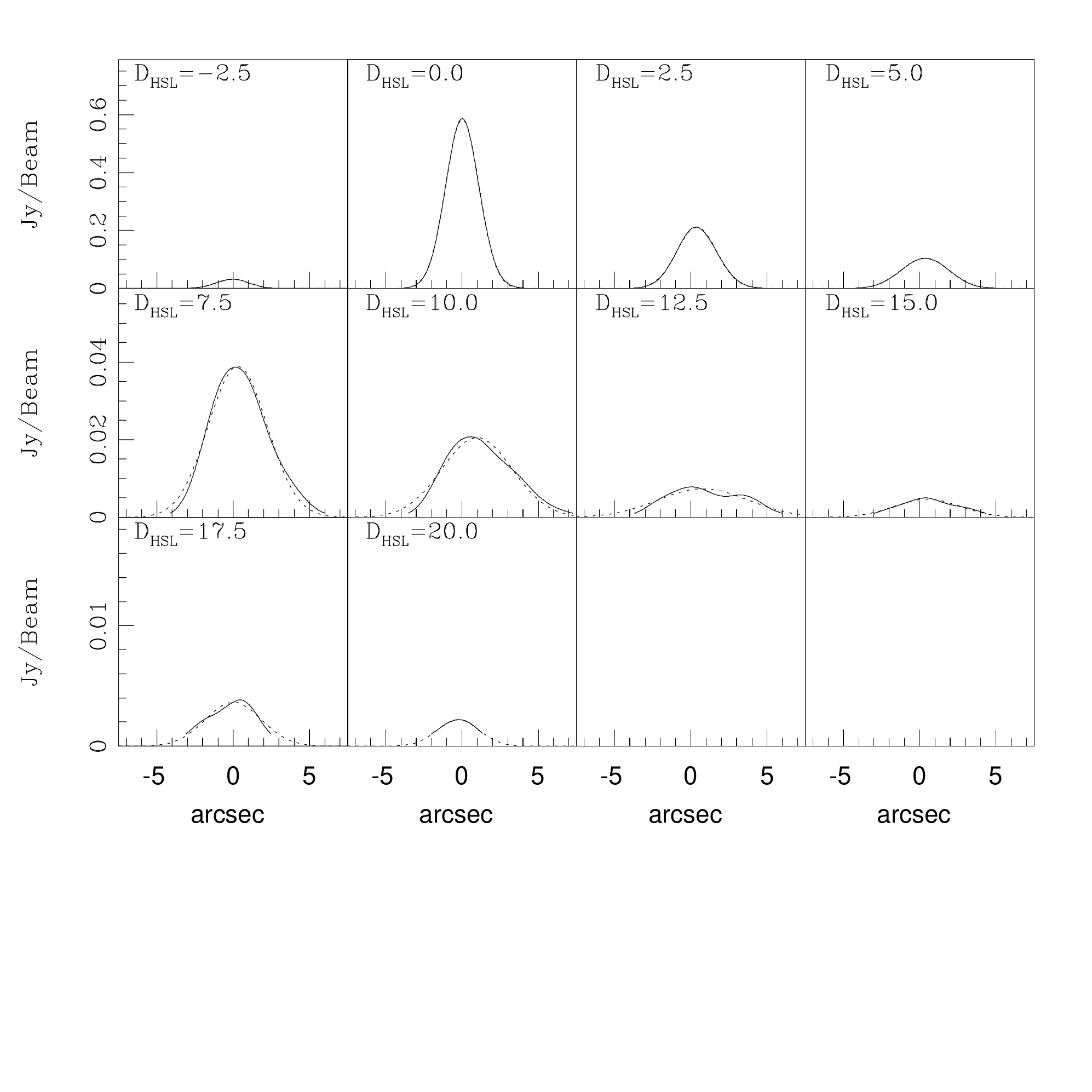}{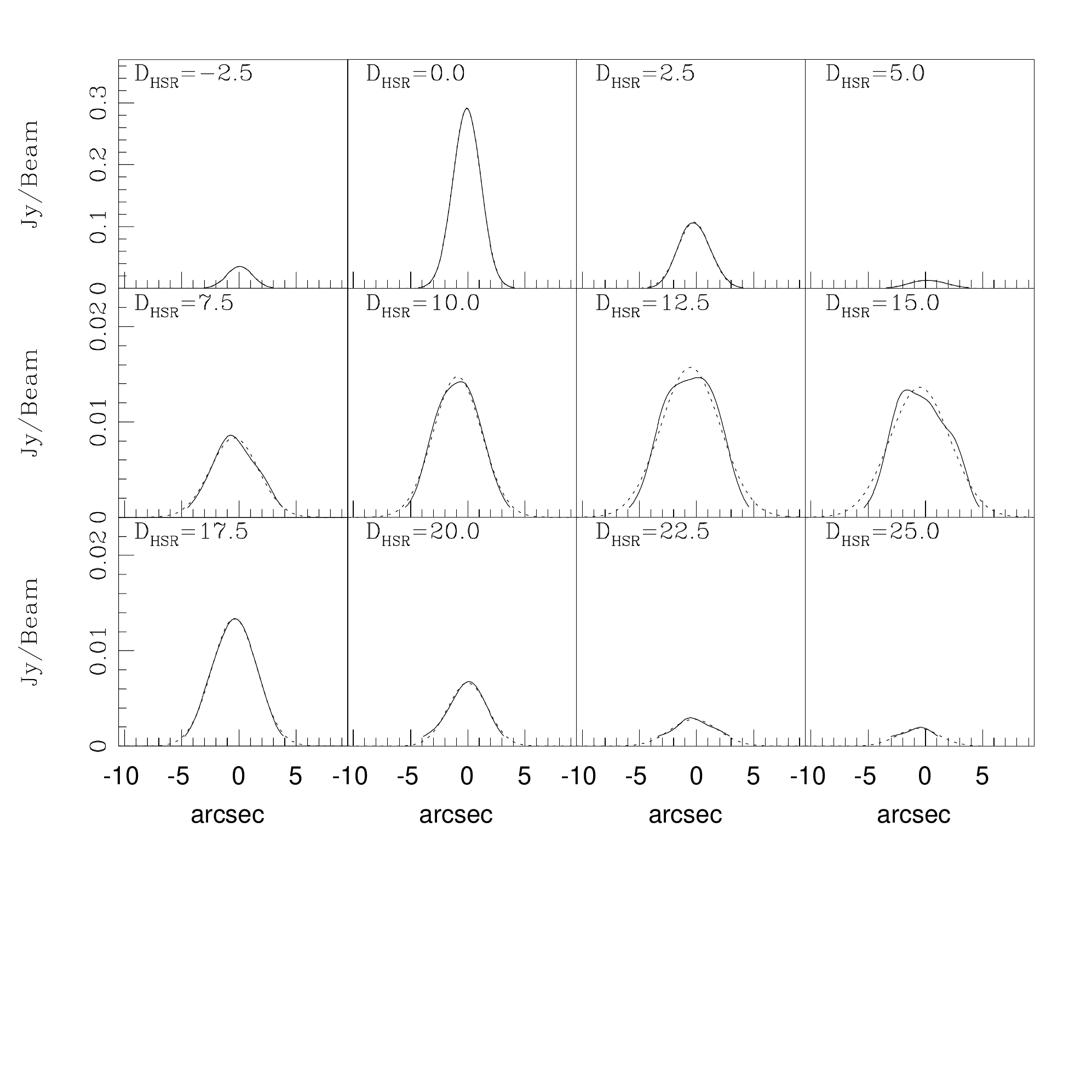}
\caption{3C54 
surface brightness (solid line) and best fit 
Gaussian (dotted line) for each cross-sectional slice of the 
radio bridge at 1.4 GHz, as in Fig. \ref{3C6.1SG}.
A Gaussian provides an excellent description of the surface brightness
profile of each cross-sectional slice. }
\label{3C54SG}
\end{figure}

\begin{figure}
\plottwo{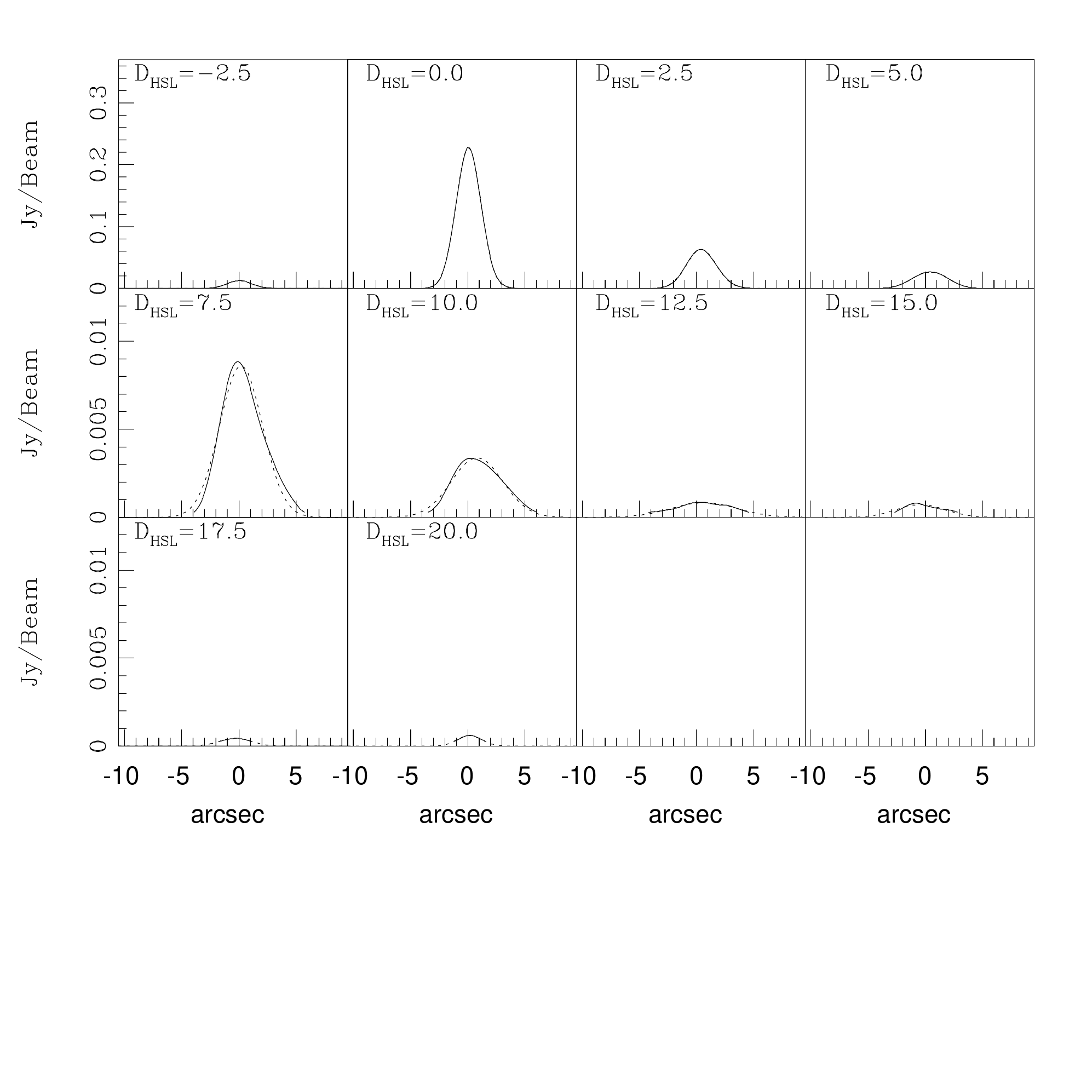}{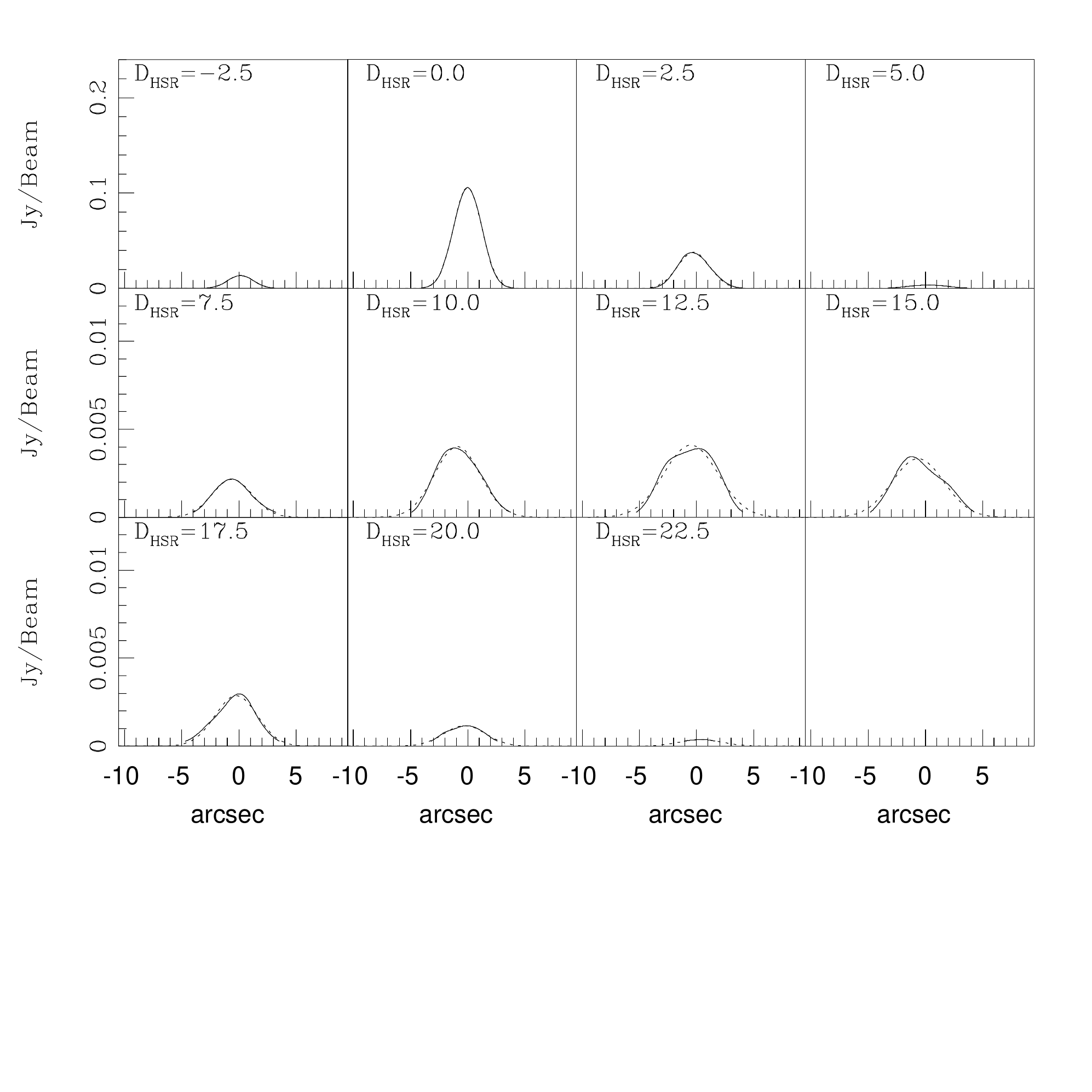}
\caption{3C54 
surface brightness (solid line) and best fit 
Gaussian (dotted line) for each cross-sectional slice of the 
radio bridge at 5 GHz, as in 
Fig. \ref{3C54SG} but at 5 GHz. 
Results obtained at 5 GHz are nearly identical to those
obtained at 1.4 GHz.}
\end{figure}

\begin{figure}
\plottwo{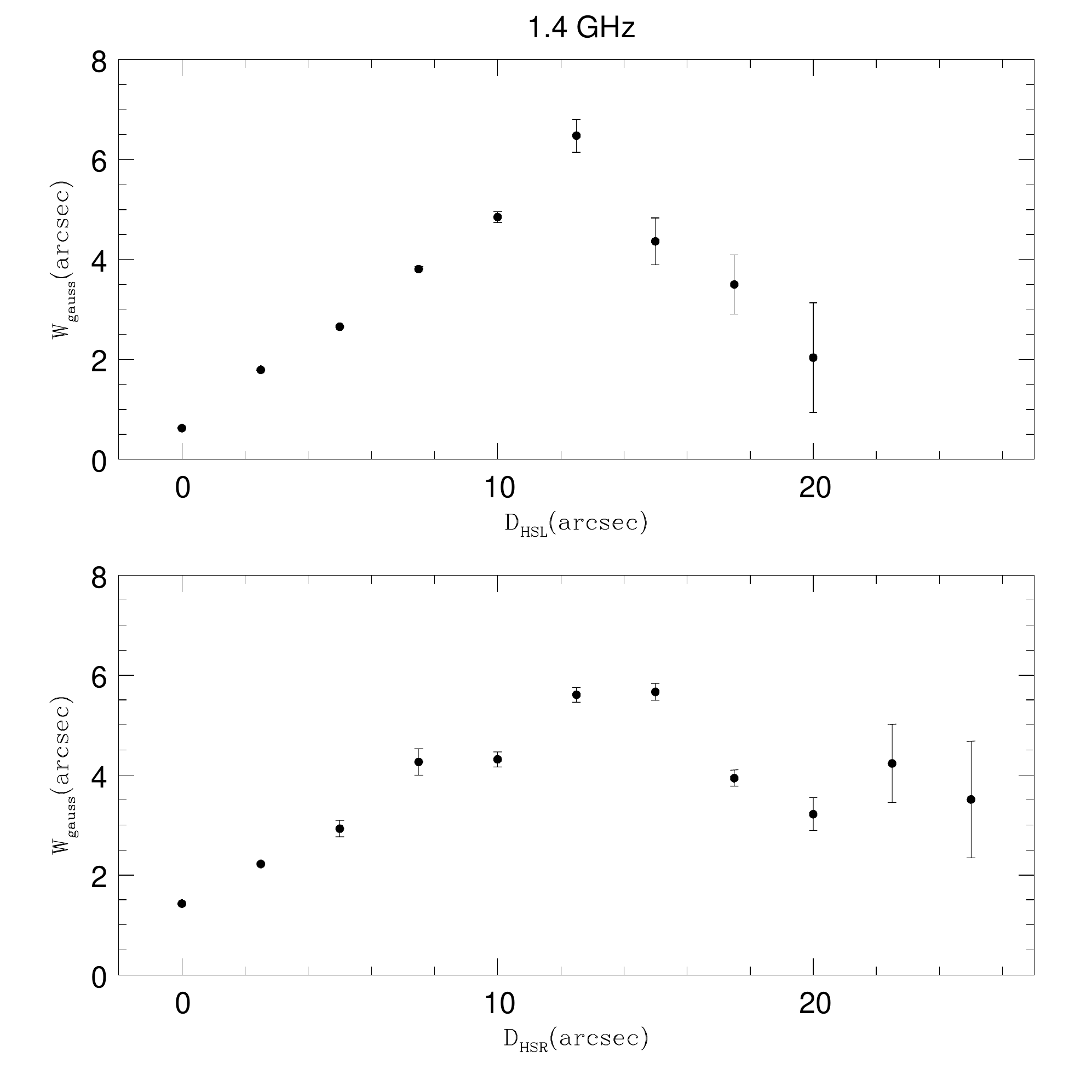}{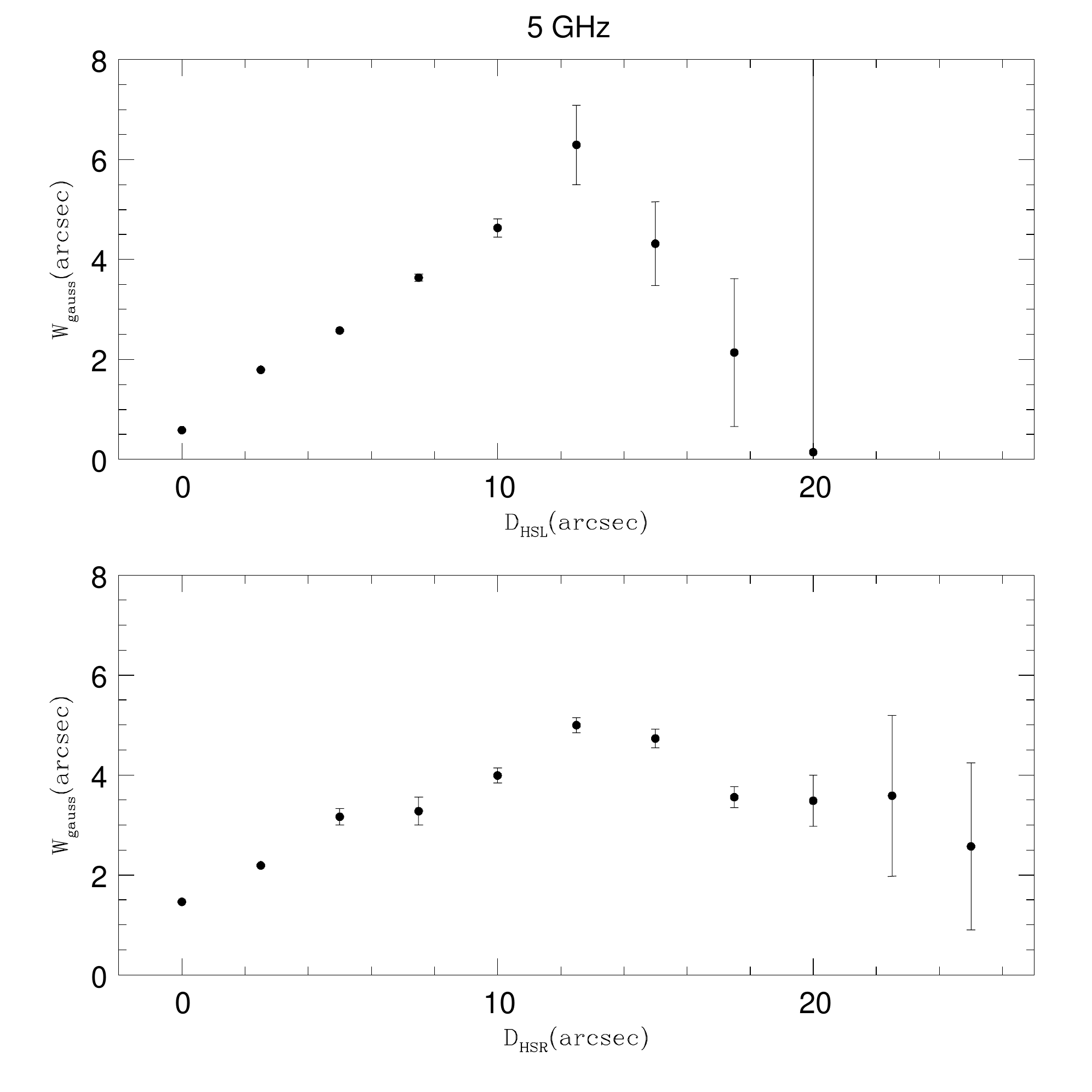}
\caption{3C54  Gaussian FWHM as a function of 
distance from the hot spot 
at 1.4 and 5 GHz (left and
right panels, respectively) for the left and right hand sides of the 
source (top and bottom panels, respectively), as in Fig. \ref{3C6.1WG}.
The right and left hand sides of the source are quite 
symmetric.  Similar results are obtained at 
1.4 and 5 GHz. }
\end{figure}

\begin{figure}
\plottwo{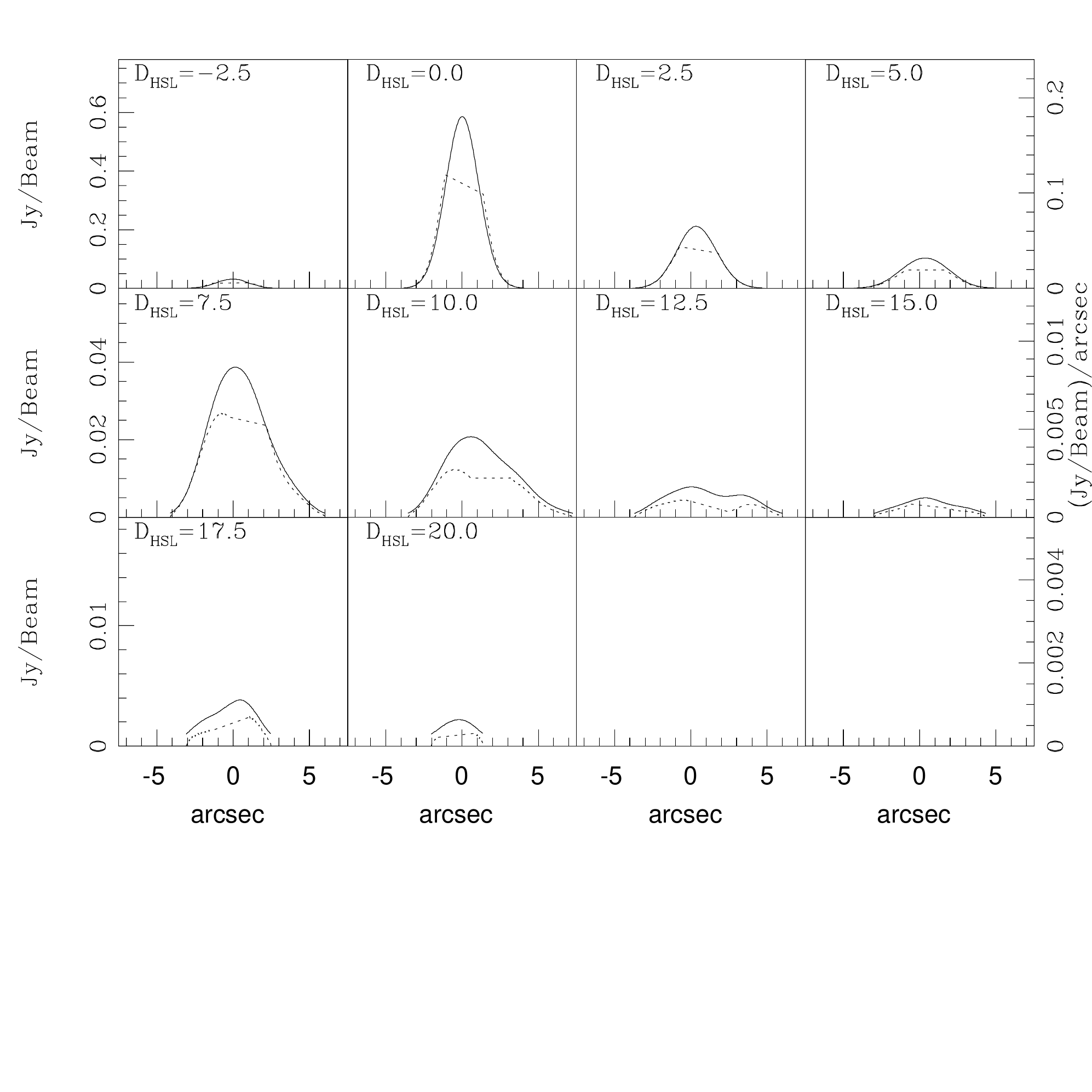}{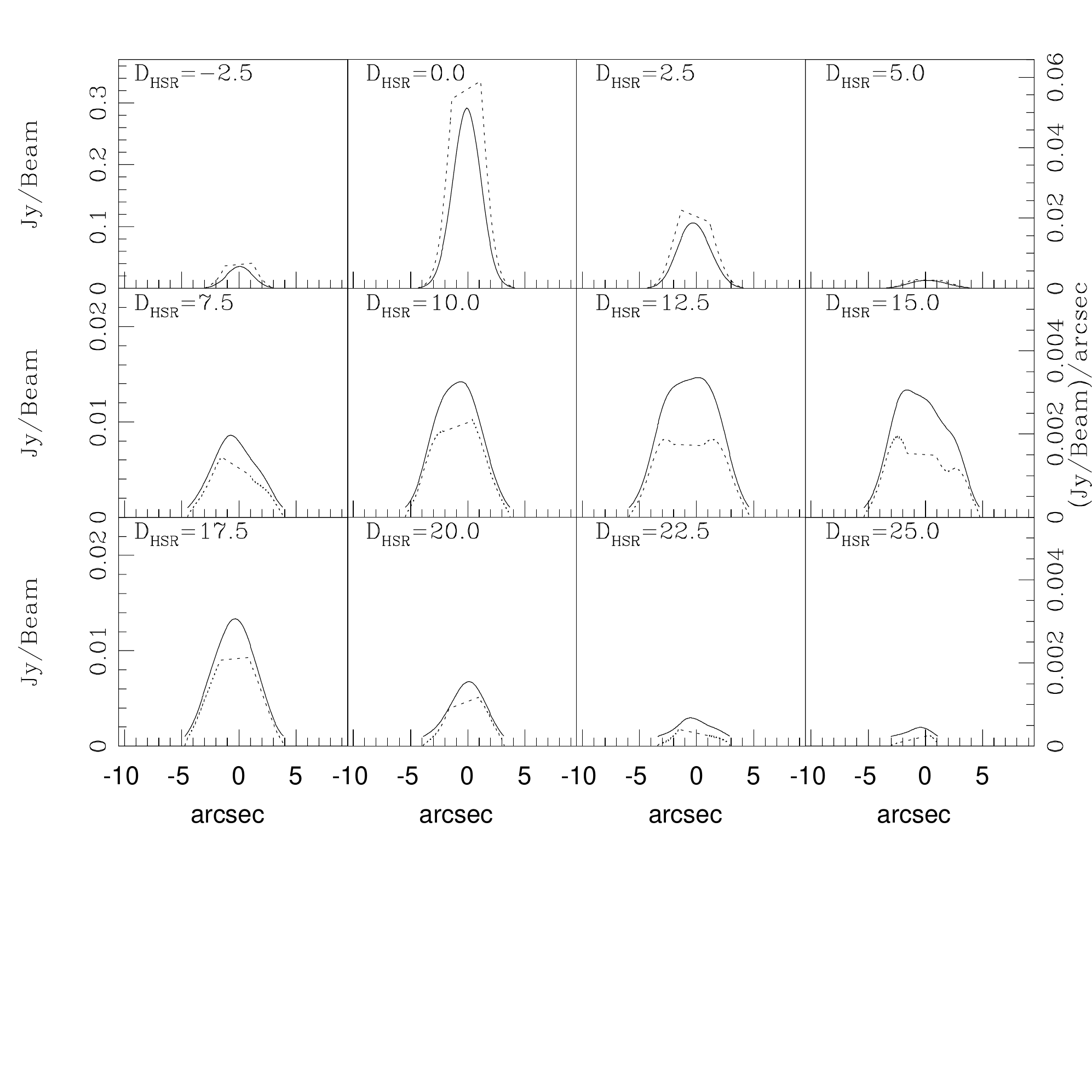}
\caption{3C54 emissivity (dotted line) and surface brightness
(solid line) for each cross-sectional slice of the radio 
bridge at 1.4 GHz, as in Fig. \ref{3C6.1SEM}. A constant volume
emissivity per slice provides a reasonable description of the 
data over most of the source.  }
\label{3C54SEM}
\end{figure}

\begin{figure}
\plottwo{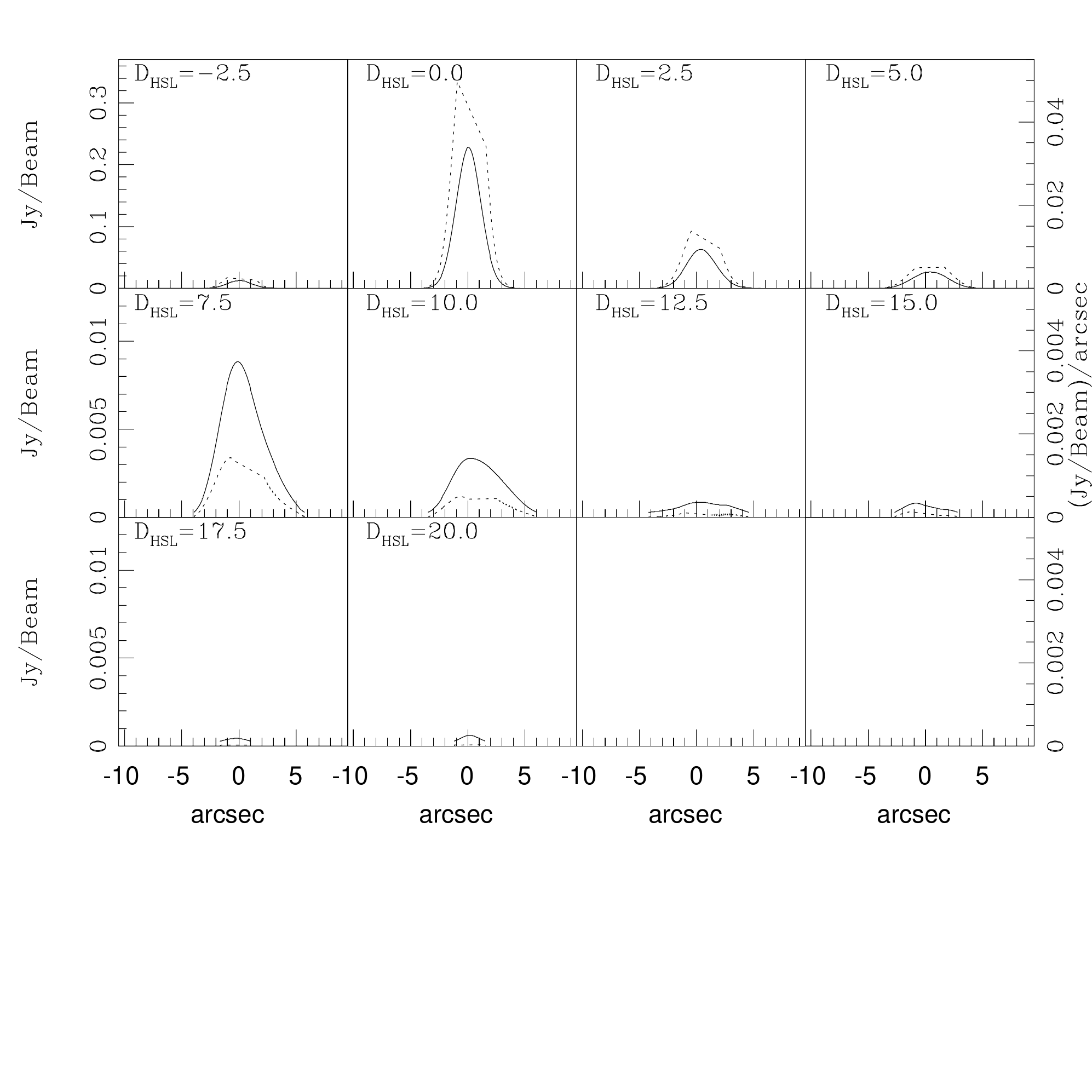}{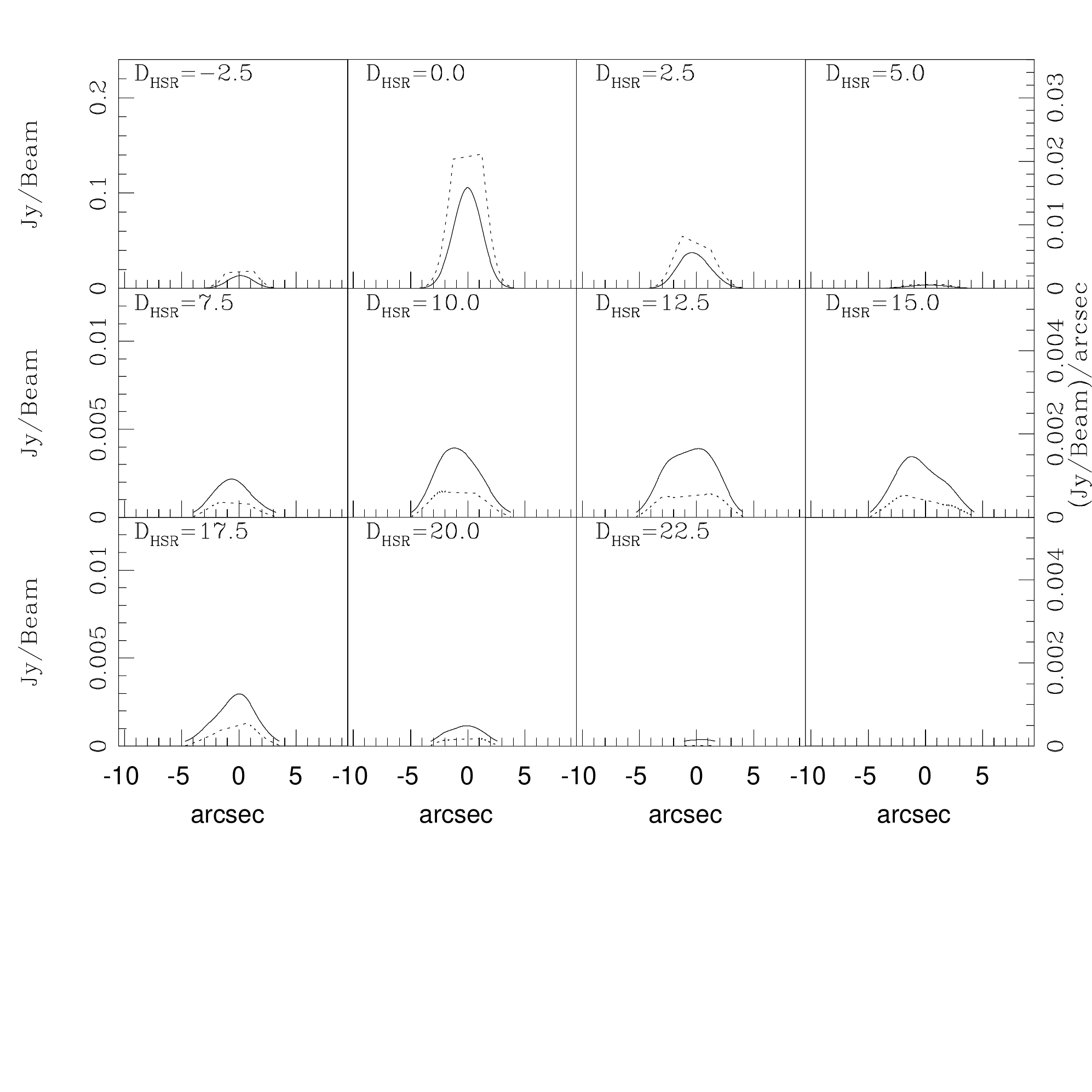}
\caption{3C54
emissivity (dotted line) and surface brightness
(solid line) for each cross-sectional slice of the radio 
bridge at 5 GHz, as in Fig. \ref{3C54SEM} but at 5 GHz.
Results obtained at 5 GHz are nearly identical to those
obtained at 1.4 GHz.}
\end{figure}

\begin{figure}
\plottwo{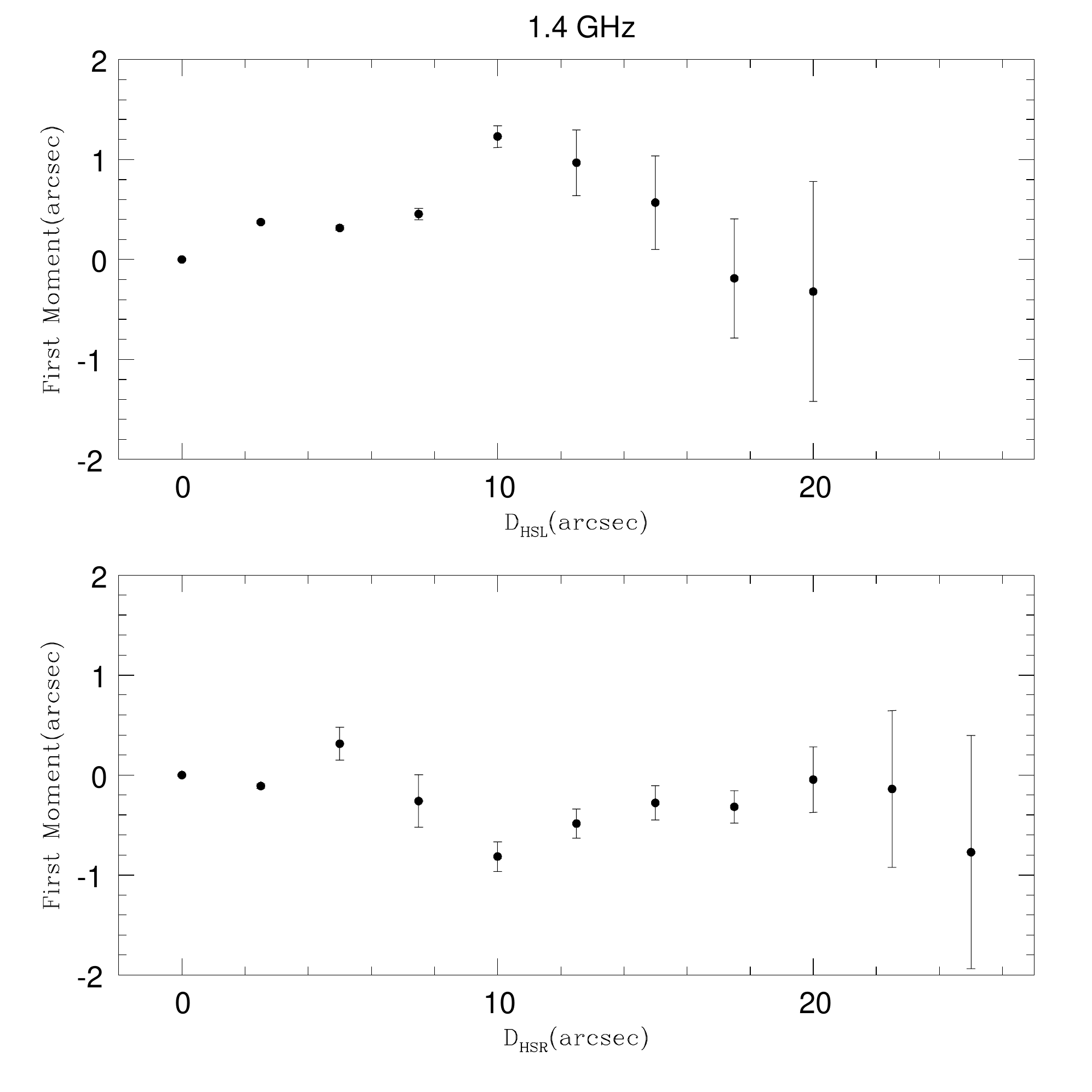}{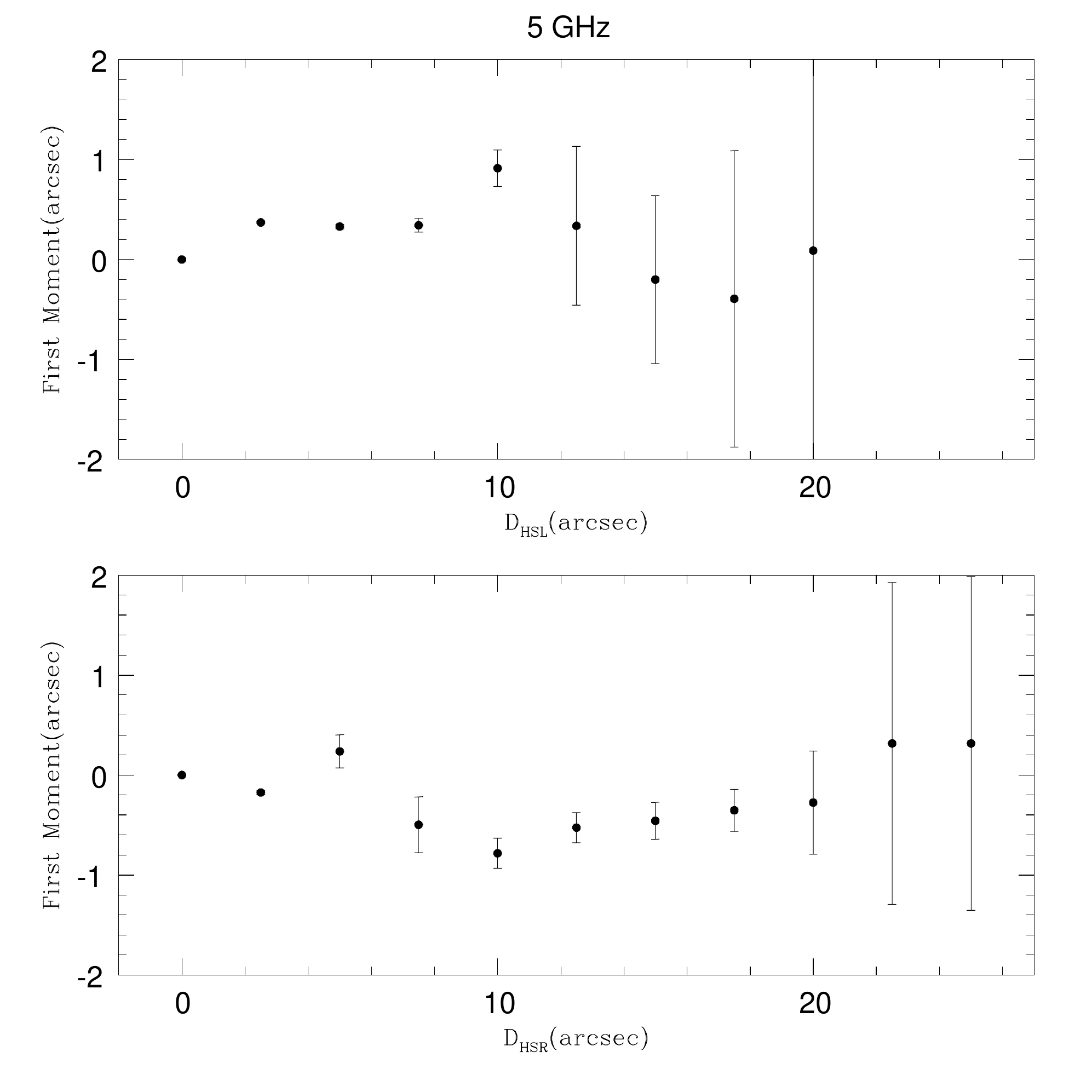}
\caption{3C54 first moment as a function of distance from 
the hot spot at 1.4 GHz and 5 GHz (left and right panels,
respectively) for the left and right hand sides of the source
(top and bottom, respectively).}
\end{figure}

\begin{figure}
\plottwo{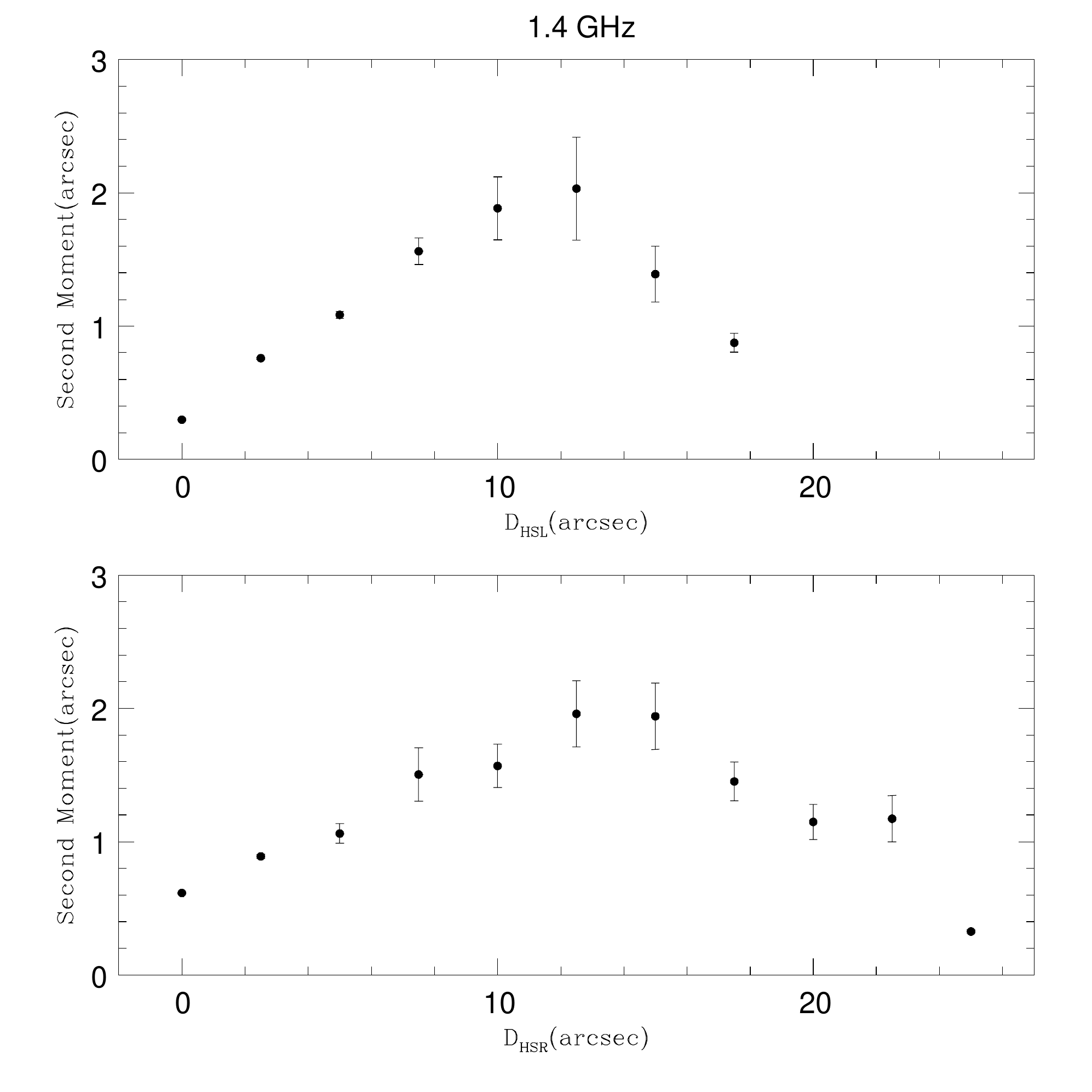}{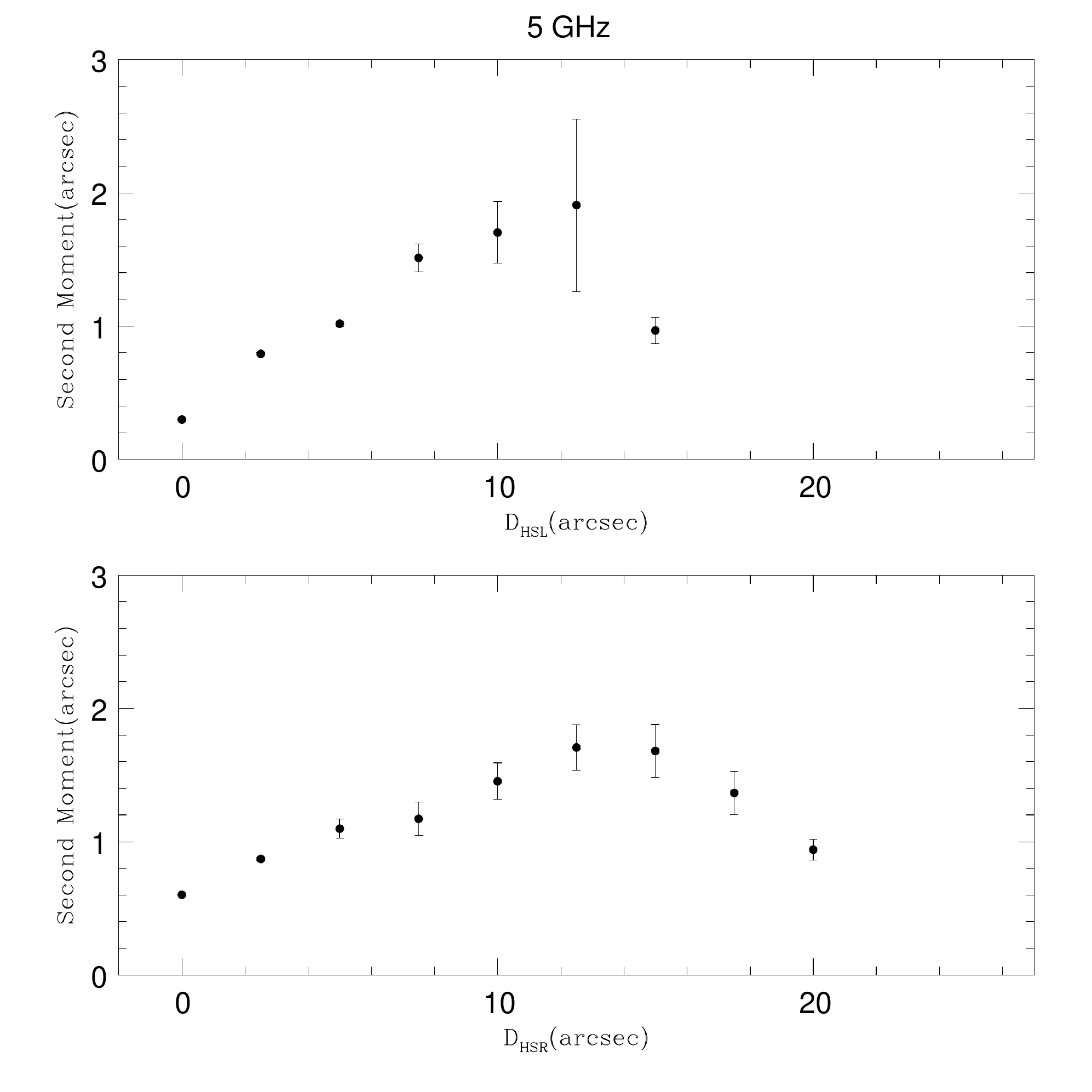}
\caption{3C54 second moment as a function of distance from the
hot spot at 1.4 GHz and 5 GHz (left and right panels, 
respectively) for the left and right sides of the source
(top and bottom panels, respectively).}
\end{figure}

\begin{figure}
\plottwo{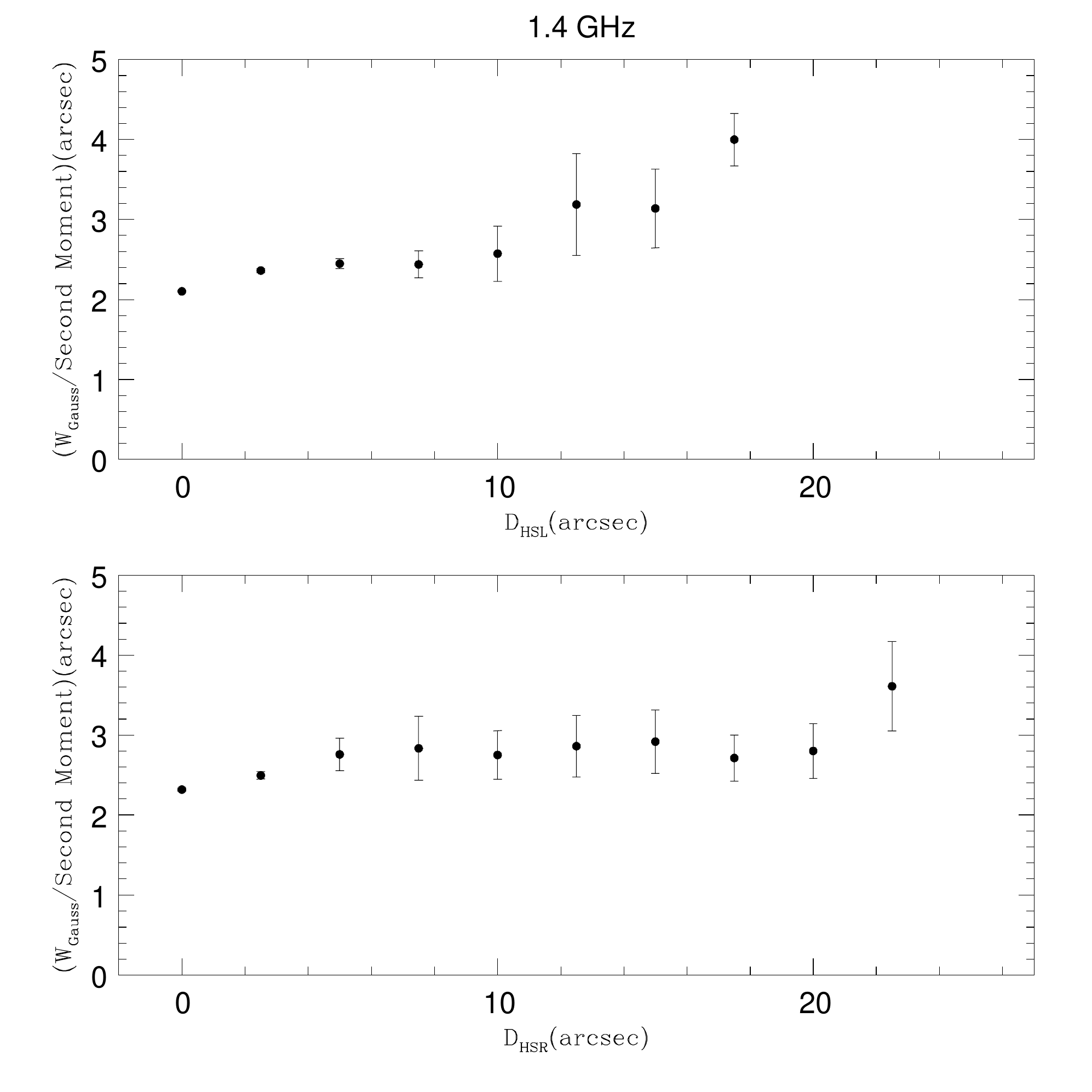}{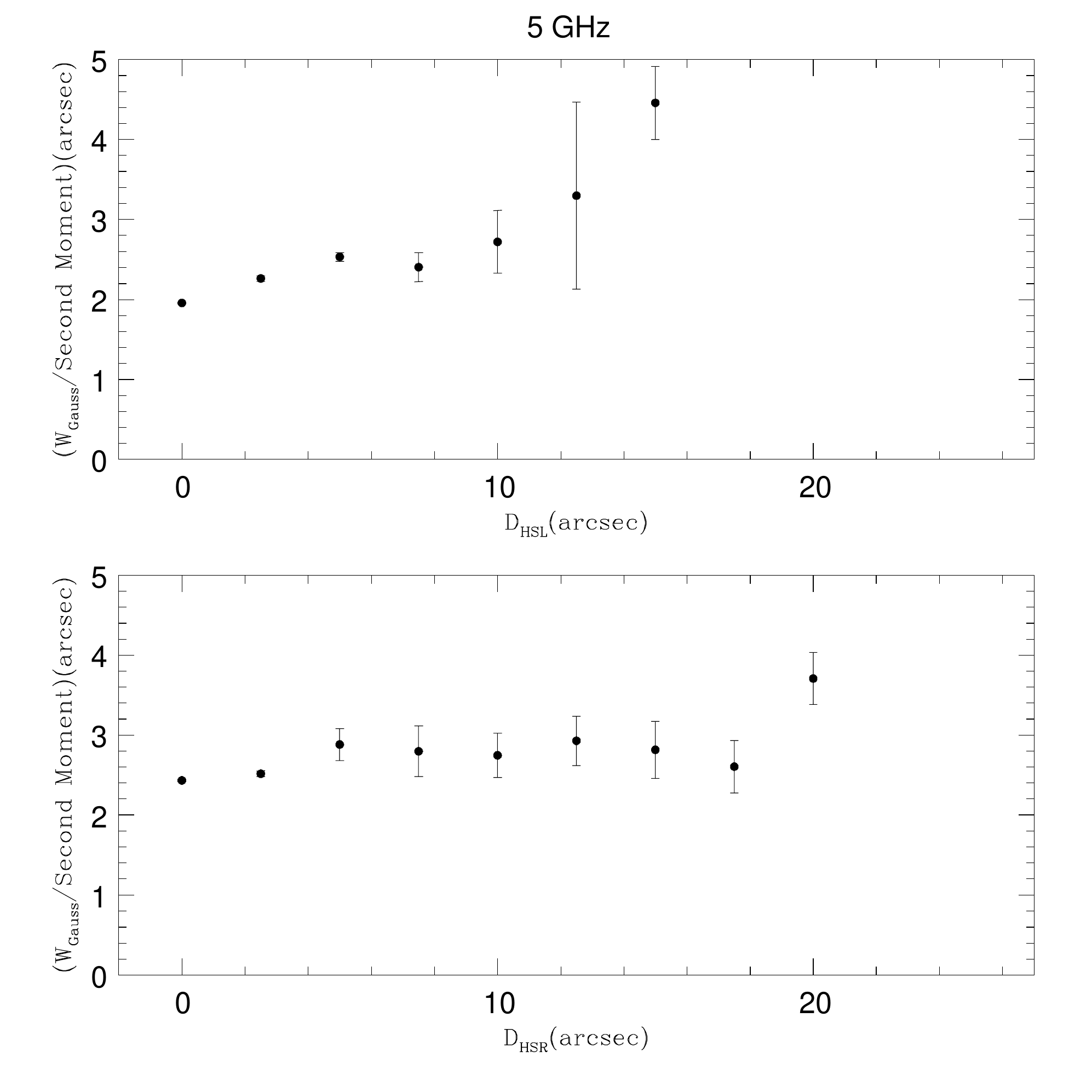}
\caption{3C54 ratio of the Gaussian FWHM to the second moment as a
function of distance from the hot spot at 1.4 and 5 GHz 
(left and right panels, respectively) for the left and right
hand sides of the source (top and bottom panels, respectively). The value of this ratio is fairly constant for each side of the source, and has a 
value of about 2 to 2.5. Similar results are obtained at 1.4 and 5 GHz.} 
\end{figure}

\begin{figure}
\plottwo{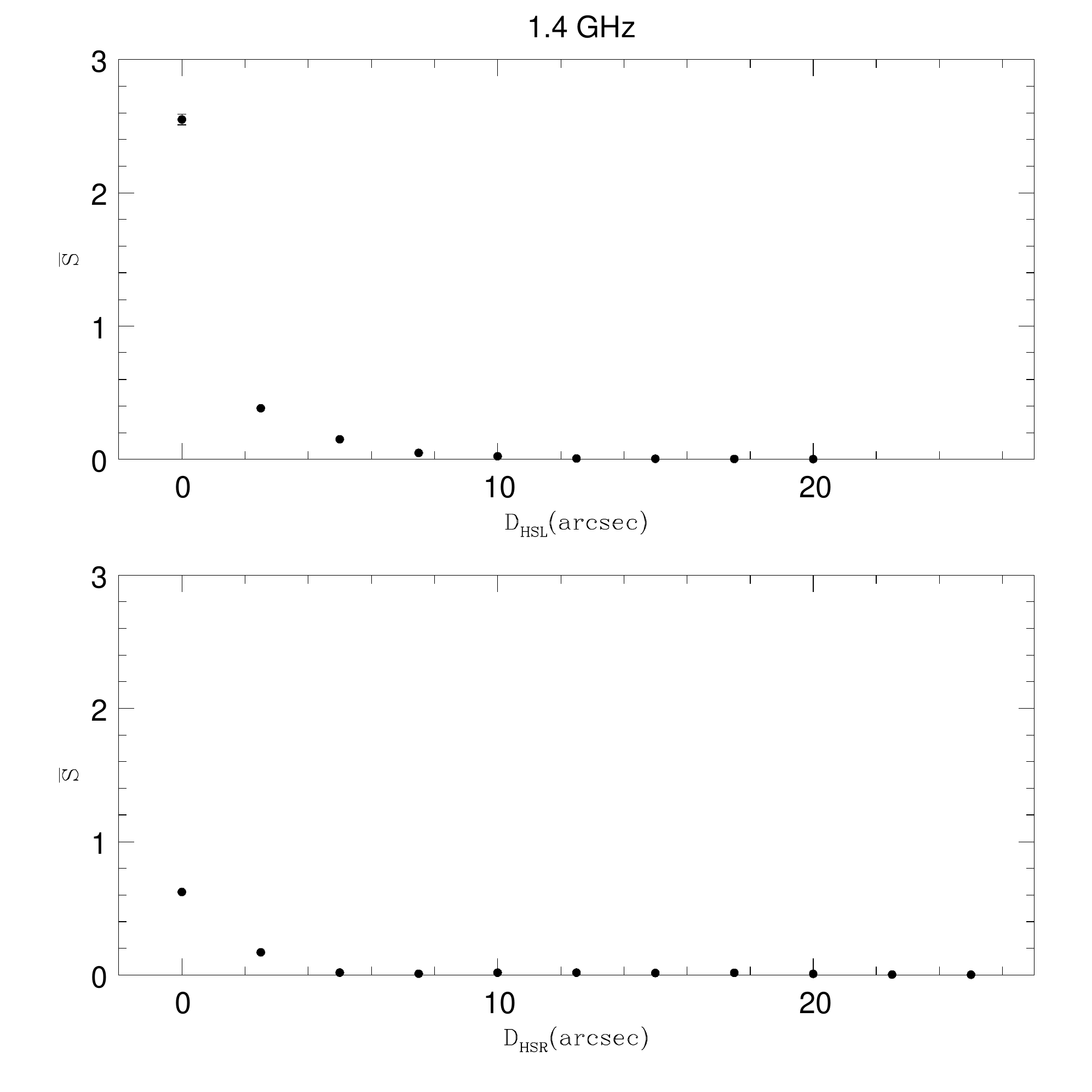}{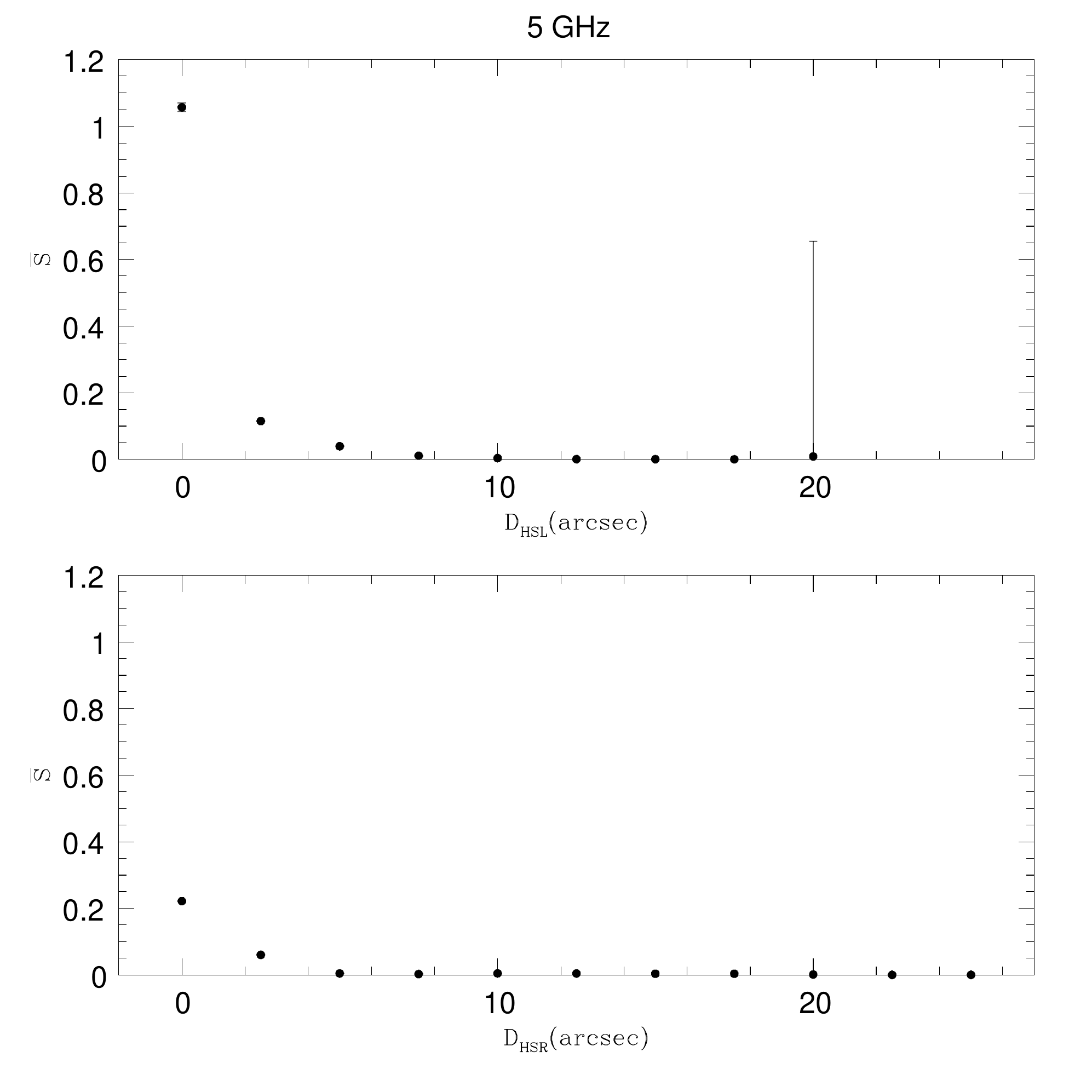}
\caption{3C54 average surface brightness in units of Jy/beam 
as a function of distance from the 
hot spot at 1.4 and 5 GHz 
(left and right panels, respectively) for the left and right
hand sides of the source (top and bottom panels, respectively).}
\end{figure}

\begin{figure}
\plottwo{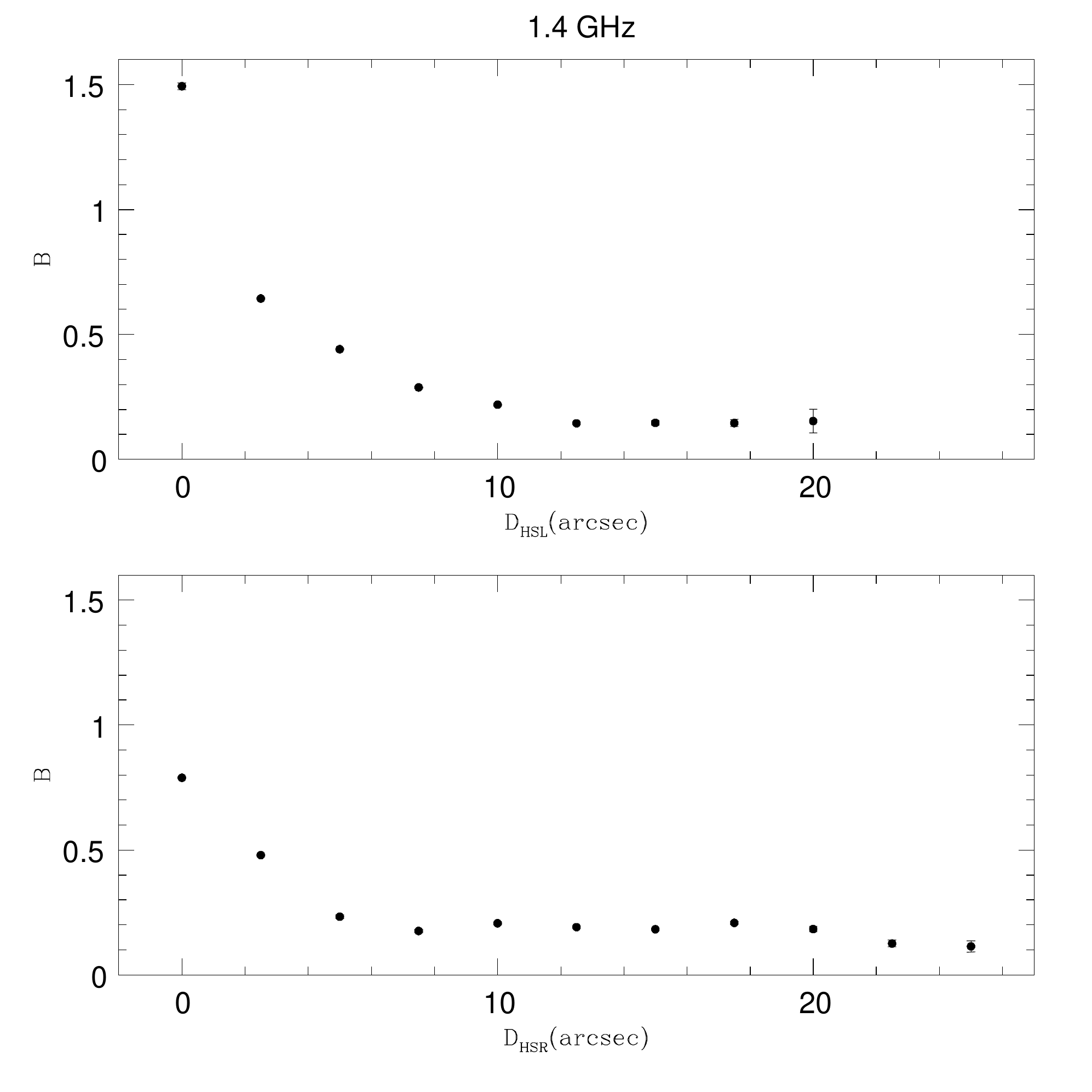}{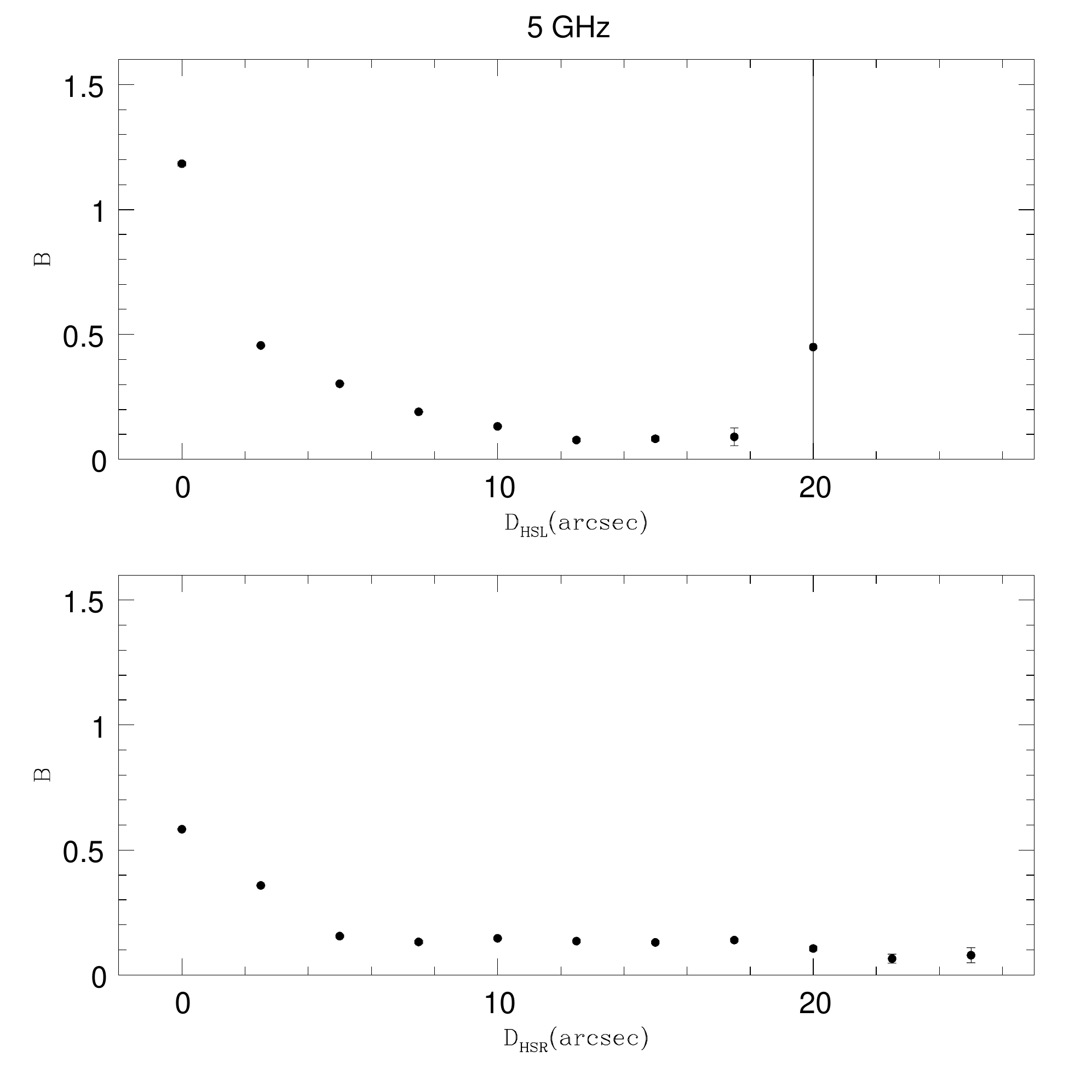}
\caption{3C54 minimum energy magnetic field strength 
as a function of distance from the 
hot spot at 1.4 and 5 GHz 
(left and right panels, respectively) for the left and right
hand sides of the source (top and bottom panels, respectively).
The normalization is given in Table 1. 
}
\end{figure}

\begin{figure}
\plottwo{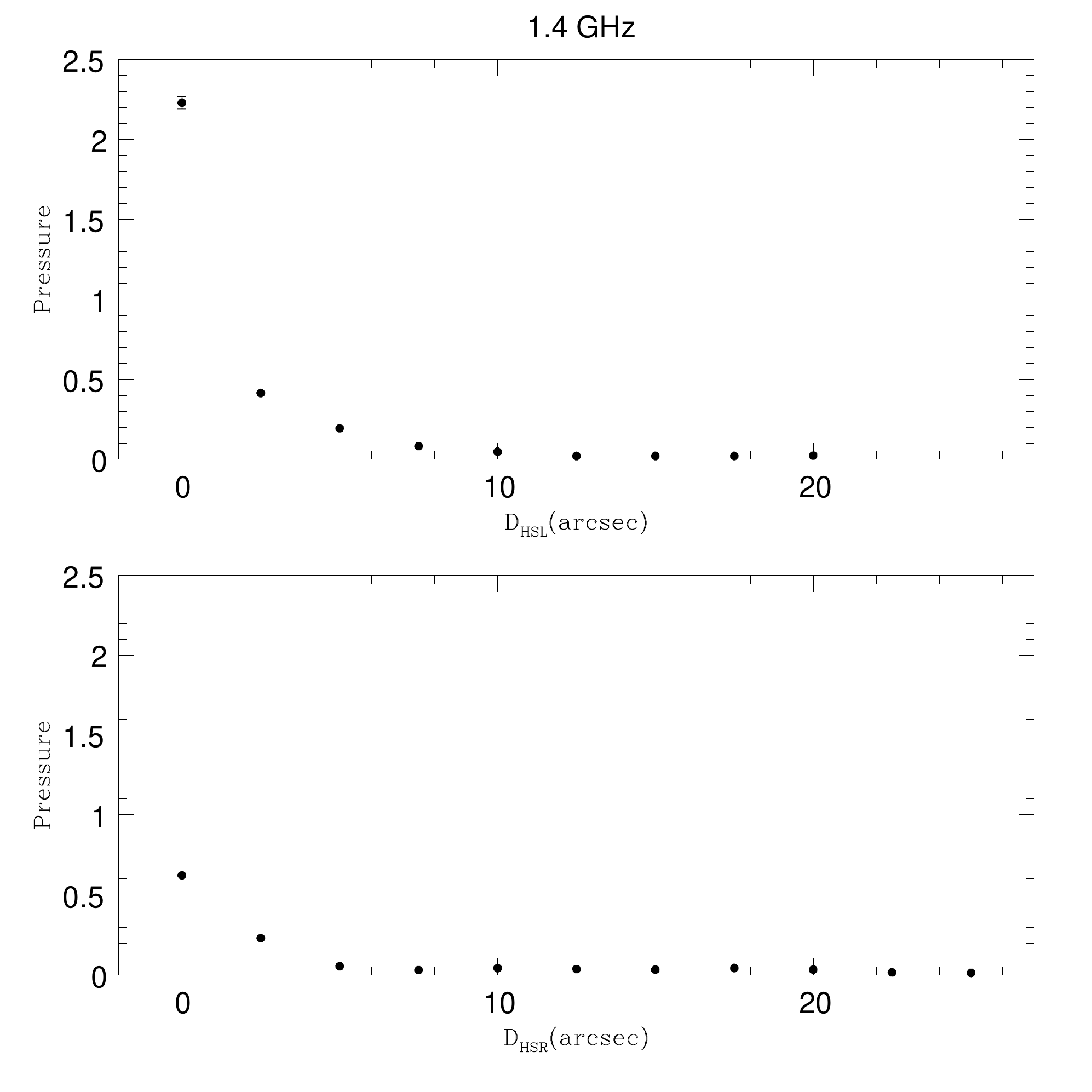}{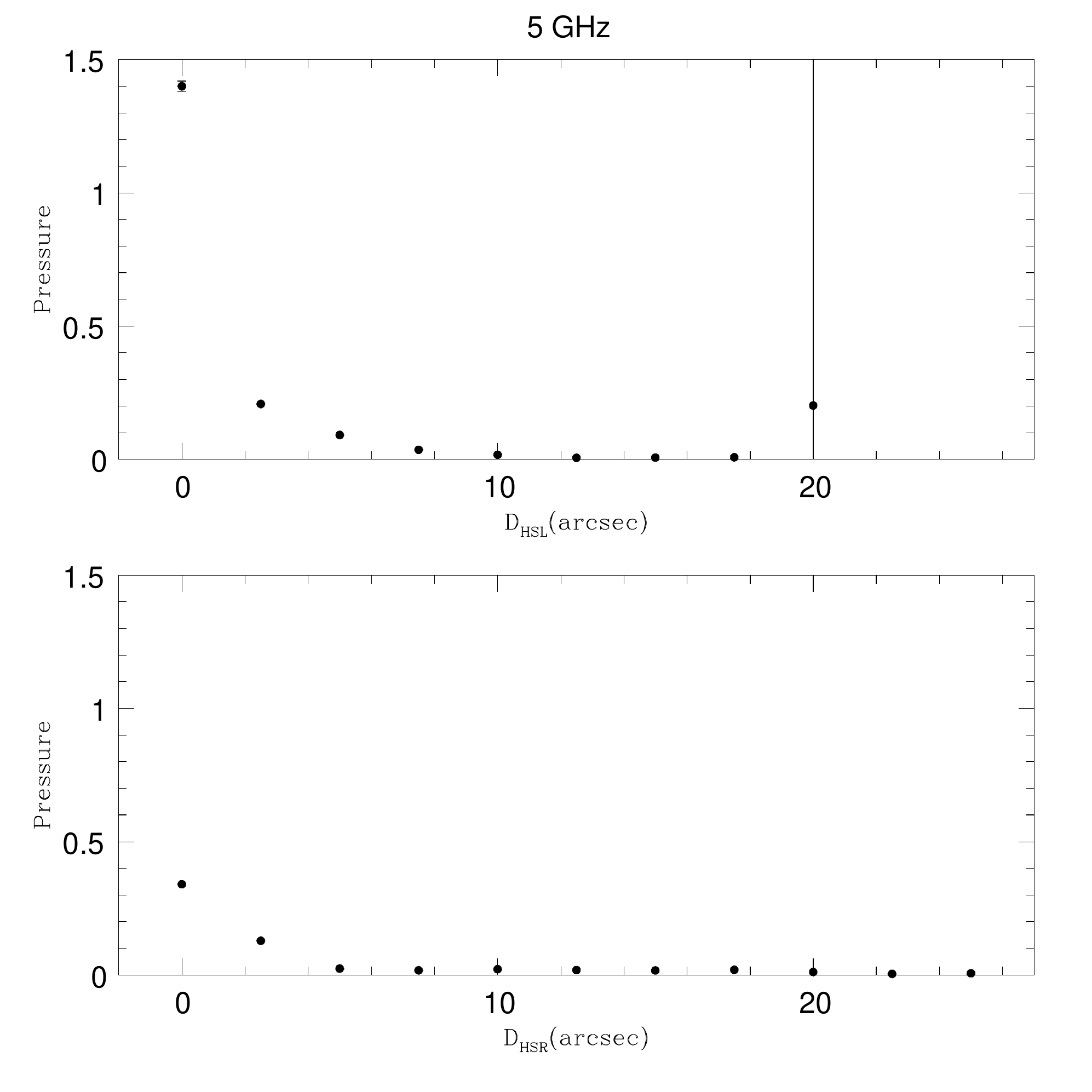}
\caption{3C54 
minimum energy pressure as a function of 
distance from the hot spot at 1.4 and 5 GHz 
(left and right panels, respectively) for the left and right
hand sides of the source (top and bottom panels, respectively), 
as in Fig. \ref{3C6.1P}. The normalization is given in Table 1.
}
\end{figure}

\clearpage
\begin{figure}
\plotone{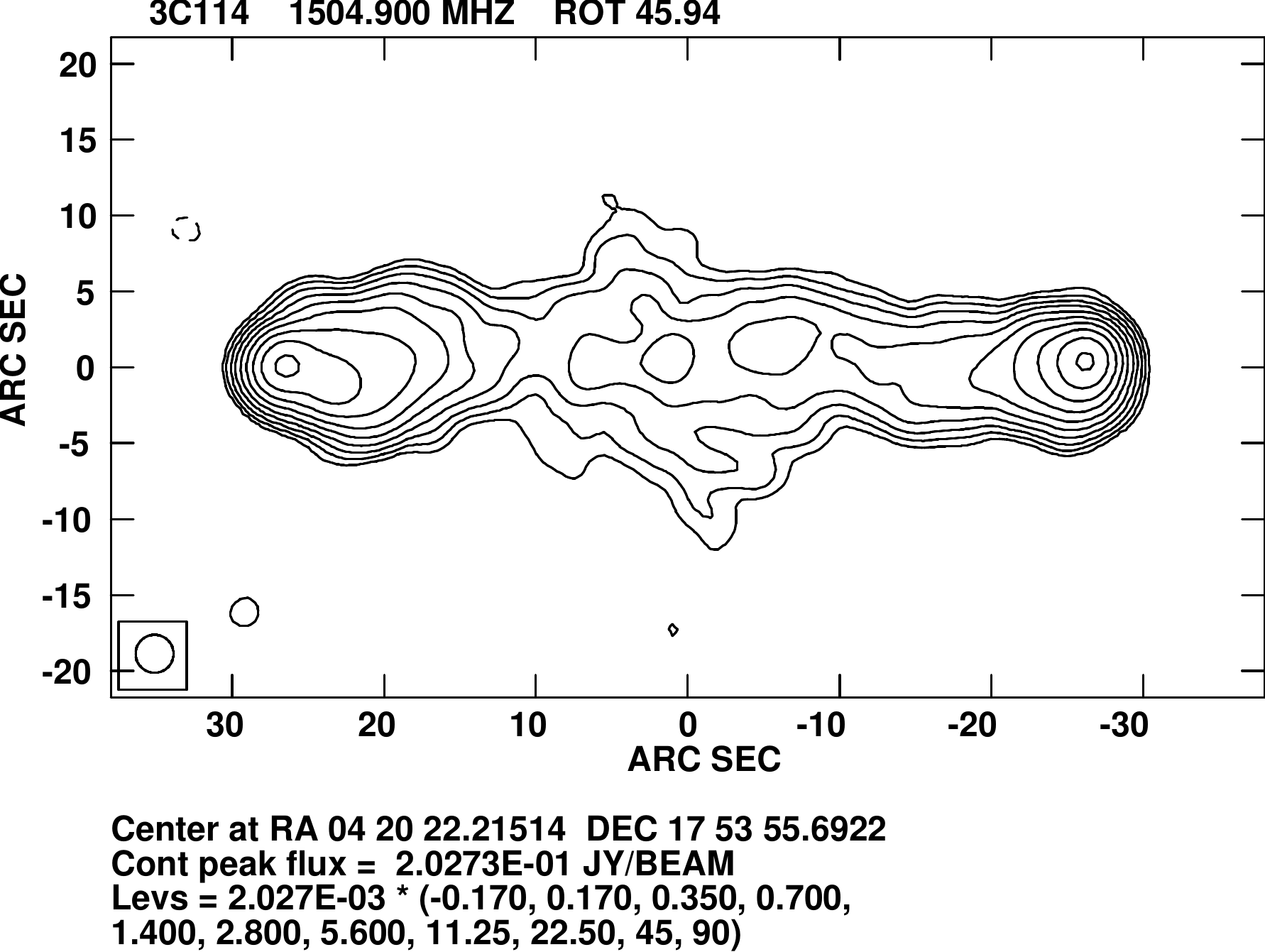}
\caption{3C114 at 1.4 GHz and 2.5'' resolution rotated by 45.94 deg 
counter-clockwise, as in Fig. \ref{3C6.1rotfig}.}
\end{figure}

\begin{figure}
\plottwo{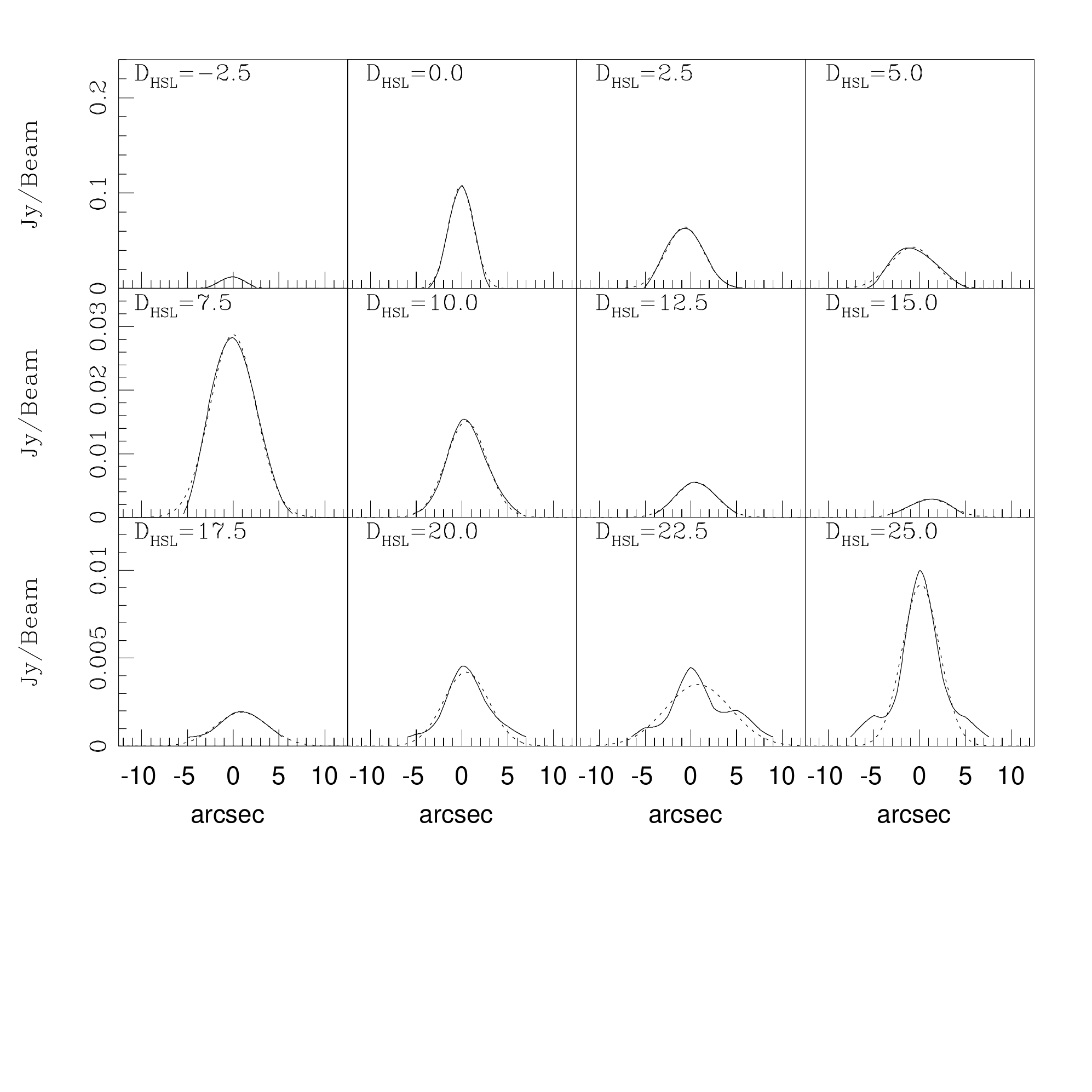}{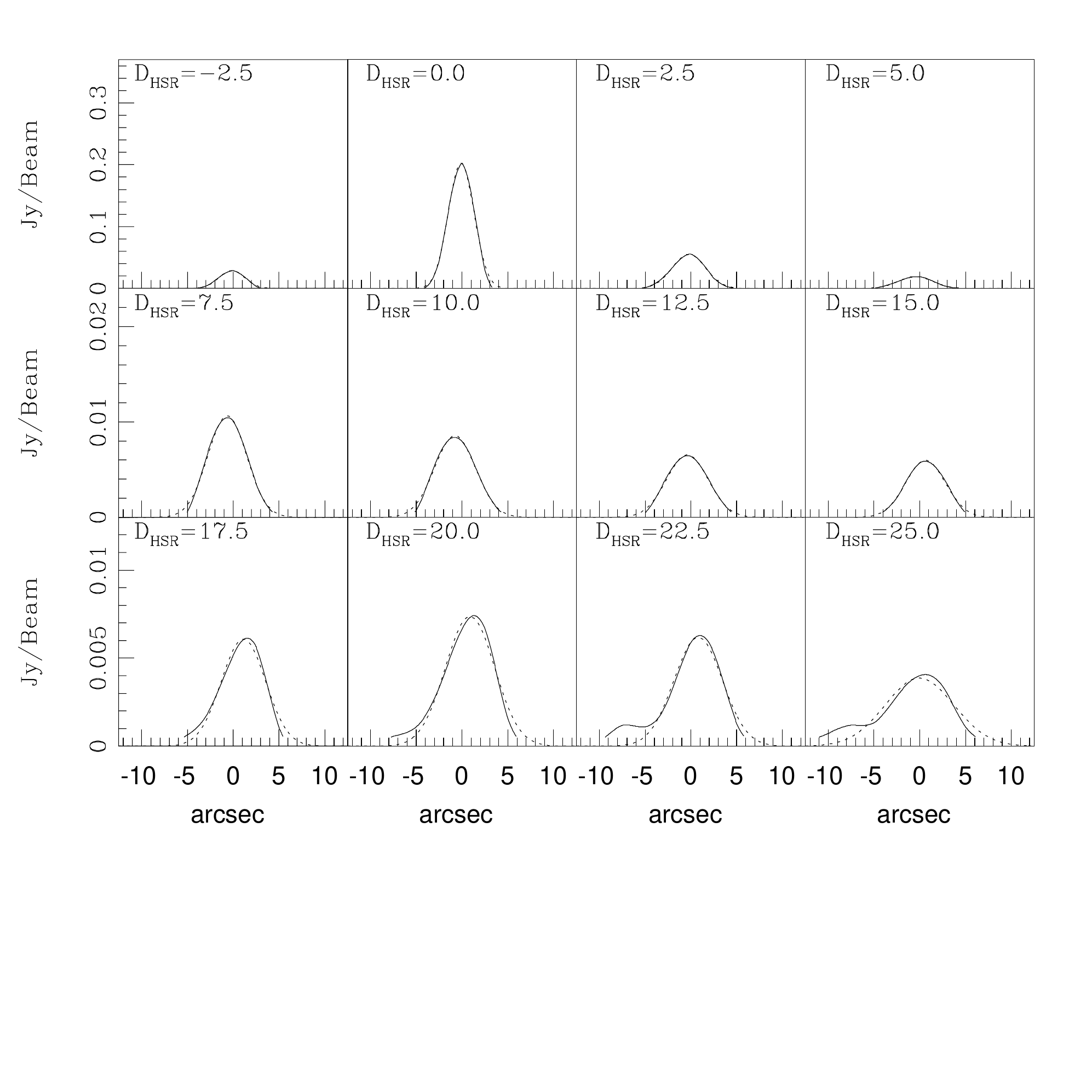}
\caption{3C114 
surface brightness (solid line) and best fit 
Gaussian (dotted line) for each cross-sectional slice of the 
radio bridge at 1.4 GHz, as in Fig. \ref{3C6.1SG}.
A Gaussian provides an excellent description of the surface brightness
profile of each cross-sectional slice. }
\label{3C114SG}
\end{figure}

\begin{figure}
\plottwo{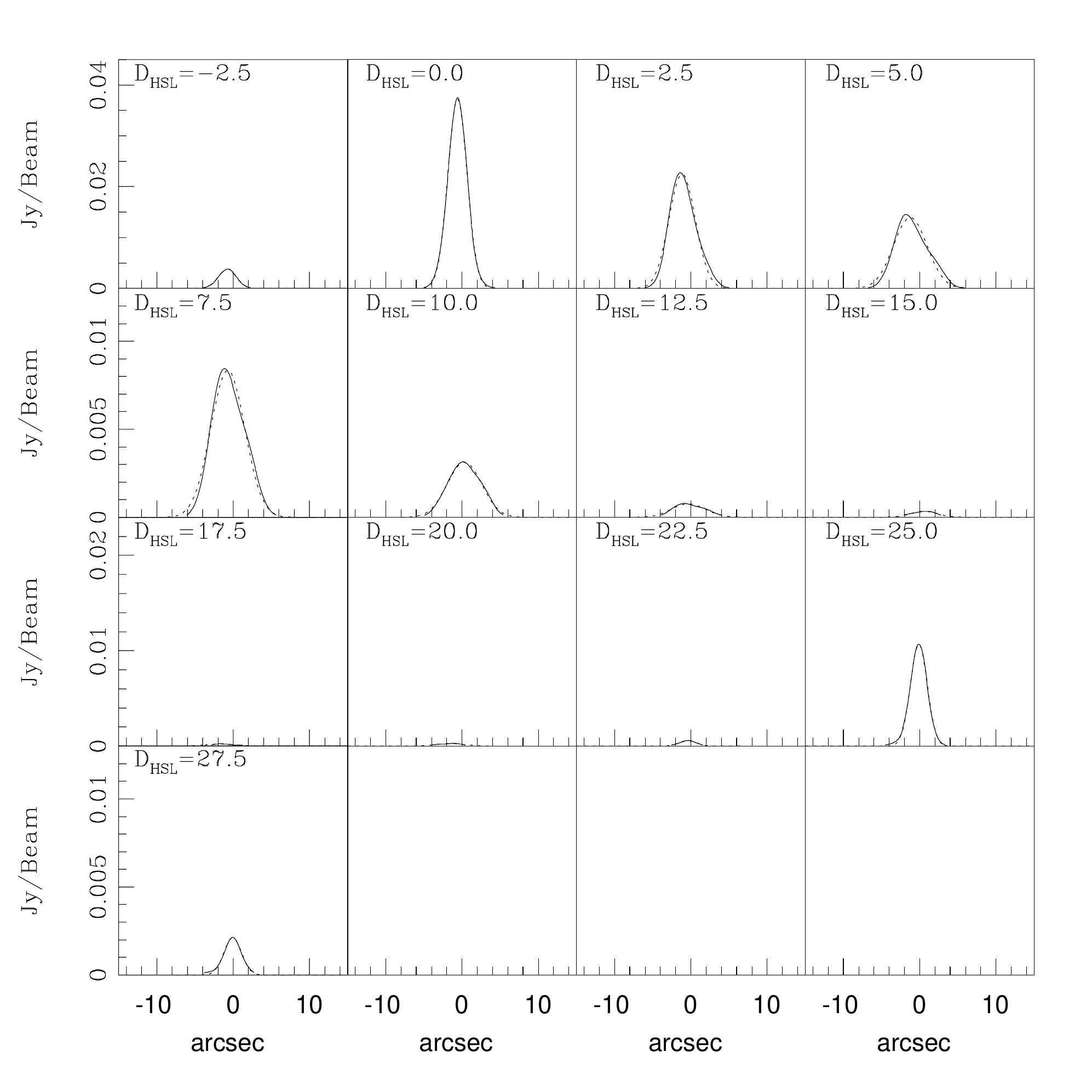}{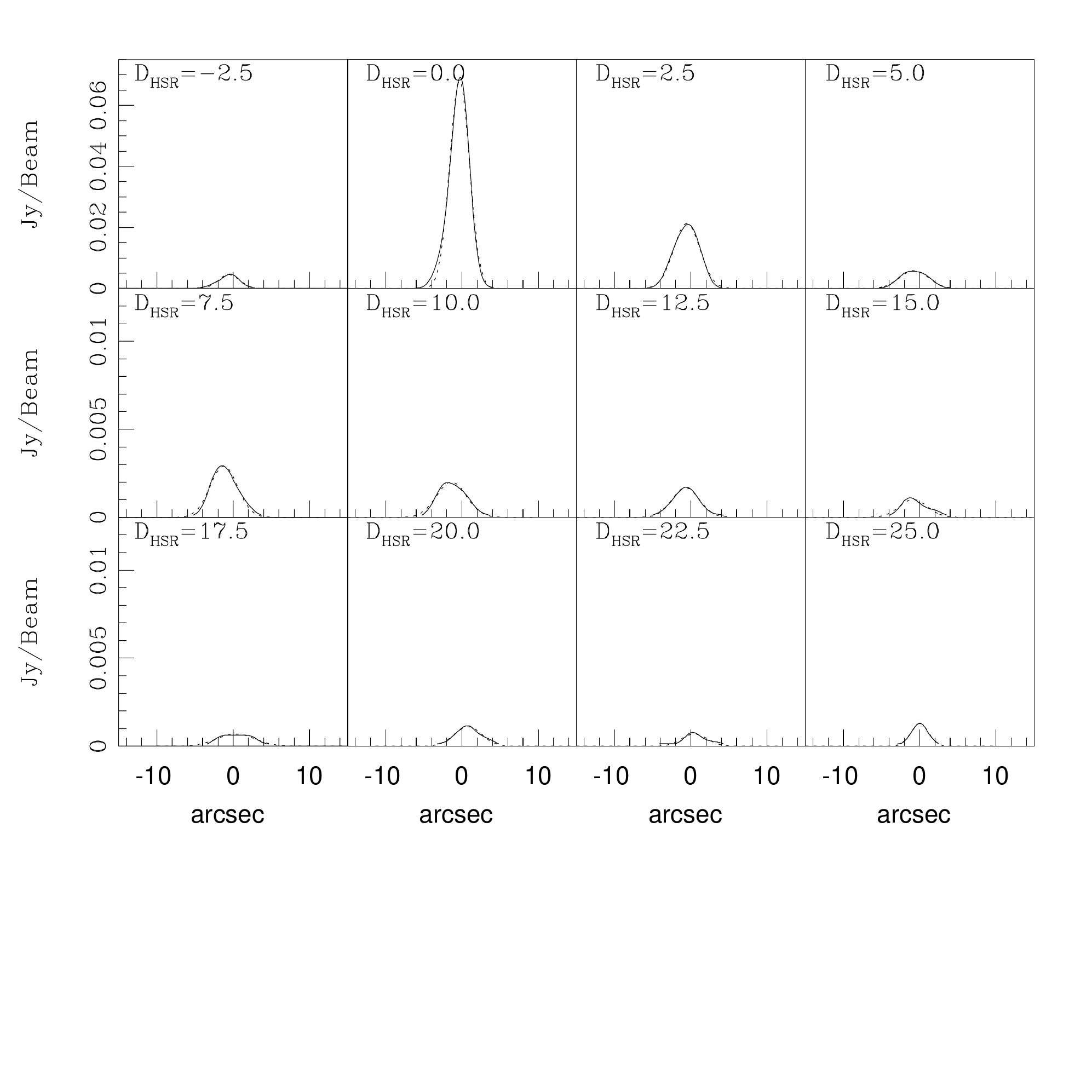}
\caption{3C114 
surface brightness (solid line) and best fit 
Gaussian (dotted line) for each cross-sectional slice of the 
radio bridge at 5 GHz, as in 
Fig. \ref{3C114SG} but at 5 GHz. 
Results obtained at 5 GHz are nearly identical to those
obtained at 1.4 GHz.}
\end{figure}

\begin{figure}
\plottwo{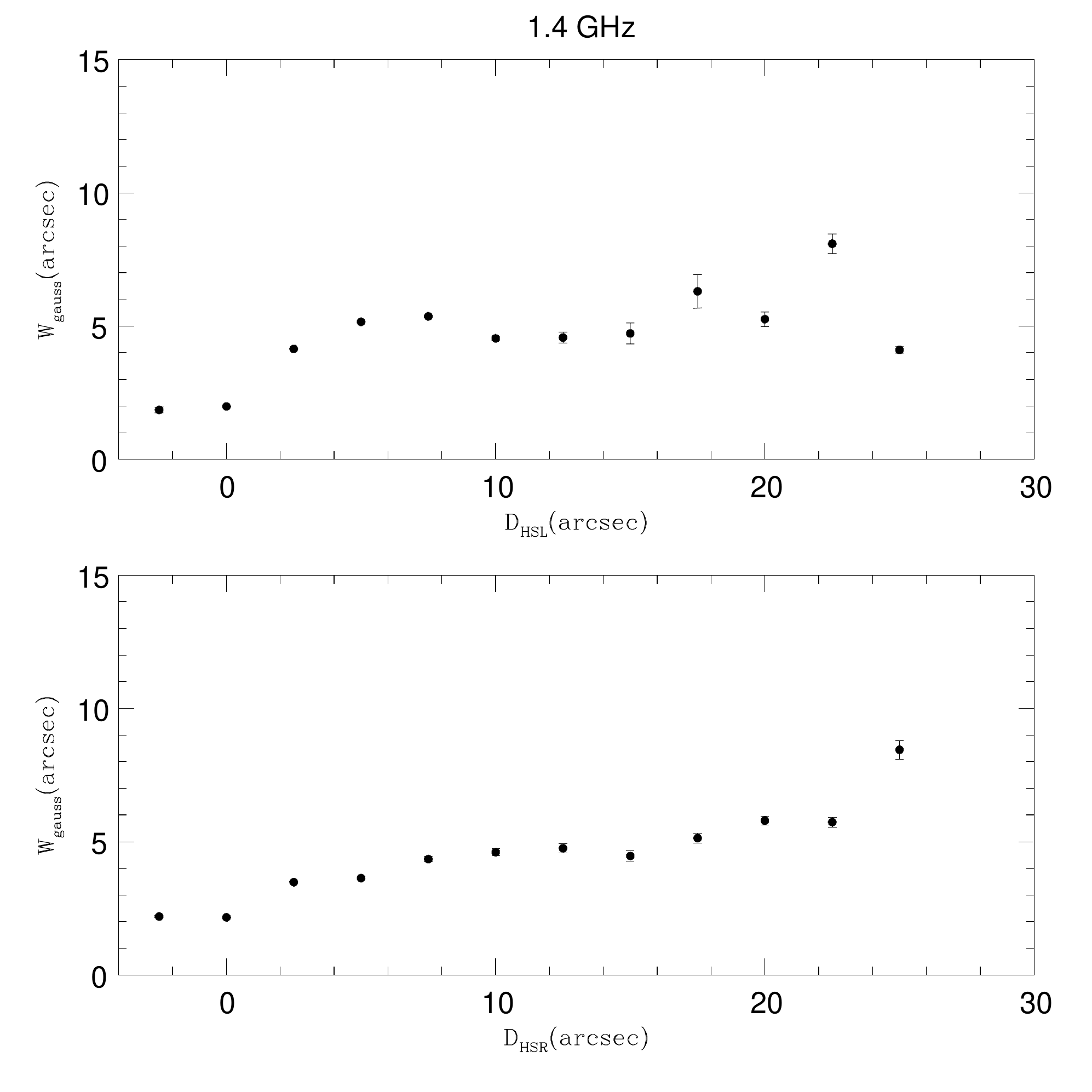}{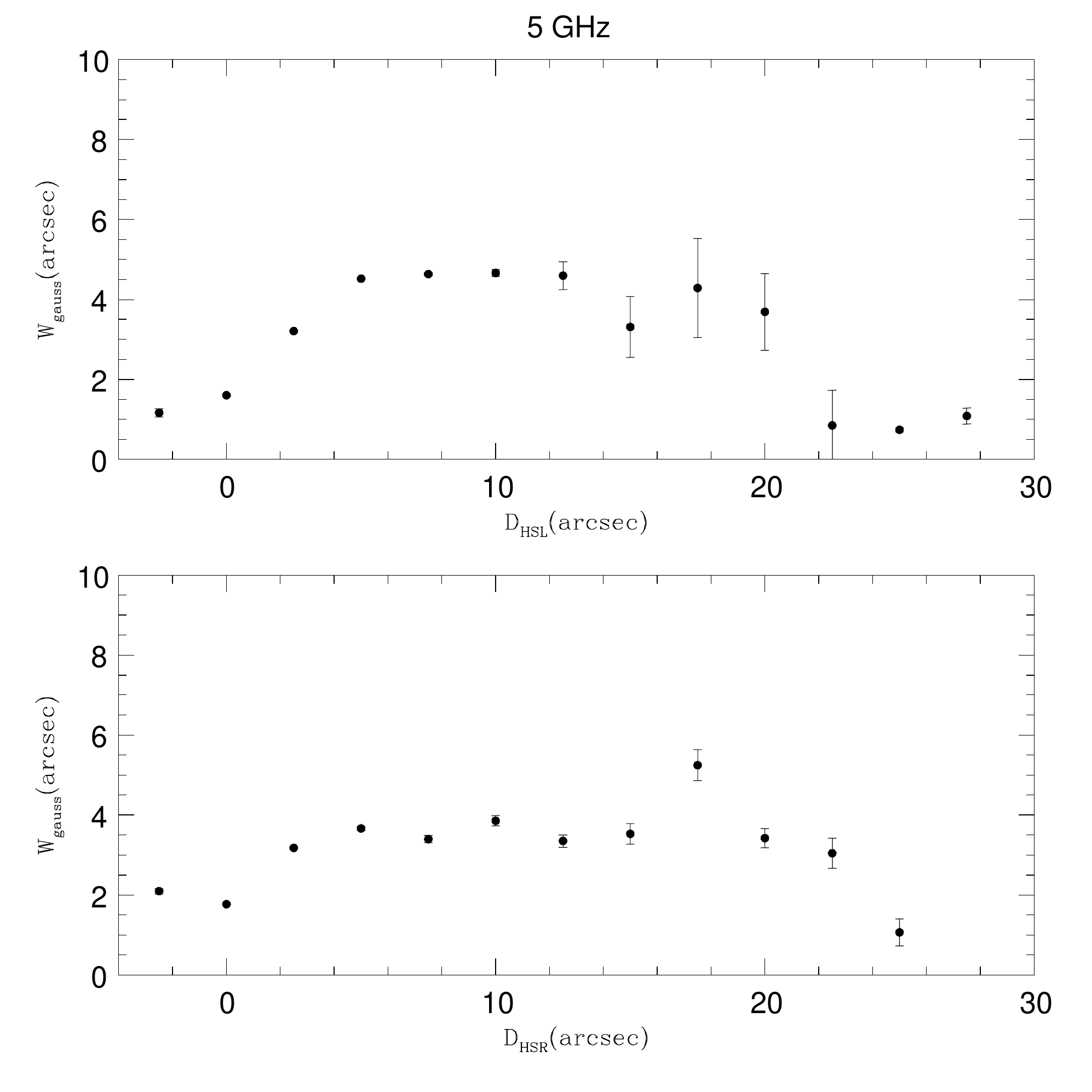}
\caption{3C114  Gaussian FWHM as a function of 
distance from the hot spot  
at 1.4 and 5 GHz (left and
right panels, respectively) for the left and right hand sides of the 
source (top and bottom panels, respectively), as in Fig. \ref{3C6.1WG}.
The right and left hand sides of the source are highly 
symmetric.  Similar results are obtained at 
1.4 and 5 GHz. }
\end{figure}

\begin{figure}
\plottwo{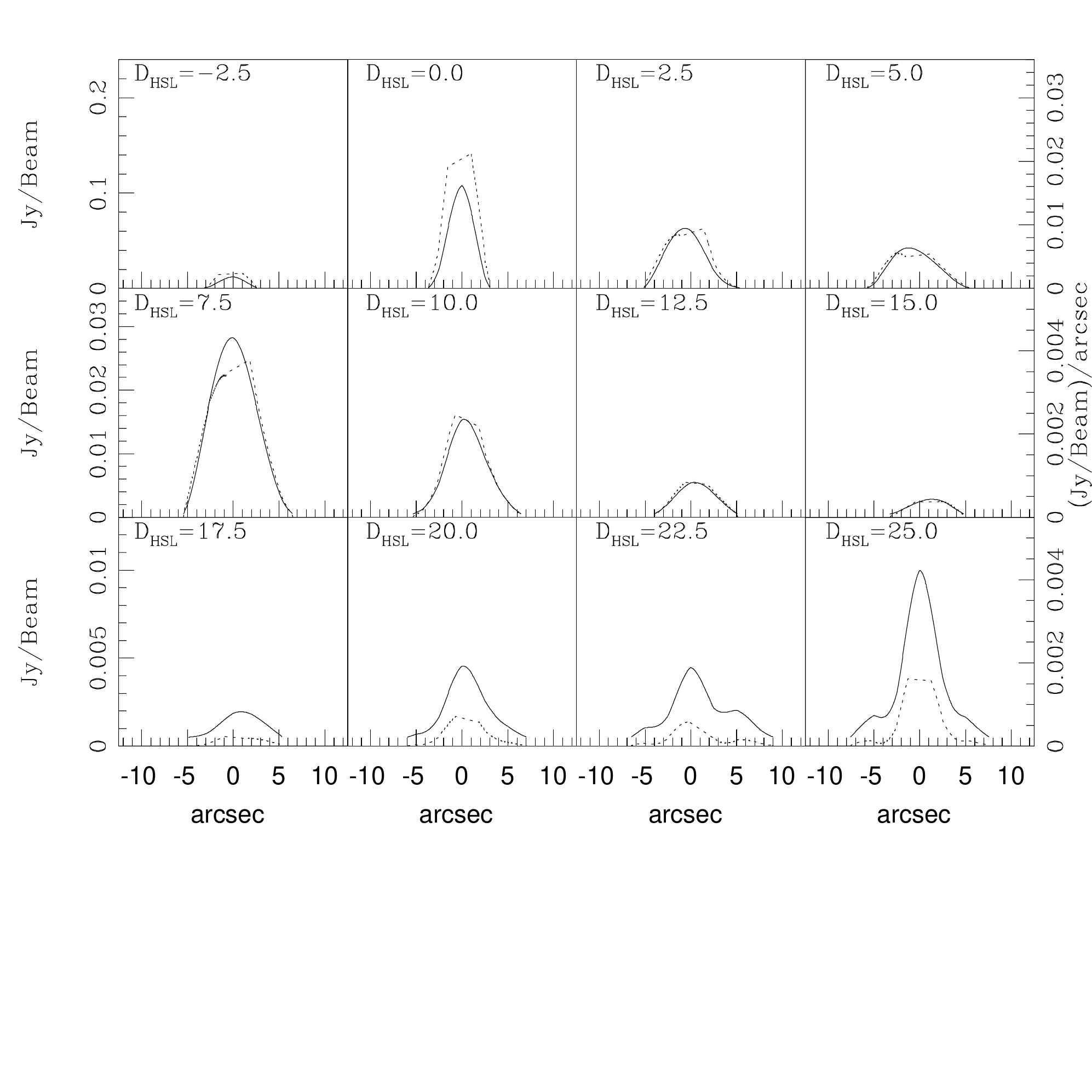}{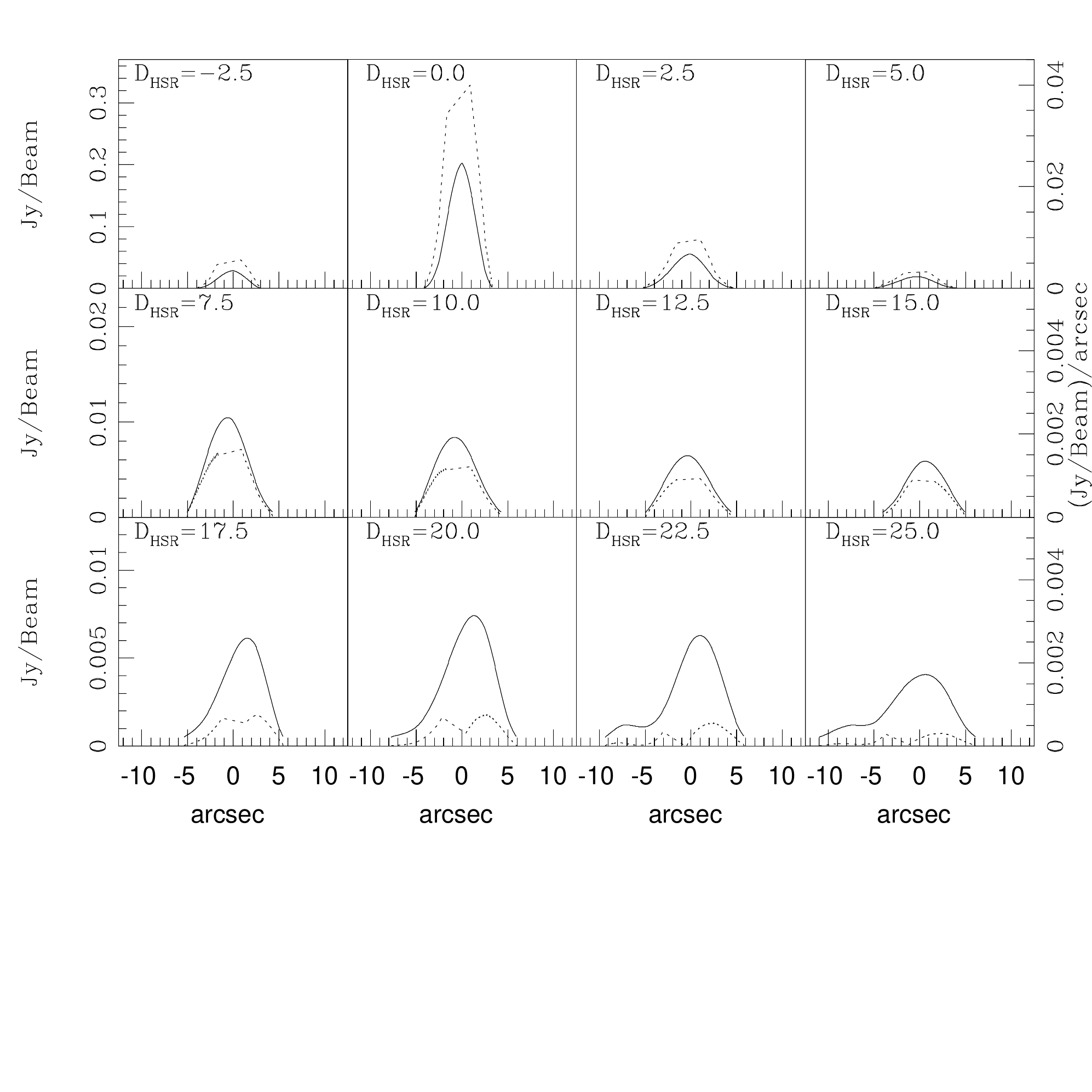}
\caption{3C114 emissivity (dotted line) and surface brightness
(solid line) for each cross-sectional slice of the radio 
bridge at 1.4 GHz, as in Fig. \ref{3C6.1SEM}. A constant volume
emissivity per slice provides a reasonable description of the 
data over most of the source. }
\label{3C114SEM}
\end{figure}

\begin{figure}
\plottwo{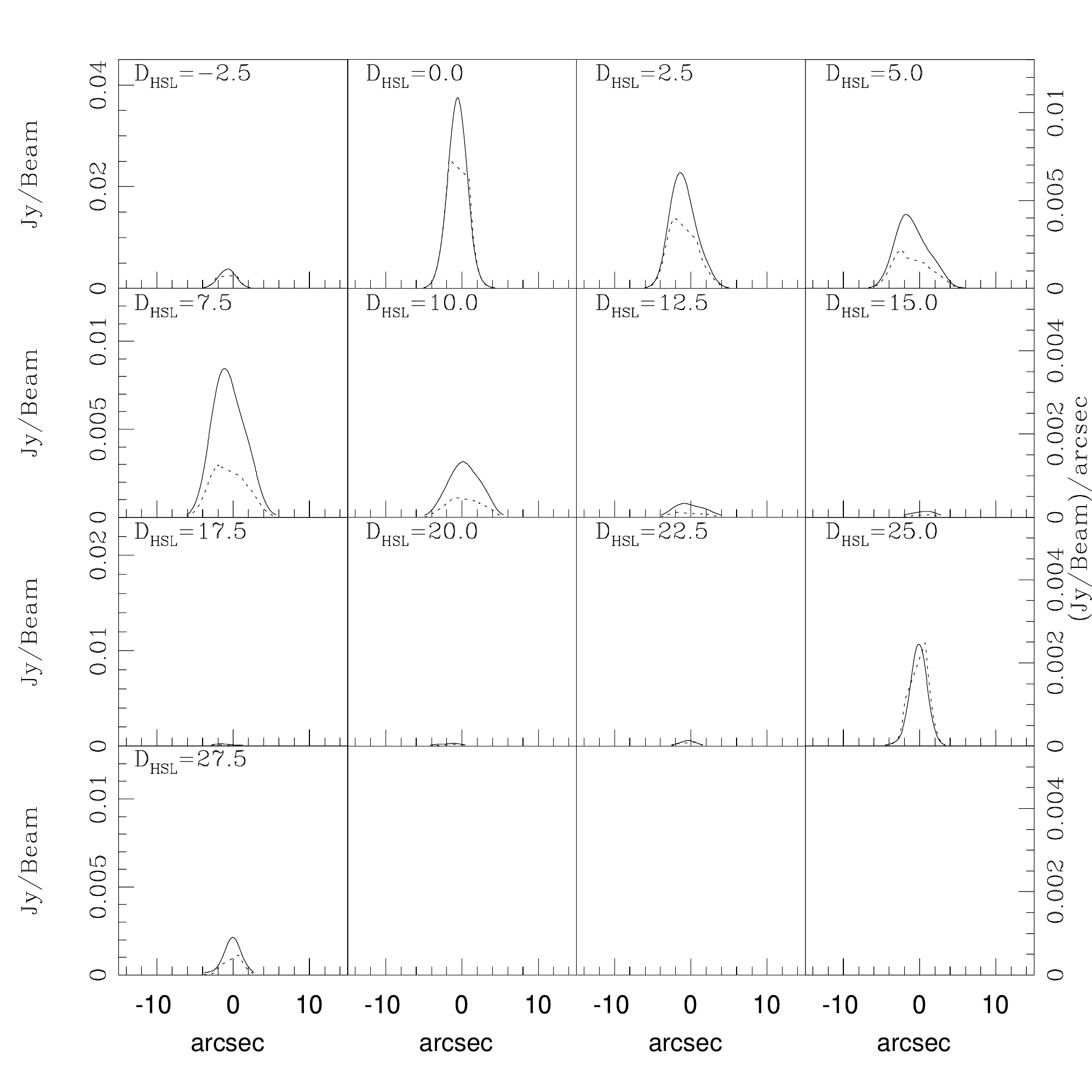}{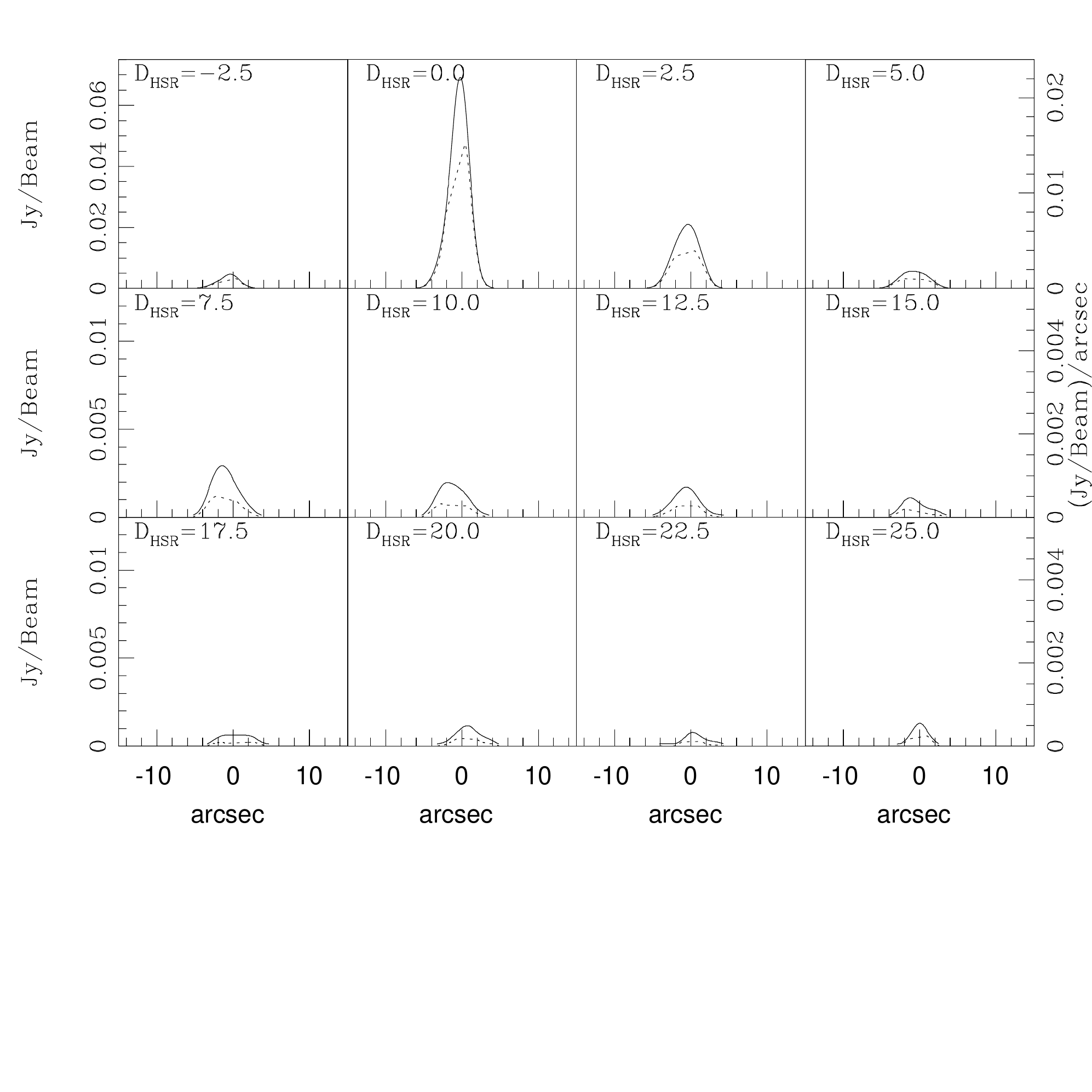}
\caption{3C114 
emissivity (dotted line) and surface brightness
(solid line) for each cross-sectional slice of the radio 
bridge at 5 GHz, as in Fig. \ref{3C114SEM} but at 5 GHz.
Results obtained at 5 GHz are nearly identical to those
obtained at 1.4 GHz.}
\end{figure}

\begin{figure}
\plottwo{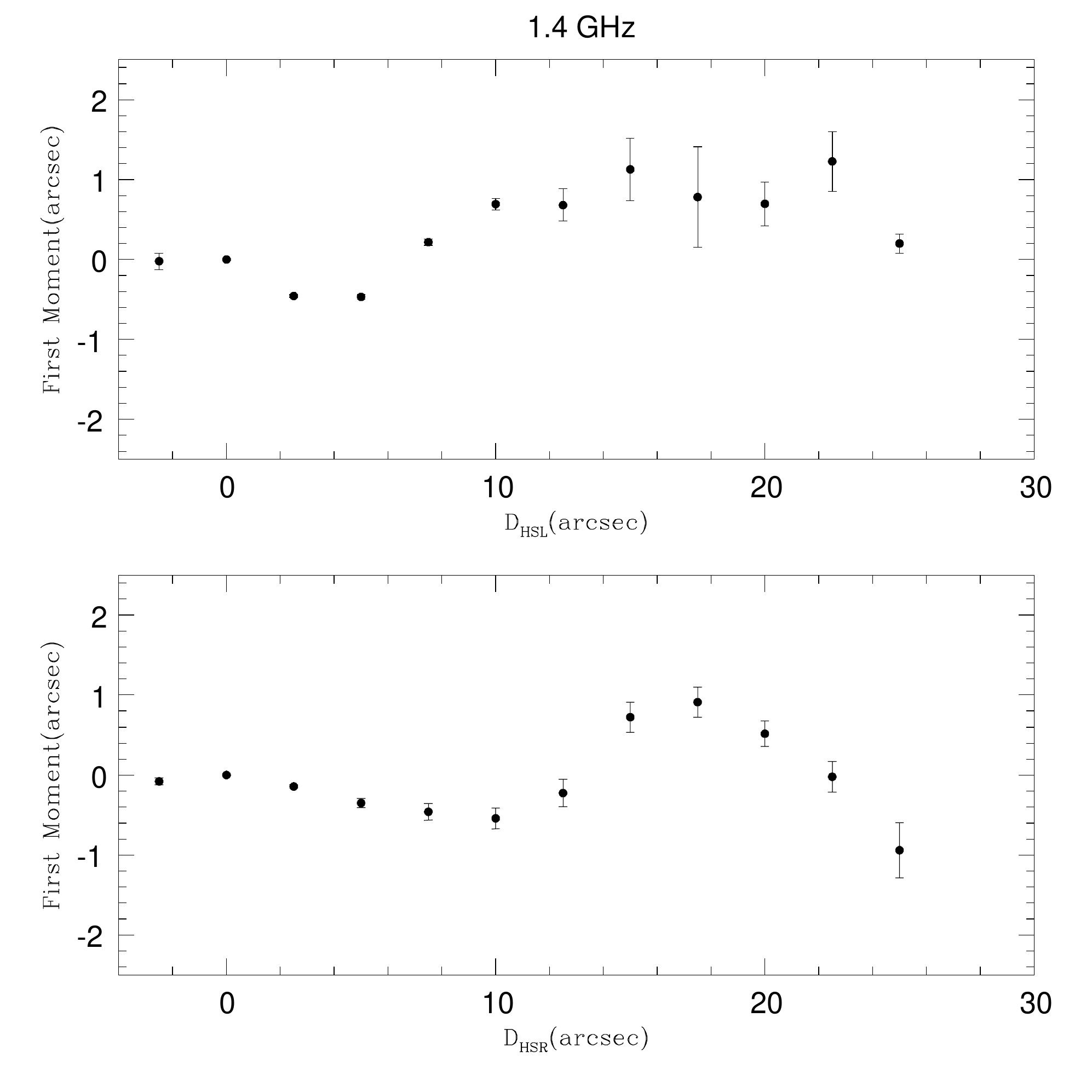}{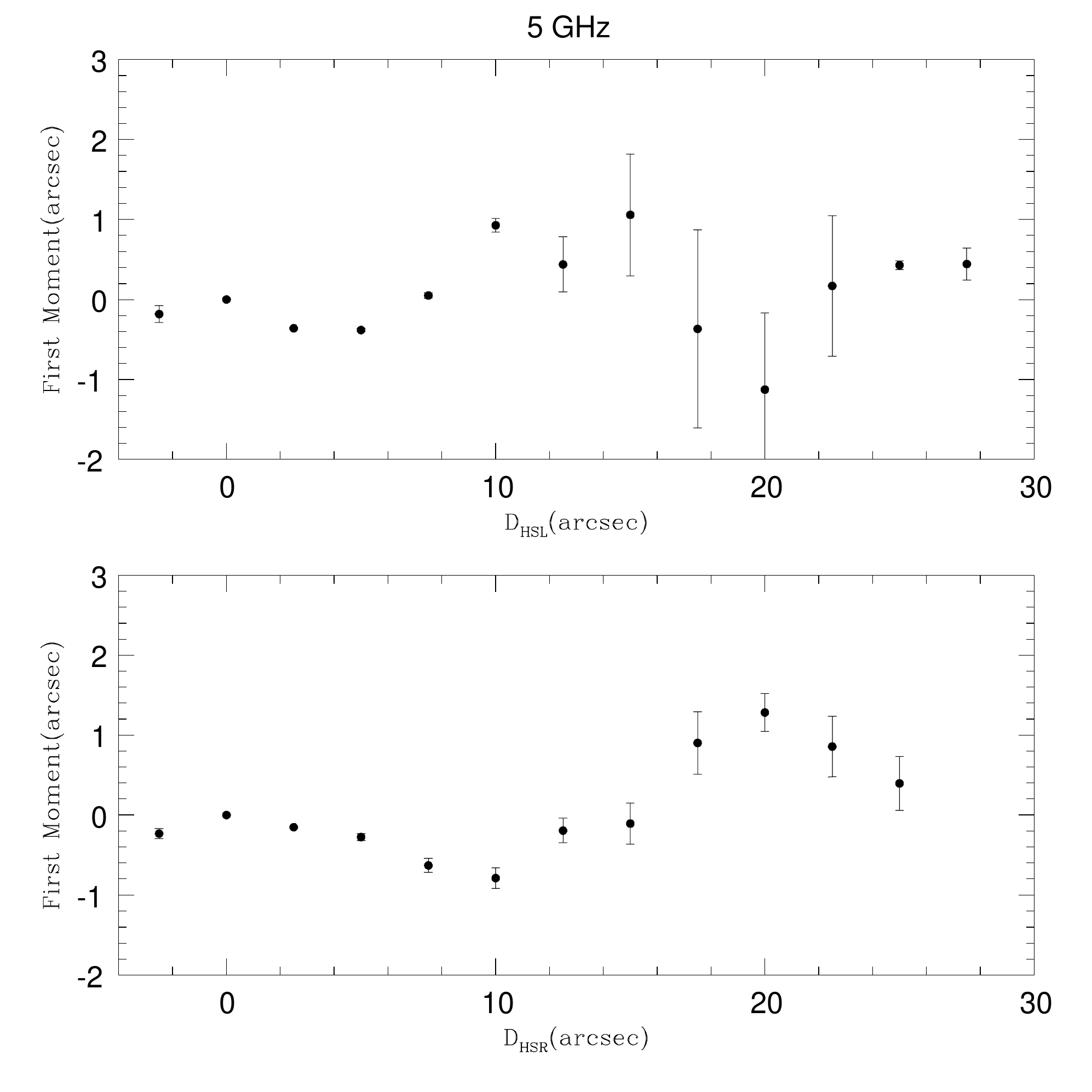}
\caption{3C114 first moment as a function of distance from 
the hot spot at 1.4 GHz and 5 GHz (left and right panels,
respectively) for the left and right hand sides of the source
(top and bottom, respectively).}
\end{figure}

\begin{figure}
\plottwo{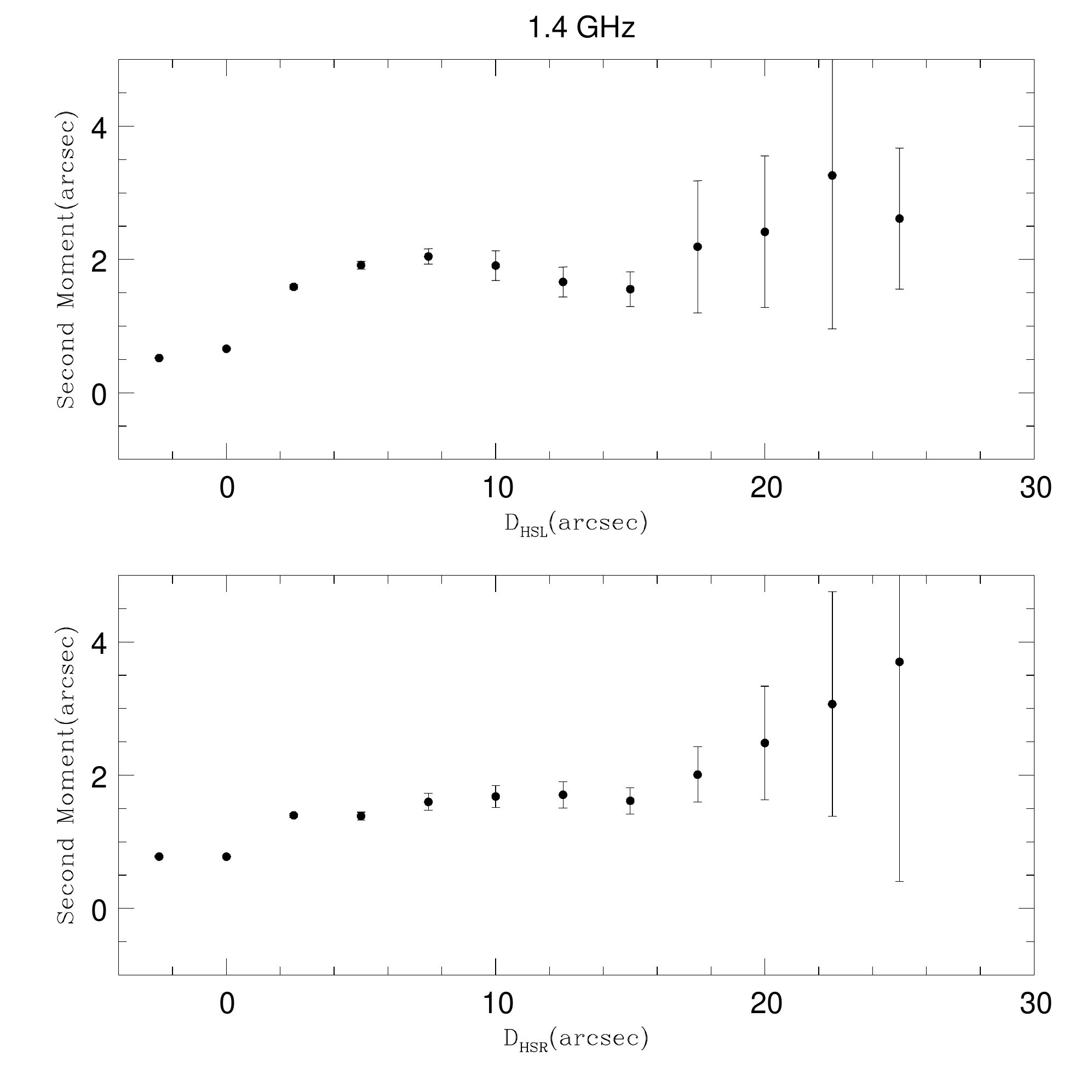}{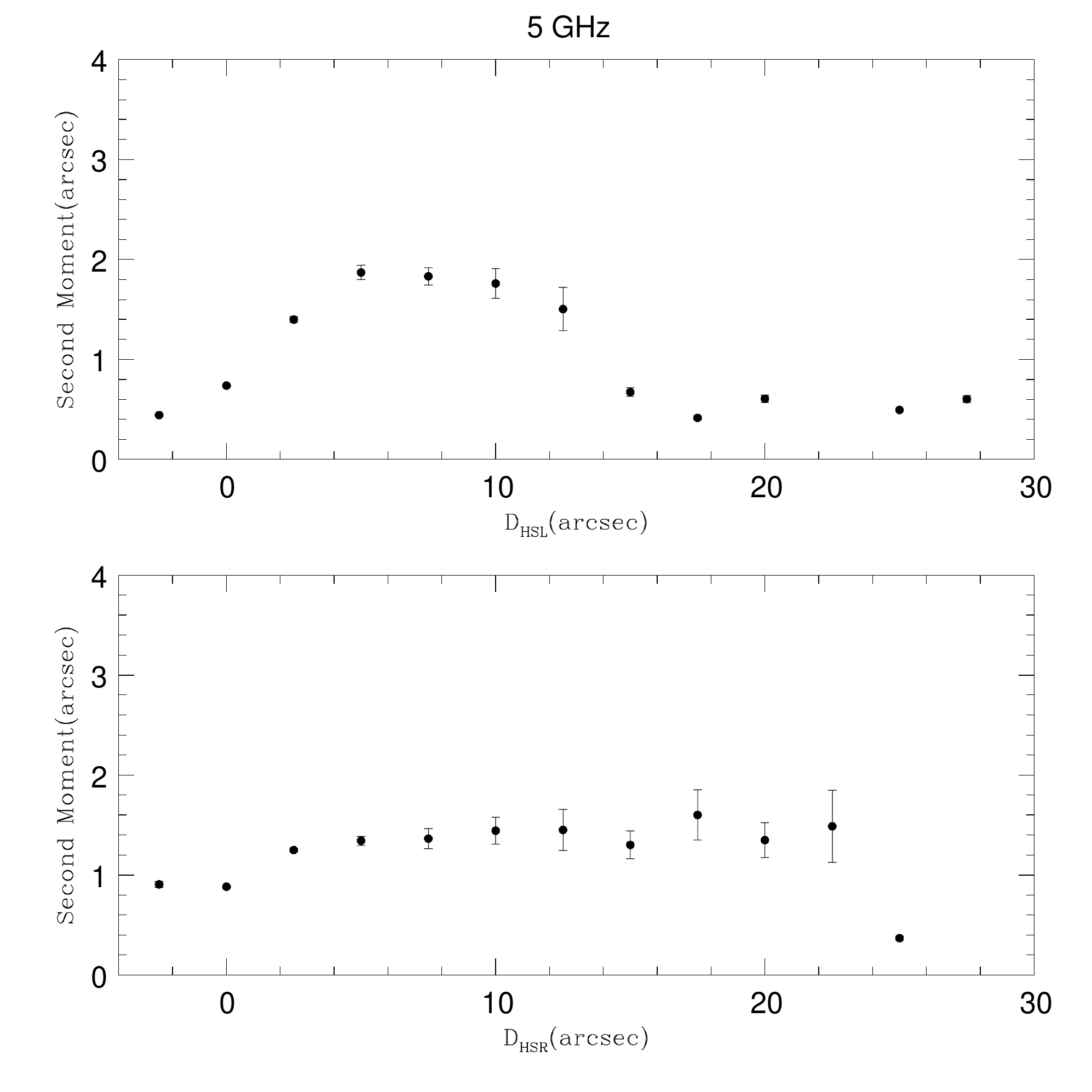}
\caption{3C114 second moment as a function of distance from the
hot spot at 1.4 GHz and 5 GHz (left and right panels, 
respectively) for the left and right sides of the source
(top and bottom panels, respectively).}
\end{figure}

\begin{figure}
\plottwo{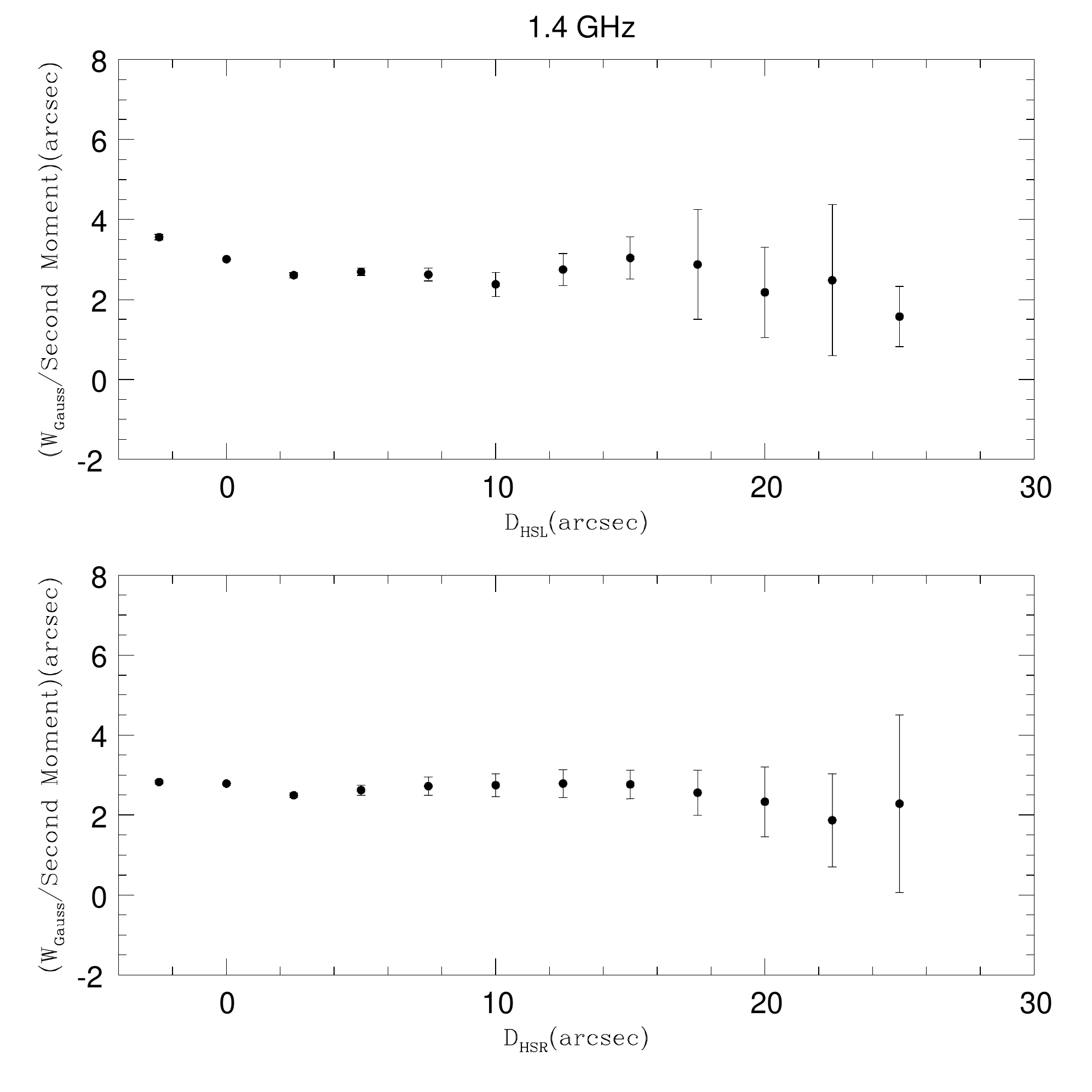}{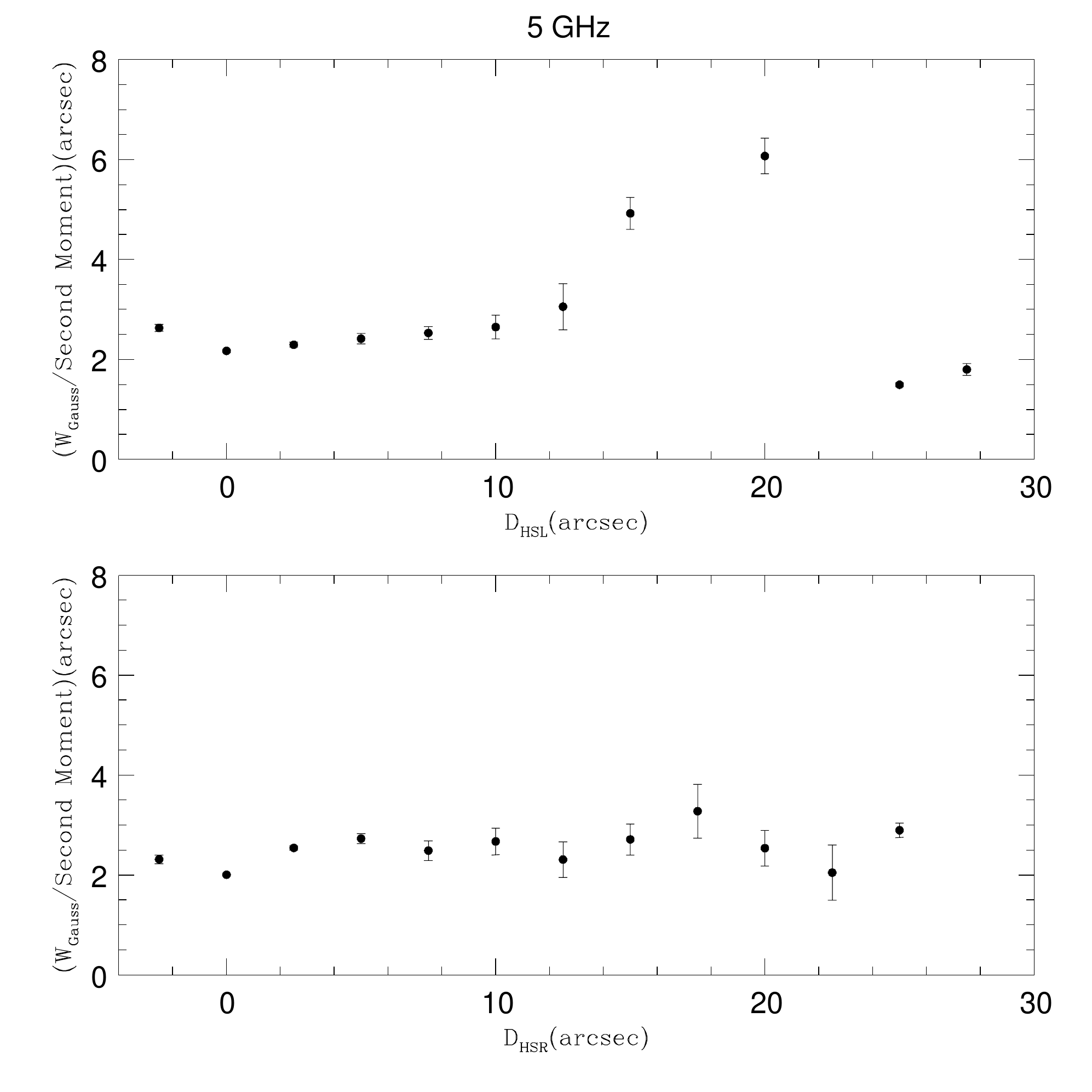}
\caption{3C114 ratio of the Gaussian FWHM to the second moment as a
function of distance from the hot spot at 1.4 and 5 GHz 
(left and right panels, respectively) for the left and right
hand sides of the source (top and bottom panels, respectively). The 
value of this ratio is fairly constant for each side of the source, and has a 
value of about 2.5 to 3. Similar results are obtained at 1.4 and 5 GHz.} 
\end{figure}

\begin{figure}
\plottwo{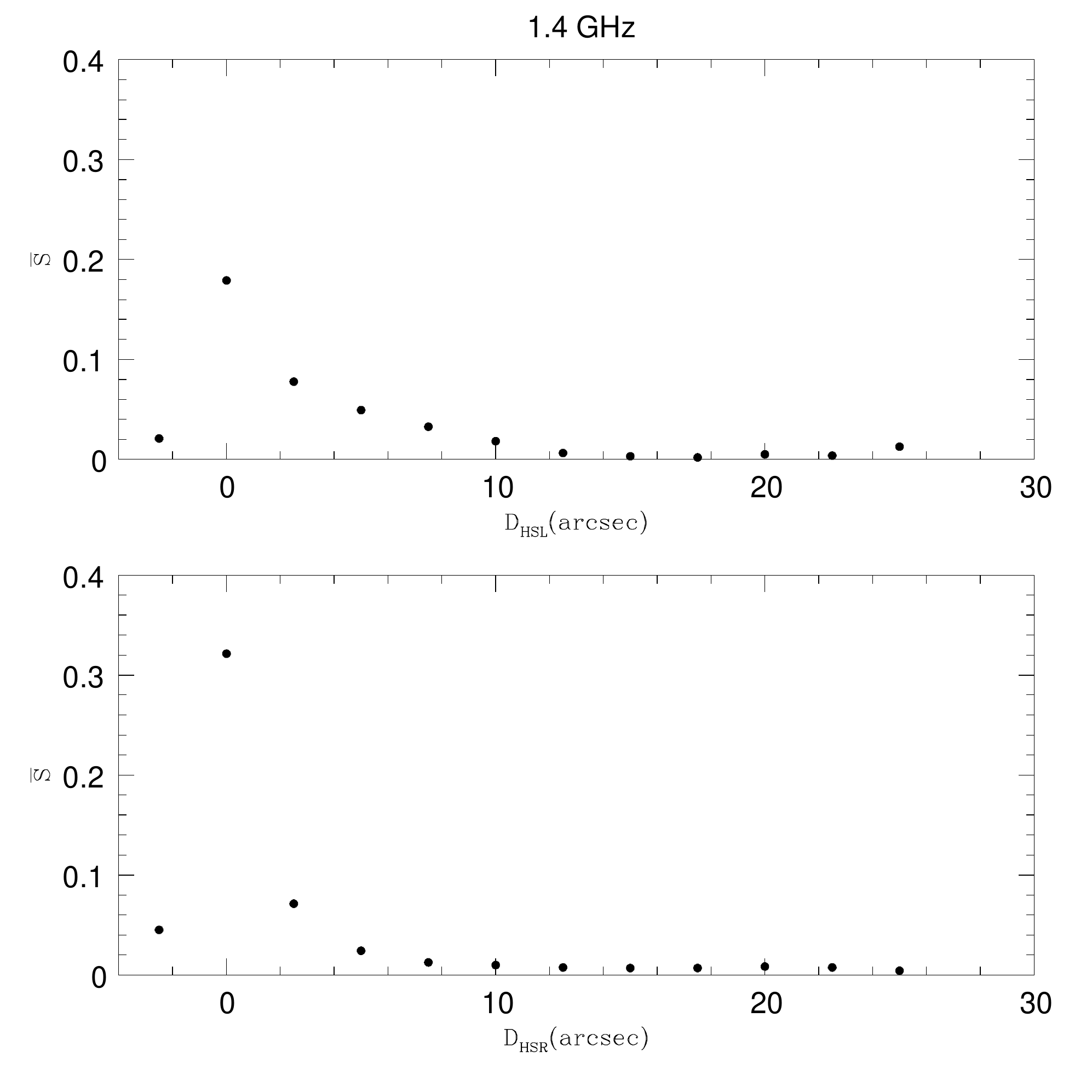}{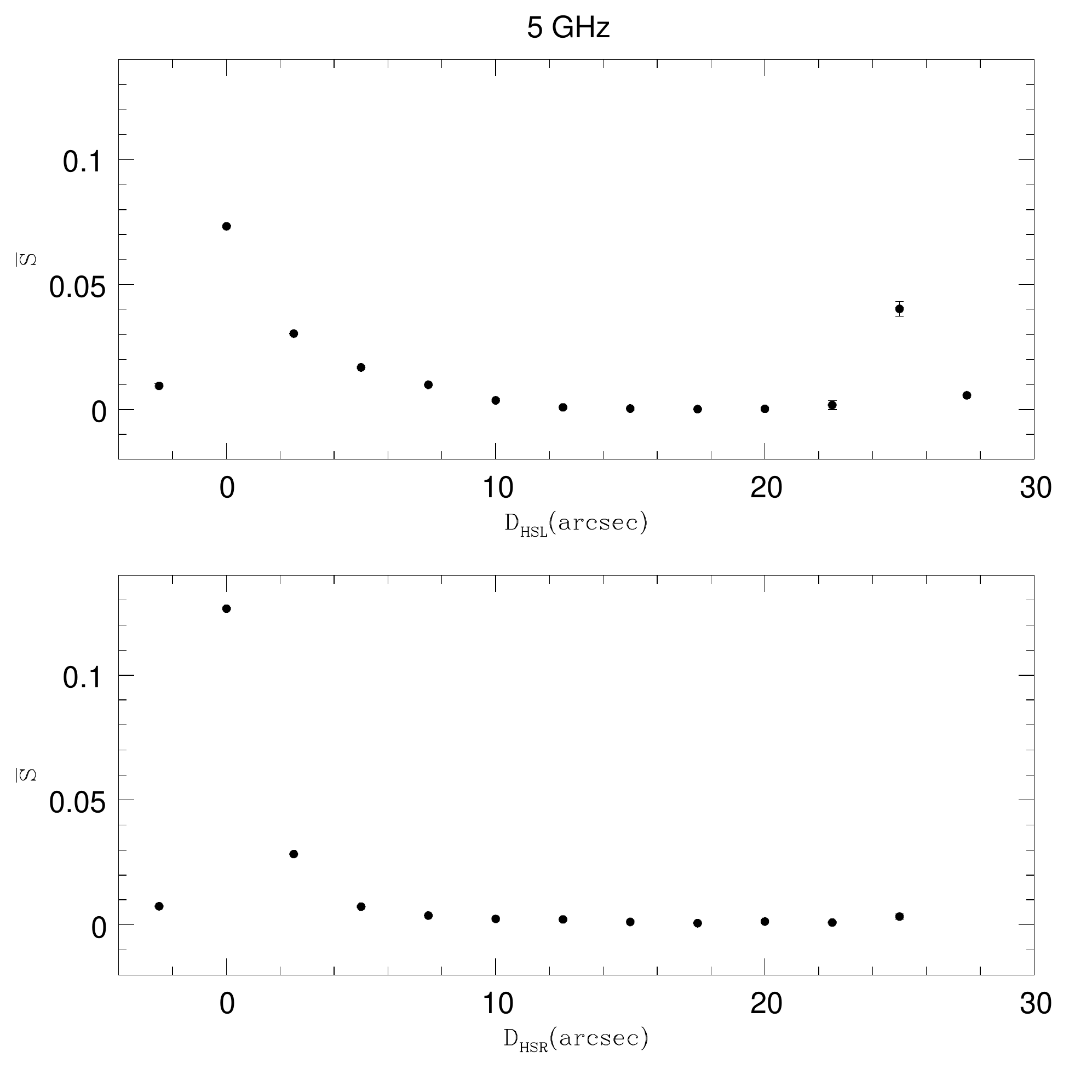}
\caption{3C114 average surface brightness in units of Jy/beam 
as a function of distance from the 
hot spot at 1.4 and 5 GHz 
(left and right panels, respectively) for the left and right
hand sides of the source (top and bottom panels, respectively).}
\end{figure}

\begin{figure}
\plottwo{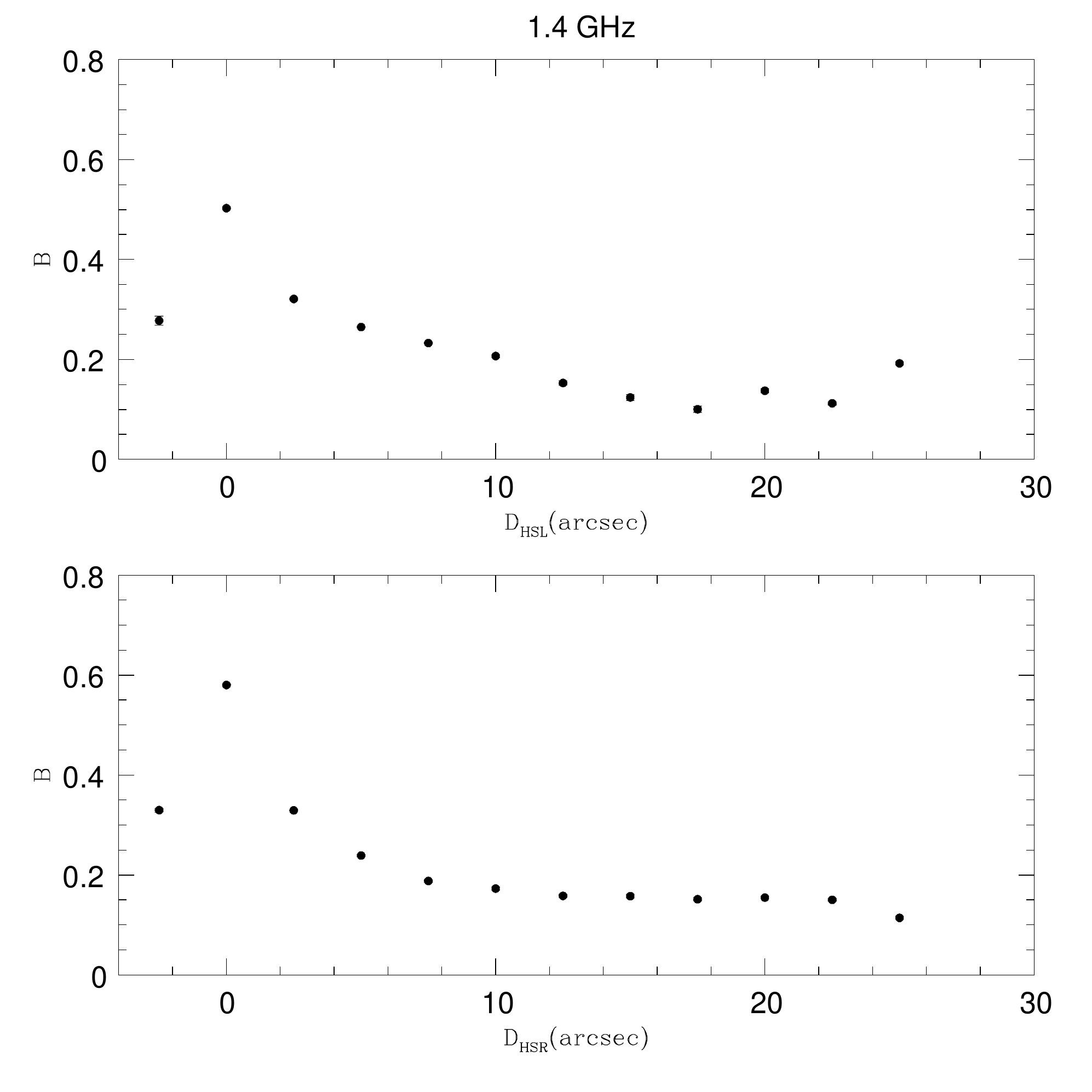}{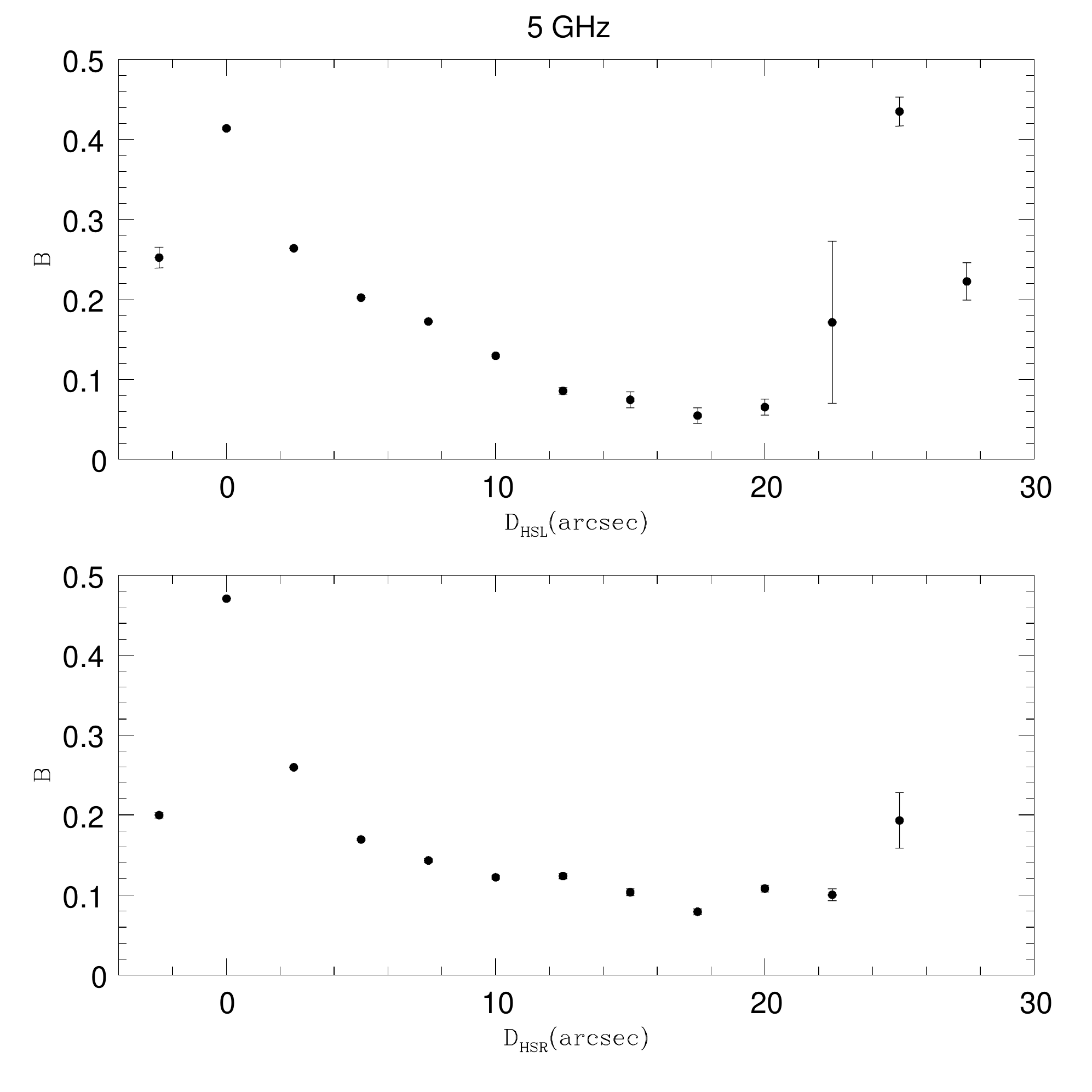}
\caption{3C114 minimum energy magnetic field strength 
as a function of distance from the 
hot spot at 1.4 and 5 GHz 
(left and right panels, respectively) for the left and right
hand sides of the source (top and bottom panels, respectively).
The normalization is given in Table 1. 
}
\end{figure}

\begin{figure}
\plottwo{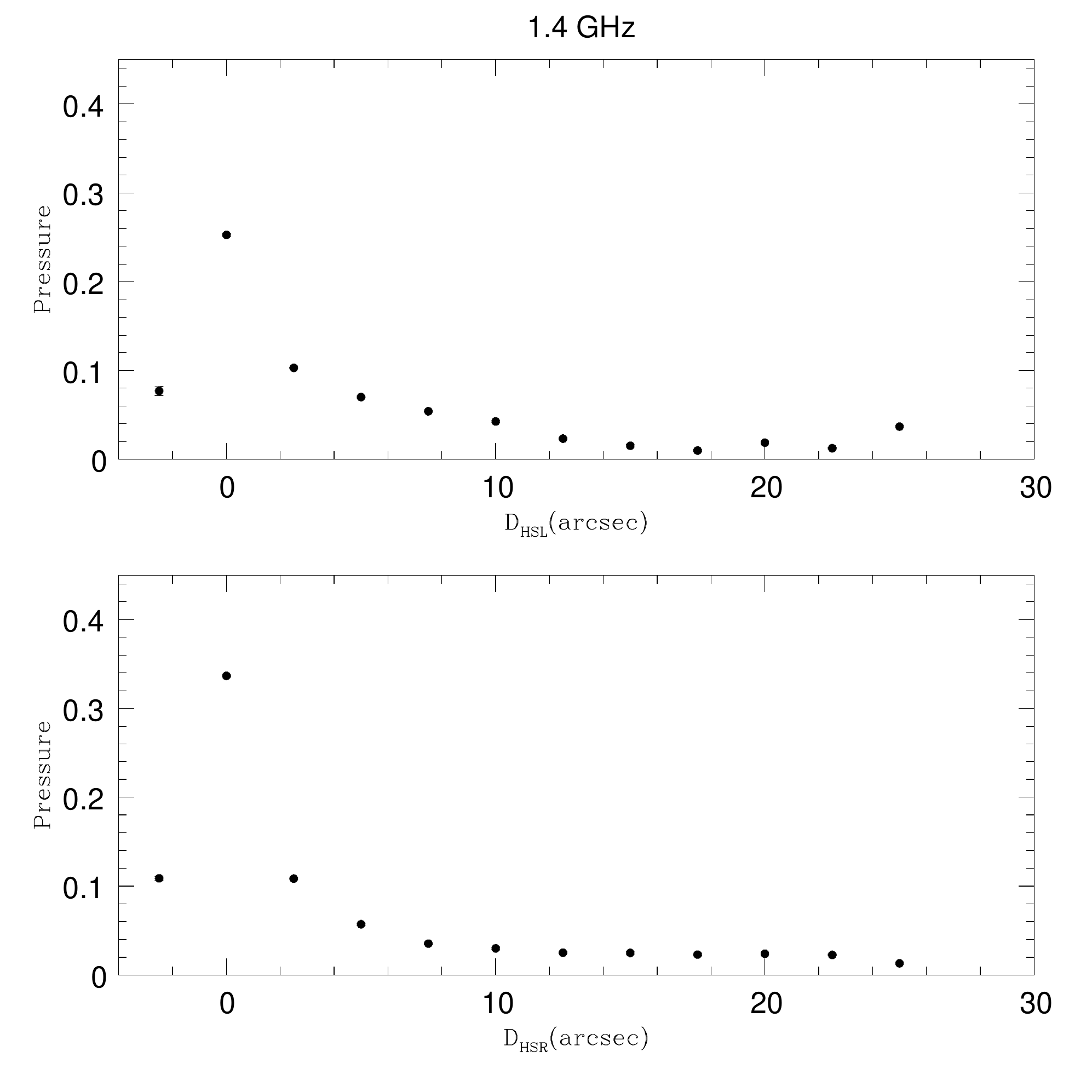}{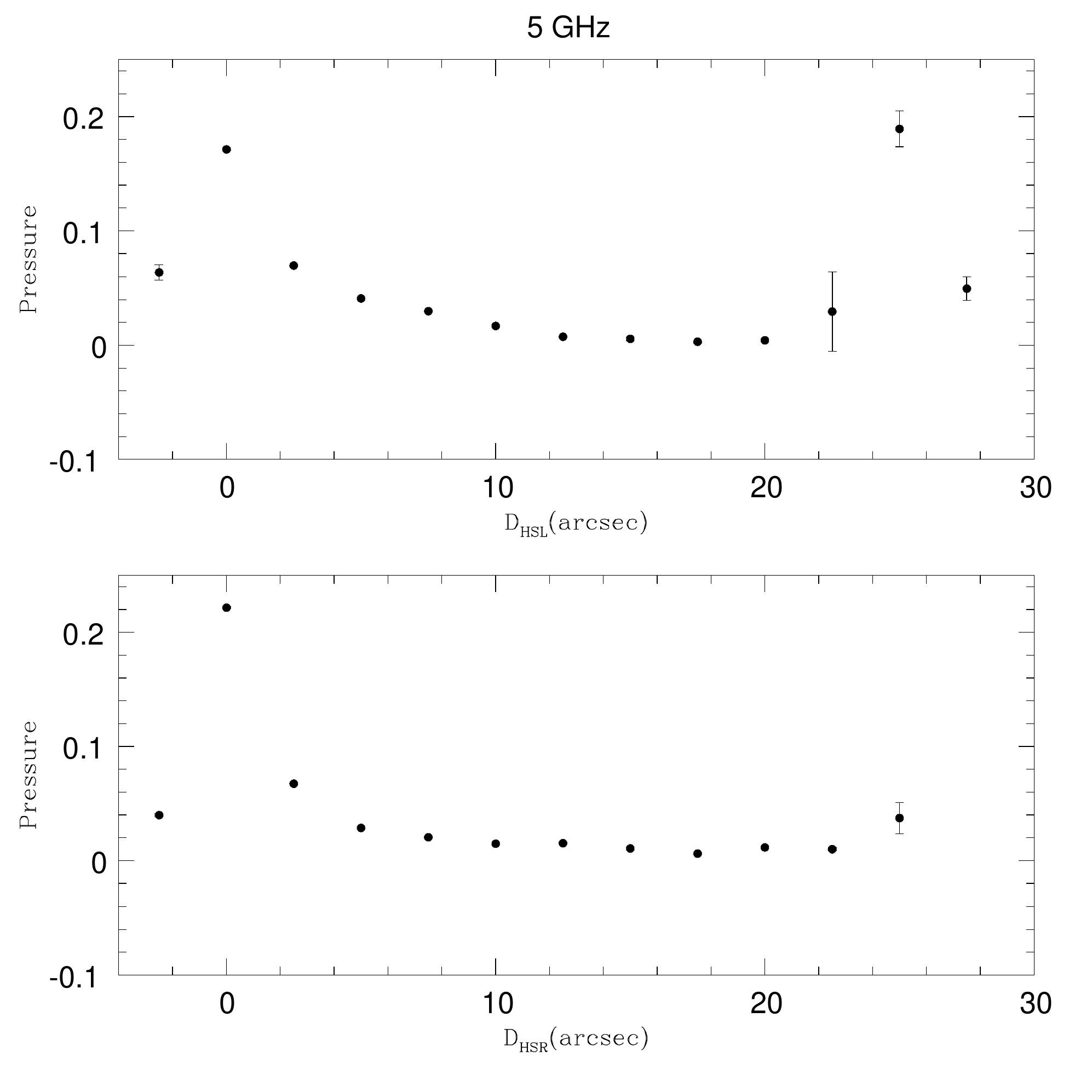}
\caption{3C114 
minimum energy pressure as a function of 
distance from the hot spot at 1.4 and 5 GHz 
(left and right panels, respectively) for the left and right
hand sides of the source (top and bottom panels, respectively), 
as in Fig. \ref{3C6.1P}. The normalization is given in Table 1.
}
\end{figure}

\clearpage
\begin{figure}
\plotone{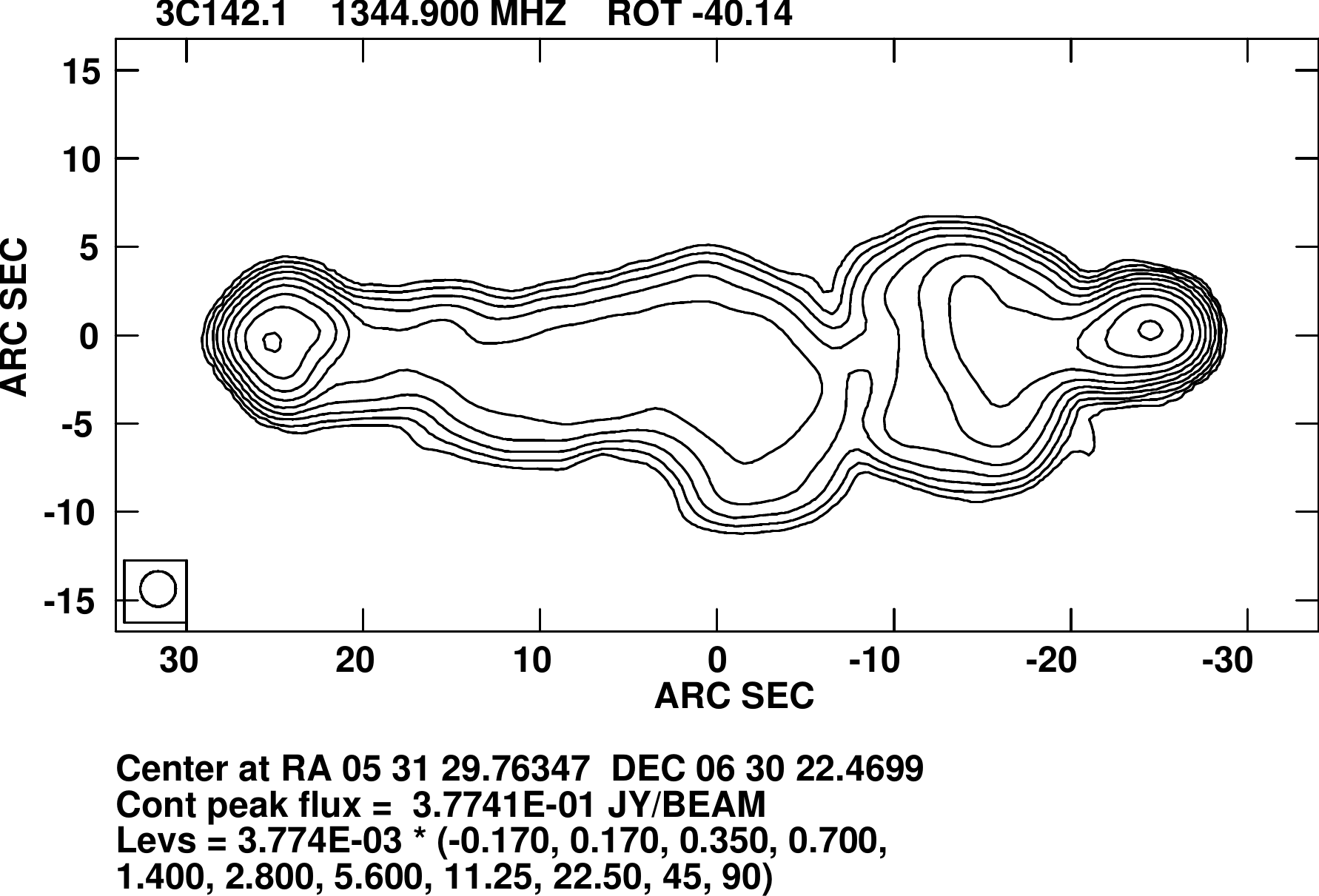}
\caption{3C142.1 at 1.4 GHz and 2'' resolution rotated by 40.14 deg 
clockwise, as in Fig. \ref{3C6.1rotfig}.}
\end{figure}

\begin{figure}
\plottwo{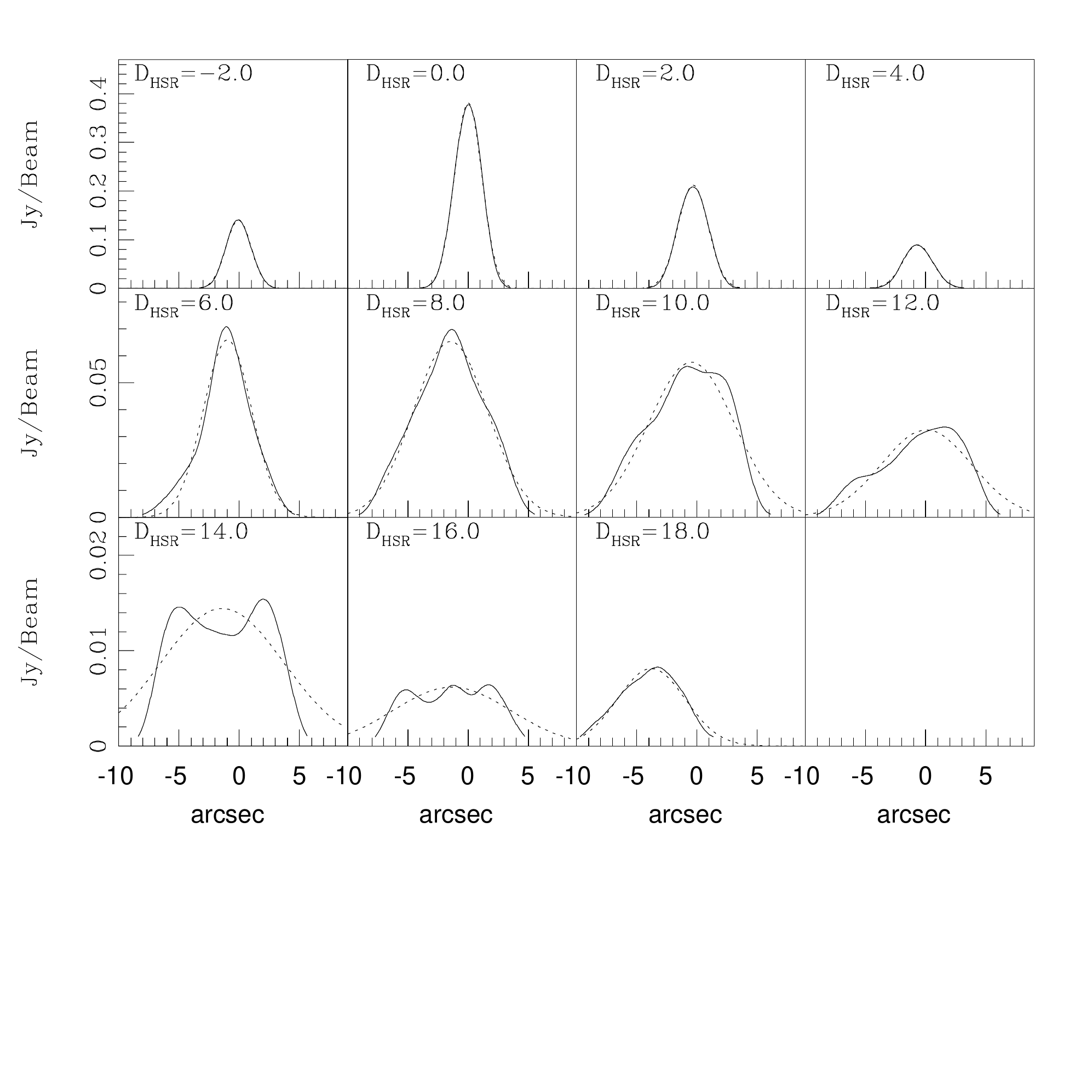}{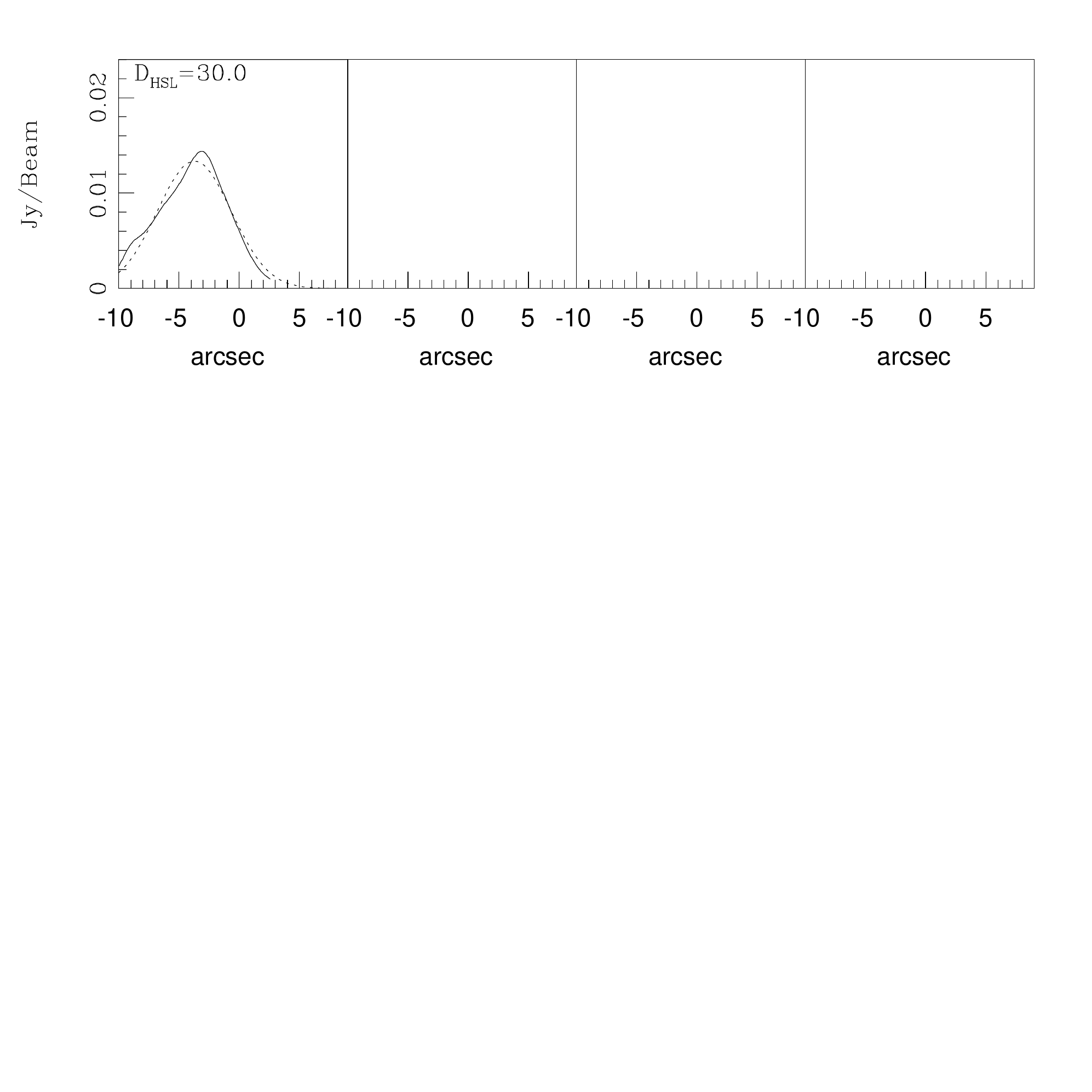}
\caption{3C142.1 
surface brightness (solid line) and best fit 
Gaussian (dotted line) for each cross-sectional slice of the 
radio bridge at 1.4 GHz, as in Fig. \ref{3C6.1SG}.
A Gaussian provides an excellent description of the surface brightness
profile of each cross-sectional slice. }
\label{3C142.1SG}
\end{figure}

\begin{figure}
\plotone{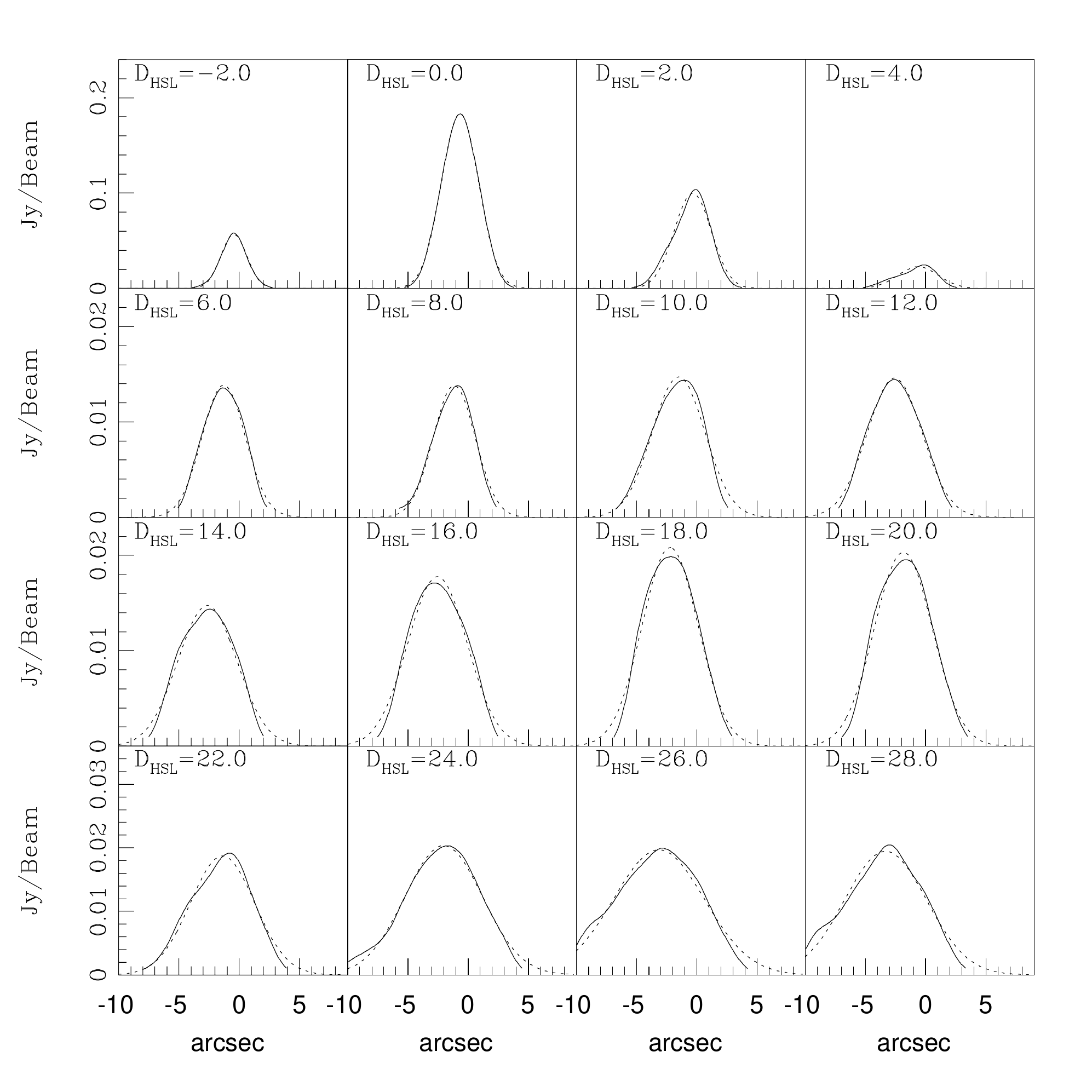}
\caption{3C142.1 
surface brightness (solid line) and best fit 
Gaussian (dotted line) for each cross-sectional slice of the 
radio bridge at 1.4 GHz, as in Fig. \ref{3C142.1SG}.
A Gaussian provides an excellent description of the surface brightness
profile of each cross-sectional slice. }
\end{figure}

\begin{figure}
\plottwo{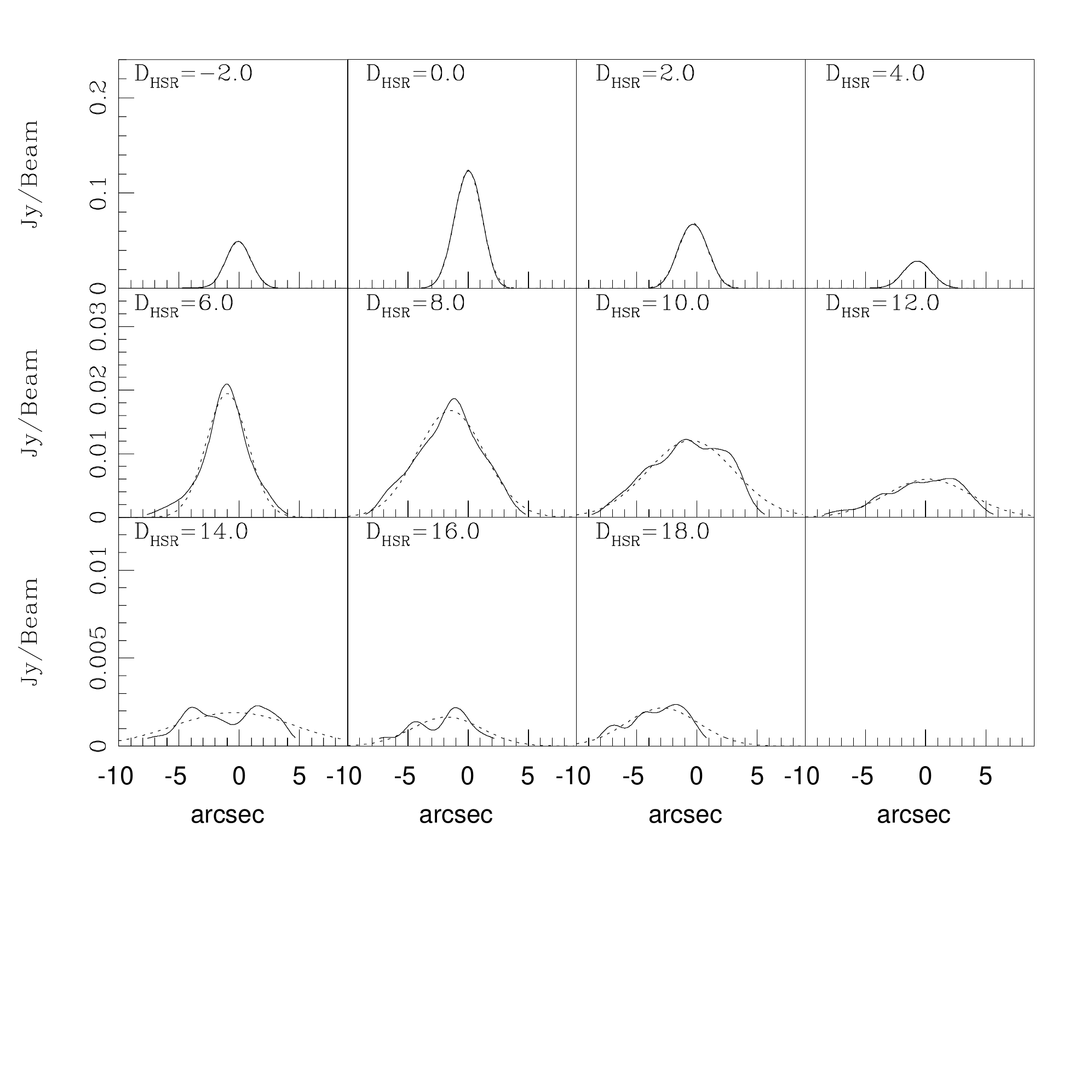}{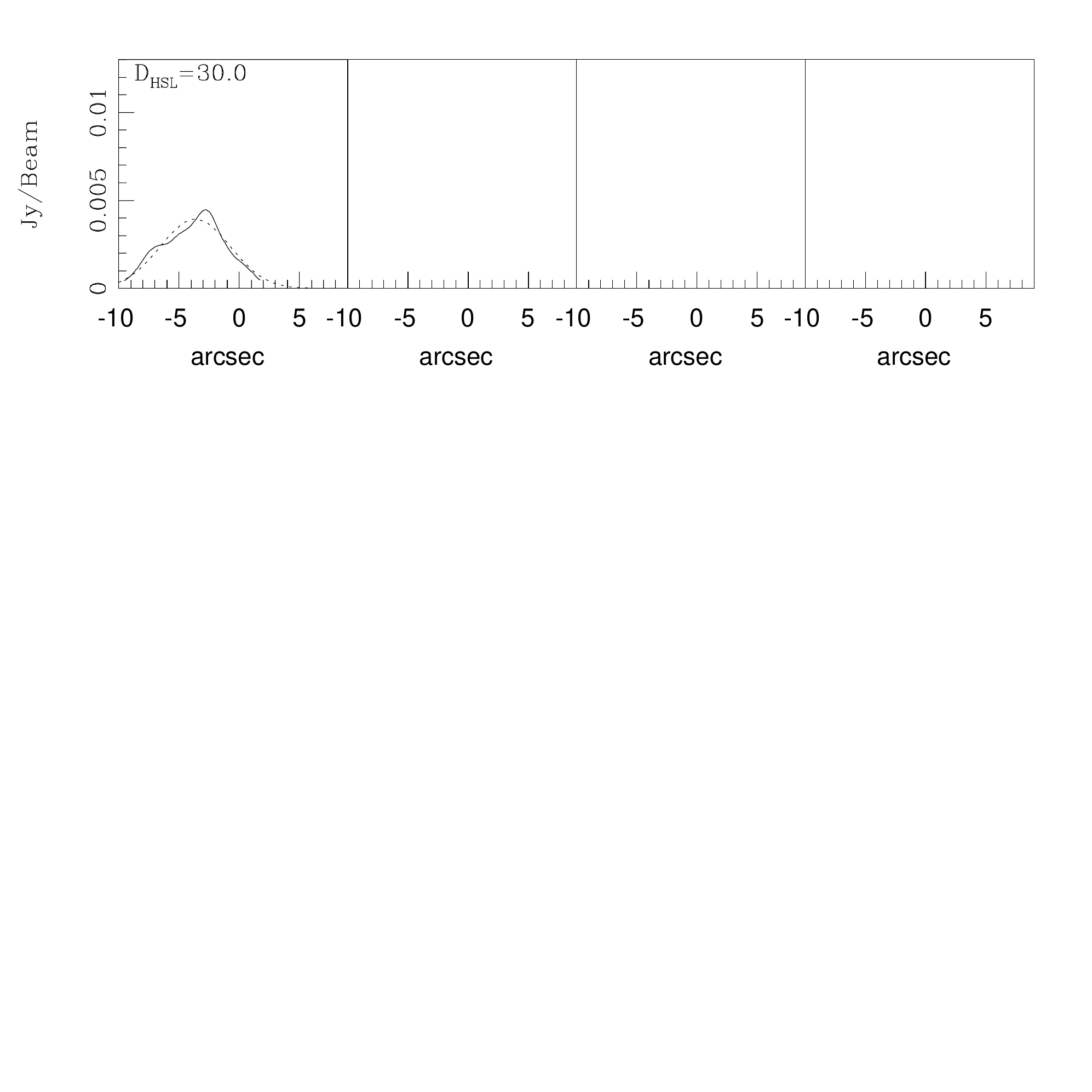}
\caption{3C142.1 
surface brightness (solid line) and best fit 
Gaussian (dotted line) for each cross-sectional slice of the 
radio bridge at 5 GHz, as in 
Fig. \ref{3C142.1SG} but at 5 GHz. 
Results obtained at 5 GHz are nearly identical to those
obtained at 1.4 GHz.}
\label{3C142.1SGA}
\end{figure}

\begin{figure}
\plotone{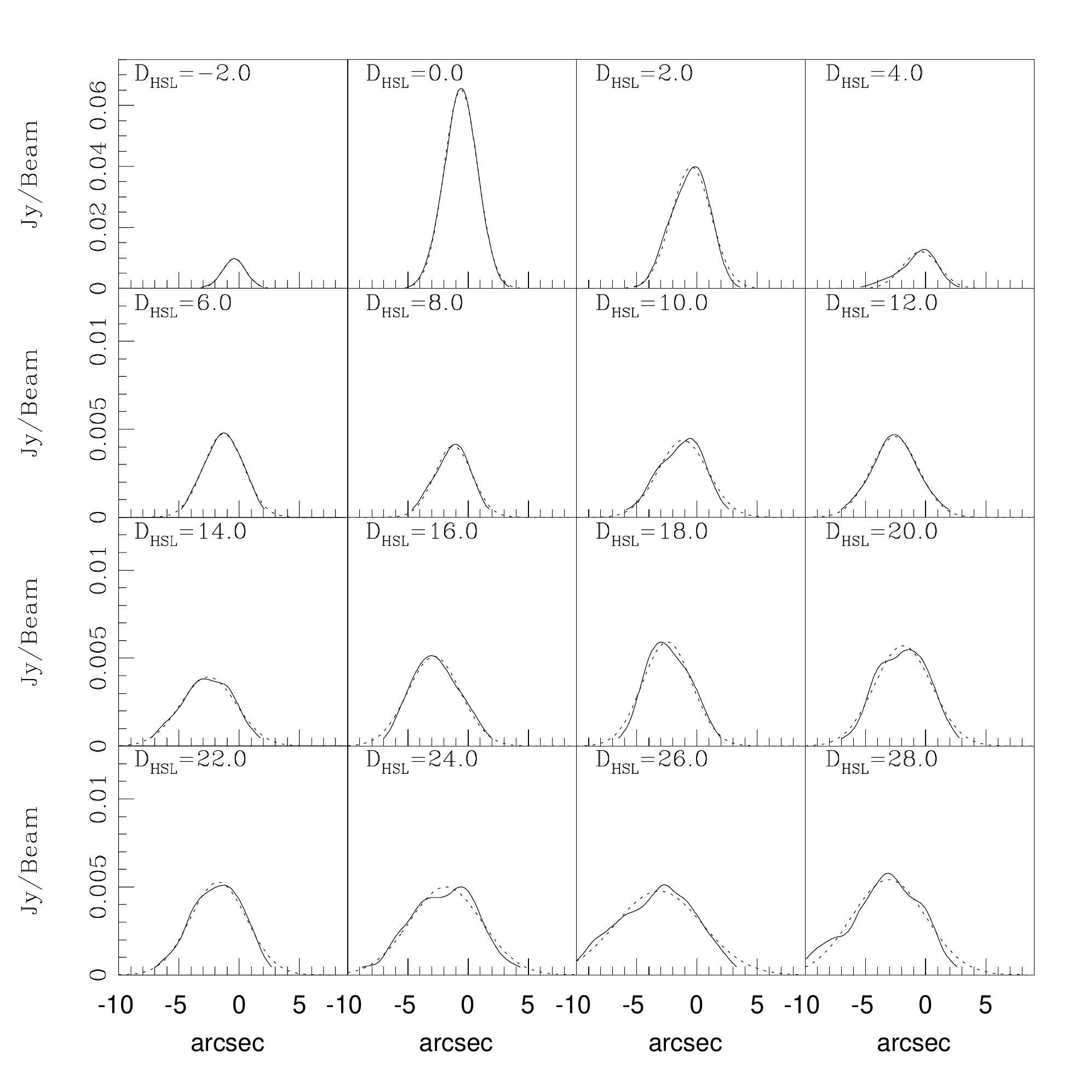}
\caption{3C142.1 
surface brightness (solid line) and best fit 
Gaussian (dotted line) for each cross-sectional slice of the 
radio bridge at 5 GHz, as in 
Fig. \ref{3C142.1SGA}. 
Results obtained at 5 GHz are nearly identical to those
obtained at 1.4 GHz.}
\end{figure}

\begin{figure}
\plottwo{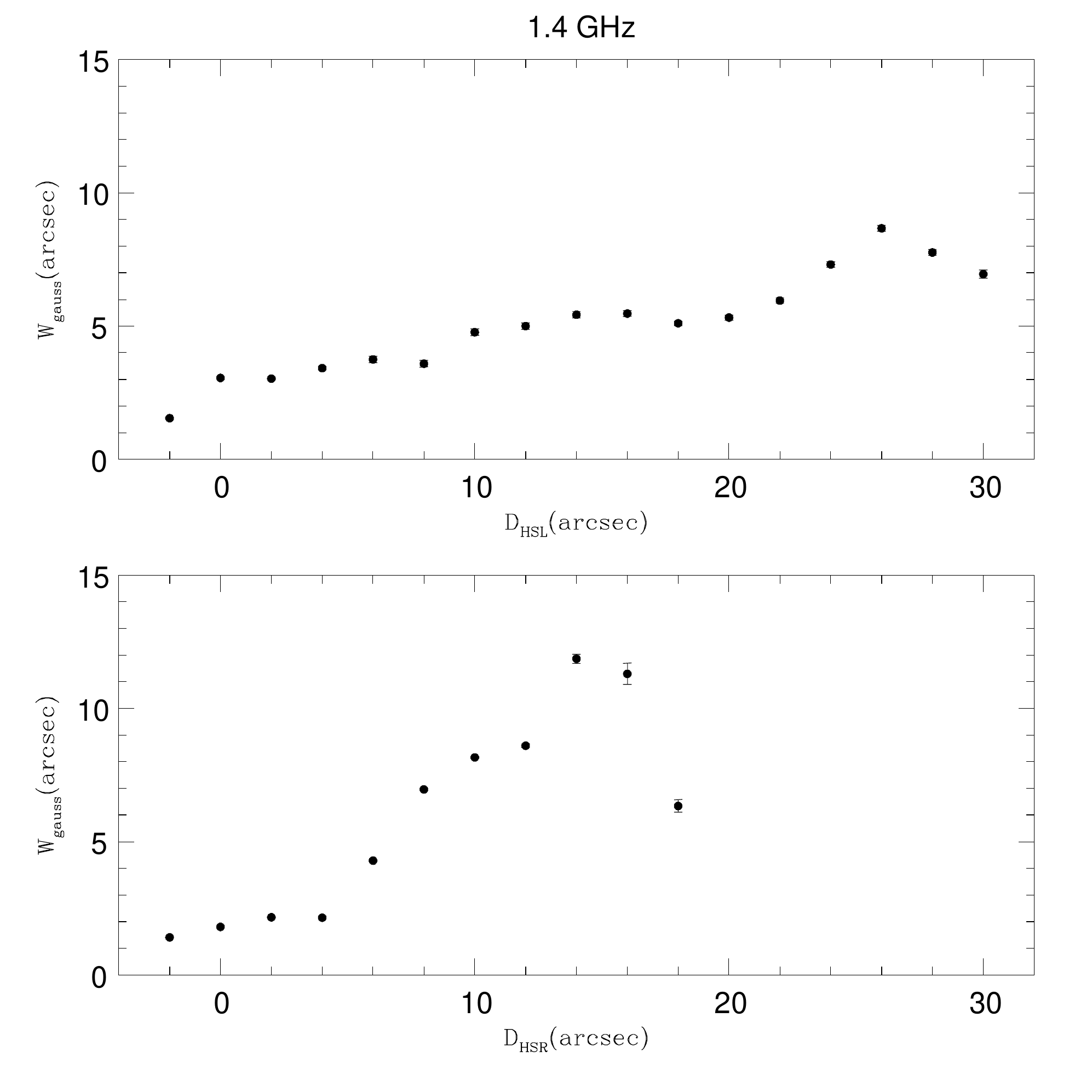}{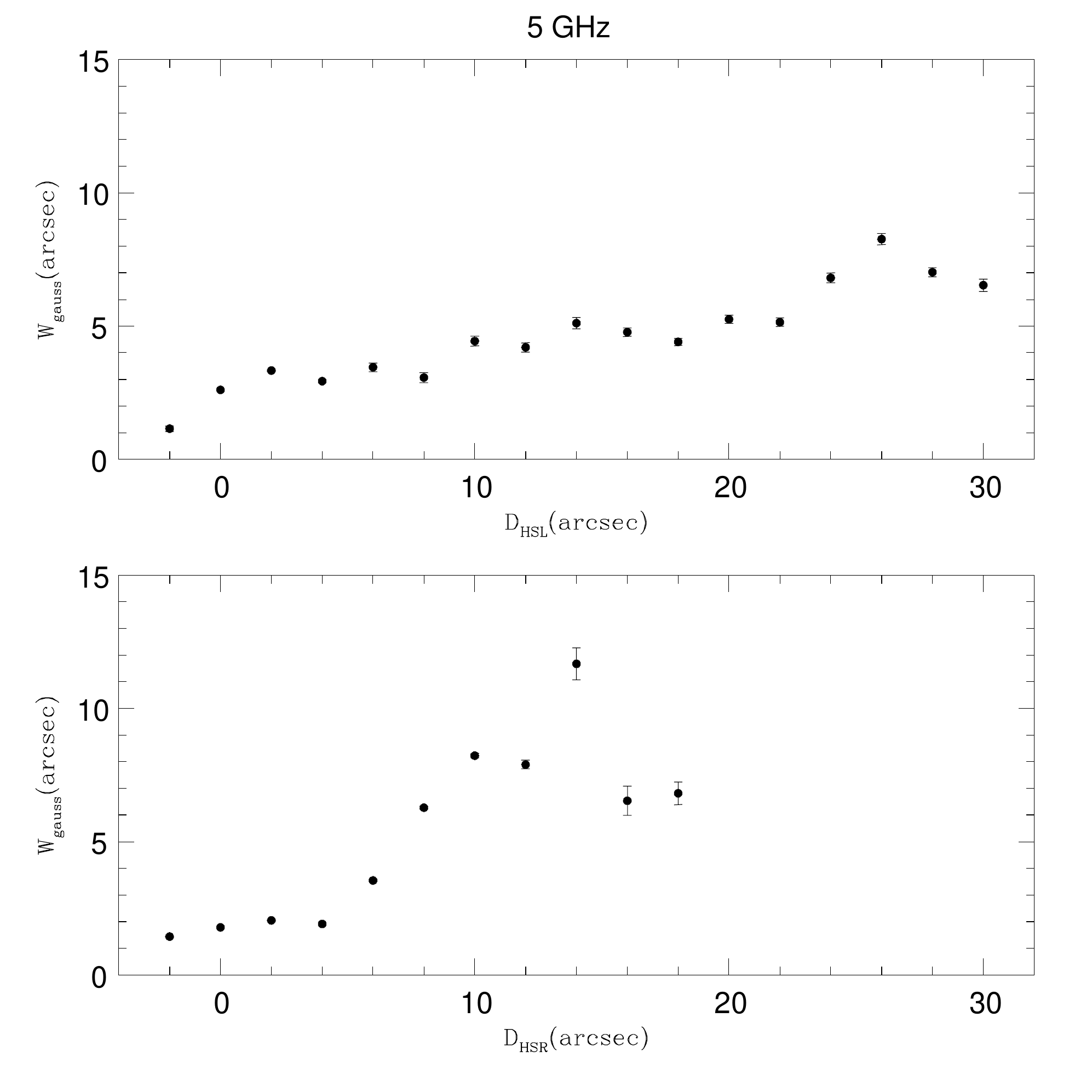}
\caption{3C142.1  Gaussian FWHM as a function of 
distance from the hot spot  
at 1.4 and 5 GHz (left and
right panels, respectively) for the left and right hand sides of the 
source (top and bottom panels, respectively), as in Fig. \ref{3C6.1WG}.
The right and left hand sides of the source are somewhat 
symmetric.  Similar results are obtained at 
1.4 and 5 GHz. }
\end{figure}

\begin{figure}
\plottwo{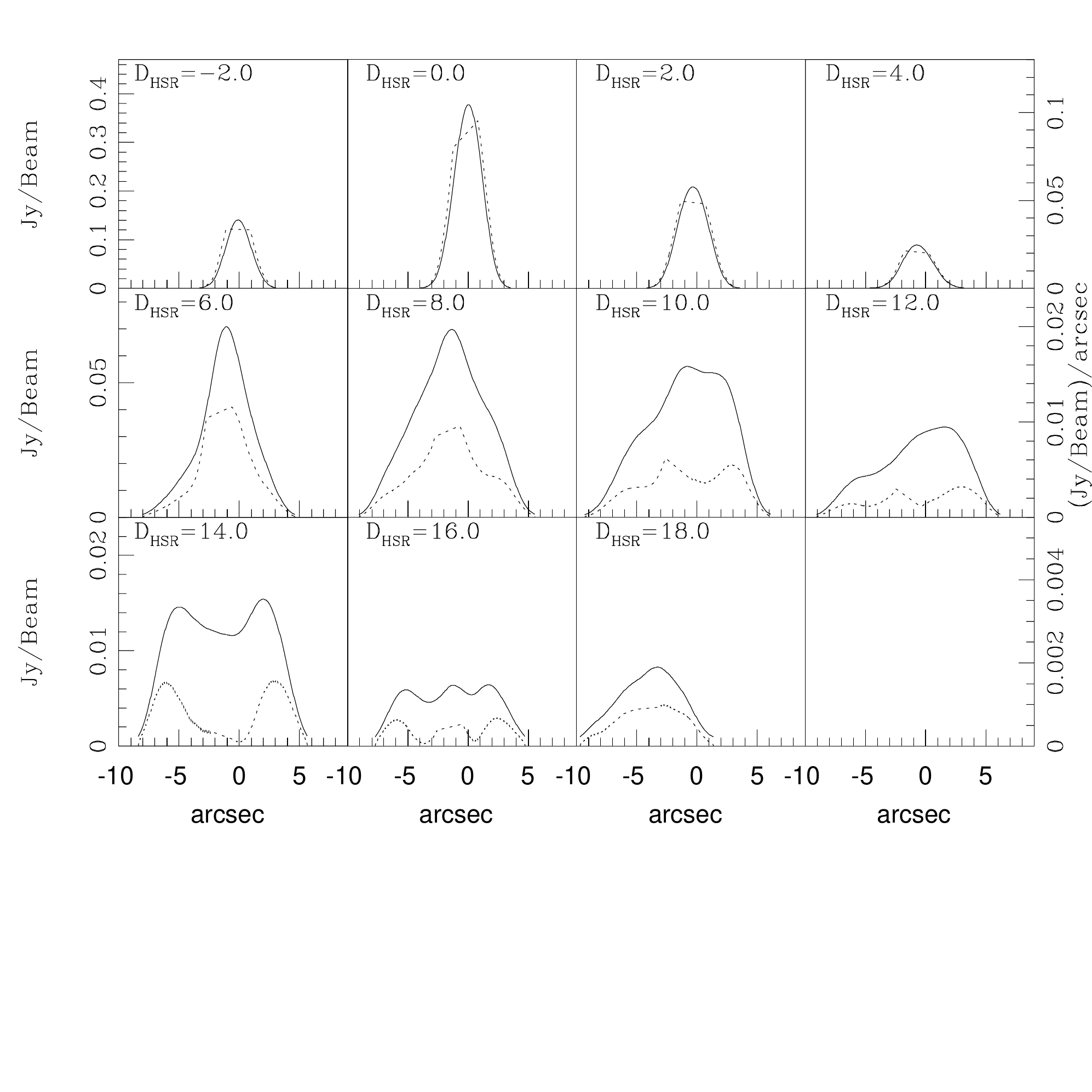}{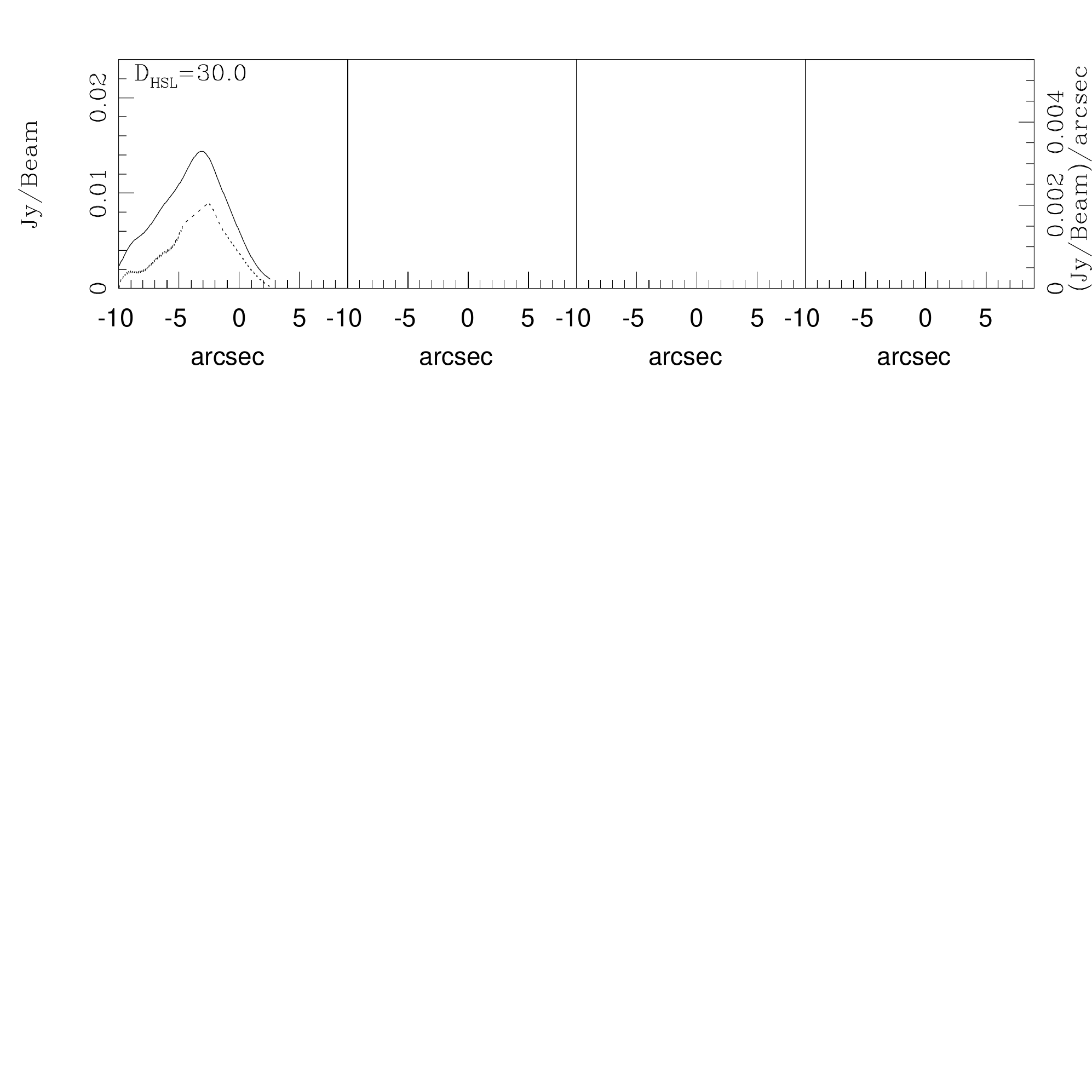}
\caption{3C142.1 emissivity (dotted line) and surface brightness
(solid line) for each cross-sectional slice of the radio 
bridge at 1.4 GHz, as in Fig. \ref{3C6.1SEM}. A constant volume
emissivity per slice provides a reasonable description of the 
data over most of the source. }
\label{3C142.1SEM}
\end{figure}

\begin{figure}
\plotone{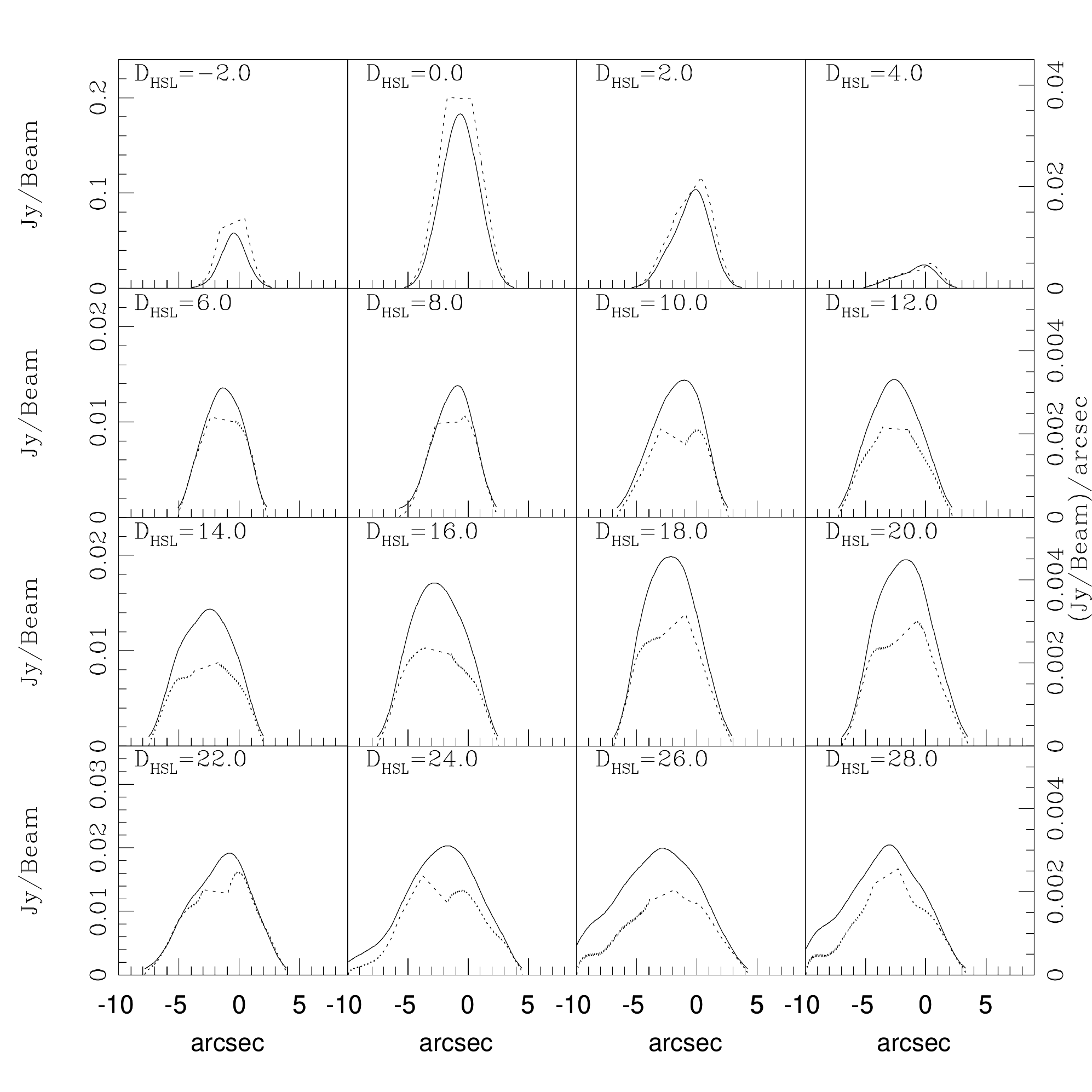}
\caption{3C142.1 emissivity (dotted line) and surface brightness
(solid line) for each cross-sectional slice of the radio 
bridge at 1.4 GHz, as in Fig. \ref{3C142.1SEM}. A constant volume
emissivity per slice provides a reasonable description of the 
data over most of the source.
}
\end{figure}

\begin{figure}
\plottwo{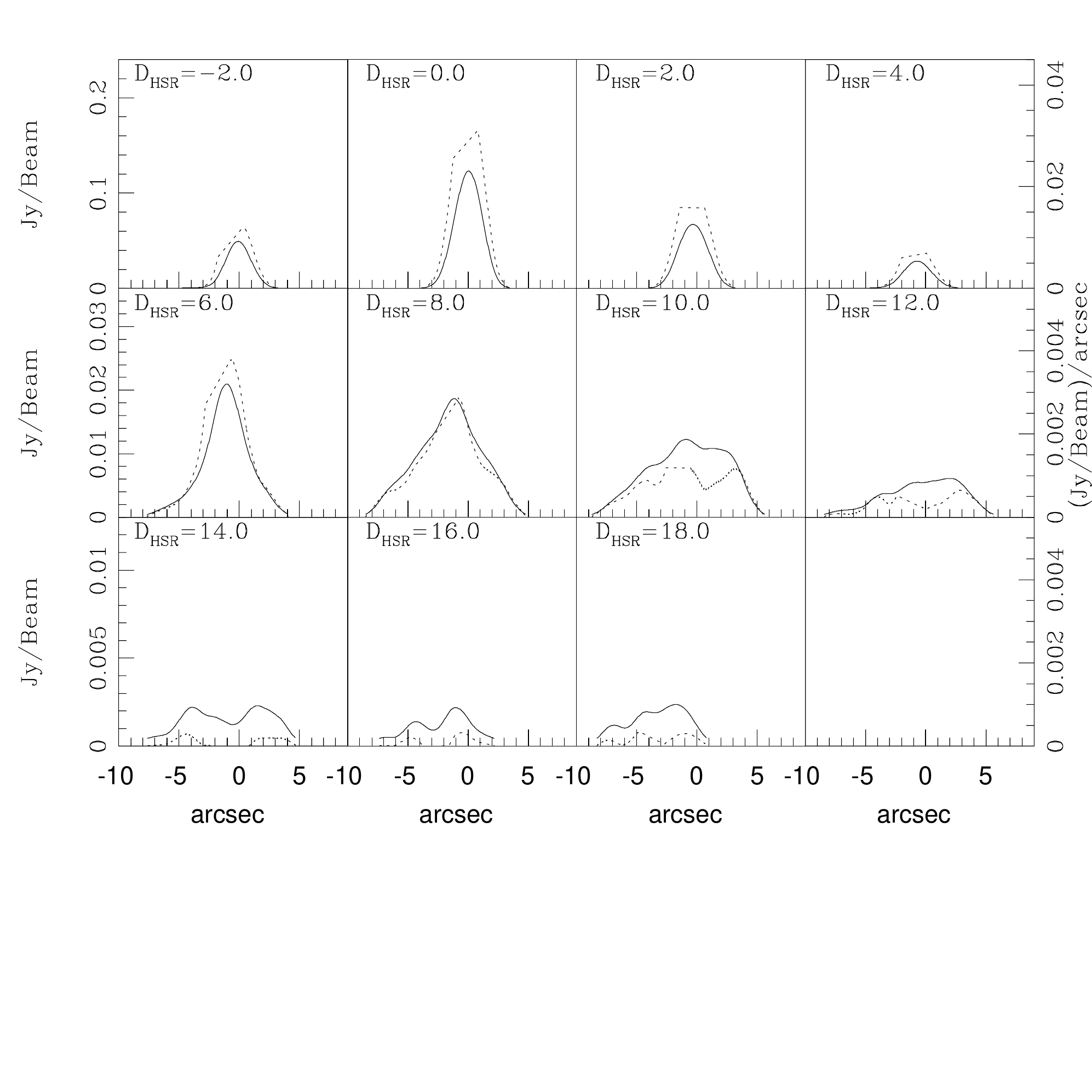}{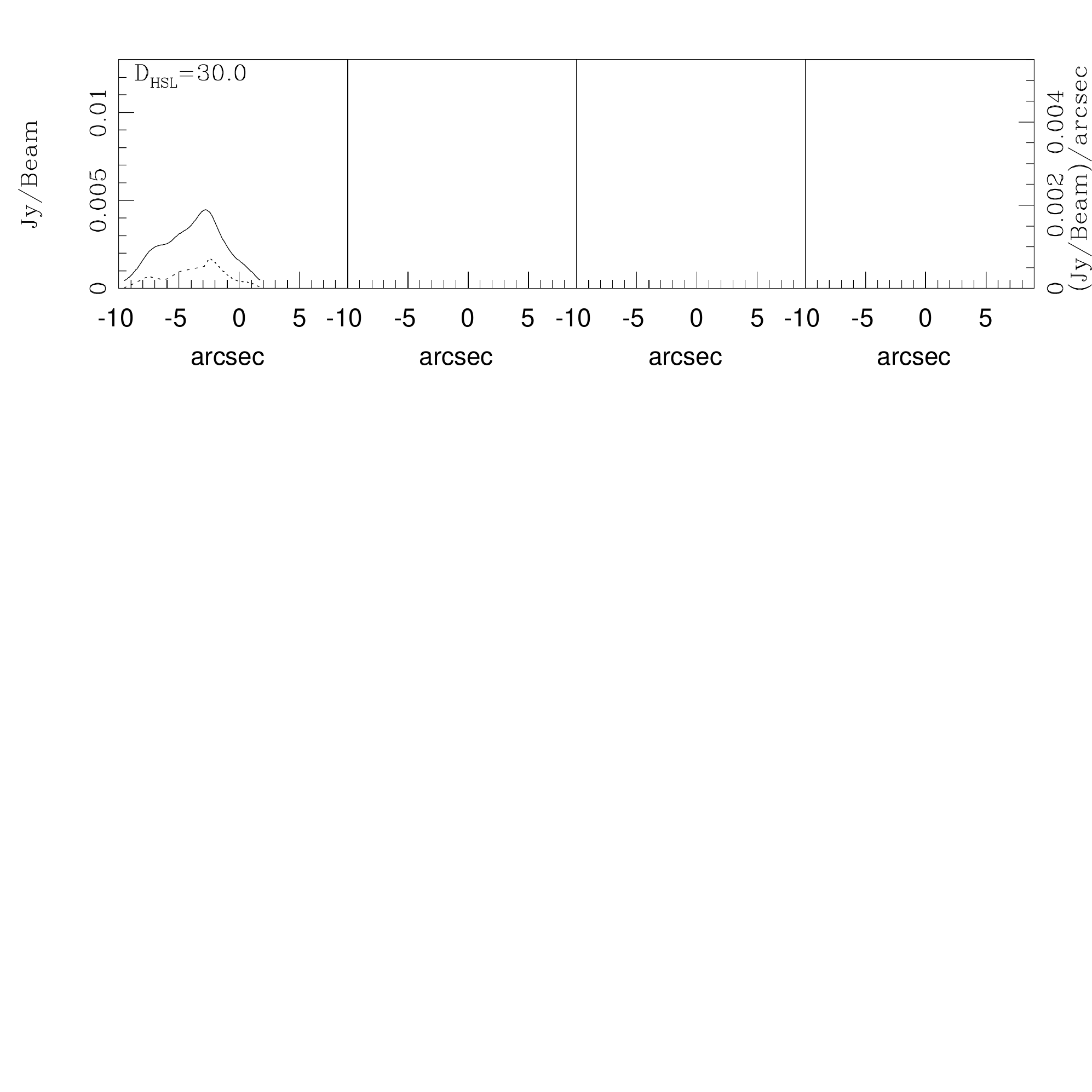}
\caption{3C142.1
emissivity (dotted line) and surface brightness
(solid line) for each cross-sectional slice of the radio 
bridge at 5 GHz, as in Fig. \ref{3C142.1SEM} but at 5 GHz.
Results obtained at 5 GHz are nearly identical to those
obtained at 1.4 GHz.}
\end{figure}

\begin{figure}
\plotone{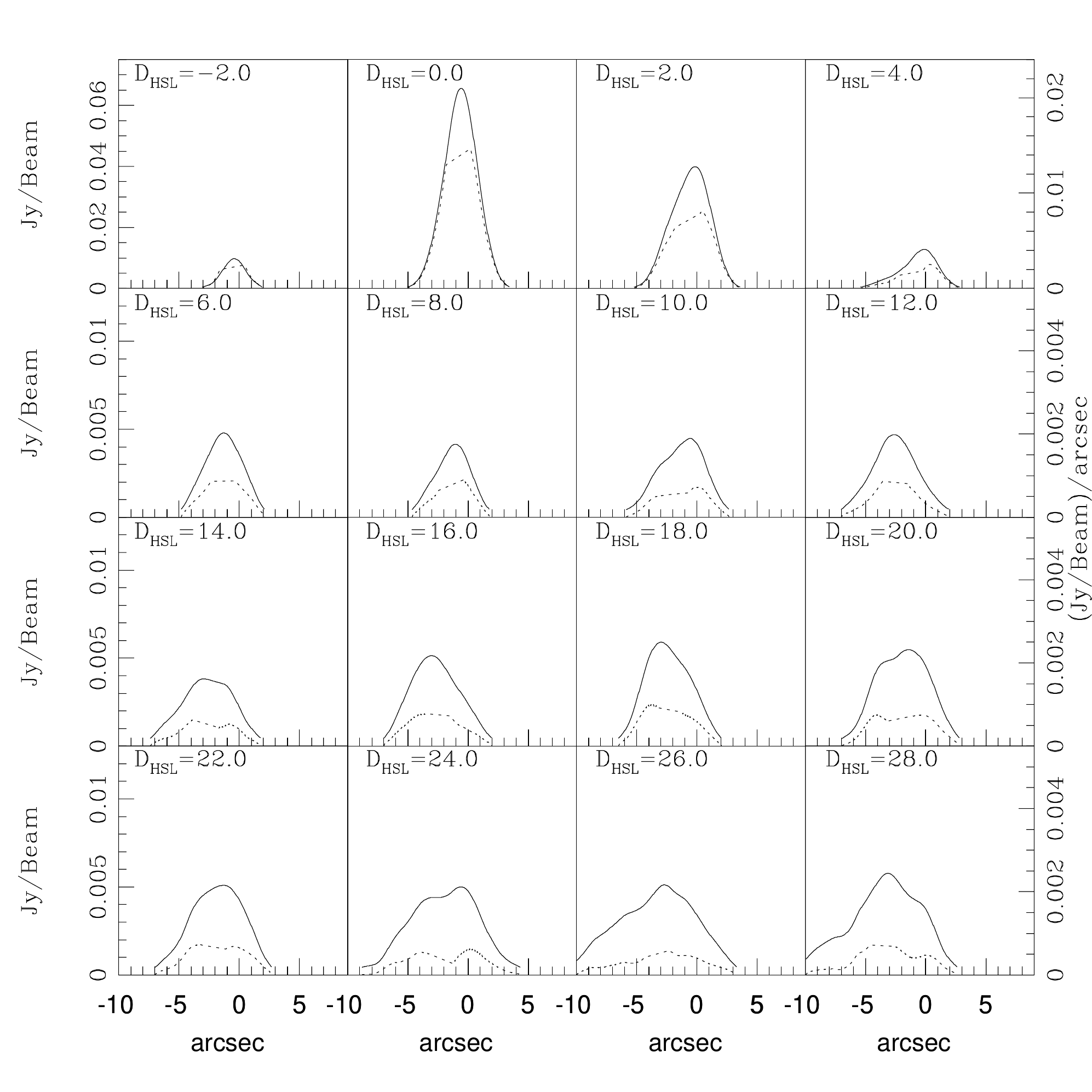}
\caption{3C142.1
emissivity (dotted line) and surface brightness
(solid line) for each cross-sectional slice of the radio 
bridge at 5 GHz, as in Fig. \ref{3C142.1SEM} but at 5 GHz.
Results obtained at 5 GHz are nearly identical to those
obtained at 1.4 GHz.
}
\end{figure}

\begin{figure}
\plottwo{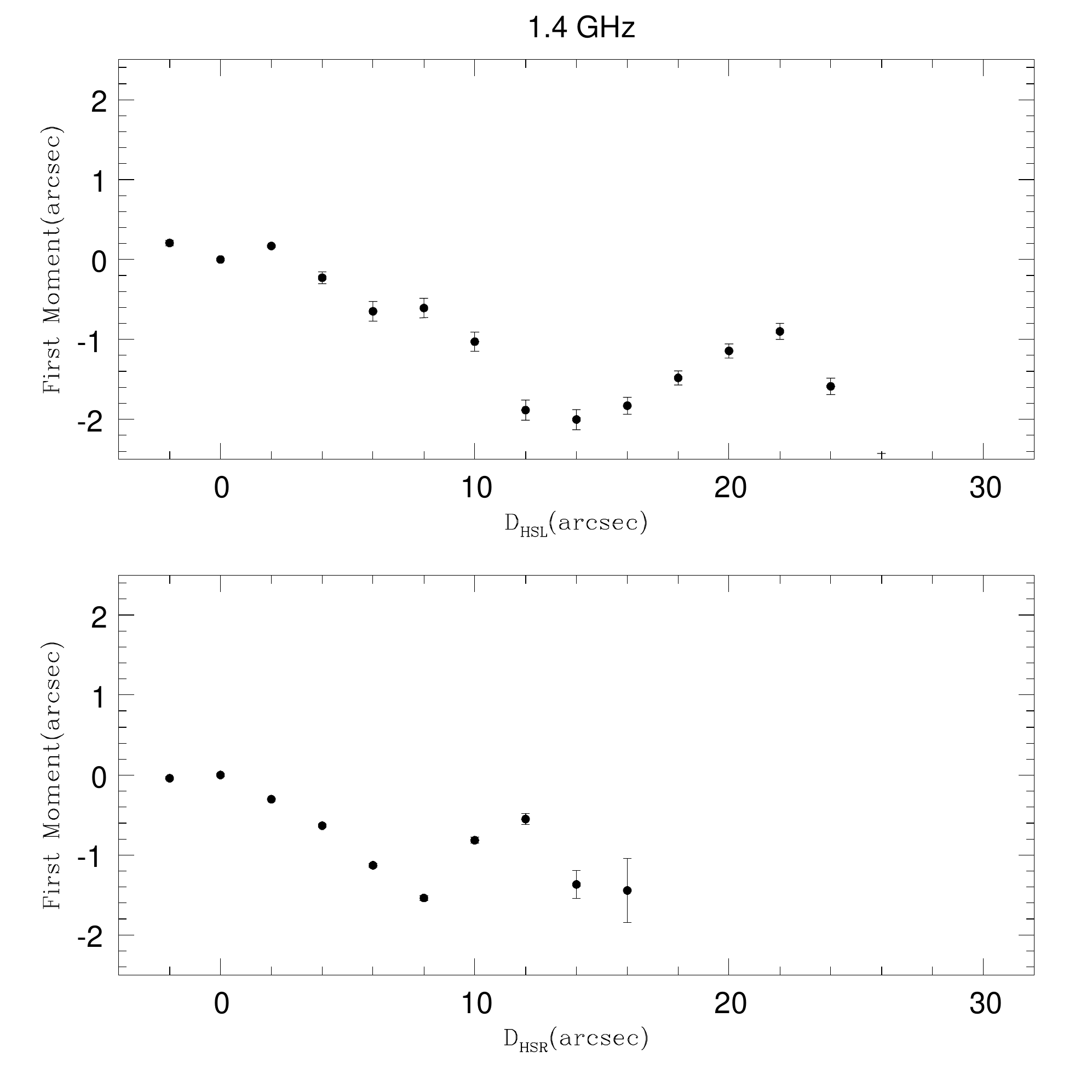}{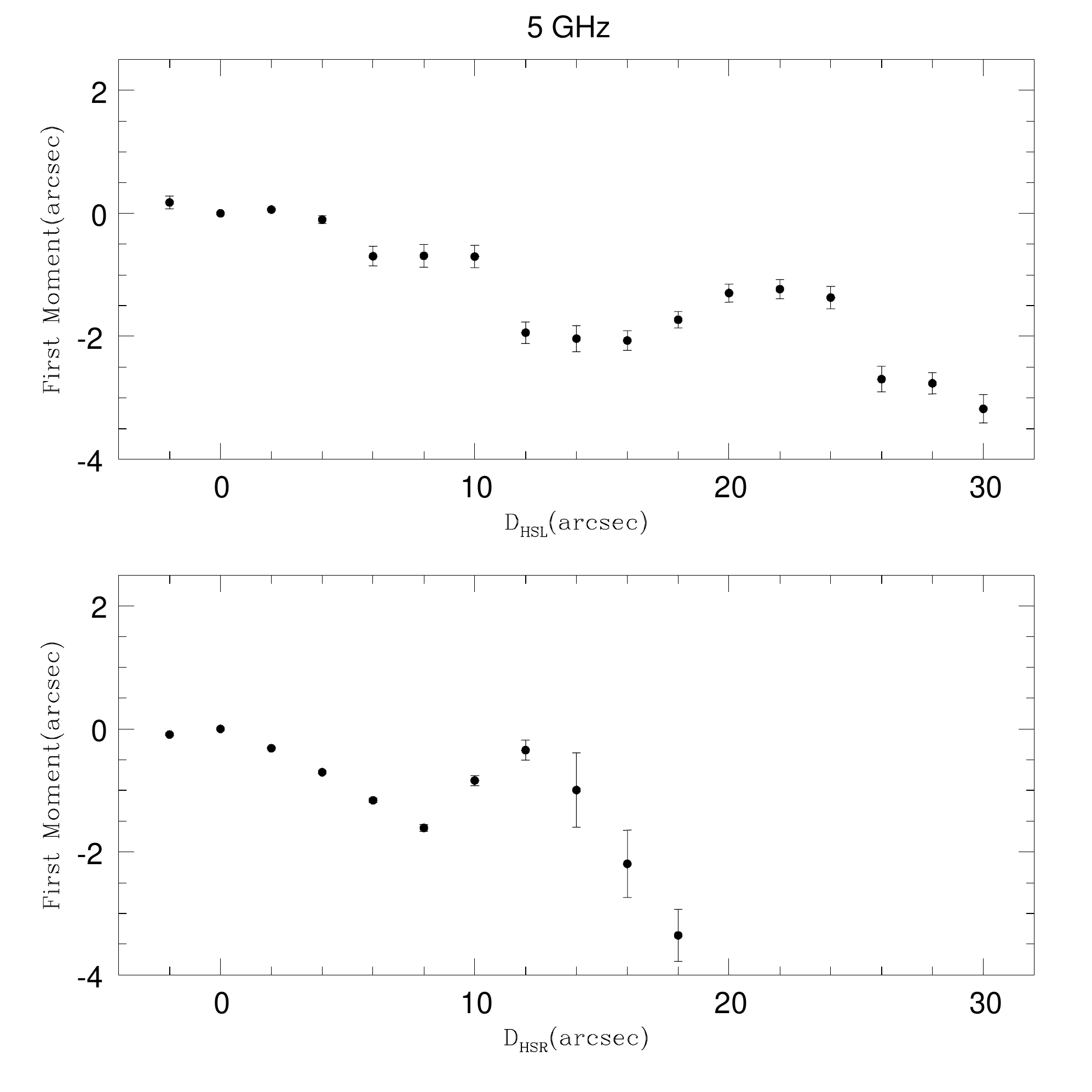}
\caption{3C142.1 first moment as a function of distance from 
the hot spot at 1.4 GHz and 5 GHz (left and right panels,
respectively) for the left and right hand sides of the source
(top and bottom, respectively).}
\end{figure}

\begin{figure}
\plottwo{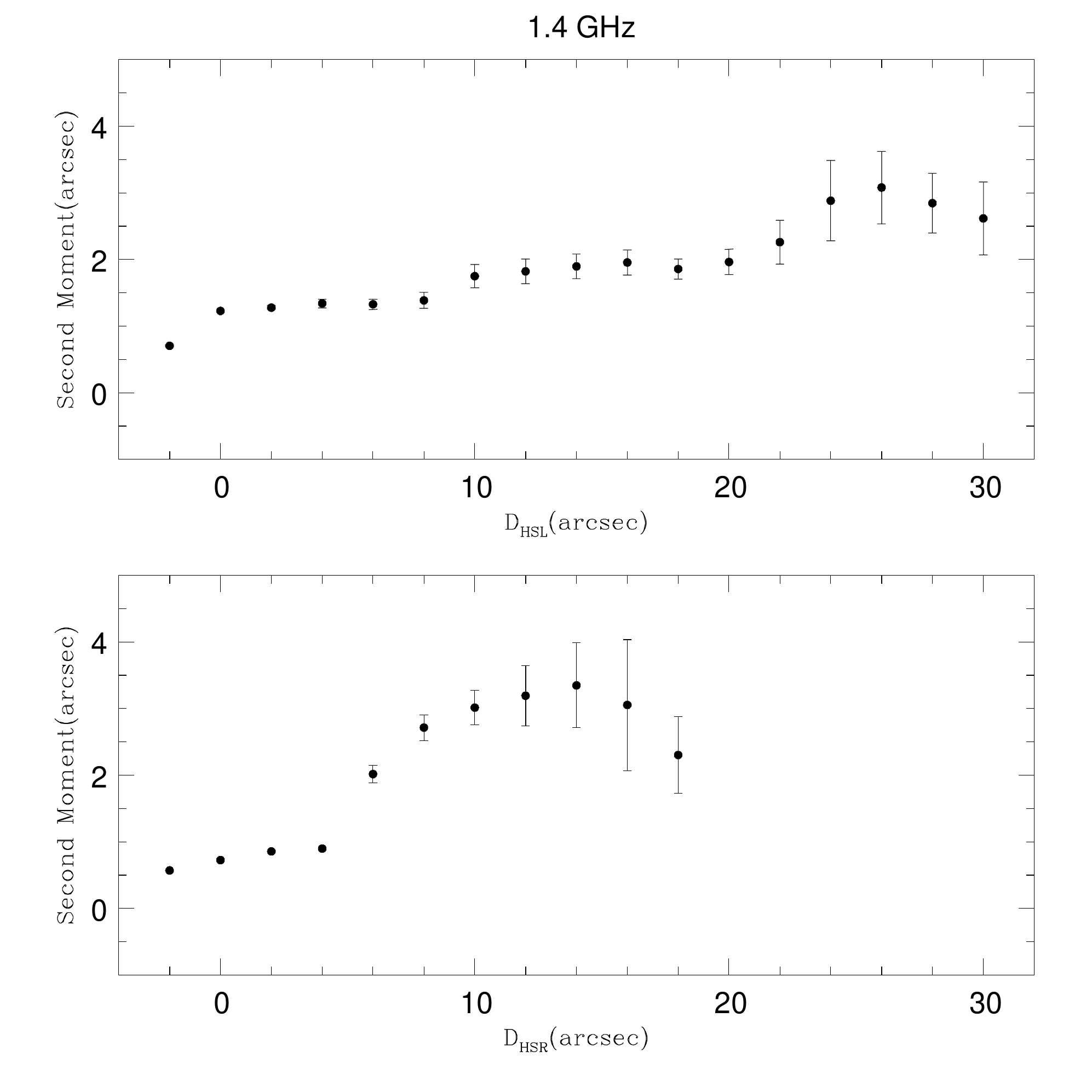}{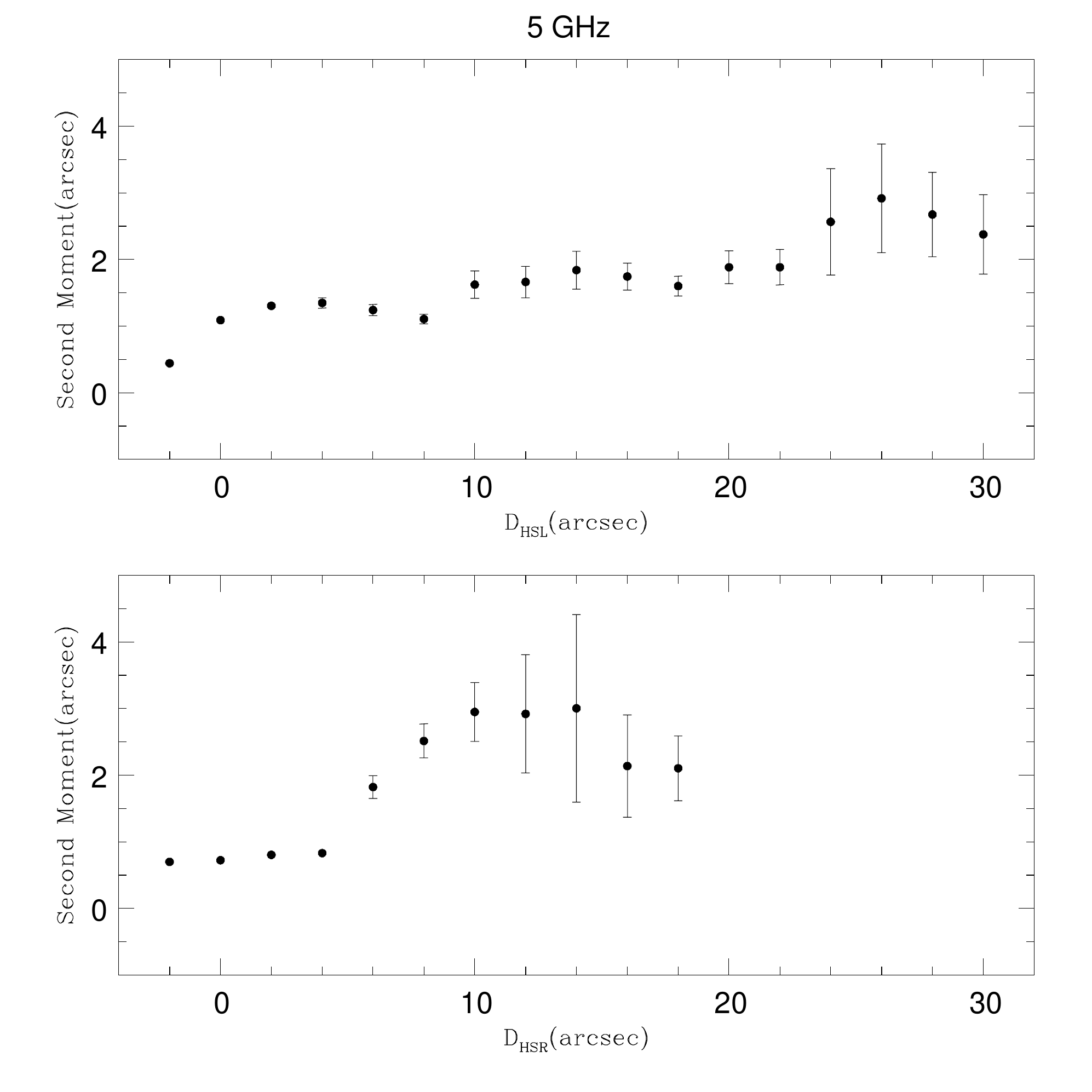}
\caption{3C142.1 second moment as a function of distance from the
hot spot at 1.4 GHz and 5 GHz (left and right panels, 
respectively) for the left and right sides of the source
(top and bottom panels, respectively).}
\end{figure}

\begin{figure}
\plottwo{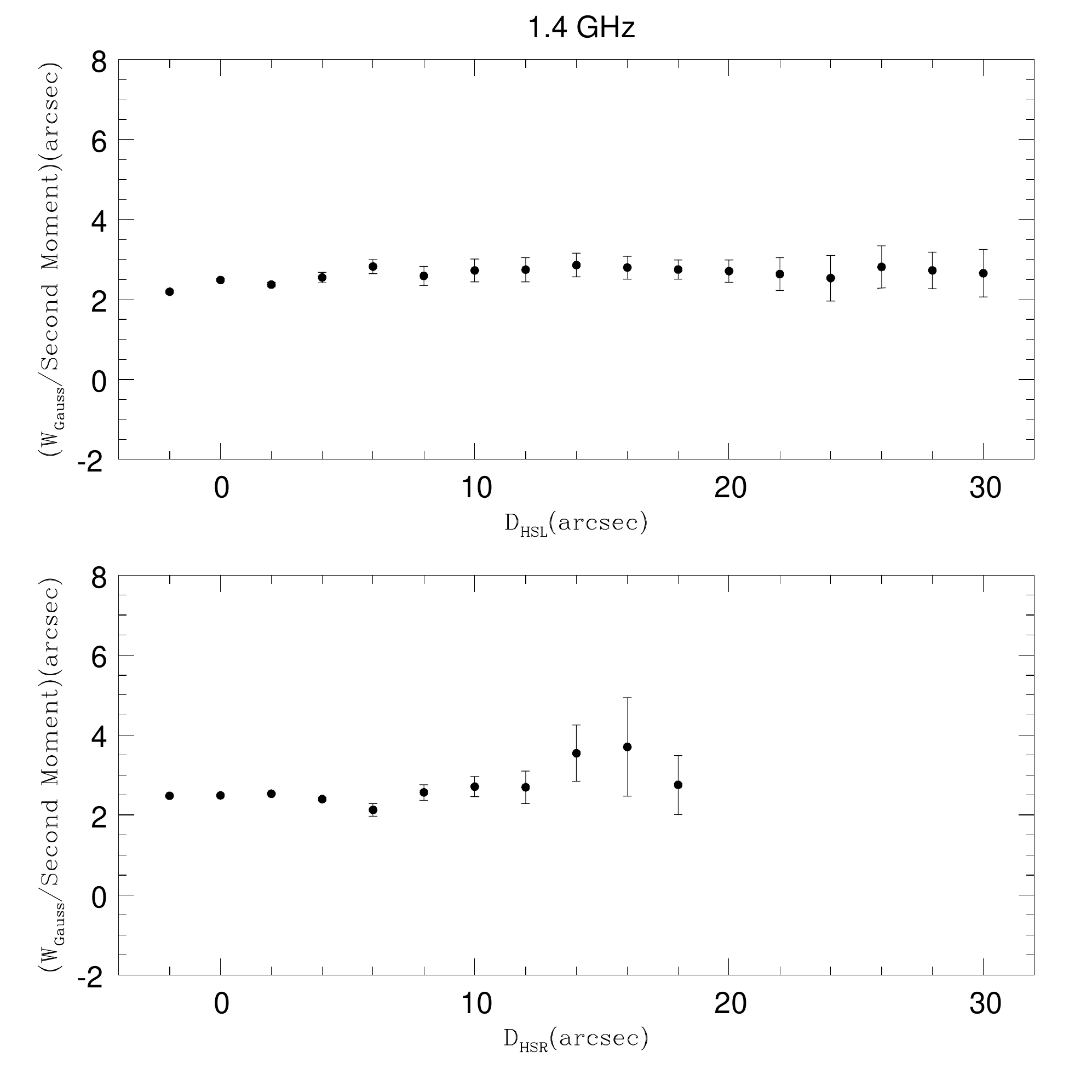}{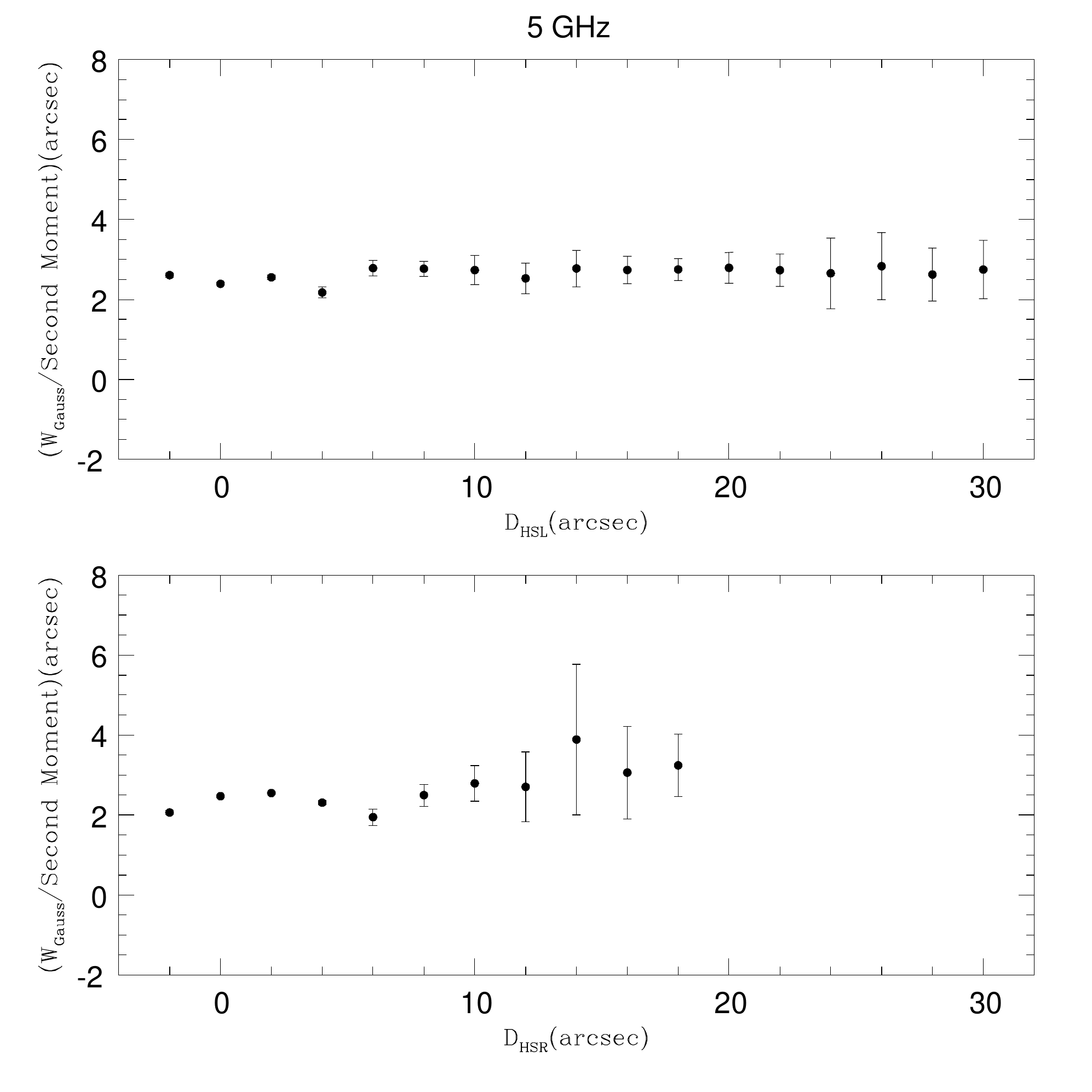}
\caption{3C142.1 ratio of the Gaussian FWHM to the second moment as a
function of distance from the hot spot at 1.4 and 5 GHz 
(left and right panels, respectively) for the left and right
hand sides of the source (top and bottom panels, respectively). The value of this ratio is fairly constant for each side of the source, and has a value of about 2. Similar results are obtained at 1.4 and 5 GHz.} 
\end{figure}

\begin{figure}
\plottwo{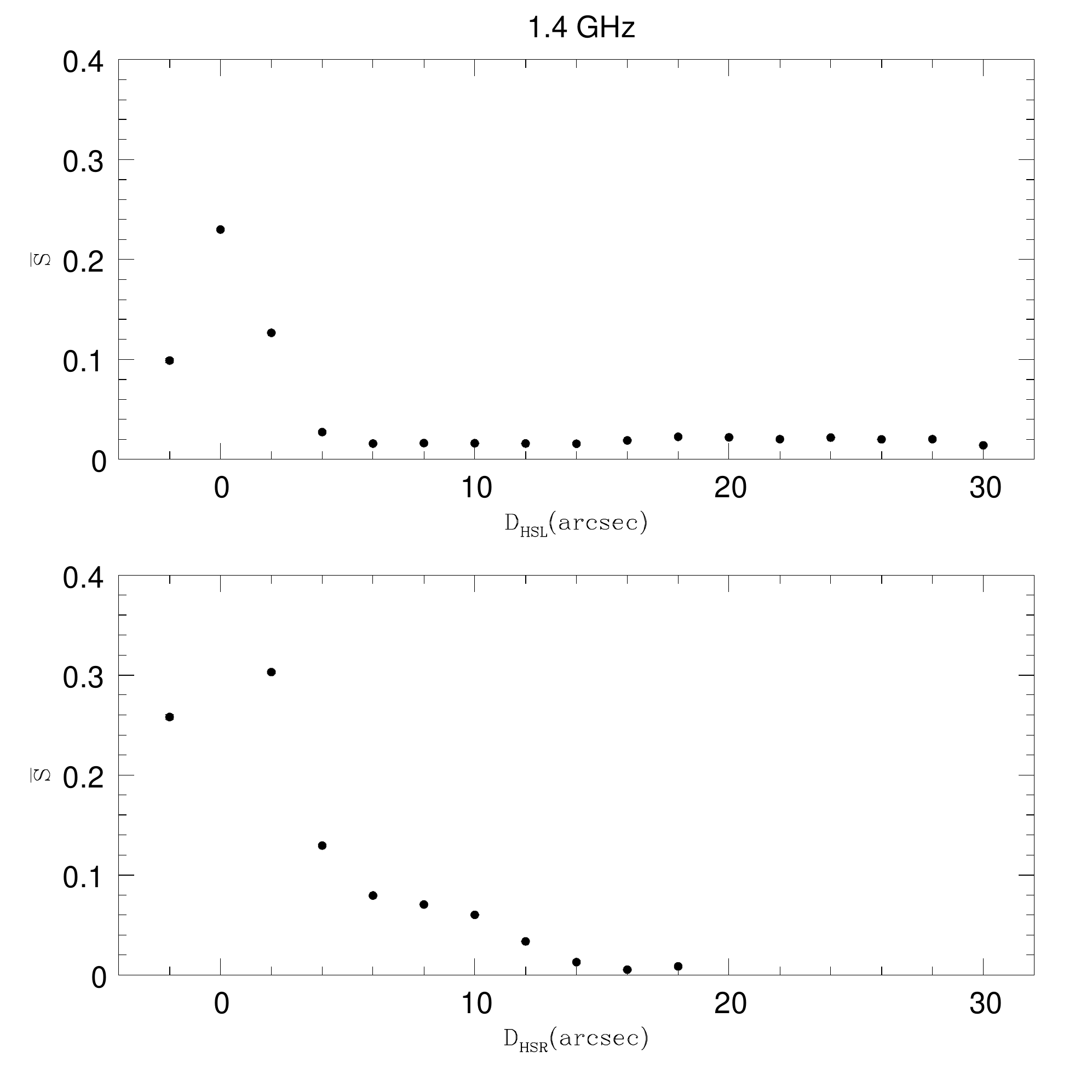}{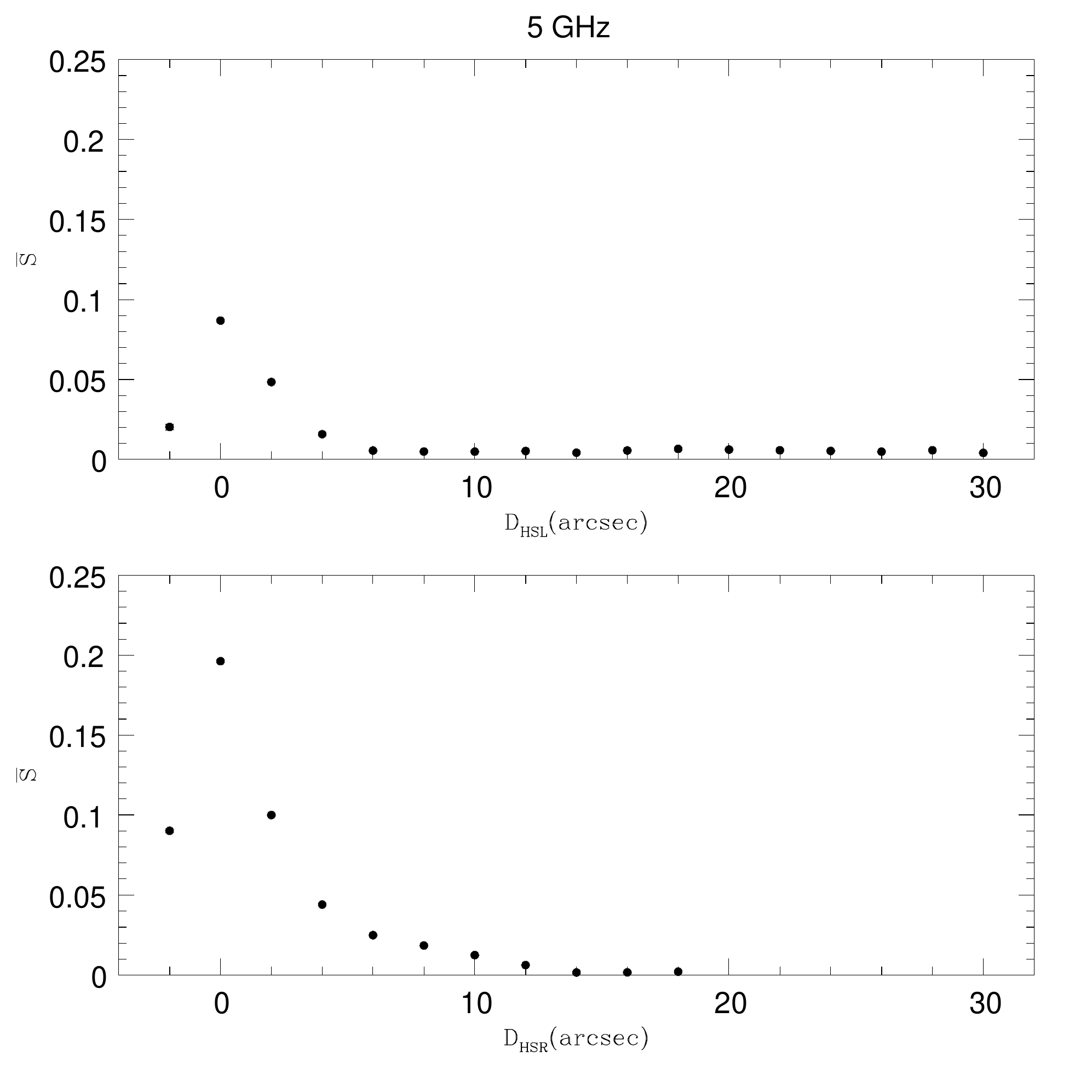}
\caption{3C142.1 average surface brightness in units of Jy/beam 
as a function of distance from the 
hot spot at 1.4 and 5 GHz 
(left and right panels, respectively) for the left and right
hand sides of the source (top and bottom panels, respectively).}
\end{figure}

\begin{figure}
\plottwo{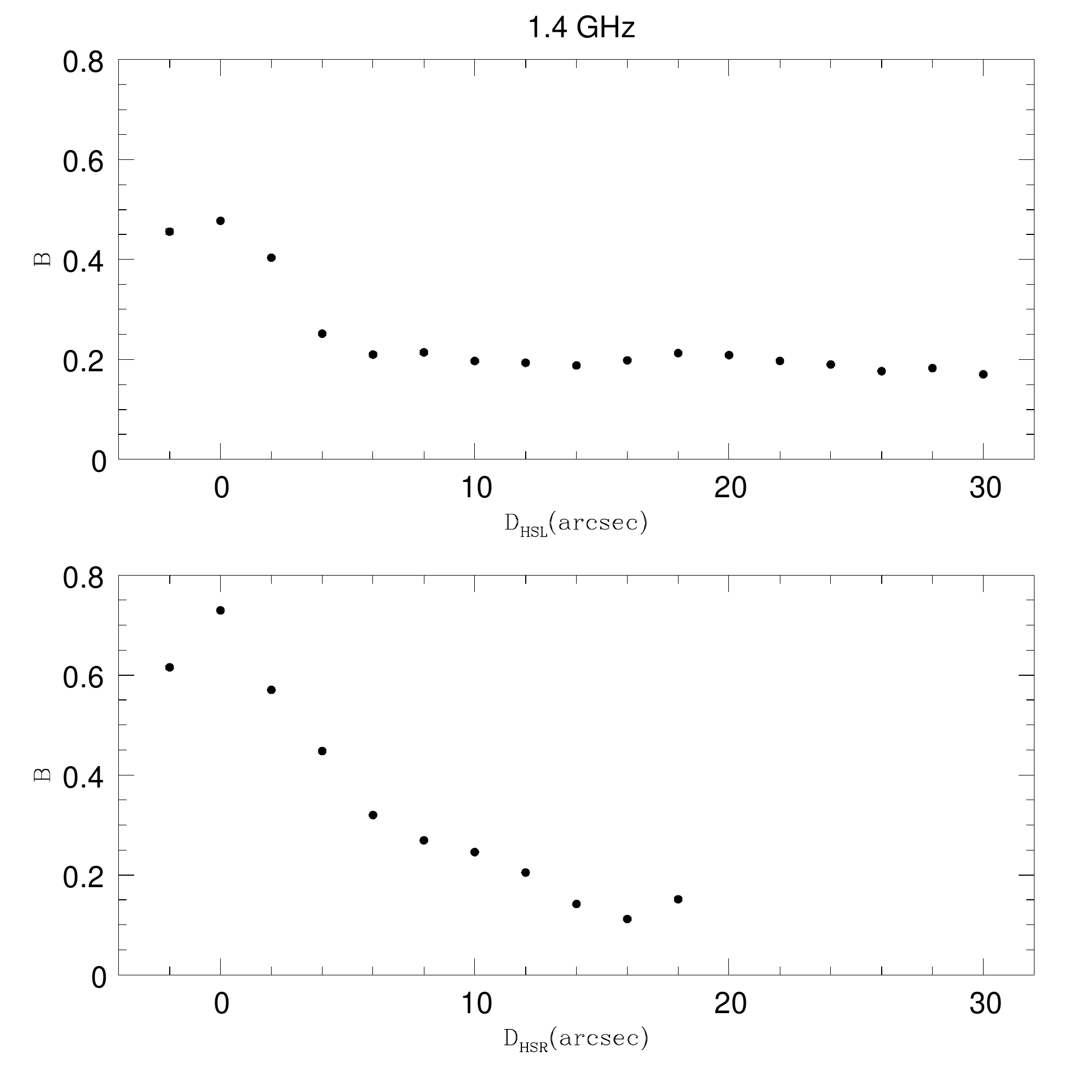}{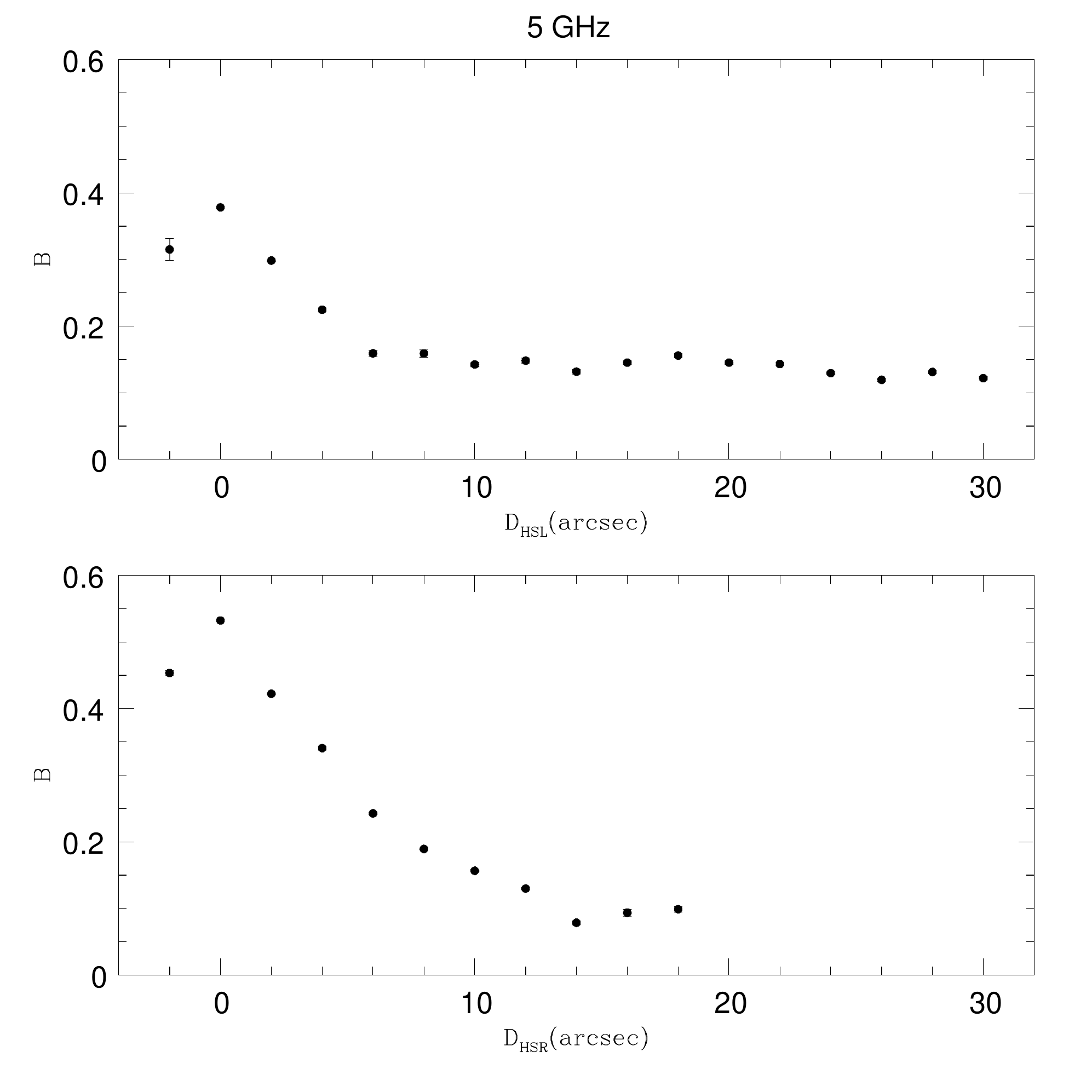}
\caption{3C142.1 minimum energy magnetic field strength 
as a function of distance from the 
hot spot at 1.4 and 5 GHz 
(left and right panels, respectively) for the left and right
hand sides of the source (top and bottom panels, respectively).
The normalization is given in Table 1. 
}
\end{figure}

\begin{figure}
\plottwo{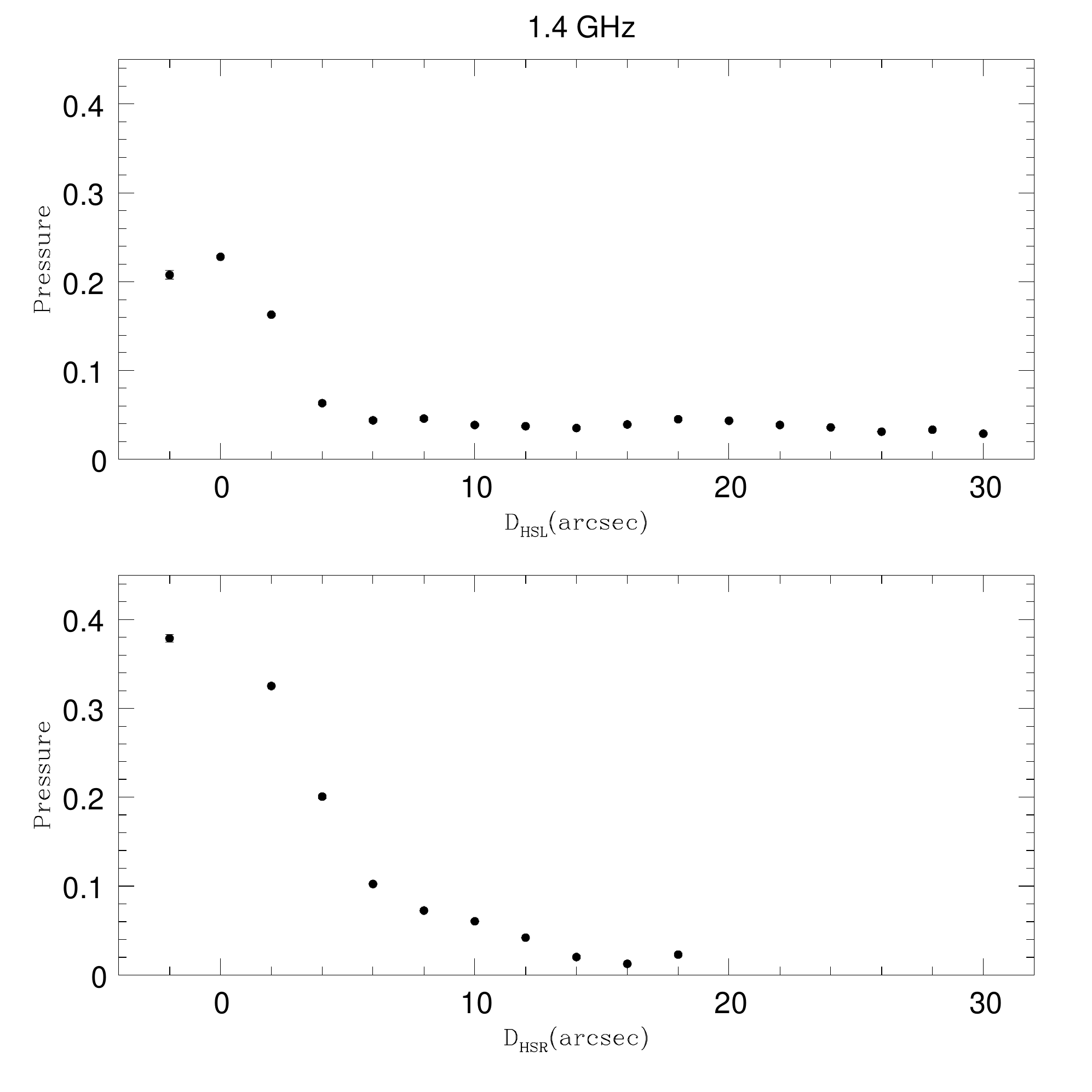}{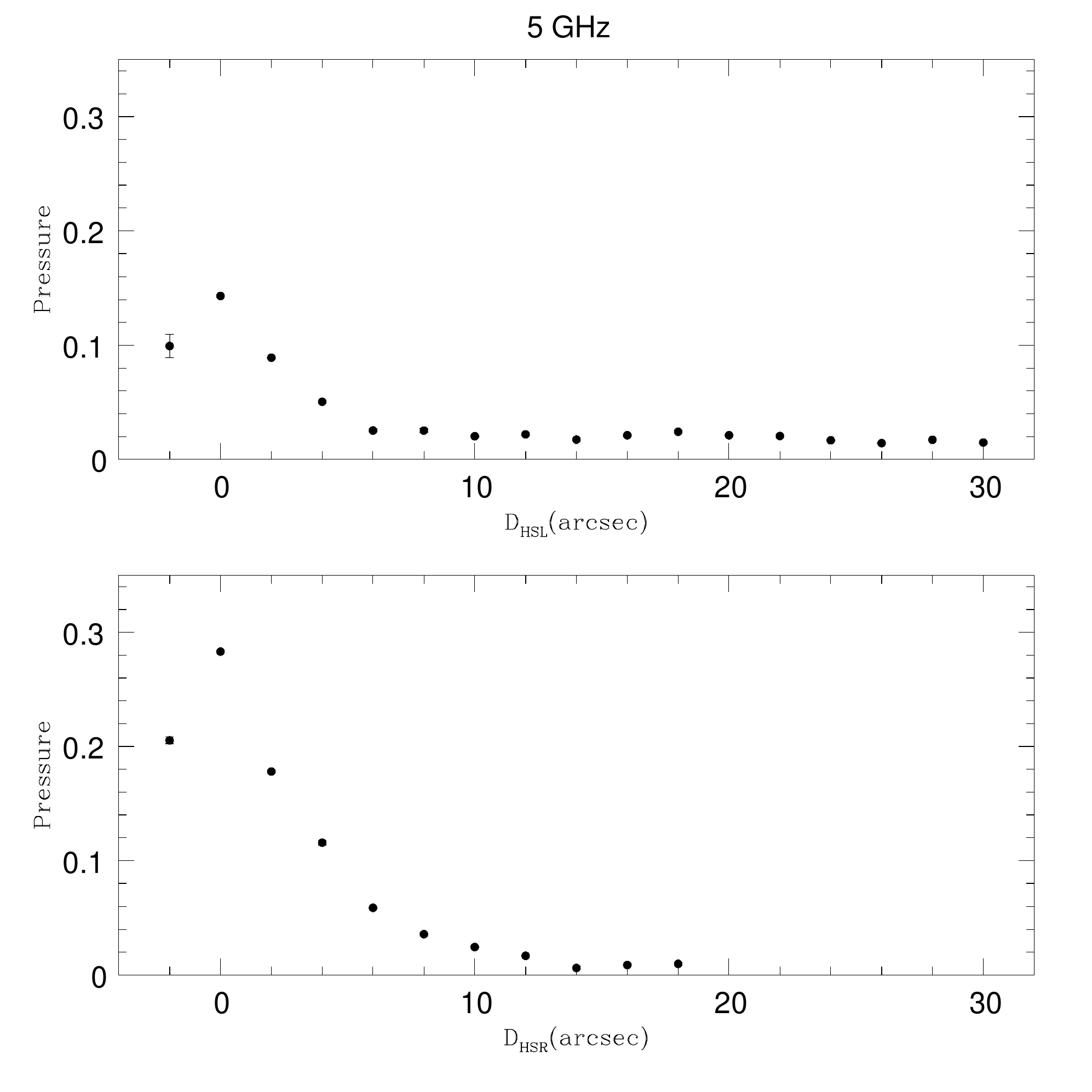}
\caption{3C142.1 
minimum energy pressure as a function of 
distance from the hot spot at 1.4 and 5 GHz 
(left and right panels, respectively) for the left and right
hand sides of the source (top and bottom panels, respectively), 
as in Fig. \ref{3C6.1P}. The normalization is given in Table 1.
}
\end{figure}

\clearpage
\begin{figure}
\plotone{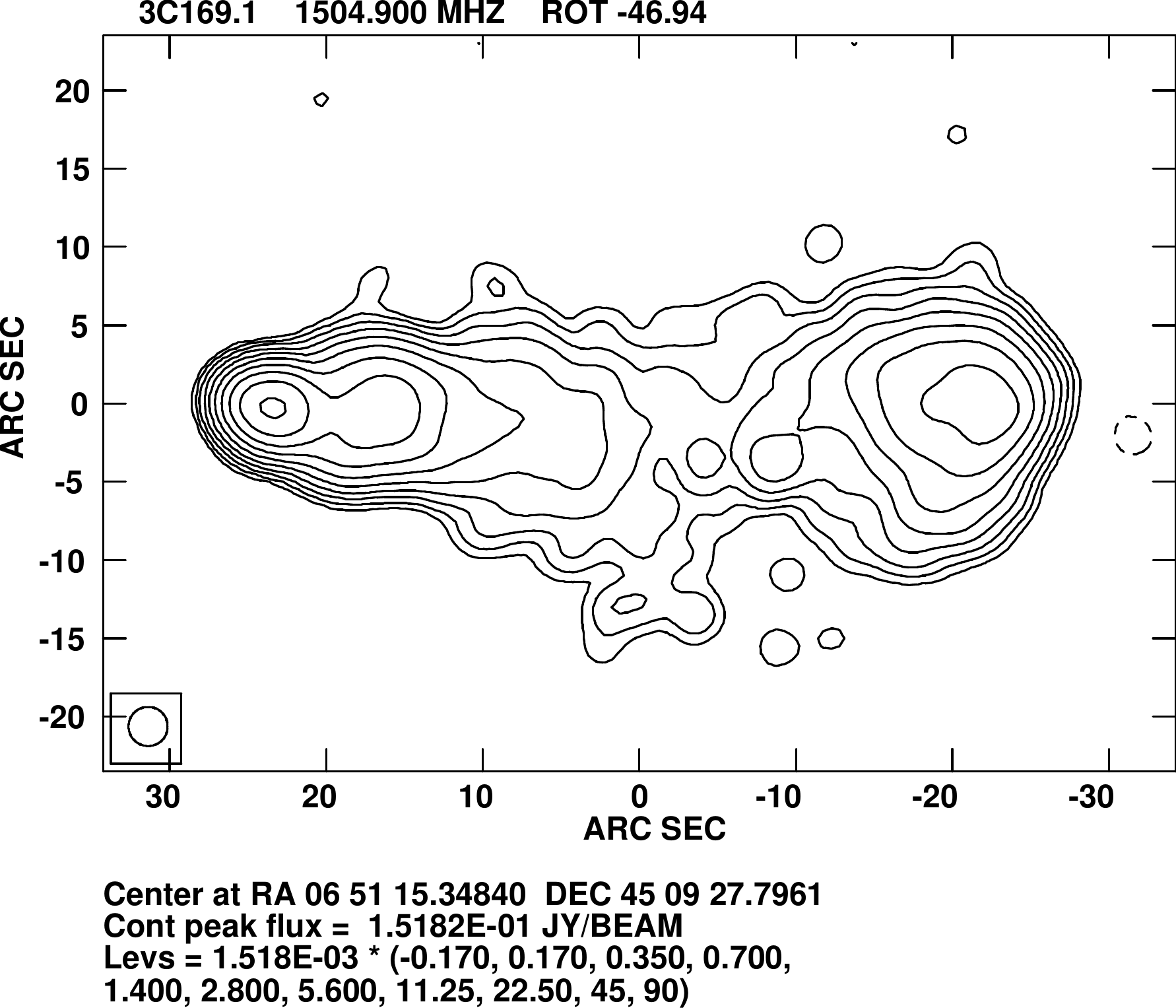}
\caption{3C169.1 at 1.4 GHz and 2.5'' resolution rotated by 46.94 deg 
clockwise, as in Fig. \ref{3C6.1rotfig}.}
\end{figure}

\begin{figure}
\plottwo{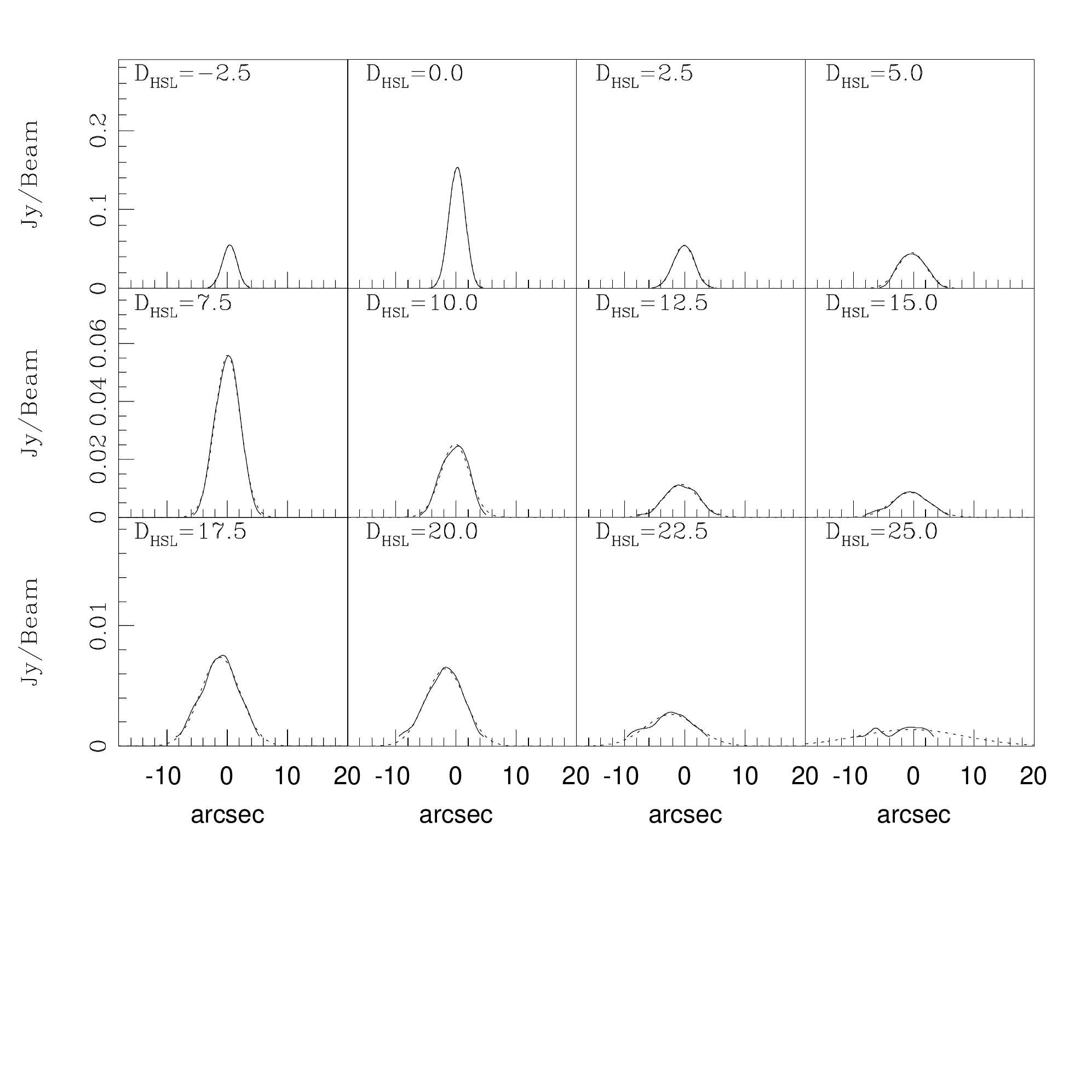}{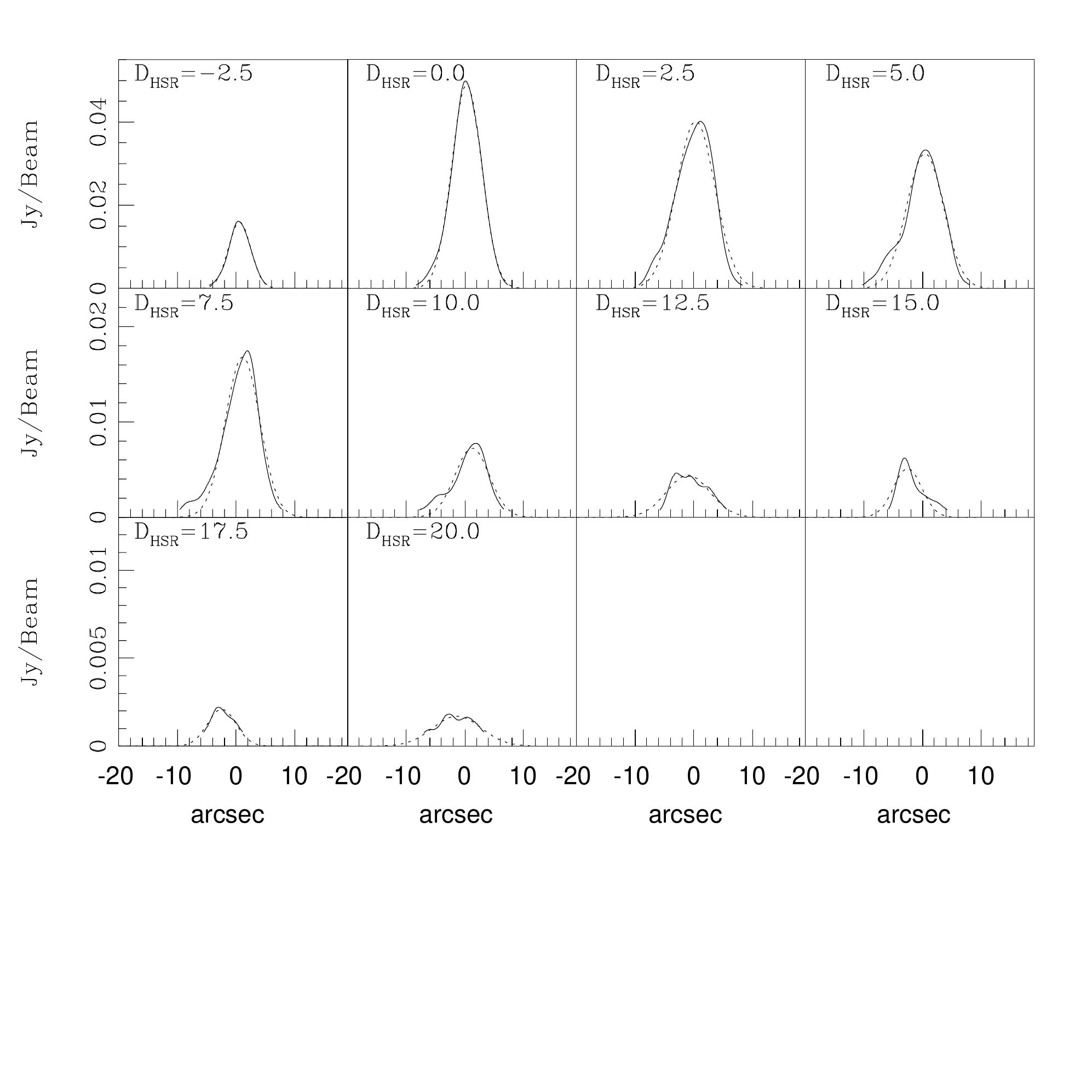}
\caption{3C169.1 
surface brightness (solid line) and best fit 
Gaussian (dotted line) for each cross-sectional slice of the 
radio bridge at 1.4 GHz, as in Fig. \ref{3C6.1SG}.
A Gaussian provides an excellent description of the surface brightness
profile of each cross-sectional slice. }
\label{3C169.1SG}
\end{figure}

\begin{figure}
\plottwo{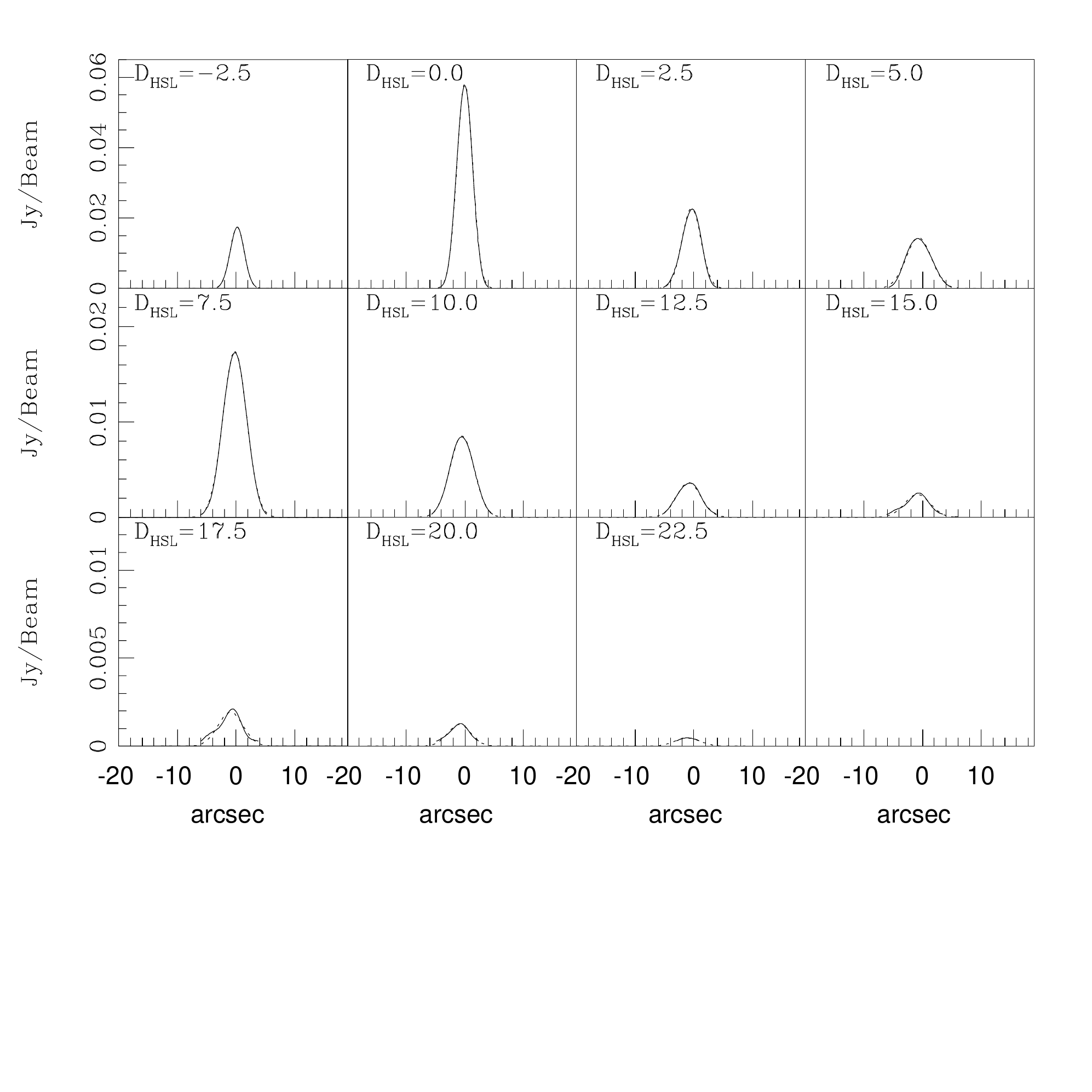}{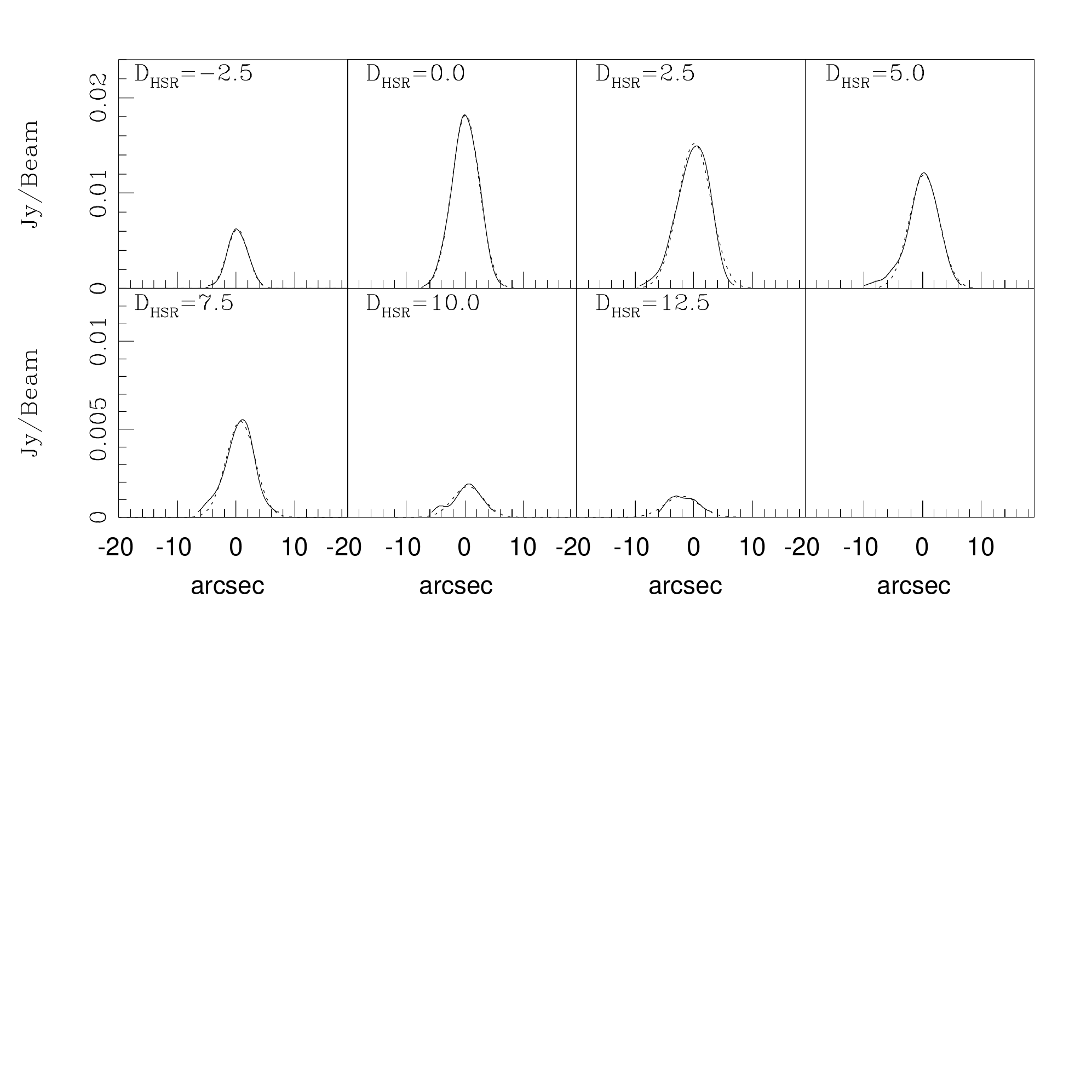}
\caption{3C169.1 
surface brightness (solid line) and best fit 
Gaussian (dotted line) for each cross-sectional slice of the 
radio bridge at 5 GHz, as in 
Fig. \ref{3C169.1SG} but at 5 GHz. 
Results obtained at 5 GHz are nearly identical to those
obtained at 1.4 GHz.}
\end{figure}

\begin{figure}
\plottwo{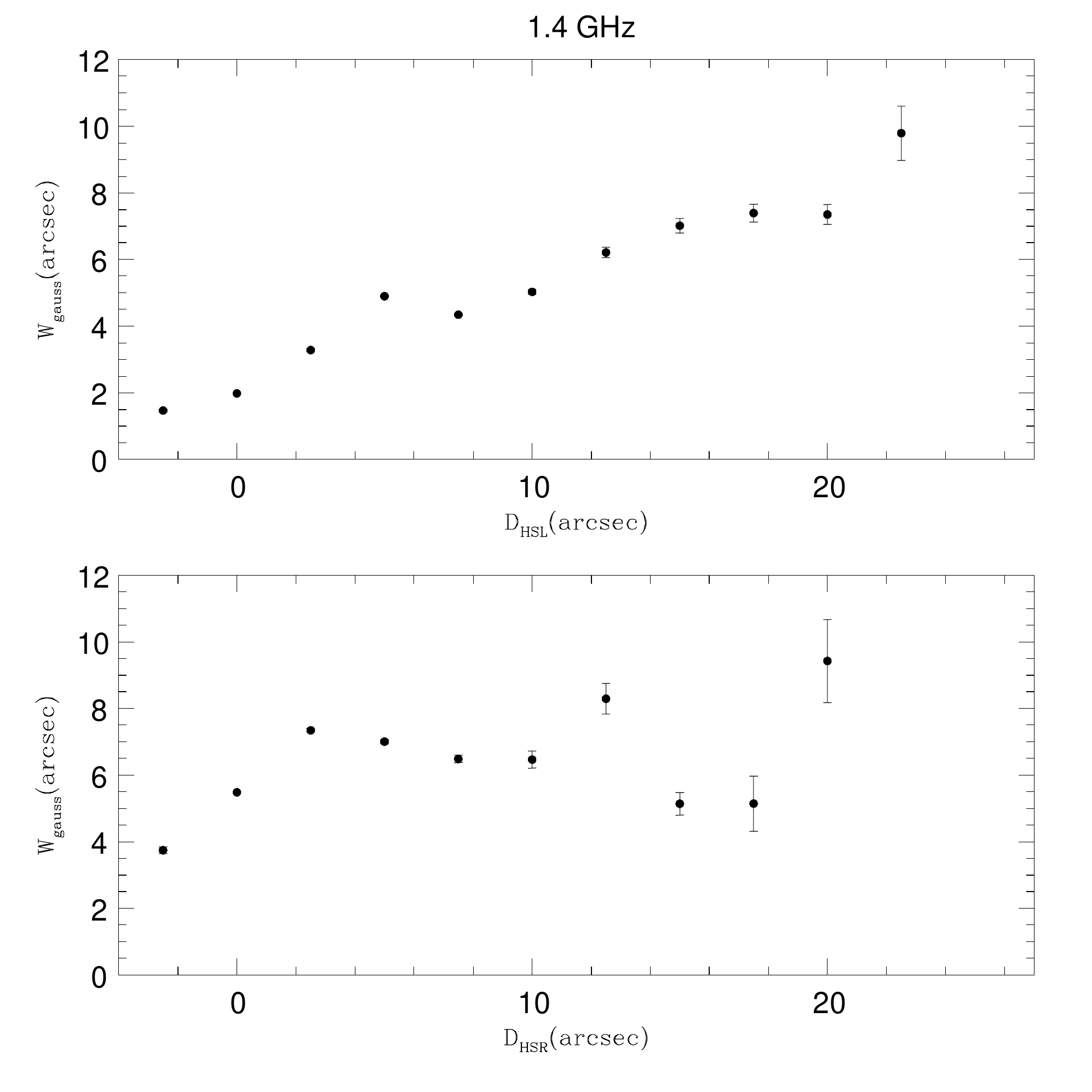}{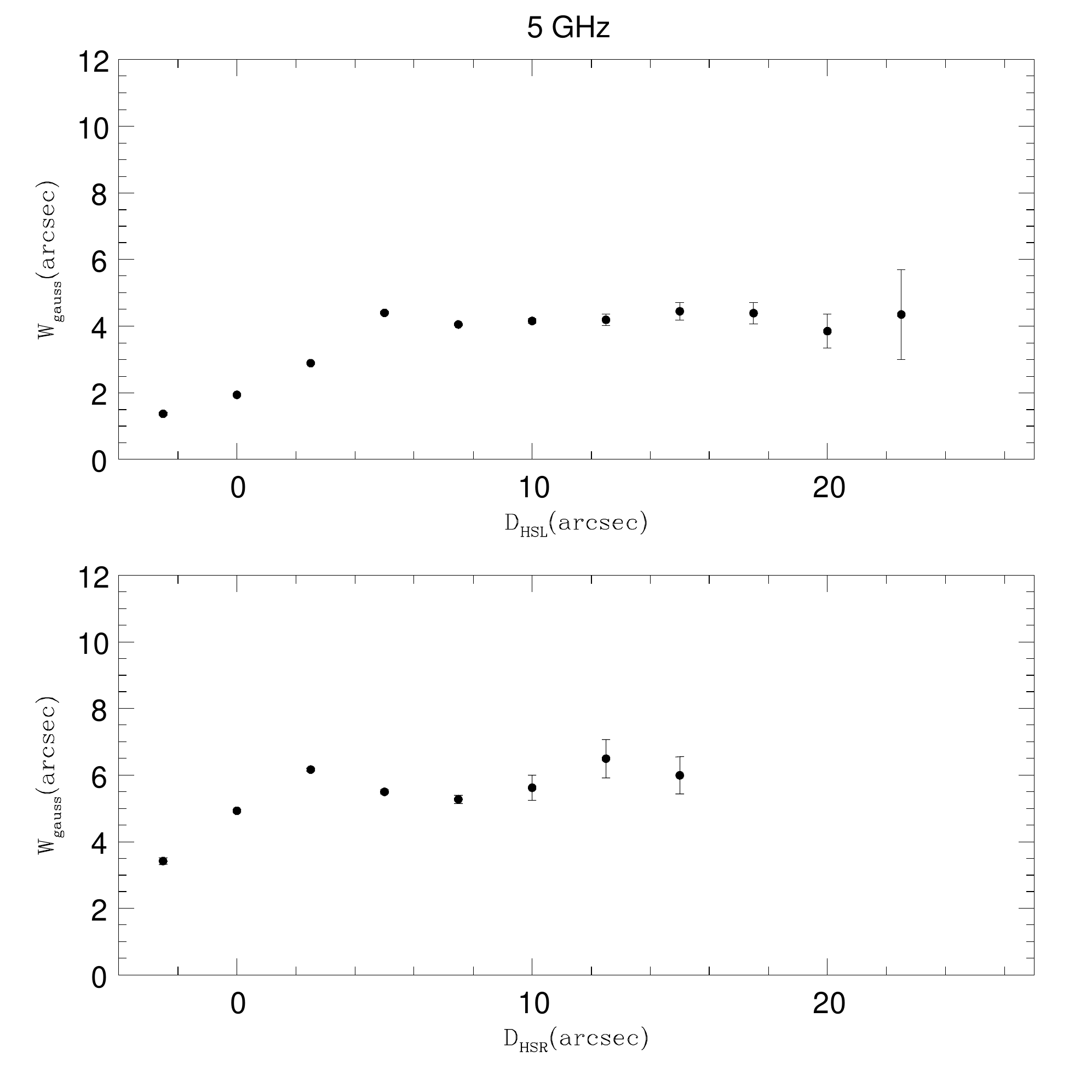}
\caption{3C169.1  Gaussian FWHM as a function of 
distance from the hot spot  
at 1.4 and 5 GHz (left and
right panels, respectively) for the left and right hand sides of the 
source (top and bottom panels, respectively), as in Fig. \ref{3C6.1WG}.
The right and left hand sides of the source are fairly 
symmetric.  Similar results are obtained at 
1.4 and 5 GHz. }
\end{figure}

\begin{figure}
\plottwo{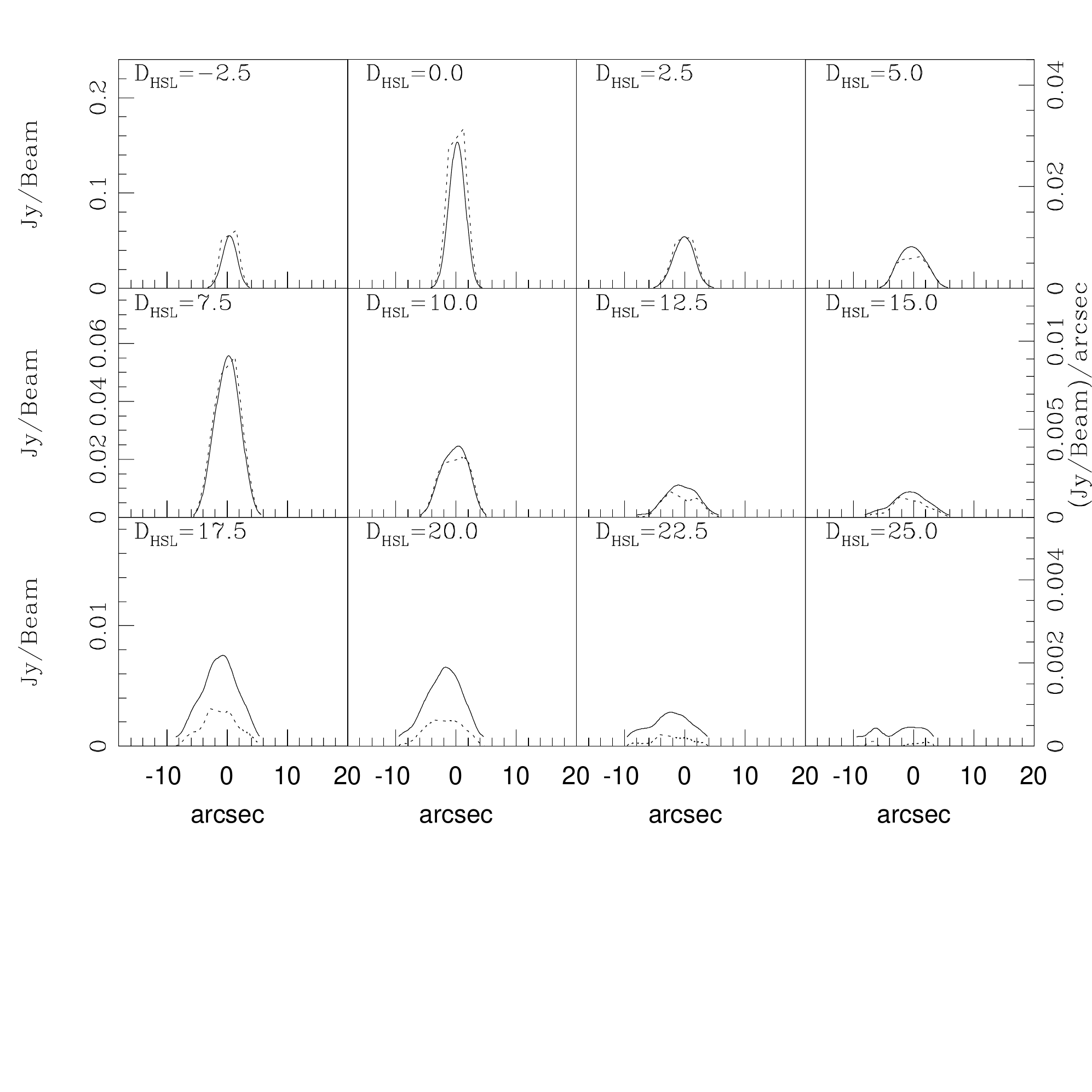}{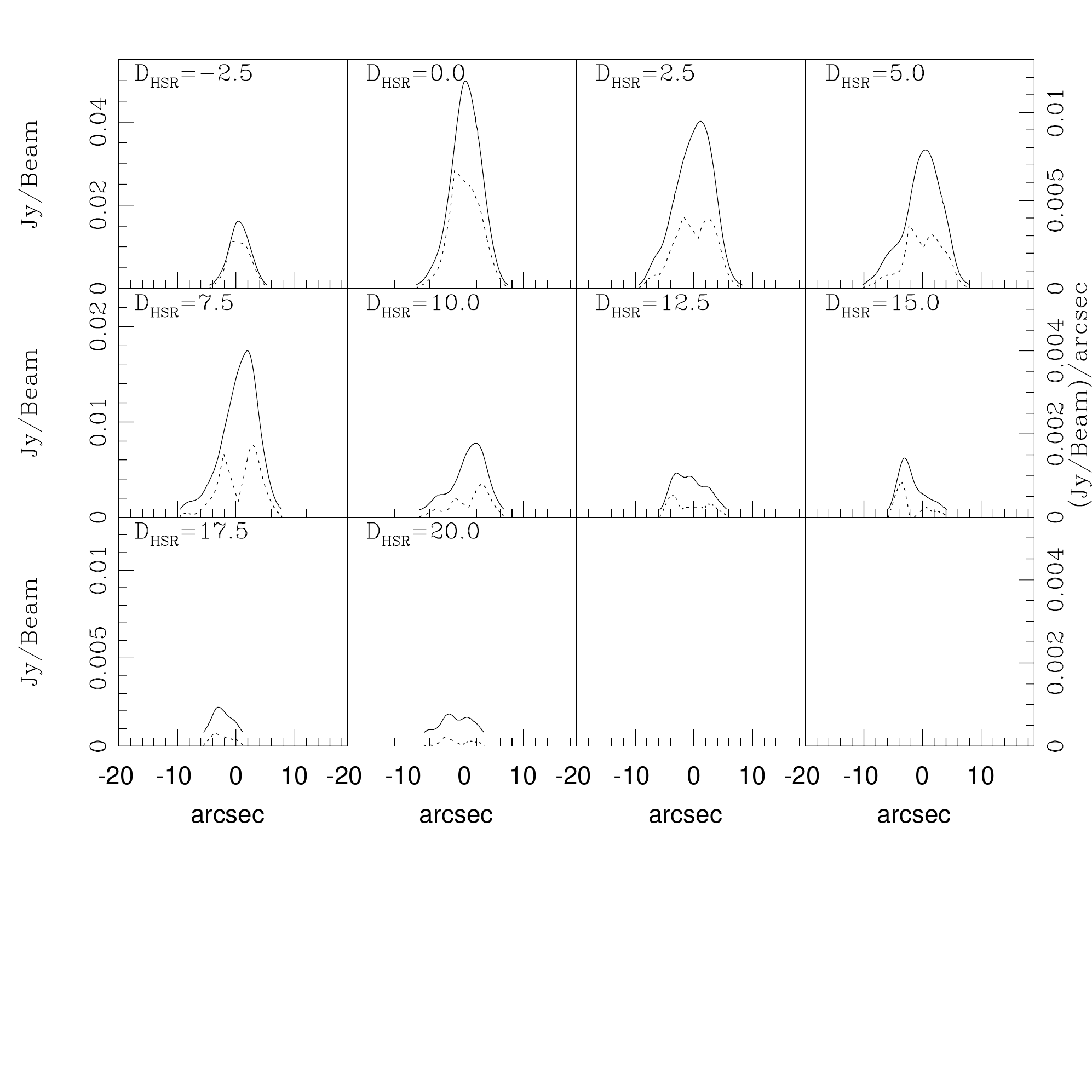}
\caption{3C169.1 emissivity (dotted line) and surface brightness
(solid line) for each cross-sectional slice of the radio 
bridge at 1.4 GHz, as in Fig. \ref{3C6.1SEM}. A constant volume
emissivity per slice provides a reasonable description of the 
data over most of the source.}
\label{3C169.1SEM}
\end{figure}

\begin{figure}
\plottwo{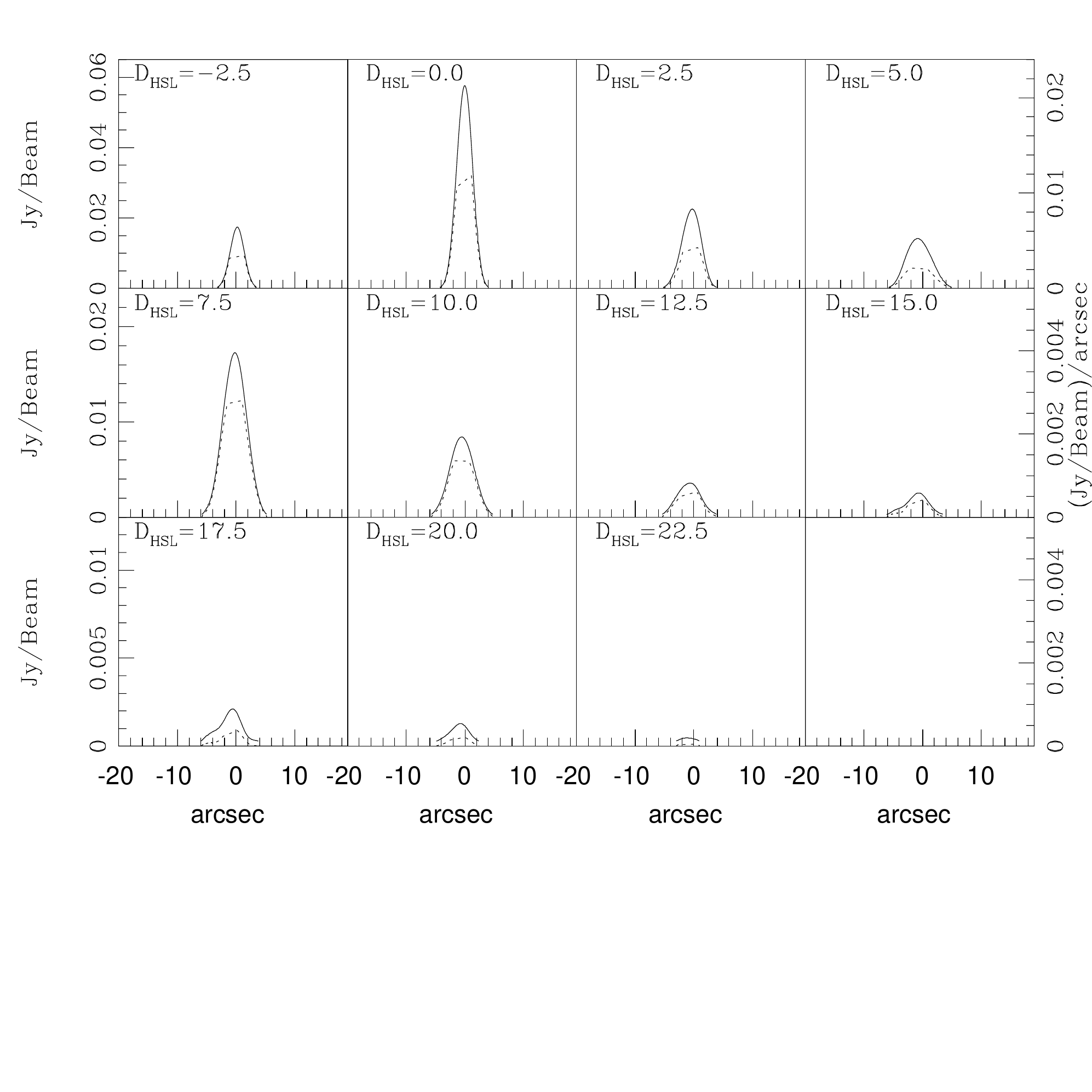}{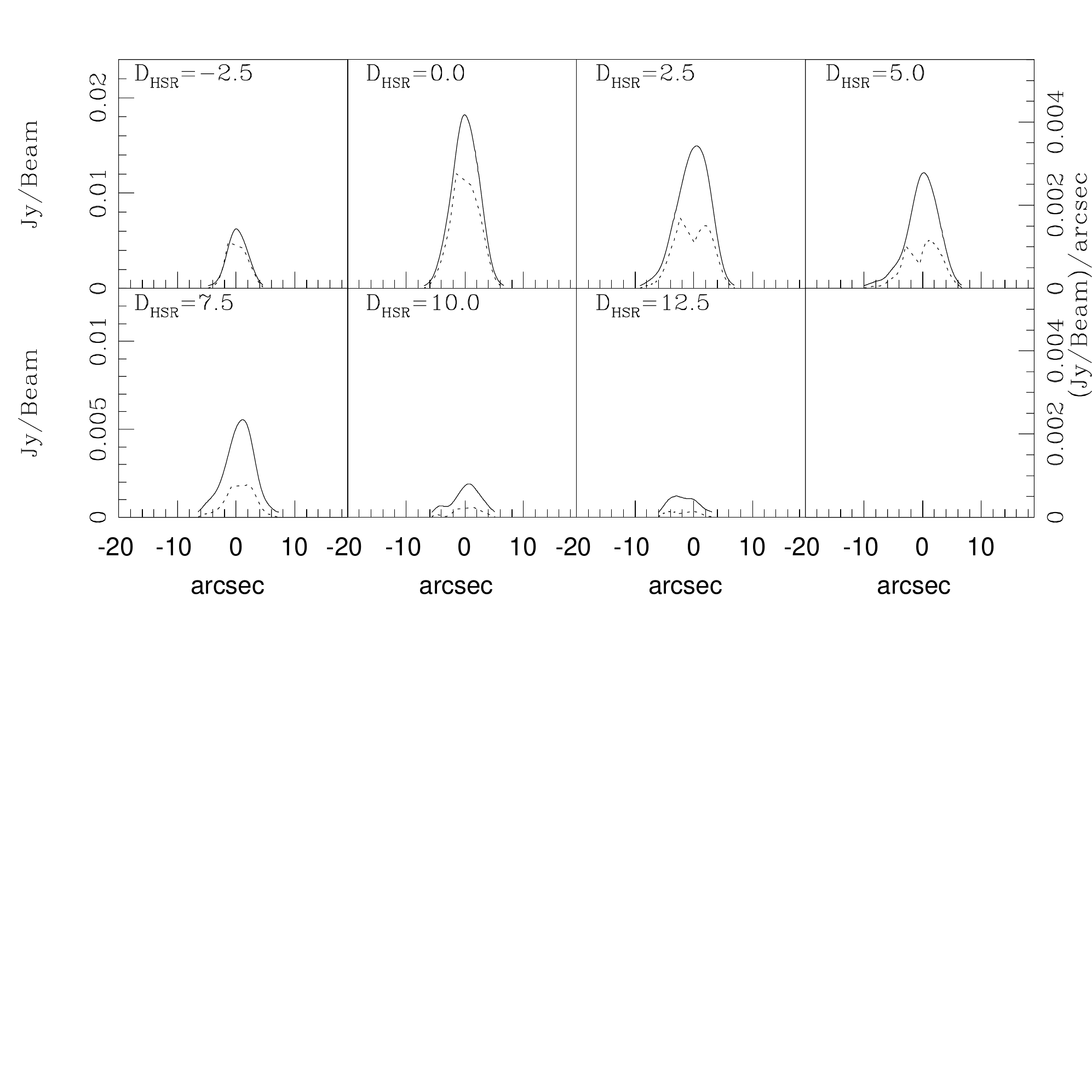}
\caption{3C169.1
emissivity (dotted line) and surface brightness
(solid line) for each cross-sectional slice of the radio 
bridge at 5 GHz, as in Fig. \ref{3C169.1SEM} but at 5 GHz.
Results obtained at 5 GHz are nearly identical to those
obtained at 1.4 GHz.}
\end{figure}

\begin{figure}
\plottwo{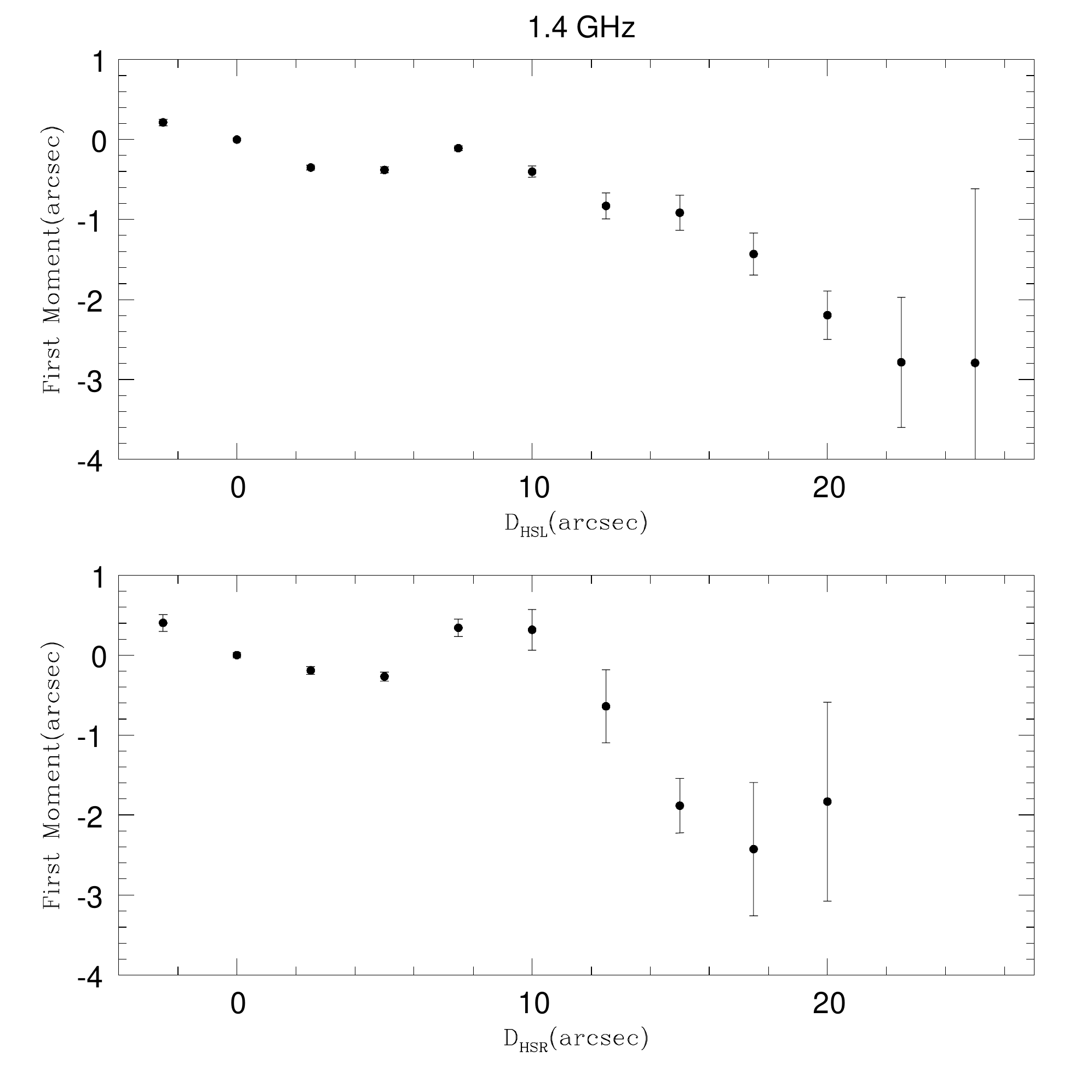}{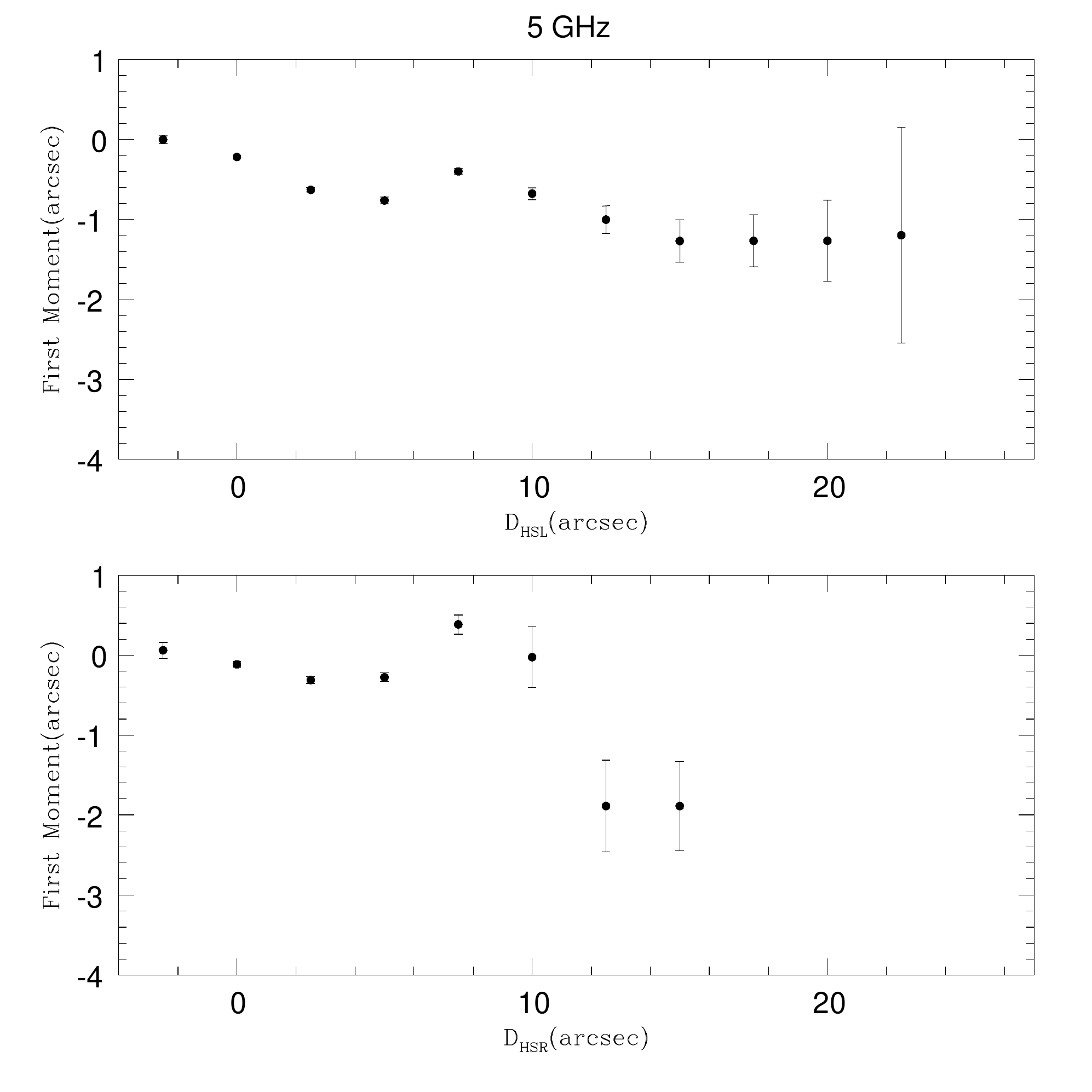}
\caption{3C169.1 first moment as a function of distance from 
the hot spot at 1.4 GHz and 5 GHz (left and right panels,
respectively) for the left and right hand sides of the source
(top and bottom, respectively).}
\end{figure}

\begin{figure}
\plottwo{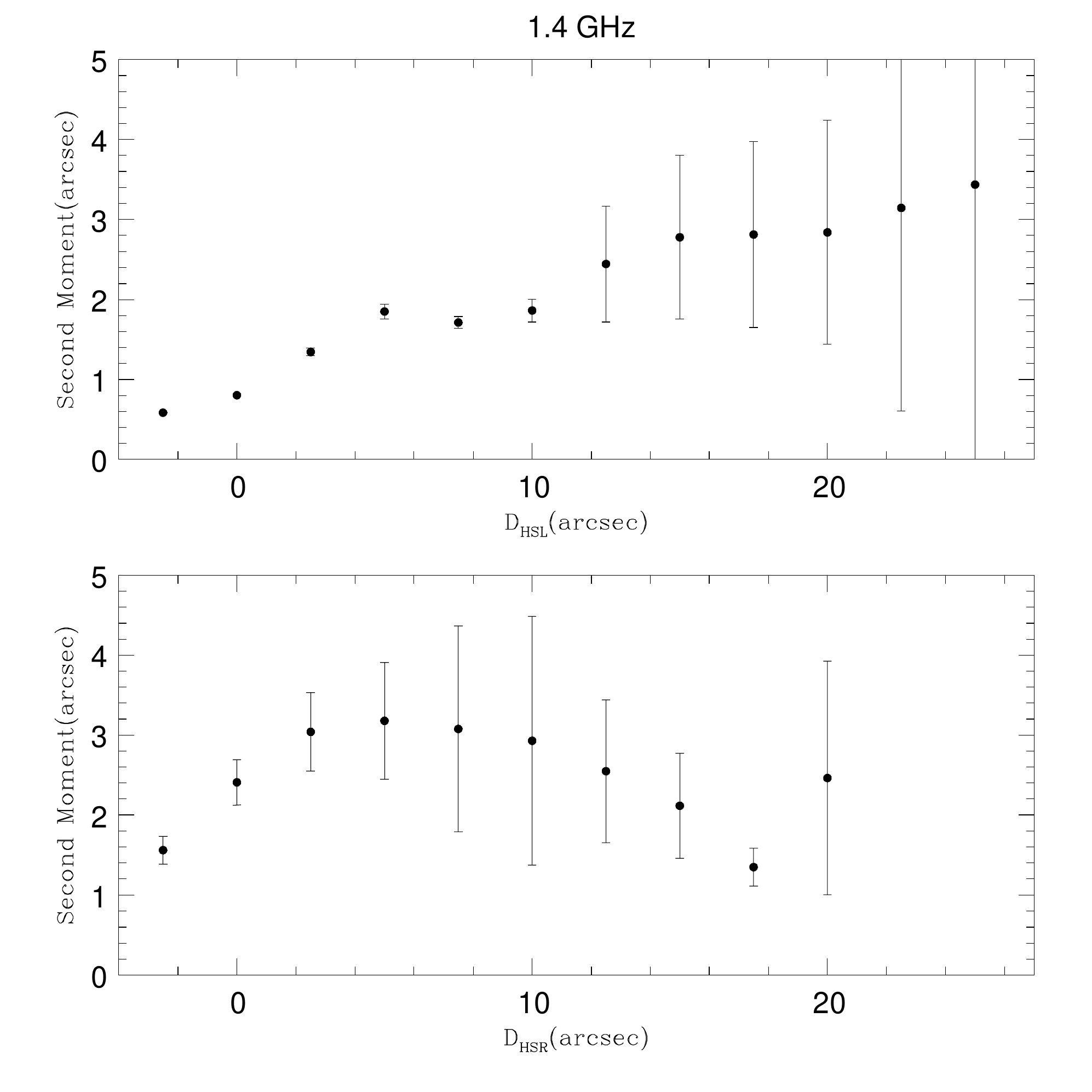}{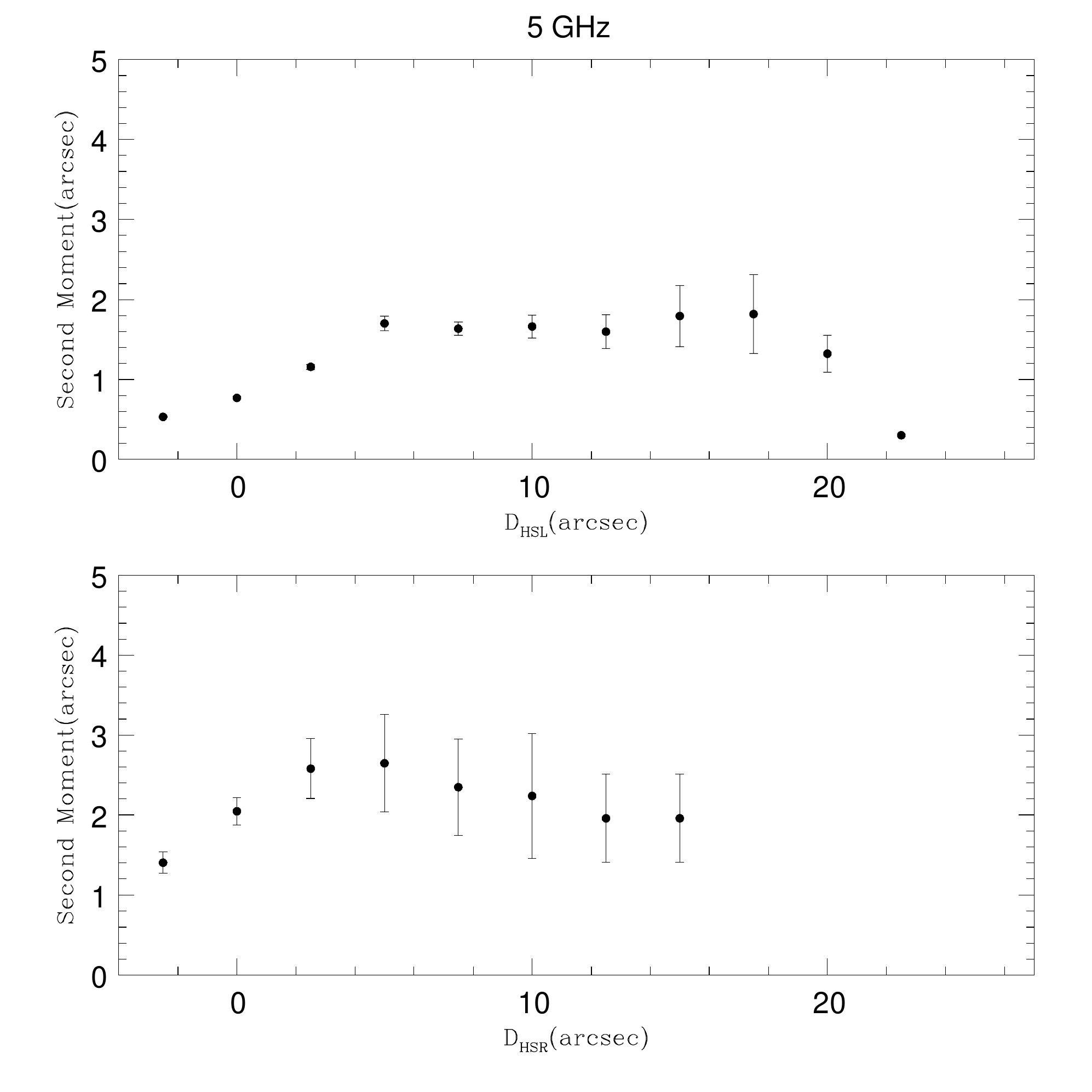}
\caption{3C169.1 second moment as a function of distance from the
hot spot at 1.4 GHz and 5 GHz (left and right panels, 
respectively) for the left and right sides of the source
(top and bottom panels, respectively).}
\end{figure}

\begin{figure}
\plottwo{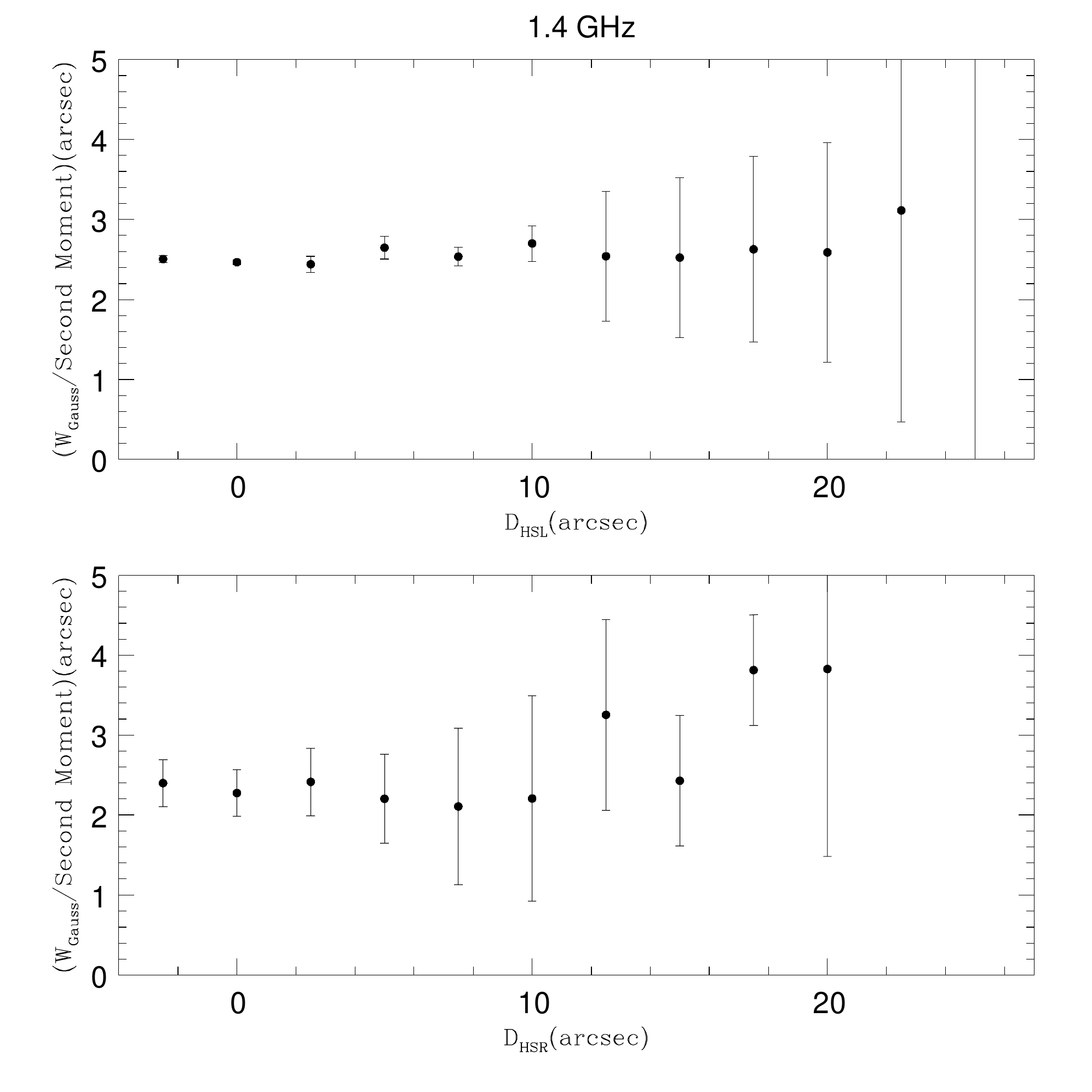}{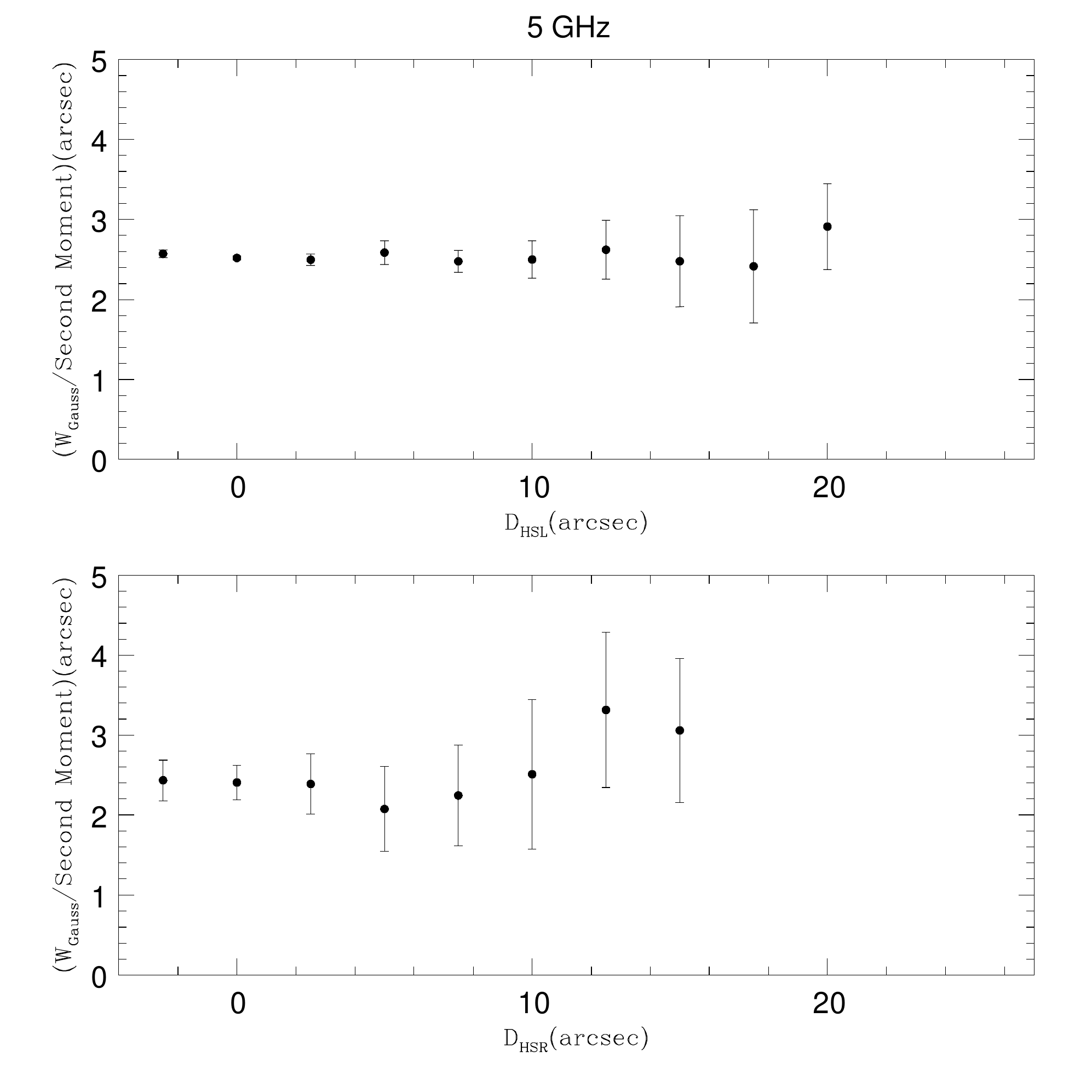}
\caption{3C169.1 ratio of the Gaussian FWHM to the second moment as a
function of distance from the hot spot at 1.4 and 5 GHz 
(left and right panels, respectively) for the left and right
hand sides of the source (top and bottom panels, respectively). The value of this ratio is fairly constant for each side of the source, and has a value of about 2.5. Similar results are obtained at 1.4 and 5 GHz.} 
\end{figure}

\begin{figure}
\plottwo{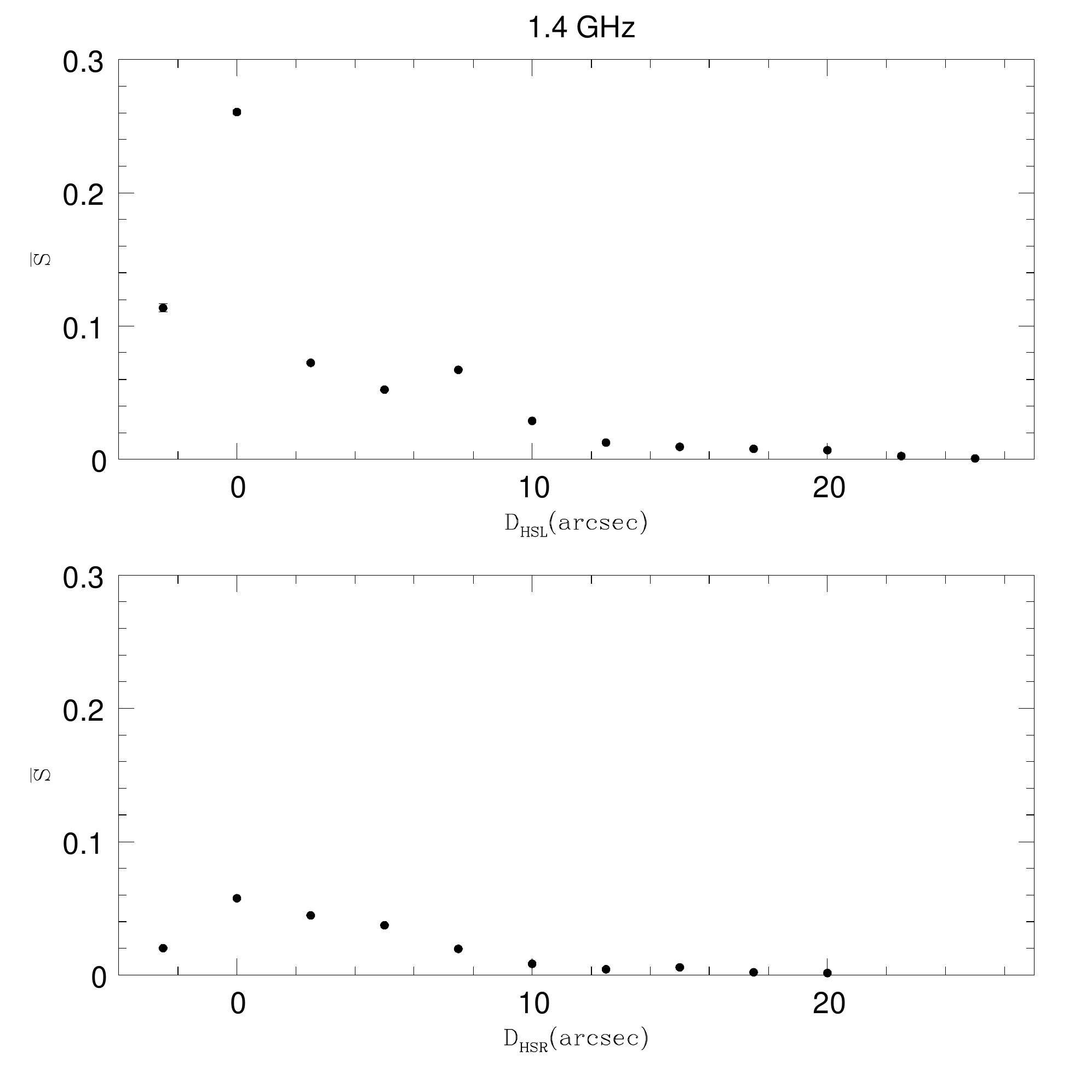}{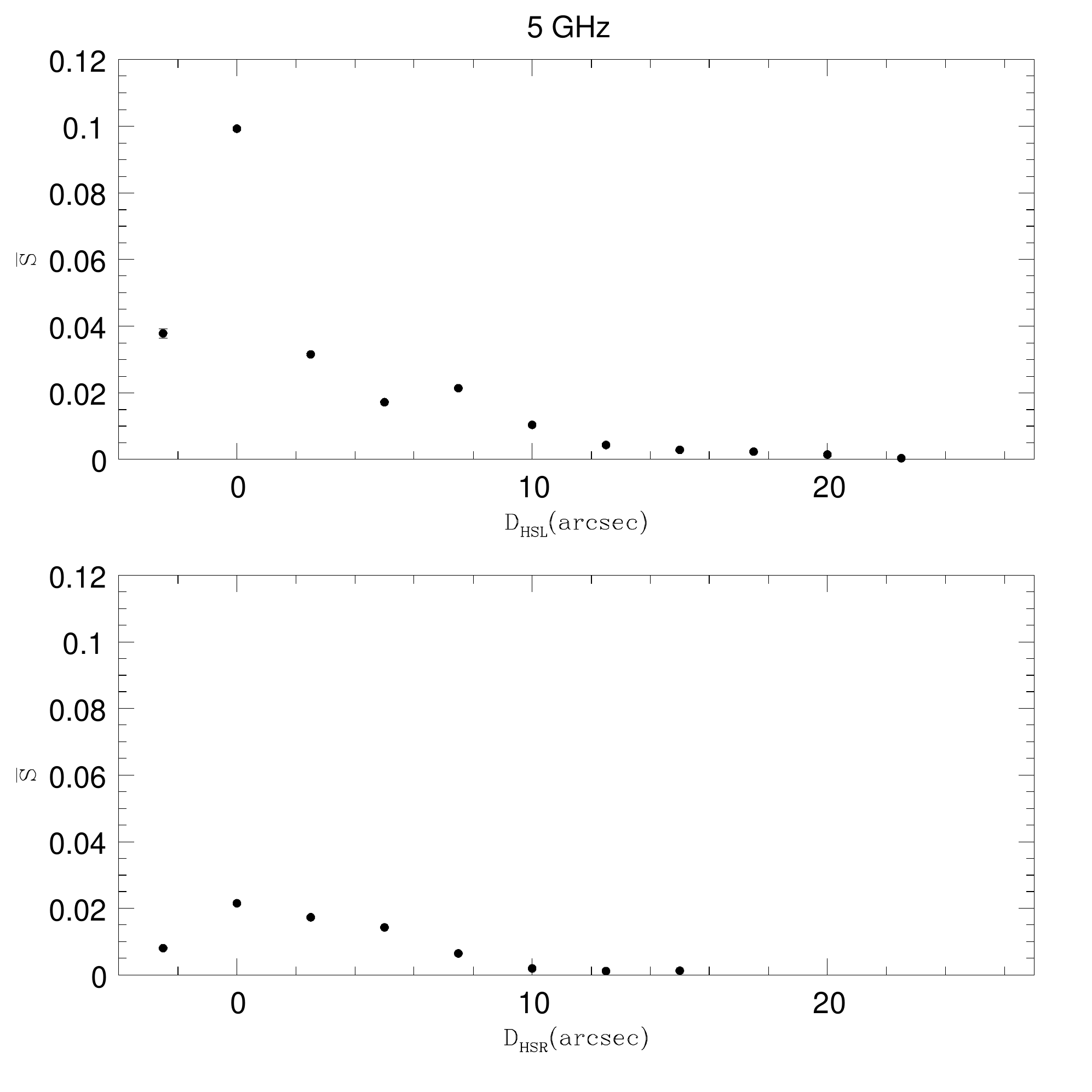}
\caption{3C169.1 average surface brightness in units of Jy/beam 
as a function of distance from the 
hot spot at 1.4 and 5 GHz 
(left and right panels, respectively) for the left and right
hand sides of the source (top and bottom panels, respectively).}
\end{figure}

\begin{figure}
\plottwo{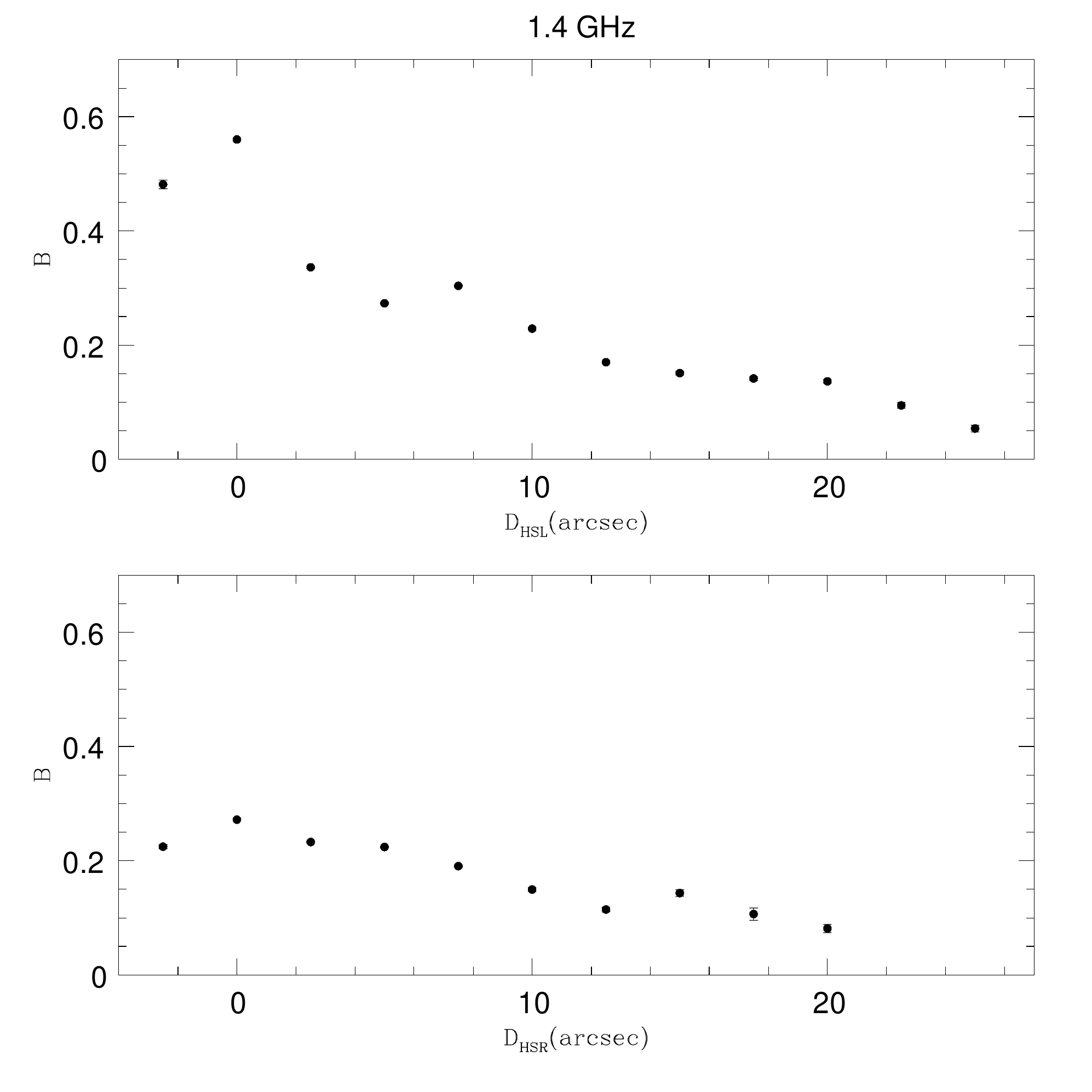}{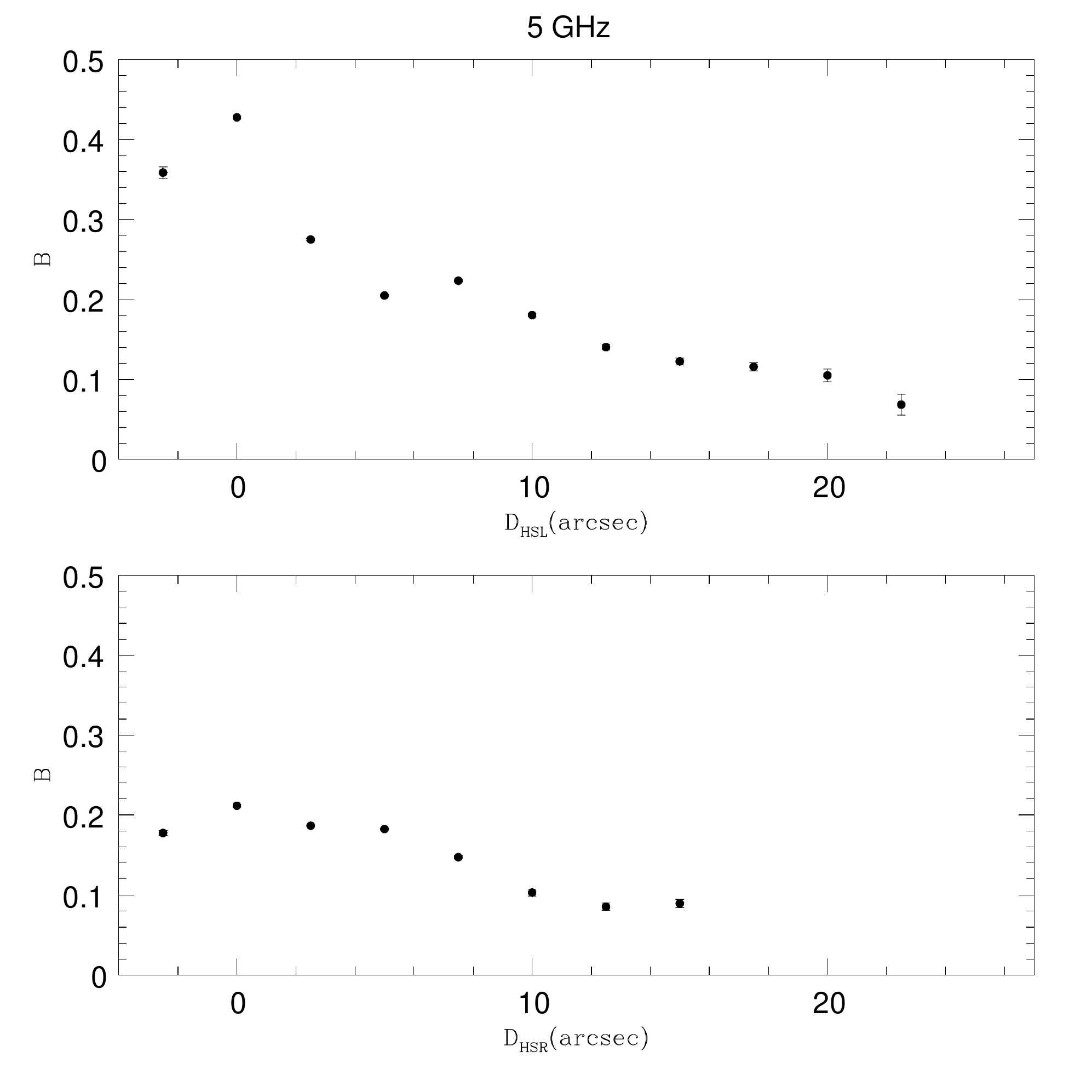}
\caption{3C169.1 minimum energy magnetic field strength 
as a function of distance from the 
hot spot at 1.4 and 5 GHz 
(left and right panels, respectively) for the left and right
hand sides of the source (top and bottom panels, respectively).
The normalization is given in Table 1. 
The flat magnetic field strength across the right hand side of 
the source is caused by the weak hot spot on that side. 
}
\end{figure}

\begin{figure}
\plottwo{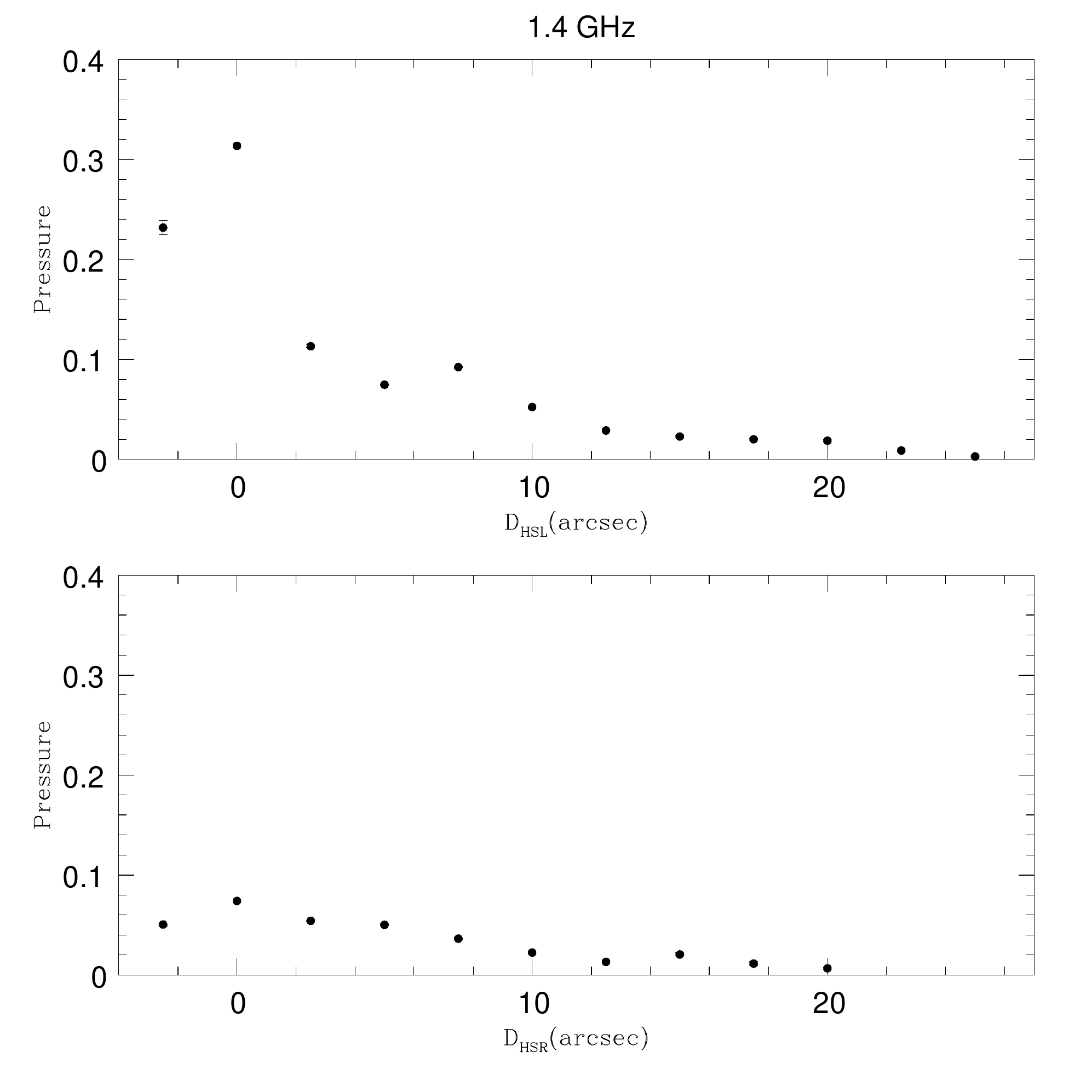}{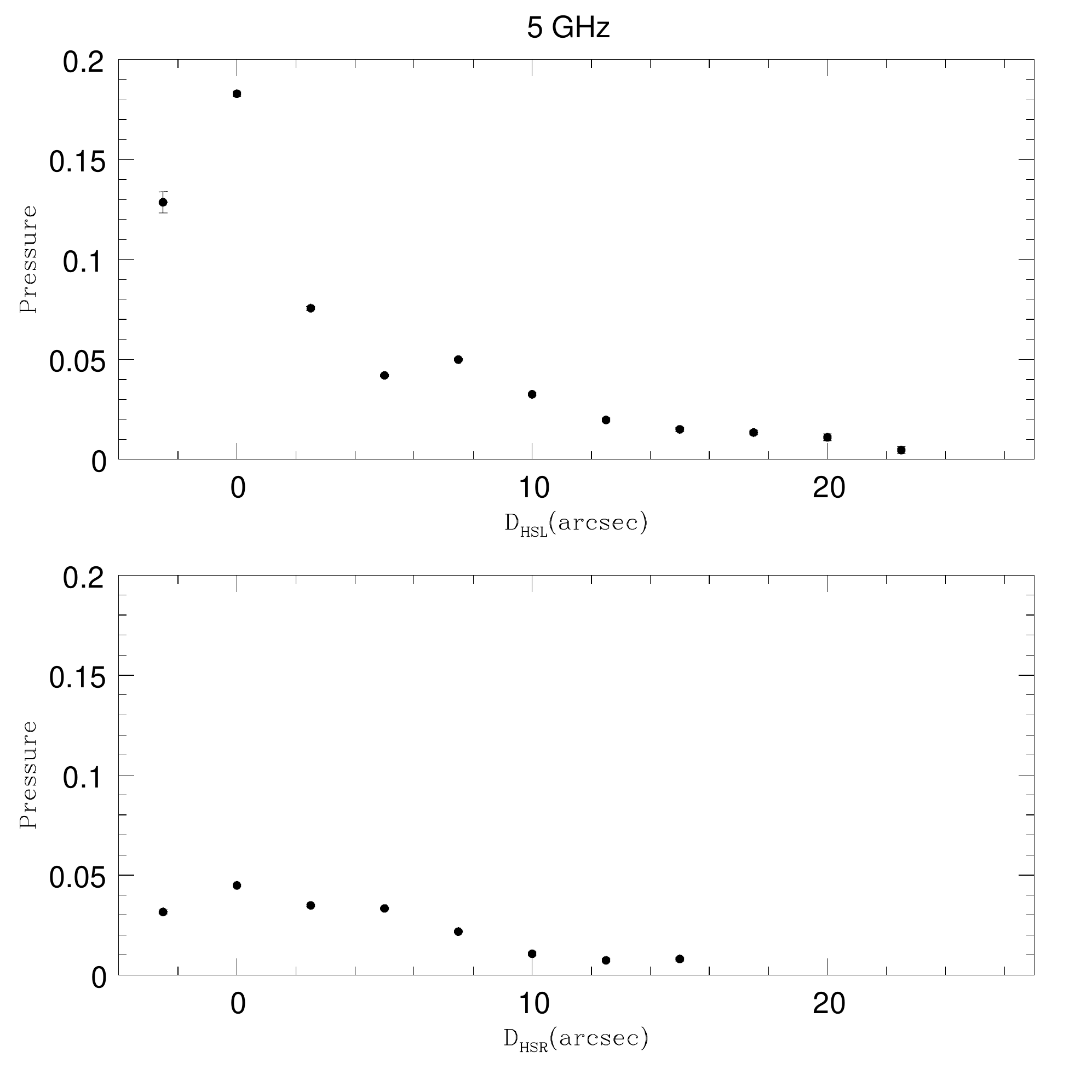}
\caption{3C169.1 
minimum energy pressure as a function of 
distance from the hot spot at 1.4 and 5 GHz 
(left and right panels, respectively) for the left and right
hand sides of the source (top and bottom panels, respectively), 
as in Fig. \ref{3C6.1P}. The normalization is given in Table 1.
The flat pressure profile across the right hand side of the source
is caused by the weak hot spot on that side of the source. 
}
\end{figure}

\clearpage
\begin{figure}
\plotone{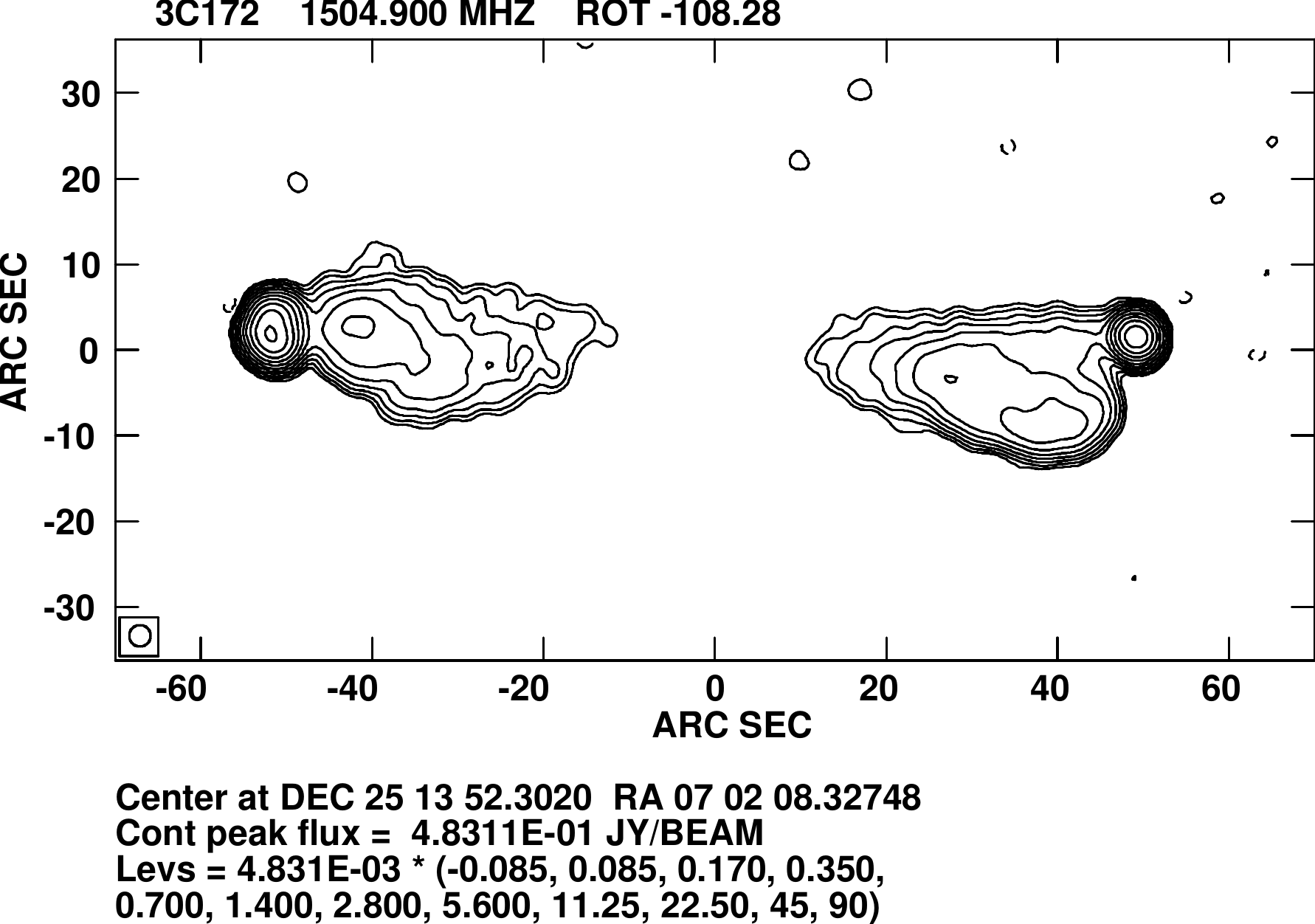}
\caption{3C172 at 1.4 GHz and 2.5'' resolution rotated by 108.28 deg 
clockwise, as in Fig. \ref{3C6.1rotfig}.}
\end{figure}

\begin{figure}
\plottwo{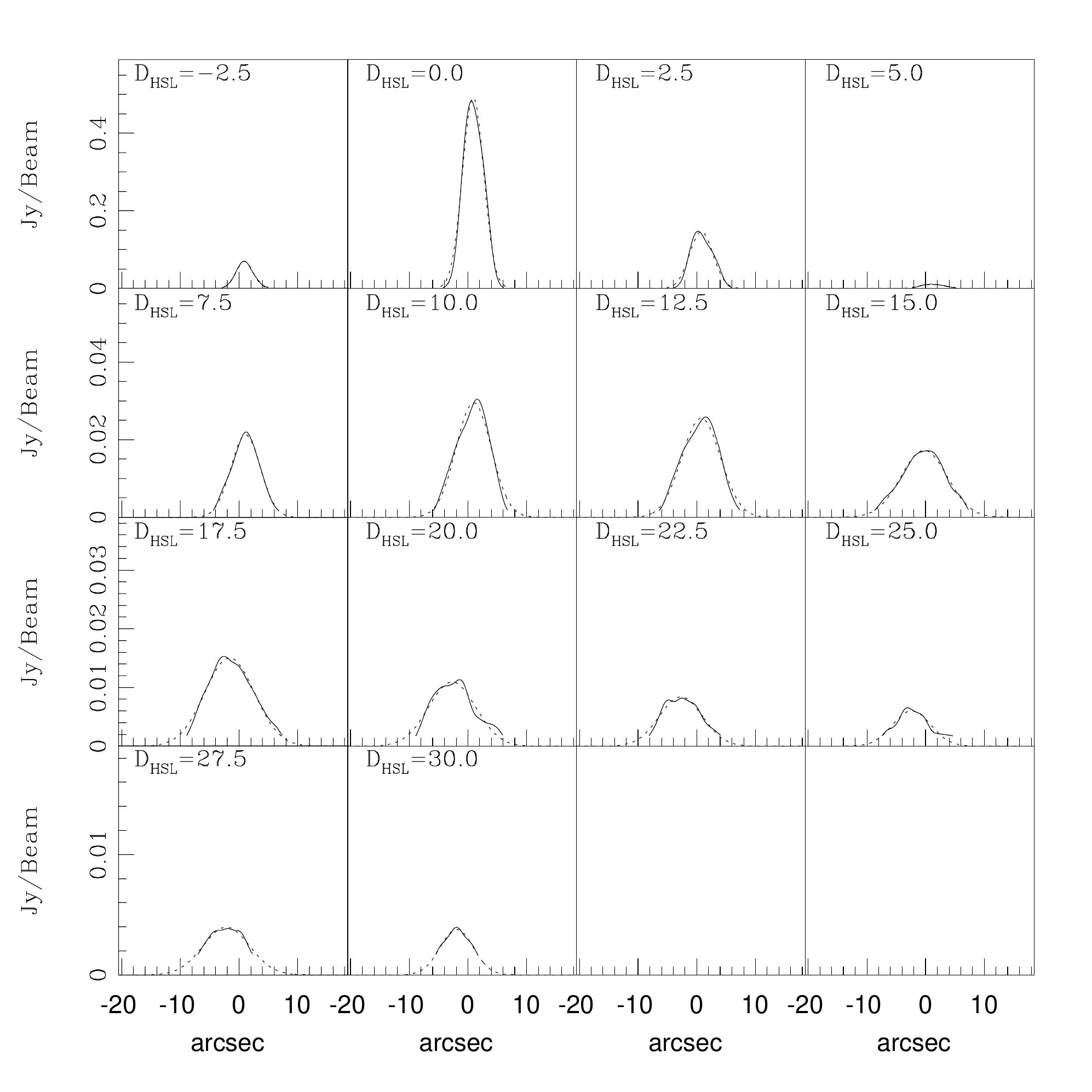}{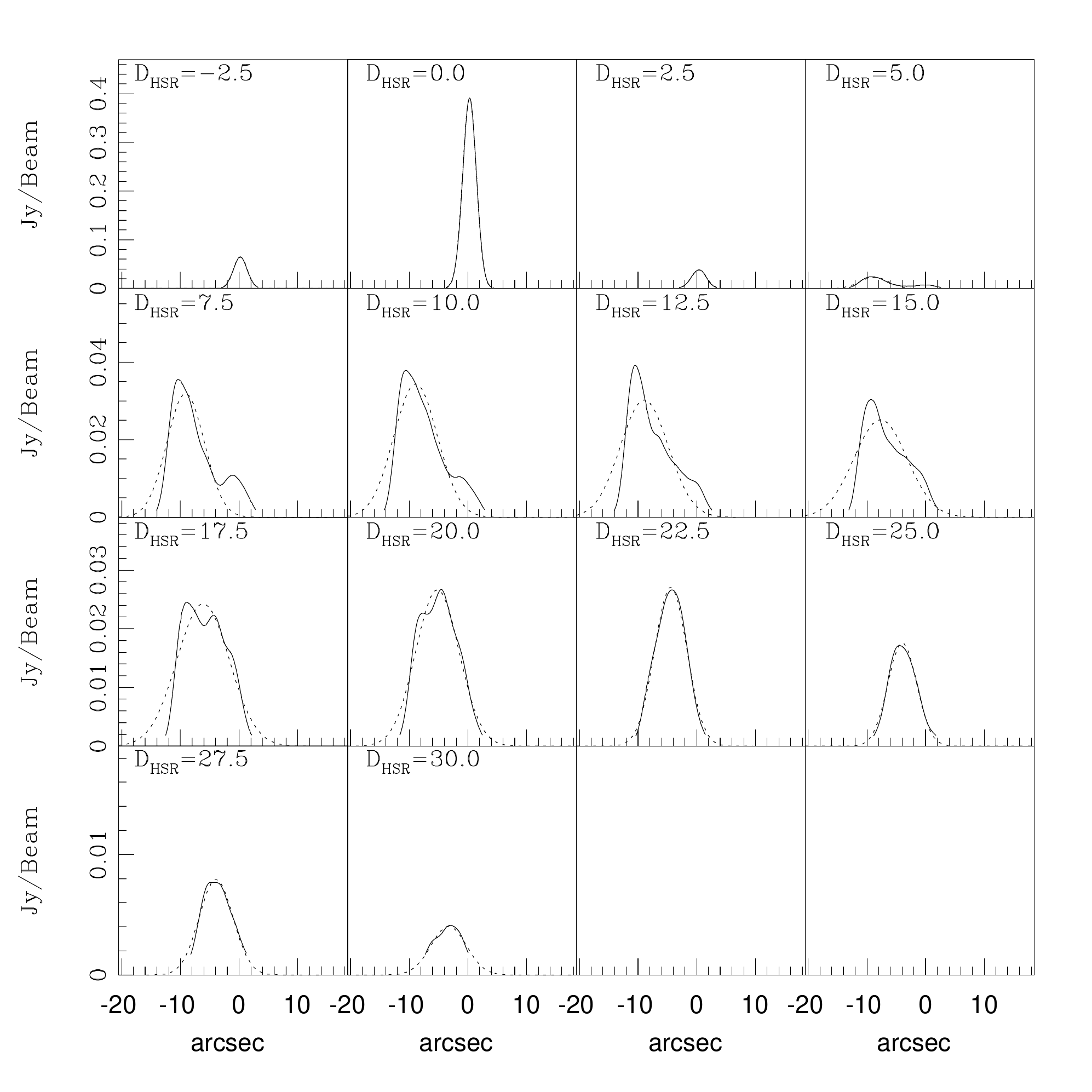}
\caption{3C172 
surface brightness (solid line) and best fit 
Gaussian (dotted line) for each cross-sectional slice of the 
radio bridge at 1.4 GHz, as in Fig. \ref{3C6.1SG}.
A Gaussian provides an excellent description of the surface brightness
profile of each cross-sectional slice. }
\label{3C172SG}
\end{figure}

\begin{figure}
\plottwo{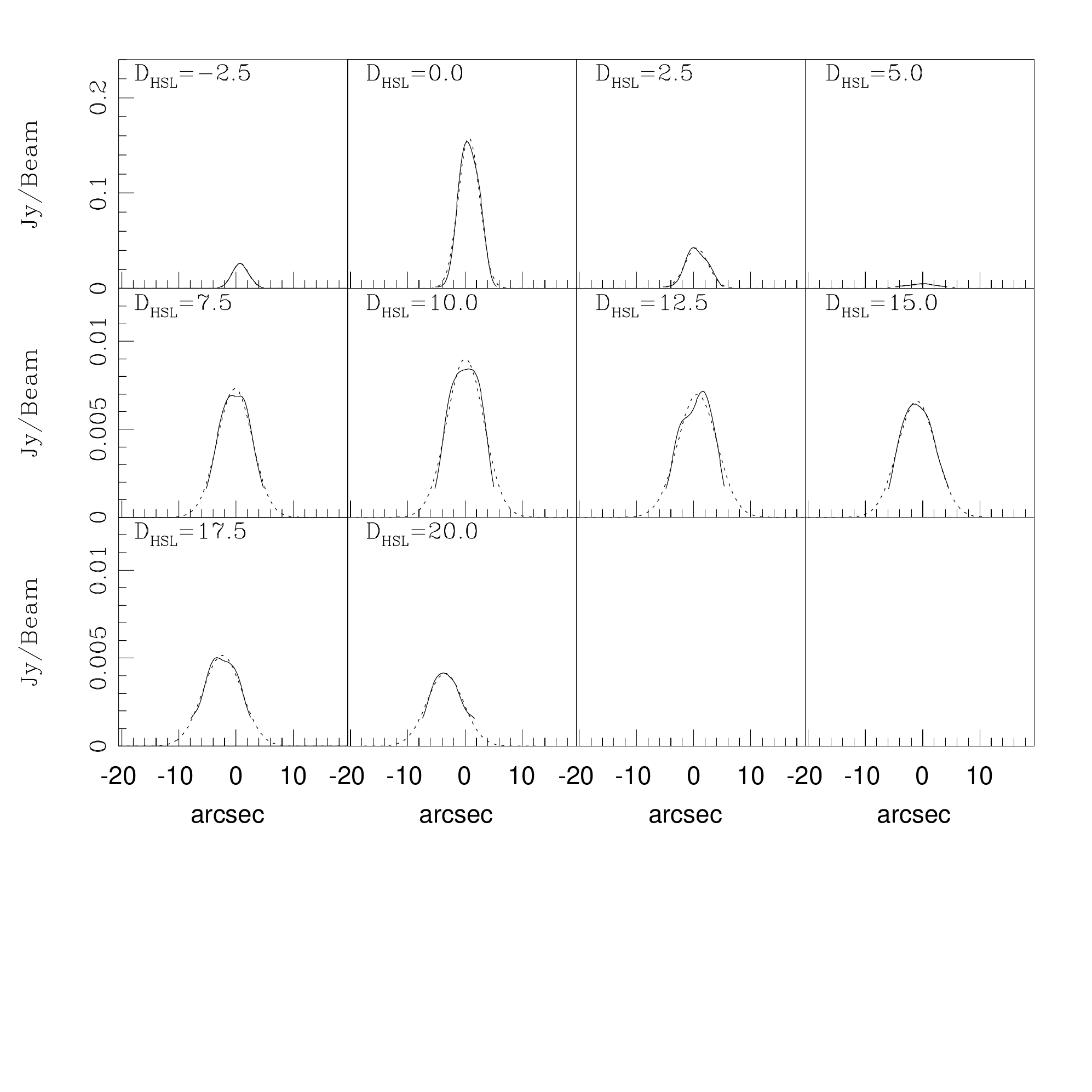}{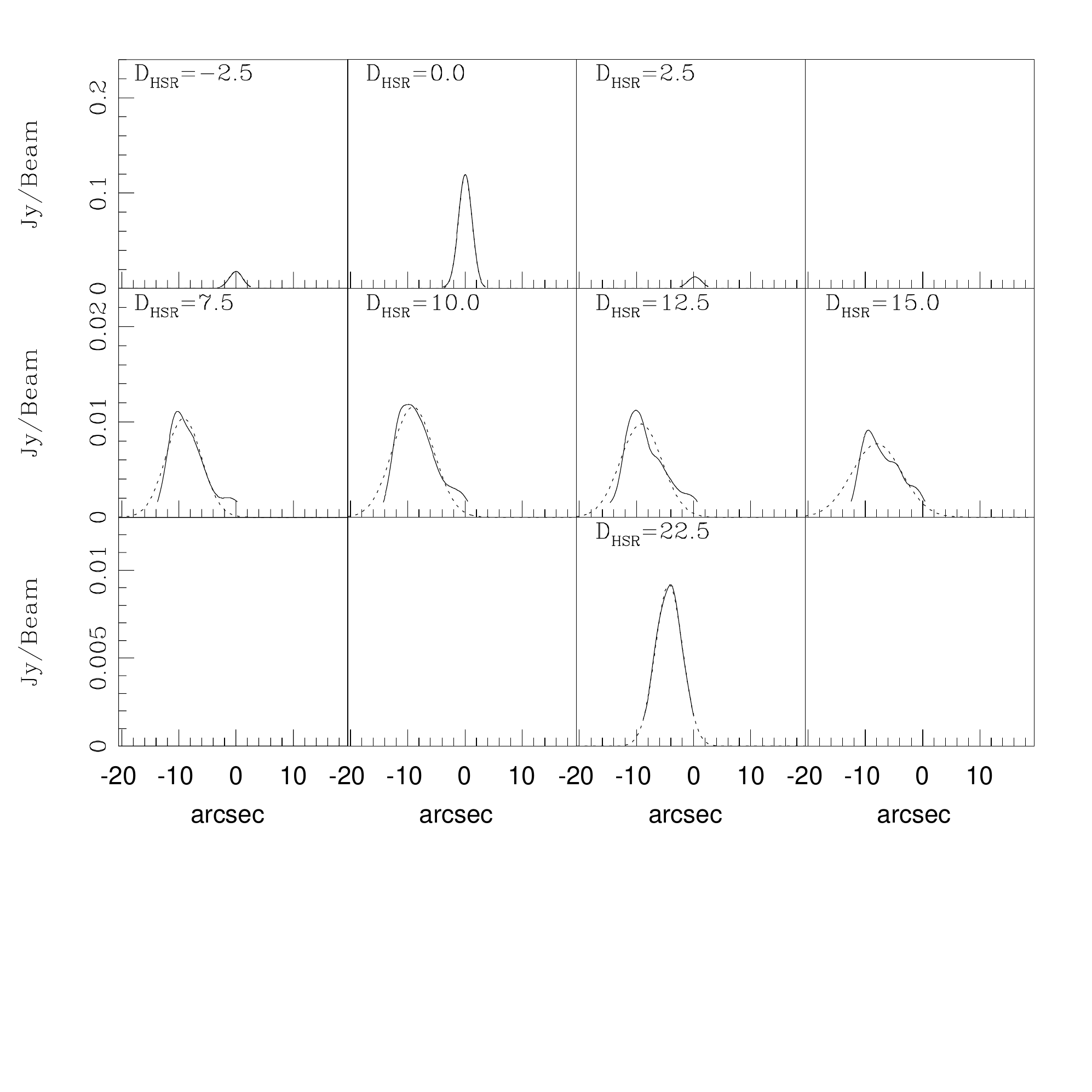}
\caption{3C172 
surface brightness (solid line) and best fit 
Gaussian (dotted line) for each cross-sectional slice of the 
radio bridge at 5 GHz, as in 
Fig. \ref{3C172SG} but at 5 GHz. 
Results obtained at 5 GHz are nearly identical to those
obtained at 1.4 GHz.}
\end{figure}

\begin{figure}
\plottwo{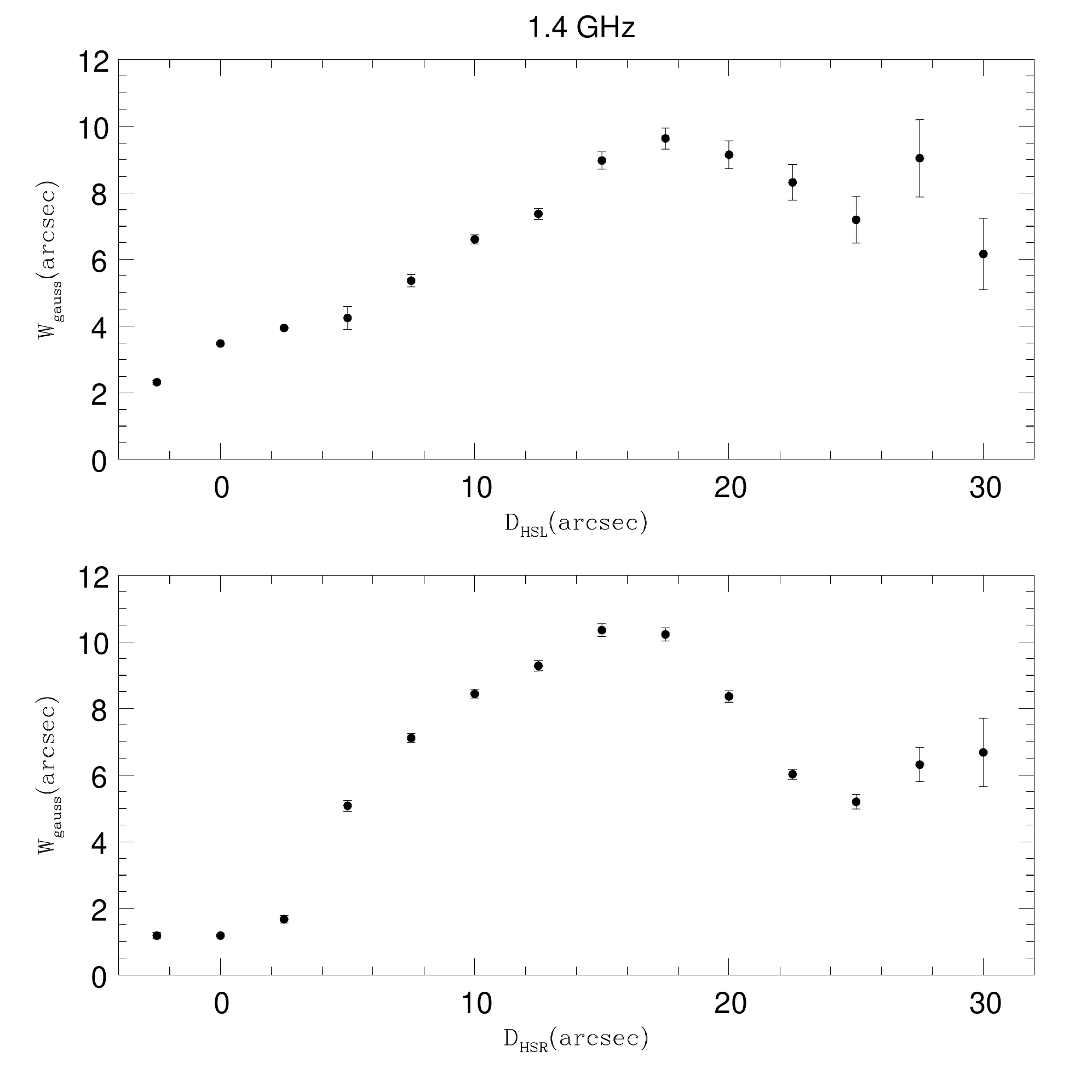}{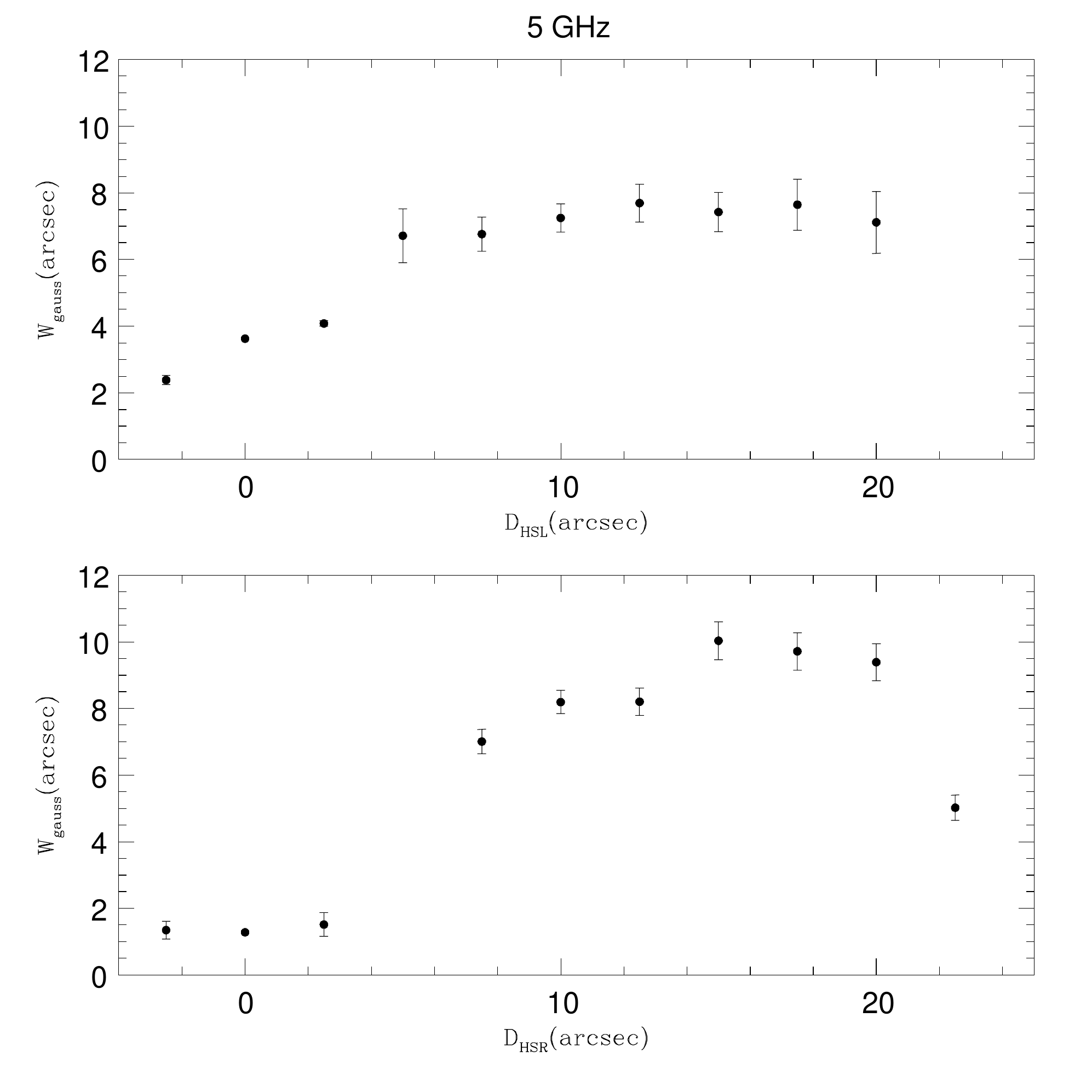}
\caption{3C172  Gaussian FWHM as a function of 
distance from the hot spot  
at 1.4 and 5 GHz (left and
right panels, respectively) for the left and right hand sides of the 
source (top and bottom panels, respectively), as in Fig. \ref{3C6.1WG}.
The right and left hand sides of the source are fairly 
symmetric.  Similar results are obtained at 
1.4 and 5 GHz. }
\end{figure}

\begin{figure}
\plottwo{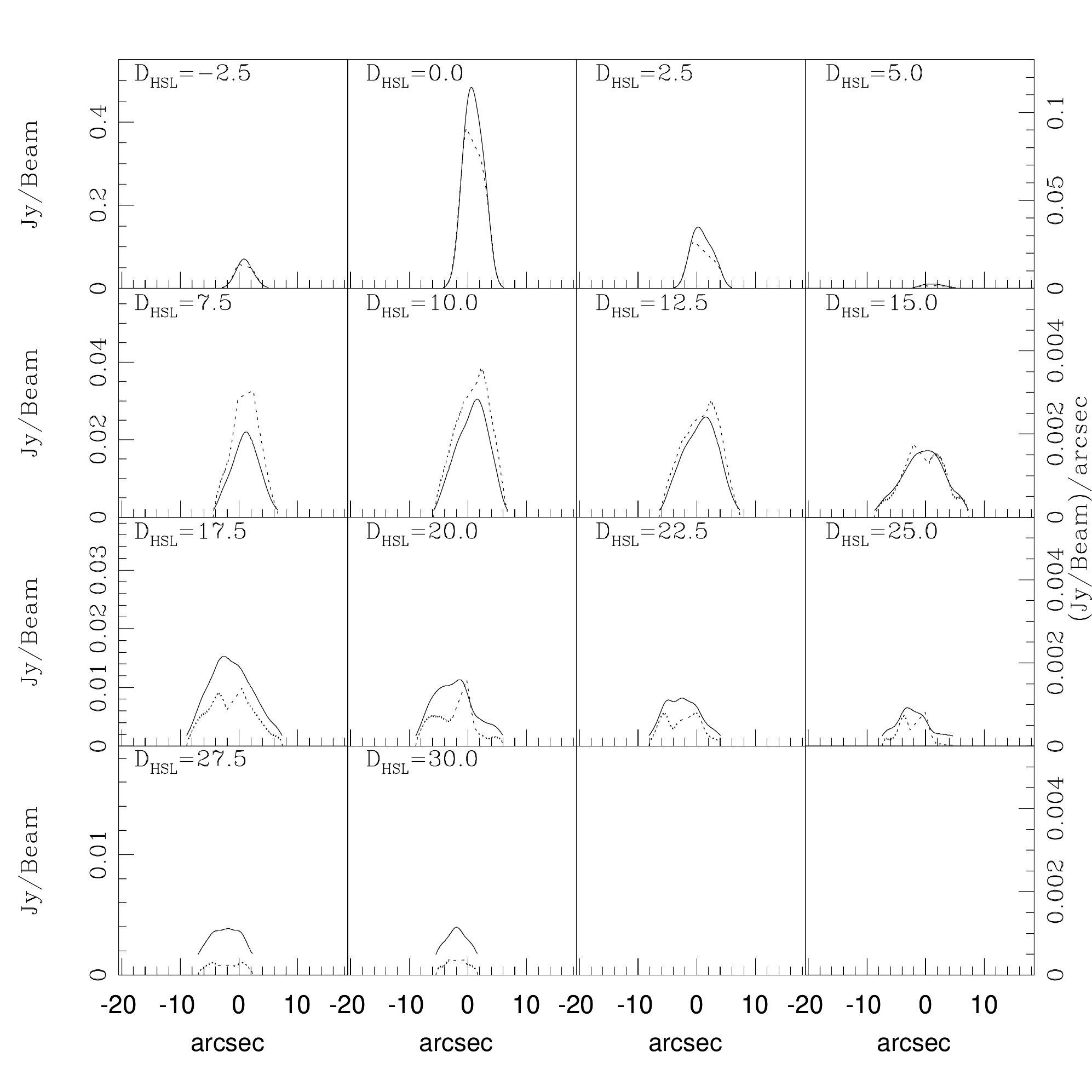}{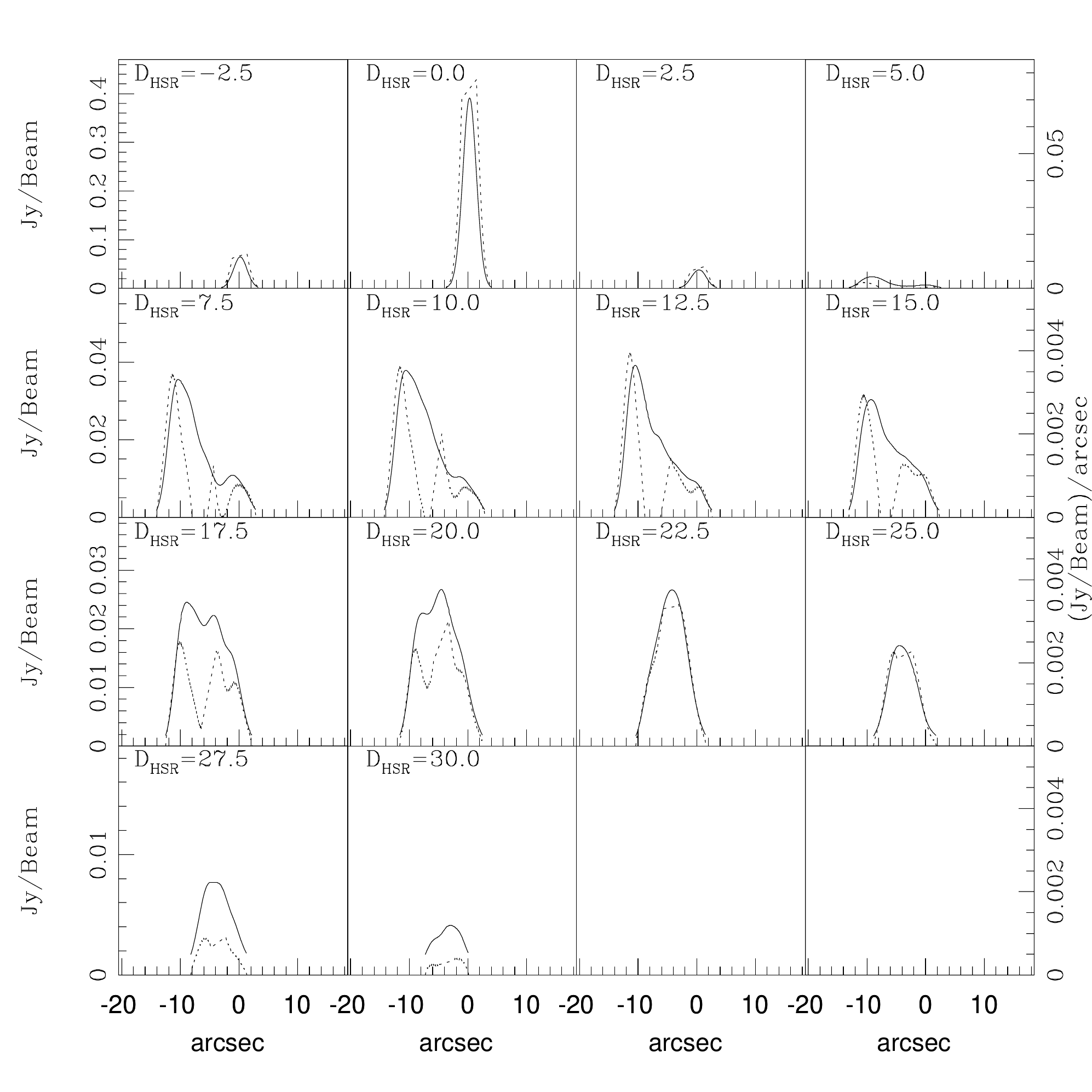}
\caption{3C172 emissivity (dotted line) and surface brightness
(solid line) for each cross-sectional slice of the radio 
bridge at 1.4 GHz, as in Fig. \ref{3C6.1SEM}. A constant volume
emissivity per slice provides a reasonable description of the 
data over most of the source. }
\label{3C172SEM}
\end{figure}

\begin{figure}
\plottwo{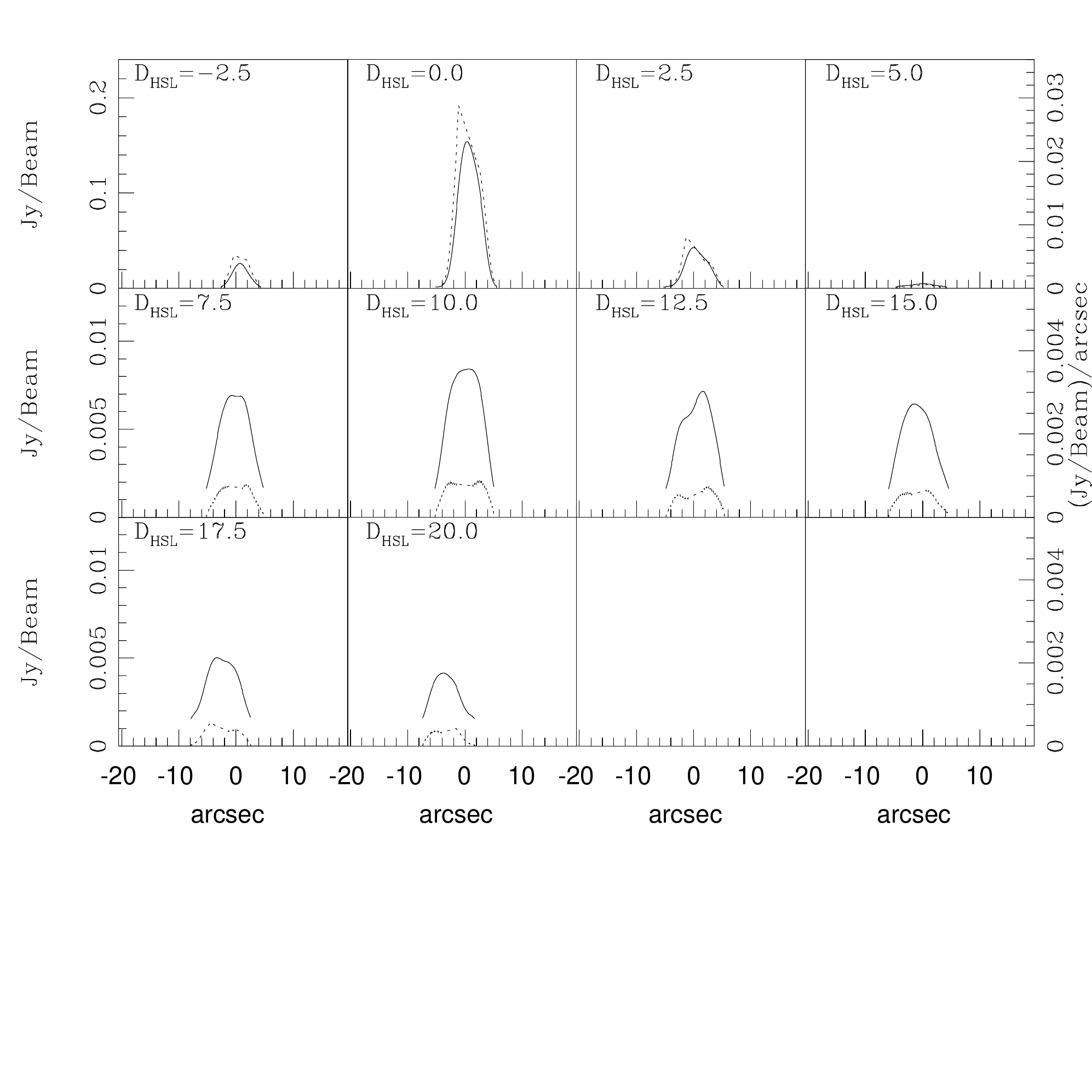}{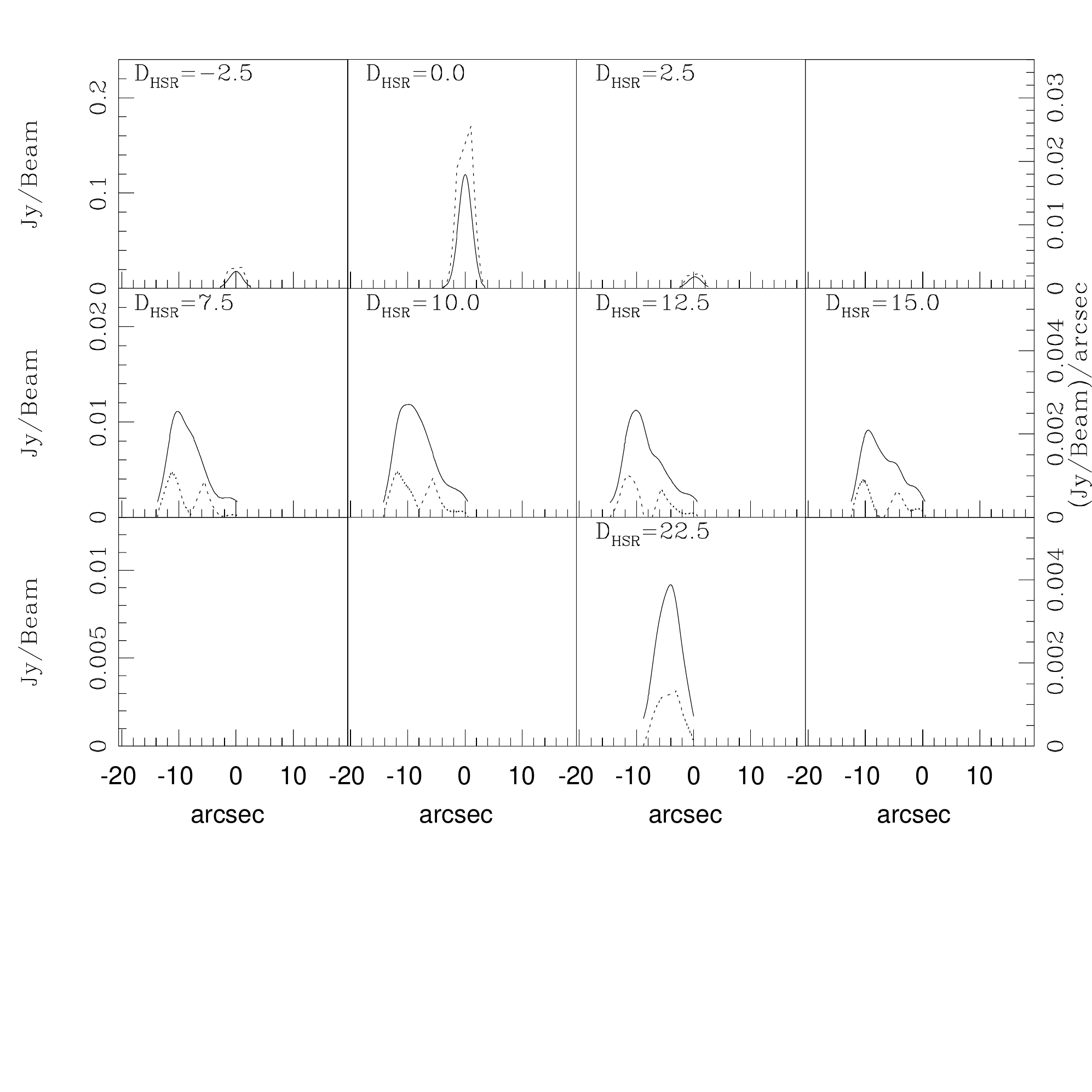}
\caption{3C172
emissivity (dotted line) and surface brightness
(solid line) for each cross-sectional slice of the radio 
bridge at 5 GHz, as in Fig. \ref{3C172SEM} but at 5 GHz.
Results obtained at 5 GHz are nearly identical to those
obtained at 1.4 GHz.}
\end{figure}

\begin{figure}
\plottwo{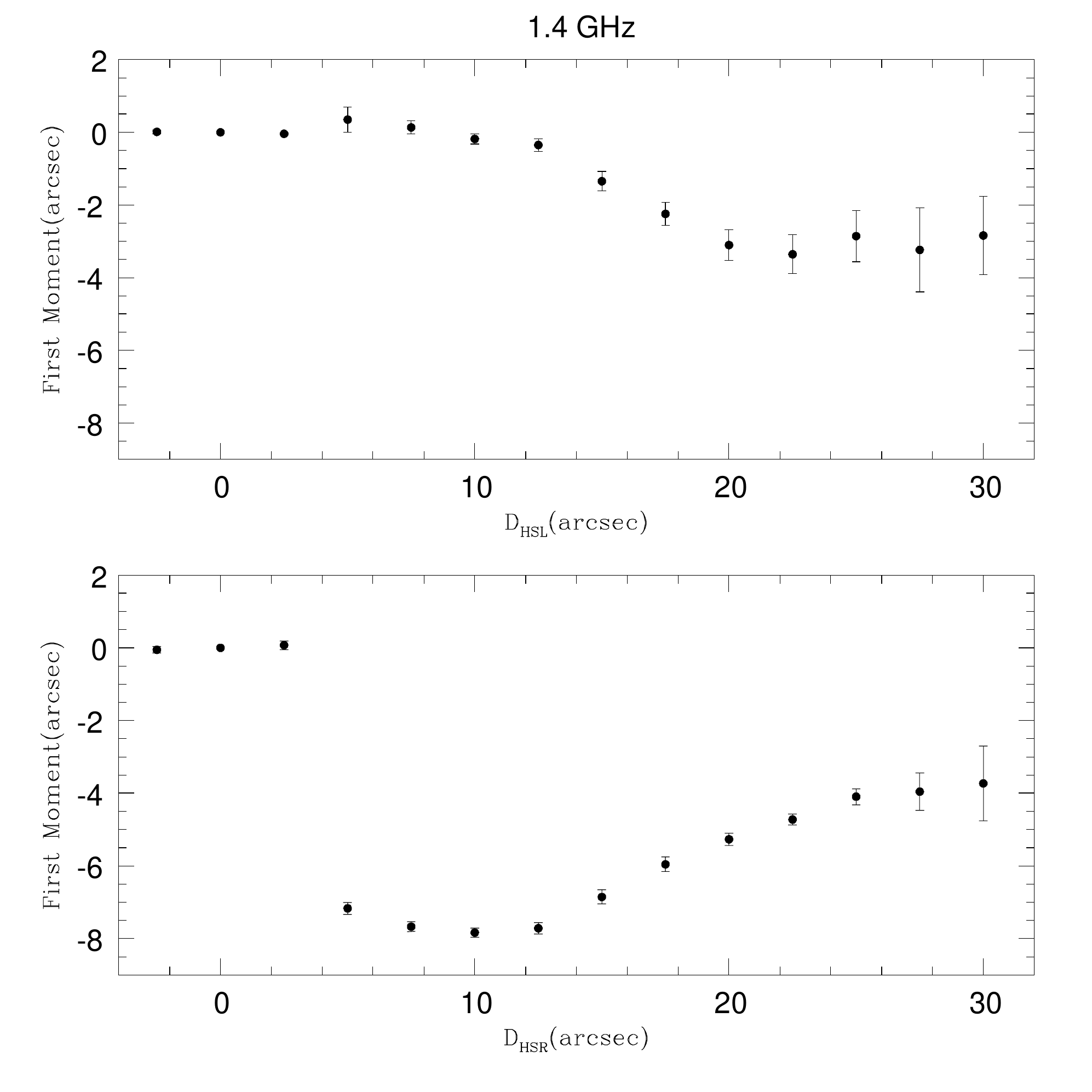}{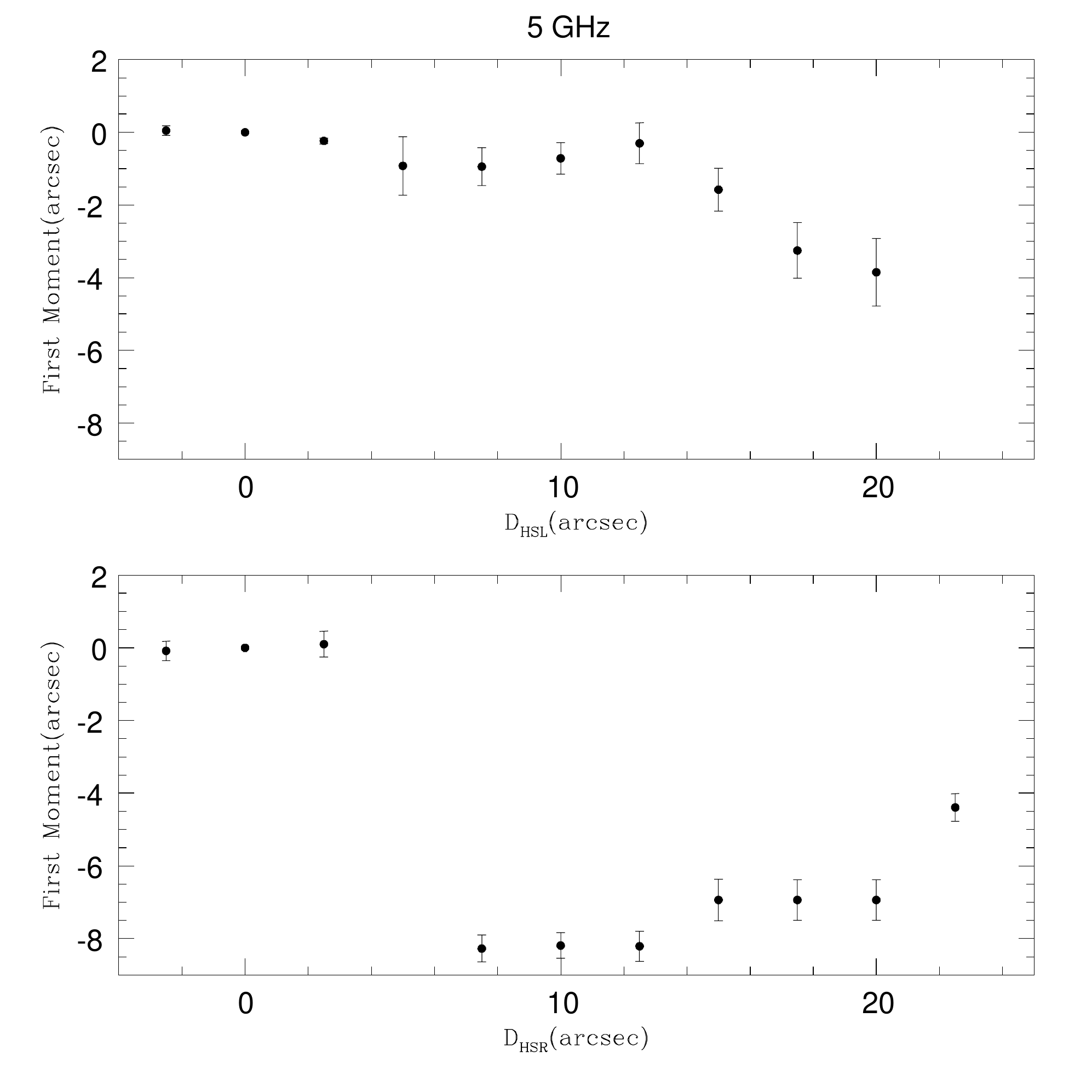}
\caption{3C172 first moment as a function of distance from 
the hot spot at 1.4 GHz and 5 GHz (left and right panels,
respectively) for the left and right hand sides of the source
(top and bottom, respectively).}
\end{figure}

\begin{figure}
\plottwo{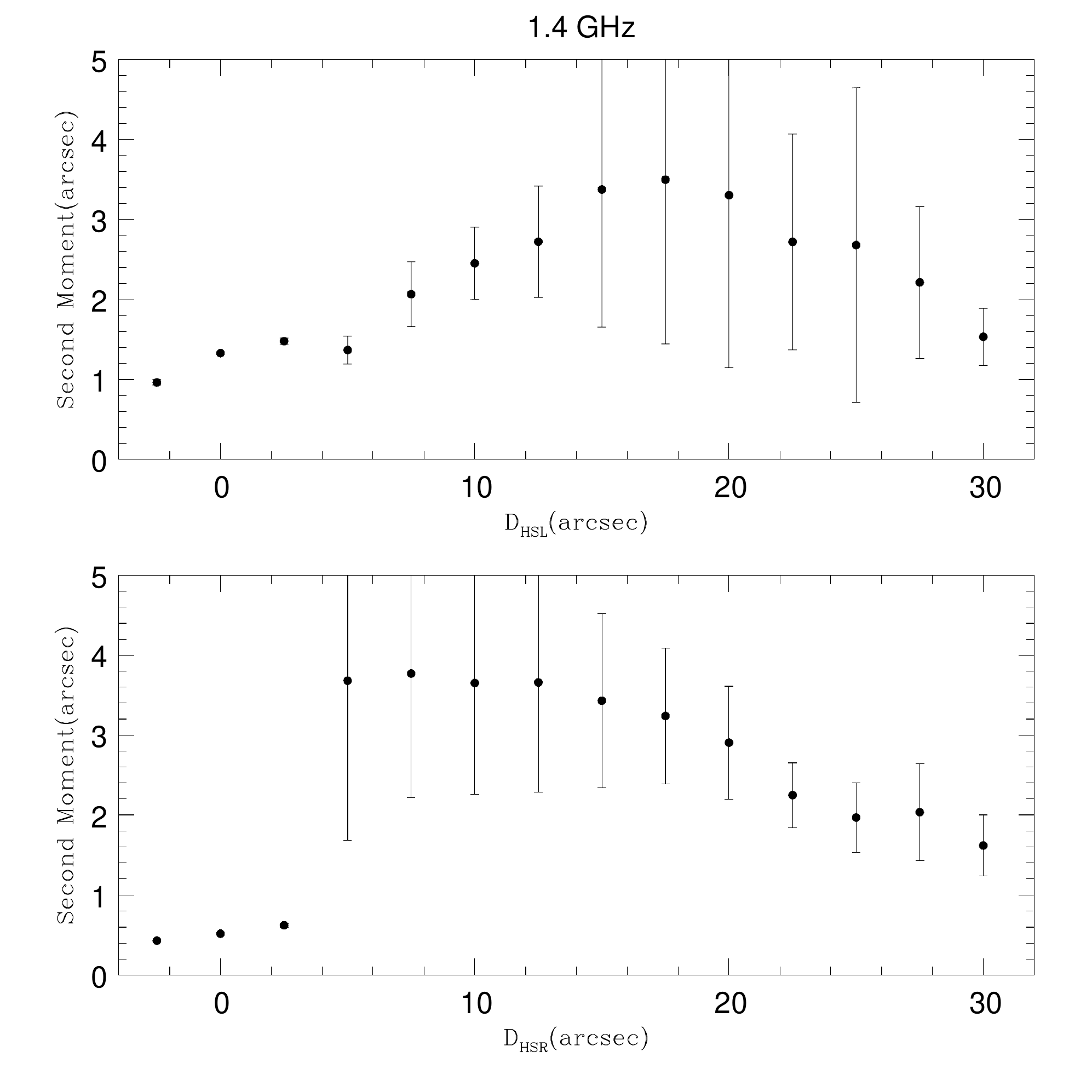}{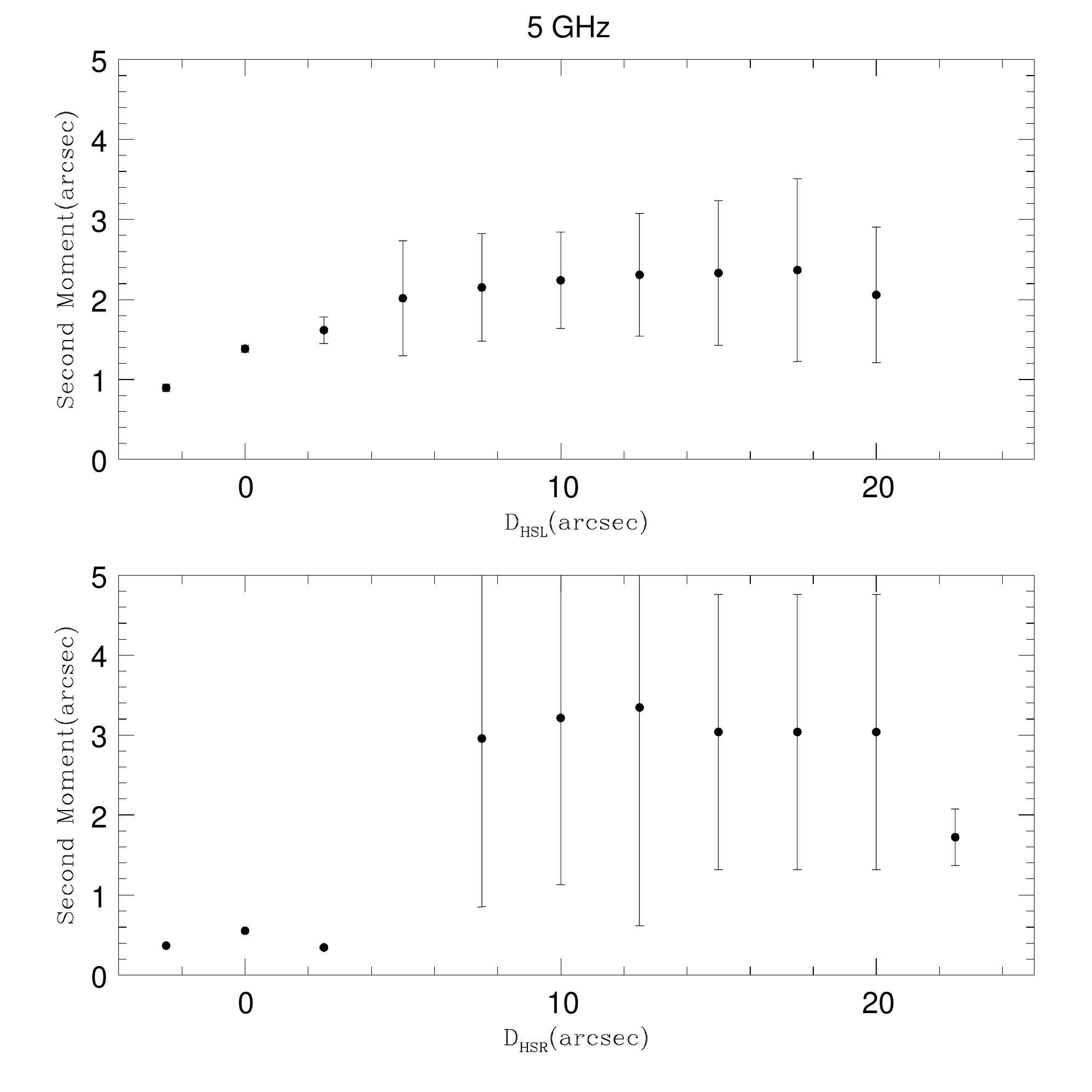}
\caption{3C172 second moment as a function of distance from the
hot spot at 1.4 GHz and 5 GHz (left and right panels, 
respectively) for the left and right sides of the source
(top and bottom panels, respectively).}
\end{figure}

\begin{figure}
\plottwo{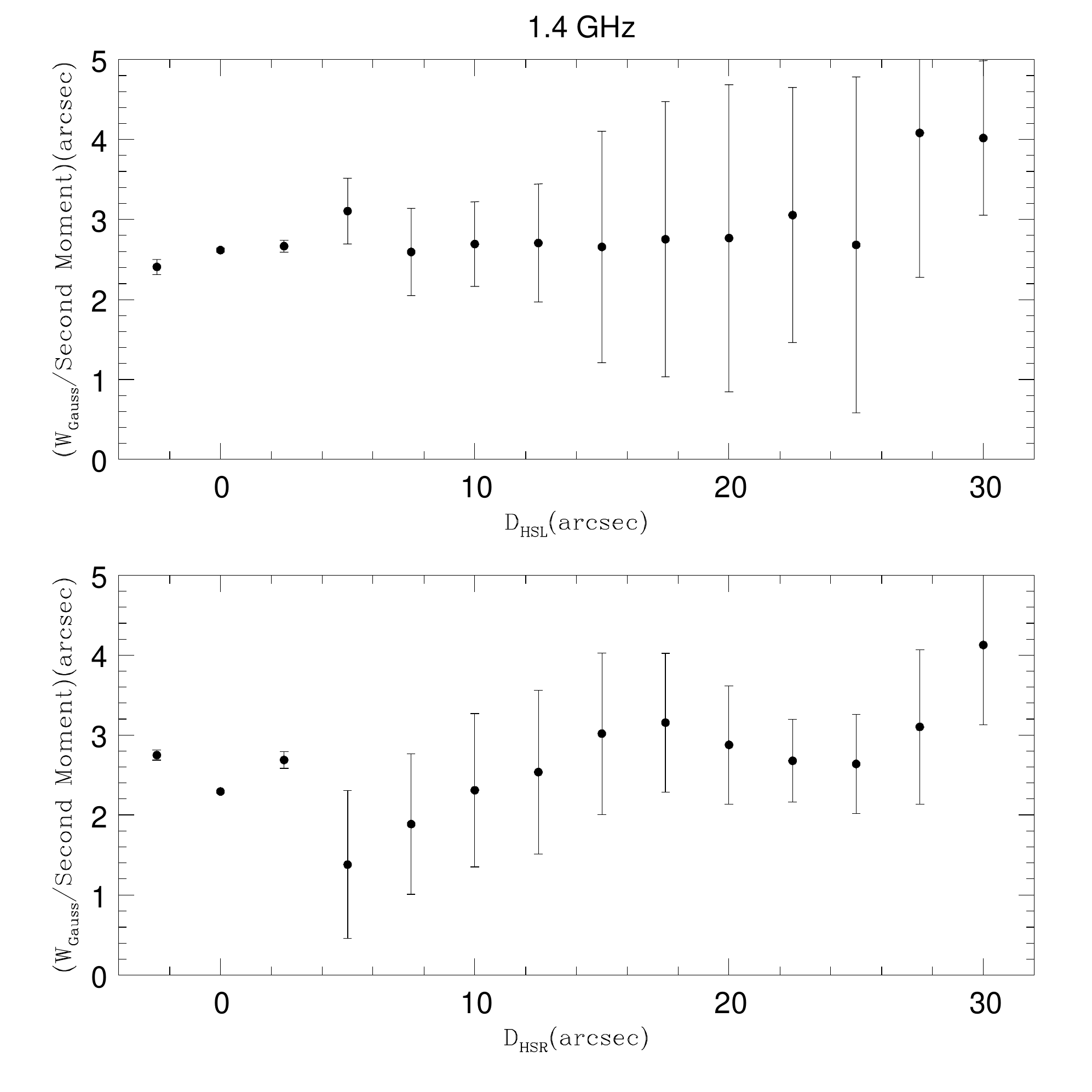}{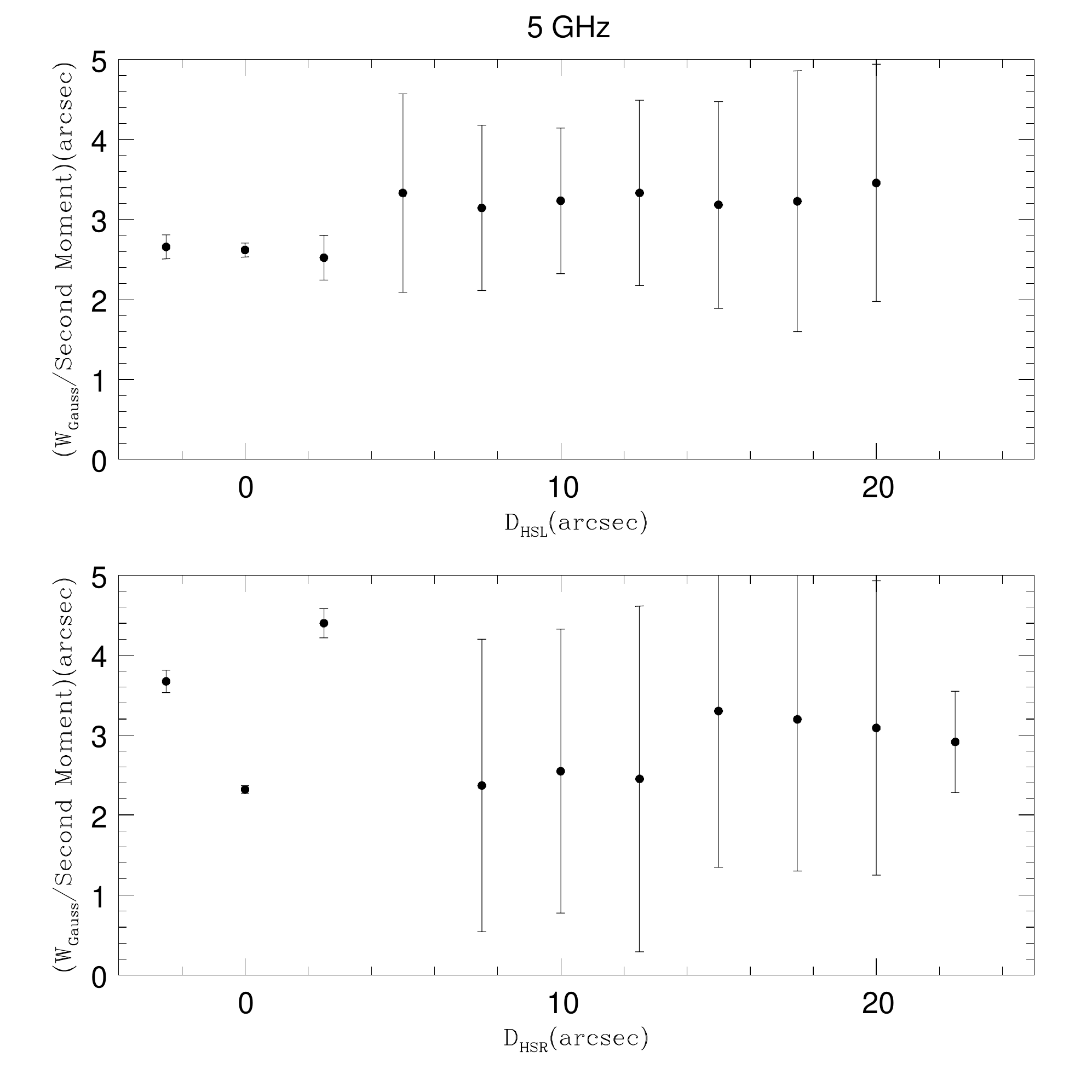}
\caption{3C172 ratio of the Gaussian FWHM to the second moment as a
function of distance from the hot spot at 1.4 and 5 GHz 
(left and right panels, respectively) for the left and right
hand sides of the source (top and bottom panels, respectively). The value of this ratio is fairly constant for each side of the source, and has a value of about 2.5. Similar results are obtained at 1.4 and 5 GHz.} 
\end{figure}

\begin{figure}
\plottwo{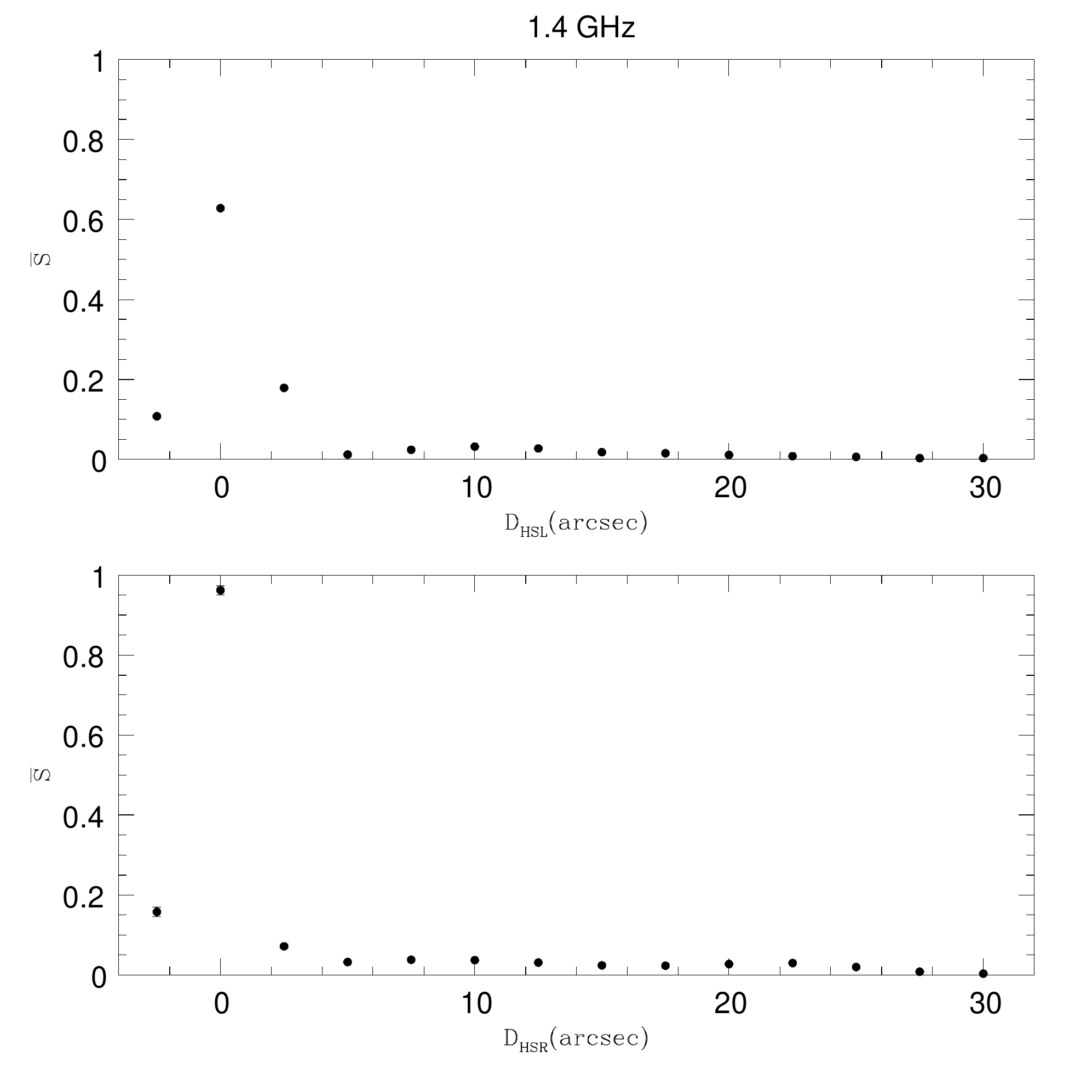}{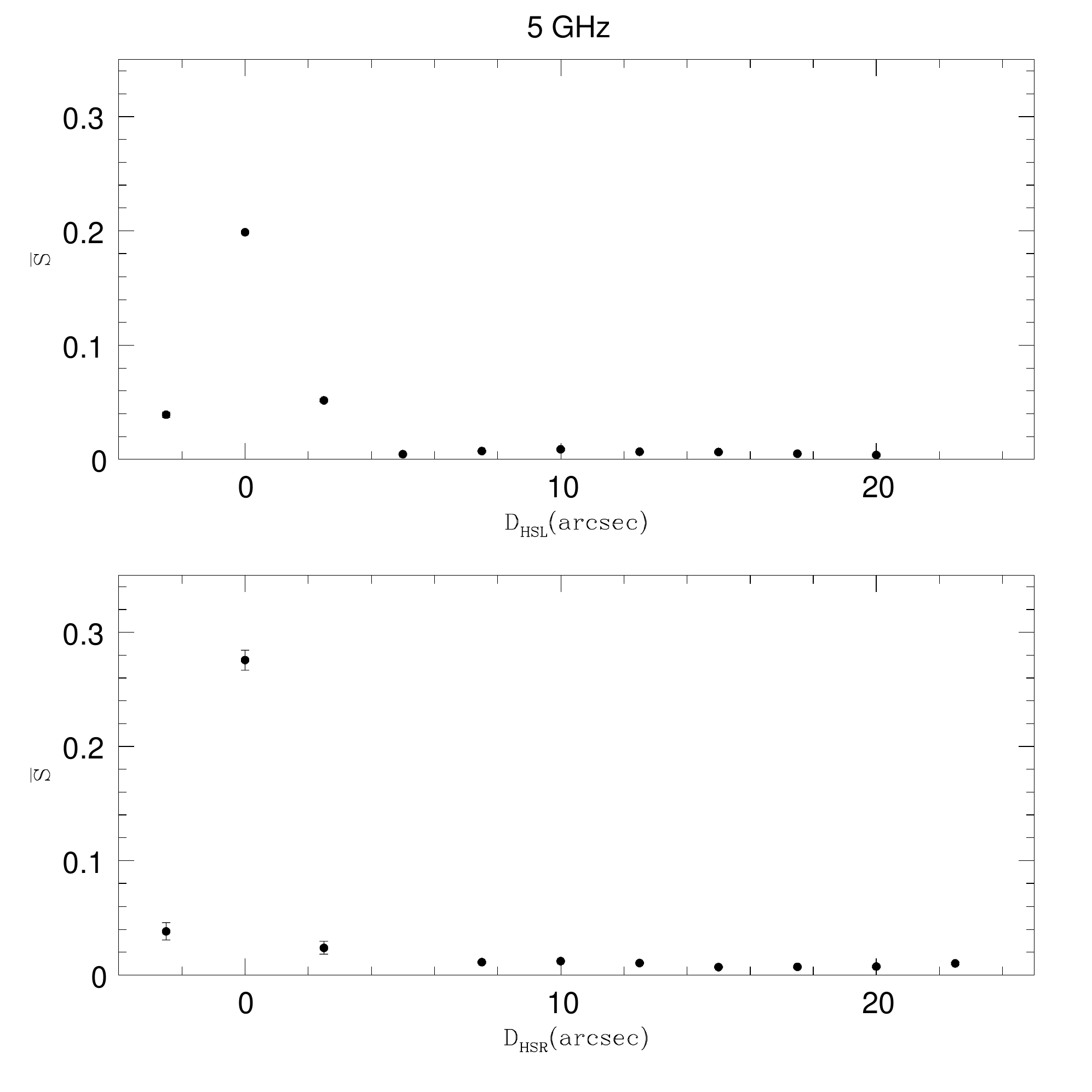}
\caption{3C172 average surface brightness in units of Jy/beam 
as a function of distance from the 
hot spot at 1.4 and 5 GHz 
(left and right panels, respectively) for the left and right
hand sides of the source (top and bottom panels, respectively).}
\end{figure}

\begin{figure}
\plottwo{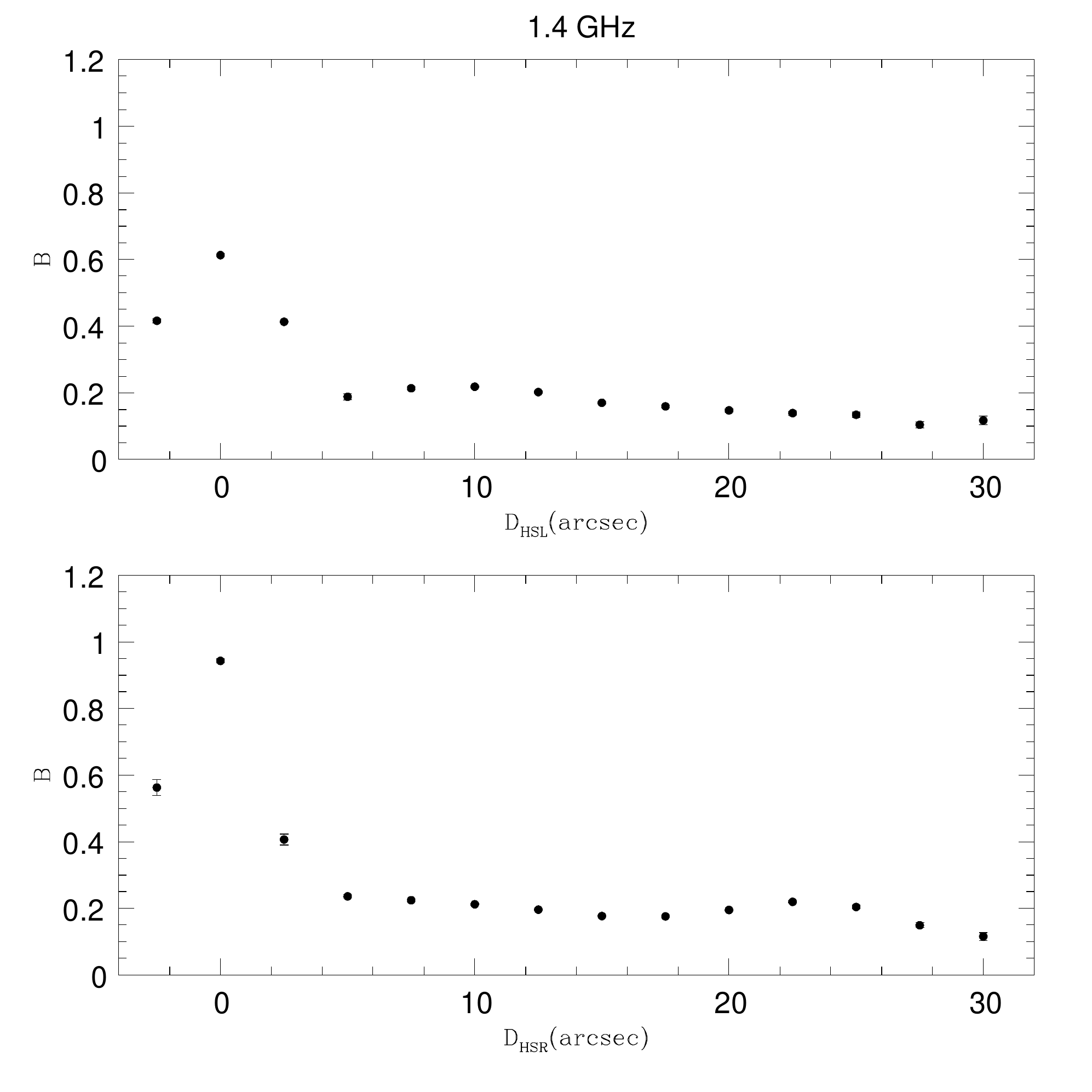}{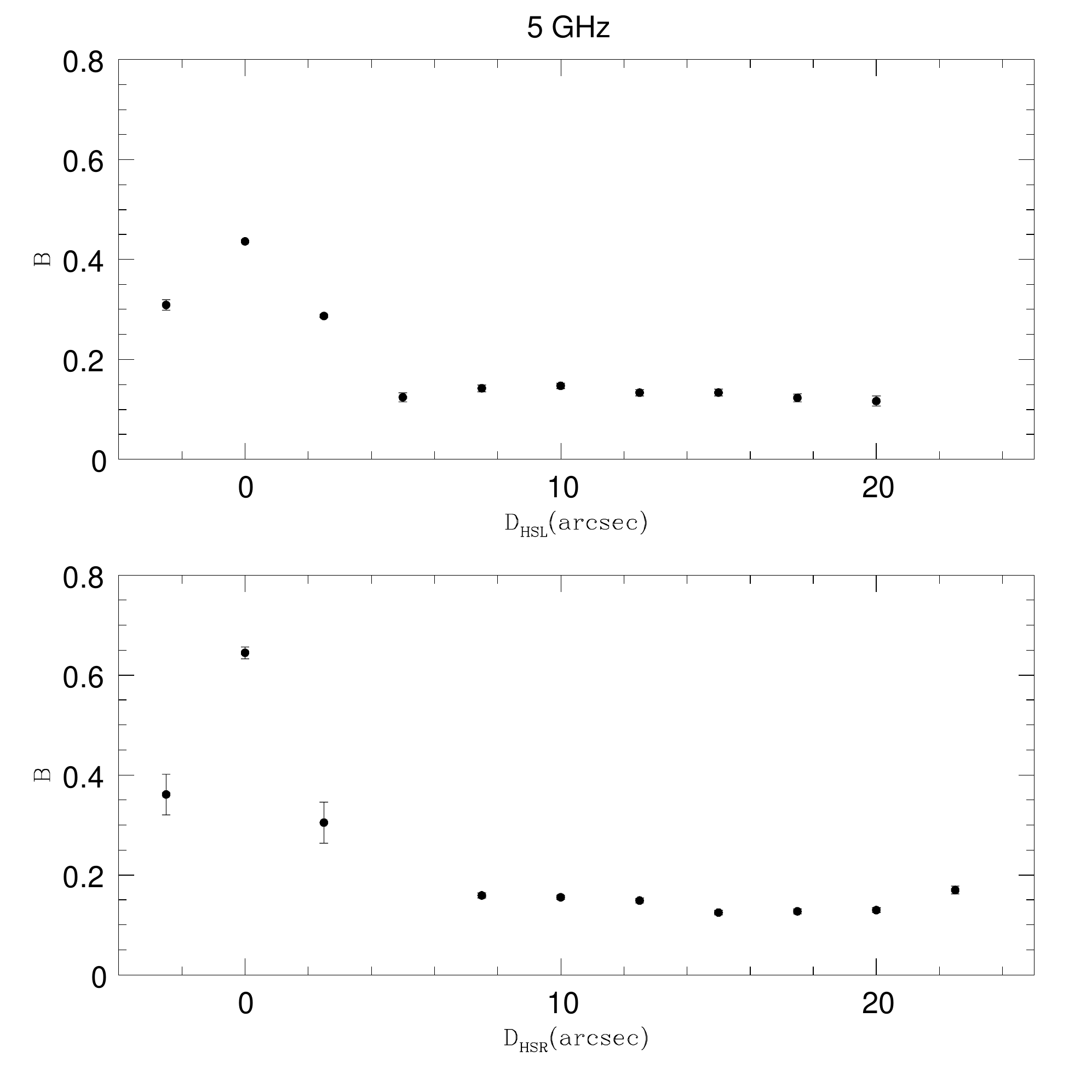}
\caption{3C172 minimum energy magnetic field strength 
as a function of distance from the 
hot spot at 1.4 and 5 GHz 
(left and right panels, respectively) for the left and right
hand sides of the source (top and bottom panels, respectively).
The normalization is given in Table 1. 
}
\end{figure}

\begin{figure}
\plottwo{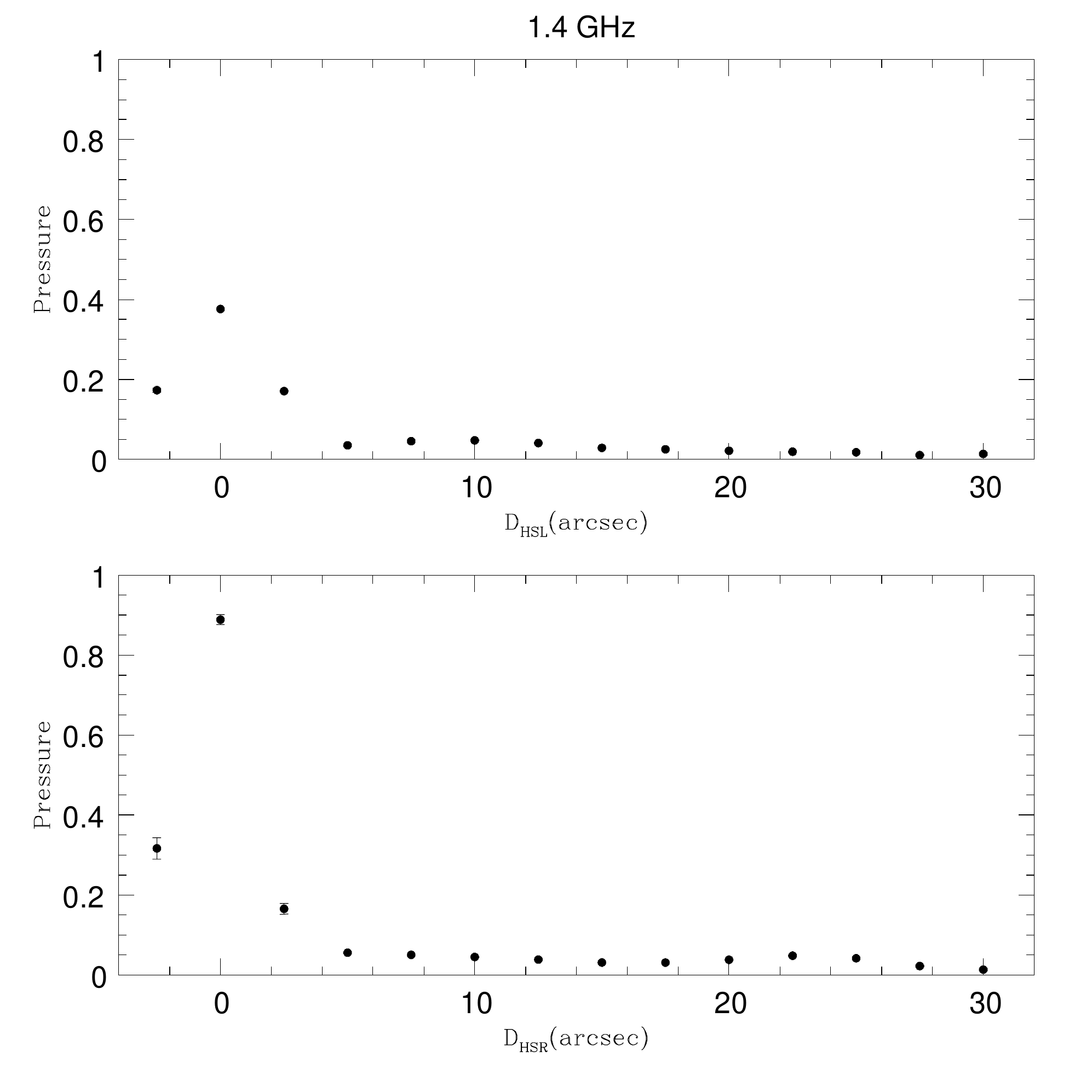}{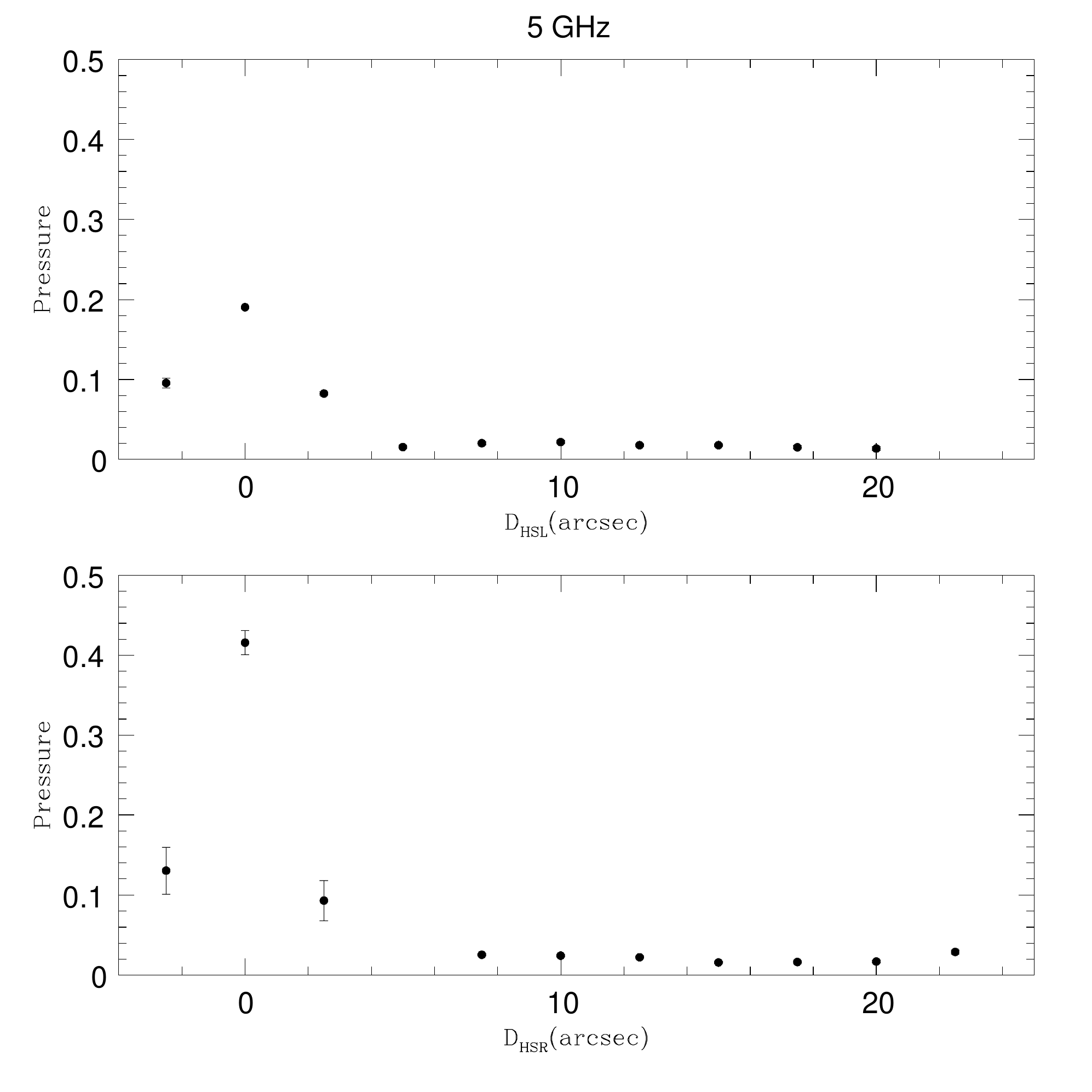}
\caption{3C172 
minimum energy pressure as a function of 
distance from the hot spot at 1.4 and 5 GHz 
(left and right panels, respectively) for the left and right
hand sides of the source (top and bottom panels, respectively), 
as in Fig. \ref{3C6.1P}. The normalization is given in Table 1.
}
\end{figure}

\clearpage
\begin{figure}
\plotone{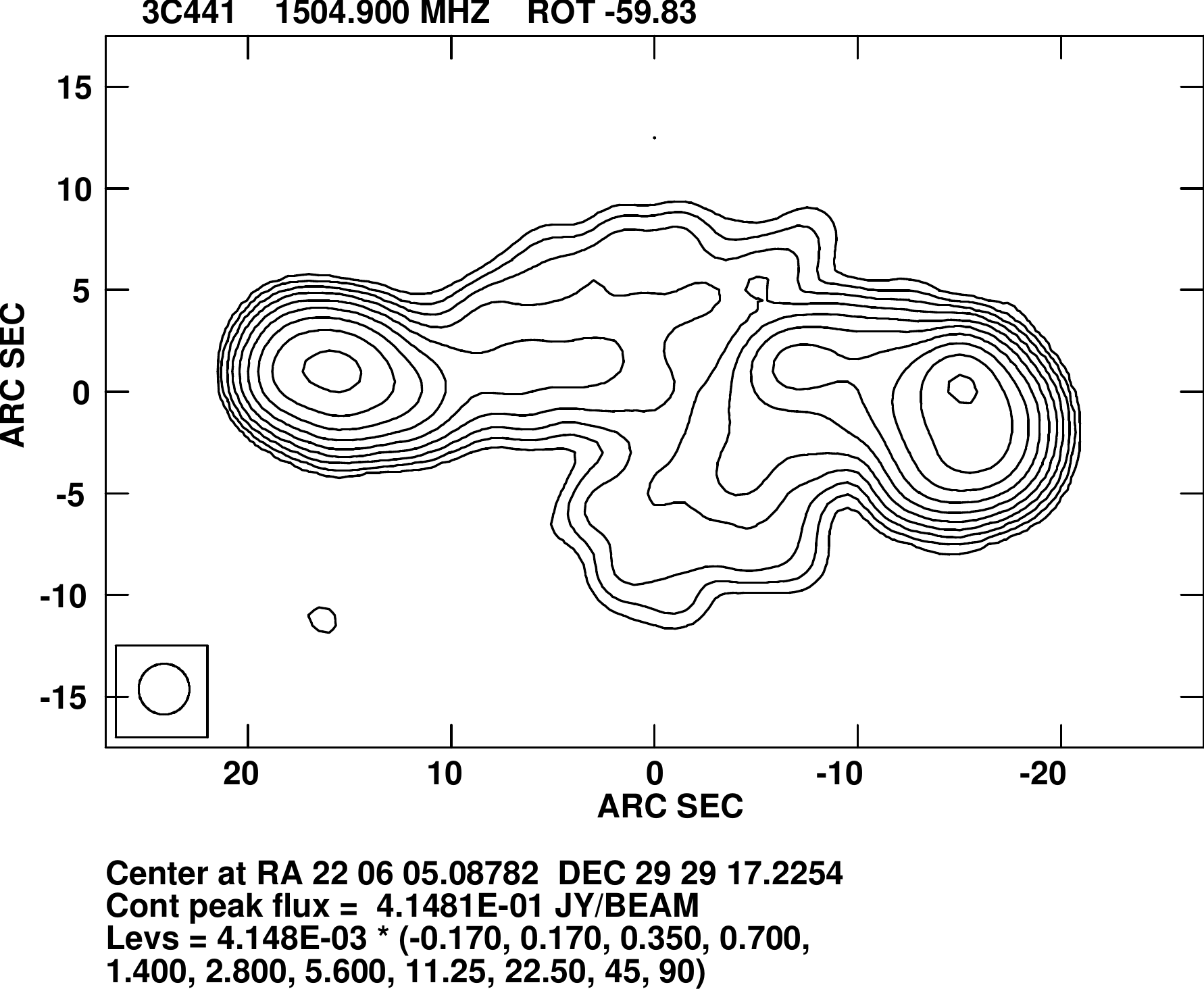}
\caption{3C441 at 1.4 GHz and 2.5'' resolution rotated by 59.83 deg 
clockwise, as in Fig. \ref{3C6.1rotfig}.}
\end{figure}

\begin{figure}
\plottwo{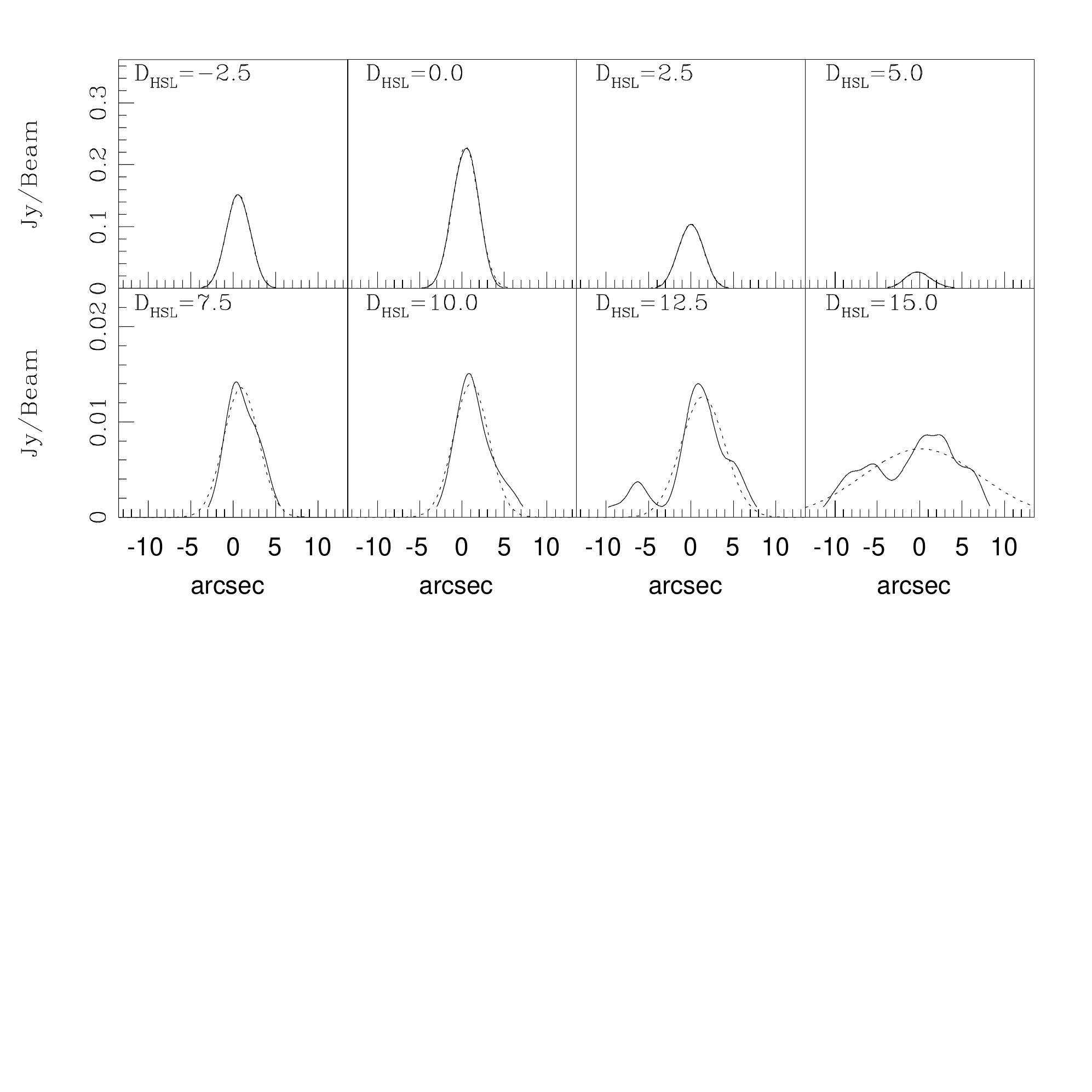}{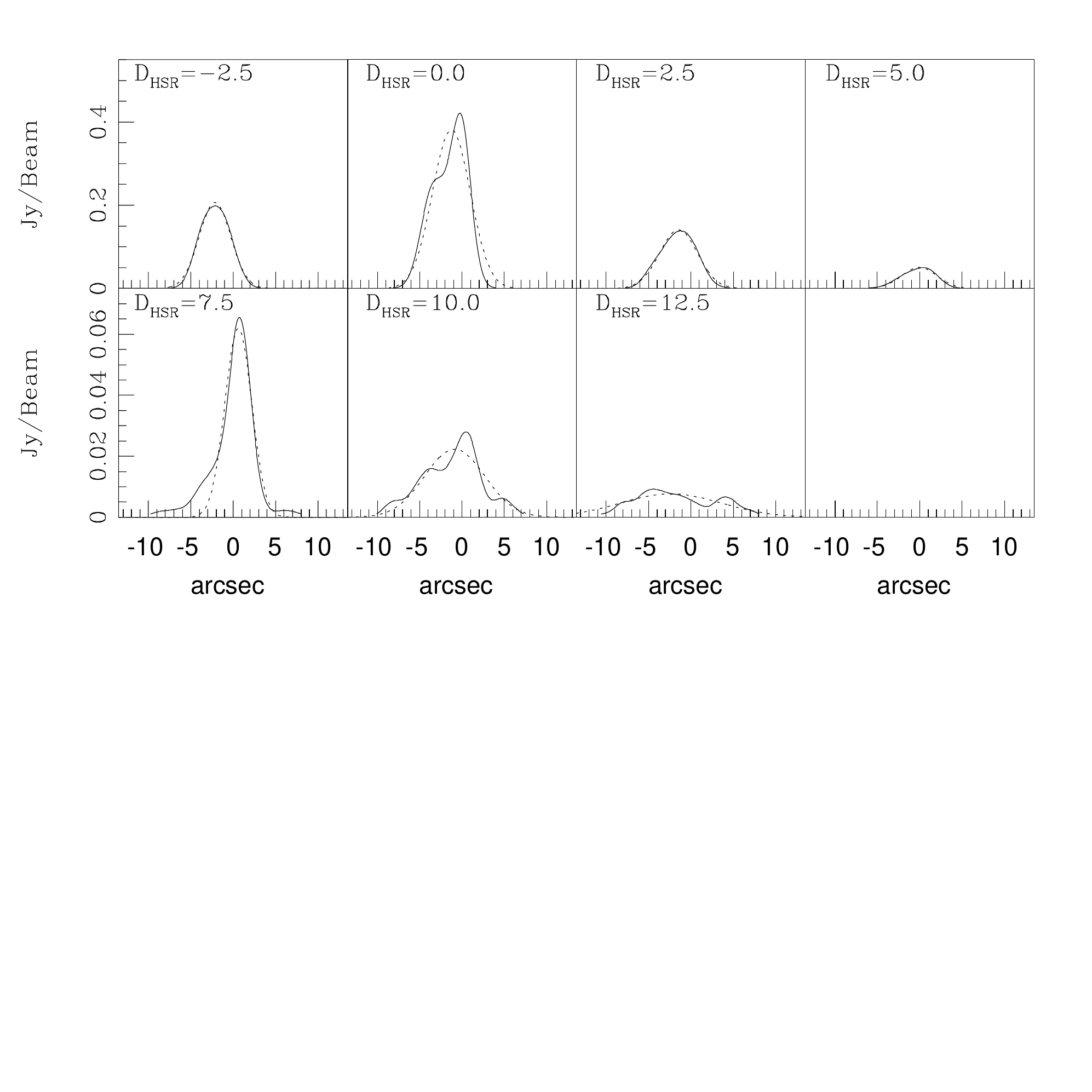}
\caption{3C441 
surface brightness (solid line) and best fit 
Gaussian (dotted line) for each cross-sectional slice of the 
radio bridge at 1.4 GHz, as in Fig. \ref{3C6.1SG}.
A Gaussian provides an excellent description of the surface brightness
profile of each cross-sectional slice. }
\label{3C441SG}
\end{figure}

\begin{figure}
\plottwo{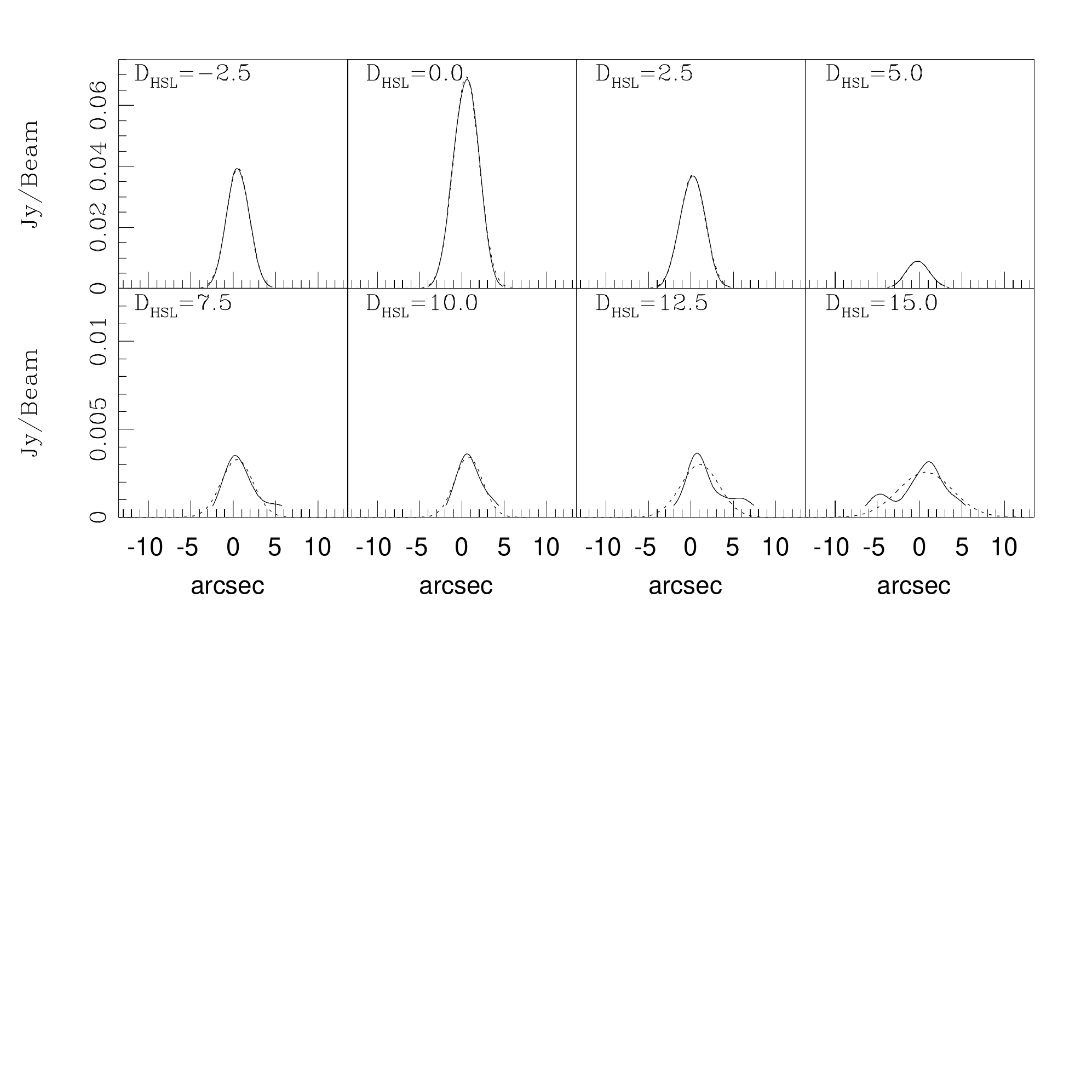}{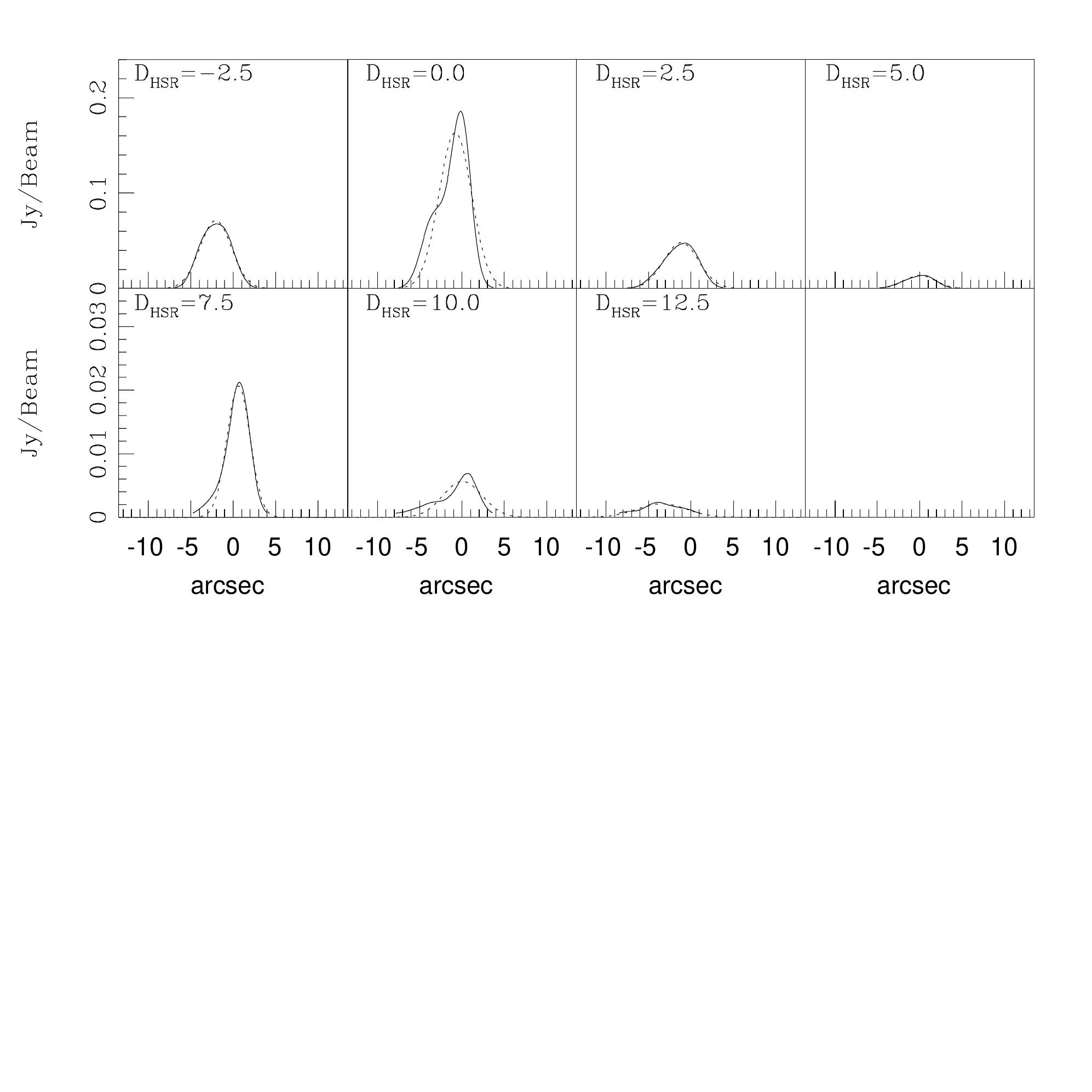}
\caption{3C441 
surface brightness (solid line) and best fit 
Gaussian (dotted line) for each cross-sectional slice of the 
radio bridge at 5 GHz, as in 
Fig. \ref{3C441SG} but at 5 GHz. 
Results obtained at 5 GHz are nearly identical to those
obtained at 1.4 GHz.}
\end{figure}

\begin{figure}
\plottwo{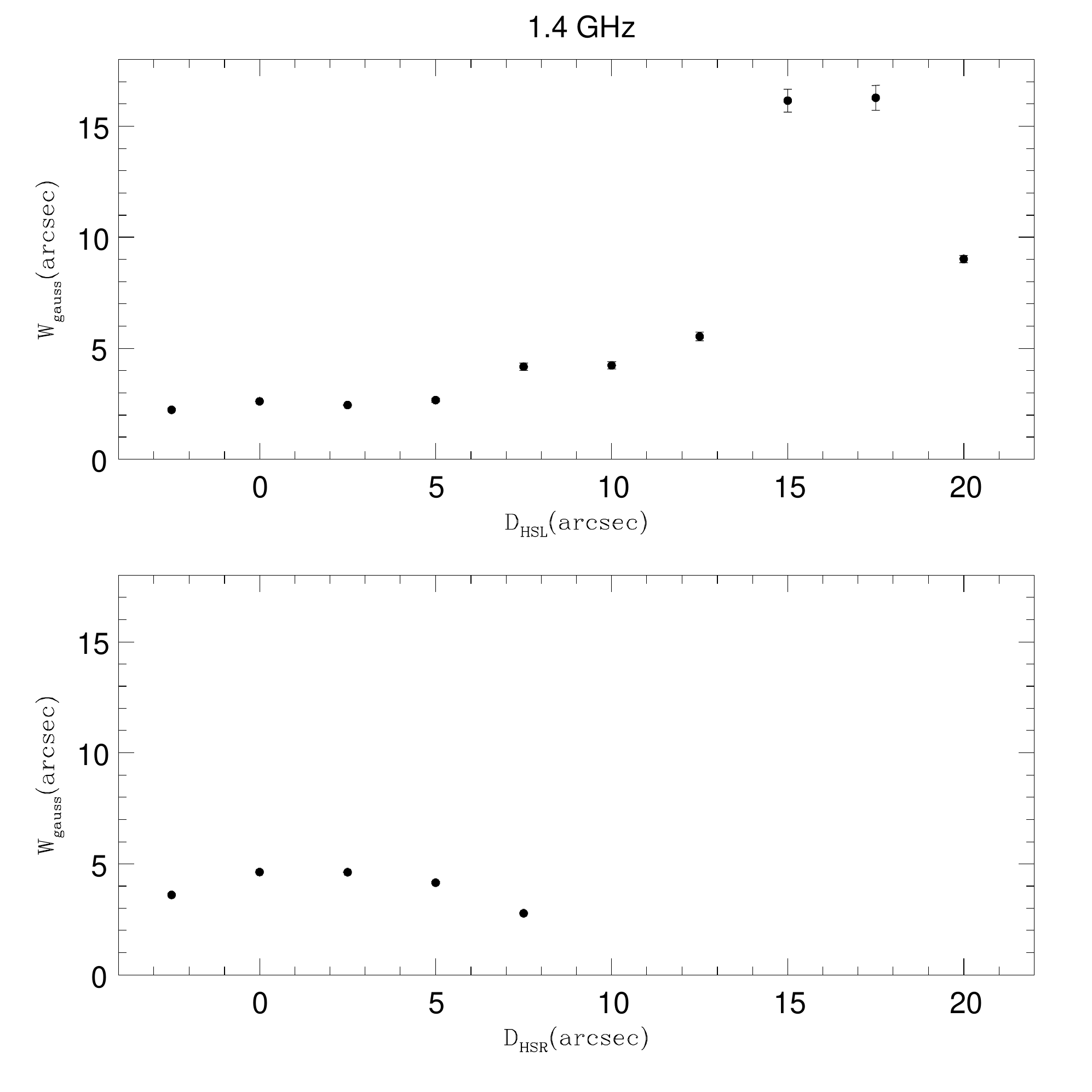}{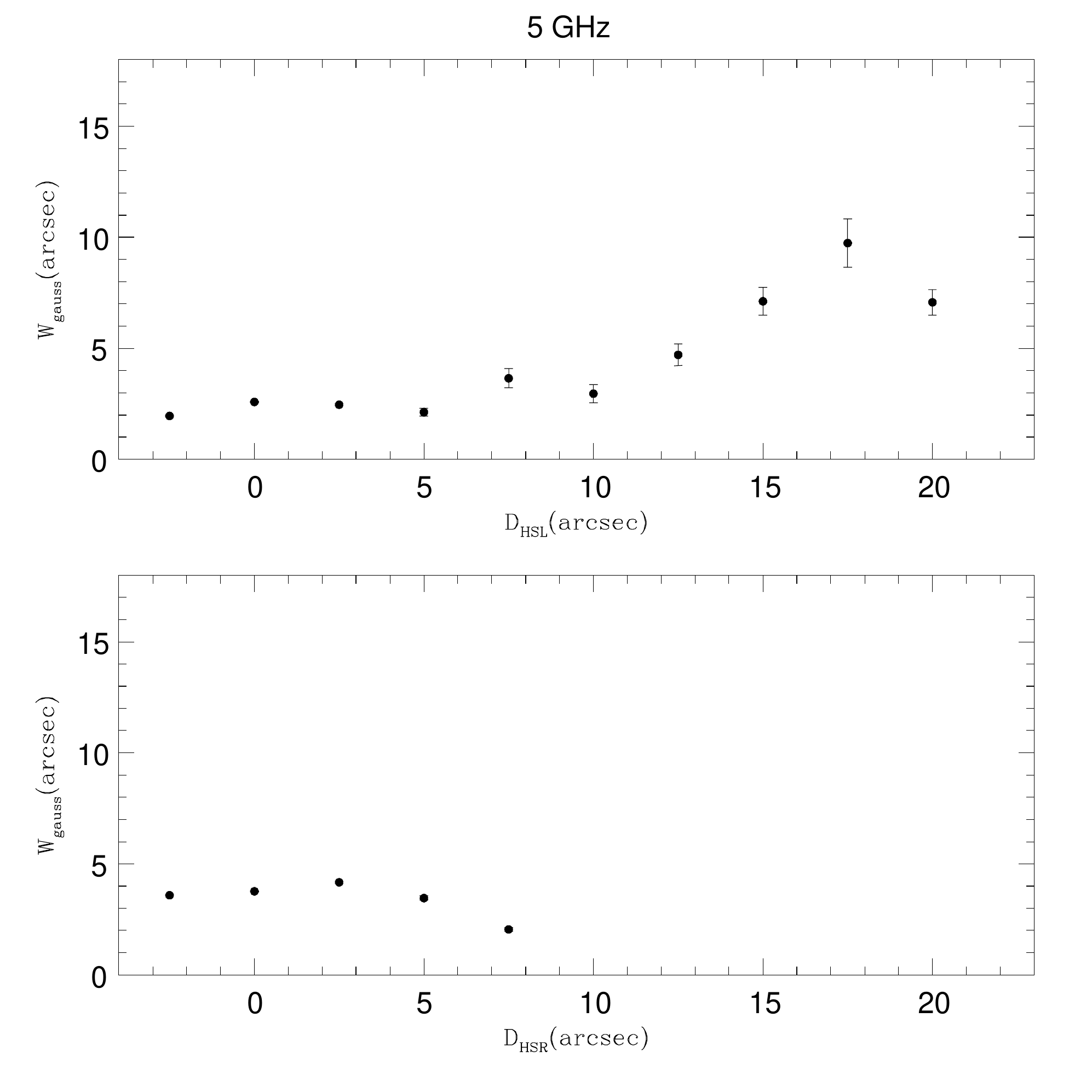}
\caption{3C441  Gaussian FWHM as a function of 
distance from the hot spot  
at 1.4 and 5 GHz (left and
right panels, respectively) for the left and right hand sides of the 
source (top and bottom panels, respectively), as in Fig. \ref{3C6.1WG}.
The right and left hand sides of the source are somewhat
symmetric.  Similar results are obtained at 
1.4 and 5 GHz. }
\end{figure}

\begin{figure}
\plottwo{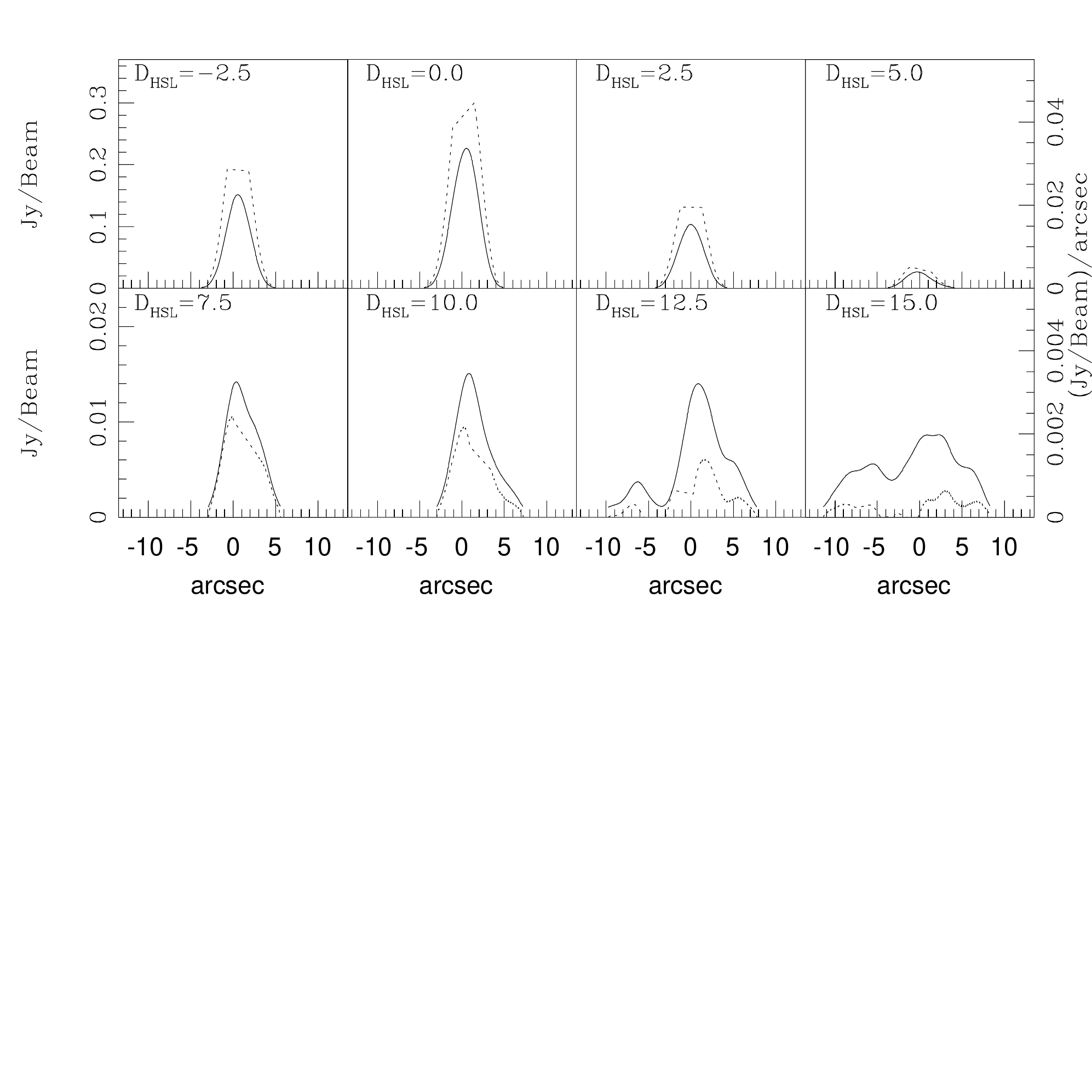}{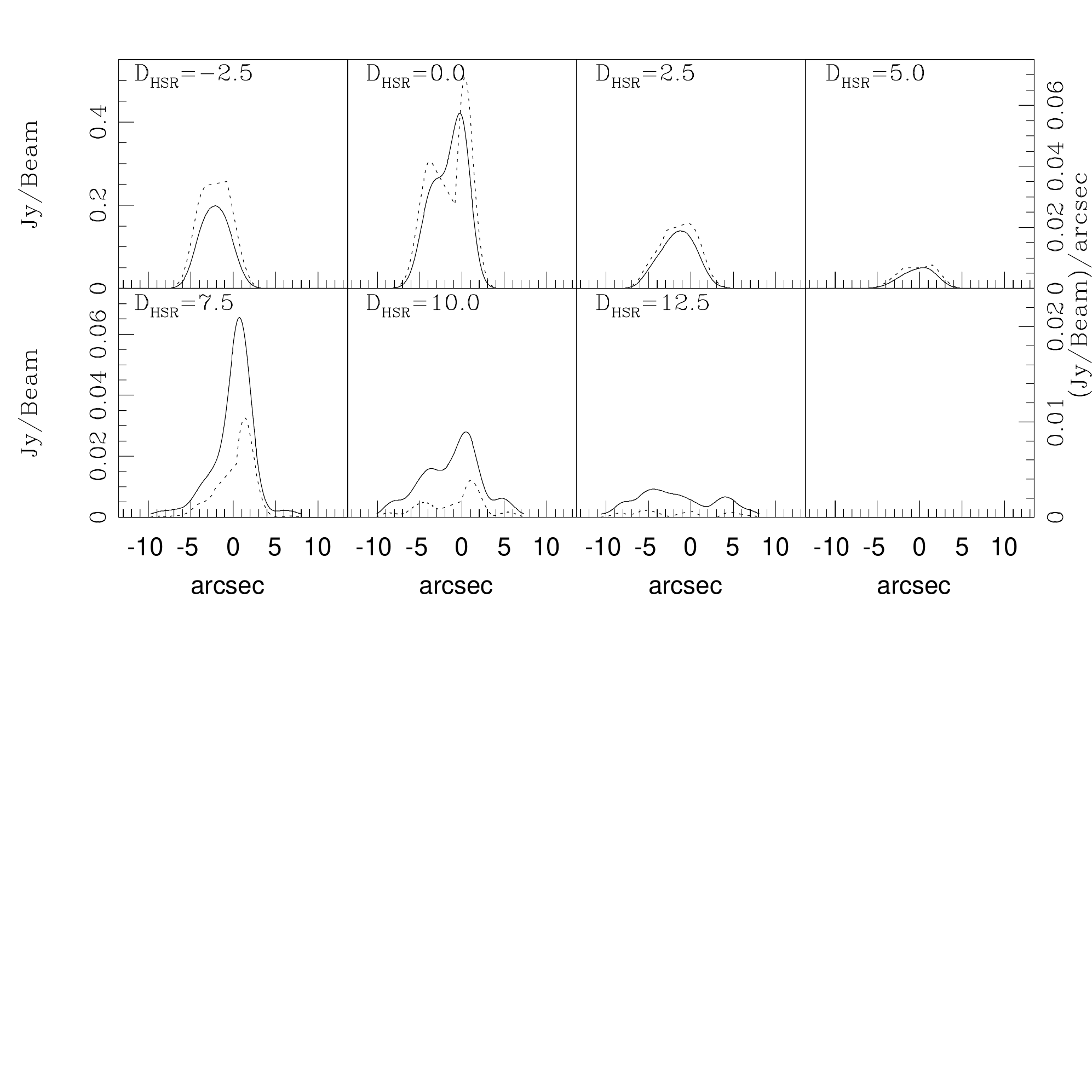}
\caption{3C441 emissivity (dotted line) and surface brightness
(solid line) for each cross-sectional slice of the radio 
bridge at 1.4 GHz, as in Fig. \ref{3C6.1SEM}. A constant volume
emissivity per slice provides a reasonable description of the 
data over most of the source. }
\label{3C441SEM}
\end{figure}

\begin{figure}
\plottwo{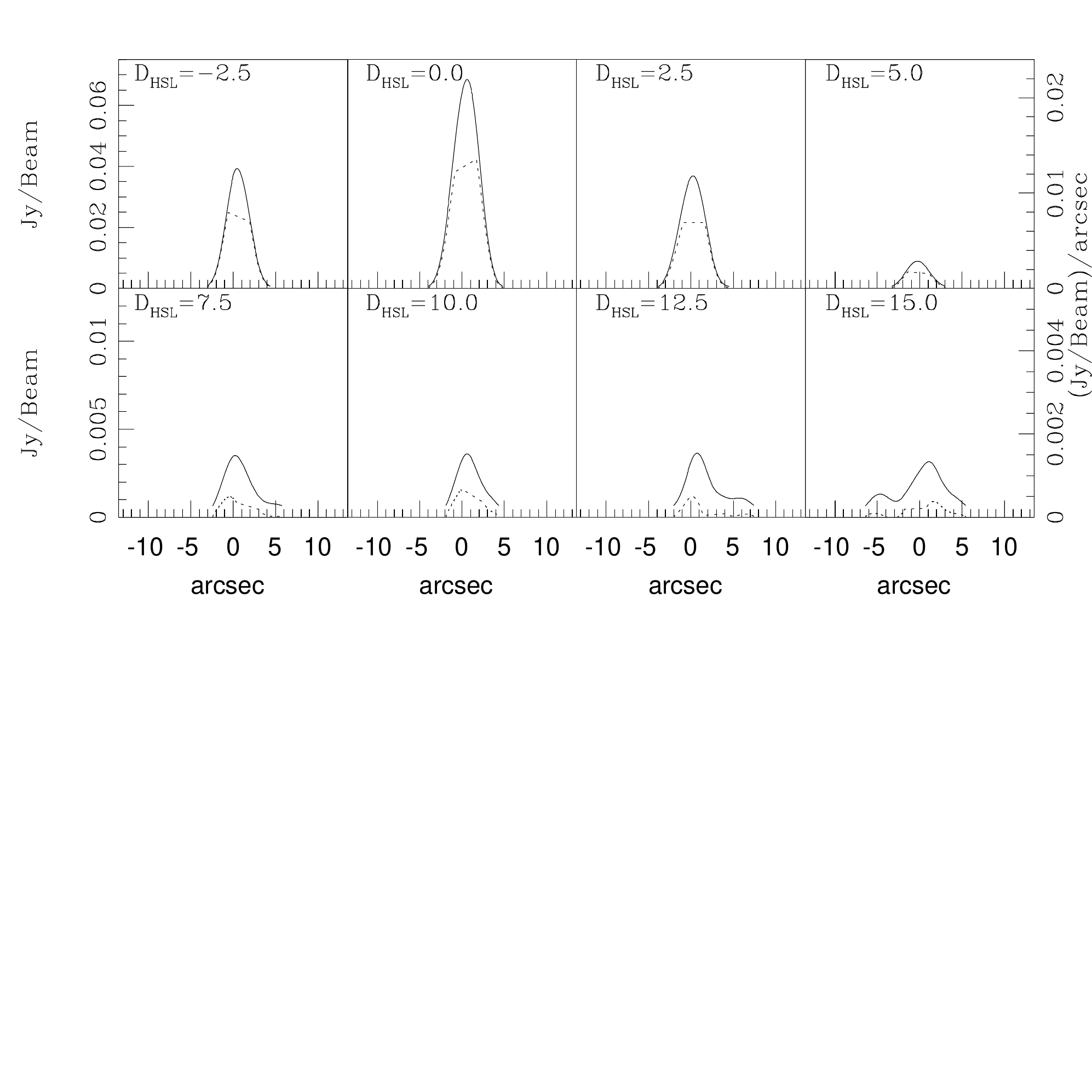}{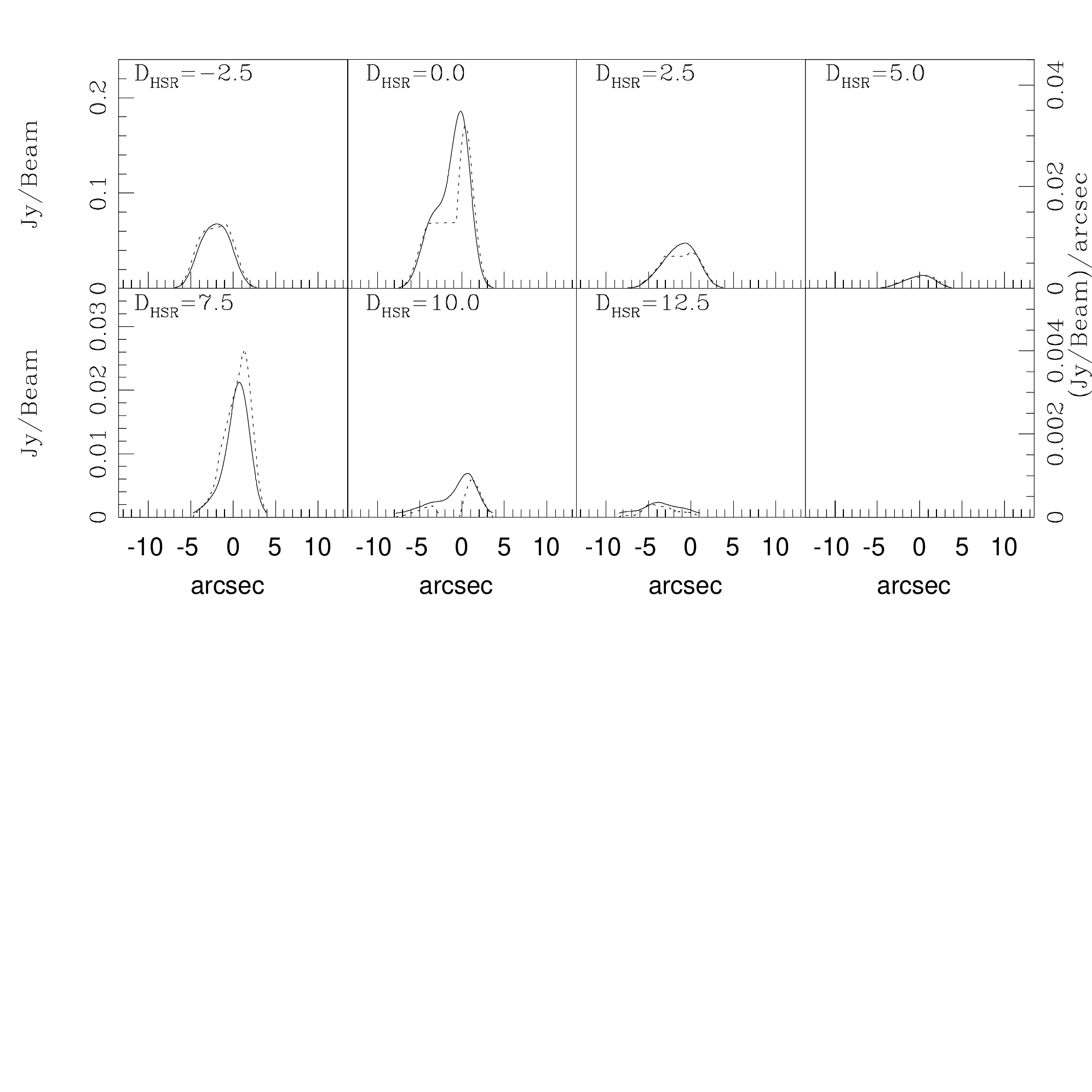}
\caption{3C441
emissivity (dotted line) and surface brightness
(solid line) for each cross-sectional slice of the radio 
bridge at 5 GHz, as in Fig. \ref{3C441SEM} but at 5 GHz.
Results obtained at 5 GHz are nearly identical to those
obtained at 1.4 GHz.}
\end{figure}

\begin{figure}
\plottwo{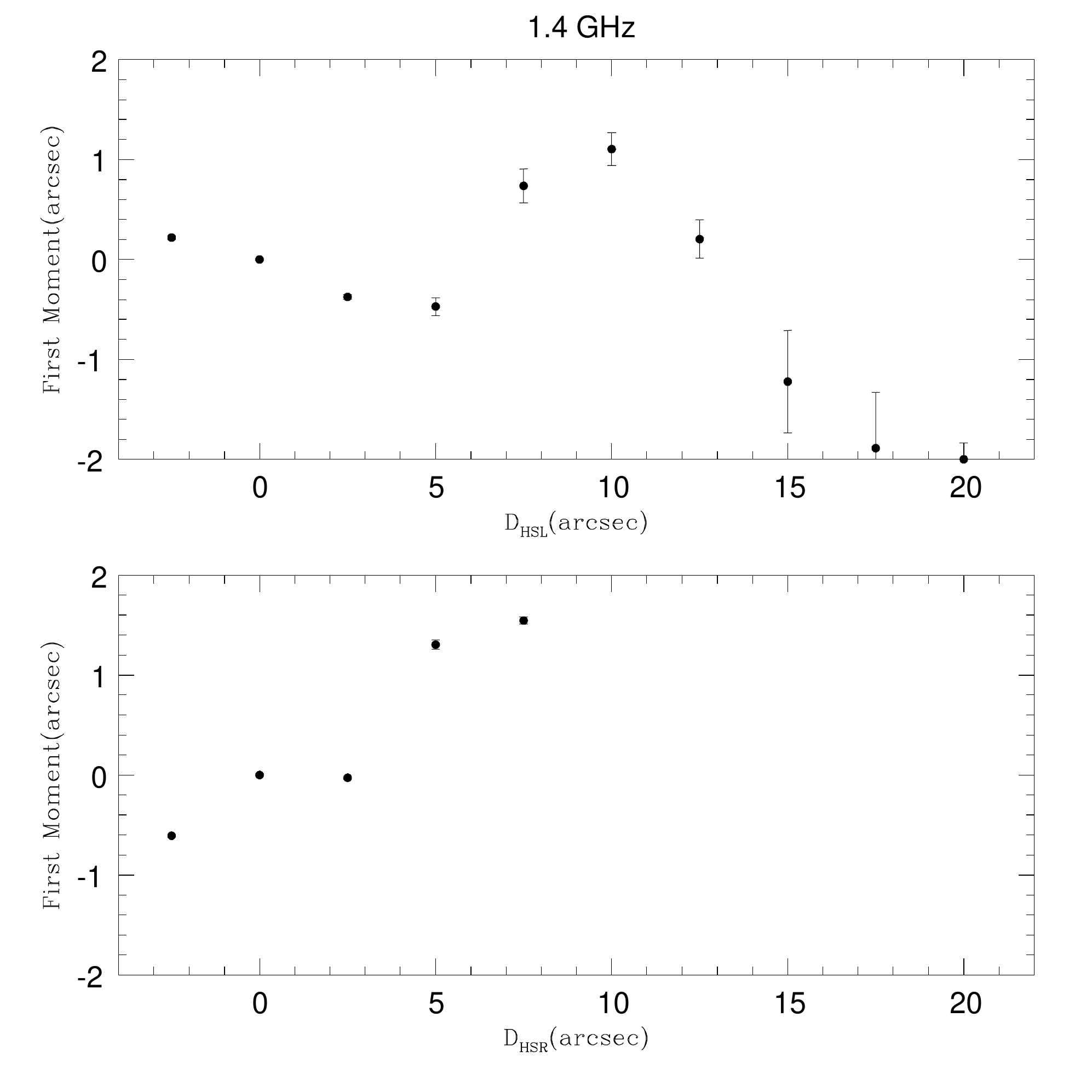}{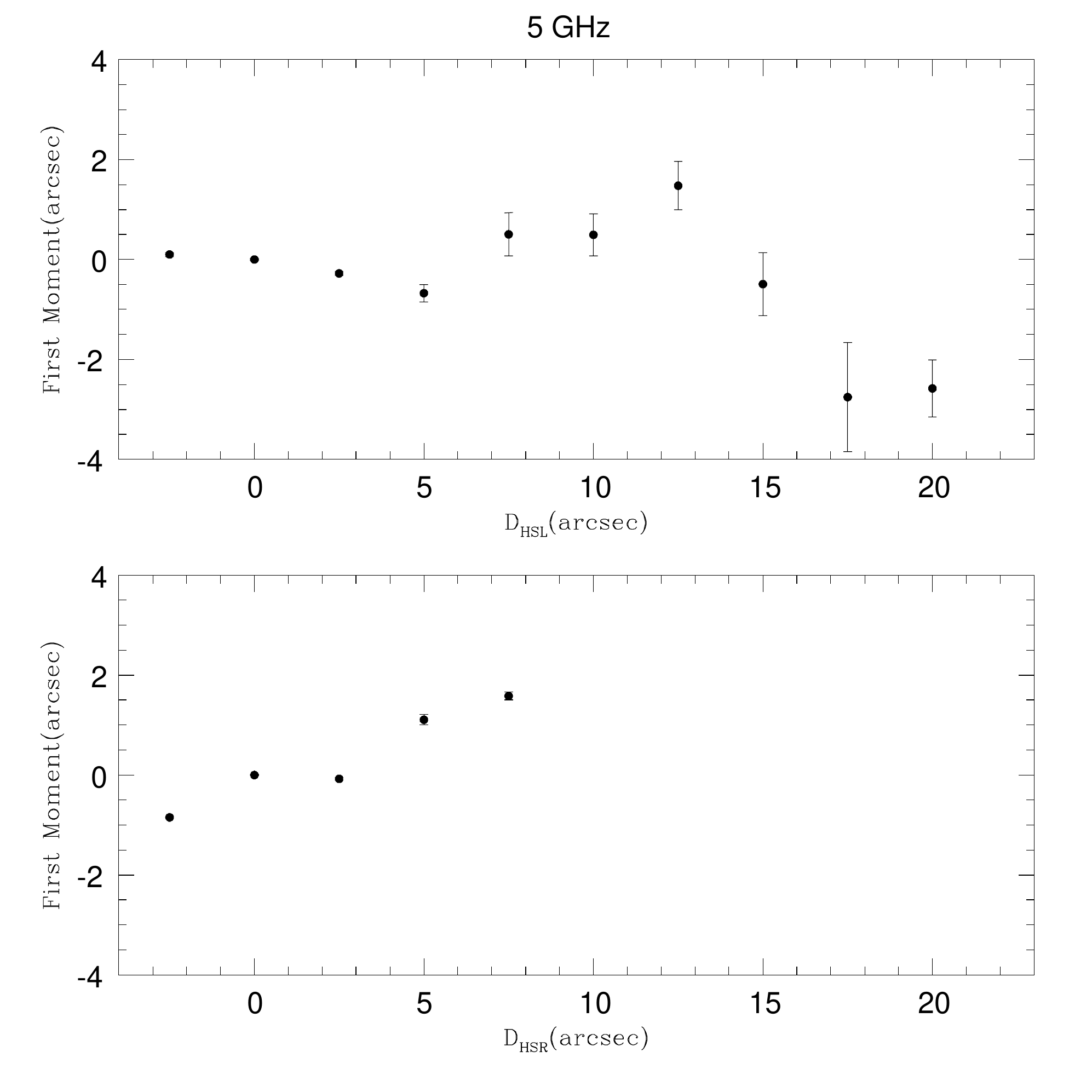}
\caption{3C441 first moment as a function of distance from 
the hot spot at 1.4 GHz and 5 GHz (left and right panels,
respectively) for the left and right hand sides of the source
(top and bottom, respectively).}
\end{figure}

\begin{figure}
\plottwo{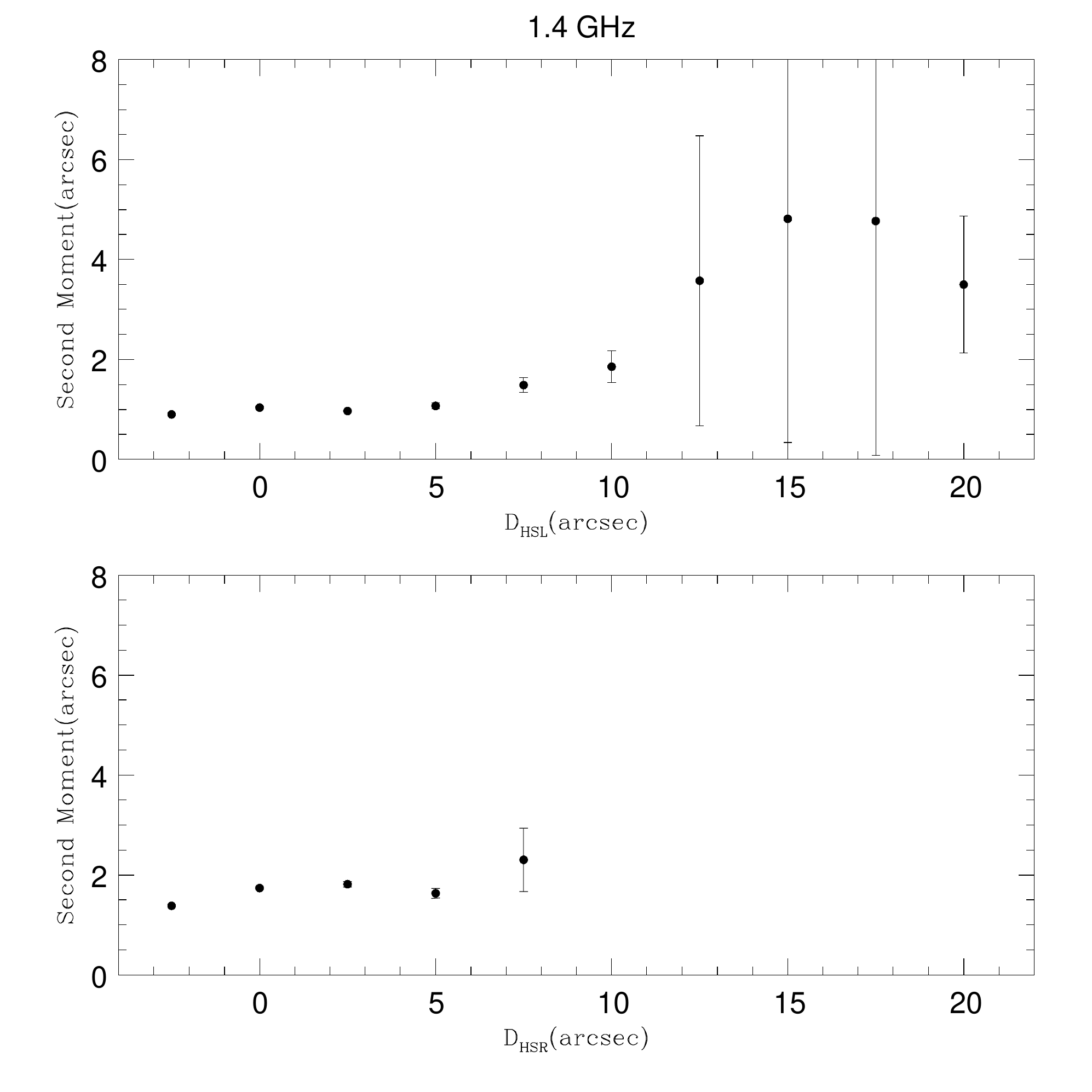}{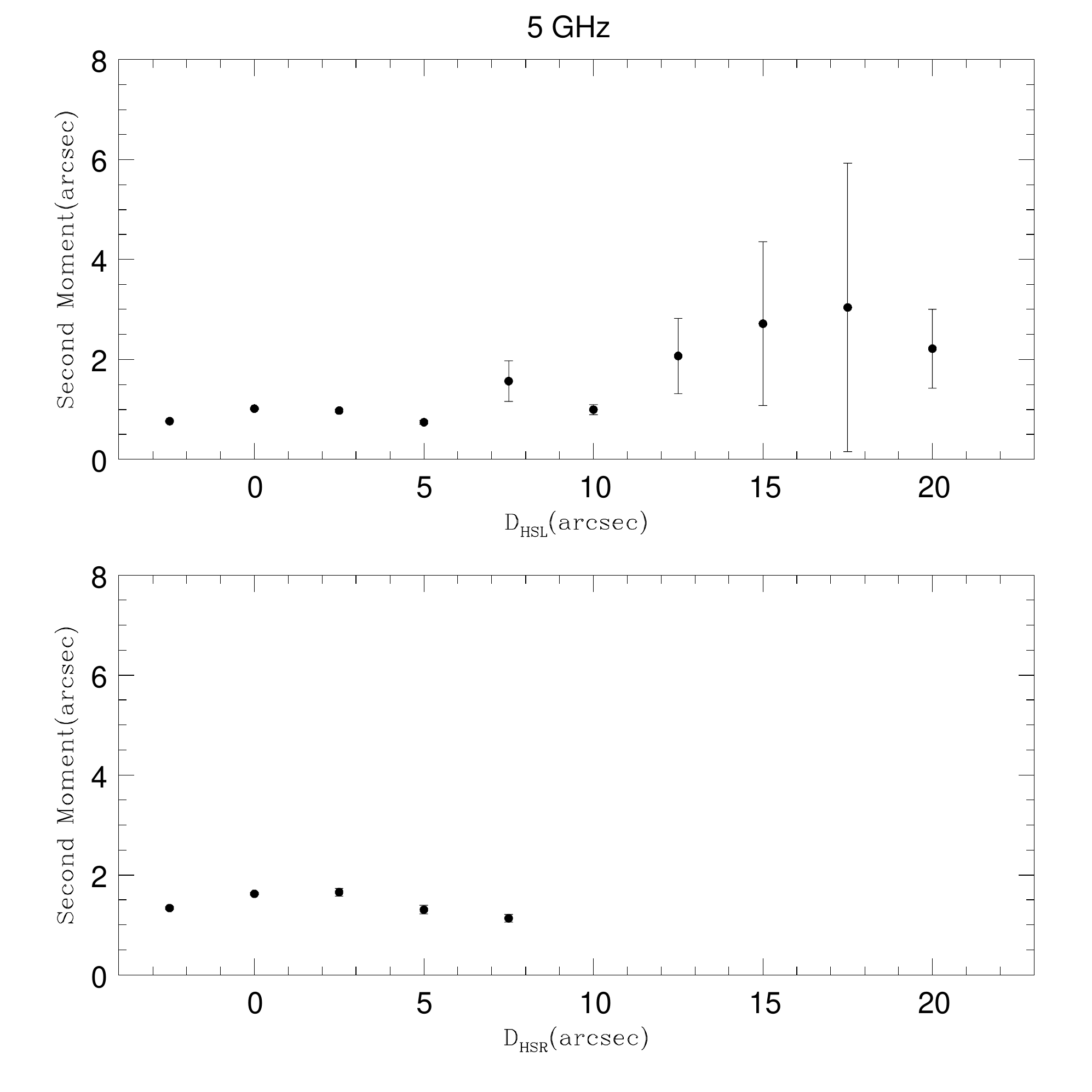}
\caption{3C441 second moment as a function of distance from the
hot spot at 1.4 GHz and 5 GHz (left and right panels, 
respectively) for the left and right sides of the source
(top and bottom panels, respectively).}
\end{figure}

\begin{figure}
\plottwo{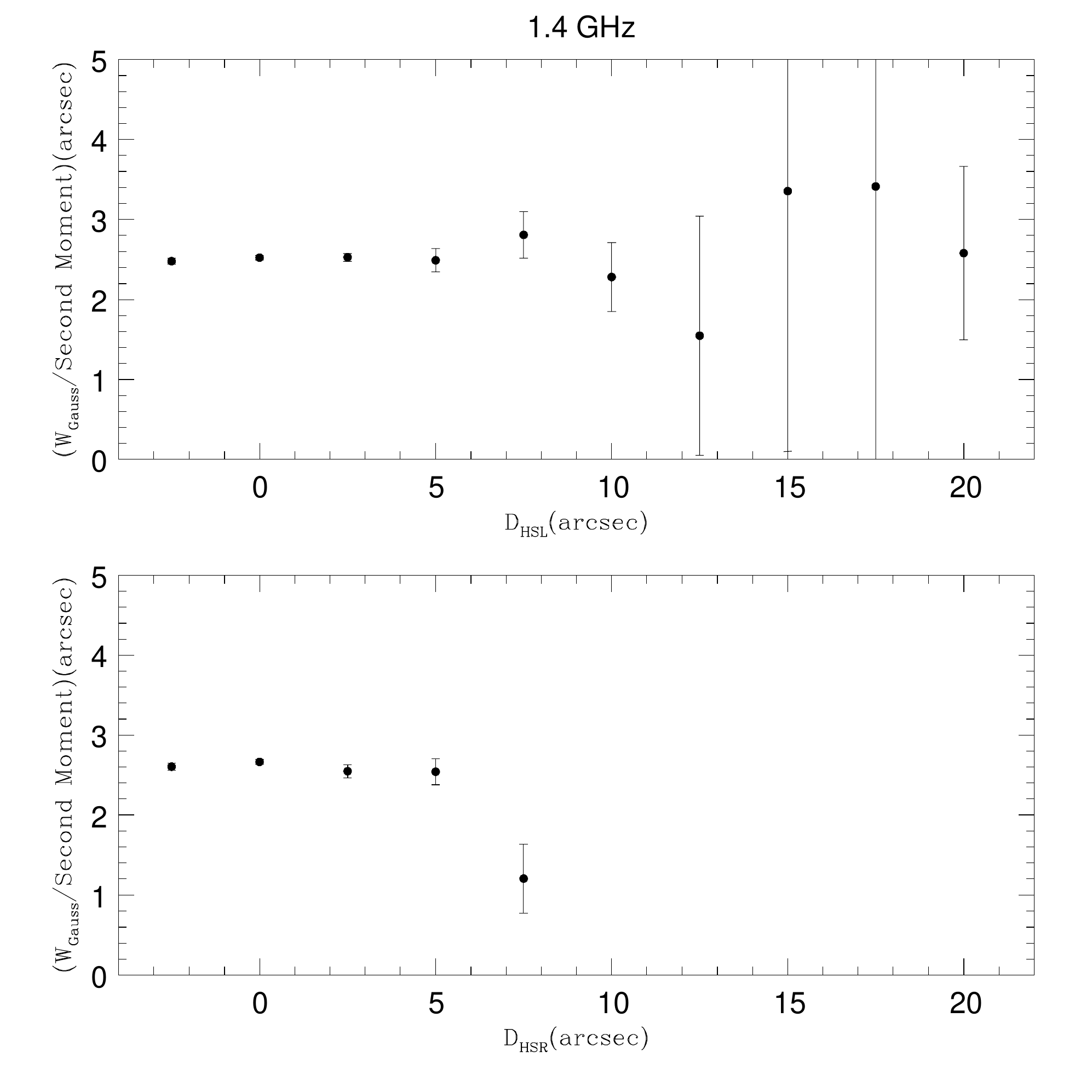}{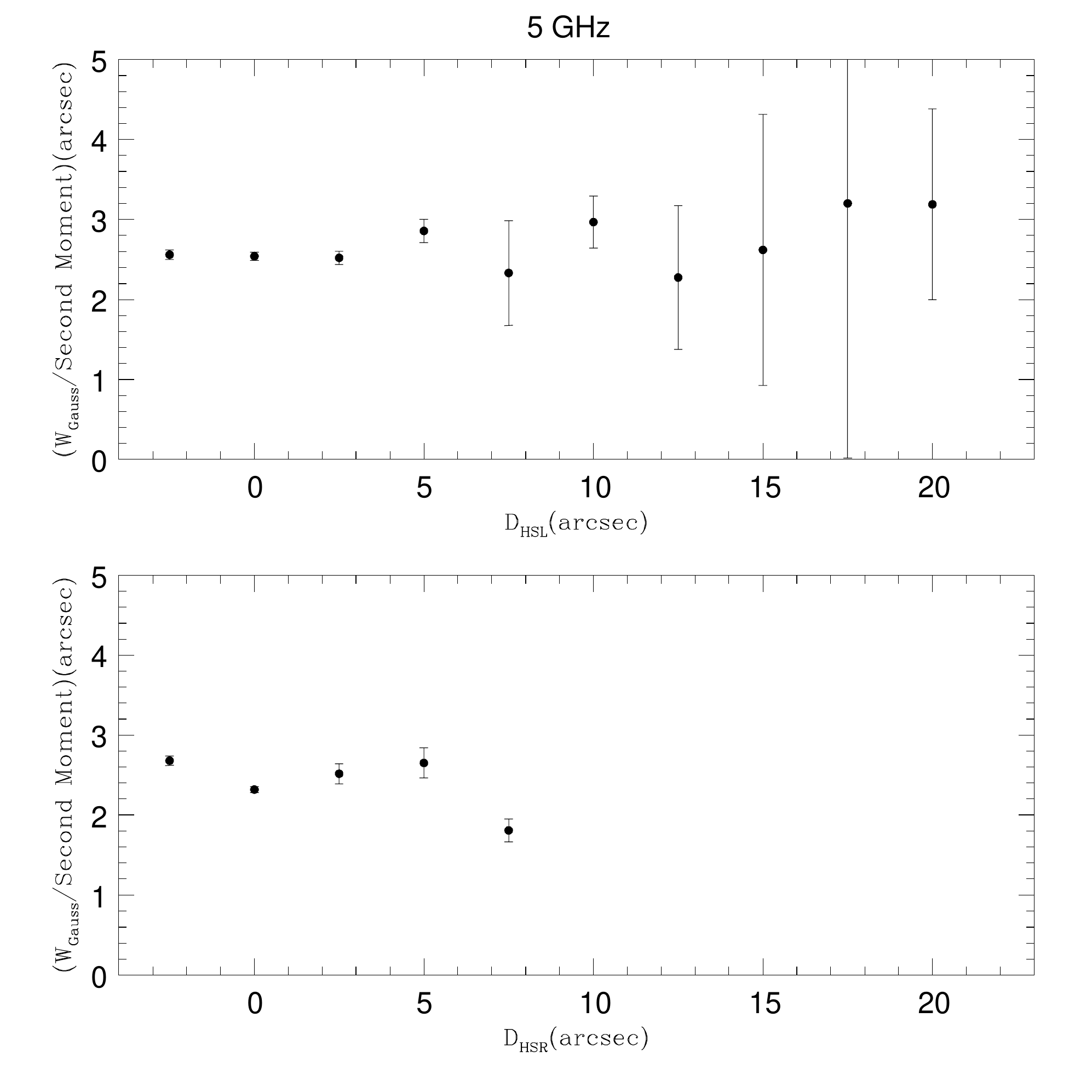}
\caption{3C441 ratio of the Gaussian FWHM to the second moment as a
function of distance from the hot spot at 1.4 and 5 GHz 
(left and right panels, respectively) for the left and right
hand sides of the source (top and bottom panels, respectively). The value of this ratio is fairly constant for each side of the source, and has a value of about 2.5. Similar results are obtained at 1.4 and 5 GHz.} 
\end{figure}

\begin{figure}
\plottwo{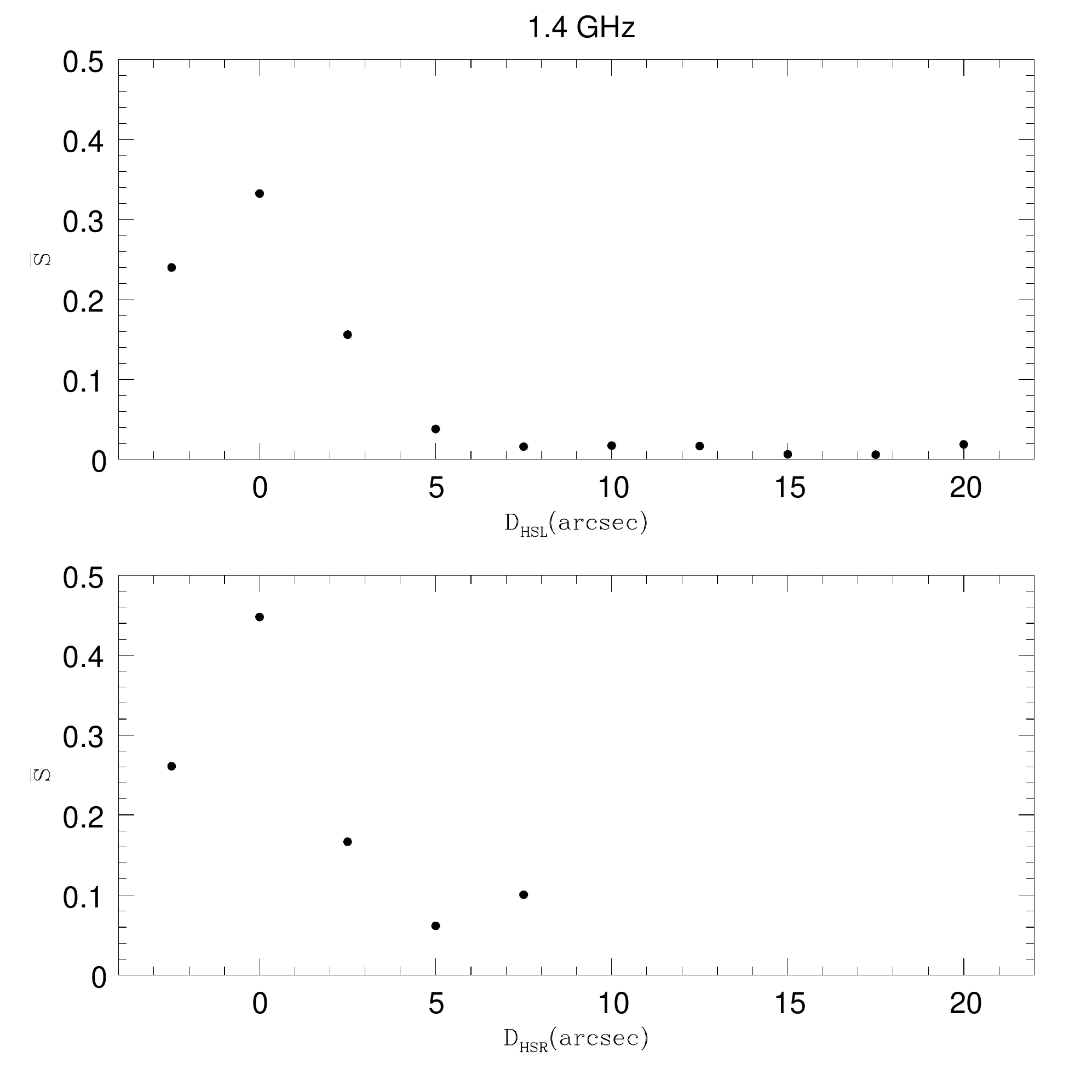}{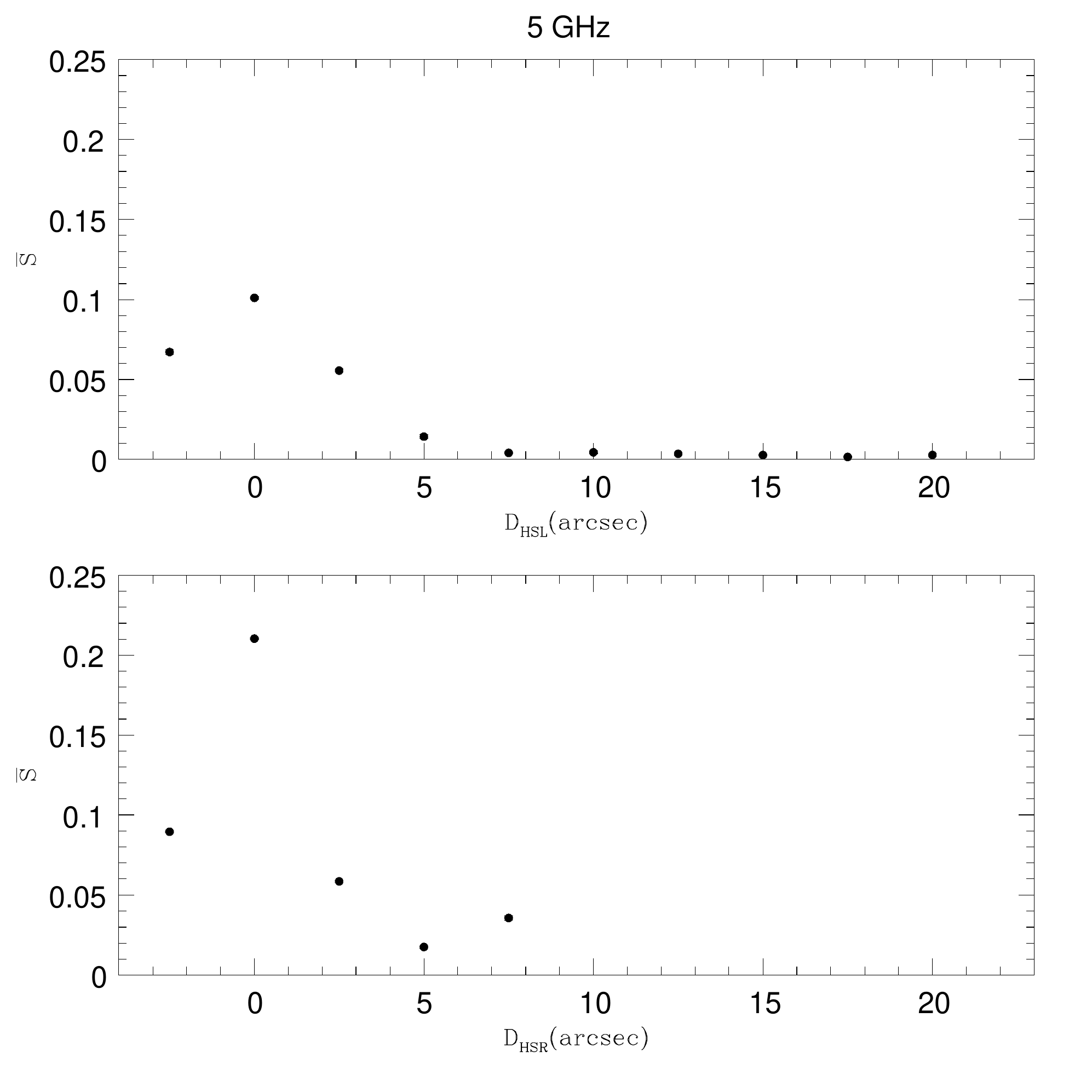}
\caption{3C441 average surface brightness in units of Jy/beam 
as a function of distance from the 
hot spot at 1.4 and 5 GHz 
(left and right panels, respectively) for the left and right
hand sides of the source (top and bottom panels, respectively).}
\end{figure}

\begin{figure}
\plottwo{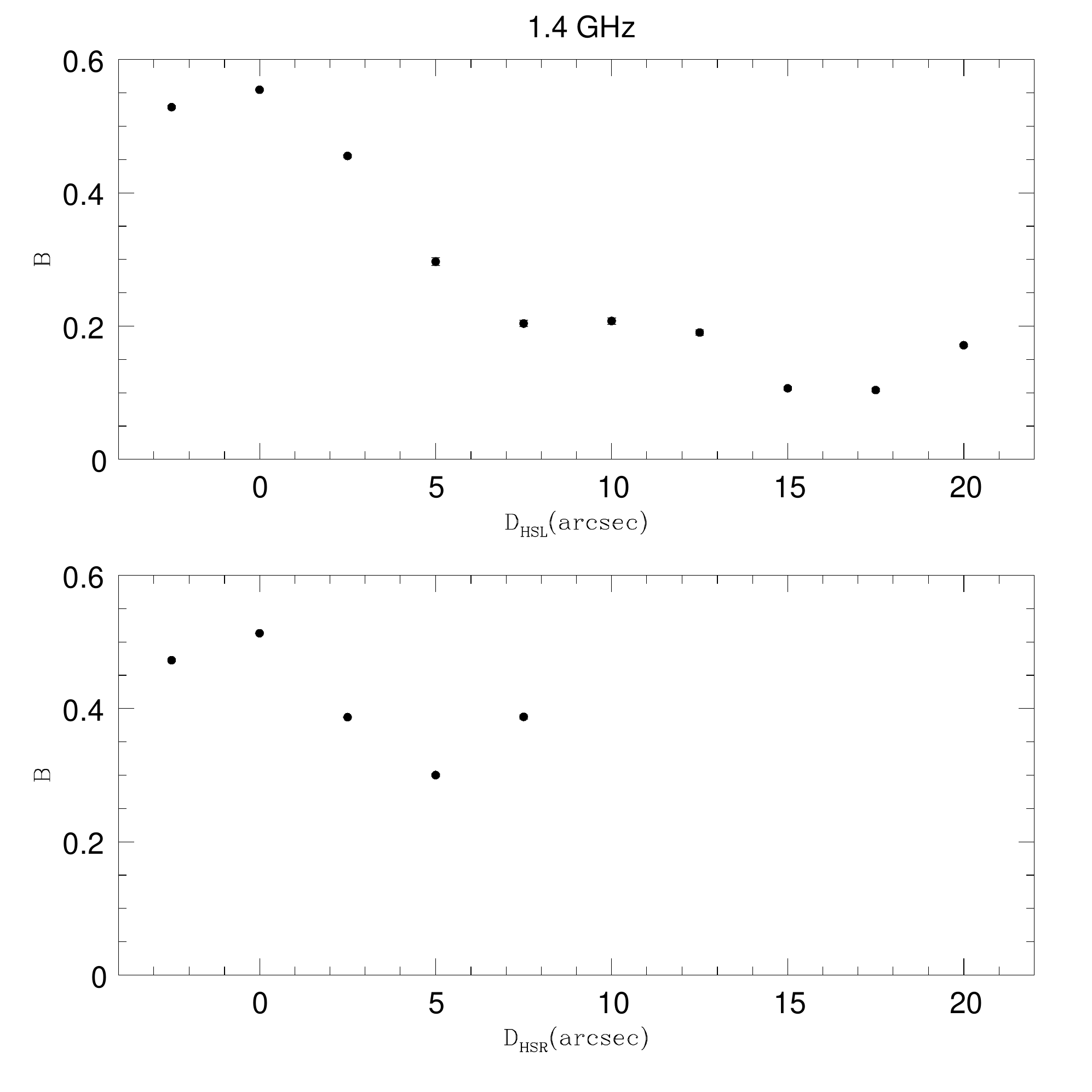}{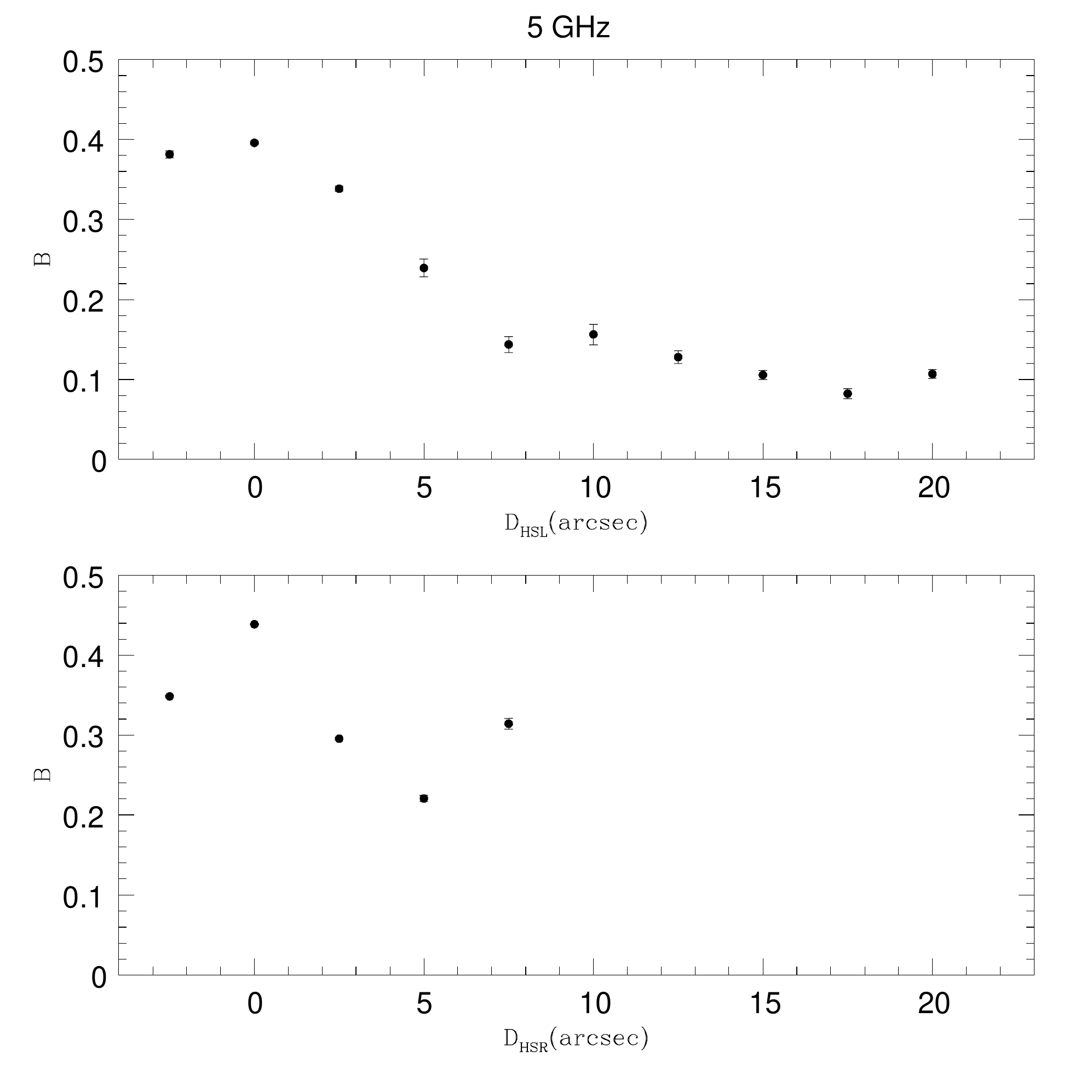}
\caption{3C441 minimum energy magnetic field strength 
as a function of distance from the 
hot spot at 1.4 and 5 GHz 
(left and right panels, respectively) for the left and right
hand sides of the source (top and bottom panels, respectively).
The normalization is given in Table 1. 
}
\end{figure}

\begin{figure}
\plottwo{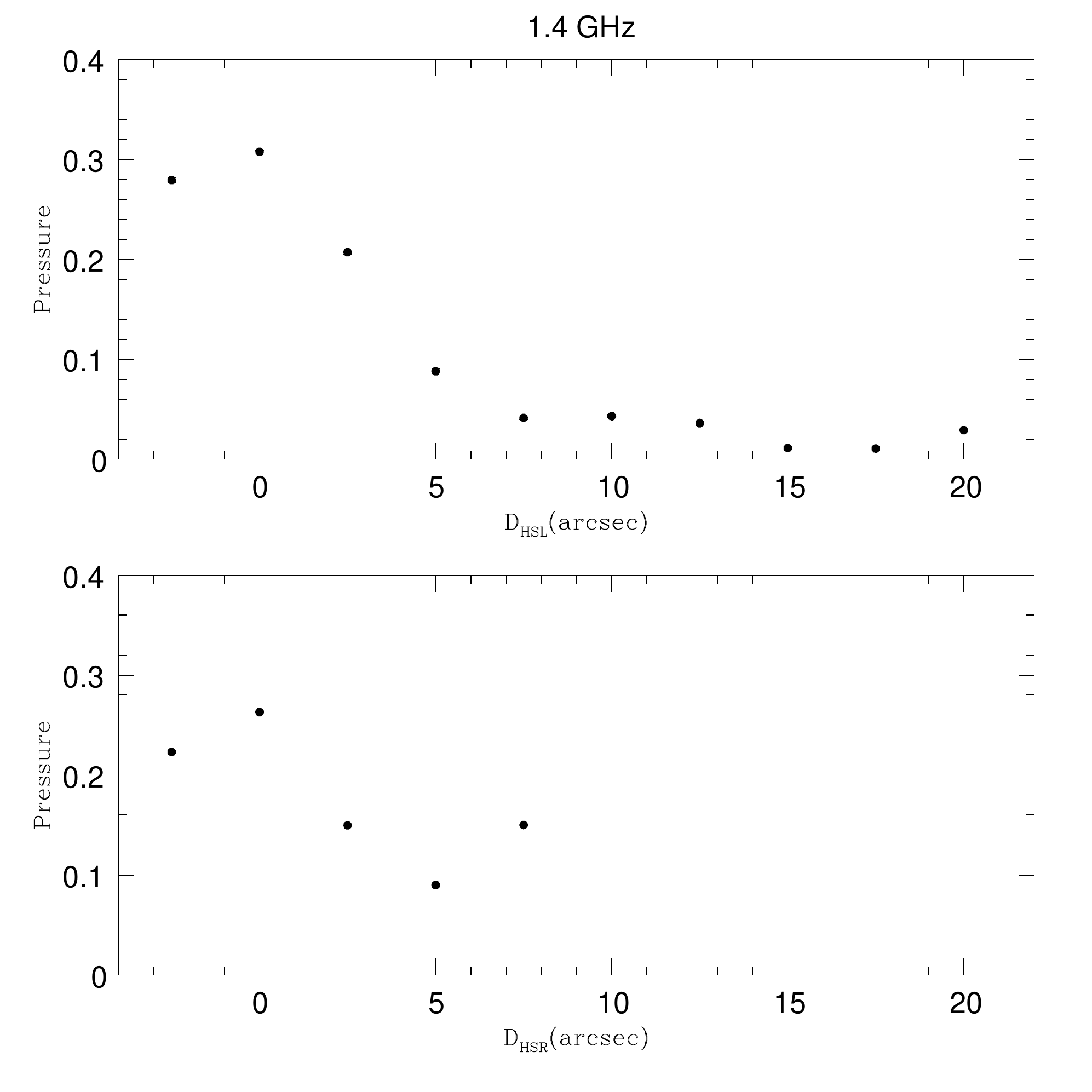}{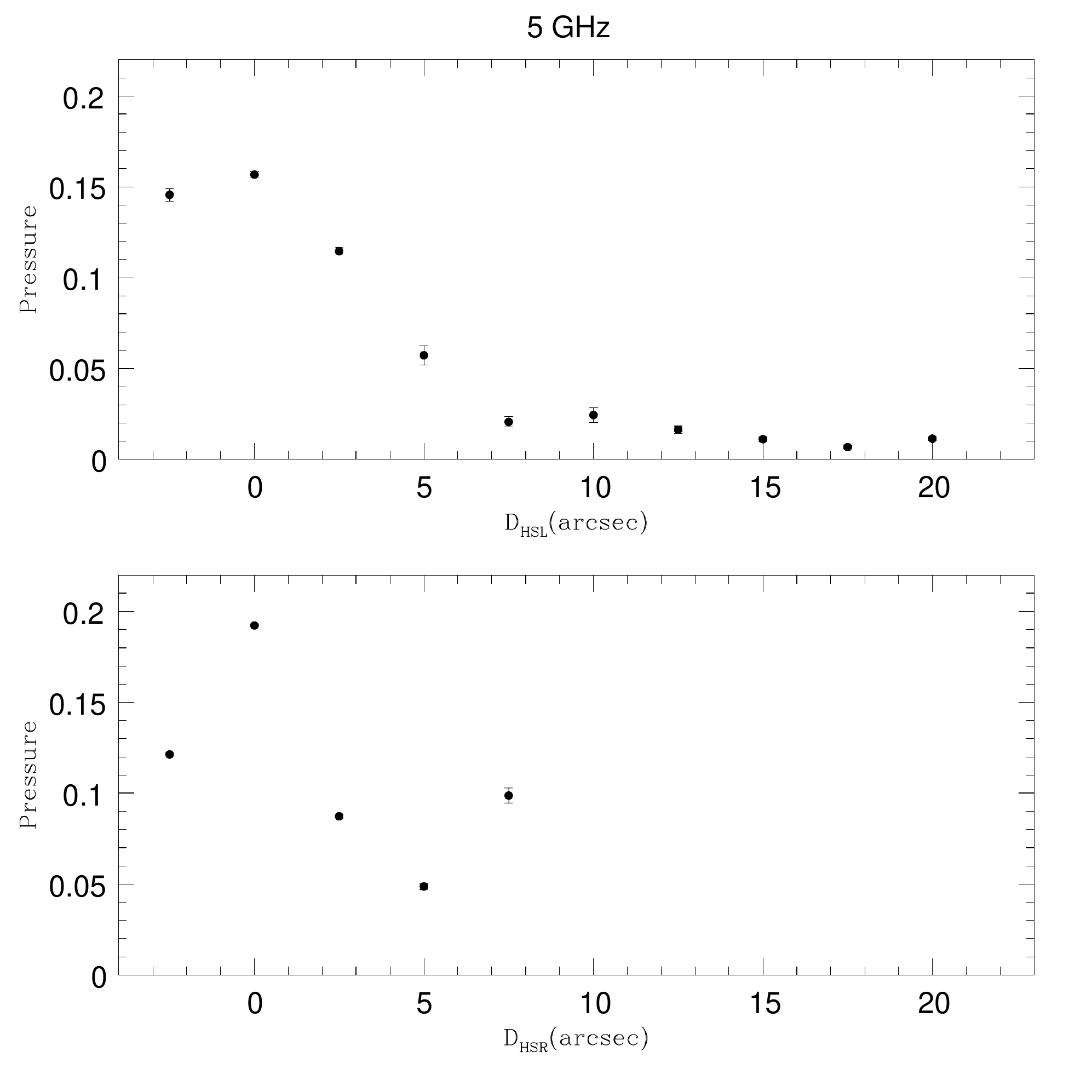}
\caption{3C441 
minimum energy pressure as a function of 
distance from the hot spot at 1.4 and 5 GHz 
(left and right panels, respectively) for the left and right
hand sides of the source (top and bottom panels, respectively), 
as in Fig. \ref{3C6.1P}. The normalization is given in Table 1.
}
\end{figure}

\clearpage
\begin{figure}
\plotone{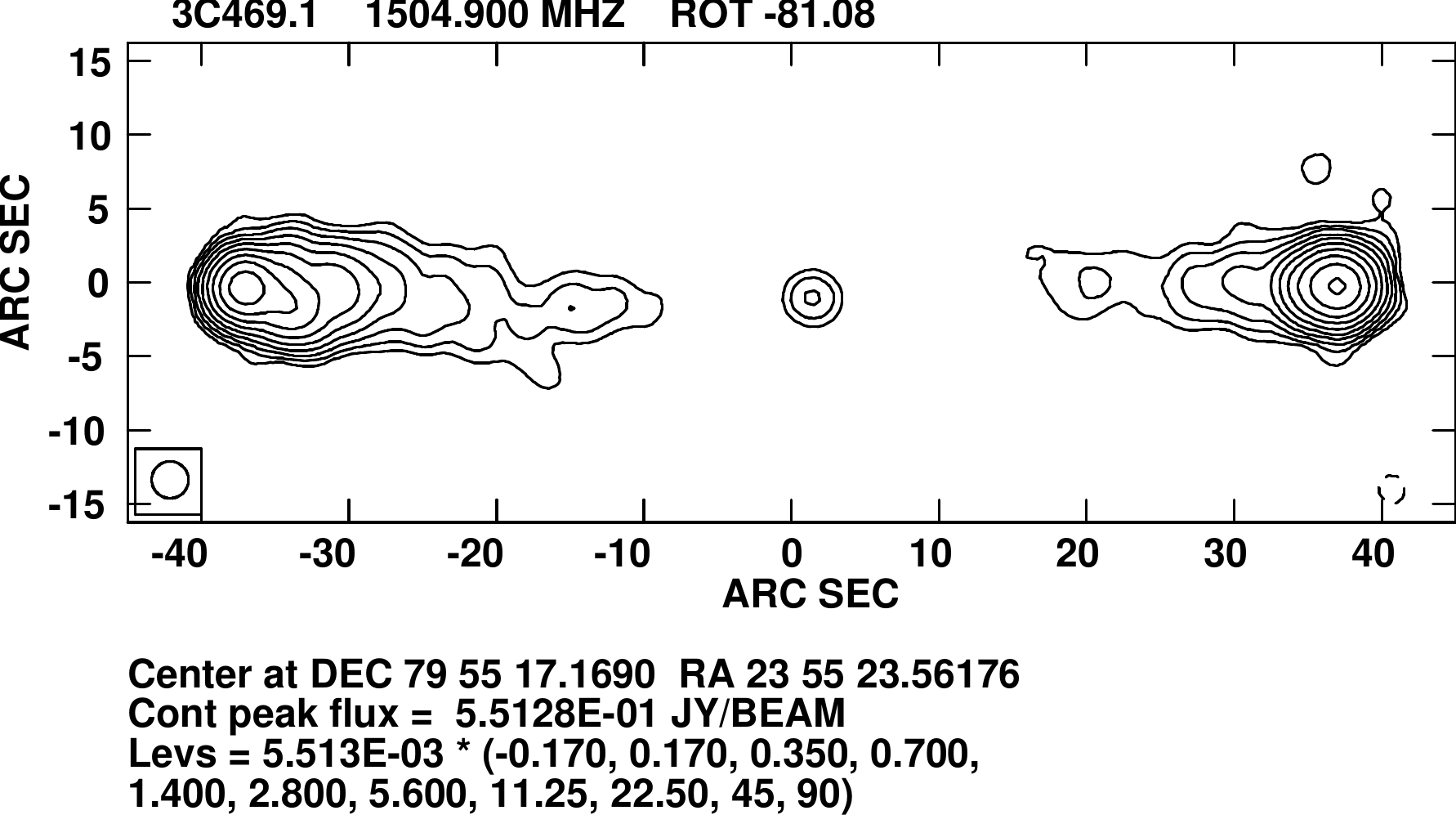}
\caption{3C469.1 at 1.4 GHz and 2.5'' resolution rotated by 81.08 deg 
clockwise, as in Fig. \ref{3C6.1rotfig}.}
\end{figure}

\begin{figure}
\plottwo{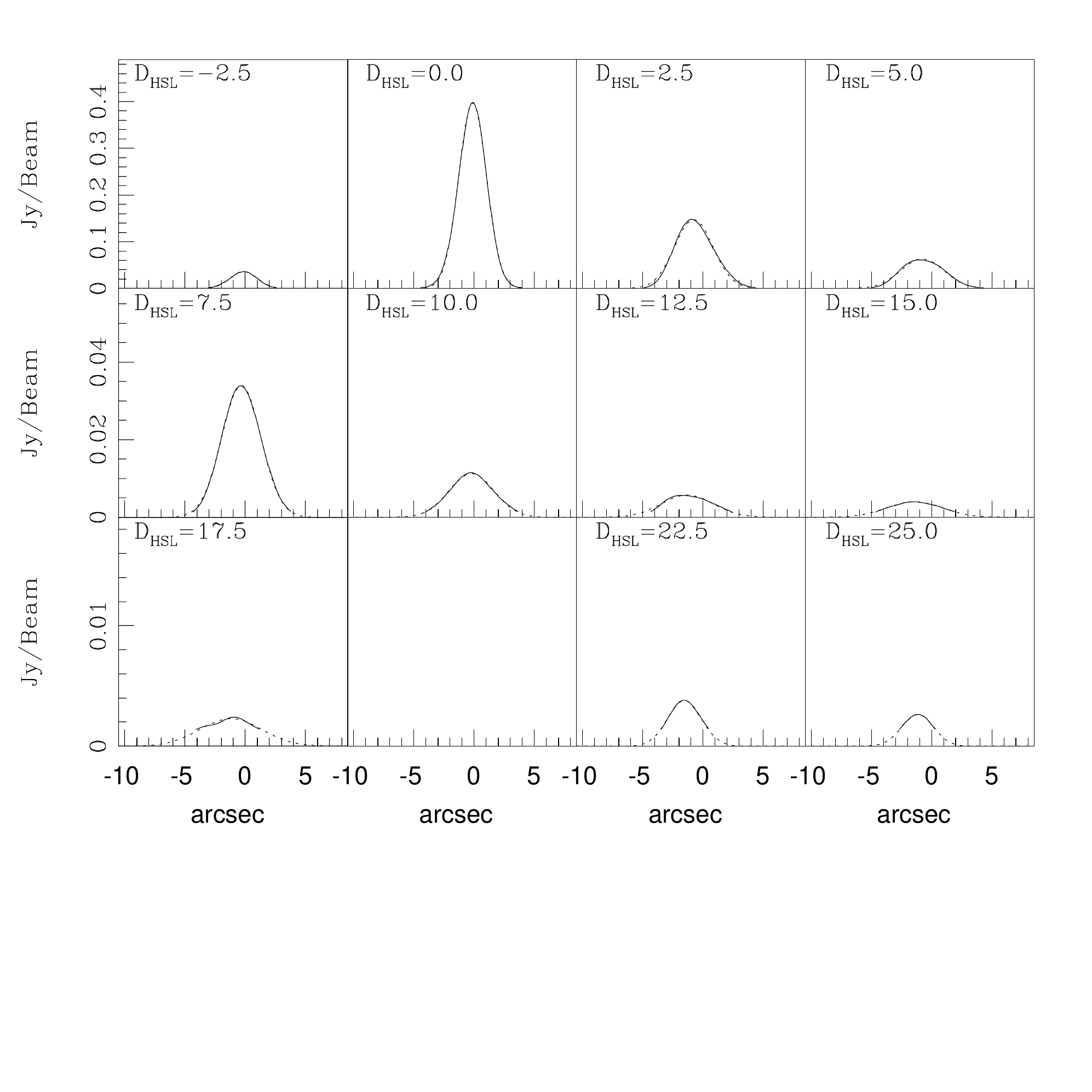}{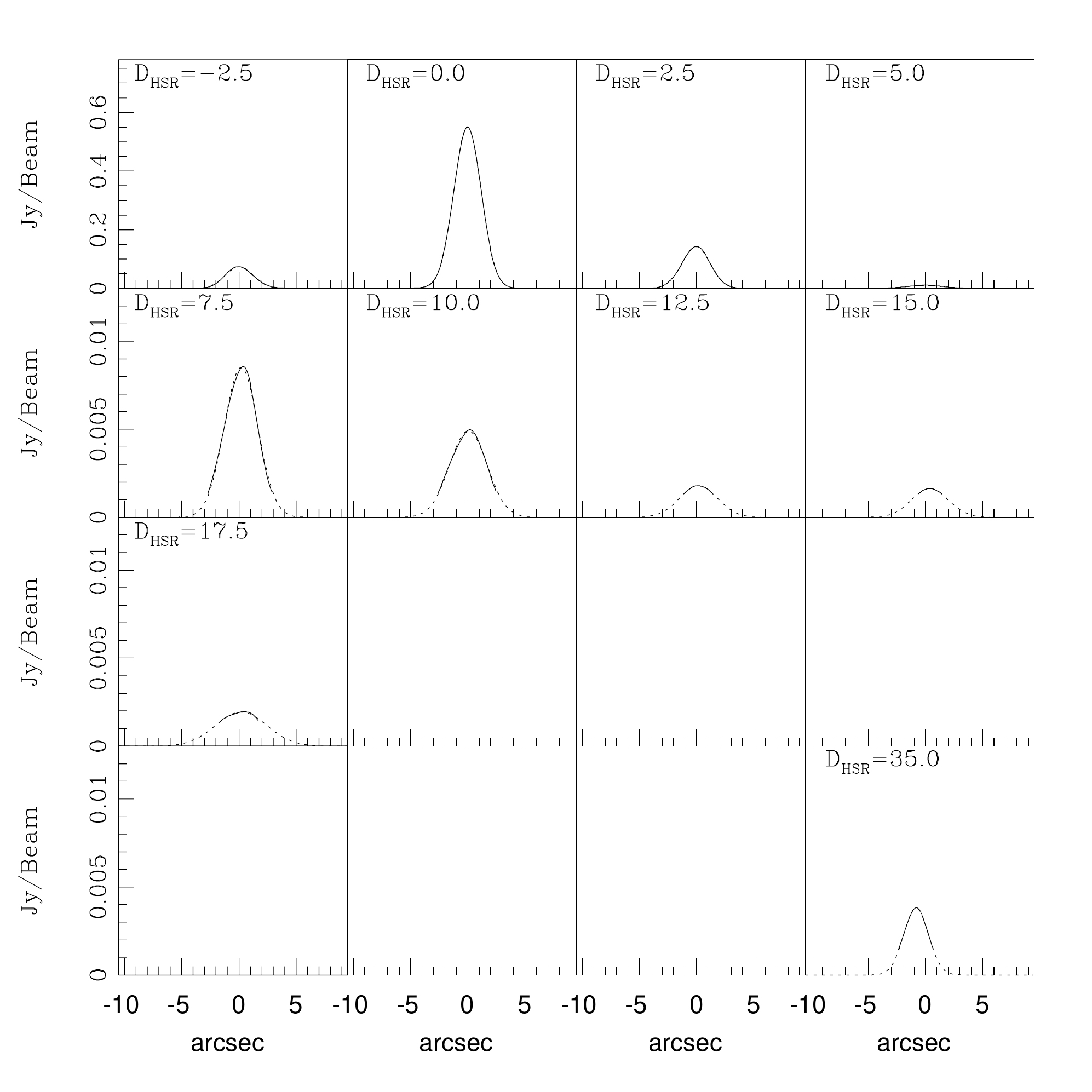}
\caption{3C469.1 
surface brightness (solid line) and best fit 
Gaussian (dotted line) for each cross-sectional slice of the 
radio bridge at 1.4 GHz, as in Fig. \ref{3C6.1SG}.
A Gaussian provides an excellent description of the surface brightness
profile of each cross-sectional slice. }
\label{3C469.1SG}
\end{figure}

\begin{figure}
\plottwo{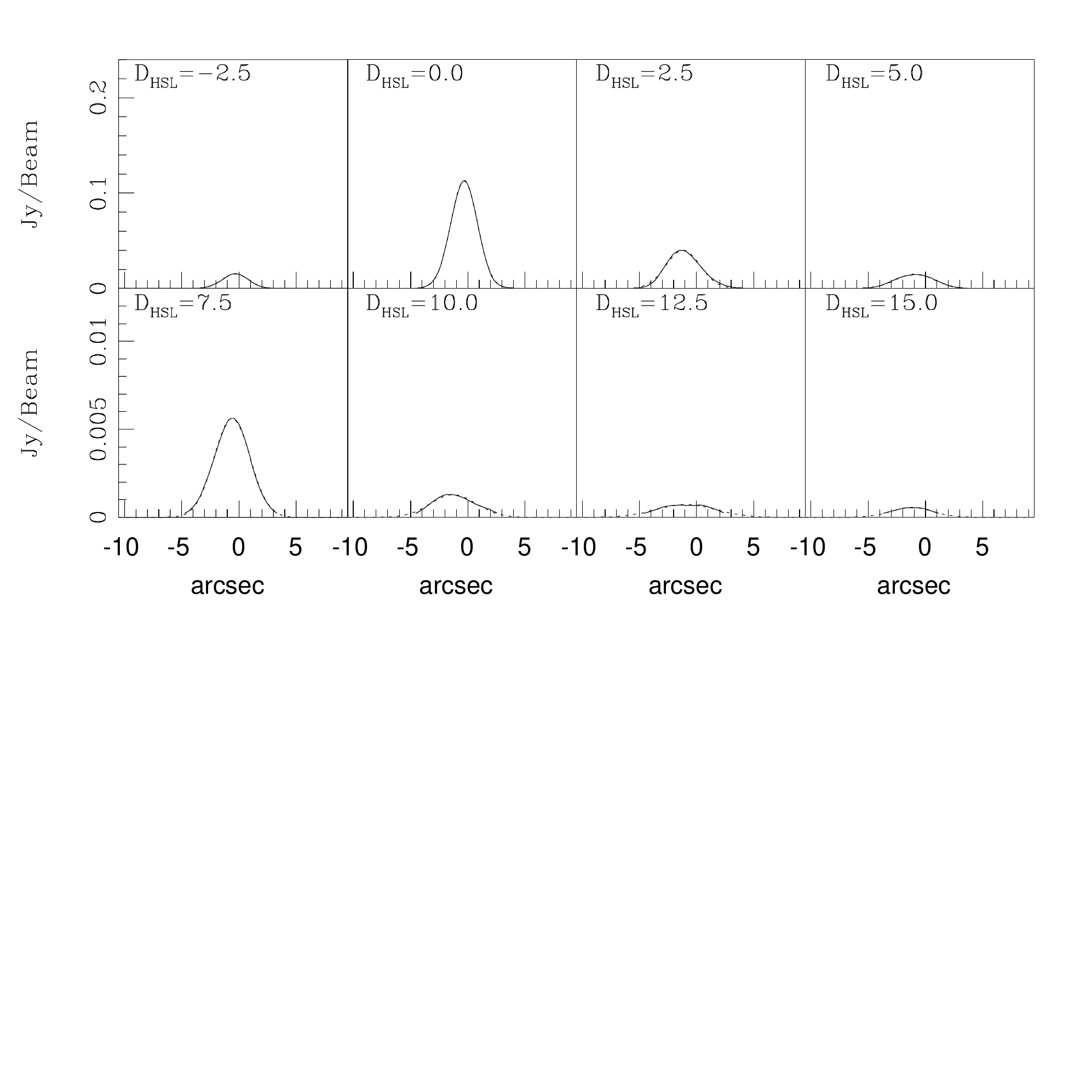}{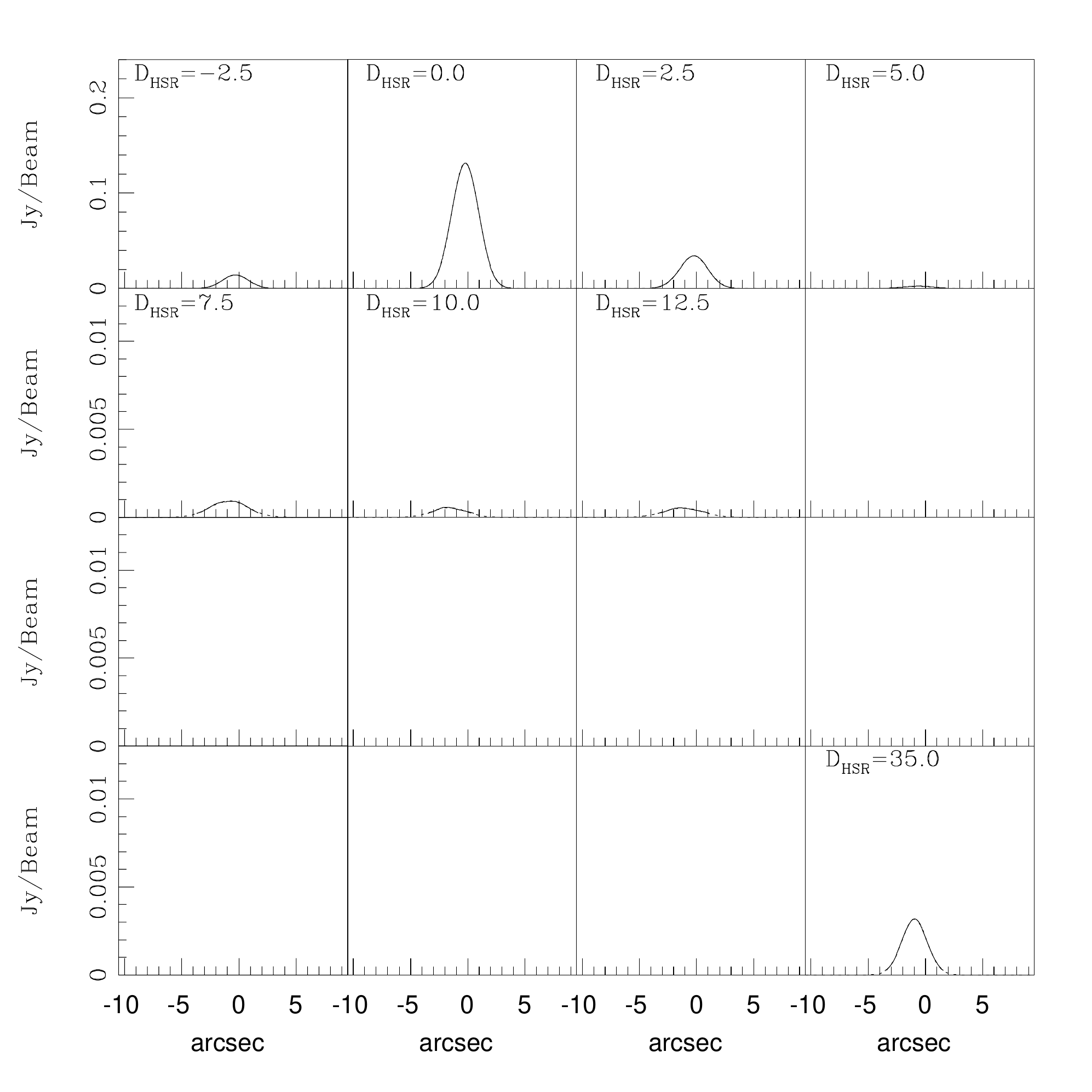}
\caption{3C469.1 
surface brightness (solid line) and best fit 
Gaussian (dotted line) for each cross-sectional slice of the 
radio bridge at 5 GHz, as in 
Fig. \ref{3C469.1SG} but at 5 GHz. 
Results obtained at 5 GHz are nearly identical to those
obtained at 1.4 GHz.}
\end{figure}

\begin{figure}
\plottwo{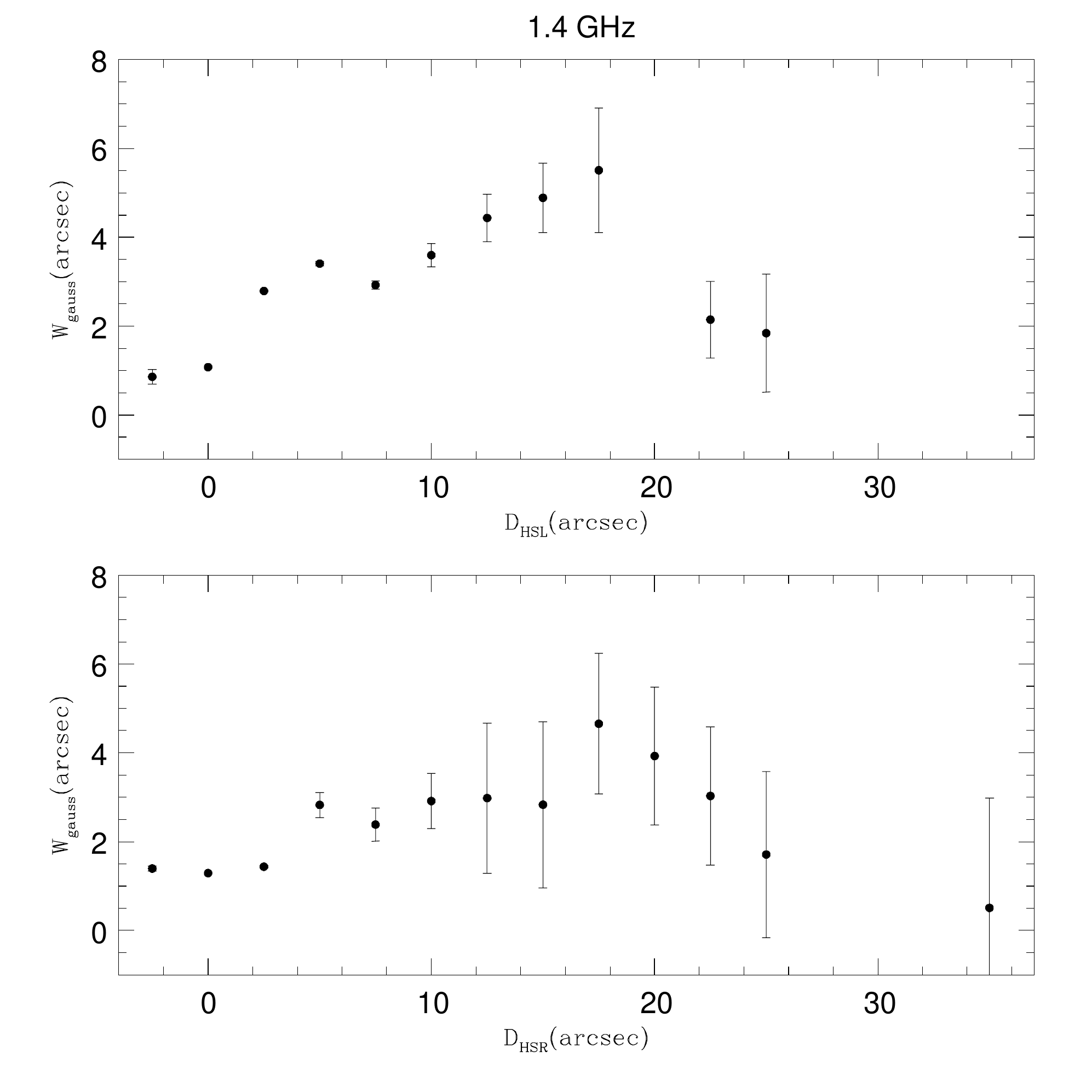}{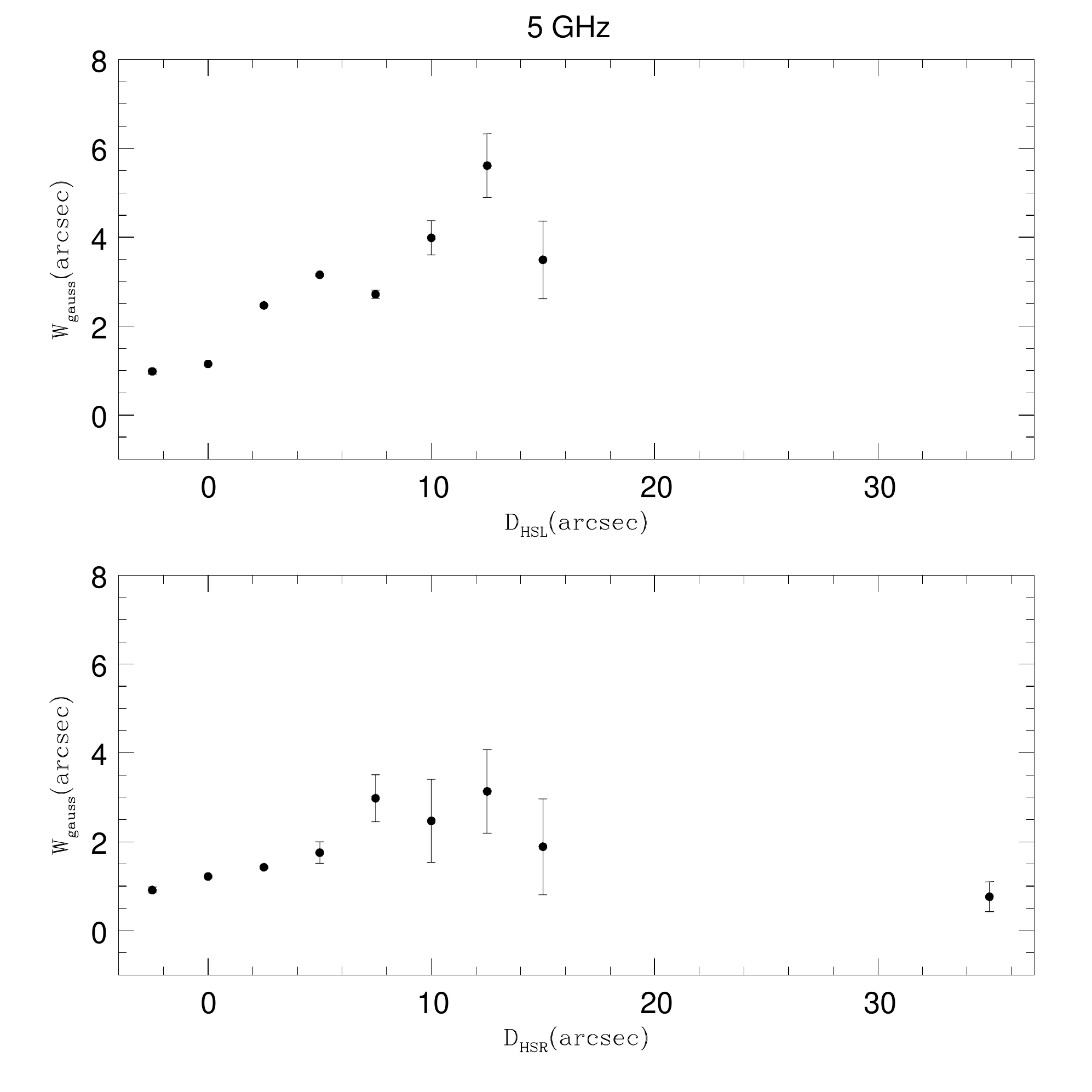}
\caption{3C469.1  Gaussian FWHM as a function of 
distance from the hot spot  
at 1.4 and 5 GHz (left and
right panels, respectively) for the left and right hand sides of the 
source (top and bottom panels, respectively), as in Fig. \ref{3C6.1WG}.
The right and left hand sides of the source are somewhat
symmetric.  Similar results are obtained at 
1.4 and 5 GHz. }
\end{figure}

\begin{figure}
\plottwo{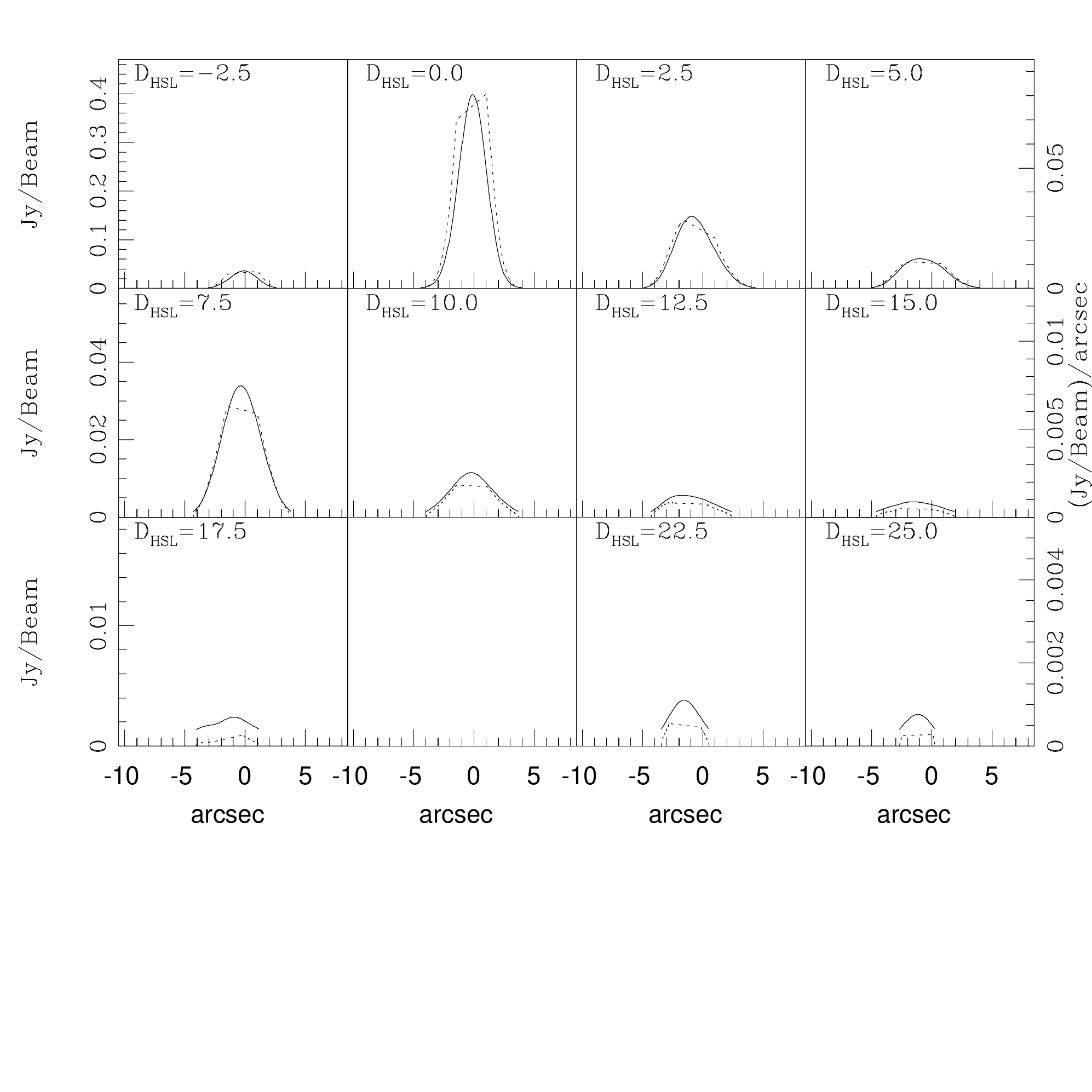}{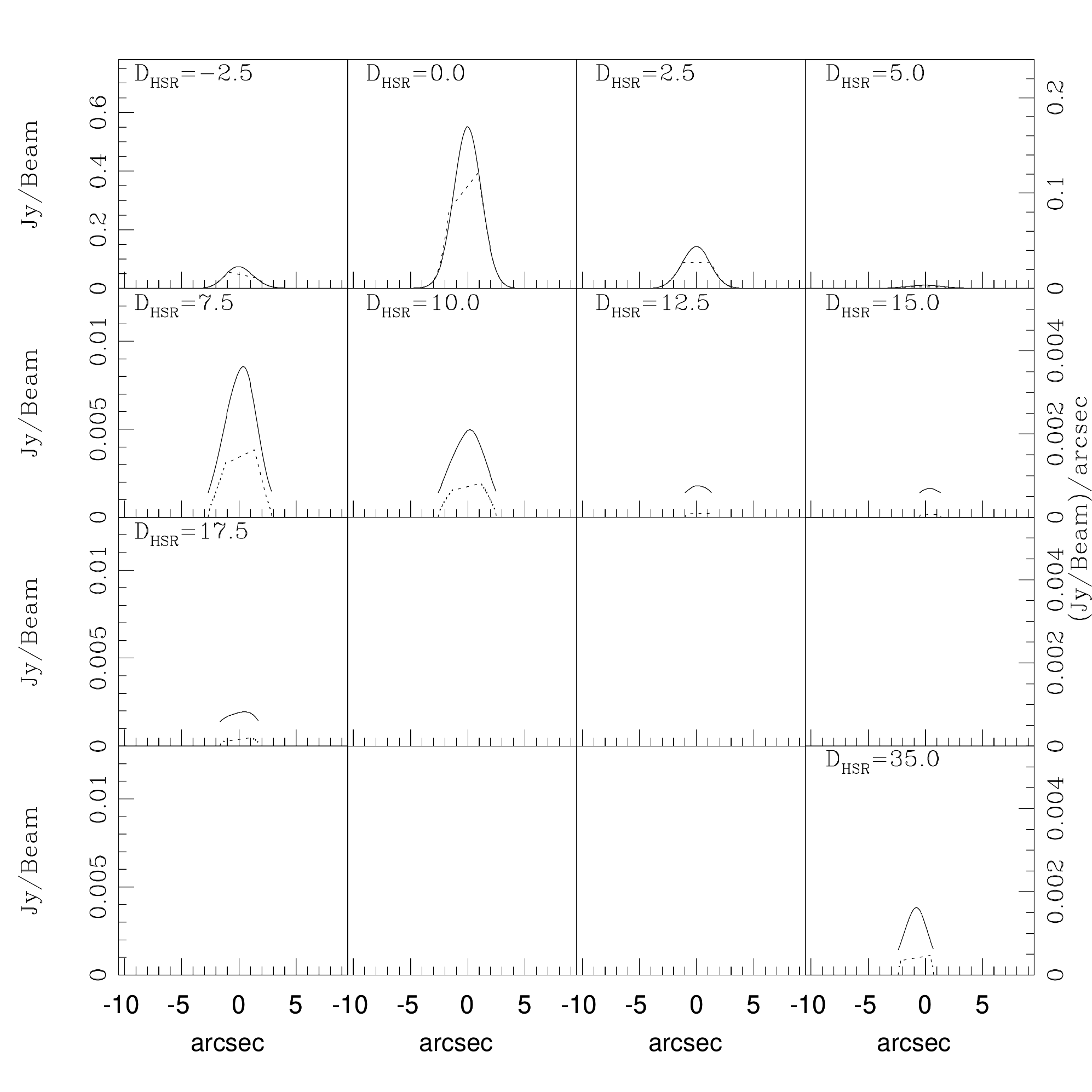}
\caption{3C469.1 emissivity (dotted line) and surface brightness
(solid line) for each cross-sectional slice of the radio 
bridge at 1.4 GHz, as in Fig. \ref{3C6.1SEM}. A constant volume
emissivity per slice provides a reasonable description of the 
data over most of the source. }
\label{3C469.1SEM}
\end{figure}

\begin{figure}
\plottwo{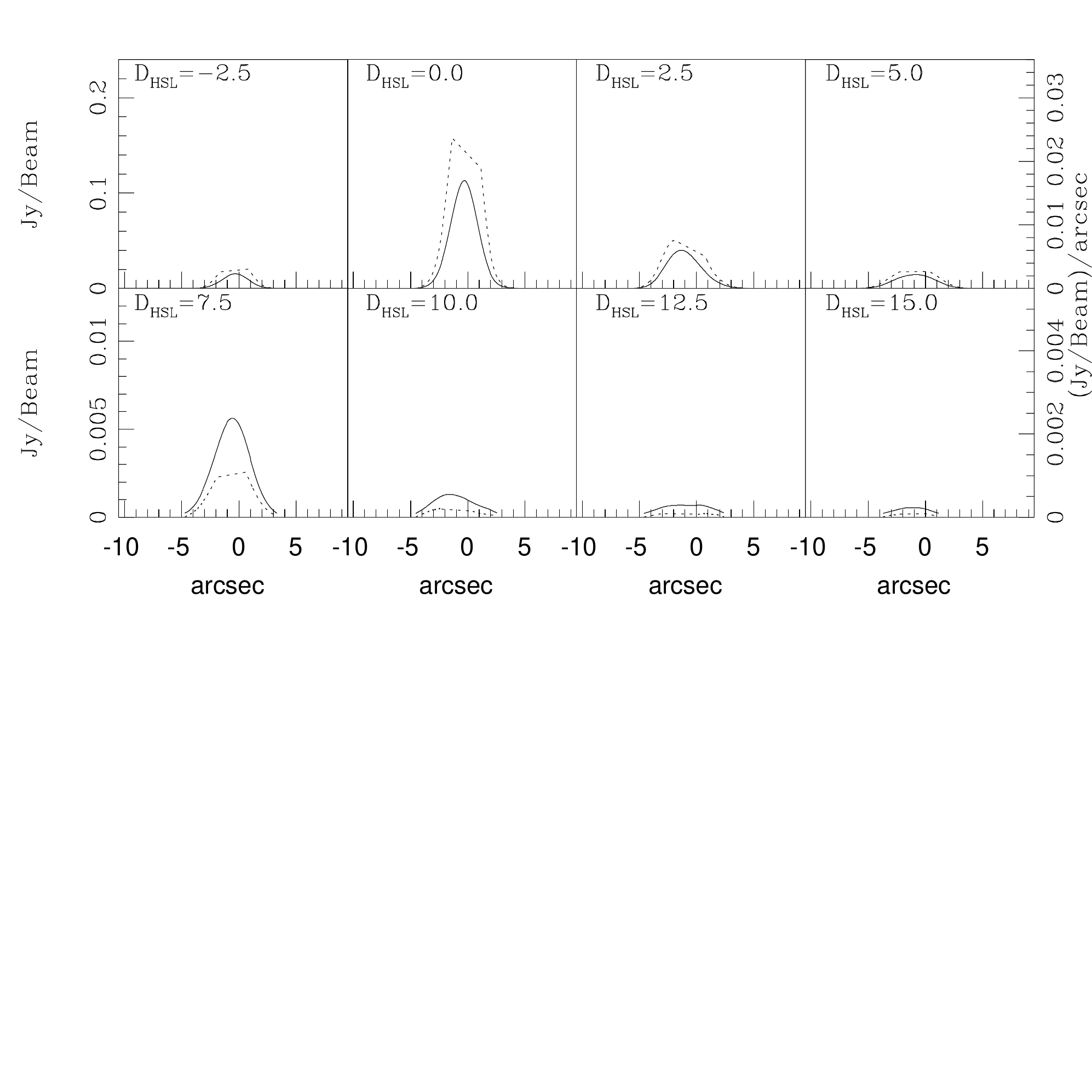}{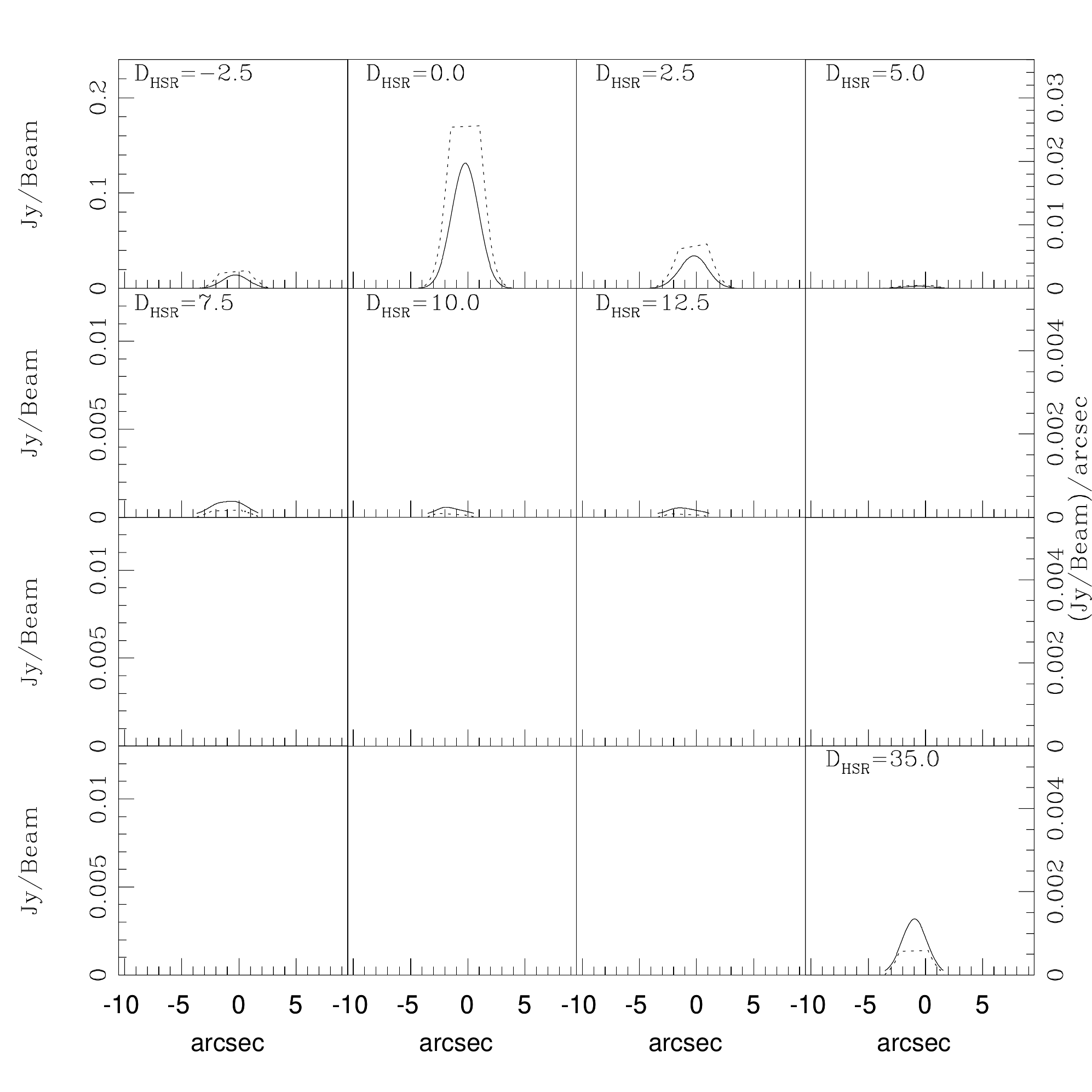}
\caption{3C469.1
emissivity (dotted line) and surface brightness
(solid line) for each cross-sectional slice of the radio 
bridge at 5 GHz, as in Fig. \ref{3C469.1SEM} but at 5 GHz.
Results obtained at 5 GHz are nearly identical to those
obtained at 1.4 GHz.}
\end{figure}

\begin{figure}
\plottwo{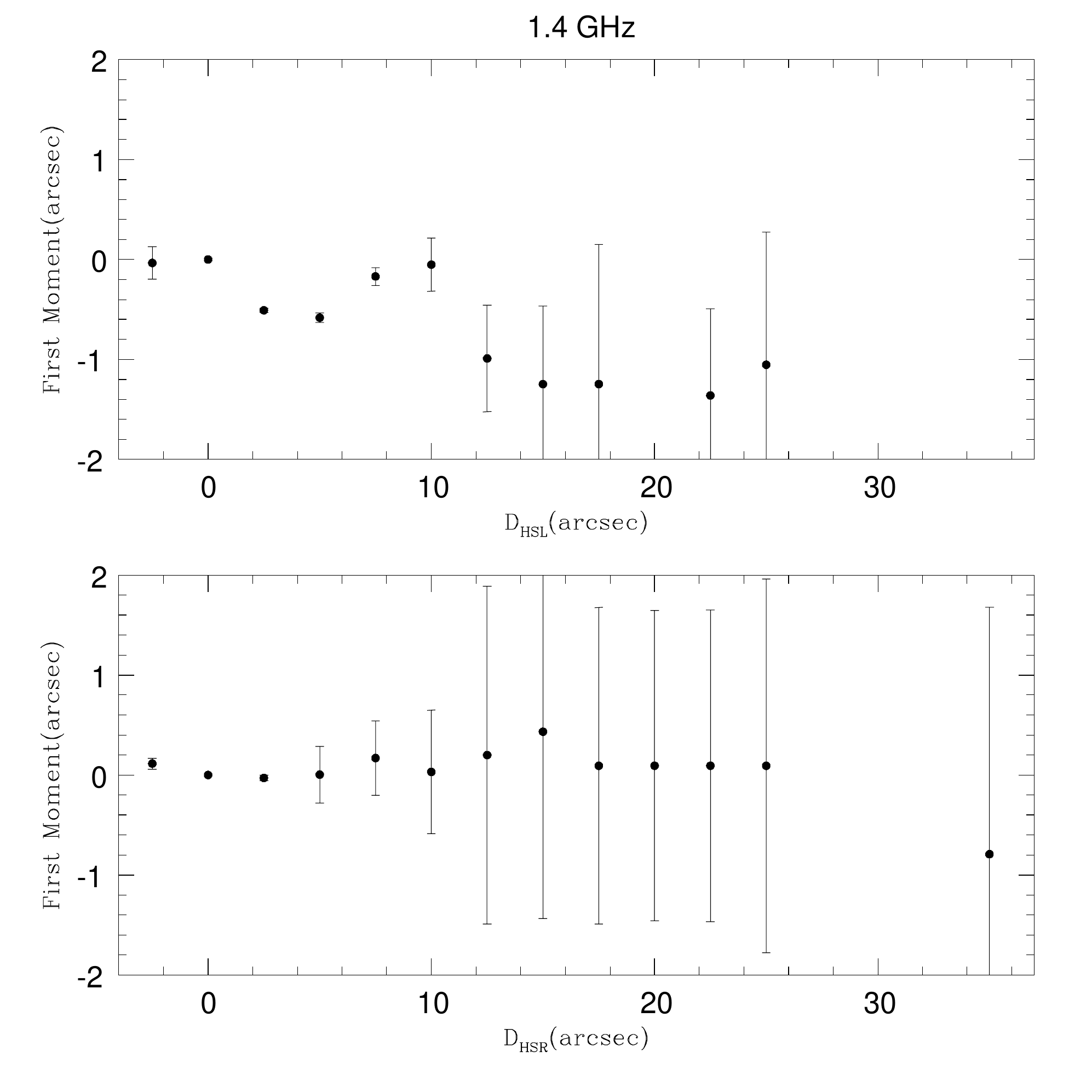}{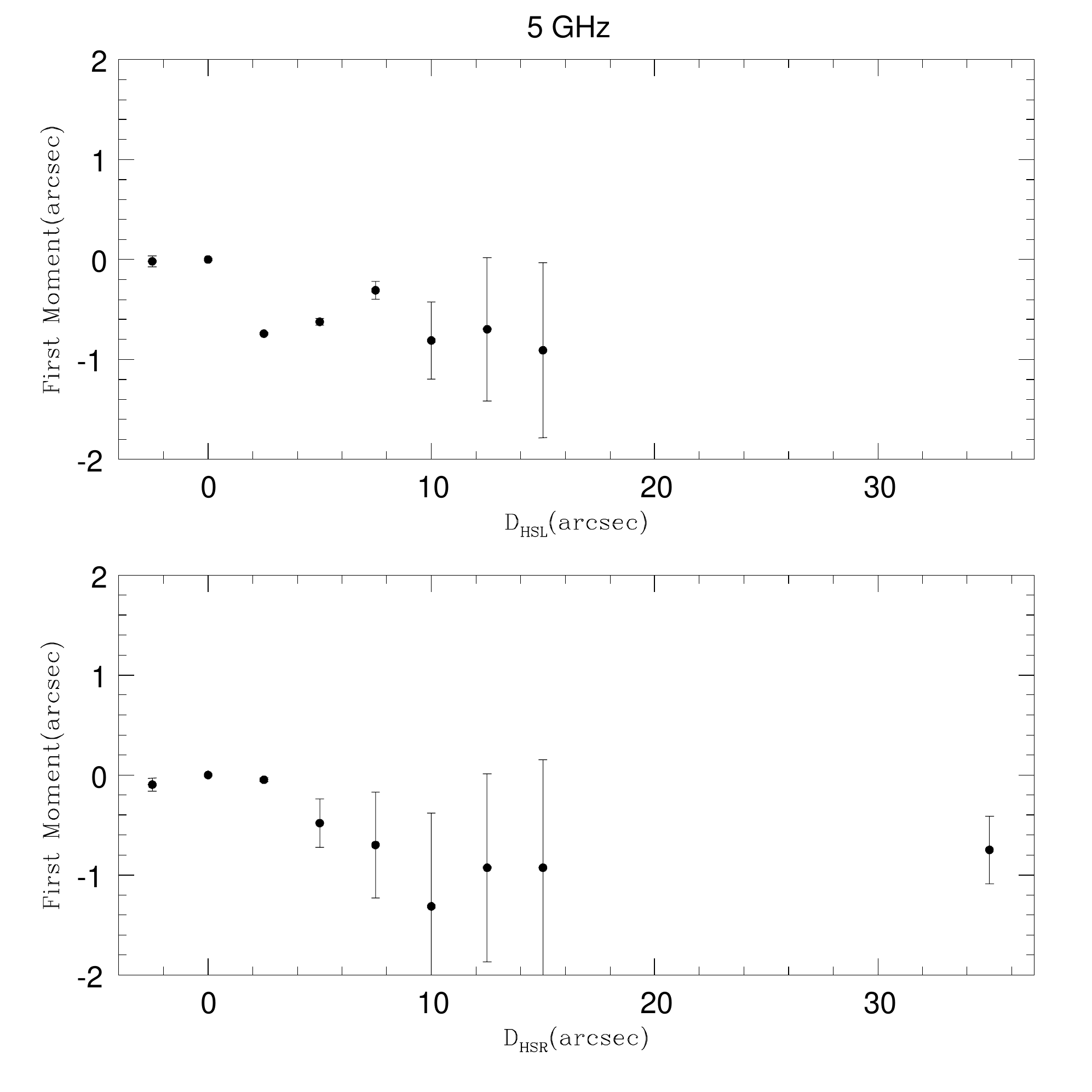}
\caption{3C469.1 first moment as a function of distance from 
the hot spot at 1.4 GHz and 5 GHz (left and right panels,
respectively) for the left and right hand sides of the source
(top and bottom, respectively).}
\end{figure}

\begin{figure}
\plottwo{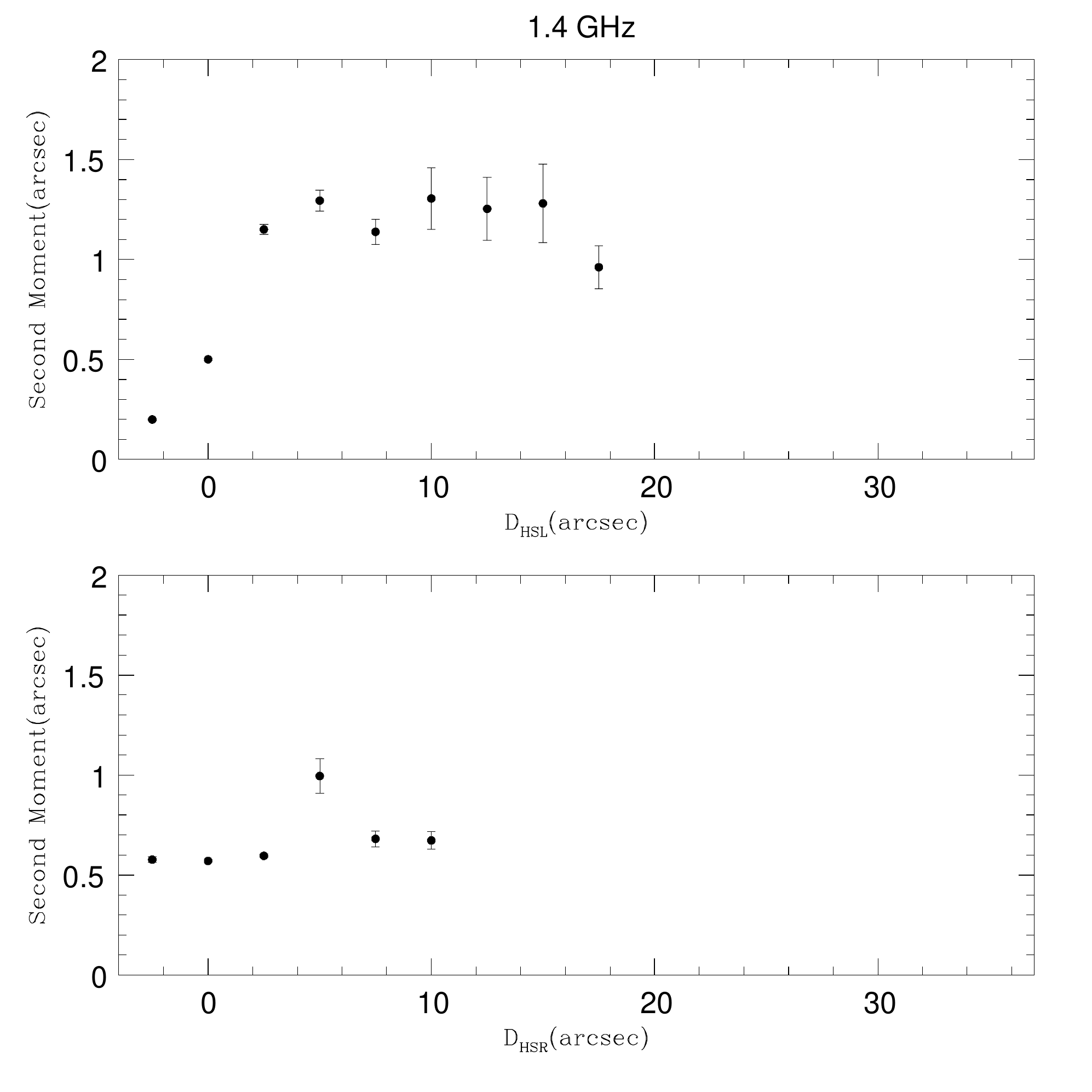}{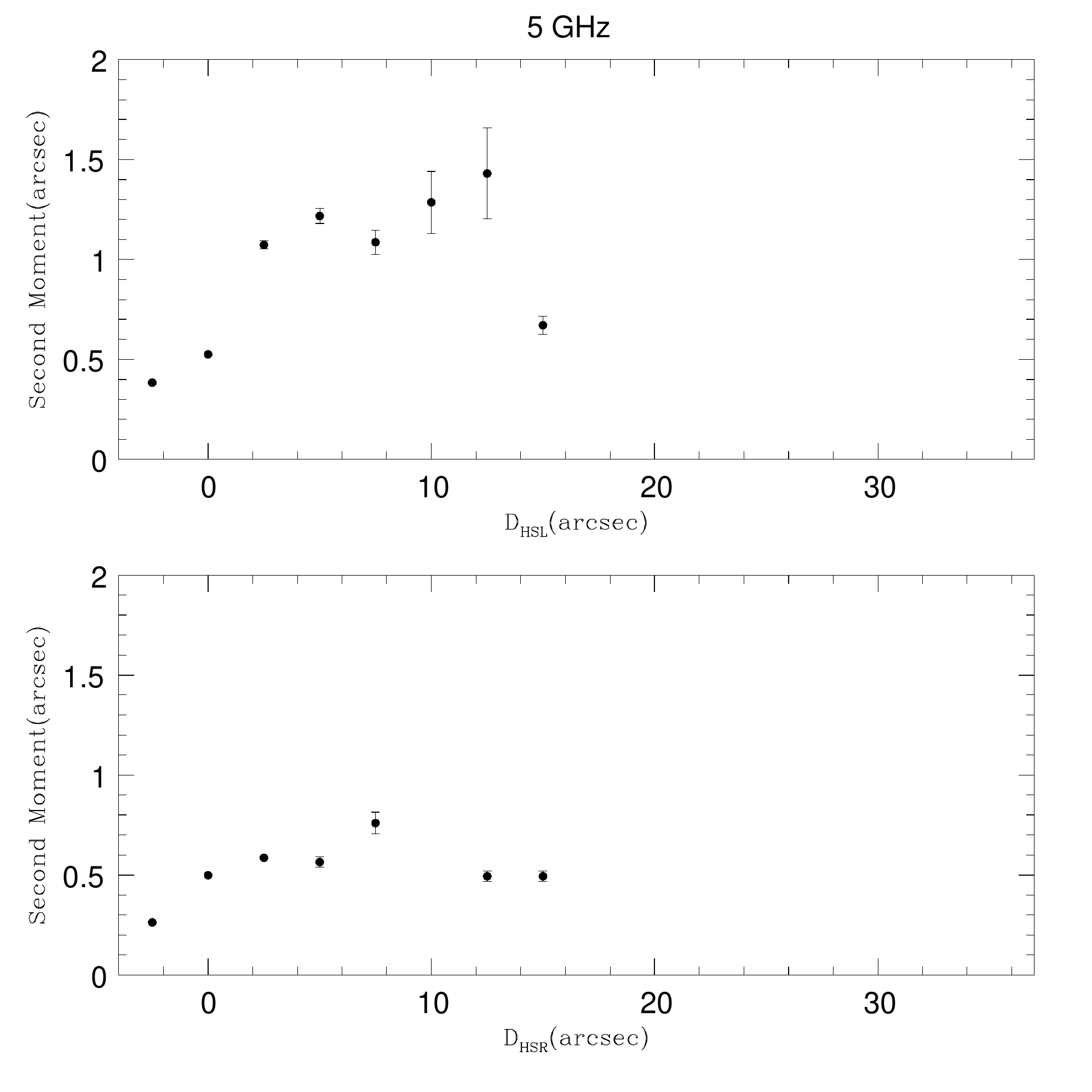}
\caption{3C469.1 second moment as a function of distance from the
hot spot at 1.4 GHz and 5 GHz (left and right panels, 
respectively) for the left and right sides of the source
(top and bottom panels, respectively).}
\end{figure}

\begin{figure}
\plottwo{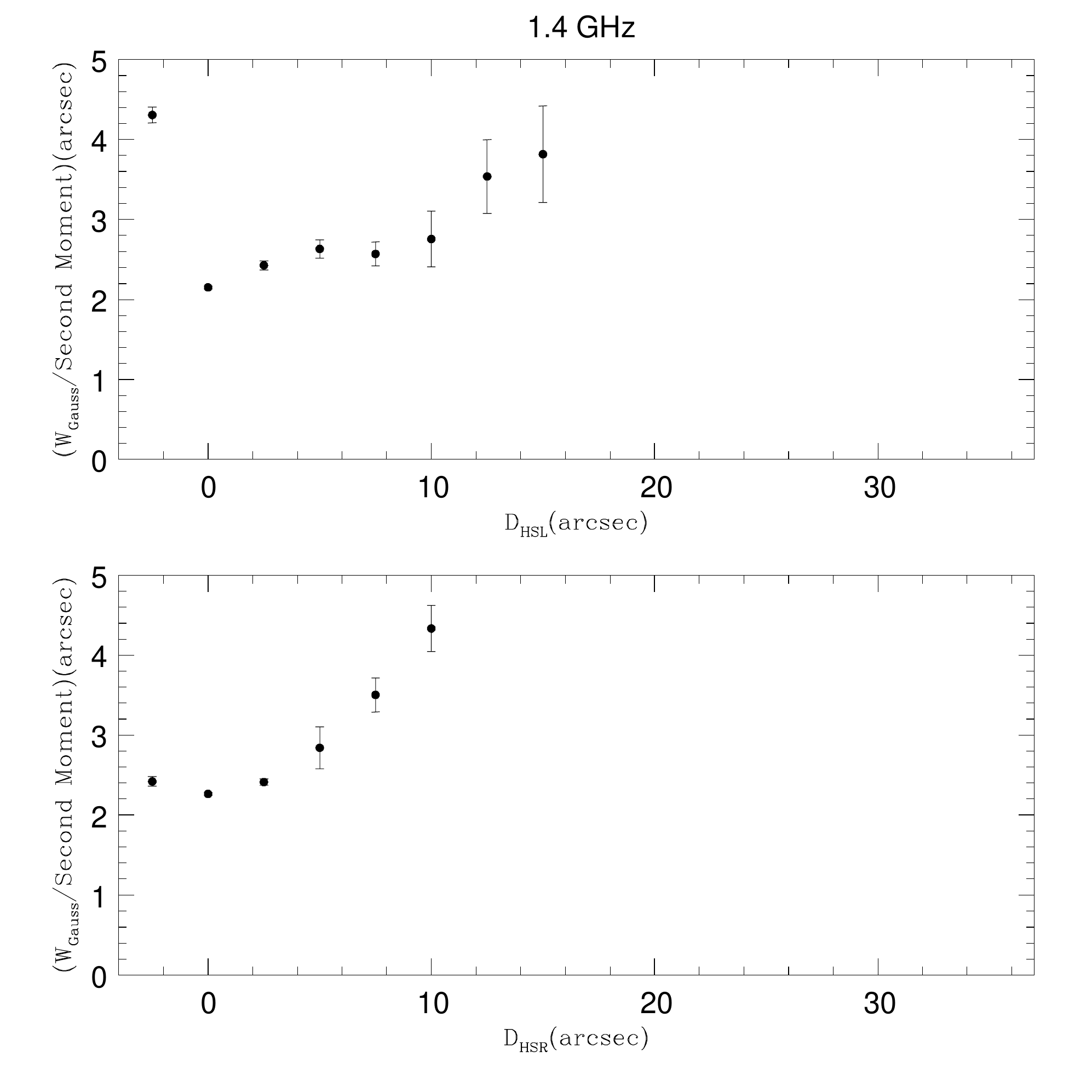}{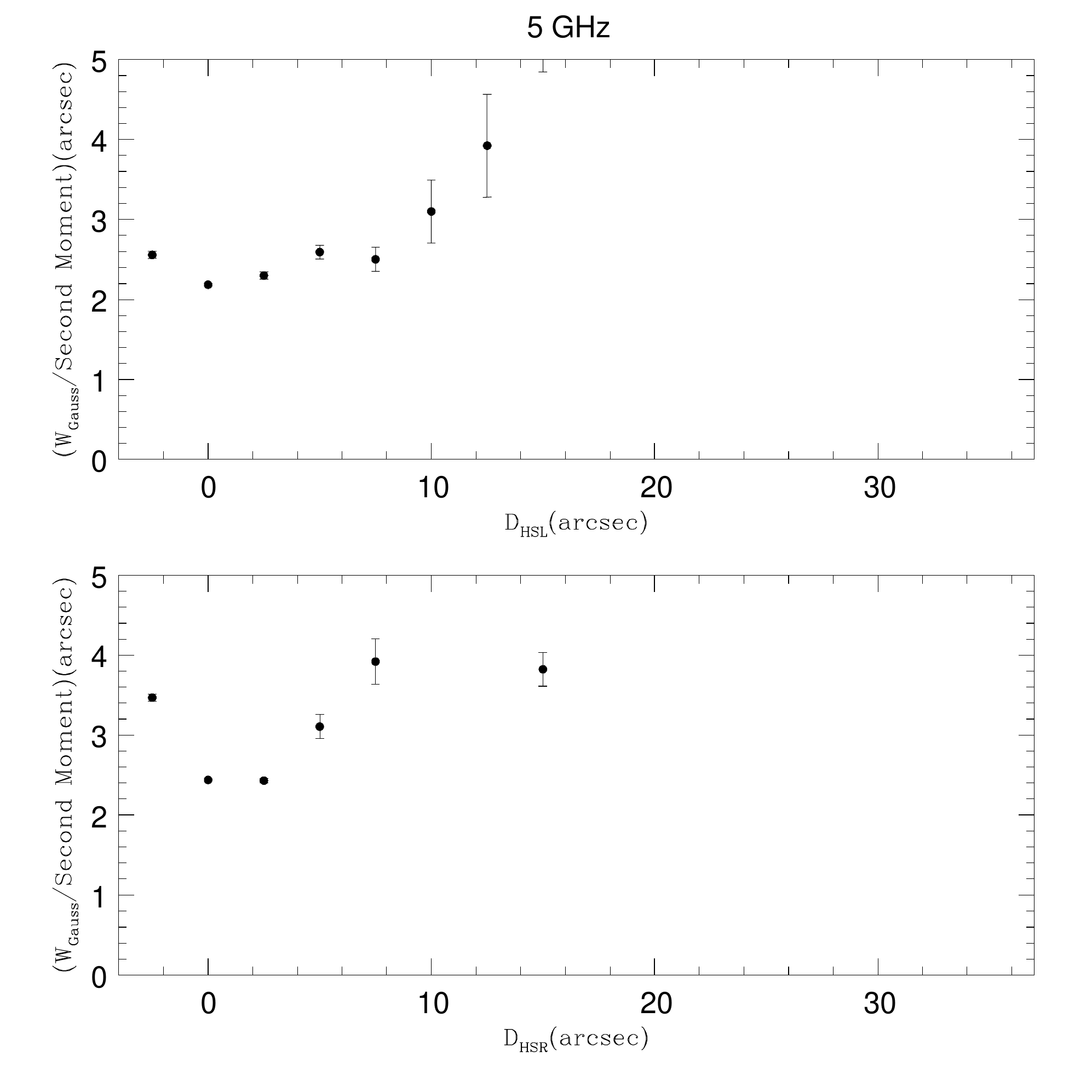}
\caption{3C469.1 ratio of the Gaussian FWHM to the second moment as a
function of distance from the hot spot at 1.4 and 5 GHz 
(left and right panels, respectively) for the left and right
hand sides of the source (top and bottom panels, respectively). 
The value of this ratio shifts across each side of the 
source, and has a value of about 2 to 3. Similar results are obtained 
at 1.4 and 5 GHz.} 
\end{figure}

\begin{figure}
\plottwo{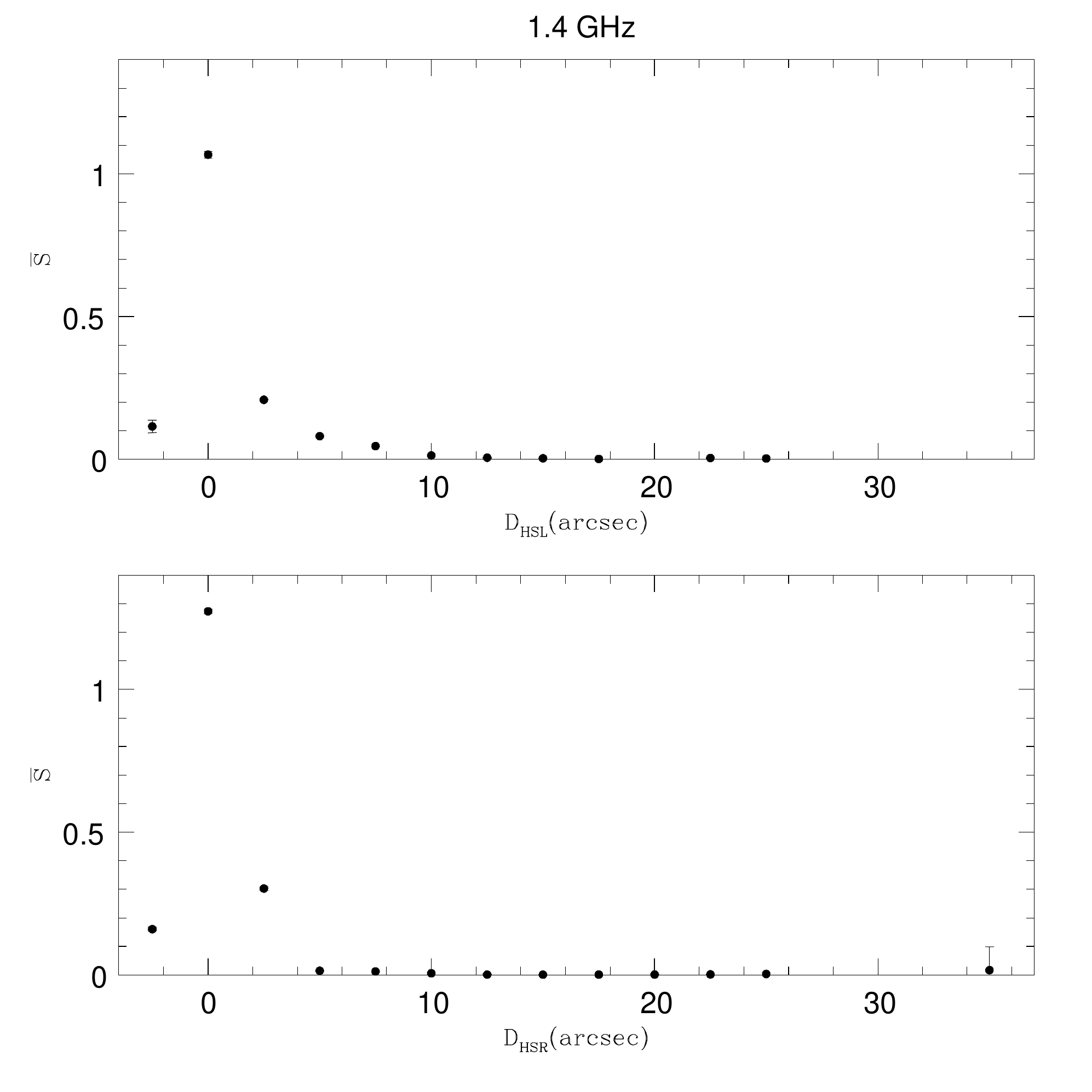}{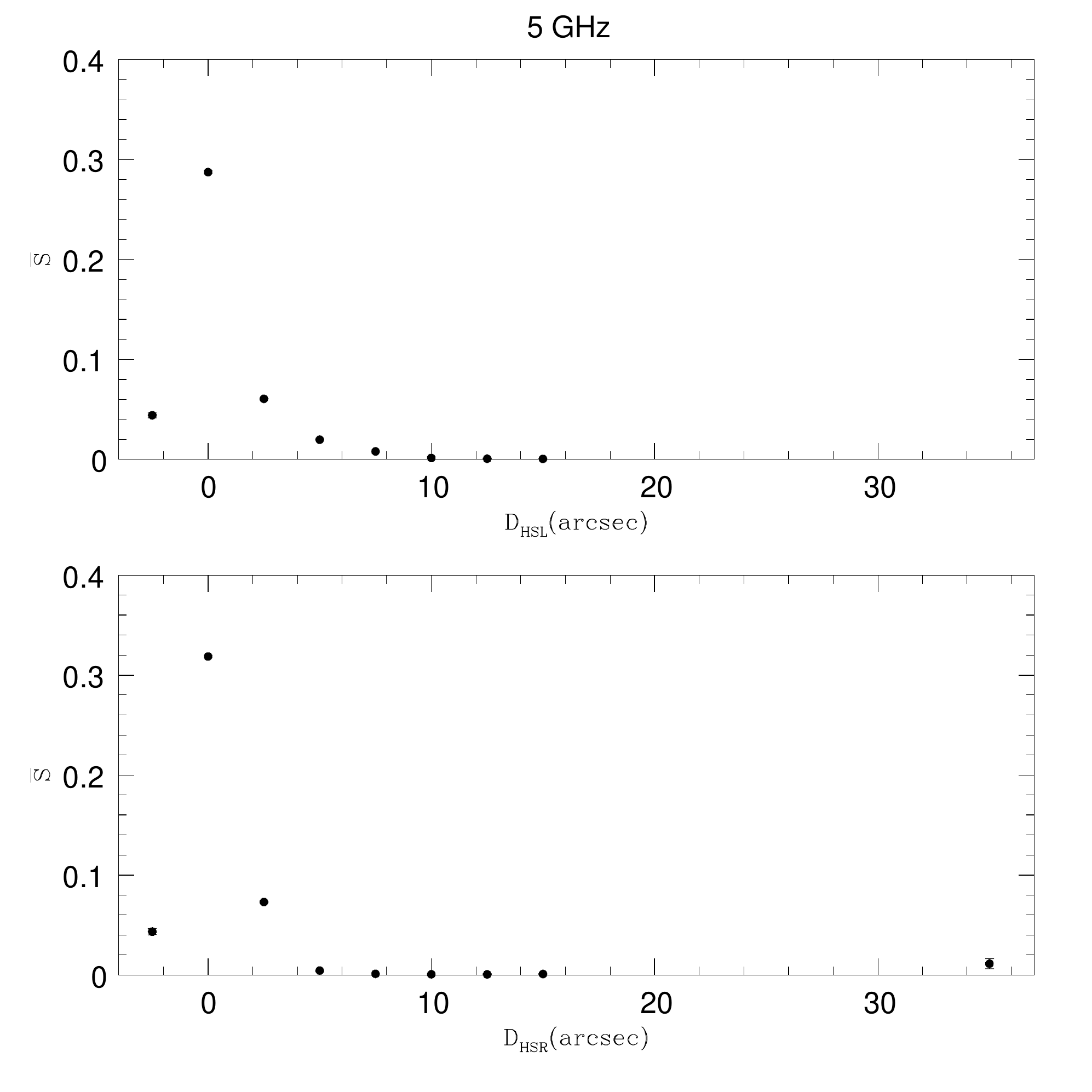}
\caption{3C469.1 average surface brightness in units of Jy/beam 
as a function of distance from the 
hot spot at 1.4 and 5 GHz 
(left and right panels, respectively) for the left and right
hand sides of the source (top and bottom panels, respectively).}
\end{figure}

\begin{figure}
\plottwo{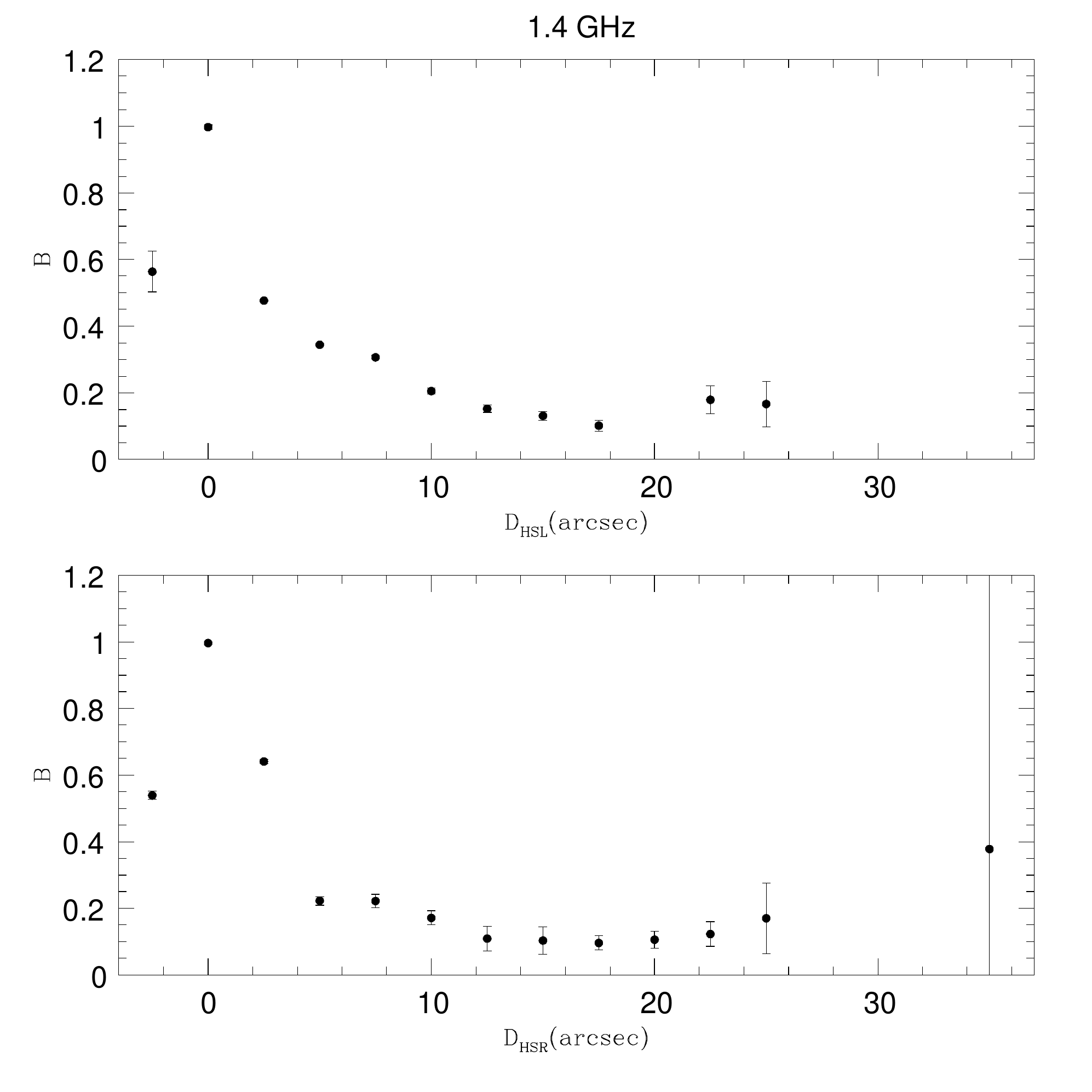}{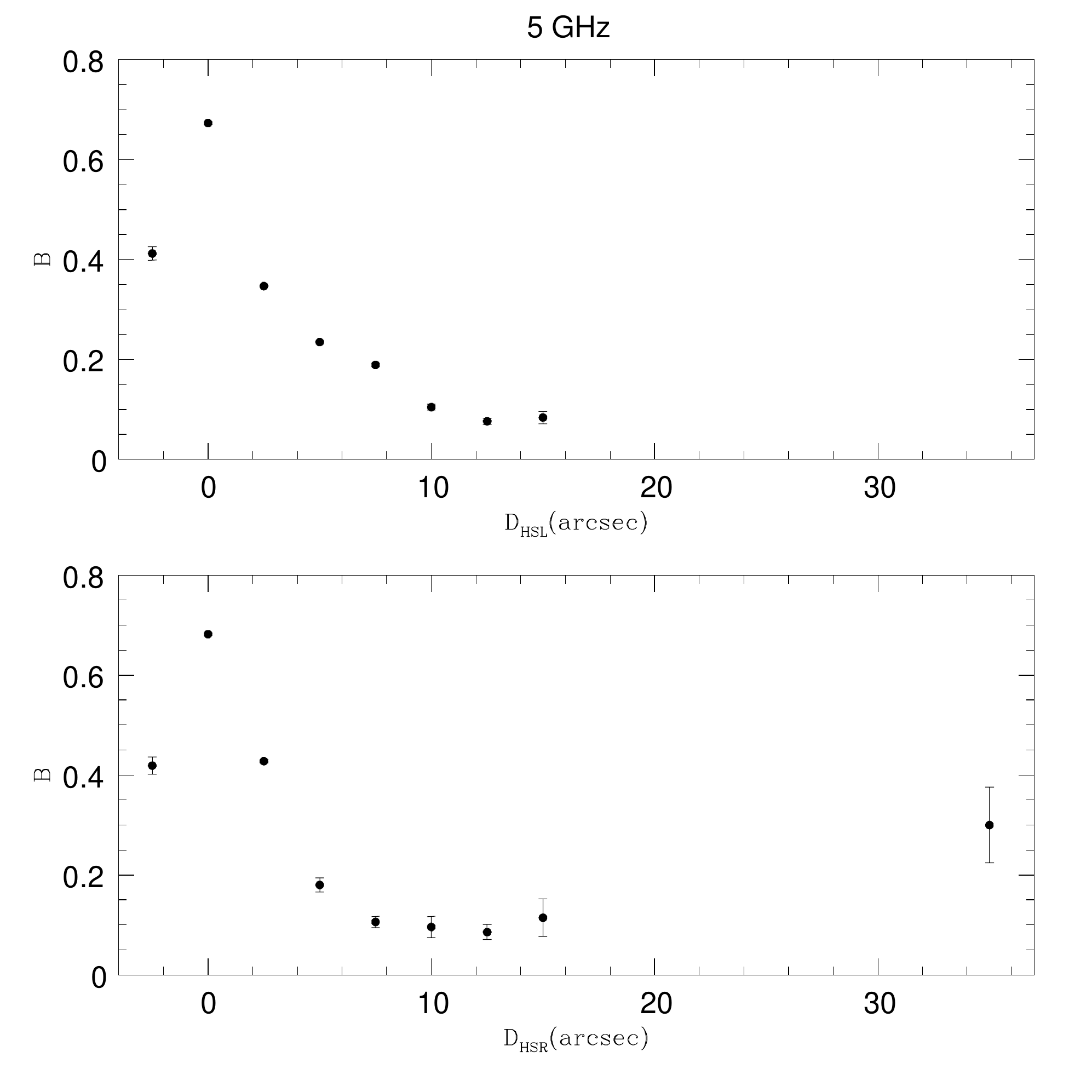}
\caption{3C469.1 minimum energy magnetic field strength 
as a function of distance from the 
hot spot at 1.4 and 5 GHz 
(left and right panels, respectively) for the left and right
hand sides of the source (top and bottom panels, respectively).
The normalization is given in Table 1.}
\end{figure}

\begin{figure}
\plottwo{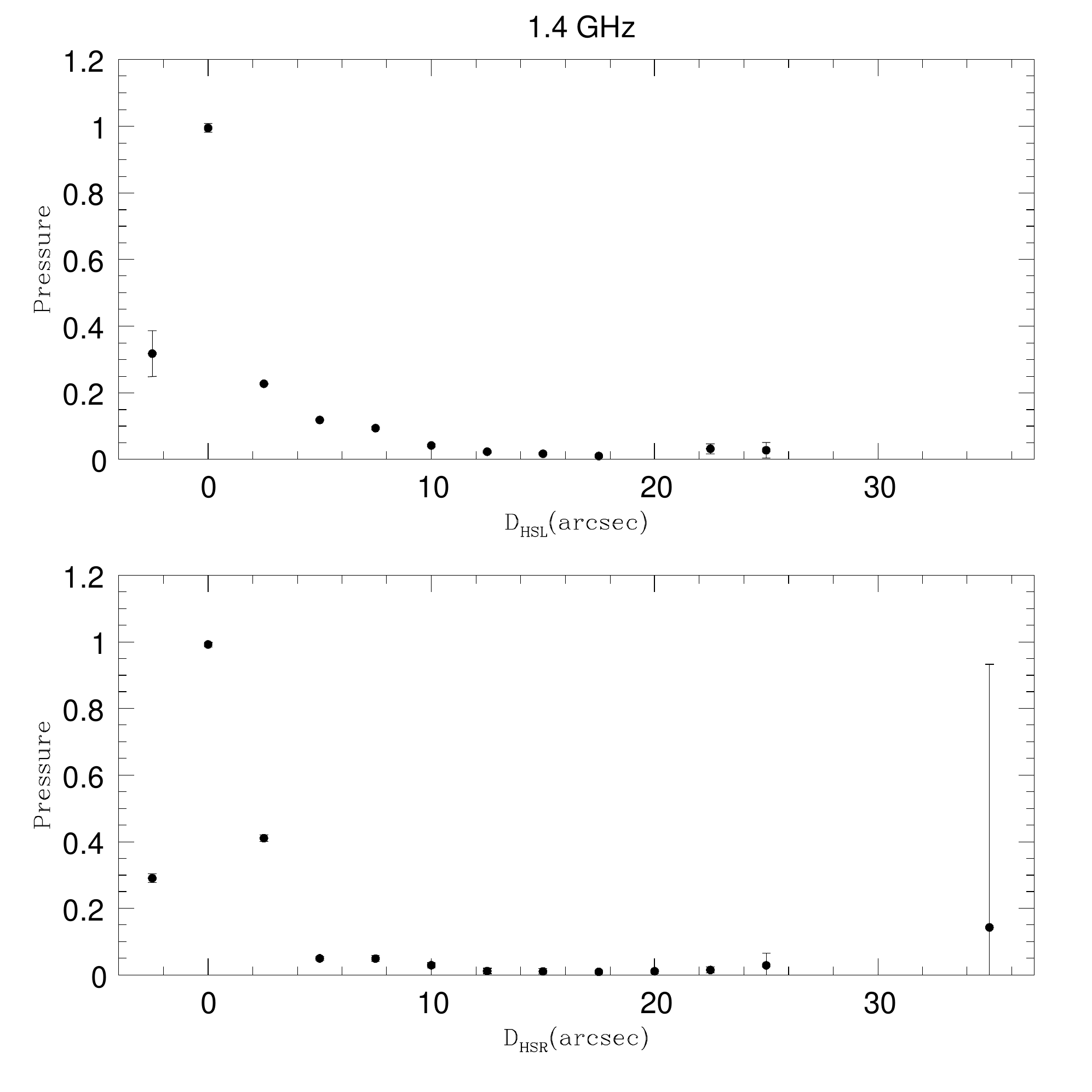}{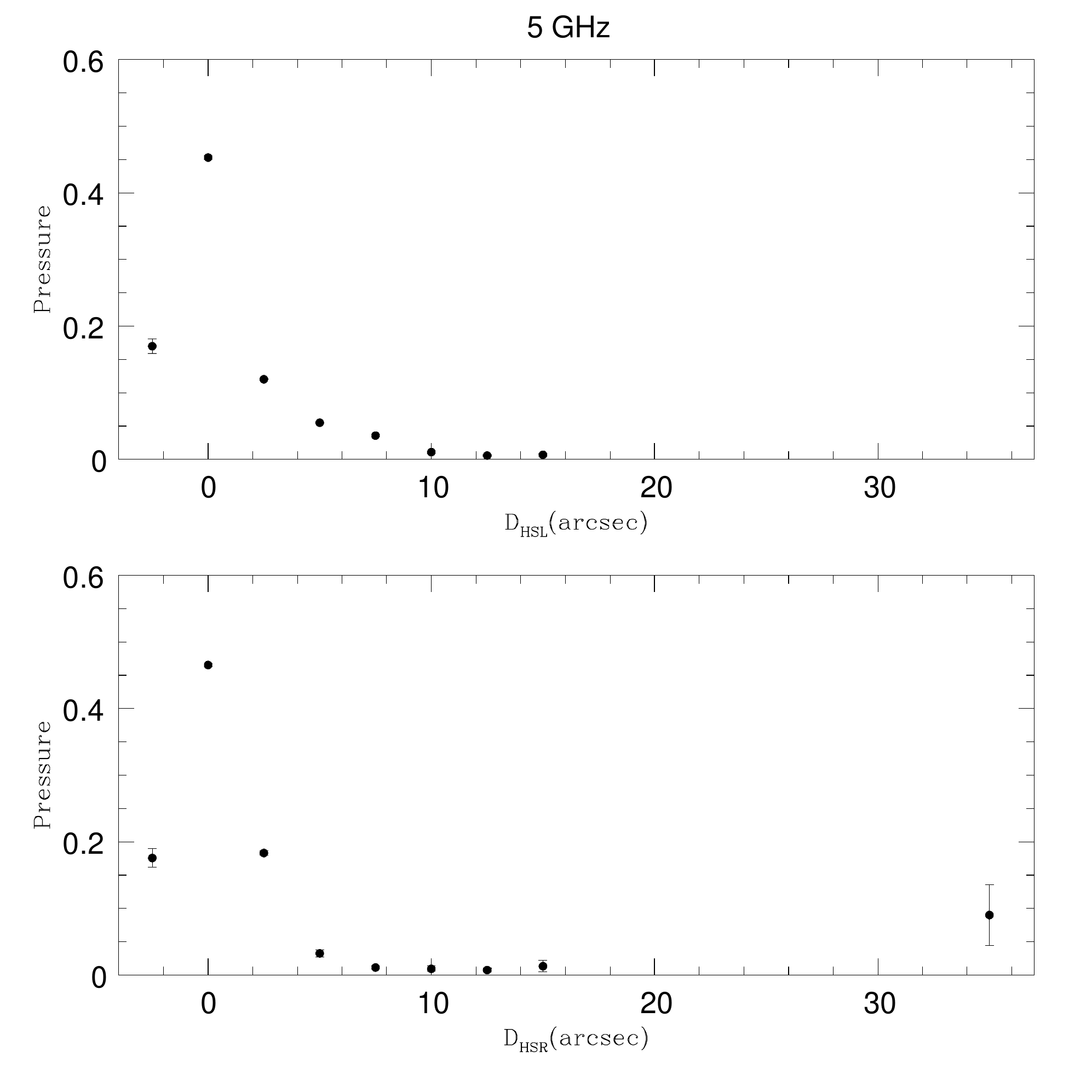}
\caption{3C469.1 
minimum energy pressure as a function of 
distance from the hot spot at 1.4 and 5 GHz 
(left and right panels, respectively) for the left and right
hand sides of the source (top and bottom panels, respectively), 
as in Fig. \ref{3C6.1P}. The normalization is given in Table 1.
}
\end{figure}


\begin{thebibliography}{11}
\expandafter\ifx\csname natexlab\endcsname\relax\def\natexlab#1{#1}\fi

\bibitem[{{Alexander}(1987)}]{Alexander87}
{Alexander}, P. 1987, \mnras, 225, 27

\bibitem[Alexander \& Leahy (1987)]{AL87}
Alexander, P., \& Leahy, J. P., 1987, \mnras, 225, 1

\bibitem[{{Best} {et~al.}(1997){Best}, {Longair}, \& {Rottgering}}]{BLR97}
{Best}, P. N., {Longair}, M. S., \& {Rottgering}, H. J. A. 1997,
\mnras, 286, 785

\bibitem[{{Blundell} {et~al.}(1999){Blundell}, {Rawlings}, \&
  {Willott}}]{Blundell99}
{Blundell}, K.~M., {Rawlings}, S., \& {Willott}, C.~J. 1999, \aj, 117, 677

\bibitem[{{Burbidge}(1956)}]{Burbidge56}
{Burbidge}, G.~R. 1956, \apj, 124, 416

\bibitem[{{Carilli} {et~al.}(1991){Carilli}, {Perley}, {Dreher}, \&
  {Leahy}}]{Carilli91}
{Carilli}, C.~L., {Perley}, R.~A., {Dreher}, J.~W., \& {Leahy}, J.~P. 1991,
  \apj, 383, 554

\bibitem[{{Carvalho} {et~al.}(2005){Carvalho}, {Daly}, {Mory}, \&
  {O'Dea}}]{Carvalho05}
{Carvalho}, J.~C., {Daly}, R.~A., {Mory}, M.~P., \& {O'Dea}, C.~P. 2005, \apj,
  620, 126

\bibitem[{Courant} \& {Friedrichs} (1948)]{CF48}
{Courant}, R., \& {Friedrichs}, K. O. 1948,
{\it Supersonic Flow and Shock Waves} (New York: Wiley Interscience)

\bibitem[{{Daly} {et~al.}(2008){Daly}, {Djorgovski},
{Freeman},{Mory},{O'Dea},{Kharb}, \& {Baum}}]{Daly08}
{Daly}, R. ~A., {Djorgovski}, S. ~G.,
{Freeman}, K.~A., {Mory}, M. ~P., {O'Dea}, C. ~P., {Kharb}, P.
\& {Baum}, S. A. 2008, ApJ, 677, 1

\bibitem[{{Daly} \& {Marscher} (1988)}]{DM88}
{Daly}, R. A., \& {Marscher}, A. P. 1988,
ApJ, 334, 539

\bibitem[{{Daly} {et~al.}(2009){Daly}, {Mory}, {O'Dea}, {Kharb},
{Baum}, {Guerra}, \& {Djorgovski}}]{Daly098}
{Daly}, R. ~A., {Mory}, M. ~P., {O'Dea}, C. ~P., {Kharb}, P.
{Baum}, S. A., {Guerra}, E. ~J., \& {Djorgovski}, S. ~G.,
2009, ApJ, 691, 1058


\bibitem[{{Eilek} {et~al.}(1997)}]{E97}
{Eilek}, J. A., {Melrose}, D. B., \& {Walker}, M. A. 1997,
ApJ, 483, 282

\bibitem[{{Fanaroff} \& {Riley}(1974)}]{FanaroffRiley74}
{Fanaroff}, B.~L., \& {Riley}, J.~M. 1974, \mnras, 167, 31P

\bibitem[{{Goodlet} {et~al.}(2004)
{Goodlet}, {Kaiser}, {Best}, \& {Dennett-Thorpe}}]{GKBD04}
{Goodlet}, J. A., {Kaiser}, C. R., {Best}, P. N., \& {Dennett-Thorpe}, J.
2004, \mnras, 347, 508

\bibitem[Guerra et al. (2000)]{GDW00}
Guerra, E. J., Daly, R. A. , \& Wan, L. 2000, ApJ, 544, 659

\bibitem[Jamrozy et al. (2008)]{J08} Jamrozy, M., Konar, C., Machalski,
J., \& Saikia, D. J. 2008, MNRAS, 385, 1286

\bibitem[Kaiser (2000)]{K00} Kaiser, C. R., 2000, A\&A, 362, 447

\bibitem[{{Kharb} {et~al.}(2008){Kharb}, {O'Dea}, {Baum}, {Daly}, {Mory},
  {Donahue}, \& {Guerra}}]{Kharb08}
{Kharb}, P., {O'Dea}, C., {Baum}, S., {Daly}, R., {Mory}, M., {Donahue}, M., \&
  {Guerra}, E. 2008, ApJS, 174, 74

\bibitem[{Konar} {et~al.} (2009)]{K09}
Konar, C., Hardcastle, M. J., Croston, J. H., \& Saikia, D. J. 2009,
MNRAS, 400, 480

\bibitem[{K\"onigl}(1980)]{K80}
K\"onigl, A. 1980, {\it Phys. Fluids}, 23, 1083

\bibitem[{Leahy}, {Muxlow}, \& Stephens (1989)]{LMS89}
{Leahy}, J.~P., {Muxlow}, T.~W.~B., \& {Stephens}, P.~W. 1989, \mnras, 239, 401

\bibitem[{{Leahy} \& {Williams}(1984)}]{LW84}
{Leahy}, J.~P., \& {Williams}, A.~G. 1984, \mnras, 210, 929

\bibitem[{{Liu} {et~al.}(1992){Liu}, {Pooley}, \& {Riley}}]
{LPR92}
{Liu}, R., {Pooley}, G., \& {Riley}, J. M. 1992, \mnras, 257, 545

\bibitem[Mack et al. (1998)]{M98} Mack, K.-H., Klein, U., O'Dea, C. P.,
Willis, A. G., \& Saripalli, L., 1998, A\&A, 329, 431

\bibitem[Machalski et al. (2007)]{M07} Machalski, J., Chyzy, K. T.,
Stawarz, L., \& Koziel, D., 2007, A\&A, 462, 43

\bibitem[Miley (1980)]{Miley1980} Miley, G. K. 1980,
ARA\&A, 18 165



\bibitem[Murgia et al. (1999)]{M99} Murgia, M., Fanti, C., Fanti, R.,
Gregorini, L.,
Klein, U., Mack, K.-H., Vigotti, M., 1999, \aap, 345, 769

\bibitem[Myers \& Spangler (1985)]{MS85} Myers, S. T., \& Spangler,
S. R.,
1985, \apj, 291, 52

\bibitem[{{Napier} {et~al.}(1983){Napier}, {Thompson}, \& {Ekers}}]{Napier83}
{Napier}, P.~J., {Thompson}, A.~R., \& {Ekers}, R.~D. 1983, IEEE Proceedings,
  71, 1295


\bibitem[{{O'Dea} {et~al.}(2009) {O'Dea}, {Daly}, {Freeman}, {Kharb},
\& {Baum}}]{O'Dea08}
{O'Dea}, C., {Daly}, R., {Freeman}, K. A., {Kharb}, P.,
\& {Baum}, S. 2009, A\&A, 494, 471

\bibitem[{Owczarek} (1964)]{O64}
{Owczarek}, J. A. 1964, {\it Fundamentals of Gas Dynamics}
(Scranton: International Textbook Co.)

\bibitem[Parma et al. (1999)]{P99} Parma, P., Murgia, M., Morganti, R.,
Capetti, A., de Ruiter, H.
R., \& Fanti, R., 1999, A\&A, 344, 7

\bibitem[{{Scheuer}(1982)}]{Scheuer82}
{Scheuer}, P.~A.~G. 1982, in IAU Symp. 97: Extragalactic Radio Sources, ed.
  D.~S. {Heeschen} \& C.~M. {Wade}, 163--165

\bibitem[{{Tregillis} {et~al.}(2004)
{Tregillis}, {Jones}, \& {Ryu}}]{TJR04}
{Tregillis}, I. L., {Jones}, T. W., \& {Ryu}, D. 2004, ApJ, 601, 778

\end{thebibliography}

\end{document}